\documentclass{article}

\usepackage[left=3cm, right=3cm, bottom=4cm]{geometry}

\usepackage{amsmath, amsthm, amssymb,amsbsy}
\usepackage{algcompatible}
\usepackage{algorithm}
\usepackage{algorithmicx}
\usepackage{algpseudocode}
\usepackage{graphicx}
\usepackage[colorlinks,citecolor=blue]{hyperref}
\usepackage{tikz}
\usetikzlibrary{calc,positioning, math, tikzmark}
\usepackage{tikzpagenodes}
\usepackage{everypage}

\usepackage{pgfplots}
\usepackage{bbm}
\usepackage{float}
\usepackage[shortlabels]{enumitem}
\usepackage{comment}
\usepackage{caption}
\usepackage{setspace}
\usepackage{tcolorbox}
\usepackage{booktabs}
\usepackage{longtable}

\usepackage{glossaries}

\newif\ifmaybe
\maybefalse

\newif\ifremove
\removefalse

\newif\ifreplace
\replacetrue

\newcommand{\editstart}{}
\newcommand{\editfinish}{}

\usepackage[caption=false,font=normalsize,labelfon
t=sf,textfont=sf]{subfig}

\algnewcommand\algorithmicinput{\textbf{Input:}}
\algnewcommand\INPUT{\item[\algorithmicinput]}
\algnewcommand\algorithmicoutput{\textbf{Output:}}
\algnewcommand\OUTPUT{\item[\algorithmicoutput]}

\algrenewcommand\algorithmicrequire{\textbf{Input:}}
\algrenewcommand\algorithmicensure{\textbf{Output:}}



\makeatletter
   \def\vhrulefill#1{\leavevmode\leaders\hrule\@height#1\hfill \kern\z@}
\makeatother
   
\newcounter{alg}
\newenvironment{myalg}[1][]%
{
  \refstepcounter{alg}
  \noindent \rule{\textwidth}{1pt}
  \textbf{Algorithm \thealg.} #1 \\[-0.3em]
  \noindent  \rule{\textwidth}{0.5pt} %
}%
{%
  \rule{\textwidth}{0.5pt} %
}

\usepackage{framed}
\usepackage{tcolorbox}
\tcbuselibrary{breakable,skins}

\newtcolorbox{mybox}{                       
   enhanced,
   colframe=blue!50,
   boxrule=1pt,
   width=1.05\textwidth,
   arc=1mm,
   breakable,                               
}

\title{A Universal Low Complexity Compression Algorithm for Sparse Marked Graphs\footnote{This paper was presented in part at 2020 IEEE International Symposium on Information Theory.}}

\author{Payam Delgosha\thanks{Department of Computer Science, University of
    Illinois Urbana-Champaign, \texttt{delgosha@illinois.edu}}
   \, and   Venkat Anantharam\thanks{Department of Electrical Engineering and Computer
    Sciences, University of California, Berkeley, \texttt{ananth@berkeley.edu}}}

\newcommand{\evwrt}[2]{\mathbb{E}_{#1} \left [ #2 \right ] }
\newcommand{\pr}[1]{\mathbb{P} \left ( #1 \right ) }
\newcommand{\prwrt}[2]{\mathbb{P}_{#1} \left ( #2 \right ) }

\newcommand{\snorm}[1]{\Vert #1 \Vert}
\newcommand{\one}[1]{\mathbbm{1} \left [ #1 \right ]}

\newcommand{\concat}{+}

\newtheorem{lem}{Lemma}
\newtheorem*{lem*}{Lemma}
\newtheorem{thm}{Theorem}
\newtheorem{definition}{Definition}
\newtheorem{prop}{Proposition}

\newtheorem{rem}{Remark}
\newtheorem{cor}{Corollary}

\newcommand{\mG}{\mathcal{G}}

\newcommand{\vnu}{\vec{\nu}}
\newcommand{\vm}{\vec{m}}
\newcommand{\vu}{\vec{u}}
\newcommand{\vd}{\vec{d}}
\newcommand{\va}{\vec{a}}
\newcommand{\vA}{\vec{A}}
\newcommand{\vb}{\vec{b}}
\newcommand{\vx}{\vec{x}}
\newcommand{\vv}{\vec{v}}
\newcommand{\vy}{\vec{y}}
\newcommand{\vs}{\vec{s}}

\newcommand{\tM}{\widetilde{M}}
\newcommand{\vmn}{\vec{m}^{(n)}}
\newcommand{\vun}{\vec{u}^{(n)}}
\newcommand{\vdn}{\vec{d}^{(n)}}
\newcommand{\mn}{m^{(n)}}

\newcommand{\un}{u^{(n)}}
\newcommand{\dn}{d^{(n)}}

\newcommand{\mA}{\mathcal{A}}
\newcommand{\mAn}{\mathcal{A}^{(n)}}
\newcommand{\mB}{\mathcal{B}}
\newcommand{\mBn}{\mathcal{B}^{(n)}}

\newcommand{\mP}{\mathcal{P}}

\newcommand{\mS}{\mathcal{S}}
\newcommand{\tmS}{\widetilde{\mathcal{S}}}
\newcommand{\mF}{\mathcal{F}}

\newcommand{\mFdeltah}{\mathcal{F}^{(\delta,h)}}
\newcommand{\mFdeltak}{\mathcal{F}^{(\delta,k)}}
\newcommand{\mFdeltakp}{\mathcal{F}^{(\delta,k+1)}}

\newcommand{\mFbardeltah}{\bar{\mathcal{F}}^{(\delta,h)}}

\newcommand{\mT}{\mathcal{T}}
\newcommand{\mTn}{\mathcal{T}^{(n)}}

\newcommand{\mC}{\mathcal{C}}

\newcommand{\mEn}{\mathcal{E}^{(n)}}

\newcommand{\mVn}{\mathcal{V}^{(n)}}

\newcommand{\mCnt}{\mathcal{C}^{(\pn)}}
\newcommand{\mCnlr}{\mathcal{C}^{(n_l, n_r)}}

\newcommand{\mCdeltah}{\mathcal{C}^{(\delta,h)}}

\newcommand{\mCbardeltah}{\bar{\mathcal{C}}^{(\delta,h)}}

\newcommand{\mGb}{\bar{\mathcal{G}}}
\newcommand{\mTb}{\bar{\mathcal{T}}}
\newcommand{\mGn}{\mathcal{G}^{(n)}}
\newcommand{\mGnt}{\mathcal{G}^{(\pn)}}
\newcommand{\mGnlr}{\mathcal{G}^{(n_l, n_r)}}

\newcommand{\sfb}{\mathsf{b}}
\newcommand{\sfa}{\mathsf{a}}
\newcommand{\vsfb}{\vec{\mathsf{b}}}
\newcommand{\vsfa}{\vec{\mathsf{a}}}
\newcommand{\sfU}{\mathsf{U}}
\newcommand{\sfW}{\mathsf{W}}
\newcommand{\sfT}{\mathsf{T}}

\newcommand{\ts}{\tilde{s}}
\newcommand{\tu}{\tilde{u}}

\newcommand{\mGnmnun}{\mathcal{G}^{(n)}_{\vmn, \vun}}

\newcommand{\Gn}{G^{(n)}}
\newcommand{\In}{I^{(n)}}
\newcommand{\hGn}{\widehat{G}^{(n)}}

\newcommand{\gn}{g^{(n)}}
\newcommand{\gnt}{g^{(\pn)}}
\newcommand{\gnlr}{g^{(n_l, n_r)}}

\newcommand{\fn}{f^{(n)}}
\newcommand{\fnt}{f^{(\pn)}}
\newcommand{\fnlr}{f^{(n_l, n_r)}}

\newcommand{\edelta}{\mathsf{E}_{\Delta}}
\newcommand{\len}{\mathsf{bits}}
\newcommand{\nat}{\mathsf{nats}}
\newcommand{\Vns}{\mathcal{V}^{(n)}_\star}
\newcommand{\Dictionary}{\mathsf{Dictionary}}
\newcommand{\Array}{\mathsf{Array}}
\newcommand{\Deg}{\mathsf{Deg}}

\newcommand{\Sn}{S^{(n)}}

\newcommand{\tN}{\widetilde{N}}
\newcommand{\tz}{\widetilde{z}}

\newcommand{\reals}{\mathbb{R}}
\newcommand{\integers}{\mathbb{Z}}
\newcommand{\nats}{\mathbb{N}}

\newcommand{\ER}{Erd\H{o}s--R\'{e}nyi }
\newcommand{\LP}{L\'{e}vy--Prokhorov }

\newcommand{\dlp}{d_\text{LP}} 
\newcommand{\bch}{ \Sigma} 
\newcommand{\bchover}{\overline{\Sigma}}
\newcommand{\bchunder}{\underbar{$\Sigma$}}
\newcommand{\condmnun}{|_{(\vmn, \vun)}} 

\newcommand{\Mul}{\textproc{Mul}} 
\newcommand{\Div}{\textproc{Div}} 

\let\oldmarginpar\marginpar
\renewcommand{\marginpar}[2][rectangle,draw,rounded corners,text width = 3cm, scale=0.7]{%
  \oldmarginpar{%
    \tikz \node at (0,0) [#1]{#2};}%
}

\newcommand{\edgemark}{\Xi} 
\newcommand{\vermark}{\Theta} 

\newcommand{\vtype}{\Pi} 
\newcommand{\vvtype}{\vec{\Pi}} 
\newcommand{\vdeg}{\vec{\deg}} 

\newcommand{\UC}{\mathsf{UC}} 
\newcommand{\vDn}{\vec{D}^{(n)}} 
\newcommand{\Dn}{D^{(n)}} 
\newcommand{\dnn}{d^{(n)}} 
\newcommand{\tD}{\widetilde{D}}

\newcommand{\vgamma}{\vec{\gamma}}

\newcommand{\gamman}{\gamma^{(n)}}
\newcommand{\vgamman}{\vec{\gamma}^{(n)}}
\newcommand{\tgamman}{\tilde{\gamma}^{(n)}}
\newcommand{\vtgamman}{\vec{\tilde{\gamma}}^{(n)}}
\newcommand{\xn}{x^{(n)}}
\newcommand{\xnp}{{x'}^{(n)}}

\newcommand{\tP}{\widetilde{P}}

\newcommand{\pn}{\tilde{n}} 

\newcommand{\pp}[1]{#1'}
\newcommand{\ppp}[1]{#1''}
\newcommand{\tp}{\pp{t}} 
\newcommand{\ip}{\pp{i}} 

\newcommand{\thetan}{\theta^{(n)}}
\newcommand{\vthetan}{\vec{\theta}^{(n)}}
\newcommand{\eln}{\ell^{(n)}}

\newcommand{\Nn}{\mathcal{N}^{(n)}}
\newcommand{\tNn}{\widetilde{\mathcal{N}}^{(n)}}
\newcommand{\T}[2]{t^{(n)}_h(#1, #2)}
\newcommand{\TT}[2]{\tilde{t}^{(n)}_{h,\delta}(#1, #2)}
\newcommand{\type}[2]{\psi^{(n)}_{h,\delta}(#1, #2)}

\allowdisplaybreaks 

\definecolor{bluenodecolor}{RGB}{62,126,176}
\definecolor{rednodecolor}{RGB}{173,61,58}

\definecolor{blueedgecolor}{RGB}{3,151,255}
\definecolor{orangeedgecolor}{RGB}{255,149,8}

\tikzstyle{nodeB} = [fill=bluenodecolor, circle, inner sep = 1.7pt]
\tikzstyle{nodeR} = [fill=rednodecolor, rectangle, inner sep = 2.3pt]
\tikzstyle{edgeB} = [very thick, blueedgecolor]
\tikzstyle{edgeO} = [very thick, orangeedgecolor, decoration = {zigzag,segment length = 0.2cm, amplitude = 0.5mm},decorate]

\newcommand{\drawedge}[4]{\draw[edge#3] (#1) -- ($(#1)!0.5!(#2)$); \draw[edge#4]
  (#2) -- ($(#2)!0.5!(#1)$);}
\newcommand{\nodelabel}[3]{\node at ($(#1)+(#2:5mm)$) {#3};}
\newcommand{\nodelabeldist}[4]{\node at ($(#1)+(#2:#3)$) {#4};}
\newcommand{\nodelabeldistalignleft}[4]{\node[anchor=west] at ($(#1)+(#2:#3)$) {#4};}
\newcommand{\nodelabeldistalignright}[4]{\node[anchor=east] at ($(#1)+(#2:#3)$)
  {#4};}

\newcommand{\tone}{\!\tikz[baseline=(one.base)]{\node[bluenodecolor] (one) at (0,0) {1}; \node[nodeB,scale=0.9] at (0,-0.25) {};}\!}

\newcommand{\ttwo}{\!\tikz[baseline=(two.base)]{\node[rednodecolor] (two) at (0,0) {2}; \node[nodeR,scale=0.9] at (0,-0.25) {};}\!}

\newcommand{\xone}{\!\tikz[baseline=(one.base)]{\node[blueedgecolor] (one) at
    (0,0) {1}; \draw[edgeB] (-0.12,-0.25) -- (0.12,-0.25);}\!}

\newcommand{\xtwo}{\!\tikz[baseline=(two.base)]{\node[orangeedgecolor] (two) at
    (0,0) {2}; \draw[edgeO] (-0.12,-0.25) -- (0.12,-0.25);}\!}

\newcommand{\stack}[2]{\!\tikz[baseline=(top.base)]{
    \node (top) at (0,0) {#1};
    \node[rotate=-90,scale=0.5,anchor=west] at (0,-0.3) {#2};
  }\!}

\newcommand{\tthree}[2][0]{\scalebox{#2}{\tikz[baseline=(p.base)]{
      \node (p) at (0,#1-0.5) {};
      \node[nodeB] (r) at (0,0) {};
      \node[nodeR] at (-0.3,-0.5) (r1) {};
      \node[nodeR] at (0.3,-0.5) (r2) {};
      \drawedge{r}{r1}{B}{B};
      \drawedge{r}{r2}{B}{O};
      \draw[edgeB] (r) -- ($(r) + (0,0.3)$);
    }}}

\newcommand{\tfour}[2][0]{\scalebox{#2}{\tikz[baseline=(p.base)]{
      \node (p) at (0,#1-0.5) {};
      \node[nodeR] (r) at (0,0) {};
      \node[nodeR] at (0,-0.5) (r1) {};
      \draw[edgeO] (r) -- (r1);
      \draw[edgeB] (r) -- ($(r) + (0,0.3)$);
    }}}

\newcommand{\tfive}[2][0]{\scalebox{#2}{\tikz[baseline=(p.base)]{
      \node (p) at (0,#1-0.5) {};
      \node[nodeR] (r) at (0,0) {};
      \node[nodeB] at (0,-0.5) (r1) {};
      \draw[edgeB] (r) -- (r1);
      \draw[edgeO] (r) -- ($(r) + (0,0.3)$);
    }}}

\newcommand{\tikzint}[1]{\pgfmathparse{#1}\pgfmathprintnumber{\pgfmathresult}}

\makeatletter
\newcommand{\checkmarkpage}[4]
{\@ifundefined{save@pt@#1}{#2}{%
  \edef\markid{\csname save@pt@#1\endcsname}%
  \edef\markpage{\csname save@pg@\markid\endcsname}%
  \ifnum\thepage<\markpage\relax #2%
  \else%
    \ifnum\thepage=\markpage\relax #3%
    \else #4%
    \fi%
  \fi}%
}
\makeatother

\newcounter{outlineid}
\newcounter{outlinedone}

  {\par\tikzmark{end\theoutlineid}\stepcounter{outlineid}\ignorespacesafterend}%

\newcommand{\drawoutline}{\checkmarkpage{begin\theoutlinedone}{}%
  {\begin{tikzpicture}[remember picture,overlay]
    \path ({pic cs:begin\theoutlinedone}-| current page text area.west)
      ++(0pt,\ht\strutbox) coordinate(A);
    \checkmarkpage{end\theoutlinedone}%
      {\path (current page text area.south west) ++(0pt,-\dp\strutbox)
         coordinate(B);}%
      {\path ({pic cs:end\theoutlinedone}-| current page text area.west)
        ++(0pt,\ht\strutbox) coordinate(B);}%
      {}
      \fill[yellow] ($(A) + (-.333em,0pt)$) rectangle ($(B) + (-1cm,0pt)$);
   \end{tikzpicture}}%
  {\begin{tikzpicture}[remember picture,overlay]
    \coordinate (A) at (current page text area.north west);
    \checkmarkpage{end\theoutlinedone}%
      {\path (current page text area.south west) ++(0pt,-\dp\strutbox)
         coordinate(B);}%
      {\path ({pic cs:end\theoutlinedone}-| current page text area.west)
        ++(0pt,\ht\strutbox) coordinate(B);}%
      {}
      \fill[yellow] ($(A) + (-.333em,0pt)$) rectangle ($(B) + (-1cm,0pt)$);
   \end{tikzpicture}}%
  \checkmarkpage{end\theoutlinedone}{}%
    {\stepcounter{outlinedone}\drawoutline}%
    {}
 }
\AddEverypageHook{\drawoutline}

\newif\ifdraft\draftfalse

\definecolor{peditcolor}{rgb}{0,0,0}
\definecolor{pcommentcolor}{RGB}{128, 128, 0}

\begin{document}

\colorlet{Cyan}{cyan}
\colorlet{Orange}{orange}
\tikzstyle{Node} = [circle,fill,inner sep=1.5pt]
\tikzstyle{Node2} = [rectangle,fill,inner sep=2pt]
\tikzstyle{Root} = [circle,fill=magenta,inner sep=1.7pt]
\tikzstyle{Root2} = [rectangle,fill=magenta,inner sep=2.5pt]
\tikzstyle{Cedge} = [Cyan, thick]
\tikzstyle{Oedge} = [Orange, densely dotted, very thick]

\maketitle



\begin{abstract}
  Many modern applications involve accessing 
  and processing graphical
  data, i.e. data that is naturally indexed by graphs. Examples come from
  internet graphs, social networks, genomics and
  proteomics, and other sources. The typically large size of such data motivates seeking efficient ways for its compression and decompression.
  The current compression  methods are usually tailored to specific models, or do not provide
  theoretical guarantees. 
  In  this paper, we introduce a low--complexity lossless compression algorithm for sparse marked graphs, i.e. graphical data indexed by sparse graphs, which is
  capable of universally achieving the optimal compression rate in a precisely defined sense. In order to
  define  universality, we employ the framework of local weak convergence,
  which allows one to make sense of a notion of stochastic processes for sparse graphs.
  Moreover, we investigate the performance of
  our algorithm through some experimental results on both synthetic and
  real--world data. 
\end{abstract}

\section{Introduction}
\label{sec:introduction}

\editstart

Nowadays, a large amount of data 
arises in a form that is indexed by
combinatorial
structures, such as graphs, rather than classical time series.  Examples include internet graphs,  social
networks, and  biological data. Such data typically
needs to be processed in practice,
for instance to infer   communities in a social network,
or to
predict whether two proteins interact in a biological network. Typically, the
size of such  graphical data is large, which argues for the need to find
efficient and close to optimal ways of compressing 
them so that they can be 
stored for subsequent data mining tasks.

The problem of graphical data compression has drawn a lot of attention in different
fields.
Some of the  existing algorithms 
 rely on  properties specific to certain types of  graphs, such as web graphs or
 social network,  which are extracted through
analyzing real--world samples of such data
\cite{boldi2004webgraph,boldi2011layered,liakos2014pushing}. Such methods usually
do not come with theoretical guarantees of optimality. On the other hand,  there 
has been some progress in studying the information content of some models of
random graphs from an information theoretic perspective 
\cite{lempel1986compression,
  choi2012compression,aldous2014entropy,abbe2016graph,asadi2017compressing,
  basu2017universal,luczak2019asymmetry,luczak2019compression, turowski2020compression}.

\subsection{Prior work}
\label{sec:prior-work}

There have been two main approaches towards the problem of graphical data
compression.
One approach is to  assume that the graphical data is generated from a
certain statistical model and the encoder aims to achieve the entropy of this
input distribution. For instance, 
Choi and Szpankowski studied the
structural entropy of the \ER model,
i.e.\ the entropy associated to the isomorphism 
  class of such graphs \cite{choi2012compression}. Moreover, they proposed a compression scheme which
  asymptotically achieves the structural entropy.
 Aldous and Ross studied the asymptotics of the
  entropy of several models of random graphs, including the sparse \ER ensemble
  \cite{aldous2014entropy}.
Abbe studied the asymptotic behavior of the entropy
  of stochastic block models, and discussed the optimal compression rate for
  such models up to the first order term \cite{abbe2016graph}. They also considered the case where
  vertices in a stochastic block model can carry data which is conditionally
  independent given their community membership. 
  {\L}uczak  et al.\ studied the asymptotics of the entropy associated to the
  preferential attachment model, both for labeled and unlabeled regimes, and
  used this to  design  optimal compression schemes
  \cite{luczak2019asymmetry}.
  Turowski et al.\ studied the information content of the duplication model
    \cite{turowski2020compression}. They analyzed the asymptotic
    behavior of the entropy of such models for both the labeled and unlabeled
    regimes, and designed compression methods to achieve such entropy bounds.


A second line of research offers an alternative approach to compress specific
types of graphical data,
such as web graphs \cite{bharat1998connectivity,silverstein1999analysis,broder2000graph,boldi2004webgraph}, social
networks 
\cite{chierichetti2009compressing, maserrat2010neighbor,boldi2011layered,maserrat2012compression}, or biological
networks \cite{dinkla2012compressed,ay2012metabolic,kamal2014memory,sheikhizadeh2016pantools,hosseini2016survey}. 
These works usually take advantage of some properties specific to a specific data
source, where such properties are usually inferred through observing real--world
data samples.
Consequently, results in this category of work usually do not come with
information theoretic guarantee of optimality.
For
instance, the web graph framework of \cite{boldi2004webgraph} employ the
\emph{locality} and \emph{similarity} properties existing in web
graphs to design an efficient compression algorithm tailored for such data. The
\emph{locality} property refers to the fact that a web page usually refers to
other web pages whose URLs have a long prefix in common, while the
\emph{similarity} property refers to the fact that if we sort web pages based on
the lexicographic order of their URLs, web pages that are close to each other
tend to have many successors in common. Therefore, techniques such as
\emph{reference encoding} and \emph{gap encoding} are useful in compressing web
graphs.  
 The reader is referred
to~\cite{besta2018survey} for a survey on graph compression methods.



\subsection{Our Contribution}
\label{sec:our-contrib}

The key property distinguishing our approach from the existing ones is
\emph{universality}. More precisely,  we introduce a scheme which is capable of
compressing graphs which come from a certain ``stochastic process'' without any
prior knowledge of this process, yet is able to achieve the optimal compression rate,
in a precisely defined sense. Additionally, in contrast to several earlier works, we assume that the graphs
are ``marked'' so that vertices and edges can carry additional information
on top of the connectivity structure of the graph. This is the key feature 
which makes our approach applicable in modeling real--world data represented  by graphs,
which we call {\em graphical data}. Our focus in this paper is  on \emph{sparse}
marked graphs, in the sense that the number of edges in the graph scales
linearly with the number of vertices.  The
motivation for this is that usually real--world graphs are less or more  sparse.

To make sense of the notion of a ``stochastic process''  for sparse
marked graphs, we
employ the framework of local weak convergence
\cite{BenjaminiSchramm01rec,aldous2004objective,aldous2007processes}. Moreover,
we employ a notion of entropy
called the \emph{marked BC
entropy} \cite{delgosha2019notion}, which is an extension of
the notion of entropy introduced by Bordenave and Caputo in \cite{bordenave2015large} and  serves
as a  counterpart of the Shannon entropy rate for this framework. This notion of
entropy is proved to give  the optimal
compression rate for sparse  graphical data \cite{delgosha2020universal}. 
The marked BC entropy is defined by analyzing the asymptotic behavior of
the size of \emph{typical} graphs. It can be seen that the logarithm of the size
of the 
set of
\color{black}
typical 
marked
\color{black}
graphs has a leading term of order $n \log n$, $n$ being the
number of vertices in the graph, and the entropy shows up at the second order
term, which scales linearly with the number of vertices. In other words, the
marked BC entropy captures the per--vertex growth rate of the 
logarithm of the 
\color{black}
size of 
the set of
\color{black}
typical
marked
\color{black}
graphs, after carefully separating out the leading term. 
The authors have already
introduced a universal compression scheme in \cite{delgosha2020universal} which shows that this
notion of entropy is indeed the optimal information theoretic threshold of
compression. 
{
In~\cite{delgosha2020universal}, the encoder needs to find the index of the input
graph among all graphs which have
{\color{black}the same
frequency of local structures as}
the input graph. 
Although this method is proved to universally achieve  the marked BC entropy in an
asymptotic sense, it is computationally intractable.
The focus of this paper is to provide a
compression algorithm which asymptotically achieves the optimal compression rate
and also is  computationally efficient.
}

There have been some attempts in the literature to address universality in the
context of graphical data compression, for instance to compress deep neural
networks \cite{basu2017universal}, \cite{basu2018universal}, or stochastic block
models \cite{bhatt2020universal}. However, such attempts usually address universality in
a limited fashion, i.e.\ the graph is generated from a class of  ensembles with
unknown parameters. Moreover, such attempts usually
consider the ratio of the codeword length to the overall ensemble entropy, as
opposed to our framework which considers the per--vertex entropy by carefully
separating out the leading term.


\subsection{Notational Conventions}
\label{sec:intro-notation}


$\nats$ denotes the set of positive integers. For a positive integer $n$, $[n]$ denotes the set of integers $\{1, 2, \dots,
n\}$. For integers $i$ and $j$, $[i:j]$ denotes the set $\{i,\dots, j\}$ if $i
\leq j$, and the empty set otherwise. Throughout the paper, $\log$ refers to the logarithm in the natural basis, while
$\log_2$ refers to logarithm in base 2. We write 
$\{0,1\}^{*+}$, $\{0,1\}^*\backslash \emptyset$, or
\color{black}
$\{0,1\}^* - \emptyset$  for
the set of nonempty binary sequences with finite length. For such a sequence $x
\in \{0,1\}^* - \emptyset$, $\len(x)$ refers to the length of $x$ in bits, while
$\nat(x)$ denotes the length of $x$ in nats, i.e.\ $\nat(x) =
\len(x) \times \log 2$. 
For two binary sequences $x, y \in \{0,1\}^* - \emptyset$,
$x \concat y$ denotes the concatenation of $x$ and $y$, e.g.\ $(0,1) \concat
(1,0,0) = (0,1,1,0,0)$.
Occasionally we write $xy$ for the concatenation of $x$ and $y$.
\color{black}

We
write $:= $ and $=: $ for equality by definition.
 For two sequence $(a_n: n \geq
1)$ and $(b_n: n \geq 1)$ of nonnegative real numbers, we write $a_n = O(b_n)$
if there exists a constant $C > 0$ such that $a_n \leq C b_n$ for $n$ large
enough.
Moreover, we write $a_n = o(b_n)$ if $a_n / b_n \rightarrow 0$ as $n
\rightarrow \infty$. Furthermore, we write $a_n = \Omega(b_n)$ if there exists a
constant $C > 0$ such that $a_n \geq C b_n$ for $n$ large enough. Also, we write $a_n = \Theta(b_n)$
if there are constants $C_1, C_2 > 0$ such that $C_1 b_n \leq a_n \leq C_2 b_n$
for $n$ large enough.

Given a totally ordered set $\mA$, we define the lexicographic order on finite
sequences from $\mA$ as follows. Assume that  two finite sequences $\va = (a_1, \dots, a_k)$ and $\vb = (b_1, \dots, b_l)$ 
with symbols in $\mA$ are given.
If $\va$ is a prefix of $\vb$, i.e. if $l > k$ and
$a_i = b_i$ for $1 \leq i \leq k$, we say that $\va$ is lexicographically
smaller than $\vb$ and write  $\va \prec \vb$. Similarly, if $\vb$
is a prefix of $\va$, we have $\vb \prec \va$. If 
neither of these cases holds,
\color{black}
and $\va \neq \vb$, there exists an index $1 \leq i \leq \min\{k,l\}$ such that
$a_i \neq b_i$. Let $i$ be the minimum such index. Then, if $a_i < b_i$, we let
$\va \prec \vb$, while if $b_i < a_i$, we let $\vb \prec \va$. For instance, if
$\mA = \nats$ is the set of positive integers, we have $(1,1,3) \prec (1,2,0)$
and $(1,1,1) \prec (2,0)$.

We denote the Elias delta code by $\edelta: \{1, 2, \dots \}
\rightarrow \{0,1\}^* - \emptyset$.
This is a prefix--free code for positive
integers \cite{elias1975universal}. 
{
  Given an integer $N \geq 1$, in order to encode $N$ using the Elias delta
  code, we first find $m:= \lfloor \log_2 N \rfloor$. Note that the binary
  representation of $N$ has $m+1$ bits. Then, we write $r:= \lfloor
  \log_2 (m+1) \rfloor$ zeros, followed by the  binary representation of
  $m+1$ using $r+1$ bits. Then, we write all but the leading bit in the binary representation of
  $N$, i.e.\ the last $m$ bits in the binary representation of $N$. Note that  since $N \geq 1$,
the leading bit  in the binary representation of $N$ is always one.}
The length of the encoded version of an
integer $N$ using the Elias delta code is $\lfloor  \log_2 N  \rfloor + O(\log \log
N)$. Moreover, we denote the Elias delta decoder
by $\edelta^{-1}$, so that $\edelta^{-1}(\edelta(n)) = n$ for all $n \geq 1$. 


For nonnegative integers $r, s \geq 0$, we define the binomial coefficient
$\binom{r}{s} := \frac{r!}{s!(r-s)!}$ if $r \geq s$, and  $\binom{r}{s}
:= 0$ if $r < s$.
For nonnegative integers $r$ and $s$, we define the falling factorial $(r)_s$
to be $r(r-1)(r-2) \dots (r-(s-1))$. In other words, if $r \geq s$, we have
$(r)_s = r! / (r-s)!$, while if $r < s$, we have $(r)_s = 0$.
We use the symbol $\div$ for integer division, i.e. for
integers $r \geq
0$ and $s > 1$, we have $r \div s := \lfloor  r / s \rfloor$.
For an even integer $k > 0$, we define
  \begin{equation*}
    (k-1)!! := \frac{k!}{2^{k/2} (k/2)!},
  \end{equation*}
  which is the number of matchings on $k$ objects. Moreover, we define $(-1)!!
  := 1$.
\color{black}
For a
probability distribution $P$ defined on a finite set, $H(P)$ denotes the Shannon entropy
of $P$.
Other notation used in this document is defined on its
first appearance.

{
Let $\Mul(n,m)$ denote the time complexity of multiplying two positive integers
whose binary representations have $n$ and $m$ bits, respectively. Furthermore,
let $\Div(n,m)$ denote the time complexity of finding the quotient and remainder
after dividing two positive integers with $n$ and $m$ bits, respectively. It can
be shown that, using fast arithmetic algorithms, integer multiplication and
division can be done so that $\Mul(n,m) = O(r \log r \log \log r)$ and
$\Div(n,m) = O(r \log r \log \log r)$, where $r = \max
\{n,m\}$.\footnote{Although the asymptotic complexities are the same, the
  constant for division is larger} The reader is referred to
\cite{brent2010modern} and \cite{crandall2006prime}
for more details on such fast arithmetic algorithms. With an abuse of notation, we
use $\Mul(n)$ and $\Div(n)$ as an upper bound for the time complexity of
multiplying and dividing two positive integers with at  most $n$ bits,
respectively, so  that $\Mul(n) = O(n \log n \log \log n)$  and  $\Div(n)  = O(n
\log n  \log \log n)$.
} 

We illustrate  algorithms in this document  using standard pseudocode notation. The
expression $a \gets b$ in a pseudocode means we evaluate $b$ and store the
result in variable $a$. A \textbf{function} in a pseudocode denotes a procedure
which receives some data as input, does some calculation, and returns 
 one or several variables. On the other hand, a
\textbf{procedure} is part of a bigger algorithm which is separated from the
main algorithm in order to
simplify the reading. 

\subsection{Structure of the document}
\label{sec:intro-structure}


The structure of the document is as follows.
In Section~\ref{sec:preliminaries}, we review some of the tools that we use,
specifically the local weak convergence framework and the marked BC entropy.
In Section~\ref{sec:Problem-statement}, we rigorously define the problem of
universal graphical data compression and state our main results in
Theorem~\ref{thm:optimality-complexity-main} which introduces our compression
algorithm and discusses its main properties, i.e.\ information theoretic
optimality and low computational complexity.
In Section~\ref{sec:algorithm-general}, we give an overview of the steps of our universal compression algorithm, without going
through the details.
In Section~\ref{sec:experiments}, we illustrate the performance of our
compression algorithm for both synthetic and real--world graphical data.
Section~\ref{sec:alg-details} discusses the details of our compression and
decompression algorithms. {
  In particular, the complexity
  analysis of the  compression and decompression algorithms (second part of
  Theorem~\ref{thm:optimality-complexity-main}) are provided in
  Sections~\ref{sec:enc-complexity} and \ref{sec:main-decode-complexity},
  respectively. Specifically, we will see that  the dependence of the computational complexity on the
  graph size for both
compression and decompression is of the form $\widetilde{O}(n+m)$ where $n$ and $m$ denote
the number of vertices and edges, respectively, and $\widetilde{O}$ ignores
logarithmic factors.}
In Section~\ref{sec:optimality-proof}, we  prove that our compression algorithm  is
universally optimal, i.e. we prove the first part of
Theorem~\ref{thm:optimality-complexity-main}.
To better organize the document, we bring the details for several components of
our compression and decompression algorithms discussed in
Section~\ref{sec:alg-details} to 
Sections~\ref{sec:MP}, \ref{sec:bipartite-compression}, and
\ref{sec:simple-graph-compression}.
Appendix~\ref{sec:app-algo-dependencies} provides a list of  the algorithms in
this paper and illustrates their dependencies. Moreover,
Appendix~\ref{sec:data-structures} reviews some of the  data structures and algorithms
used in this paper.
The remaining appendices contain some technical proofs. 

\editfinish


\section{Preliminaries}
\label{sec:preliminaries}

\subsection{Graphs}
\label{sec:prelim-graphs}

A simple graph $G$ consists of a set of vertices, denoted by $V(G)$, and a set
of edges, 
\ifreplace
without multiple edges or self loops.
\color{black}
\fi
\ifreplace
We use the terms ``vertex'' and ``node'' interchangeably.
\color{black}
\fi
For vertices $v$ and $w$ in $V(G)$, we write $v \sim_G w$ to denote
that there is an edge in $G$ between $v$ and $w$
\ifreplace
i.e. $v$ and $w$ are adjacent in $G$.
\color{black}
\fi
\ifreplace
All graphs in this document are
simple.
\color{black}
\fi

\begin{figure}
  \centering
  \begin{tikzpicture}[scale=0.7]
    \begin{scope}[yshift=0.7cm]
    \node[nodeB] (n1) at (0,2) {};
    \node[nodeB] (n2) at (-1,1) {};
    \node[nodeB] (n3) at (1,1) {};
    \node[nodeR] (n4) at (0,0) {};
  \end{scope}
  \begin{scope}[yshift=-0.7cm]
    \node[nodeR] (n5) at (0,0) {};
    \node[nodeB] (n6) at (-1,-1) {};
    \node[nodeB] (n7) at (1,-1) {};
    \node[nodeB] (n8) at (0,-2) {};
  \end{scope}
  \nodelabel{n1}{90}{1};
  \nodelabel{n2}{180}{2};
  \nodelabel{n3}{0}{3};
  \nodelabel{n4}{180}{4};
  \nodelabel{n5}{180}{5};
  \nodelabel{n6}{180}{6};
  \nodelabel{n7}{0}{7};
  \nodelabel{n8}{270}{8};

  \drawedge{n1}{n2}{B}{O}
  \drawedge{n1}{n3}{B}{O}
  \drawedge{n8}{n6}{B}{O}
  \drawedge{n8}{n7}{B}{O}

  \drawedge{n2}{n4}{B}{B}
  \drawedge{n3}{n4}{B}{B}
  \drawedge{n5}{n6}{B}{B}
  \drawedge{n5}{n7}{B}{B}
  \draw[edgeO] (n4) -- (n5);
  \end{tikzpicture}
  

\caption[A marked graph]{\label{fig:marked-graph} A marked graph $G$ on the vertex set
  $\{1,\dots,8\}$ where edges carry marks from $\edgemark =
  \{\text{\color{blueedgecolor} Blue (solid)}, \text{\color{orangeedgecolor}
    Orange (wavy)} \}$ (e.g.\ $\xi_G(1,2)= \text{\color{orangeedgecolor} Orange}$
  while $\xi_G(2,1) = \text{\color{blueedgecolor} Blue}$; also, $\xi_G(2,4) =
  \xi_G(4,2) = \text{\color{blueedgecolor} Blue}$) and vertices carry marks from
$\vermark = \{\tikz{\node[nodeB] at (0,0) {};}, \tikz{\node[nodeR] at (0,0)
  {};}\}$ (e.g. $\tau_G(3) = \tikz{\node[nodeB] at (0,0) {};}$).
}
\end{figure}
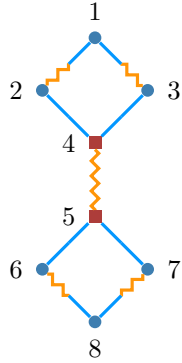

A (simple) marked graph is a simple graph such that every vertex carries a mark
coming from a finite vertex mark set $\vermark$, and also every edge carries two marks, one
towards each of its endpoints, coming from a finite edge mark set $\edgemark$.
The mark of vertex $v \in V(G)$ is denoted by $\tau_G(v)$, and the mark of an
edge between vertices $v$ and $w$ towards vertex $w$ is denoted by $\xi_G(v,w)$.
See Figure~\ref{fig:marked-graph} for an example. 
Given a marked graph $G$, the edge mark count vector of $G$ is defined to be the
vector $\vm_G := (m_G(x,x'): x, x' \in \edgemark)$ such that for
$x, x' \in \edgemark$, $m_G(x,x')$ is the number of edges in $G$ with mark $x$
towards one endpoint and mark $x'$ towards the other endpoint. Note that by
definition, we have $m_G(x,x') = m_G(x',x)$ for all $x, x' \in \edgemark$.
Furthermore, the vertex mark count vector is defined to be the vector $\vu_G :=
(u_G(\theta): \theta \in \vermark)$ where $u_G(\theta)$ for $\theta \in
\vermark$ is the number of vertices in $G$ with mark $\theta$. 
A marked tree is a marked
graph $T$  where the underlying graph is a tree. 

A \emph{walk}
\color{black}
between two vertices $v$ and $w$ in a marked or an unmarked graph
$G$ is a sequence of distinct vertices $v = v_0, v_1, \dots, v_k = w$  such that
for all $1 \leq i \leq k$, we have $v_i \sim_G v_{i-1}$. The length of 
such a walk
\color{black}
is defined to be $k$. The distance between vertices $v$ and $w$ in a marked
or an unmarked graph $G$ is defined
to be the length of 
the shortest walk
\color{black}
connecting them, or infinity if 
such a walk
\color{black}
does not exist. The \emph{connected component} of a vertex $v \in V(G)$ in
a marked or an unmarked graph $G$,
which is denoted by $G_v$, is the subgraph comprised of vertices in $G$ with a
finite distance from $v$. Note that $G_v$ by definition is connected.

Given a marked or an unmarked graph $G$, for a vertex $v \in V(G)$, the degree
of $v$, which is denoted by $\deg_G(v)$, is defined to be the number of edges
connected to $v$. Moreover, for $x, \pp{x} \in \edgemark$,
$\deg_G^{x,\pp{x}}(v)$ denotes the number of neighbors $w$ of $v$ such that
$\xi_G(w,v) = x$ and $\xi_G(v,w) = \pp{x}$. $G$ is called locally finite if the
degree of all the vertices is finite. 
All graphs in this document are assumed to be locally finite.
\color{black}

\begin{figure}
  \centering
  \begin{tikzpicture}
    \newcommand{\ucxdist}{0.7}
    \newcommand{\ucydist}{0.8}

    \begin{scope}[xshift=-4cm]
      \node[nodeB] (n1) at (0,0) {};
      \node[nodeR] (n2) at (-1,-1) {};
      \node[nodeR] (n3) at (1,-1) {};
      \node[nodeB] (n4) at (0,-2) {};
      \drawedge{n1}{n2}{B}{O}
      \drawedge{n1}{n3}{B}{O}
      \drawedge{n2}{n4}{B}{O}
      \drawedge{n3}{n4}{B}{O}

      \nodelabeldist{n1}{90}{4mm}{1};
      \nodelabeldist{n2}{180}{4mm}{2};
      \nodelabeldist{n3}{0}{4mm}{3};
      \nodelabeldist{n4}{270}{4mm}{4};

      \node at (0,-5*\ucydist) {$(a)$};
    \end{scope}

    \begin{scope}[xshift=4cm]
      \node[nodeB] (c1) at (0,0) {};
      \nodelabeldist{c1}{90}{4mm}{(1)};

      \node[nodeR] (c12) at (-\ucxdist, -\ucydist) {};
      \nodelabeldistalignright{c12}{180}{2mm}{(1,2)};
      \node[nodeR] (c13) at (\ucxdist, -\ucydist) {};
      \nodelabeldistalignleft{c13}{0}{2mm}{(1,3)};
      \drawedge{c1}{c12}{B}{O}
      \drawedge{c1}{c13}{B}{O}

      \node[nodeB] (c124) at (-2*\ucxdist, -2*\ucydist) {};
      \nodelabeldistalignright{c124}{180}{2mm}{(1,2,4)};
      \node[nodeB] (c134) at (2*\ucxdist, -2*\ucydist) {};
      \nodelabeldistalignleft{c134}{0}{2mm}{(1,3,4)};
      \drawedge{c12}{c124}{B}{O}
      \drawedge{c13}{c134}{B}{O}

      \node[nodeR] (c1243) at (-3*\ucxdist, -3*\ucydist) {};
      \nodelabeldistalignright{c1243}{180}{2mm}{(1,2,4,3)};
      \node[nodeR] (c1342) at (3*\ucxdist, -3*\ucydist) {};
      \nodelabeldistalignleft{c1342}{0}{2mm}{(1,3,4,2)};
      \drawedge{c124}{c1243}{O}{B}
      \drawedge{c134}{c1342}{O}{B}

      \node[nodeB] (c12431) at (-4*\ucxdist, -4*\ucydist) {};
      \nodelabeldistalignright{c12431}{180}{2mm}{(1,2,4,3,1)};
      \node[nodeB] (c13421) at (4*\ucxdist, -4*\ucydist) {};
      \nodelabeldistalignleft{c13421}{0}{2mm}{(1,3,4,2,1)};
      \drawedge{c1243}{c12431}{O}{B}
      \drawedge{c1342}{c13421}{O}{B}

      \node[rotate=48.814] at (-4.5*\ucxdist, -4.5*\ucydist) {\dots};
      \node[rotate=-48.814] at (4.5*\ucxdist, -4.5*\ucydist) {\dots};

      \node at (0,-5*\ucydist) {$(b)$};
    \end{scope}
  \end{tikzpicture}
  \caption{The universal cover $\UC_1(G)$ for the simple marked graph $G$
    depicted in $(a)$ is illustrated in $(b)$. By definition, every vertex in
    $\UC_1(G)$ corresponds to a non-backtracking walk in $G$ starting at vertex
    $1$. The walk associated to every vertex in $\UC_1(G)$ is shown adjacent to
    that vertex in $(b)$. Note that although $G$ is a finite graph in this
    example, $\UC_1(G)$ is an infinite graph. Moreover, the universal cover is
    always a tree. \label{fig:universal-cover-example}}
\end{figure}
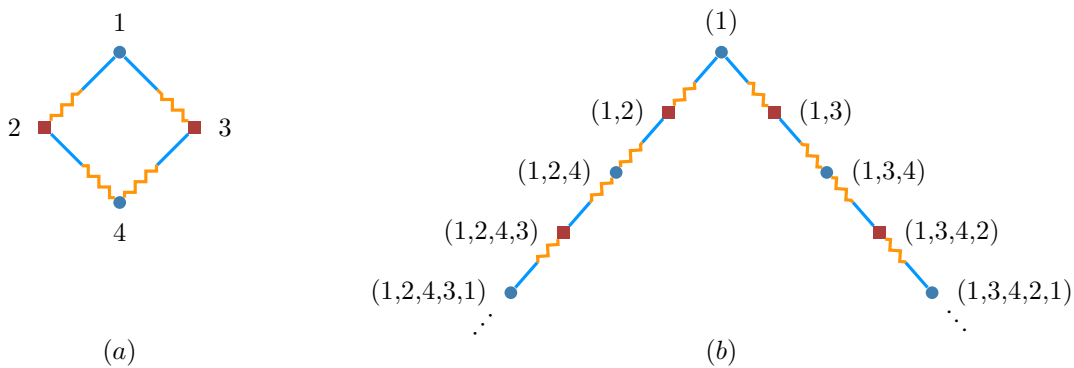

Given a 
(simple, locally finite)
\color{black}
marked graph $G$ together with a vertex $v \in V(G)$, we define the
``universal cover of $G$ at $v$'', denoted by $\UC_v(G)$, as follows. Every
vertex in $\UC_v(G)$ is in correspondence with a non--backtracking walk starting
at $v$, i.e.\ a sequence of vertices $v = v_0, v_1, \dots, v_k$ for some $k \geq
0$ such that for $1
\leq i \leq k$, we have $v_i \sim_G v_{i-1}$ and $v_i \neq v_{i-1}$. Moreover,
for $1 \leq i \leq k-1$, we require that $v_{i-1} \neq v_{i+1}$, i.e.\ the walk
is non--backtracking. 
A vertex in $\UC_v(G)$ corresponding to a walk $(v_i: 0
\leq i \leq k)$ is given the mark $\tau_G(v_k)$. Moreover, for each
non--backtracking walk $v = v_0, \dots, v_k$ for which $k > 1$, if $\tilde{v}$
and $\tilde{w}$  denote the vertices in $\UC_v(G)$ corresponding to the walks
$v_0, \dots, v_k$ and $v_0, \dots, v_{k-1}$, respectively, we place an edge
in $\UC_v(G)$ between the vertices $\tilde{v}$ and $\tilde{w}$ with mark
$\xi_G(v_{k-1}, v_k)$ towards $\tilde{v}$ and mark $\xi_G(v_k, v_{k-1})$ towards
$\tilde{w}$. In fact, $\UC_v(G)$ is the universal cover associated to the
connected component of $v$ in $G$. It can be easily verified that $\UC_v(G)$ is
a marked tree. 
With an abuse of notation, we denote the vertex in $\UC_v(G)$ associated to the
walk $(v)$  
with length 0
\color{black}
by $v$. Likewise, for a vertex $w \sim_G v$, we denote
the vertex in $\UC_v(G)$ associated to the walk $(v,w)$ by $w$. 
See Figure~\ref{fig:universal-cover-example} for an example.  See, for instance, \cite{sunada2012topological}, for more
discussion on universal covers. 

\subsection{The Framework of Local Weak Convergence}
\label{sec:prelim-lwc}

In this section, we give brief introduction to the framework of local weak
convergence, or the so-called objective method. This is the framework we use in
order to make sense of stochastic processes for the space of marked sparse
graphs. This is crucial in formulating our problem of universal compression of
graphical data. The reader is referred to \cite{BenjaminiSchramm01rec},
\cite{aldous2004objective},
\cite{aldous2007processes} for more details on this framework. Here, we
assume that the vertex and edge mark sets, $\vermark$ and $\edgemark$, are fixed
and finite.

A rooted marked graph is defined to be a marked graph $G$ with a distinguished
vertex $v \in V(G)$. We denote such a rooted graph by $(G,v)$. Here, $G$ can have a finite or a countably infinite set of
vertices. We say that two rooted marked  graphs $(G,o)$ and $(G',o')$
are isomorphic, and we  write $(G,o) \equiv (G',o')$, if there is a vertex
bijection between the vertices in $G_o$  and $G'_{o'}$, that preserves
adjacency structures, vertex marks, edge marks, and also maps $o$ to $o'$. This
imposes an equivalence relation on the space of rooted marked graphs. Note that
in order to determine whether two rooted marked graphs are isomorphic in the
above sense, we only need to compare the connected components of the root in
them. Let $[G,o]$ denote the equivalence class corresponding to a rooted marked 
graph $(G,o)$. 

For a rooted marked graph $(G, o)$ and integer $h\geq 1$, let $(G, o)_h$ denote 
the subgraph of  $G$ rooted at $o$ induced by  vertices with distance no more
than $h$ from $o$. If $h=0$,  $(G, o)_h$ is defined to  be the isolated root $o$ with mark $\tau_{G}(o)$. 
Moreover, let $[G, o]_h$ denote the  equivalence class corresponding to $(G, o)_h$, i.e.\ $[G, o]_h := [(G, o)_h]$.
Note that $[G, o]_h$ depends only on  $[G, o]$.

Note that all definitions above have natural parallels for unmarked graphs, which
are obtained by simply going over the above definitions with each of the mark
sets $\edgemark$ and $\vermark$ being of cardinality 1.

Let $\mGb_*$ denote the space of equivalence classes $[G,o]$ corresponding to
(simple, locally finite) rooted marked graphs $(G,o)$. As was mentioned above, all that
matters in defining $[G,o]$ is the connected component of the root in $(G,o)$.
We equip $\mGb_*$ with the metric $\bar{d}_*$ defined as follows. Given $[G,o]$ and
$[G',o']$, let $\hat{h}$ be the supremum over all $h  \geq 0$ such that $(G,o)_h
\equiv (G',o')_h$. Here, $(G,o)$ and $(G',o')$ are arbitrary members in the
equivalence classes $[G,o]$ and $[G',o']$ respectively.  Since all such members
are rooted isomorphic, the above definition is invariant under the specific
choice of members inside equivalence classes. If such $h$ does not exist, we
define $\hat{h} = 0$. With this, we define $\bar{d}_*([G,o], [G',o']):=
1/(1+\hat{h})$.
One can check that $\bar{d}_*$ is a metric; in particular, it satisfies the triangle inequality.
Moreover, 
$\mGb_*$ together with this metric is a Polish space, i.e. a 
complete separable metric space \cite{aldous2007processes}.
Let $\mTb_*$ denote the subset of $\mGb_*$ comprised of the 
equivalence classes $[G, o]$ arising from some $(G,o)$ where
the graph underlying $G$ is a tree. 
In the sequel we will 
think of $\mGb_*$ as a Polish space with the metric $\bar{d}_*$
defined above, rather than just a set. Note that $\mTb_*$ is a closed subset of $\mGb_*$.

For a Polish space 
$\Omega$, let $\mP(\Omega)$ denote the set of Borel probability measures on
$\Omega$. We say that a sequence of measures $\mu_n$ on $\Omega$ converges
weakly to $\mu \in \mP(\Omega)$ and write $\mu_n \Rightarrow \mu$, if for any
bounded continuous function on $\Omega$, we have $\int f d \mu_n \rightarrow
\int f d \mu$.
It can be shown that it suffices to verify this condition only for
uniformly continuous and bounded functions \cite{billingsley2013convergence}.
For a Borel set $B \subset \Omega$, the $\epsilon$--extension of
$B$, denoted by $B^\epsilon$, is defined as the union of the open balls with
radius $\epsilon$ centered around the points in $B$. For two probability
measures $\mu$ and $\nu$ in $\mP(\Omega)$, the \LP distance $\dlp(\mu, \nu)$ is
defined to be the infimum of all $\epsilon>0$ such that for all Borel sets $B
\subset \Omega$ we have $\mu(B) \leq \nu(B^\epsilon) + \epsilon$ and  $\nu(B) \leq \mu(B^\epsilon) + \epsilon$.
It is known that the \LP distance metrizes the topology of weak convergence 
on the space of probability distributions on a Polish space
(see, for instance, \cite{billingsley2013convergence}).
For $x \in \Omega$, let $\delta_x$ be the Dirac measure at $x$.

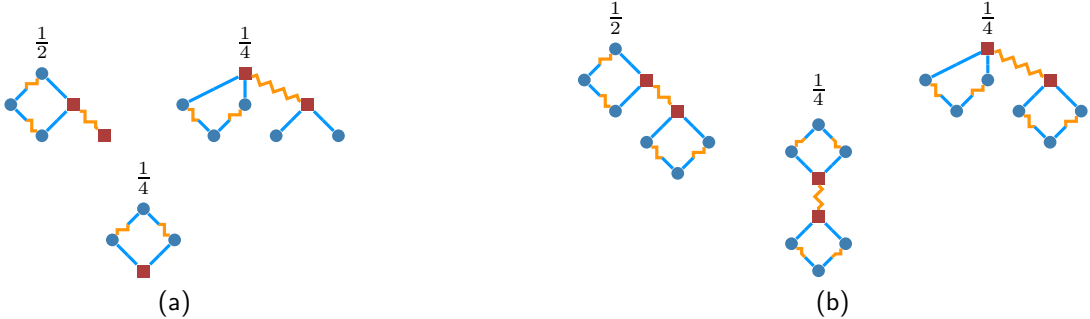
\begin{figure}
  \centering
\hfill
\subfloat[]{
\centering
\begin{tikzpicture}[scale=0.9]
  \begin{scope}[yshift=2cm,xshift=1.5cm, scale=0.23]
    \node[nodeR] (n1) at (0,0) {};
    \node[nodeB] (n2) at (-4,-2) {};
    \node[nodeB] (n3) at (0,-2) {};
    \node[nodeR] (n4) at (4,-2) {};
    \node[nodeB] (n5) at (-2,-4) {};
    \node[nodeB] (n6) at (2,-4) {};
    \node[nodeB] (n7) at (6,-4) {};

    \drawedge{n1}{n2}{B}{B}
    \drawedge{n1}{n3}{B}{B}
    \draw[edgeO] (n1) -- (n4);
    \drawedge{n2}{n5}{O}{B}
    \drawedge{n3}{n5}{O}{B}
    \drawedge{n4}{n6}{B}{B}
    \drawedge{n4}{n7}{B}{B}

    \node at (0,2) {$\frac{1}{4}$};

  \end{scope}

  \begin{scope}[yshift=2cm,xshift=-1.5cm,scale=0.23]
    \node[nodeB] (n1) at (0,0) {};
    \node[nodeB] (n2) at (-2,-2) {};
    \node[nodeR] (n3) at (2,-2) {};
    \node[nodeB] (n4) at (0,-4) {};
    \node[nodeR] (n5) at (4,-4) {};
    \drawedge{n1}{n2}{O}{B}
    \drawedge{n1}{n3}{B}{B}
    \drawedge{n2}{n4}{B}{O}
    \drawedge{n3}{n4}{B}{B}
    \draw[edgeO] (n3) -- (n5);
    
    \node at (0,2) {$\frac{1}{2}$};

  \end{scope}

  \begin{scope}[scale=0.23]
    \node[nodeB] (n1) at (0,0) {};
    \node[nodeB] (n2) at (-2,-2) {};
    \node[nodeB] (n3) at (2,-2) {};
    \node[nodeR] (n4) at (0,-4) {};
    
    \drawedge{n1}{n2}{B}{O}
    \drawedge{n1}{n3}{B}{O}
    \drawedge{n2}{n4}{B}{B}
    \drawedge{n3}{n4}{B}{B}
    
    \node at (0,2) {$\frac{1}{4}$};
  \end{scope}
\end{tikzpicture}
\label{fig:U-d2}
}
\hfill\hfill\hfill
\subfloat[]{
\centering
\begin{tikzpicture}[scale=0.9]
  \begin{scope}[yshift=1cm,xshift=2.5cm,scale=0.23]
        \node[nodeR] (n1) at (0,0) {};
    \node[nodeB] (n2) at (-4,-2) {};
    \node[nodeB] (n3) at (0,-2) {};
    \node[nodeR] (n4) at (4,-2) {};
    \node[nodeB] (n5) at (-2,-4) {};
    \node[nodeB] (n6) at (2,-4) {};
    \node[nodeB] (n7) at (6,-4) {};
    \node[nodeB] (n8) at (4,-6) {};
    
    \drawedge{n1}{n2}{B}{B}
    \drawedge{n1}{n3}{B}{B}
    \draw[edgeO] (n1) -- (n4);
    \drawedge{n2}{n5}{O}{B}
    \drawedge{n3}{n5}{O}{B}
    \drawedge{n4}{n6}{B}{B}
    \drawedge{n4}{n7}{B}{B}
    \drawedge{n6}{n8}{O}{B}
    \drawedge{n7}{n8}{O}{B}
    
    \node at (0,2) {$\frac{1}{4}$};

  \end{scope}

  \begin{scope}[yshift=1cm,xshift=-3cm,scale=0.23]

    \node[nodeB] (n1) at (0,0) {};
    \node[nodeB] (n2) at (-2,-2) {};
    \node[nodeR] (n3) at (2,-2) {};
    \node[nodeB] (n4) at (0,-4) {};
    \node[nodeR] (n5) at (4,-4) {};
    \node[nodeB] (n6) at (2,-6) {};
    \node[nodeB] (n7) at (6,-6) {};
    \node[nodeB] (n8) at (4,-8) {};
    
    \drawedge{n1}{n2}{O}{B}
    \drawedge{n1}{n3}{B}{B}
    \drawedge{n2}{n4}{B}{O}
    \drawedge{n3}{n4}{B}{B}
    \drawedge{n5}{n6}{B}{B}
    \drawedge{n5}{n7}{B}{B}
    \drawedge{n6}{n8}{O}{B}
    \drawedge{n7}{n8}{O}{B}
    \draw[edgeO] (n3) -- (n5);

    \node at (0,2) {$\frac{1}{2}$};

  \end{scope}

  \begin{scope}[scale=0.4,yshift=-3cm]
        \begin{scope}[yshift=0.7cm]
    \node[nodeB] (n1) at (0,2) {};
    \node[nodeB] (n2) at (-1,1) {};
    \node[nodeB] (n3) at (1,1) {};
    \node[nodeR] (n4) at (0,0) {};
  \end{scope}
  \begin{scope}[yshift=-0.7cm]
    \node[nodeR] (n5) at (0,0) {};
    \node[nodeB] (n6) at (-1,-1) {};
    \node[nodeB] (n7) at (1,-1) {};
    \node[nodeB] (n8) at (0,-2) {};
  \end{scope}

  \drawedge{n1}{n2}{B}{O}
  \drawedge{n1}{n3}{B}{O}
  \drawedge{n8}{n6}{B}{O}
  \drawedge{n8}{n7}{B}{O}

  \drawedge{n2}{n4}{B}{B}
  \drawedge{n3}{n4}{B}{B}
  \drawedge{n5}{n6}{B}{B}
  \drawedge{n5}{n7}{B}{B}
  \draw[edgeO] (n4) -- (n5);

    \node at (0,4) {$\frac{1}{4}$};
  \end{scope}
\end{tikzpicture}
\label{fig:U-U}
}
\hfill
\caption{\label{fig:UG} 
With $G$ being the graph from Figure~\ref{fig:marked-graph}, 
(a) illustrates $U_2(G)$, which is a probability distribution on rooted marked graphs of depth at most 2 and
(b) depicts $U(G)$, which is a probability distribution on $\mGb_*$. 
}
\end{figure}

For a finite marked graph $G$, define $U(G) \in \mP(\mGb_*)$ as 
\begin{equation}
  \label{eq:UG}
  U(G) := \frac{1}{|V(G)|} \sum_{o \in V(G)} \delta_{[G, o]}.
\end{equation}
Note that $U(G) \in \mP(\mGb_*)$. In creating $U(G)$
from $G$, we have created a probability distribution on 
rooted marked graphs from the given marked graph $G$ 
by rooting the graph at a vertex chosen uniformly at random. 
Furthermore, for an integer $h \geq 1$, let 
\begin{equation}
\label{eq:UkG}
  U_h(G) := \frac{1}{|V(G)|} \sum_{o \in V(G)} \delta_{[G, o]_h}.
\end{equation}
We then have $U_h(G) \in \mP(\mGb_*)$.
See Figure~\ref{fig:UG} for an example.

We say that a probability distribution $\mu$ on $\mGb_*$ is the {\em local weak limit} of a sequence of finite marked graphs $\{G_n\}_{n=1}^\infty$ when $U(G_n)$ converges weakly to $\mu$ 
(with respect to the topology induced by the metric $\bar{d}_*$).
This turns out to be equivalent to the condition that, for any finite depth $h \ge 0$, the structure of $G_n$ from the point of view of a root chosen uniformly at random and then looking around it only to depth $h$, converges in distribution to $\mu$ truncated up to depth $h$. This description of what is being captured by the definition justifies the term ``local weak convergence''.

In fact, $U_h(G)$ could be thought of as the ``depth $h$ empirical distribution'' of the marked graph $G$. On the other hand, a probability distribution $\mu \in \mP(\mGb_*)$ that arises as a local weak limit plays the role of a stochastic process on graphical data, and a sequence of marked graphs $\{G_n\}_{n=1}^\infty$ could be thought of as being 
asymptotically distributed like
this process when $\mu$ is the local weak limit of the sequence.

The degree of a probability measure $\mu \in \mP(\mGb_*)$,
denoted by $\deg(\mu)$, is defined as
\begin{equation*}
  \deg(\mu) := \evwrt{\mu}{ \deg_{G}(o)} = \int \deg_{G}(o) d \mu([G, o]),
\end{equation*}
which is the expected degree of the root.
Similarly, for $\mu \in \mP(\mGb_*)$ and $x, x' \in \Xi$, let $\deg_{x,x'}(\mu)$ be defined as 
\begin{equation*}
  \deg_{x,x'}(\mu) := \int \deg_{G}^{x,x'}(o) d \mu([G, o]),
\end{equation*}
which is the expected number of edges  connected to the root with mark $x$
towards the root and mark $x'$ towards the other endpoint. 
We use the notation $\vdeg(\mu) := (\deg_{x,x'}(\mu): x, x' \in \edgemark)$.
Moreover, for $\mu \in \mP(\mGb_*)$ and $\theta \in \vermark$, let
\begin{equation}
\label{eq:vertype-probability}
  \vtype_\theta(\mu) := \mu(\{[G, o] \in \mGb_*: \tau_{G}(o) = \theta \}),
\end{equation}
which is  the probability under $\mu$ that the mark of the root is $\theta$. 
We use the notation $\vvtype(\mu) := (\vtype_\theta(\mu): \theta \in \vermark)$. 

All the preceding definitions and concepts have the obvious parallels in the 
case of unmarked graphs. These can be arrived at by simply walking through
the definitions while restricting the mark sets $\vermark$ and $\edgemark$
to be of cardinality $1$. It is convenient, however, to sometimes use
the special
notation for the unmarked case
that matches the one currently in use in the literature. 
We will therefore write $\mG_*$ for the set of rooted isomorphism classes of unmarked graphs. This is just the set $\mGb_*$ in the case where 
both $\edgemark$ and $\vermark$ are sets of cardinality $1$. 
We will also denote the metric on $\mG_*$ by $d_*$, which is
just $\bar{d}_*$ when both $\edgemark$ and $\vermark$ are sets of cardinality $1$. 

Since each vertex in $G_n$ has the same chance of being chosen as 
the root in the definition of $U(G_n)$, this should manifest itself as some
kind of stationarity property of the limit $\mu$, with respect
to changes of the root.
A probability distribution $\mu \in \mP(\mGb_*)$ is called {\em sofic} 
if there exists a sequence of finite graphs $G_n$ with local weak limit $\mu$. 
The definition of unimodularity is made in an attempt to understand
what it means for a Borel probability distribution on $\mGb_*$
to be sofic.

To define unimodularity, let $\mGb_{**}$ be the set of isomorphism classes
$[G,o,v]$ where $G$ is a marked connected graph with two distinguished vertices
$o$ and $v$ in $V(G)$ (ordered, but not necessarily distinct). Here, isomorphism
is defined by an adjacency preserving vertex bijection which preserves
vertex and edge marks, and also maps the two distinguished vertices of one
object to the respective ones of the other. A measure $\mu \in \mP(\mGb_*)$ is
said to be unimodular if, for all measurable functions $f:
\mGb_{**} \rightarrow \reals_+$, we have
\begin{equation}
  \label{eq:unim-integral}
  \int \sum_{v \in V(G)} f([G,o,v]) d\mu([G,o]) = \int \sum_{v \in V(G)} f([G,v,o]) d\mu([G,o]).
\end{equation}
Here, the summation is taken over all vertices $v$ which are in the same
connected component of $G$ as $o$. It can be seen that it suffices to check the
above condition for a function $f$ such that $f([G,o,v]) = 0$ unless $v \sim_G
o$. This is called \emph{involution invariance} \cite{aldous2007processes}.
Let $\mP_u(\mGb_*)$ denote the set of unimodular probability measures on
$\mGb_*$. Also, since $\mTb_* \subset \mGb_*$, we can define the set of
unimodular probability measures on $\mTb_*$ and denote it by $\mP_u(\mTb_*)$. 
A sofic probability measure is unimodular. Whether the other direction also
holds is unknown.

\subsection{The Marked BC Entropy}
\label{sec:marked-BC-entropy}

In order to make sense of information theoretic optimality of our compression
algorithm, we need to define a notion of entropy for stochastic processes on
graphs. As we discussed in the previous subsection, we employ the framework of
local weak convergence to make sense of stochastic processes on the space of
rooted graphs. Motivated by this, we look for a notion of entropy defined for
probability measures on the space of rooted marked graphs, i.e.\ $\mGb_*$, defined
above. Throughout our discussion, we assume that the set of vertex and edge
marks, i.e.\ $\edgemark$ and $\vermark$, are fixed and finite sets. The notion
of entropy we employ here is a generalization of the notion of entropy defined
by Bordenave and Caputo in \cite{bordenave2015large} to the marked framework
discussed above. This generalization is due to us, and the reader is referred to
\cite{delgosha2019notion} for more details.


An edge mark
count vector is defined to be a vector of nonnegative integers $\vm := (m(x, x'): x, x' \in \edgemark)$ such that
$m(x,x') = m(x',x)$ for all $x, x' \in \edgemark$. 
A vertex mark count
vector is defined to be a vector of nonnegative integers $\vu := (u(\theta): \theta \in
\Theta)$. Since $\edgemark$ is finite, we may assume it is an ordered set.
We define $\snorm{\vm}_1 := 
\sum_{x \leq x' \in \edgemark} 
m(x, x')$
and $\snorm{\vu}_1 :=
\sum_{\theta \in \vermark} u(\theta)$.

For an integer $n \in \nats$ and edge mark and vertex mark count vectors $\vm$
and $\vu$, define $\mGn_{\vm, \vu}$ to be the set of 
marked
graphs on the vertex set
$[n]$ such that $\vm_G = \vm$ and $\vu_G = \vu$.  Note that
$\mGn_{\vm, \vu}$ is empty unless $\snorm{\vm}_1 \leq \binom{n}{2}$ and
$\snorm{\vu}_1 = n$. 

An {\em average degree vector} is defined to be a vector of nonnegative reals $\vd = (d_{x,x'} : x, x' \in \edgemark)$ such
  that for all $x, x' \in \edgemark$,  we have  $d_{x,x'} =
  d_{x',x}$. Moreover, we require that $\sum_{x,x' \in \edgemark} d_{x,x'} > 0$.



\begin{definition}
\label{def:deg-seq-adapt}  
  Given an average degree vector $\vd$ and a probability distribution $Q = (q_\theta: \theta \in \vermark)$, we say that a sequence 
$(\vmn,\vun)$ of edge mark count vectors and vertex mark count vectors $\vmn$ and $\vun$ is adapted to $(\vd, Q)$, if the following conditions hold:
  \begin{enumerate}
  \item For each $n$, we have $\snorm{\vmn}_1 \leq \binom{n}{2}$ and $\snorm{\vun}_1 = n$;
  \item For $x \in \edgemark$, we have $\mn(x,x) / n \rightarrow d_{x,x}/2$;
  \item For $x \neq x' \in \edgemark$, we have $\mn(x, x') / n \rightarrow d_{x,x'} = d_{x',x}$;
  \item For $\theta \in \vermark$, we have $\un(\theta) / n \rightarrow q_\theta$;
  \item For $x, x' \in \edgemark$, $d_{x,x'} = 0$ implies $\mn(x, x') = 0$ for all $n$;
  \item For $\theta \in \vermark$, $q_\theta  = 0$ implies $\un(\theta) = 0$ for all $n$.
  \end{enumerate}
\end{definition}


If $\vmn$ and $\vun$ are sequences such that $(\vmn,\vun)$ is adapted to $(\vd,
Q)$ then, using Stirling's approximation, it can be shown that
  \begin{equation}
    \label{EQ:LOG-MGNMNUN-STIRLING}
    \log | \mGn_{\vmn, \vun} | =  \snorm{\vmn}_1 \log n + n H(Q)  + n \sum_{x,x' \in \edgemark} s(d_{x,x'}) + o(n),
  \end{equation}
  where 
  \begin{equation*}
    s(d) :=
    \begin{cases}
      \frac{d}{2} - \frac{d}{2} \log d & d > 0, \\
      0 & d = 0.
    \end{cases}
  \end{equation*}

The key idea to define the  BC entropy is to count the number of ``typical''
graphs. More precisely, given $\mu \in \mP(\mGb_*)$ such that $0 < \deg(\mu) <
\infty$, for $\epsilon > 0$ and edge and vertex mark count vectors $\vmn$ and
$\vun$, respectively, define
\begin{equation*}
  \mGnmnun(\mu, \epsilon) := \{G \in \mGn_{\vmn, \vun}: \dlp(U(G), \mu) < \epsilon\}, 
\end{equation*}
where $\dlp$ refers to the \LP metric on $\mP(\mGb_*)$. In fact, one can
interpret $\mGnmnun(\mu, \epsilon)$ as the set of $\epsilon$--typical graphs
with respect to $\mu$. It turns out that, roughly speaking,  the number of
$\epsilon$--typical graphs scales as follows:
\begin{equation*}
  |\mGnmnun(\mu, \epsilon)| \approx \exp(\snorm{\vmn}_1 \log n + n \bch(\mu) + o(n)), 
\end{equation*}
where $\bch(\mu)$ is the marked BC entropy of $\mu$ which will be defined below.
In order to make this precise, we make the following definition:

\begin{definition}
  \label{def:BC-entropy}
Assume $\mu \in \mP(\mGb_*)$ is given, with $0 < \deg(\mu) < \infty$. For $\epsilon>0$, and edge and vertex mark count vectors
$\vm$ and $\vu$, let $\mGn_{\vm, \vu}(\mu, \epsilon)$ be as defined above. 
  Fix an average degree vector $\vd$ and a probability distribution $Q = (q_\theta:
  \theta \in \vermark)$, and also fix sequences of edge and vertex mark
  count vectors $\vmn$ and $\vun$ such that $(\vmn,\vun)$ is adapted to $(\vd, Q)$. With these, define
  \begin{equation*}
    \bchover_{\vd, Q}(\mu, \epsilon)\condmnun := \limsup_{n \rightarrow \infty} \frac{\log |\mGn_{\vmn, \vun}(\mu, \epsilon)| - \snorm{\vmn}_1 \log n}{n},
  \end{equation*}
which we call the $\epsilon$--upper BC entropy. Since this is increasing
in $\epsilon$, we can define the {\em upper BC entropy} as 
  \begin{equation*}
    \bchover_{\vd, Q}(\mu)\condmnun := \lim_{\epsilon \downarrow 0} \bchover_{\vd, Q}(\mu, \epsilon)\condmnun.
  \end{equation*}
We may define the $\epsilon$--lower BC entropy $\bchunder_{\vd, Q}(\mu,
\epsilon)\condmnun$ similarly as
\begin{equation*}
    \bchunder_{\vd, Q}(\mu, \epsilon)\condmnun := \liminf_{n \rightarrow \infty} \frac{\log |\mGn_{\vmn, \vun}(\mu, \epsilon)| - \snorm{\vmn}_1 \log n}{n}.
  \end{equation*}
Since this is increasing
in $\epsilon$, we can define the {\em lower BC entropy} $\bchunder_{\vd, Q}(\mu)\condmnun$ as
\begin{equation*}
    \bchunder_{\vd, Q}(\mu)\condmnun := \lim_{\epsilon \downarrow 0} \bchunder_{\vd, Q}(\mu, \epsilon)\condmnun.
  \end{equation*}
 
    \end{definition}

    The following Theorem~\ref{thm:badcases} shows that certain conditions must 
be met for the marked BC entropy to be of interest.

\begin{thm}[Theorem~1 in \cite{delgosha2019notion}]
  \label{thm:badcases}
  Let an average degree vector $\vd = (d_{x,x'} : x,x' \in \edgemark)$ and a
  probability distribution $Q = (q_\theta: \theta \in \vermark)$ be given. Suppose $\mu \in \mP(\mGb_*)$ with
 $0 < \deg(\mu) < \infty$ satisfies any one of the following conditions:
 \begin{enumerate}
    \item $\mu$ is not unimodular;
    \item $\mu$ is not supported on $\mTb_*$;
    \item $\deg_{x,x'}(\mu) \neq d_{x,x'}$ for some $x,x' \in \edgemark$, or $\vtype_\theta(\mu) \neq q_\theta$ for some $\theta \in \vermark$.
    \end{enumerate}
    Then, for any choice of the
    sequences $\vmn$ and $\vun$ such that $(\vmn,\vun)$ is adapted to $(\vd, Q)$, we have $\bchover_{\vd, Q}(\mu)\condmnun = -\infty$. 
  \end{thm}

  A consequence of Theorem~\ref{thm:badcases}
is that the only case of interest in the discussion of marked
BC entropy is when $\mu \in \mP_u(\mTb_*)$,
$\vd = \vdeg(\mu)$, $Q = \vvtype(\mu)$,
and the
sequences $\vmn$ and $\vun$ are such that $(\vmn,\vun)$ is adapted to
$(\vdeg(\mu), \vvtype(\mu))$.
Namely, the only upper and lower marked BC entropies of interest are 
$\bchover_{\vdeg(\mu), \vvtype(\mu)}(\mu)\condmnun$ and $\bchunder_{\vdeg(\mu), \vvtype(\mu)}(\mu)\condmnun$ respectively.

The following Theorem~\ref{thm:bch-properties} establishes that the upper and lower
marked BC entropies do not depend on the 
choice of the defining pair of sequences 
$(\vmn,\vun)$. Further, 
this theorem establishes that
the upper marked BC entropy 
is always equal to the lower marked BC entropy.

\begin{thm}[Theorem 2 in \cite{delgosha2019notion}]
  \label{thm:bch-properties}
  Assume that an average degree vector $\vd = (d_{x,x'} : x,x' \in \edgemark)$ together with a
  probability distribution $Q = (q_\theta: \theta \in \vermark)$ are given. For
  any  $\mu \in \mP(\mGb_*)$ such that
 $0 < \deg(\mu) < \infty$, we have 
  \begin{enumerate}
  \item \label{thm:BC-invariant} The values of $\bchover_{\vd, Q}(\mu)\condmnun$ and
    $\bchunder_{\vd, Q}(\mu)\condmnun$ are invariant under the specific choice of the
    sequences $\vmn$ and $\vun$ such that $(\vmn,\vun)$ is adapted to $(\vd, Q)$. With this,
    we may simplify the notation and unambiguously write $\bchover_{\vd, Q}(\mu)$ and
    $\bchunder_{\vd, Q}(\mu)$. 
  \item \label{thm:BC-well} 
  $\bchover_{\vd, Q}(\mu) = \bchunder_{\vd, Q}(\mu)$. 
  We may therefore unambiguously write $\bch_{\vd, Q}(\mu)$ 
for this common value,
and call it the {\em marked BC entropy} of 
$\mu \in \mP(\mGb_*)$ for the 
average degree vector $\vd$ and a probability distribution $Q = (q_\theta:
  \theta \in \vermark)$.
  Moreover, $\bch_{\vd, Q}(\mu) \in [-\infty, s(\vd) + H(Q)]$. Here, $s(\vd):=
  \sum_{x,x' \in \edgemark} s(d_{x,x'})$. 
  \end{enumerate}
\end{thm}

From Theorem~\ref{thm:badcases} we conclude that unless 
$\vd = \vdeg(\mu)$, $Q = \vvtype(\mu)$, and $\mu$
  is a unimodular measure on $\mTb_*$, we have 
  $\bch_{\vd, Q}(\mu) = -\infty$. 
  In view of this, for $\mu \in \mP(\mGb_*)$
  with $0 < \deg(\mu) < \infty$, we write $\bch(\mu)$
  for  $\bch_{\vdeg(\mu), \vvtype(\mu)}(\mu)$. Likewise, we may write
  $\bchunder(\mu)$ and $\bchover(\mu)$ for $\bchunder_{\vdeg(\mu),
    \vvtype(\mu)}(\mu)$ and $\bchover_{\vdeg(\mu),
    \vvtype(\mu)}(\mu)$, respectively. 
    Note that, unless $\mu \in \mP_u(\mTb_*)$, 
    we have $\bchover(\mu) = \bchunder(\mu) = \bch(\mu) = -\infty$.
    
We are now in a position to define the marked BC entropy.

\begin{definition}
  \label{def:BC-entropy-new}
  
  For $\mu \in \mP(\mGb_*)$
  with $0 < \deg(\mu) < \infty$, the marked BC entropy of $\mu$ is defined to be $\bch(\mu)$.
 

\end{definition}

Next, we discuss how to approximate the marked BC entropy of a probability
distribution $\mu \in \mP_u(\mTb_*)$ in terms of the finite depth truncation of
$\mu$. 
Given $\mu \in \mP(\mGb_*)$ and an integer $h \geq 0$, we define the depth $h$
truncation of $\mu$, denoted by  $\mu_h$, to be
the law of $[G,o]_h$ when $[G,o]$ has law $\mu$. 
Let $\mGb_*^h$ denote the subset of $\mGb_*$ consisting of isomorphism classes of marked rooted graphs with depth no more
than $h$. Likewise, define $\mTb_*^h$ to be the subset of $\mTb_*$ consisting of
isomorphism classes of marked rooted trees with depth no more than $h$. 
Note that for $\mu \in \mP(\mGb_*)$, $\mu_h$ is a
probability measure on $\mGb_*^h$. Likewise, for $\mu \in \mP(\mTb_*)$, we have
$\mu_h \in \mP(\mTb_*^h)$. 

{
  For a marked graph
$G$, on a finite or countably infinite vertex set, and adjacent vertices $u$ and $v$ in $G$, we define $G(u,v)$ to be the pair
$(\xi_G(u,v), (G',v))$ where  $G'$ is the connected component of $v$ in the graph
obtained from $G$ by removing the edge between $u$ and $v$. Similarly, for
 $h \geq 0$, $G(u,v)_h$ is defined as $(\xi_G(u,v), (G',v)_h)$.
}
  For a rooted marked graph $(G, o)$, integer $h \ge 1$, and $g, g' \in \Xi \times \mGb_*^{h-1}$, we define
  \begin{equation}
    \label{eq:Eh-g-g'}
    E_h(g, g')(G, o) := |\{v \sim_G o: G(v,o)_{h-1} \equiv g, G(o,v)_{h-1} \equiv g'  \}|.
  \end{equation}
Also, for $[G,o] \in \mGb_*$, we can write $E_h(g,g')([G,o])$ for
$E_h(g,g')(G,o)$, where $(G,o)$ is an arbitrary member of $[G,o]$.
This notation is well-defined, since $E_h(g, g')(G,o)$,
thought of as a function of $(G,o)$ for fixed integer $h \ge 1$
and $g, g' \in \Xi \times \mGb_*^{h-1}$, is invariant under rooted isomorphism.

  For  $h \geq 1$, $P \in \mP(\mGb_*^h)$, and $g, g' \in \Xi \times \mGb_*^{h-1}$, define
  \begin{equation}
            \label{eq:esubPdefinition}
    e_P(g, g') := \evwrt{P}{E_h(g, g')(G, o)}.
  \end{equation}
  Here, $(G, o)$ is a member of the isomorphism class $[G, o]$ that has law $P$.
This notation is well-defined
for the same reason as above.
{
  When it is clear from the context, we may simplify the notation and
write $E_h(g,g')$  instead of $E_h(g,g')(G,o)$.}

  \begin{definition}
    \label{def:admissible}
    
      Let $h \ge 1$. A probability distribution $P \in \mP(\mGb_*^h)$ is called
  {\em admissible} if $\evwrt{P}{\deg_G(o)} < \infty$ and $e_P(g, g') = e_P(g', g)$
  for all $g, g' \in \edgemark \times \mGb_*^{h-1}$. 
\end{definition}

It can be seen that for a unimodular measure $\mu \in \mP(\mGb_*)$ and $h \geq
1$, it holds that $\mu_h$ is admissible \cite[Lemma~1]{delgosha2019notion}. 

For $h \ge 1$ and admissible $P \in \mP(\mTb_*^h)$ such that $d:= \evwrt{P}{\deg_T(o)} > 0$, let $\pi_P$ denote the probability
distribution on $(\edgemark \times \mTb_*^{h-1}) \times (\edgemark \times
\mTb_*^{h-1})$ defined as
\begin{equation*}
  \pi_P(t, t') := \frac{e_P(t, t')}{d}.
\end{equation*}
Since for each $[T,
o] \in \mTb_*$ we have $\deg_T(o) = \sum_{t, t' \in \edgemark \times
  \mTb_*^{h-1}} E_h(t, t')(T, o)$, 
we have $d = \sum_{t, t' \in \edgemark
  \times \mTb_*^{h-1}} e_P(t, t')$. Consequently, $\pi_P$ is indeed a
probability distribution. 

 For $h \ge 1$ and admissible $P \in \mP(\mTb_*^h)$ with $H(P) < \infty$ and $\evwrt{P}{\deg_T(o)}
 > 0$, define
    \begin{equation}
      \label{eq:Jh-def}
      J_h(P) := -s(d) + H(P) - \frac{d}{2} H(\pi_P) - \sum_{t, t' \in \edgemark \times \mTb_*^{h-1}} \evwrt{P}{\log E_h(t, t')!},
    \end{equation}
    where $d := \evwrt{P}{\deg_T(o)}$ is the average degree at the root and
    $s(d) = \frac{d}{2} - \frac{d}{2} \log d$. 
  Note that $s(d)$ is finite, since $d < \infty$. Also, $H(P) < \infty$,
   $H(\pi_P) \geq 0$, and for each $t, t' \in \edgemark \times
  \mTb_*^{h-1}$, $\evwrt{P}{\log E_h(t, t')!} \geq 0$. Thereby, $J_h(P)$ is
  well-defined and is in the range $[-\infty, \infty)$.

Now, we are ready to state the following result which approximates the marked BC
entropy of $\mu$ in terms of $J_h(\mu_h)$ for $h \geq 1$. 

  \begin{thm}[Theorem 3 in \cite{delgosha2019notion}]
  \label{thm:Jh}
  Let $\mu \in \mP_u(\mTb_*)$ be a unimodular probability measure with 
  $0 < \deg(\mu) < \infty$. 
  Then, 
  \begin{enumerate}
  \item \label{thm:Jh--infty} If $\evwrt{\mu}{\deg_T(o) \log \deg_T(o)} = \infty$, then $\bchunder(\mu)
    = \bchover(\mu) = \bch(\mu) = -\infty$.
  \item  \label{thm:Jh-lim} If $\evwrt{\mu}{\deg_T(o) \log \deg_T(o)} < \infty$, then, for each $h \ge 1$, the probability measure $\mu_h$ is admissible, and $H(\mu_h) < \infty$. Furthermore, the sequence $(J_h(\mu_h): h \geq 1)$ is
    nonincreasing, and
      \begin{equation*}
    \bchunder(\mu) = \bchover(\mu) = \bch(\mu) = \lim_{h \rightarrow \infty} J_h(\mu_h). 
  \end{equation*}
  \end{enumerate}
\end{thm}



\editstart 

\section{Problem Statement and Main Results}
\label{sec:Problem-statement}

Let $\mGb_n$  denote the set of simple marked graphs on the vertex set $[n]$ with
edge and vertex marks coming from fixed and finite sets $\edgemark$ and
$\vermark$ respectively. Without loss of generality, we may
assume that $\edgemark = \{1, \dots, |\edgemark|\}$ and $\vermark = \{1, \dots,
|\vermark|\}$. These mark sets are fixed and known to  both the
encoder and the decoder.
Our goal is to design a 
lossless
\color{black}
compression algorithm which
maps simple marked graphs in $\mGb_n$ to $\{0,1\}^* - \emptyset$ in an one-to-one manner. Our algorithm uses two
integers $h \geq 1$ and $\delta \geq 1$ as hyperparameters. Therefore, for each $n \geq 1$, we will introduce an
encoding function $\fn_{h, \delta} : \mGb_n \rightarrow \{0,1\}^* - \emptyset$ together with
a decoding function $\gn_{h, \delta}$ such that $\gn_{h, \delta} \circ \fn_{h,
  \delta}(G) = G$ for all $G \in \mGb_n$. We require $\fn_{h, \delta}$ to satisfy
the standard prefix--free condition. 
In addition to this, we want $\fn_{h,
  \delta}$ to be universally optimal from an information theoretic perspective. More precisely, if a sequence
of simple marked graphs $\Gn$ is given which converges to a limit $\mu \in
\mP(\mG_*)$ in the local weak sense as was described in Section~\ref{sec:prelim-lwc}, i.e.\ $U(G_n) \Rightarrow \mu$, then we want the asymptotic
codeword length $\nat(\fn_{h, \delta}(\Gn))$, after normalization
{
  and sending $n$, $\delta$, and $h$ to infinity in that order}, to be no more
than the marked BC entropy of the limit $\mu$, as was defined in Section~\ref{sec:marked-BC-entropy}. 
Motivated by the definition of the marked BC entropy in
Section~\ref{sec:marked-BC-entropy}, the correct normalization is to consider
$(\nat(\fn_{h, \delta}(\Gn)) - \mn \log n)/n$,
where $\mn$ is the total number of edges in $\Gn$. 
Here, in order to address universality, we assume that the encoder does not know
the limit $\mu$, and only gets to see the simple marked graph $\Gn$ as its input.

We assume that a simple marked graph $\Gn$ is given to the encoder by
representing $(1)$ its vertex mark sequence $\vthetan = (\thetan_v: v \in [n])$,
where $\thetan_v \in \vermark$ is the mark of the vertex $v$ in $\Gn$, i.e.\
$\thetan_v = \tau_{\Gn}(v)$, and $(2)$ the
list of 
marked
\color{black}
edges in $\Gn$ of the form $(v_i, w_i, x_i, x'_i)$ for $1 \leq i \leq
\mn$, where $\mn$ denotes the total number of edges in $\Gn$, and for $1 \leq i
\leq \mn$, the tuple $(v_i, w_i, x_i, x'_i)$ represents an edge between the
vertices $v_i$ and $w_i$, with mark $x_i$ towards $v_i$ and mark $x'_i$ towards
$w_i$, i.e.\ $\xi_{\Gn}(w_i, v_i) = x_i$ and $\xi_{\Gn}(v_i, w_i) = x'_i$. We
assume that each 
marked
\color{black}
edge is represented only once in this list, but  in an arbitrary
orientation, i.e.\ either $v_i < w_i$ or $w_i < v_i$. 
We call this form of representing $\Gn$ the \emph{``edge list''}
representation.

Our main contribution in this paper is to introduce  a compression
algorithm, which  $(1)$ is universally optimal from an information theoretic
perspective,
and $(2)$ has a low computational complexity
for both encoding and decoding.
\color{black}
Theorem~\ref{thm:optimality-complexity-main} below formally states 
what this means.
\color{black}
Before
stating the theorem, we need a few definitions.  

For a marked graph
$G$, on a finite or countably infinite vertex set, and adjacent vertices $u$ and $v$ in $G$, 
recall that
\color{black}
we define $G(u,v)$ to be the pair
$(\xi_G(u,v), (G',v))$ where  $G'$ is the connected component of $v$ in the graph
obtained from $G$ by removing the edge between $u$ and $v$. Similarly, for
 $h \geq 0$, $G(u,v)_h$ is defined as $(\xi_G(u,v), (G',v)_h)$.
See Figure~\ref{fig:Guv} for an example.
Let $G[u,v]$ denote the pair $(\xi_G(u,v), [G',v])$, so that $G[u,v] \in \edgemark \times \mGb_*$. Likewise, for $h \geq 0$, let $G[u,v]_h$ denote
$(\xi_G(u,v), [G',v]_h)$, so that $G[u,v]_h \in \edgemark \times \mGb_*^h$.

\begin{figure}
  \centering
  \begin{tikzpicture}
  \begin{scope}[xshift=-5cm]
  \node[nodeR,label={[label distance=1mm]90:1}] (n1) at (0,0) {};
  \node[nodeB,label={[label distance=1mm]180:2}] (n2) at (-1.5,-1.5) {};
  \node[nodeB,label={[label distance=1mm]0:3}] (n3) at (1.5,-1.5) {};
  \node[nodeR,label={[label distance=1mm]180:4}] (n4) at (-1.5,-3.7) {};
  \node[nodeR,label={[label distance=1mm]0:5}] (n5) at (1.5,-3.7) {};

  \drawedge{n1}{n3}{B}{O}
  
  \drawedge{n1}{n2}{B}{O}

  \drawedge{n2}{n4}{B}{O}

  \drawedge{n4}{n5}{B}{B}

  \drawedge{n3}{n5}{B}{B}

  \drawedge{n2}{n5}{O}{B}

  \node at (0,-4.7) {$(a)$};
\end{scope}

\begin{scope}
  \node[nodeB] (n3) at (0,0) {};
  \nodelabel{n3}{0}{3};
  \node[nodeR] (n5) at (0,-1.5) {};
  \nodelabel{n5}{0}{5};
  \node[nodeB] (n2) at (-0.9,-2.6) {};
  \nodelabel{n2}{180}{2};
  \node[nodeR] (n4) at (0.9,-2.6) {};
  \nodelabel{n4}{0}{4};
  \node[nodeR] (n1) at (0,-3.7) {};
  \nodelabel{n1}{0}{1};

  \drawedge{n3}{n5}{B}{B}
  \drawedge{n5}{n2}{B}{O}
  \drawedge{n5}{n4}{B}{B}
  \drawedge{n2}{n1}{O}{B}
  \drawedge{n2}{n4}{B}{O}

  \draw[edgeO] ($(n3)+(0,0.6)$) -- (n3);
  \node at (0,-4.7) {$(b)$};
\end{scope}

\begin{scope}[xshift=5cm]
  \node[nodeB] (n3) at (0,0) {};
  \nodelabel{n3}{0}{3};
  \node[nodeR] (n5) at (0,-1.5) {};
  \nodelabel{n5}{0}{5};
  \node[nodeB] (n2) at (-0.9,-2.6) {};
  \nodelabel{n2}{180}{2};
  \node[nodeR] (n4) at (0.9,-2.6) {};
  \nodelabel{n4}{0}{4};

  \drawedge{n3}{n5}{B}{B}
  \drawedge{n5}{n2}{B}{O}
  \drawedge{n5}{n4}{B}{B}
  \drawedge{n2}{n4}{B}{O}

  \draw[edgeO] ($(n3)+(0,0.6)$) -- (n3);
  \node at (0,-4.7) {$(c)$};
\end{scope}
  
\end{tikzpicture}

  \caption[Marked Graph]{$(a)$ A marked graph $G$ on the vertex set $\{1, \dots, 5\}$
    with vertex mark set $\vermark = \{\tikz{\node[nodeB] at (0,0) {};}, \tikz{\node[nodeR] at (0,0)
  {};}\}$ and
    edge mark set $\edgemark =
  \{\text{\color{blueedgecolor} Blue (solid)}, \text{\color{orangeedgecolor}
    Orange (wavy)} \}$. In $(b)$, $G(1,3)$ is
    illustrated where the first component $\xi_G(1,3)$ is depicted as a half--edge with the corresponding mark going towards the root $3$,  and
  $(c)$ illustrates $G(1,3)_2$. Note that $G(1,3)$ can be interpreted as cutting
the edge between $1$ and $3$ and leaving the half--edge connected to $3$ in
place. Note that, in constructing $G(u,v)$, although we are removing
the edge between $u$ and $v$, it might be the case that $u$ is still reachable
from $v$ 
through another walk,
\color{black}
as is the case in the above example in $(b)$.}
  \label{fig:Guv}
\end{figure}

For $g \in \edgemark \times \mGb_*$, we call the $\edgemark$ component of $g$
its {\em mark component} and denote it by $g[m]$. Moreover, we call the $\mGb_*$
component of $g$ its {\em subgraph component} and denote it by $g[s]$.
Given a marked graph $G$ and adjacent vertices $u$ and $v$ in $G$, and
for $g \in \edgemark \times \mGb_*$,  we write $G(u,v) \equiv g$ to denote that
$\xi_G(u,v) = {g[m]}$ and also $(G',v)$ falls in the isomorphism class {$g[s]$}.
We define the expression $G(u,v)_h \equiv g$ for $g \in \edgemark \times
\mGb_*^h$ in a similar fashion. 

For positive integers $h$ and $\delta$, by treating $\mTb_*^h$ as a subset of
$\mGb_*^h$, we define $\mFdeltah$ to be the set of $t = (t[m], t[s]) \in \edgemark \times
\mTb_*^{h-1}$ such that in the subtree component of $t$, i.e.\ $t[s]$, the
degree of the root is strictly less than $\delta$, and the degree of all other
vertices is at most $\delta$. Moreover, for $[T,o] \in \mTb_*$, we define
\begin{equation}
\label{eq:deg-star-deg}
  \deg_\star^{(h, \delta)}([T,o]) := |\{ v \sim_T o: T[v,o]_{h-1} \notin \mFdeltah \text{ or } T[o,v]_{h-1} \notin \mFdeltah \text{ or } \deg_T(o) > \delta \text{ or } \deg_T(v) > \delta\}|.
\end{equation}

Note that the degree at the root for any $[T,o] \in \mTb_*^0$ is zero, meaning
that $\mF^{(\delta, 1)} = \edgemark \times \mTb_*^0$ for any $\delta\geq 1$.
The condition
``$\deg_T(o) > \delta \text{ or } \deg_T(v) > \delta$"
in the definition of 
$\deg_\star^{(h, \delta)}([T,o])$ therefore plays a nontrivial role when $h=1$.
\color{black}


\begin{thm}
  \label{thm:optimality-complexity-main}
There exists a compression/decompression algorithm with encoding and decoding
functions $\fn_{h,\delta}$ and $\gn_{h, \delta}$ as defined above, which has the
following properties:
\begin{enumerate}
  \item \textbf{(Optimality)}
  Assume that $\mu \in \mP_u(\mTb_*)$ with $\deg(\mu) \in (0,\infty)$ is given such
  that $\evwrt{\mu}{\deg_T(o) \log \deg_T(o)} < \infty$ and $\bch(\mu) >
  -\infty$. Also, assume that 
   a sequence $\Gn$ of simple marked graphs is given such that $U(\Gn) \Rightarrow
  \mu$, and with $\mn$ being the total number of edges in $\Gn$,  we have $\mn / n \rightarrow
  \deg(\mu)/2$. Then, for $h\geq 1$ and $\delta \geq 1$, we have 
  \begin{equation}
    \label{eq:opt-statement-eta-0-bound}
    \limsup_{n \rightarrow \infty} \frac{\nat(\fn_{h, \delta}(\Gn)) - \mn \log n}{n} \leq J_h(\mu_h) + \eta_0(\mu; h, \delta),
  \end{equation}
  where
  \begin{align*}
    \eta_0(\mu; h, \delta) &:= \eta_1(\mu; h, \delta)(|\edgemark|^2 \log 2 + \log |\vermark| + 2 - 2\log \eta_1(\mu; h, \delta))
                             + \eta_2(\mu; h, \delta) (1+\log 4) \\
                           &\qquad + \eta_3(\mu; h, \delta) + (1+2(|\edgemark| \times |\vermark|)^2)\eta_4(\mu; h, \delta), \\
    \eta_1(\mu; h, \delta) &:= \prwrt{\mu}{\deg_\star^{(h, \delta)}([T,o]) > 0}, \\
    \eta_2(\mu; h, \delta) &:= \frac{1}{2} \evwrt{\mu}{\deg_\star^{(h, \delta)}([T,o])}, \\
    \eta_3(\mu; h, \delta) &:=
                             \begin{cases}
                               \frac{1}{2} \evwrt{\mu}{\deg_\star^{(h, \delta)}([T,o]) \log \frac{1}{\mu_h([T,o]_h)}},  & \text{if } h > 1; \\
                               \frac{1}{2} \evwrt{\mu}{\deg_\star^{(h, \delta)}([T,o]) \log \frac{1}{\mu_2([T,o]_2)}}, & \text{if } h = 1,
                             \end{cases}
                              \\
    \eta_4(\mu; h, \delta) &:=  \evwrt{\mu}{\deg_\star^{(h, \delta)}([T,o]) \log \deg_T(o)}. 
  \end{align*}
  Moreover, with fixed $h$, as $\delta \rightarrow \infty$, $\eta_i(\mu; h,
  \delta) \rightarrow 0$ for $i \in \{0, 1, 2, 3, 4\}$, and we have 
  \begin{equation}
    \label{eq:optimality-thm-statement-delta-n}
    \limsup_{\delta \rightarrow \infty} \limsup_{n \rightarrow \infty} \frac{\nat(\fn_{h, \delta}(\Gn)) - \mn \log n}{n} \leq J_h(\mu_h).
  \end{equation}
  Furthermore,
  \begin{equation}
    \label{eq:optimality-thm-statement}
    \limsup_{h \rightarrow \infty} \limsup_{\delta \rightarrow \infty} \limsup_{n \rightarrow \infty} \frac{\nat(\fn_{h, \delta}(\Gn)) - \mn \log n}{n} \leq \bch(\mu).
  \end{equation}
\item \textbf{(Computational Complexity)}  The time complexity of encoding a
  graph $\Gn$ with $\mn$ many edges using our compression algorithm is $O((\mn +n ) h \delta \log n (\log n +
\log(|\edgemark| + |\vermark|)) + n \delta^2 \log^4 n \log \log n +
|\edgemark|^2 (n+\mn (\log n + \log |\edgemark|)))$. Also, the
memory complexity of the compression phase is $O((\mn + \delta)(\log n+ \log(|\edgemark| + |\vermark|)) + n \delta \log
|\edgemark| + n \delta^2 \log n)$. Moreover, the time complexity of decoding for
this graph using our decompression algorithm is
$O(\mn(\log |\edgemark| +
\log n ) + n \delta \log \delta \log |\edgemark| + n \delta^2 \log^5 n \log \log n +
|\edgemark|^2 n + \mn \log^2 \mn)$.
Furthermore, the memory
complexity of the decompression phase is  $O((\mn + n \delta)
(\log |\edgemark| + \log n ) + \mn \log \mn + n \delta^2 \log n)$.
\end{enumerate}
\end{thm}

{
The proof of the first part of
  Theorem~\ref{thm:optimality-complexity-main} (optimality) is given in
  Section~\ref{sec:optimality-proof}, and the complexity (second part of the
  theorem) is discussed in Section~\ref{sec:alg-details}. More specifically, the
complexity of  encoding is discussed in Section~\ref{sec:enc-complexity}, and
the complexity of decoding is discussed in
Section~\ref{sec:main-decode-complexity}.} 

The following corollary simplifies the computational complexity associated to
our algorithm in the special case that $\mn = \Theta(n)$ and all the other parameters are
constant and do not grow in $n$. This is to study the effect of $n$, the number
of vertices, on the complexity of our algorithm. 

\begin{cor}
  \label{cor:complexity-simplified}
  If $\mn = \Theta(n)$, and $h$, $\delta$, $|\vermark|$, and $|\edgemark|$ are all
  constants not growing with $n$, then the time and memory complexities of our compression algorithm are
  $O(n \log^4 n \log \log n)$ and $O(n \log n)$, respectively. Moreover, the
  time and memory complexities of our decompression algorithm are $O(n \log^5 n
  \log \log n)$ and $O(n \log n)$, respectively. 
\end{cor}

\begin{rem}
  \label{rem:optimal-up-to-logarithms}
  Note that since the graph is given to our algorithm in its edge list form, the time
  complexity of reading the graph from the input is $\Omega(\mn \log n)$. Therefore,
  motivated by Corollary~\ref{cor:complexity-simplified},
  the dependence of the complexity of our algorithm in $n$, the number of
  vertices, is optimal up to logarithmic factors. 
\end{rem}

First, in Section~\ref{sec:algorithm-general} below, we explain the steps of our
algorithm without going into the details. The detailed steps of the algorithm are
presented in Section~\ref{sec:alg-details} together with its complexity
analysis.
More precisely, Sections \ref{sec:enc-alg} and \ref{sec:enc-complexity} discuss
the details of the compression algorithm and its complexity respectively.
Also, Sections \ref{sec:dec-alg} and \ref{sec:main-decode-complexity} discuss
the details of the decompression algorithm and its complexity respectively. 


\editfinish


\editstart

\section{Steps of the Universal Compression Algorithm}
\label{sec:algorithm-general}

\ifreplace
In this section, we give an overview of the steps of our universal compression
algorithm.
\color{black}
\fi
The details of the algorithm will
be presented in Section~\ref{sec:alg-details}.
Throughout this section, we use the graph in
Figure~\ref{fig:sample-graph-for-compression} as a sample input  to illustrate
the steps of our
algorithm. Recall from Section~\ref{sec:Problem-statement} that we assume edge
and vertex marks that are integer values, i.e. $\edgemark = \{1, \dots,
|\edgemark|\}$ and $\vermark = \{1, \dots, |\vermark|\}$. Following this convention, in the
graph of Figure~\ref{fig:sample-graph-for-compression}, we have $\edgemark =
\{\xone, \xtwo\}$ and $\vermark = \{\tone, \ttwo\}$. 



\begin{figure}
  \centering
  \begin{tikzpicture}
    \node[nodeR, label={[label distance=1mm]54:1}] (n1) at (0,0) {};
    \node[nodeB, label={[label distance=1mm]90:2}] (n2) at (90:2) {};
    \node[nodeB, label={[label distance=1mm]90:3}] (n3) at (162:2) {};
    \node[nodeB, label={[label distance=1mm]90:4}] (n4) at (234:2) {};
    \node[nodeB, label={[label distance=1mm]90:5}] (n5) at (306:2) {};
    \node[nodeB, label={[label distance=1mm]90:6}] (n6) at (18:2) {};
    \foreach \i in {2,...,6}
    {
      \drawedge{n1}{n\i}{O}{B};
      \begin{scope}[shift={(n\i)}]
        \begin{scope}[rotate={90+72*(\i-2)}]
          \node[nodeR] (n1\i) at (45:1) {};
          \node at (45:1.4) {\pgfmathparse{2*\i+4} \pgfmathprintnumber{\pgfmathresult}};
          \node[nodeR] (n2\i) at (-45:1) {};
          \node at (-45:1.4) {\pgfmathparse{2*\i+3} \pgfmathprintnumber{\pgfmathresult}};
          \drawedge{n\i}{n1\i}{B}{B};
          \drawedge{n\i}{n2\i}{B}{B};
          \draw[edgeO] (n1\i) -- (n2\i);
        \end{scope}
      \end{scope}
    }
  \end{tikzpicture}
  \caption[Sample graph for demonstrating our compression algorithm]{The simple marked graph we use to demonstrate the  steps of our compression
    algorithm. Here, $\edgemark =
  \{\text{\color{blueedgecolor} Blue (solid)}, \text{\color{orangeedgecolor}
    Orange (wavy)} \} = \{\xone, \xtwo\}$ and $\vermark = \{\tikz{\node[nodeB] at (0,0) {};}, \tikz{\node[nodeR] at (0,0)
  {};}\} = \{\tone, \ttwo\}$.}
  \label{fig:sample-graph-for-compression}
\end{figure}

\subsection{Preprocessing}
\label{sec:alg-high-level-preprocessing}
Recall from Section~\ref{sec:Problem-statement} that the input graph $\Gn$ is given in its edge
list form, i.e.\ is represented by its
vertex mark sequence $\vthetan$ and a list of marked edges. 
First, we go through a  preprocessing step to convert the edge list  representation to the
following form which is convenient for our algorithm. We call this
representation the \emph{``neighbor list''} representation of $\Gn$, which
consists of the following components:
\begin{enumerate}
\item The sequence $\vthetan = (\thetan_v: v \in [n])$, where $\thetan_v$
  denotes the vertex mark of $v$. 
\item A sequence $\vdn = (\dnn_v: v \in [n])$, such that $\dnn_v$ for $1 \leq v
  \leq n$ is the degree of vertex $v$ in $\Gn$. 
\item For $1 \leq v \leq n$, the neighbors of vertex $v$ in $\Gn$  in
  an increasing order of the form $1 \leq \gamman_{v,1} < \gamman_{v,2} < \dots <
  \gamman_{v,\dnn_v} \leq n$. Furthermore, for $1 \leq v \leq n$ and $1 \leq i \leq
  d_v$, $\xn_{v,i}$ and $\xnp_{v,i}$ denote the two edge marks corresponding to the
  edge connecting $v$ to $\gamman_{v,i}$, so that $\xn_{v,i} =
  \xi_{\Gn}(\gamman_{v,i}, v)$ and $\xnp_{v,i} = \xi_{\Gn}(v, \gamman_{v,i})$.
{\color{peditcolor}Moreover, let $\vec{\gamma}^{(n)} := (\gamman_{v,i}: 1 \leq v
  \leq n, 1 \leq i \leq \dnn_v)$.}
  
  
\item For $1 \leq v \leq n$ and $1 \leq i \leq \dn_v$, $\tgamman_{v,i}$ denotes
  the index of $v$ among the neighbors of $\gamman_{v,i}$, so that
  $\gamman_{\gamman_{v,i}, \tgamman_{v,i}} = v$.  {\color{peditcolor}Moreover, let $\vec{\tilde{\gamma}}^{(n)} := (\tgamman_{v,i}: 1 \leq v
  \leq n, 1 \leq i \leq \dnn_v)$.}
\end{enumerate}
For instance, for the graph in Figure~\ref{fig:sample-graph-for-compression},
the neighbor list for vertex $5$ is $(\gamma^{(16)}_{5,i}: i \in [3]) = (1,13,14)$.
Moreover,
$(x^{(16)}_{5,i}: i \in [3]) = (\xone, \xone, \xone)$,
while
$({x'}^{(16)}_{5,i}:i \in [3]) = (\xtwo, \xone, \xone)$.
Also, $(\tilde{\gamma}^{(16)}_{5,i}: i \in [3]) = (4,1,1)$.

\editfinish

\editstart

\subsection{Definition of Edge Types}
\label{sec:alg-high-level-edge-type-def}

Our algorithm uses two integers $h \geq 1$ and $\delta \geq 1$ as hyperparameters.
\ifreplace
We
assume that they are fixed. 
\color{black}
\fi
With this, let $\mFdeltah \subset
\edgemark \times \mTb_*^{h-1}$ be the set consisting of all $t = (t[m],
  t[s]) \in \edgemark \times \mTb_*^{h-1}$ such that  in the subtree
  component of $t$, i.e.\ $t[s]$, the degree of the root is strictly less than $\delta$, and
  the degree of all other vertices is at most $\delta$. 
Moreover, for $x \in \edgemark$,
  let $\star_x$ be fictitious distinct elements not present  in $\mFdeltah$, and
  define $\mFbardeltah := \mFdeltah \cup \{ \star_x: x \in \edgemark\}$.
Note that $\star_x$ for $x \in \edgemark$ are auxiliary objects,
    and are not of the form of a pair of a mark and a rooted tree.
  Also,
  define $\mCdeltah := \mFdeltah \times \mFdeltah$ and $\mCbardeltah := \mFbardeltah
  \times \mFbardeltah$.

  For adjacent vertices $v \sim_{\Gn} w$ in $\Gn$, we define
  \begin{equation}
    \label{eq:t-v-w}
    \T{v}{w} := (\UC_v(\Gn))[w,v]_{h-1},
  \end{equation}
  which is indeed a member of $\edgemark \times \mTb_*^{h-1}$. Moreover, define
  \begin{equation}
    \label{eq:tilde-t-v-w}
    \TT{v}{w} :=
    \begin{cases}
      \T{v}{w} & \T{v}{w} \in \mFdeltah, \T{w}{v} \in \mFdeltah, \deg_{\Gn}(v) \leq \delta, \text{ and } \deg_{\Gn}(w) \leq \delta \\
      \star_{\xi_{\Gn}(w,v)} & \text{otherwise}
    \end{cases}
  \end{equation}
If $h > 1$, $\T{v}{w} \in \mFdeltah$ automatically implies that
$\deg_{\Gn}(v) \leq \delta$.
The reason is that $\deg_{\Gn}(v)$ is one plus the degree at the root in the
subgraph component of $(\UC_v(\Gn))[w,v]_{h-1}$.
However, this is not the case for $h = 1$. This is
why the two conditions $\deg_{\Gn}(v) \leq \delta$ and $\deg_{\Gn}(w)
\leq \delta$ in the above definition are not degenerate. 
In fact, the degree at the root for any $[T,o] \in \mTb_*^0$ is zero, meaning
that $\mF^{(\delta, 1)} = \edgemark \times \mTb_*^0$ for any $\delta\geq 1$.

  Observe that, by definition, $\TT{v}{w} \in \mFbardeltah$. 
  Furthermore, we define
  \begin{equation}
    \label{eq:type-v-w}
    \type{v}{w} := (\TT{v}{w} , \TT{w}{v})  \in \mCbardeltah,
  \end{equation}
  and call it the ``type'' of the edge $(v,w)$. Note that we have $\type{v}{w} =
  \overline{\type{w}{v}}$, where $\overline{(a,b)} := (b,a)$. The notion of edge
  types plays
  a crucial role in our compression algorithm.

  For instance, for the graph in Figure~\ref{fig:sample-graph-for-compression},
  with $h=2$, we have 
  \begin{equation*}
    t^{(16)}_2(1,6) =
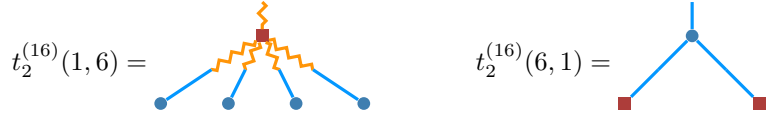
  \begin{tikzpicture}[scale=0.9,baseline=(pivot.base)]
    \node[nodeR] (r) at (0,0) {};
    \node (pivot) at (0,-0.5) {};
    \node[nodeB] (n1) at (-1.5,-1) {};
    \node[nodeB] (n2) at (-0.5,-1) {};
    \node[nodeB] (n3) at (0.5,-1) {};
    \node[nodeB] (n4) at (1.5,-1) {};
    \drawedge{r}{n1}{O}{B};
    \drawedge{r}{n2}{O}{B};
    \drawedge{r}{n3}{O}{B};
    \drawedge{r}{n4}{O}{B};
    \draw[edgeO] ($(r)+(0,0.5)$) -- (r);
  \end{tikzpicture}
  \qquad \qquad 
  t^{(16)}_2(6,1) =
  \begin{tikzpicture}[scale=0.9,baseline=(pivot.base)]
    \node[nodeB] (r) at (0,0) {};
    \node (pivot) at (0,-0.5) {};
    \node[nodeR] (n1) at (-1,-1) {};
    \node[nodeR] (n2) at (1,-1) {};
    \drawedge{r}{n1}{B}{B};
    \drawedge{r}{n2}{B}{B};
    \draw[edgeB] ($(r)+(0,0.5)$) -- (r);
  \end{tikzpicture}
  \end{equation*}
If $\delta = 4$, then $t^{(16)}_2(1,6) \notin \mF^{(4,2)}$. Therefore, we have
$\tilde{t}^{(16)}_{2,4}(1,6) = \star_{\xtwo}$. Note that although $t^{(16)}_2(6,1)
\in \mF^{(4,2)}$, we have $\tilde{t}^{(16)}_{2,4}(6,1) = \star_{\xone}$.

  Note that, based on~\eqref{eq:type-v-w}, both $\UC_v(\Gn)$ and $\UC_w(\Gn)$
  appear in the definition of $\type{v}{w}$. However, using the following lemma,
  we can give an equivalent expression for $\type{v}{w}$ which only depends on
  $\UC_v(\Gn)$ and is more convenient to work with.  The proof of the following Lemma~\ref{lem:UCv-UCu} is given in Appendix~\ref{sec:uc-lemma-proof}.

  \begin{lem}
    \label{lem:UCv-UCu}
    Given a simple marked graph $G$ and adjacent vertices $u \sim_G v$, we
    have $\UC_v(G)[v,u] = \UC_u(G)[v,u]$.
  \end{lem}
  Using this, with $t = \UC_v(\Gn)[w,v]_{h-1}$ and $t' = \UC_v(\Gn)[v,w]_{h-1}$,
  we have
  \begin{equation}
    \label{eq:type-v-w-equivalent}
    \type{v}{w} = 
    \begin{cases}
      (t,t') & t \in \mFdeltah,  t' \in \mFdeltah, \deg_{\Gn}(v) \leq \delta, \text{ and } \deg_{\Gn}(w) \leq \delta \\
      (\star_{\xi_{\Gn}(w,v)}, \star_{\xi_{\Gn}(v,w)}) & \text{otherwise}
    \end{cases}
  \end{equation}
This equivalent form will be useful in our future analysis. 


  \subsection{Finding Edge Types}
\label{sec:main-find-edge-type}

Our next step is to find $\type{v}{w}$ for adjacent
  vertices $v$ and $w$ in $\Gn$. 
\ifreplace
The algorithm returns an
array $\vec{c} = (c_{v,i}: v \in [n], i \in [\dn_v])$ where $c_{v,i}$ is a pair of
integers
representing
  $\TT{v}{\gamman_{v,i}}$ and $\TT{\gamman_{v,i}}{v}$ respectively.
\color{black}
\fi
In
  addition to the array $\vec{c}$, our algorithm returns the following:
  \begin{enumerate}
  \item $\mathsf{TCount}$: the range of integers showing up in $\vec{c}$, so
    that for $v \in [n]$ and $i \in [\dn_v]$, $c_{v,i}$ is a pair of integers,
    each in the range $\{1, \dots, \mathsf{TCount}\}$. 
\ifreplace
In Section~\ref{sec:MP} we will see that
\color{black}
\fi
    there is a one to one mapping from a subset of
    $\mFbardeltah$ to the set of
    integers $\{1, \dots, \mathsf{TCount}\}$.
    More precisely, with
    \begin{equation}
      \label{eq:mTn-deg}
      \begin{aligned}
        \mTn &:= \{\TT{v}{w}: v \sim_{\Gn} w \} \cup \{\T{v}{w} : (v,w) \in \mBn\},
        \end{aligned}
      \end{equation}
      and
      \begin{align*}
        \mBn &:= \{(v,w): v \sim_{\Gn} w, \T{v}{w} \in \mFdeltah \text{ and } (\T{w}{v} \notin \mFdeltah \text{ or } \deg_{\Gn}(v) > \delta \text{ or } \deg_{\Gn}(w) > \delta)\},
      \end{align*}
    there exists a one to one mapping $J_n: \mTn \rightarrow \{1, \dots,
    \mathsf{TCount}\}$. With this, for all $v \in [n]$ and $i \in [\dn_v]$, we have
    \begin{equation}
      \label{eq:mp-cvi-Jn-high-level-section}
    c_{v,i} = (J_n(\TT{v}{\gamman_{v,i}}), J_n(\TT{\gamman_{v,i}}{v})).
    \end{equation}
    See Lemma~\ref{lem:message-passing-c-J-Tn-relation} in Section~\ref{sec:MP} for a proof of the
    existence of this mapping. Moreover, as we will see in this
    Lemma~\ref{lem:message-passing-c-J-Tn-relation}, we have $\mathsf{TCount}
    \leq 4\mn$.

  \item An array $\mathsf{TIsStar} = (\mathsf{TIsStar}(i): 1 \leq i \leq
    \mathsf{TCount})$, where $\mathsf{TIsStar}(i)$ for $1 \leq i \leq
    \mathsf{TCount}$ is 1 if the member of $\mFbardeltah$ corresponding to $i$,
    i.e.\ $J_n^{-1}(i)$,  is
    of the form $\star_x$ for some $x \in \edgemark$, and 0 otherwise.
  \item An array $\mathsf{TMark} = (\mathsf{TMark}(i): 1 \leq i \leq
    \mathsf{TCount})$, where  for $1 \leq i \leq
    \mathsf{TCount}$, with $t = J_n^{-1}(i)$ being the member in $\mFbardeltah$ corresponding
    to $i$, if $t$ is of the form $\star_x$ for some $x \in \edgemark$, we have
    $\mathsf{TMark}(i) = x$; otherwise, we have $\mathsf{TMark}(i) = t[m]$, i.e.\
    the mark component associated to $t \in \mFdeltah$. 
  \end{enumerate}
  

Note that since we want to find integer representations for $\TT{v}{w}$ for $v
\sim_{\Gn} w$, we ideally would expect the set $\mTn$ above to only consist of
$\TT{v}{w}$ for $v \sim_{\Gn} w$. However, it turns out our algorithms
encounters some extra objects in the process of exploring edge types, and these
extra objects are precisely $\T{v}{w}$ for $(v,w) \in \mBn$.

  To summarize, we prove the following result in Section~\ref{sec:MP} by
  introducing Algorithm~\ref{alg:type-extract-message-passing-new} therein.

    \begin{prop}
    \label{prop:MP}
    There is an algorithm that given a simple marked graph $\Gn$ 
and
    parameters $\delta \ge 1$ and $h \ge 1$,
\color{black}
finds integer representations of
    $\type{v}{w}$ for all $v \sim_{\Gn} w $, i.e.\ $\vec{c} = (c_{v,i}: v \in
    [n], i\in [\dn_v])$ satisfying~\eqref{eq:mp-cvi-Jn-high-level-section}, together with 
    $\mathsf{TCount}$, $\mathsf{TIsStar}$, and $\mathsf{TMark}$ as explained
    above.  The algorithm runs in $O((\mn+n) h \delta \log n (\log n + \log(|\edgemark| + |\vermark|)))$
 time and requires $O((\mn +\delta)( \log n +  \log (|\edgemark| + |\vermark|)))$ bits of  memory, where $\mn$ denotes the
    number of edges in  $\Gn$.
  \end{prop}

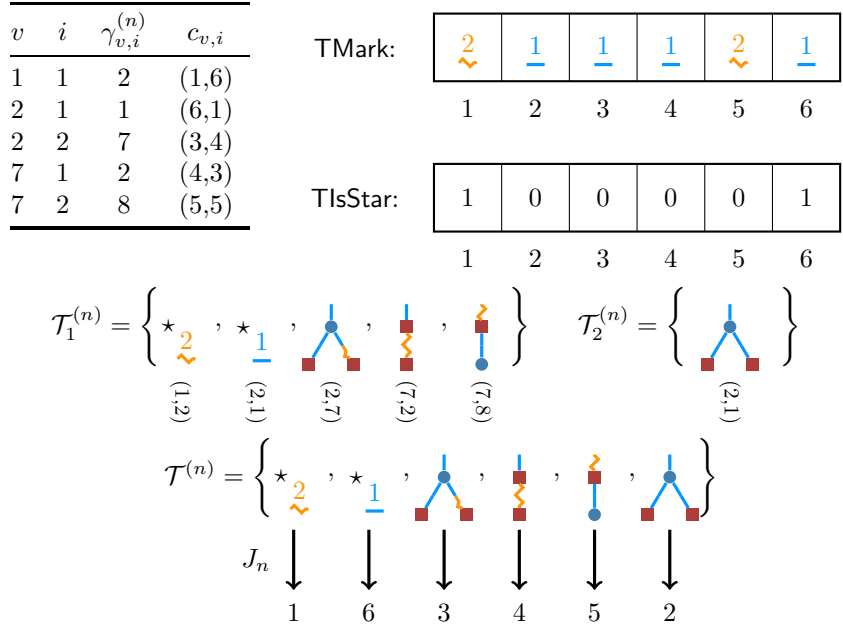
\begin{figure}
  \centering
      \begin{tabular}[t]{c@{\qquad}c}
  \begin{tabular}{@{}lcccr@{}}
\toprule
 $v$& $i$   &$\gamman_{v,i}$  &  $c_{v,i}$ \\ \midrule
 1 &  1 &  2 &  (1,6)\\
    2 &  1 &  1 &  (6,1)\\
    2 & 2 & 7 &   (3,4)\\
    7 & 1 & 2 &  (4,3)\\
    7 & 2 & 8 &  (5,5)\\
    \bottomrule
  \end{tabular}%
    &%
      \begin{tikzpicture}[baseline=(pivot.center)]
        \node (pivot) at (0,-1) {};
        \begin{scope}
          \node at (0,0) {$\mathsf{TMark}$:};
          \begin{scope}[xshift=1.5cm,scale=0.9]
            \draw[thick] (-0.5,-0.5) rectangle (5.5,0.5);
            \draw (0.5,-0.5) -- (0.5,0.5) (1.5,-0.5) -- (1.5,0.5) (2.5,-0.5) -- (2.5,0.5) (3.5,-0.5) -- (3.5,0.5) (4.5,-0.5) -- (4.5,0.5);
            \node at (0,0) {\xtwo};
            \node at (0,-0.9) {1};
            \node at (1,0) {\xone};
            \node at (1,-0.9) {2};
            \node at (2,0) {\xone};
            \node at (2,-0.9) {3};
            \node at (3,0) {\xone};
            \node at (3,-0.9) {4};
            \node at (4,0) {\xtwo};
            \node at (4,-0.9) {5};
            \node at (5,0) {\xone};
            \node at (5,-0.9) {6};
          \end{scope}
        \end{scope}
        \begin{scope}[yshift=-2cm]
          \node at (0,0) {$\mathsf{TIsStar}$:};
          \begin{scope}[xshift=1.5cm,scale=0.9]
            \draw[thick] (-0.5,-0.5) rectangle (5.5,0.5);
            \draw (0.5,-0.5) -- (0.5,0.5) (1.5,-0.5) -- (1.5,0.5) (2.5,-0.5) -- (2.5,0.5) (3.5,-0.5) -- (3.5,0.5) (4.5,-0.5) -- (4.5,0.5);
            \node at (0,0) {1};
            \node at (0,-0.9) {1};
            \node at (1,0) {0};
            \node at (1,-0.9) {2};
            \node at (2,0) {0};
            \node at (2,-0.9) {3};
            \node at (3,0) {0};
            \node at (3,-0.9) {4};
            \node at (4,0) {0};
            \node at (4,-0.9) {5};
            \node at (5,0) {1};
            \node at (5,-0.9) {6};
          \end{scope}
        \end{scope}

      \end{tikzpicture}

\end{tabular}

\begin{tikzpicture}

      \begin{scope}[xshift=-5cm]
      \node at (0,0) {$\mTn_1 =\Bigg \{$};
      \begin{scope}[xshift=1cm]
        \node at (0,-0.2) {$\star_{\xtwo}$};
      \end{scope}
      \begin{scope}[xshift=2cm]
        \node at (0,-0.2) {$\star_{\xone}$};
      \end{scope}
      \begin{scope}[xshift=3cm]
        \node[nodeB] (r) at (0,0) {};
        \node[nodeR] (r1) at (-0.3,-0.5) {};
        \node[nodeR] (r2) at (0.3,-0.5) {};
        \draw[edgeB] (r) -- ($(r)+(0,0.3)$);
        \drawedge{r}{r1}{B}{B}
        \drawedge{r}{r2}{B}{O}
      \end{scope}
      \begin{scope}[xshift=4cm]
        \node[nodeR] (r) at (0,0) {};
        \node[nodeR] (r1) at (0,-0.5) {};
        \draw[edgeB] (r) -- ($(r)+(0,0.3)$);
        \draw[edgeO] (r) -- (r1);
      \end{scope}
      \begin{scope}[xshift=5cm]
        \node[nodeR] (r) at (0,0) {};
        \node[nodeB] (r1) at (0,-0.5) {};
        \draw[edgeO] (r) -- ($(r)+(0,0.3)$);
        \drawedge{r}{r1}{B}{B}
      \end{scope}
      \foreach \x in {1.5,2.5,...,4.5}
      \node at (\x,0) {,};
      \node[rotate=-90,scale=0.8] at (1,-1) {(1,2)};
      \node[rotate=-90,scale=0.8] at (2,-1) {(2,1)};
      \node[rotate=-90,scale=0.8] at (3,-1) {(2,7)};
      \node[rotate=-90,scale=0.8] at (4,-1) {(7,2)};
      \node[rotate=-90,scale=0.8] at (5,-1) {(7,8)};
      
      \node at (5.5,0) {$\Bigg \}$};

    \end{scope}

    \begin{scope}[xshift=2cm]
      \node at (0,0) {$\mTn_2 =\Bigg \{$};
      \begin{scope}[xshift=1.3cm]
        \node[nodeB] (r) at (0,0) {};
        \node[nodeR] (r1) at (-0.3,-0.5) {};
        \node[nodeR] (r2) at (0.3,-0.5) {};
        \draw[edgeB] (r) -- ($(r)+(0,0.3)$);
        \drawedge{r}{r1}{B}{B}
        \drawedge{r}{r2}{B}{B}
      \end{scope}
      \node[rotate=-90,scale=0.8] at (1.3,-1) {(2,1)};
      
      \node at (2.1,0) {$\Bigg \}$};

    \end{scope}

  
    \begin{scope}[xshift=-3.5cm,yshift=-2cm]
      \node at (0,0) {$\mTn =\Bigg \{$};
      \begin{scope}[xshift=1cm]
        \node at (0,-0.2) {$\star_{\xtwo}$};
      \end{scope}
      \begin{scope}[xshift=2cm]
        \node at (0,-0.2) {$\star_{\xone}$};
      \end{scope}
      \begin{scope}[xshift=3cm]
        \node[nodeB] (r) at (0,0) {};
        \node[nodeR] (r1) at (-0.3,-0.5) {};
        \node[nodeR] (r2) at (0.3,-0.5) {};
        \draw[edgeB] (r) -- ($(r)+(0,0.3)$);
        \drawedge{r}{r1}{B}{B}
        \drawedge{r}{r2}{B}{O}
      \end{scope}
      \begin{scope}[xshift=4cm]
        \node[nodeR] (r) at (0,0) {};
        \node[nodeR] (r1) at (0,-0.5) {};
        \draw[edgeB] (r) -- ($(r)+(0,0.3)$);
        \draw[edgeO] (r) -- (r1);
      \end{scope}
      \begin{scope}[xshift=5cm]
        \node[nodeR] (r) at (0,0) {};
        \node[nodeB] (r1) at (0,-0.5) {};
        \draw[edgeO] (r) -- ($(r)+(0,0.3)$);
        \drawedge{r}{r1}{B}{B}
      \end{scope}
      \begin{scope}[xshift=6cm]
        \node[nodeB] (r) at (0,0) {};
        \node[nodeR] (r1) at (-0.3,-0.5) {};
        \node[nodeR] (r2) at (0.3,-0.5) {};
        \draw[edgeB] (r) -- ($(r)+(0,0.3)$);
        \drawedge{r}{r1}{B}{B}
        \drawedge{r}{r2}{B}{B}
      \end{scope}

      \foreach \x in {1.5,2.5,...,5.5}
      \node at (\x,0) {,};
      \node at (6.5,0) {$\Bigg \}$};

      \foreach \x in {1,...,6}
      \draw[->,very thick] (\x,-0.7) -- (\x,-1.5);
      \node at (0.5,-1.1) {$J_n$};
      \node at (1,-1.8) {1};
      \node at (2,-1.8) {6};
      \node at (3,-1.8) {3};
      \node at (4,-1.8) {4};
      \node at (5,-1.8) {5};
      \node at (6,-1.8) {2};
    \end{scope}
  \end{tikzpicture}
  \caption{The result of running the message passing algorithm to find edge
    types (i.e.\ Algorithm~\ref{alg:type-extract-message-passing-new} in
    Section~\ref{sec:MP}) for the graph of
    Figure~\ref{fig:sample-graph-for-compression} with parameters $h =2$ and
    $\delta = 4$. The array $\vec{c}$ is
    depicted on the top left table. Due to the symmetry in the graph, we have only
    presented a subset of $(v,i)$ pairs, e.g.\ $v=1, i \in \{2,\dots, 5\}$ is
    identical to $v=1,i=1$.
    The set $\mTn$ together with the mapping $J_n$ are shown.
Here, $\mTn_1$ denotes the set
  $\{\tilde{t}^{(n)}_{h,\delta}(v,w): v \sim_{\Gn} w\}$, while $\mTn_2$ denotes the
set $\{t^{(n)}_h(v,w): (v,w) \in \mBn\}$, so that $\mTn = \mTn_1 \cup \mTn_2$. For each element in $\mTn_1$ or
$\mTn_2$, we write one corresponding pair $(v,w)$ below that element (note that
there might be more than one pair resulting in that element). In this
example, we have $\mBn = \{(v,1): 2 \leq v \leq 6\}$. Also, we have $\mathsf{TCount} = 6$. The
    $\mathsf{TMark}$ and $\mathsf{TIsStar}$ arrays are also illustrated.
 \label{fig:message-passing-example-summary}}
\end{figure}

After we find the quantities $\mathsf{TCount}$, $\vec{c}$, $\mathsf{TIsStar}$,
and $\mathsf{TMark}$, we write them to the output bit sequence so that
the decoder can use them later during the decompression phase. Figure~\ref{fig:message-passing-example-summary} illustrates the result of running this algorithm
on the graph of Figure~\ref{fig:sample-graph-for-compression}.

  \subsection{Encoding Star Vertices}
\label{sec:main-star-vertices}

  We separately encode the edges in $\Gn$ which have a  type of the form $(\star_x,
  \star_{x'})$ for some $x, x' \in \edgemark$. But before that, we first
  encode $\Vns$, which is defined to be the set of vertices $v \in [n]$ such that for at least
  one of their neighbors $w \sim_{\Gn} v$, we have $\type{v}{w} =  (\star_x,
  \star_{x'})$ where $x = \xi_{\Gn}(w,v)$ and $\pp{x} = \xi_{\Gn}(v,w)$. 
  We call an edge $(v,w)$ with type $(\star_x, \star_{x'})$ for some $x, x' \in
  \edgemark$ a \emph{``star edge''}. Likewise, we call a  vertex $v \in \Vns$ a
  \emph{``star vertex''}. 
  Note that with our
  discussion above, a vertex $v \in [n]$ is in $\Vns$ iff for some $1 \leq i \leq
  d_v$, with $c_{v,i} = (t,t')$, we have either $\mathsf{TIsStar}(t) = 1$ or
  $\mathsf{TIsStar}(t') = 1$.
For instance, for the graph of Figure~\ref{fig:sample-graph-for-compression},
with $h=2$ and $\delta = 4$,
\color{black}
based on the edge types shown in Figure~\ref{fig:message-passing-example-summary}, the edges $(1,i)$ for $2 \leq
i \leq 6$ are star edges, and $\Vns = \{1, \dots, 6\}$.

  In order to encode $\Vns$, we first represent $\Vns$ using a bit sequence $\vy =
  (y_i: i \in [n])$ of size $n$,
  where $y_i = 1$ if $i \in \Vns$ and zero
  otherwise. Then we encode this bit sequence. Note that in principle, we can
  use classical methods such as Lempel--Ziv coding \cite{cover2012elements} to encode $\vy$. 
  However, here we introduce a different scheme based on one of our results that
  will be discussed later in the document.
  We choose this alternative method to simplify our analysis, since this method allows us to give a bound
  on the number of bits used to compress a sequence as well as its time and
  memory complexities. In the following Proposition~\ref{prop:encode-sequence}, we discuss a compression
  scheme in a slightly more general setting where each coordinate of the
  sequence can be an arbitrary nonnegative integer, a special case being a  
  zero-one
  sequence, such as the sequence $\vy$ representing $\Vns$ as above. This more
  general setup will be useful later when we discuss encoding vertex types in
  Section~\ref{sec:main-encode-vertex-types} below. We prove
  Proposition~\ref{prop:encode-sequence} in Section~\ref{sec:alg-details}. 

  \begin{prop}
  \label{prop:encode-sequence}
  There exists a compression algorithm that,
  given $n$ and
  a sequence $\vy = (y_1, \dots,
  y_{n})$ of nonnegative integers, produces a prefix--free bit sequence of size
  at most
  \begin{equation}
    \label{eq:encode-sequence-length-bound}
    \log_2 \left( 1 + \left\lceil \frac{n!}{\prod_{j=0}^{K-1} \phi_j !}\right \rceil \right) + O(K \log n).
  \end{equation}
  Here, $K = 1 + \max \{y_i: i \in [n]\}$  and $\phi_j,
  0 \leq j \leq K-1$, denotes the frequency of the integer $j$ in $\vy$.
  Furthermore, the time and memory complexities of this algorithm are $O(n
  \log^4 n\log \log n)$ and
  $O(n\log n)$ respectively. The corresponding decoding algorithm has the same
  bounds on the time and
  memory complexities.
\end{prop}


\subsection{Encoding Star Edges}
\label{sec:main-star-edges}

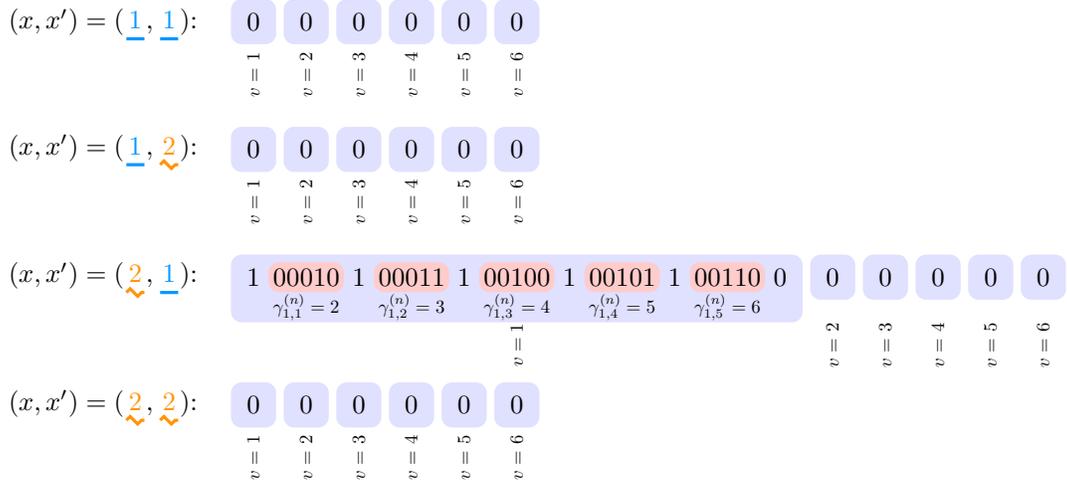
\begin{figure}
  \centering
  \begin{tikzpicture}
    \begin{scope}
      \node at (0,0) {$(x,\pp{x}) = (\xone, \xone)$:};
      \begin{scope}[xshift=2cm]
        \foreach \x/\i in {0/1,0.7/2,1.4/3,2.1/4,2.8/5,3.5/6}{
          \begin{scope}[xshift=\x cm]
            \fill[blue!20,rounded corners,opacity=0.6] (-0.3,-0.3) rectangle (0.3,0.3);
            \node at (0,0) {0};
            \node[scale=0.7,rotate=90] at (0,-0.7) {$v=$ \i};
          \end{scope}
        }
      \end{scope}
    \end{scope} 
    \begin{scope}[yshift=-1.7cm]
      \node at (0,0) {$(x,\pp{x}) = (\xone, \xtwo)$:};
      \begin{scope}[xshift=2cm]
        \foreach \x/\i in {0/1,0.7/2,1.4/3,2.1/4,2.8/5,3.5/6}{
          \begin{scope}[xshift=\x cm]
            \fill[blue!20,rounded corners,opacity=0.6] (-0.3,-0.3) rectangle (0.3,0.3);
            \node at (0,0) {0};
            \node[scale=0.7,rotate=90] at (0,-0.7) {$v=$ \i};
          \end{scope}
        }
      \end{scope}
    \end{scope} 

    \begin{scope}[yshift=-3.4cm]
      \node at (0,0) {$(x,\pp{x}) = (\xtwo, \xone)$:};
      \begin{scope}[xshift=2cm]
        \fill[blue!20,rounded corners,opacity=0.6] (-0.3,-0.6) rectangle (7.3,0.3);
        \begin{scope}[xshift=0.7cm]
          \fill[red!20, rounded corners] (-0.5,-0.2) rectangle (0.5,0.2);
          \node[scale=0.7] at (0,-0.4) {$\gamman_{1,1} = 2$};
        \end{scope}
        \begin{scope}[xshift=2.1cm]
          \fill[red!20, rounded corners] (-0.5,-0.2) rectangle (0.5,0.2);
          \node[scale=0.7] at (0,-0.4) {$\gamman_{1,2} = 3$};
        \end{scope}
        \begin{scope}[xshift=3.5cm]
          \fill[red!20, rounded corners] (-0.5,-0.2) rectangle (0.5,0.2);
          \node[scale=0.7] at (0,-0.4) {$\gamman_{1,3} = 4$};
        \end{scope}
        \begin{scope}[xshift=4.9cm]
          \fill[red!20, rounded corners] (-0.5,-0.2) rectangle (0.5,0.2);
          \node[scale=0.7] at (0,-0.4) {$\gamman_{1,4} = 5$};
        \end{scope}
        \begin{scope}[xshift=6.3cm]
          \fill[red!20, rounded corners] (-0.5,-0.2) rectangle (0.5,0.2);
          \node[scale=0.7] at (0,-0.4) {$\gamman_{1,5} = 6$};
        \end{scope}

        \node at (0,0) {1};
        \node at (0.7,0) {00010}; 
        \node at (1.4,0) {1};
        \node at (2.1,0) {00011}; 
        \node at (2.8,0) {1};
        \node at (3.5,0) {00100}; 
        \node at (4.2,0) {1};
        \node at (4.9,0) {00101}; 
        \node at (5.6,0) {1};
        \node at (6.3,0) {00110}; 
        \node at (7,0) {0};

        \node[scale=0.7,rotate=90] at (3.5,-0.9) {$v=1$};

        \foreach \x/\i in {7.7/2,8.4/3,9.1/4,9.8/5,10.5/6}{
          \begin{scope}[xshift=\x cm]
            \fill[blue!20,rounded corners,opacity=0.6] (-0.3,-0.3) rectangle (0.3,0.3);
            \node at (0,0) {0};
            \node[scale=0.7,rotate=90] at (0,-0.9) {$v=$ \i};
          \end{scope}
        }
      \end{scope}
    \end{scope} 

    \begin{scope}[yshift=-5.1cm]
      \node at (0,0) {$(x,\pp{x}) = (\xtwo, \xtwo)$:};
      \begin{scope}[xshift=2cm]
        \foreach \x/\i in {0/1,0.7/2,1.4/3,2.1/4,2.8/5,3.5/6}{
          \begin{scope}[xshift=\x cm]
            \fill[blue!20,rounded corners,opacity=0.6] (-0.3,-0.3) rectangle (0.3,0.3);
            \node at (0,0) {0};
            \node[scale=0.7,rotate=90] at (0,-0.7) {$v=$ \i};
          \end{scope}
        }
      \end{scope}
    \end{scope} 
    
  \end{tikzpicture}
  \caption[Example of encoding star edges]{Encoding star edges for the graph of
    Figure~\ref{fig:sample-graph-for-compression} with parameters $h=2$ and
    $\delta = 4$. Each line illustrates the bit sequence corresponding to a pair
    $(x,\pp{x}) \in \edgemark \times \edgemark$. For each such pair, we go over
    the vertices in $\Vns = \{1,2,\dots, 6\}$ in an increasing order. Note that all star edges in
    this example are encoded when we encounter the mark pair $(\xone, \xtwo)$. The reason is
    that, as we have discussed in Section~\ref{sec:main-star-edges}, in order to
    avoid representing edges twice, we encode
    a star edge $(v,\gamman_{v,i})$ for $v \in \Vns$ and $i \in [\dn_v]$ only when
    $v < \gamman_{v,i}$. Therefore, all star edges are encoded for $v = 1$ and
    $1 \leq i \leq 5$. As we can see, the index of the endpoint of a star edge,
    i.e.\ $\gamman_{v,i}$, is encoded using $1 + \lfloor \log_2 n \rfloor$ bits,
    which is 5 bits in this example. Once we finish encoding star edges
    connected to a vertex $v$ with mark pair $(x,\pp{x})$, we write a bit with value zero and
    move to the next vertex. \label{fig:encode-star-edge-example}}
\end{figure}

Next, we encode the star edges, i.e.\ those edges with type $(\star_x,
\star_{x'})$ for some $x, x' \in \edgemark$. Note that, due to the definition of
$\Vns$, both endpoints of a star edge are in $\Vns$. Motivated by this, for each pair of
edge marks $x,x' \in\edgemark$, we go through the vertices $v$ in $\Vns$ in an
increasing order and encode the edges connected to $v$ with type $(\star_x,
\star_{x'})$. In order to make sure that we do not represent an edge twice, we
only consider edges connected to $v$ such that the index of the other endpoint
is greater than $v$.  More precisely, for fixed $x, x' \in \edgemark$, and for
$v \in \Vns$, for each $1 \leq i \leq \dn_v$, if we have
$\type{v}{\gamman_{v,i}} = (\star_{x}, \star_{x'})$ and $\gamman_{v,i}
> v$, we write a bit with value 1 to the output
together with the binary representation of $\gamman_{v,i}$ using $1 +
\lfloor \log_2 n \rfloor$ bits. Note that in our notation we have
$\type{v}{\gamman_{v,i}} = (\star_{x}, \star_{x'})$ iff, with $c_{v,i}= (t,\tp)$, we
have $\mathsf{TIsStar}(t) = 1$, $\mathsf{TIsStar}(\tp) = 1$, $\xn_{v,i} = x$,
and $\xnp_{v,i} = x'$. Finally, when we finish checking all the neighbors
of $v$, we write a bit with value 0 to the output. This ensures a
prefix--free representation of star edges. Note that since the decoder
knows $\Vns$ from the previous step, it can unambiguously reconstruct the star
edges. Figure~\ref{fig:encode-star-edge-example} illustrates this procedure for
the graph of Figure~\ref{fig:sample-graph-for-compression}. 

\subsection{Encoding Vertex Types}
\label{sec:main-encode-vertex-types}

For $1 \leq v \leq n$, define $\Dn(v) := (\Dn_{t,t'}(v): t, t' \in \mFdeltah)$ where,
for $t, t' \in \mFdeltah$, we write
\begin{equation}
  \label{eq:Dn-def}
  \Dn_{t,t'}(v) := | \{ w \sim_{\Gn} v: \type{v}{w} = (t,t')\} |.
\end{equation}
See Figure~\ref{fig:Dn-example} for the value of 
$D^{(16)}_{t,\pp{t}}(v)$
\color{black}
for several
$v, t$, and $\pp{t}$ in the graph of
Figure~\ref{fig:sample-graph-for-compression},
where $h=2$ and $\delta = 4$.
\color{black}

\begin{figure}
\begin{center}
\scalebox{0.8}{
  \begin{tikzpicture}
    \begin{scope}[xshift=-5cm]
      \node[scale=1.5] at (0,0) {$D^{(16)}_{\tthree{0.5},\, \tfour{0.5}}(2) = 2$};
    \end{scope}
    \begin{scope}[xshift=0cm]
      \node[scale=1.5] at (0,0) {$D^{(16)}_{\tfour{0.5},\, \tthree{0.5}}(7) = 1$};
    \end{scope}
    \begin{scope}[xshift=5cm]
      \node[scale=1.5] at (0,0) {$D^{(16)}_{\tfive{0.5},\, \tfive{0.5}}(7) = 1$};
    \end{scope}
  \end{tikzpicture}
  }
\end{center}
\caption{$D^{(16)}_{t,\pp{t}}(v)$ for several values of $v$, $t$, and $\pp{t}$ in the
  graph of Figure~\ref{fig:sample-graph-for-compression} with parameters $h = 2$
and $\delta = 4$. Note that due to symmetry, $D^{(16)}(v)$ are the same for $v \in
\{2,\dots, 6\}$, and also for $v \in \{7, \dots, 16\}$. Moreover, since all
edges connected to vertex $v = 1$ are star edges, we have $D^{(16)}_{t,\pp{t}}(1) =
0$ for all $t, \pp{t} \in \mFdeltah$. \label{fig:Dn-example}}
\end{figure}



{
  Note that if for a vertex $v$, we have $\Dn_{t,t'}(v) > 0$
  for some $t, t' \in \mFdeltah$, then we have $\deg_{\Gn}(v) \leq \delta$. The
  reason is that $\Dn_{t,t'}(v) > 0$ means there exists $w \sim_{\Gn} v$ such
  that $\TT{v}{w} = t$ and $\TT{w}{v} = t'$. Since $t,t' \in \mFdeltah$,
  recalling~\eqref{eq:tilde-t-v-w}, this in particular means that we must have $\TT{v}{w} = \T{v}{w}$ and
  $\deg_{\Gn}(v) \leq \delta$.}
This implies that 
  \begin{equation}
    \label{eq:sum-Dn-ttp-delta}
    \sum_{t, \pp{t} \in \mFdeltah} \Dn_{t,\pp{t}}(v)  \leq \delta \qquad \forall v \in [n].
  \end{equation}
  Note that if for a vertex $v \in [n]$, we have $\Dn_{t,\pp{t}}(v) = 0$ for all
   $t,\pp{t} \in \mFdeltah$, then the above 
   inequality automatically holds for that vertex. 
  In particular, \eqref{eq:sum-Dn-ttp-delta} implies that 
  \begin{equation}
    \label{eq:Dn-bounded-delta}
    \Dn_{t,t'}(v) \leq  \delta \qquad \forall v \in [n], t,t' \in \mFdeltah.
  \end{equation}
We define the ``type'' of a vertex $v \in [n]$  to be the pair $(\thetan_v,
\Dn(v))$. 
    The next step in our compression algorithm is to encode vertex types, i.e.\ jointly encode the
  sequences $\vthetan$ and $\vDn = (\Dn(v): v \in [n])$. In order to do so, we
  will construct a sequence $\vy = (y_v: v \in [n])$ consisting of positive
  integers such that $y_v = y_w$ iff $\thetan_v = \thetan_w$ and $\Dn(v) =
  \Dn(w)$. Then, we use the algorithm of Proposition~\ref{prop:encode-sequence}
  to encode the sequence $\vy$. The details will be given in
  Section~\ref{sec:alg-details}. 
For instance, in the graph of Figure~\ref{fig:sample-graph-for-compression},
with parameters $h = 2$ and $\delta = 4$,  $(\thetan_v,\Dn(v))$ are the same for $2 \leq v
\leq 6$, and also  $(\thetan_v, \Dn(v))$ are the same for $7 \leq v \leq 16$.
Therefore, as we will see in Section~\ref{sec:encode-vertex-types},
the sequence
$\vec{y}$ in this example will be of them form
\begin{equation*}
  \vec{y} = (1,2,2,2,2,2,3,3,3,3,3,3,3,3,3,3).
\end{equation*}

\subsection{Encoding Partition Graphs}
\label{sec:main-encode-partition}

Our next step is to partition the remaining edges in the graph, i.e.\ those
edges which are not
star edges. In order to do so, we partition such edges based on their types.  This will result in a number of unmarked graphs
which will be encoded separately. More precisely, let $\mEn$  denote the set of all edge types that appear in the graph,
excluding star edges,  i.e.\
\begin{equation}
  \label{eq:set-of-types}
  \mEn := \{(t, t') \in \mCdeltah: \exists w \sim_{\Gn} v, \type{v}{w} = (t,t')\}.
\end{equation}
For $(t, t') \in \mEn$, let $\mVn_{t, t'}$ denote the set of vertices $v \in [n]$
such that $\Dn_v(t,t') > 0$. In fact, $\mVn_{t,t'}$ is the set of vertices with
at least one edge with type $(t,t')$ connected to them. Note that if $v \in \mVn_{t,t'}$, and $w
\sim_{\Gn} v $ with $\type{v}{w} = (t,t')$, then $w \in \mVn_{t',t}$. Let $\Nn_{t,t'}:= |\mVn_{t,t'}|$ denote
the number of vertices in $\mVn_{t,t'}$.  For $(t,t') \in \mEn$, define the mapping
$\In_{t,t'} : \mVn_{t,t'} \rightarrow \{1, \dots, \Nn_{t,t'}\}$
 so that for $v \in \mVn_{t,t'}$,  $\In_{t,t'}(v)$ denotes the
index of $v$ in $\mVn_{t,t'}$ when the elements of $\mVn_{t,t'}$ are sorted in
an increasing order.
Figure~\ref{fig:mEn-mVn-In} illustrates the set $\mEn$ together with
$\mVn_{t,\pp{t}}$ and $\In_{t,\pp{t}}$ for $(t,\pp{t}) \in \mEn$ for the graph in
Figure~\ref{fig:sample-graph-for-compression},
where again $h=2$ and $\delta = 4$.
\color{black}

\begin{figure}
\begin{center}
  \scalebox{0.7}{
  \begin{tikzpicture}
    \begin{scope}
      \node[scale=1.9] at (0,0) {$\mEn = \left\{ (\,\tthree[0.2]{0.45}\,,\,\tfour[0.2]{0.45}\,), (\,\tfour[0.2]{0.5}\,,\, \tthree[0.2]{0.45}\,), (\,\tfive[0.2]{0.5}\,,\, \tfive[0.2]{0.45}\,)\right\}$};
    \end{scope}
    \begin{scope}[xshift=-5cm,yshift=-2cm]
      \node[scale=1.9] at (0,0) {$\mVn_{\tthree{0.5}\,,\, \tfour{0.5}} = \{2, 3, 4, 5, 6\}$};
      \foreach \i/\x in {0/1,1/2,2/3,3/4,4/5}{
        \draw[->,very thick, >=latex] (0.63*\i,-0.3) -- (0.63*\i,-1.5);
        \node[scale=1.5] at (0.63*\i, -1.8) {\x};
      }
      \node[scale=1.5] at (3.7,-0.9) {$\In_{\tthree{0.5}\,,\,\tfour{0.5}}$};
    \end{scope}
    \begin{scope}[xshift=5cm,yshift=-2cm]
      \node[scale=1.9] at (0,0) {$\mVn_{\tfour{0.5}\,,\, \tthree{0.5}} = \{7,8,\dots,16\}$};
      \draw[->,very thick, >=latex] (-0.1,-0.3) -- (-0.1,-1.5);
      \node[scale=1.5] at (-0.1,-1.8) {1};
      \draw[->,very thick, >=latex] (0.53,-0.3) -- (0.53,-1.5);
      \node[scale=1.5] at (0.53,-1.8) {2};
      \draw[->,very thick, >=latex] (2.6,-0.3) -- (2.6,-1.5);
      \node[scale=1.5] at (2.6,-1.8) {10};
      \node[scale=1.5] at (3.8,-0.9) {$\In_{\tfour{0.5}\,,\,\tthree{0.5}}$};
    \end{scope}
    \begin{scope}[xshift=0cm,yshift=-5.5cm]
      \node[scale=1.9] at (0,0) {$\mVn_{\tfive{0.5}\,,\, \tfive{0.5}} = \{7,8,\dots,16\}$};
      \draw[->,very thick, >=latex] (-0.3,-0.3) -- (-0.3,-1.5);
      \node[scale=1.5] at (-0.3,-1.8) {1};
      \draw[->,very thick, >=latex] (0.33,-0.3) -- (0.33,-1.5);
      \node[scale=1.5] at (0.33,-1.8) {2};
      \draw[->,very thick, >=latex] (2.4,-0.3) -- (2.4,-1.5);
      \node[scale=1.5] at (2.4,-1.8) {10};
      \node[scale=1.5] at (3.8,-0.9) {$\In_{\tfive{0.5}\,,\,\tfive{0.5}}$};
    \end{scope}
  \end{tikzpicture}
  }
\end{center}
\caption{The set $\mEn$ for the graph in
  Figure~\ref{fig:sample-graph-for-compression} with parameters $h=2$ and
  $\delta = 4$. Also, $\mVn_{t,\pp{t}}$ and
  $\In_{t,\pp{t}}$ for $(t,\pp{t}) \in \mEn$ are illustrated.\label{fig:mEn-mVn-In}}
\end{figure}

For $(t,t') \in \mEn$, we construct the \emph{``partition graph''} $\Gn_{t,t'}$
associated to the pair 
$(t,t')$ as follows: 
\begin{itemize}
\item If $t = t'$, let $\Gn_{t,t}$ be a simple unmarked graph on $\Nn_{t,t}$
  nodes $\{1, \dots, \Nn_{t,t}\}$. The nodes in $\Gn_{t,t}$ are in a one
  to one correspondence with the nodes in $\mVn_{t,t}$ so that $v \in
  \mVn_{t,t}$ corresponds to the node $\In_{t,t}(v)$ in $\Gn_{t,t}$. Moreover, 
 for each edge $(v,w)$ in $\Gn$ such that $\type{v}{w} =
  (t,t)$, we place an edge in $\Gn_{t,t}$ between the nodes $\In_{t,t}(v)$ and
  $\In_{t,t}(w)$.
\item If $t \neq t'$ , let $\Gn_{t,t'}$ be a simple unmarked bipartite graph
  with $\Nn_{t,t'}$ left nodes $\{1, \dots, \Nn_{t,t'}\}$  and $\Nn_{t',t}$
  right nodes $\{1, \dots, \Nn_{t',t}\}$. The left nodes in $\Gn_{t,t'}$ are in
  a one to one correspondence with the nodes in $\mVn_{t,t'}$, so that $v \in \mVn_{t,t'}$
  corresponds to the left node $\In_{t,t'}(v)$ in $\Gn_{t,\pp{t}}$. Likewise, the right nodes in
  $\Gn_{t,t'}$ are in a one to one correspondence with the nodes in $\mVn_{t',t}$, so that $w
  \in \mVn_{t',t}$ corresponds to the right node $\In_{t',t}(w)$ in $\Gn_{t,t'}$.
  Moreover, for each edge $(v,w)$ in $\Gn$ such that $\type{v}{w} = (t,t')$, we
  place an edge in $\Gn_{t,t'}$ between the left node $\In_{t,t'}(v)$ and the
  right node $\In_{t',t}(w)$.
\end{itemize}


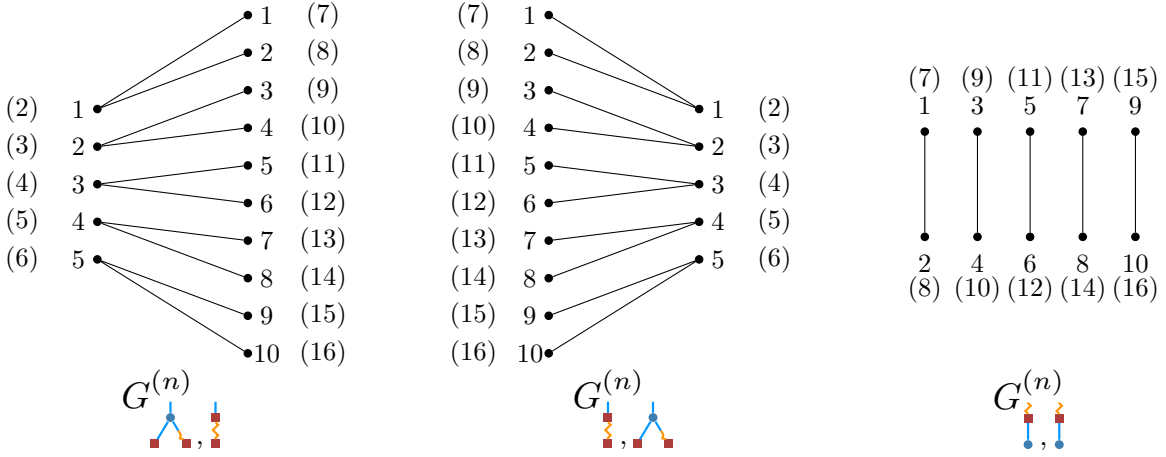
\begin{figure}
\begin{center}
  \begin{tikzpicture}
    \begin{scope}[scale=0.5,xshift=-6cm]
      \begin{scope}[xshift=-2cm]
        \foreach \y/\i in {2/1,1/2,0/3,-1/4,-2/5}{
          \node[fill, draw, circle, inner sep=1pt] at (0,\y) (l\i) {};
          \nodelabel{l\i}{180}{\i}
          \node at (-2,\y) {(\tikzint{4-\y})};
        }
      \end{scope}
      \begin{scope}[xshift=2cm]
        \foreach \y/\i in {-4.5/10,-3.5/9,-2.5/8,-1.5/7,-0.5/6,0.5/5,1.5/4,2.5/3,3.5/2,4.5/1}{
          \node[fill, draw, circle, inner sep=1pt] at (0,\y) (r\i) {};
          \nodelabel{r\i}{0}{\i}
          \node at (2,\y) {(\tikzint{6+\i})};
        }
      \end{scope}
      \draw
      (r1) -- (l1) -- (r2)
      (r3) -- (l2) -- (r4)
      (r5) -- (l3) -- (r6)
      (r7) -- (l4) -- (r8)
      (r9) -- (l5) -- (r10);
      \node[scale=1.4] at (0,-6) {$\Gn_{\tthree{0.5}\,,\, \tfour{0.5}}$};
    \end{scope}
    \begin{scope}[scale=0.5,xshift=6cm]
      \begin{scope}[xshift=-2cm]
        \foreach \y/\i in {-4.5/10,-3.5/9,-2.5/8,-1.5/7,-0.5/6,0.5/5,1.5/4,2.5/3,3.5/2,4.5/1}{
          \node[fill, draw, circle, inner sep=1pt] at (0,\y) (l\i) {};
          \nodelabel{l\i}{180}{\i}
          \node at (-2,\y) {(\tikzint{6+\i})};
        }
      \end{scope}
      \begin{scope}[xshift=2cm]
        \foreach \y/\i in {2/1,1/2,0/3,-1/4,-2/5}{
          \node[fill, draw, circle, inner sep=1pt] at (0,\y) (r\i) {};
          \nodelabel{r\i}{0}{\i}
          \node at (2,\y) {(\tikzint{4-\y})};
        }
      \end{scope}
      \draw
      (l1) -- (r1) -- (l2)
      (l3) -- (r2) -- (l4)
      (l5) -- (r3) -- (l6)
      (l7) -- (r4) -- (l8)
      (l9) -- (r5) -- (l10);
      \node[scale=1.4] at (0,-6) {$\Gn_{\tfour{0.5}\,,\,\tthree{0.5}}$};
    \end{scope}
    \begin{scope}[scale=0.7, xshift=12cm]
      \foreach \x/\i in {-2/1,-1/2,0/3,1/4,2/5}{
        \node[fill, draw, circle, inner sep = 1pt] (t\i) at (\x,1) {};
        \node[fill, draw, circle, inner sep = 1pt] (b\i) at (\x,-1) {};
        \draw (t\i) -- (b\i);
        \nodelabel{t\i}{90}{\tikzint{2*\i-1}};
        \nodelabel{b\i}{-90}{\tikzint{2*\i}};
        \node at (\x,2) {(\tikzint{5+2*\i})};
        \node at (\x,-2) {(\tikzint{6+2*\i})};
      }
      \node[scale=1.4] at (0,-4.3) {$\Gn_{\tfive{0.5}\,,\,\tfive{0.5}}$};
    \end{scope}
  \end{tikzpicture}
\end{center}
  \caption{Partition graphs for the graph of
    Figure~\ref{fig:sample-graph-for-compression} with parameters $h = 2$ and
    $\delta = 4$. See Figure~\ref{fig:message-passing-example-summary} for
    edge types, and also Figure~\ref{fig:mEn-mVn-In} for $\mEn$,
    $\mVn_{t,\pp{t}}$, and $\In_{t,\pp{t}}$ for $(t,\pp{t}) \in \mEn$ in this
    example. Next to each vertex in each partition graph, the index of its
    corresponding  vertex
    in $\Gn$ is written in parentheses. \label{fig:partition-graphs-example}}
\end{figure}


Figure~\ref{fig:partition-graphs-example} illustrates the partition graphs for the
graph of Figure~\ref{fig:sample-graph-for-compression} with parameters $h=2$ and
$\delta = 4$. 
The next step in our compression algorithm is to encode these partition graphs. 
Observe that for $(t,t') \in \mEn$ such that $t \neq t'$, the partition graphs
$\Gn_{t,t'}$ and $\Gn_{t',t}$ are basically the same, and one is obtained from
the other by flipping the right and the left nodes (this is evident for the
graph in Figure~\ref{fig:sample-graph-for-compression} by looking at 
Figure~\ref{fig:partition-graphs-example}). Let $\mEn_{\leq}$ be the set
of $(t,t') \in \mEn$ such that with $\tilde{t}= J_n(t)$ and $\tilde{t}' = J_n(\pp{t})$
being the integer representations of $t$ and $t'$ respectively,
we have  $\tilde{t} \leq \tilde{t}'$.
With the above discussion, we may only consider
those partition graphs $\Gn_{t,t'}$ such that $(t,t') \in \mEn_{\leq}$.
Note that each edge in $\Gn$ which is not an star edge appears in precisely one
of the partition graphs $\{\Gn_{t,t'}: (t,t') \in \mEn_{\leq}\}$, which
justifies our terminology ``partition graph''. 
As an example, for the graph of Figure~\ref{fig:sample-graph-for-compression}
with parameters $h=2$ and $\delta = 4$, from
Figure~\ref{fig:message-passing-example-summary}, we realize that
\begin{equation*}
  J_n(\tthree{0.5}) = 3 < 4 = J_n(\tfour{0.5}),
\end{equation*}
which implies that
\begin{equation*}
  \mEn_{\leq} = \left\{ (\tthree{0.5}\,,\, \tfour{0.5}), (\tfive{0.5}\,,\, \tfive{0.5}) \right\}.
\end{equation*}

Before discussing how to encode these partition graphs, we explain how this can
be helpful in reconstructing the original marked graph $\Gn$ at the decoder. Recall from
Section~\ref{sec:main-encode-vertex-types} that we have already encoded
$\vthetan$, the sequence of vertex marks. On the other hand, in
Section~\ref{sec:main-star-edges}, we have encoded all the star edges. Hence, it
remains to encode the remaining edges, which is precisely what is being done in
this step. As we discussed above, every edge in $\Gn$ that is not a star edge
appears in exactly one partition graph. On the other hand, for $(t, t') \in
\mEn_{\leq}$ such that $t \neq t'$, if there is an edge between the left node
$i$ and right node $j$ in $\Gn_{t,t'}$, we realize that there must be a corresponding edge in
$\Gn$ between the vertices $(\In_{t,t'})^{-1}(i)$ and $(\In_{t',t})^{-1}(j)$ with mark
$t[m]$ towards $(\In_{t,t'})^{-1}(i)$ and mark $t'[m]$ towards $(\In_{t',t})^{-1}(j)$.
Here, $(\In_{t,t'})^{-1}$ denotes the inverse of the mapping $\In_{t,t'}$ defined
above, and $(\In_{t,t'})^{-1}(i)$ is the vertex in $\Gn$ corresponding to the left
vertex $i$ in $\Gn_{t,t'}$. Likewise, $(\In_{t',t})^{-1}(j)$ is the vertex in $\Gn$
corresponding to the right vertex $j$  in $\Gn_{t,t'}$.
Recalling our discussion in Section~\ref{sec:main-find-edge-type}, if
$\tilde{t} = J_n(t)$ and $\tilde{t'} = J_n(\pp{t})$ are the integer representations of $t$ and $t'$
respectively, then $t[m] = \mathsf{TMark}(\tilde{t})$ and $t'[m] =
\mathsf{TMark}(\tilde{t'})$. Consequently, for $(t,t') \in \mEn_{\leq}$ with $t
\neq t'$, for each edge between a left node $i$
and a right node $j$ in $\Gn_{t,t'}$, at the decoder, we place an edge between vertices
$(\In_{t,t'})^{-1}(i)$ and $(\In_{t',t})^{-1}(j)$ with mark $\mathsf{TMark}(\tilde{t})$
towards $(\In_{t,t'})^{-1}(i)$ and mark $\mathsf{TMark}(\tilde{t}')$ towards
$(\In_{t',t})^{-1}(j)$ in the decoded graph.
On the other hand, for $(t,t) \in \mEn_{\leq}$, if there is an edge between vertices $i$
and $j$ in $\Gn_{t,t}$, we realize that there is an edge in $\Gn$ between the vertices
$(\In_{t,t})^{-1}(i)$ and $(\In_{t,t})^{-1}(j)$ with marks $t[m]$ towards both
endpoints. 
Consequently, for $(t,t) \in \mEn_{\leq}$, for each edge between vertices $i$
and $j$ in $\Gn_{t,t}$, we place an edge between the vertices $(\In_{t,t})^{-1}(i)$
and $(\In_{t,t})^{-1}(j)$ with mark $\mathsf{TMark}(\tilde{t})$ towards both
endpoints in the decoded graph, where $\tilde{t} = J_n(t)$ is the integer representation
of $t$. 
It is easy to see that since we have already encoded $(\Dn(v): v
\in [n])$ as in Section~\ref{sec:main-encode-vertex-types}, the decoder can
reconstruct $\In_{t,\tp}$ for $(t,\tp) \in \mEn$.
Hence, we can reconstruct $\Gn$ given the
partition graphs and the previous steps.

Now we explain how we encode the partition graphs. Recall from Section~\ref{sec:main-encode-vertex-types} above that we have transmitted the sequence $\vDn$ to the decoder. We claim
that this is enough for the decoder to infer the degree sequences of the
partition graphs. Note that, by construction, for $(t,t) \in \mEn_\leq$ and $v \in
\mVn_{t,t}$, the degree of $\In_{t,t}(v)$ in $\Gn_{t,t}$ is $\Dn_{t,t}(v)$.
Thereby, for $(t,t) \in \mEn_\leq$, the degree of a node $1 \leq i \leq
\Nn_{t,t}$ in $\Gn_{t,t}$ is precisely $\Dn_{t,t}((\In_{t,t})^{-1}(i))$.
Similarly, for $(t,t') \in \mEn_{\leq}$ such that $t \neq t'$, in $\Gn_{t,t'}$, the degrees of a
left node $1 \leq i \leq \Nn_{t,t'}$ and a right node $1 \leq j \leq \Nn_{t',t}$
are $\Dn_{t,t'}((\In_{t,t'})^{-1}(i))$ and $\Dn_{t',t}((\In_{t',t})^{-1}(j))$
respectively.

Motivated by this, it suffices to design compression schemes to encode simple
unmarked graphs and simple unmarked bipartite graphs with given degree
sequences. In the following, we discuss two such compression schemes, one for
simple unmarked bipartite graphs, and one for simple unmarked graphs.  Note that
these two schemes can be of independent interest.

\subsubsection{Encoding a  Simple Unmarked Bipartite Graph given its Degree
  Sequence}
\label{sec:main-bip-comp}

\ifreplace
Let $\mGnlr_{\va, \vb}$ be the set of simple unmarked bipartite graphs 
such that the degree of a left node $1 \leq i \leq n_l$ is
$a_i$, and the degree of a right node $1 \leq j \leq n_r$ is $b_j$. 
\color{black}
\fi
Here, we
assume that the degree sequences $\va = (a_i: i \in [n_l])$ and $\vb = (b_j: j
\in [n_r])$ are fixed and graphical. In Section~\ref{sec:bipartite-compression},
we will introduce a coding
scheme which is capable of encoding elements in $\mGnlr_{\va, \vb}$, assuming that
both the encoder and the decoder know the degree sequences $\va$ and $\vb$. This
is motivated by our discussion above for compressing partition graphs of the
form $\Gn_{t,t'}$ for $(t,t') \in \mEn_{\leq}$ such that $t \neq t'$. 

We assume that a
graph $G  \in \mGnlr_{\va, \vb}$ is represented via its adjacency list
$\vgamma^G = (\vgamma^G_v: v \in [n_l])$, where for $1 \leq v \leq n_l$,
$\vgamma^G_v = (\gamma^G_{v,1} <  \dots < \gamma^G_{v,a_v})$ are the right nodes
adjacent to the left node $v$. The following proposition summarizes the result
we prove in Section~\ref{sec:bipartite-compression}.

\begin{prop}
  \label{prop:bipartite-color-graph-compress}
  With the above setup,
  given $n_l, n_r \geq 1$ and a graphical sequence pair $\va, \vb$, 
there exists a compression algorithm that maps every
  simple unmarked bipartite graph $G \in \mGnlr_{\va, \vb}$ to an integer
  $\fnlr_{\va, \vb}(G)$ such that
  \begin{equation*}
    0 \leq \fnlr_{\va, \vb}(G) \leq \left\lceil \frac{S!}{\prod_{i=1}^{n_l} a_i! \prod_{j=1}^{n_r} b_j! }\right \rceil,
  \end{equation*}
  where $S := \sum_{i=1}^{n_l} a_i = \sum_{j=1}^{n_r} b_j$. Moreover, there
  exists a decompression algorithm  $\gnlr_{\va,
    \vb}$ such that for all $G \in \mGnlr_{\va, \vb}$, we have $\gnlr_{\va, \vb}
  \circ \fnlr_{\va, \vb} (G) = G$. 
  Further, if for some $\delta \leq
  n_r$, we have $a_i \leq \delta$ for all $1 \leq i \leq n_l$, the time
  and memory complexities of the compression algorithm are $O(\delta \pn \log^4
  \pn \log \log \pn)$ and $O(\delta \pn \log \pn)$ respectively, where $\pn =
  \max\{n_l, n_r\}$. Furthermore, the
  decompression algorithm has the same time and memory complexities. 
\end{prop}

Note that, due to~\eqref{eq:Dn-bounded-delta}, all the degrees in a partition
graph $\Gn_{t,\tp}$ for $(t,\tp) \in \mEn_\leq$, $t \neq \tp$, are bounded by
the parameter $\delta$.



\subsubsection{Encoding a Simple Unmarked Graph given its Degree Sequence}
\label{sec:main-self-comp}

With $\va = (a_1, \dots, a_{\pn})$ being a graphical degree sequence, let
$\mGnt_{\va}$ denote the set of simple unmarked graphs on the vertex set $[\pn]$
such that the degree of a vertex $1 \leq v \leq \pn$ is $a_v$. Here, we use
$\pn$ for the number of vertices to avoid confusion with the number of vertices
in $\Gn$ which was denoted by $n$. 
We may assume that $\pn \geq 2$, since a partition graph of the form $\Gn_{t,t}$
for some $(t,t) \in \mEn_\leq$ has at least one edge, thereby has at least two
vertices. 
In
Section~\ref{sec:simple-graph-compression}, we introduce a coding scheme
which is capable of encoding elements in $\mGnt_{\va}$, assuming that both the
encoder and the decoder know the degree sequence $\va$. This is motivated by our
discussion above for compressing partition graphs of the form $\Gn_{t,t}$ for
$(t,t) \in \mEn_{\leq}$. 

For $G \in
\mGnt_{\va}$ and a vertex $1 \leq v \leq \pn$, we denote by $\hat{a}^G_v$ the
number of vertices $w$ adjacent to $v$ in $G$ such that $w > v$, and call it the
``forward degree'' of vertex $v$ in $G$. Also, use the notation $v < \gamma^G_{v,1}< \dots< \gamma^G_{v,\hat{a}^G_v}$ to denote
  these neighbors in an increasing order. We call $\vgamma^G_v = (\gamma^G_{v,
    1}, \dots, \gamma^G_{v,\hat{a}^G_v})$ the ``forward adjacency list'' of
  vertex $v$. We assume that a graph $G \in \mGnt_{\va}$ is presented  to our compression
  algorithm by providing its forward adjacency list $\vgamma^G = (\vgamma^G_v: v
  \in [\pn])$. 

  Similar to the scheme of Section~\ref{sec:bipartite-compression} which we
  introduced above, our algorithm generates an integer $\fnt_{\va}(G)$ given $G
  \in \mGnt_{\va}$. However, in addition to this, the encoder generates a
  sequence $\vec{\tilde{f}}^{(\pn)}_{\va}(G)$ of nonnegative integers of length
  $\lfloor 16 \pn / \log^2 \pn \rfloor$. As we will see later in
  Section~\ref{sec:simple-graph-compression}, this is useful during the
  decompression phase to ensure
  efficiency. 
  We summarize the result in the following proposition.
  \color{black}

  \begin{prop}
    \label{prop:equal-color-graph-compress}
    With the above setup,
    given $\pn \geq 2$ and a graphical sequence $\va$,
 there exists a compression algorithm that maps every
    simple unmarked graph $G \in \mGnt_{\va}$ to an integer $\fnt_{\va}(G)$
    together with a sequence $\vec{\tilde{f}}^{(\pn)}_{\va}(G) =
    (\tilde{f}^{(\pn)}_{\va, i}(G): 1 \leq i \leq \lfloor 16 \pn / \log^2 \pn
    \rfloor)$ of nonnegative integers such that
    \begin{equation*}
      0 \leq \fnt_{\va}(G) \leq \left\lceil \frac{(S-1)!!}{\prod_{v=1}^{\pn}a_v! } \right \rceil,
    \end{equation*}
    where $S := \sum_{v=1}^{\pn} a_v$,
    \ifreplace
  and we recall that for an even integer $k > 0$, we define
$(k-1)!! := \frac{k!}{2^{k/2} (k/2)!}$,
  which is the number of matchings on $k$ objects, with $(-1)!!
  := 1$ by convention.
  \color{black}
  \fi
Also, for $1 \leq i \leq \lfloor 16 \pn / \log^2 \pn
    \rfloor$, we have  $0 \leq \tilde{f}^{(\pn)}_{\va, i}(G) \leq S$. Moreover,
    there exists a decompression algorithm with a decompression map $\gnt_{\va}$
    such that for all $G \in \mGnt_{\va}$, we have $\gnt_{\va}(\fnt_{\va}(G),
    \vec{\tilde{f}}^{(\pn)}_{\va}(G)) = G$. On the other hand, if for some
    $\delta \leq \pn$, we have $a_i \leq \delta$ for all $1 \leq i \leq \pn$,
    the time and memory complexities of the compression algorithm are $O(\delta
    \pn \log^4 \pn \log \log \pn)$ and $O(\delta \pn \log \pn)$ respectively.
    Furthermore, with the same assumption, the time and memory complexities of the decompression algorithm
    are $O(\delta \pn \log^5 \pn \log \log \pn)$ and $O(\delta \pn \log \pn)$
    respectively. 
  \end{prop}

Note that due to~\eqref{eq:Dn-bounded-delta}, all the degrees in a partition
graph $\Gn_{t,t}$ for $(t,t) \in \mEn_\leq$ are bounded by
the parameter $\delta$.

\editfinish


\section{Experiments}
\label{sec:experiments}

In this section, we illustrate the performance of our algorithm for three types
of datasets. First, in Section~\ref{sec:synt-exp}, we consider some synthetic
data. The advantage of such datasets is that, since the local weak limit is known
for them, we can compare the performance of the
algorithm with what Theorem~\ref{thm:optimality-complexity-main} predicts. 
Then, in Section~\ref{sec:exp-tree}, we consider some real--world datasets which do not have
large cycles, or roughly speaking they are locally tree--like. The motivation
for considering such datasets is  that the
optimality guarantee of Theorem~\ref{thm:optimality-complexity-main} requires
the local weak limit to be supported on rooted trees, i.e.\ $\mu \in
\mP_u(\mTb_*)$. Indeed, recall from Theorem~\ref{thm:badcases} that the BC
entropy is not of interest unless $\mu \in \mP_u(\mTb_*)$. Finally, in
Section~\ref{sec:data-social}, we test the performance of our algorithm for
datasets which go beyond the theoretical requirements of
Theorem~\ref{thm:optimality-complexity-main}. More precisely, we consider some
real--world social network datasets in this section. Social networks often have small cycles and are
not locally tree-like. Hence, they fail to have local weak limits supported on
$\mTb_*$. Consequently, the assumptions of
Theorem~\ref{thm:optimality-complexity-main} do not hold for such datasets,
and the theoretical guarantee no longer exists. Nonetheless, it is interesting
to see that our algorithm has a relatively good performance for such datasets,
and its performance is comparable to the state of the art in most cases. 


\subsection{Synthetic data}
\label{sec:synt-exp}

\begin{figure}
  \begin{center}
    \scalebox{0.85}{
\begin{tikzpicture}

\definecolor{color0}{rgb}{0.12156862745098,0.466666666666667,0.705882352941177}
\definecolor{color1}{rgb}{1,0.498039215686275,0.0549019607843137}
\definecolor{color2}{rgb}{0.172549019607843,0.627450980392157,0.172549019607843}
\definecolor{color3}{rgb}{0.83921568627451,0.152941176470588,0.156862745098039}
\definecolor{color4}{rgb}{0.580392156862745,0.403921568627451,0.741176470588235}

\begin{axis}[
legend cell align={left},
legend style={fill opacity=0.8, draw opacity=1, text opacity=1, draw=white!80.0!black},
log basis x={10},
tick align=outside,
tick pos=both,
x grid style={white!69.01960784313725!black},
xlabel={\(\displaystyle n\) (number of vertices)},
xmajorgrids,
xmin=56.2341325190349, xmax=17782794.1003892,
xmode=log,
xtick style={color=black},
y grid style={white!69.01960784313725!black},
ylabel={\(\displaystyle l_n\) (nats)},
ymajorgrids,
ymin=1.51061079317472, ymax=22.7656733433308,
ytick style={color=black}
]
\addplot [thick, color0, mark=*, mark size=1, mark options={solid}]
table {%
100 10.0437579542089
1000 9.2588445426093
10000 10.1653947263528
100000 10.1232708618084
1000000 9.53539492094777
4000000 9.53278490861169
8000000 9.53494273507699
10000000 10.7250036030234
};
\addlegendentry{$\delta=2$}
\addplot [thick, color1, mark=*, mark size=1, mark options={solid}]
table {%
100 14.7017070075717
1000 10.8336749368415
10000 10.7132582578674
100000 10.4091802108458
1000000 9.80182960680012
4000000 9.78516256964229
8000000 9.78487497285457
10000000 10.8750472323871
};
\addlegendentry{$\delta=5$}
\addplot [thick, color2, mark=*, mark size=1, mark options={solid}]
table {%
100 21.7995341365056
1000 16.8002858671015
10000 11.4784927452056
100000 7.58579766321457
1000000 5.71558834796314
4000000 5.32981211939096
8000000 5.24248187225725
10000000 5.09856047405952
};
\addlegendentry{$\delta=10$}
\addplot [thick, color3, mark=*, mark size=1, mark options={solid}]
table {%
100 21.134112843168
1000 16.0905031542081
10000 10.7221305417786
100000 6.02006135999132
1000000 3.62328199461211
4000000 3.03425837716743
8000000 2.87031595846078
10000000 2.81536693235973
};
\addlegendentry{$\delta=20$}
\addplot [very thick, color4]
table {%
100 2.47675
10000000 2.47675
};
\addlegendentry{BC Entropy}
\end{axis}

\end{tikzpicture}
      }
 \end{center}
 \caption{Synthetic data results. 
   Note that for large $\delta$ the asymptotic performance converges to the
   actual BC entropy. \label{fig:exp-synthetic}}
\end{figure}
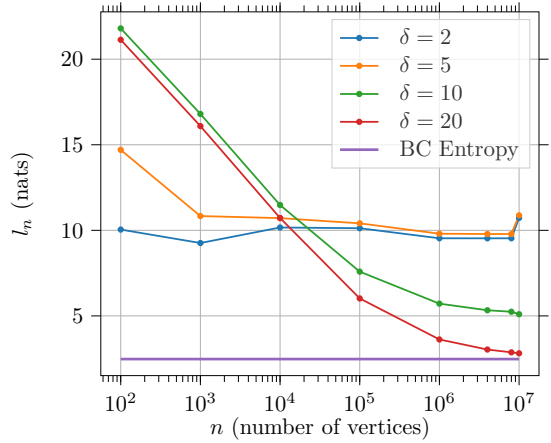

We generate a random graph $\Gn$ on $n$ vertices as follows.
At each vertex $v \in [n]$ we generate a Poisson random variable
$d_v$ with mean $3$ and pick $d_v$ many 
vertices uniformly at random from $[n] \setminus \{v\}$ without replacement,
Then, we  connect $v$ to each of these $d_v$ chosen vertices. We do this for all
nodes $1 \leq v \leq w$. If for two nodes $v \neq w$, $v$ decides to connect to
$w$, and $w$ also decides to connects to $v$, we treat this as a
single edge between $v$ and $w$. Therefore, the resulting graph is simple. We
also add a mark from $\vermark = \{1,2\}$ to each vertex independently.
{
  Moreover, we add two
independent edge marks to each edge from $\edgemark = \{1,2\}$. The choice of edge
and vertex marks is done independently throughout the graph, conditioned on the
unmarked realization of the graph.}
It can be seen  that the local weak
limit of this model is a Poisson Galton--Watson tree with mean degree 6 and
independent vertex and edge marks.
Since the limit distribution is completely
characterized by the depth 1 neighborhood distribution at the root, we choose
$h=1$, and run the algorithm with different values of $\delta$. See
Figure~\ref{fig:exp-synthetic} for the behavior of 
\begin{equation*}
 l_n := (\nat(\fn_{h,  \delta}(\Gn) ) - m_n \log n) / n,
\end{equation*}
where $m_n$ is the number of edges in
$\Gn$. As we see, for large values of $\delta$, $l_n$ converges to the marked BC
entropy of the limit as $n$ gets large, which is consistent with Theorem~\ref{thm:optimality-complexity-main}.
However, the interesting observation from Figure~\ref{fig:exp-synthetic} is that
when $n$ is small, smaller values of $\delta$ might result in smaller $l_n$.
Note that this is not a contradiction with
Theorem~\ref{thm:optimality-complexity-main}, since it only predicts the
asymptotic behavior of $l_n$ as $n$ goes to infinity. When $n$ is small,
increasing $\delta$ might result in an increased overhead of certain parts of
the compressed sequence, such as vertex types as discussed in
Section~\ref{sec:main-encode-vertex-types}, which results in increasing $l_n$. 


\subsection{Locally tree--like data}
\label{sec:exp-tree}

Recall from Theorem~\ref{thm:optimality-complexity-main} that our theoretical
guarantee holds when the limit $\mu$ is supported on marked rooted trees.
Motivated by this, we test our algorithm on the following two real--world
locally tree--like datasets. We also 
compare the compression results with the ones reported in
\cite{liakos2014pushing}, which, to the best of our knowledge, are the best
results in the 
literature for these datasets. These datasets are collected from \cite{snapnets}.
\begin{itemize}[leftmargin=*]
\item \emph{roadnet-CA}: the graph of the road network of California, consisting
  of 1,965,206 vertices and 5,533,214 edges.
\item \emph{roadnet-PA}: the graph of the road network of Pennsylvania,
  consisting of 1,088,092 vertices and 3,083,796 edges.
\end{itemize}

In order to measure whether these graphs are locally tree--like, we use a
measure called the \emph{average clustering coefficient} which is defined as
follows. 
For a simple undirected graph $G$ on the vertex set $[n]$, the clustering of a
  node $v$ in $G$ is defined as
  \begin{equation*}
    C_v := \frac{|\{(u,w): u \sim_G v, w \sim_G v, u \sim_G w\}|}{\deg_G(v) (\deg_G(v) - 1) / 2},
  \end{equation*}
  which is obtained by looking at pairs of neighbors of $v$ which can form a
  triangle, and calculating the fraction of them which actually form a triangle.
  The average clustering coefficient of $G$ is defined to be
  \begin{equation*}
    C := \frac{1}{n} \sum_{v=1}^n C_v.
  \end{equation*}
  Note that $C \in [0,1]$, and the bigger the clustering coefficient of a graph,
  the more short cycles it has, and so the less tree--like it is.
The average clustering coefficients of both the roadnet-CA and the roadnet-PA datasets
are 0.046, which is relatively small. 

Following the
convention in the literature, we report the compression ratios in \emph{bits per link} (BPL). 
Figures~\ref{fig:roadnet-bpl}, \ref{fig:roadnet-enc-time}, and
\ref{fig:roadnet-dec-time} illustrate the compression ratio, compression time,
and decompression time of our algorithm for these datasets, respectively, for
different values of $h$ and $\delta$. 
 Table~\ref{tab:roadnet-first} compares the best compression ratios of our algorithm, which, as we can see,
are more than 40\% better than the ones in \cite{liakos2014pushing}. 
As we can see from Table~\ref{tab:roadnet-first}, the value of $\delta$ in the
pair $(h, \delta)$ which yields the best result is small. Comparing this to
Figure~\ref{fig:exp-synthetic} and the discussion in Section~\ref{sec:synt-exp},
one possible explanation for this can be that $n$ is small and hence we are not
yet in the asymptotic regime. In order to address this, in
Table~\ref{tab:roadnet-20}, we force $\delta$ to be chosen large enough so that
at most $20\%$ of the edges become star edges, and find the value of $(h, \delta)$
which achieves the best compression ratio given this constraint. As we can see
in Table~\ref{tab:roadnet-20}, even in this case, our compression ratios are
still better compared to those reported in
\cite{liakos2014pushing}.

\begin{figure}
  \centering
  \hfill
  \subfloat[\emph{roadnet-CA}]{
    \scalebox{0.7}{
\begin{tikzpicture}

\definecolor{color0}{rgb}{0.12156862745098,0.466666666666667,0.705882352941177}
\definecolor{color1}{rgb}{1,0.498039215686275,0.0549019607843137}
\definecolor{color2}{rgb}{0.172549019607843,0.627450980392157,0.172549019607843}

\begin{axis}[
legend cell align={left},
legend style={fill opacity=0.8, draw opacity=1, text opacity=1, at={(0.03,0.97)}, anchor=north west, draw=white!80.0!black},
tick align=outside,
tick pos=both,
x grid style={white!69.01960784313725!black},
xlabel={\(\displaystyle \delta\)},
xmin=-0.334166666666667, xmax=3.90083333333333,
xtick style={color=black},
xtick={0.283333333333333,1.28333333333333,2.28333333333333,3.28333333333333},
xticklabels={2,5,10,20},
y grid style={white!69.01960784313725!black},
ylabel={bits per link},
ymin=0, ymax=26.3597866990144,
ytick style={color=black}
]
\draw[fill=color0,draw opacity=0,fill opacity=0.8] (axis cs:-0.141666666666667,0) rectangle (axis cs:0.141666666666667,6.08714428901539);
\addlegendimage{ybar,ybar legend,fill=color0,draw opacity=0,fill opacity=0.8};
\addlegendentry{h=2}

\draw[fill=color0,draw opacity=0,fill opacity=0.8] (axis cs:0.858333333333333,0) rectangle (axis cs:1.14166666666667,10.2622511979475);
\draw[fill=color0,draw opacity=0,fill opacity=0.8] (axis cs:1.85833333333333,0) rectangle (axis cs:2.14166666666667,10.2471800295452);
\draw[fill=color0,draw opacity=0,fill opacity=0.8] (axis cs:2.85833333333333,0) rectangle (axis cs:3.14166666666667,10.2471395467444);
\draw[fill=color1,draw opacity=0,fill opacity=0.8] (axis cs:0.141666666666667,0) rectangle (axis cs:0.425,5.95104689607161);
\addlegendimage{ybar,ybar legend,fill=color1,draw opacity=0,fill opacity=0.8};
\addlegendentry{h=3}

\draw[fill=color1,draw opacity=0,fill opacity=0.8] (axis cs:1.14166666666667,0) rectangle (axis cs:1.425,10.4674585150692);
\draw[fill=color1,draw opacity=0,fill opacity=0.8] (axis cs:2.14166666666667,0) rectangle (axis cs:2.425,10.5538198956339);
\draw[fill=color1,draw opacity=0,fill opacity=0.8] (axis cs:3.14166666666667,0) rectangle (axis cs:3.425,10.5540685756958);
\draw[fill=color2,draw opacity=0,fill opacity=0.8] (axis cs:0.425,0) rectangle (axis cs:0.708333333333333,5.93284120223798);
\addlegendimage{ybar,ybar legend,fill=color2,draw opacity=0,fill opacity=0.8};
\addlegendentry{h=4}

\draw[fill=color2,draw opacity=0,fill opacity=0.8] (axis cs:1.425,0) rectangle (axis cs:1.70833333333333,23.7298958616095);
\draw[fill=color2,draw opacity=0,fill opacity=0.8] (axis cs:2.425,0) rectangle (axis cs:2.70833333333333,25.1033153606566);
\draw[fill=color2,draw opacity=0,fill opacity=0.8] (axis cs:3.425,0) rectangle (axis cs:3.70833333333333,25.1045587609661);
\end{axis}

\end{tikzpicture}
    }
  }
  \hfill\hfill\hfill
  \subfloat[\emph{roadnet-PA}]{
    \scalebox{0.7}{
\begin{tikzpicture}

\definecolor{color0}{rgb}{0.12156862745098,0.466666666666667,0.705882352941177}
\definecolor{color1}{rgb}{1,0.498039215686275,0.0549019607843137}
\definecolor{color2}{rgb}{0.172549019607843,0.627450980392157,0.172549019607843}

\begin{axis}[
legend cell align={left},
legend style={fill opacity=0.8, draw opacity=1, text opacity=1, at={(0.03,0.97)}, anchor=north west, draw=white!80.0!black},
tick align=outside,
tick pos=both,
x grid style={white!69.01960784313725!black},
xlabel={\(\displaystyle \delta\)},
xmin=-0.334166666666667, xmax=3.90083333333333,
xtick style={color=black},
xtick={0.283333333333333,1.28333333333333,2.28333333333333,3.28333333333333},
xticklabels={2,5,10,20},
y grid style={white!69.01960784313725!black},
ylabel={bits per link},
ymin=0, ymax=10.7869018573213,
ytick style={color=black}
]
\draw[fill=color0,draw opacity=0,fill opacity=0.8] (axis cs:-0.141666666666667,0) rectangle (axis cs:0.141666666666667,6.04212211183879);
\addlegendimage{ybar,ybar legend,fill=color0,draw opacity=0,fill opacity=0.8};
\addlegendentry{h=1}

\draw[fill=color0,draw opacity=0,fill opacity=0.8] (axis cs:0.858333333333333,0) rectangle (axis cs:1.14166666666667,9.84724021952166);
\draw[fill=color0,draw opacity=0,fill opacity=0.8] (axis cs:1.85833333333333,0) rectangle (axis cs:2.14166666666667,9.84722984270036);
\draw[fill=color0,draw opacity=0,fill opacity=0.8] (axis cs:2.85833333333333,0) rectangle (axis cs:3.14166666666667,9.84722984270036);
\draw[fill=color1,draw opacity=0,fill opacity=0.8] (axis cs:0.141666666666667,0) rectangle (axis cs:0.425,6.0498839741669);
\addlegendimage{ybar,ybar legend,fill=color1,draw opacity=0,fill opacity=0.8};
\addlegendentry{h=2}

\draw[fill=color1,draw opacity=0,fill opacity=0.8] (axis cs:1.14166666666667,0) rectangle (axis cs:1.425,9.82261602259034);
\draw[fill=color1,draw opacity=0,fill opacity=0.8] (axis cs:2.14166666666667,0) rectangle (axis cs:2.425,9.80477826678548);
\draw[fill=color1,draw opacity=0,fill opacity=0.8] (axis cs:3.14166666666667,0) rectangle (axis cs:3.425,9.80477826678548);
\draw[fill=color2,draw opacity=0,fill opacity=0.8] (axis cs:0.425,0) rectangle (axis cs:0.708333333333333,5.93788694193779);
\addlegendimage{ybar,ybar legend,fill=color2,draw opacity=0,fill opacity=0.8};
\addlegendentry{h=3}

\draw[fill=color2,draw opacity=0,fill opacity=0.8] (axis cs:1.425,0) rectangle (axis cs:1.70833333333333,10.1568741901215);
\draw[fill=color2,draw opacity=0,fill opacity=0.8] (axis cs:2.425,0) rectangle (axis cs:2.70833333333333,10.2732398641155);
\draw[fill=color2,draw opacity=0,fill opacity=0.8] (axis cs:3.425,0) rectangle (axis cs:3.70833333333333,10.2732398641155);
\end{axis}

\end{tikzpicture}
    }
  }
  \hfill
  \caption{\label{fig:roadnet-bpl} Compression ratios of our algorithm for
    different values of the parameters $h$ and $\delta$ for the datasets $(a)$
    \emph{roadnet-CA}, and $(b)$ \emph{roadnet-PA}.}
\end{figure}
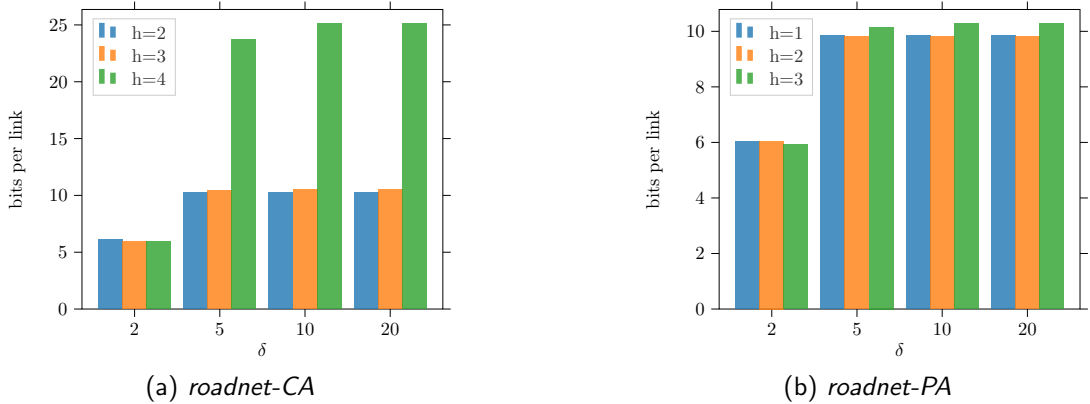

\begin{figure}
  \centering
  \hfill
  \subfloat[\emph{roadnet-CA}]{
    \scalebox{0.7}{
\begin{tikzpicture}

\definecolor{color0}{rgb}{0.12156862745098,0.466666666666667,0.705882352941177}
\definecolor{color1}{rgb}{1,0.498039215686275,0.0549019607843137}
\definecolor{color2}{rgb}{0.172549019607843,0.627450980392157,0.172549019607843}

\begin{axis}[
legend cell align={left},
legend style={fill opacity=0.8, draw opacity=1, text opacity=1, at={(0.03,0.97)}, anchor=north west, draw=white!80.0!black},
tick align=outside,
tick pos=both,
x grid style={white!69.01960784313725!black},
xlabel={\(\displaystyle \delta\)},
xmin=-0.334166666666667, xmax=3.90083333333333,
xtick style={color=black},
xtick={0.283333333333333,1.28333333333333,2.28333333333333,3.28333333333333},
xticklabels={2,5,10,20},
y grid style={white!69.01960784313725!black},
ylabel={Encode time (s)},
ymin=0, ymax=44.1141393,
ytick style={color=black}
]
\draw[fill=color0,draw opacity=0,fill opacity=0.8] (axis cs:-0.141666666666667,0) rectangle (axis cs:0.141666666666667,15.626207);
\addlegendimage{ybar,ybar legend,fill=color0,draw opacity=0,fill opacity=0.8};
\addlegendentry{h=2}

\draw[fill=color0,draw opacity=0,fill opacity=0.8] (axis cs:0.858333333333333,0) rectangle (axis cs:1.14166666666667,23.469049);
\draw[fill=color0,draw opacity=0,fill opacity=0.8] (axis cs:1.85833333333333,0) rectangle (axis cs:2.14166666666667,21.162642);
\draw[fill=color0,draw opacity=0,fill opacity=0.8] (axis cs:2.85833333333333,0) rectangle (axis cs:3.14166666666667,20.335192);
\draw[fill=color1,draw opacity=0,fill opacity=0.8] (axis cs:0.141666666666667,0) rectangle (axis cs:0.425,15.630996);
\addlegendimage{ybar,ybar legend,fill=color1,draw opacity=0,fill opacity=0.8};
\addlegendentry{h=3}

\draw[fill=color1,draw opacity=0,fill opacity=0.8] (axis cs:1.14166666666667,0) rectangle (axis cs:1.425,25.140905);
\draw[fill=color1,draw opacity=0,fill opacity=0.8] (axis cs:2.14166666666667,0) rectangle (axis cs:2.425,23.417921);
\draw[fill=color1,draw opacity=0,fill opacity=0.8] (axis cs:3.14166666666667,0) rectangle (axis cs:3.425,22.823032);
\draw[fill=color2,draw opacity=0,fill opacity=0.8] (axis cs:0.425,0) rectangle (axis cs:0.708333333333333,15.74689);
\addlegendimage{ybar,ybar legend,fill=color2,draw opacity=0,fill opacity=0.8};
\addlegendentry{h=4}

\draw[fill=color2,draw opacity=0,fill opacity=0.8] (axis cs:1.425,0) rectangle (axis cs:1.70833333333333,41.777058);
\draw[fill=color2,draw opacity=0,fill opacity=0.8] (axis cs:2.425,0) rectangle (axis cs:2.70833333333333,42.013466);
\draw[fill=color2,draw opacity=0,fill opacity=0.8] (axis cs:3.425,0) rectangle (axis cs:3.70833333333333,40.202698);
\end{axis}

\end{tikzpicture}
    }
  }
  \hfill\hfill\hfill
  \subfloat[\emph{roadnet-PA}]{
    \scalebox{0.7}{
\begin{tikzpicture}

\definecolor{color0}{rgb}{0.12156862745098,0.466666666666667,0.705882352941177}
\definecolor{color1}{rgb}{1,0.498039215686275,0.0549019607843137}
\definecolor{color2}{rgb}{0.172549019607843,0.627450980392157,0.172549019607843}

\begin{axis}[
legend cell align={left},
legend style={fill opacity=0.8, draw opacity=1, text opacity=1, at={(0.03,0.97)}, anchor=north west, draw=white!80.0!black},
tick align=outside,
tick pos=both,
x grid style={white!69.01960784313725!black},
xlabel={\(\displaystyle \delta\)},
xmin=-0.334166666666667, xmax=3.90083333333333,
xtick style={color=black},
xtick={0.283333333333333,1.28333333333333,2.28333333333333,3.28333333333333},
xticklabels={2,5,10,20},
y grid style={white!69.01960784313725!black},
ylabel={Encode time (s)},
ymin=0, ymax=12.81126,
ytick style={color=black}
]
\draw[fill=color0,draw opacity=0,fill opacity=0.8] (axis cs:-0.141666666666667,0) rectangle (axis cs:0.141666666666667,7.163105);
\addlegendimage{ybar,ybar legend,fill=color0,draw opacity=0,fill opacity=0.8};
\addlegendentry{h=1}

\draw[fill=color0,draw opacity=0,fill opacity=0.8] (axis cs:0.858333333333333,0) rectangle (axis cs:1.14166666666667,11.244538);
\draw[fill=color0,draw opacity=0,fill opacity=0.8] (axis cs:1.85833333333333,0) rectangle (axis cs:2.14166666666667,9.628854);
\draw[fill=color0,draw opacity=0,fill opacity=0.8] (axis cs:2.85833333333333,0) rectangle (axis cs:3.14166666666667,9.969542);
\draw[fill=color1,draw opacity=0,fill opacity=0.8] (axis cs:0.141666666666667,0) rectangle (axis cs:0.425,7.323825);
\addlegendimage{ybar,ybar legend,fill=color1,draw opacity=0,fill opacity=0.8};
\addlegendentry{h=2}

\draw[fill=color1,draw opacity=0,fill opacity=0.8] (axis cs:1.14166666666667,0) rectangle (axis cs:1.425,11.013995);
\draw[fill=color1,draw opacity=0,fill opacity=0.8] (axis cs:2.14166666666667,0) rectangle (axis cs:2.425,9.791376);
\draw[fill=color1,draw opacity=0,fill opacity=0.8] (axis cs:3.14166666666667,0) rectangle (axis cs:3.425,9.864647);
\draw[fill=color2,draw opacity=0,fill opacity=0.8] (axis cs:0.425,0) rectangle (axis cs:0.708333333333333,7.363001);
\addlegendimage{ybar,ybar legend,fill=color2,draw opacity=0,fill opacity=0.8};
\addlegendentry{h=3}

\draw[fill=color2,draw opacity=0,fill opacity=0.8] (axis cs:1.425,0) rectangle (axis cs:1.70833333333333,12.2012);
\draw[fill=color2,draw opacity=0,fill opacity=0.8] (axis cs:2.425,0) rectangle (axis cs:2.70833333333333,10.873475);
\draw[fill=color2,draw opacity=0,fill opacity=0.8] (axis cs:3.425,0) rectangle (axis cs:3.70833333333333,11.122111);
\end{axis}

\end{tikzpicture}
    }
  }
  \hfill
  \caption{\label{fig:roadnet-enc-time} Compression times of our algorithm in
    seconds for
    different values of the parameters $h$ and $\delta$ for the datasets $(a)$
    \emph{roadnet-CA}, and $(b)$ \emph{roadnet-PA}.}
\end{figure}

\begin{figure}
  \centering
  \hfill
  \subfloat[\emph{roadnet-CA}]{
    \scalebox{0.7}{
\begin{tikzpicture}

\definecolor{color0}{rgb}{0.12156862745098,0.466666666666667,0.705882352941177}
\definecolor{color1}{rgb}{1,0.498039215686275,0.0549019607843137}
\definecolor{color2}{rgb}{0.172549019607843,0.627450980392157,0.172549019607843}

\begin{axis}[
legend cell align={left},
legend style={fill opacity=0.8, draw opacity=1, text opacity=1, at={(0.03,0.97)}, anchor=north west, draw=white!80.0!black},
tick align=outside,
tick pos=both,
x grid style={white!69.01960784313725!black},
xlabel={\(\displaystyle \delta\)},
xmin=-0.334166666666667, xmax=3.90083333333333,
xtick style={color=black},
xtick={0.283333333333333,1.28333333333333,2.28333333333333,3.28333333333333},
xticklabels={2,5,10,20},
y grid style={white!69.01960784313725!black},
ylabel={Decode time (s)},
ymin=0, ymax=109.83024675,
ytick style={color=black}
]
\draw[fill=color0,draw opacity=0,fill opacity=0.8] (axis cs:-0.141666666666667,0) rectangle (axis cs:0.141666666666667,59.483799);
\addlegendimage{ybar,ybar legend,fill=color0,draw opacity=0,fill opacity=0.8};
\addlegendentry{h=2}

\draw[fill=color0,draw opacity=0,fill opacity=0.8] (axis cs:0.858333333333333,0) rectangle (axis cs:1.14166666666667,104.600235);
\draw[fill=color0,draw opacity=0,fill opacity=0.8] (axis cs:1.85833333333333,0) rectangle (axis cs:2.14166666666667,99.113686);
\draw[fill=color0,draw opacity=0,fill opacity=0.8] (axis cs:2.85833333333333,0) rectangle (axis cs:3.14166666666667,94.443367);
\draw[fill=color1,draw opacity=0,fill opacity=0.8] (axis cs:0.141666666666667,0) rectangle (axis cs:0.425,59.96981);
\addlegendimage{ybar,ybar legend,fill=color1,draw opacity=0,fill opacity=0.8};
\addlegendentry{h=3}

\draw[fill=color1,draw opacity=0,fill opacity=0.8] (axis cs:1.14166666666667,0) rectangle (axis cs:1.425,102.898979);
\draw[fill=color1,draw opacity=0,fill opacity=0.8] (axis cs:2.14166666666667,0) rectangle (axis cs:2.425,93.350807);
\draw[fill=color1,draw opacity=0,fill opacity=0.8] (axis cs:3.14166666666667,0) rectangle (axis cs:3.425,89.605606);
\draw[fill=color2,draw opacity=0,fill opacity=0.8] (axis cs:0.425,0) rectangle (axis cs:0.708333333333333,58.242535);
\addlegendimage{ybar,ybar legend,fill=color2,draw opacity=0,fill opacity=0.8};
\addlegendentry{h=4}

\draw[fill=color2,draw opacity=0,fill opacity=0.8] (axis cs:1.425,0) rectangle (axis cs:1.70833333333333,96.052292);
\draw[fill=color2,draw opacity=0,fill opacity=0.8] (axis cs:2.425,0) rectangle (axis cs:2.70833333333333,92.196243);
\draw[fill=color2,draw opacity=0,fill opacity=0.8] (axis cs:3.425,0) rectangle (axis cs:3.70833333333333,90.206581);
\end{axis}

\end{tikzpicture}
    }
  }
  \hfill\hfill\hfill
  \subfloat[\emph{roadnet-PA}]{
    \scalebox{0.7}{
\begin{tikzpicture}

\definecolor{color0}{rgb}{0.12156862745098,0.466666666666667,0.705882352941177}
\definecolor{color1}{rgb}{1,0.498039215686275,0.0549019607843137}
\definecolor{color2}{rgb}{0.172549019607843,0.627450980392157,0.172549019607843}

\begin{axis}[
legend cell align={left},
legend style={fill opacity=0.8, draw opacity=1, text opacity=1, draw=white!80.0!black},
tick align=outside,
tick pos=both,
x grid style={white!69.01960784313725!black},
xlabel={\(\displaystyle \delta\)},
xmin=-0.334166666666667, xmax=3.90083333333333,
xtick style={color=black},
xtick={0.283333333333333,1.28333333333333,2.28333333333333,3.28333333333333},
xticklabels={2,5,10,20},
y grid style={white!69.01960784313725!black},
ylabel={Decode time (s)},
ymin=0, ymax=60.65309355,
ytick style={color=black}
]
\draw[fill=color0,draw opacity=0,fill opacity=0.8] (axis cs:-0.141666666666667,0) rectangle (axis cs:0.141666666666667,28.240034);
\addlegendimage{ybar,ybar legend,fill=color0,draw opacity=0,fill opacity=0.8};
\addlegendentry{h=1}

\draw[fill=color0,draw opacity=0,fill opacity=0.8] (axis cs:0.858333333333333,0) rectangle (axis cs:1.14166666666667,57.764851);
\draw[fill=color0,draw opacity=0,fill opacity=0.8] (axis cs:1.85833333333333,0) rectangle (axis cs:2.14166666666667,52.245644);
\draw[fill=color0,draw opacity=0,fill opacity=0.8] (axis cs:2.85833333333333,0) rectangle (axis cs:3.14166666666667,51.284737);
\draw[fill=color1,draw opacity=0,fill opacity=0.8] (axis cs:0.141666666666667,0) rectangle (axis cs:0.425,29.718044);
\addlegendimage{ybar,ybar legend,fill=color1,draw opacity=0,fill opacity=0.8};
\addlegendentry{h=2}

\draw[fill=color1,draw opacity=0,fill opacity=0.8] (axis cs:1.14166666666667,0) rectangle (axis cs:1.425,51.494617);
\draw[fill=color1,draw opacity=0,fill opacity=0.8] (axis cs:2.14166666666667,0) rectangle (axis cs:2.425,46.4632);
\draw[fill=color1,draw opacity=0,fill opacity=0.8] (axis cs:3.14166666666667,0) rectangle (axis cs:3.425,46.466679);
\draw[fill=color2,draw opacity=0,fill opacity=0.8] (axis cs:0.425,0) rectangle (axis cs:0.708333333333333,28.885796);
\addlegendimage{ybar,ybar legend,fill=color2,draw opacity=0,fill opacity=0.8};
\addlegendentry{h=3}

\draw[fill=color2,draw opacity=0,fill opacity=0.8] (axis cs:1.425,0) rectangle (axis cs:1.70833333333333,49.570271);
\draw[fill=color2,draw opacity=0,fill opacity=0.8] (axis cs:2.425,0) rectangle (axis cs:2.70833333333333,44.647488);
\draw[fill=color2,draw opacity=0,fill opacity=0.8] (axis cs:3.425,0) rectangle (axis cs:3.70833333333333,44.4744);
\end{axis}

\end{tikzpicture}
    }
  }
  \hfill
  \caption{\label{fig:roadnet-dec-time} Decompression times of our algorithm in seconds for
    different values of the parameters $h$ and $\delta$ for the datasets $(a)$
    \emph{roadnet-CA}, and $(b)$ \emph{roadnet-PA}.}
\end{figure}

\begin{table}
  \centering
  \begin{tabular}{@{}lcccr@{}}
    \toprule
    Dataset & \cite{liakos2014pushing} & best BPL [relative\%] & value of  &   encode/decode \\
    & (BPL)  &  & ($h, \delta$) & time (sec) \\
    \midrule
    \emph{roadnet-CA} & 10.58 & 5.93 [+44\%] & (4,2) & 15.75/58.2 \\
    \emph{roadnet-PA} & 10.07 & 5.94 [+41\%] & (3,2) & 7.4/28.9 \\
    \bottomrule
  \end{tabular}
  \caption{\label{tab:roadnet-first} Comparing the compression ratios of our algorithm
  with those in \cite{liakos2014pushing} for road networks. In the third column, the best ratio of
our algorithm together with the relative improvement over the best results in
the literature are given. 
Here BPL stands for bits per link.
\color{black}
In the fourth column, the corresponding values of $h$
and $\delta$ are reported. In the fifth
column, the corresponding encoding/decoding times in seconds are given. As we can
see, for both datasets, the value of the parameter $\delta$ in the best case is
small. Motivated by Figure~\ref{fig:exp-synthetic}, one possible explanation for
this observation could be that we are not yet in the asymptotic regime. We
further address this in the following Table~\ref{tab:roadnet-20}. }
\end{table}

\begin{table}
  \centering
  \begin{tabular}{@{}lcccr@{}}
    \toprule
    Dataset & \cite{liakos2014pushing} &  best 20\% BPL [relative\%] & value of  & encode/decode\\
    & (BPL)  &  & ($h, \delta$) & time (sec) \\ 
    \midrule
    \emph{roadnet-CA} & 10.58 &  10.25 [+3\%] & (4,20) & 20.24/94.4\\
    \emph{roadnet-PA} & 10.07 &  9.80 [+2.7\%]&  (2,10) & 9.8/46.5\\
    \bottomrule
  \end{tabular}
  \caption{\label{tab:roadnet-20} 
Comparing the compression ratios of our algorithm with those in \cite{liakos2014pushing}
assuming that the value of $\delta$ is chosen so that at most 20\% of the edges
are allowed to become star edges. The third column illustrates the best
compression ratio together with its improvement over \cite{liakos2014pushing};
here BPL stands for bits per link.
\color{black}
The fourth column reports the corresponding values of $(h,\delta)$ for the
results shown in the third column. Furthermore, the fifth column reports the
encoding and decoding times in seconds for the reported values of $h$ and
$\delta$. As we can see, even in this case,
our compression ratios are better compared to those in \cite{liakos2014pushing}.}
\end{table}

\subsection{Social Networks}
\label{sec:data-social}

In this section, we consider the following social network graphs. These datasets
are accessed via the Laboratory of Web Algorithms (LAW)
website\footnote{http://law.di.unimi.it/\label{foot:law}} \cite{boldi2004webgraph, boldi2011layered}.

\begin{itemize}
\item \verb+dblp-2010+: an undirected simple graph consisting of 326,186 vertices
  and 1,615,400 edges, where each vertex represents a scientist, and two
  vertices are connected if they have a joint article. This graph is extracted
 from
  the DBLP bibliography service.
\item \verb+hollywood-2009+: a simple undirected graph,   consisting of 1,139,905
  vertices and 113,891,327 edges, where each vertex represents an actor, and two
  vertices are connected by an edge if the corresponding actors have appeared
  in a movie together.
\item \verb+amazon-2008+: a simple undirected graph consisting of 735,323 vertices and
  5,158,388 edges, describing similarity among books as reported by the Amazon store.
\item \verb+ljournal-2008+: collected by \cite{chierichetti2009compressing}, is based on the social
  website LiveJournal started in 1999. This dataset consists of 5,363,260
  vertices and 79,023,142 edges, where  vertices represent
  individuals, and there is a directed edge from $x$ to $y$ if $x$ registers $y$
  as their friend. In this case, $x$ does not need to ask for permission from
  $y$ to register them. Therefore, the edges are not necessarily symmetric.
\end{itemize}
These datasets are summarized in Table~\ref{tab:social-graphs}. As we can see in
this table, the average clustering coefficients for these datasets are far away
from zero, meaning that they are not locally tree--like, hence they do not
satisfy the conditions of our Theorem~\ref{thm:optimality-complexity-main}.
Nonetheless, it would be interesting to investigate the performance of our
algorithm on these datasets. 

As some of these datasets are directed, we need to represent them in a way which
is compatible with our setup here, i.e.\ convert each of them to a simple marked graph.
We do this by appropriately employing edge marks. More precisely, if there is a
directed edge from a node $u$ to a node $v$ in the original graph, but there is
not a directed edge from node $v$ to node $u$, we model this by a single marked
edge between nodes $u$ and $v$, with edge mark $1$ towards node $v$ and edge
mark $0$ towards node $u$. Furthermore, if there are two directed edges one from
node $u$ towards node $v$, and another from node $v$ towards node $u$, we model
this by a single marked edge between nodes $u$ and $v$ with edge marks $1$
towards both endpoints. Indeed, this representation preserves the information
regarding directional edges in the original datasets. 

\begin{table}
  \centering
  \begin{tabular}{@{}lccr@{}}
    \toprule
    Dataset & vertices & edges & average clustering \\
    & & & coefficient \\
    \midrule
    \verb+dblp-2010+ & 326,186 & 1,615,400 & 0.568 \\
    \verb+hollywood-2009+ & 1,139,905 & 113,891,327 & 0.766 \\
    \verb+amazon-2008+ & 735,323 & 5,158,388 & 0.389 \\
    \verb+ljournal-2008+ & 5,363,260 & 79,023,142 & 0.266 \\
    \bottomrule
  \end{tabular}
    \caption{\label{tab:social-graphs} Social network datasets and their properties.
}
\end{table}

\begin{table}
  \centering
  \begin{tabular}{@{}lcccr@{}}
    \toprule
    Dataset & \cite{boldi2011layered} & best BPL [relative\%] &  value of & encode/decode \\
    & (BPL) & & $(h, \delta)$  & time (sec) \\
    \midrule
    \emph{dblp-2010} & 6.78 & 5.23 [23\%] & (4,2) & 2.25/7.62 \\
    \emph{amazon-2008} & 9.12 & 8.31 [9\%] & (4,2) & 6.42/20.57 \\
    \emph{hollywood-2009} & 5.14 & 4.67 [9\%] & (4,2) & 33.67/41.88 \\
    \emph{ljournal-2008} & 10.90 & 8.73  [20\%] & (4,2) & 74.76/210.54 \\
    \bottomrule
  \end{tabular}
  \caption{\label{tab:social-first} Comparing the compression ratios of our algorithm
  with those in \cite{boldi2011layered} for social networks. We report the best
  compression ratio of our algorithm in the third column together with relative
  comparison with the results in \cite{boldi2011layered}. The corresponding
  value of the $(h, \delta)$ parameters that achieve the result in the value on the
  third column is given in the fourth column. Finally, 
  the corresponding encoding/decoding times are given in the fifth column. 
  }
\end{table}

\begin{figure}
  \centering
  \hfill
  \subfloat[\emph{dblp-2010}]{
    \scalebox{0.7}{
\begin{tikzpicture}

\definecolor{color0}{rgb}{0.12156862745098,0.466666666666667,0.705882352941177}
\definecolor{color1}{rgb}{1,0.498039215686275,0.0549019607843137}
\definecolor{color2}{rgb}{0.172549019607843,0.627450980392157,0.172549019607843}
\definecolor{color3}{rgb}{0.83921568627451,0.152941176470588,0.156862745098039}

\begin{axis}[
legend cell align={left},
legend style={fill opacity=0.8, draw opacity=1, text opacity=1, at={(0.03,0.97)}, anchor=north west, draw=white!80.0!black},
tick align=outside,
tick pos=both,
x grid style={white!69.01960784313725!black},
xlabel={\(\displaystyle \delta\)},
xmin=-0.44875, xmax=7.08625,
xtick style={color=black},
xtick={0.31875,1.31875,2.31875,3.31875,4.31875,5.31875,6.31875},
xticklabels={2,5,10,20,50,100,200},
y grid style={white!69.01960784313725!black},
ylabel={bits per link},
ymin=0, ymax=37.950632165408,
ytick style={color=black}
]
\draw[fill=color0,draw opacity=0,fill opacity=0.8] (axis cs:-0.10625,0) rectangle (axis cs:0.10625,5.36226816887458);
\addlegendimage{ybar,ybar legend,fill=color0,draw opacity=0,fill opacity=0.8};
\addlegendentry{h=1}

\draw[fill=color0,draw opacity=0,fill opacity=0.8] (axis cs:0.89375,0) rectangle (axis cs:1.10625,6.20000495233379);
\draw[fill=color0,draw opacity=0,fill opacity=0.8] (axis cs:1.89375,0) rectangle (axis cs:2.10625,7.12773554537576);
\draw[fill=color0,draw opacity=0,fill opacity=0.8] (axis cs:2.89375,0) rectangle (axis cs:3.10625,7.83235359663241);
\draw[fill=color0,draw opacity=0,fill opacity=0.8] (axis cs:3.89375,0) rectangle (axis cs:4.10625,8.22447938591061);
\draw[fill=color0,draw opacity=0,fill opacity=0.8] (axis cs:4.89375,0) rectangle (axis cs:5.10625,8.21140522471214);
\draw[fill=color0,draw opacity=0,fill opacity=0.8] (axis cs:5.89375,0) rectangle (axis cs:6.10625,8.17652098551442);
\draw[fill=color1,draw opacity=0,fill opacity=0.8] (axis cs:0.10625,0) rectangle (axis cs:0.31875,5.37676860220379);
\addlegendimage{ybar,ybar legend,fill=color1,draw opacity=0,fill opacity=0.8};
\addlegendentry{h=2}

\draw[fill=color1,draw opacity=0,fill opacity=0.8] (axis cs:1.10625,0) rectangle (axis cs:1.31875,6.28356072799307);
\draw[fill=color1,draw opacity=0,fill opacity=0.8] (axis cs:2.10625,0) rectangle (axis cs:2.31875,7.36915191283892);
\draw[fill=color1,draw opacity=0,fill opacity=0.8] (axis cs:3.10625,0) rectangle (axis cs:3.31875,8.48412034171103);
\draw[fill=color1,draw opacity=0,fill opacity=0.8] (axis cs:4.10625,0) rectangle (axis cs:4.31875,9.29303454252817);
\draw[fill=color1,draw opacity=0,fill opacity=0.8] (axis cs:5.10625,0) rectangle (axis cs:5.31875,9.44305063761298);
\draw[fill=color1,draw opacity=0,fill opacity=0.8] (axis cs:6.10625,0) rectangle (axis cs:6.31875,9.49764516528414);
\draw[fill=color2,draw opacity=0,fill opacity=0.8] (axis cs:0.31875,0) rectangle (axis cs:0.53125,5.23558747059552);
\addlegendimage{ybar,ybar legend,fill=color2,draw opacity=0,fill opacity=0.8};
\addlegendentry{h=3}

\draw[fill=color2,draw opacity=0,fill opacity=0.8] (axis cs:1.31875,0) rectangle (axis cs:1.53125,5.52060418472205);
\draw[fill=color2,draw opacity=0,fill opacity=0.8] (axis cs:2.31875,0) rectangle (axis cs:2.53125,6.46756964219388);
\draw[fill=color2,draw opacity=0,fill opacity=0.8] (axis cs:3.31875,0) rectangle (axis cs:3.53125,9.35547356691841);
\draw[fill=color2,draw opacity=0,fill opacity=0.8] (axis cs:4.31875,0) rectangle (axis cs:4.53125,18.6339878667822);
\draw[fill=color2,draw opacity=0,fill opacity=0.8] (axis cs:5.31875,0) rectangle (axis cs:5.53125,28.5670422186455);
\draw[fill=color2,draw opacity=0,fill opacity=0.8] (axis cs:6.31875,0) rectangle (axis cs:6.53125,35.8358202302835);
\draw[fill=color3,draw opacity=0,fill opacity=0.8] (axis cs:0.53125,0) rectangle (axis cs:0.74375,5.23091246750031);
\addlegendimage{ybar,ybar legend,fill=color3,draw opacity=0,fill opacity=0.8};
\addlegendentry{h=4}

\draw[fill=color3,draw opacity=0,fill opacity=0.8] (axis cs:1.53125,0) rectangle (axis cs:1.74375,5.48474928810202);
\draw[fill=color3,draw opacity=0,fill opacity=0.8] (axis cs:2.53125,0) rectangle (axis cs:2.74375,6.09640212950353);
\draw[fill=color3,draw opacity=0,fill opacity=0.8] (axis cs:3.53125,0) rectangle (axis cs:3.74375,7.31152655688993);
\draw[fill=color3,draw opacity=0,fill opacity=0.8] (axis cs:4.53125,0) rectangle (axis cs:4.74375,11.2774743097685);
\draw[fill=color3,draw opacity=0,fill opacity=0.8] (axis cs:5.53125,0) rectangle (axis cs:5.74375,18.6251727126408);
\draw[fill=color3,draw opacity=0,fill opacity=0.8] (axis cs:6.53125,0) rectangle (axis cs:6.74375,36.1434592051504);
\end{axis}

\end{tikzpicture}
    }
  }
  \hfill\hfill\hfill
  \subfloat[\emph{hollywood-2009}]{
    \scalebox{0.7}{
\begin{tikzpicture}

\definecolor{color0}{rgb}{0.12156862745098,0.466666666666667,0.705882352941177}
\definecolor{color1}{rgb}{1,0.498039215686275,0.0549019607843137}
\definecolor{color2}{rgb}{0.172549019607843,0.627450980392157,0.172549019607843}
\definecolor{color3}{rgb}{0.83921568627451,0.152941176470588,0.156862745098039}

\begin{axis}[
legend cell align={left},
legend style={fill opacity=0.8, draw opacity=1, text opacity=1, draw=white!80.0!black},
tick align=outside,
tick pos=both,
x grid style={white!69.01960784313725!black},
xlabel={\(\displaystyle \delta\)},
xmin=-0.44875, xmax=7.08625,
xtick style={color=black},
xtick={0.31875,1.31875,2.31875,3.31875,4.31875,5.31875,6.31875},
xticklabels={2,5,10,20,50,100,200},
y grid style={white!69.01960784313725!black},
ylabel={bits per link},
ymin=0, ymax=5.66519815859201,
ytick style={color=black}
]
\draw[fill=color0,draw opacity=0,fill opacity=0.8] (axis cs:-0.10625,0) rectangle (axis cs:0.10625,4.66990937773515);
\addlegendimage{ybar,ybar legend,fill=color0,draw opacity=0,fill opacity=0.8};
\addlegendentry{h=1}

\draw[fill=color0,draw opacity=0,fill opacity=0.8] (axis cs:0.89375,0) rectangle (axis cs:1.10625,4.67843989560329);
\draw[fill=color0,draw opacity=0,fill opacity=0.8] (axis cs:1.89375,0) rectangle (axis cs:2.10625,4.69939966543721);
\draw[fill=color0,draw opacity=0,fill opacity=0.8] (axis cs:2.89375,0) rectangle (axis cs:3.10625,4.75592765724821);
\draw[fill=color0,draw opacity=0,fill opacity=0.8] (axis cs:3.89375,0) rectangle (axis cs:4.10625,4.93771217539681);
\draw[fill=color0,draw opacity=0,fill opacity=0.8] (axis cs:4.89375,0) rectangle (axis cs:5.10625,5.16701174269398);
\draw[fill=color0,draw opacity=0,fill opacity=0.8] (axis cs:5.89375,0) rectangle (axis cs:6.10625,5.39542681770667);
\draw[fill=color1,draw opacity=0,fill opacity=0.8] (axis cs:0.10625,0) rectangle (axis cs:0.31875,4.66995629965748);
\addlegendimage{ybar,ybar legend,fill=color1,draw opacity=0,fill opacity=0.8};
\addlegendentry{h=2}

\draw[fill=color1,draw opacity=0,fill opacity=0.8] (axis cs:1.10625,0) rectangle (axis cs:1.31875,4.67847108322831);
\draw[fill=color1,draw opacity=0,fill opacity=0.8] (axis cs:2.10625,0) rectangle (axis cs:2.31875,4.69783775545964);
\draw[fill=color1,draw opacity=0,fill opacity=0.8] (axis cs:3.10625,0) rectangle (axis cs:3.31875,4.74582764322344);
\draw[fill=color1,draw opacity=0,fill opacity=0.8] (axis cs:4.10625,0) rectangle (axis cs:4.31875,4.87609273355819);
\draw[fill=color1,draw opacity=0,fill opacity=0.8] (axis cs:5.10625,0) rectangle (axis cs:5.31875,5.0170166864418);
\draw[fill=color1,draw opacity=0,fill opacity=0.8] (axis cs:6.10625,0) rectangle (axis cs:6.31875,5.21550344215412);
\draw[fill=color2,draw opacity=0,fill opacity=0.8] (axis cs:0.31875,0) rectangle (axis cs:0.53125,4.66882680188633);
\addlegendimage{ybar,ybar legend,fill=color2,draw opacity=0,fill opacity=0.8};
\addlegendentry{h=3}

\draw[fill=color2,draw opacity=0,fill opacity=0.8] (axis cs:1.31875,0) rectangle (axis cs:1.53125,4.67078909353651);
\draw[fill=color2,draw opacity=0,fill opacity=0.8] (axis cs:2.31875,0) rectangle (axis cs:2.53125,4.67198574304082);
\draw[fill=color2,draw opacity=0,fill opacity=0.8] (axis cs:3.31875,0) rectangle (axis cs:3.53125,4.6734223405791);
\draw[fill=color2,draw opacity=0,fill opacity=0.8] (axis cs:4.31875,0) rectangle (axis cs:4.53125,4.6777242660453);
\draw[fill=color2,draw opacity=0,fill opacity=0.8] (axis cs:5.31875,0) rectangle (axis cs:5.53125,4.686278666329);
\draw[fill=color2,draw opacity=0,fill opacity=0.8] (axis cs:6.31875,0) rectangle (axis cs:6.53125,4.70889390901557);
\draw[fill=color3,draw opacity=0,fill opacity=0.8] (axis cs:0.53125,0) rectangle (axis cs:0.74375,4.66882567800795);
\addlegendimage{ybar,ybar legend,fill=color3,draw opacity=0,fill opacity=0.8};
\addlegendentry{h=4}

\draw[fill=color3,draw opacity=0,fill opacity=0.8] (axis cs:1.53125,0) rectangle (axis cs:1.74375,4.67048480346532);
\draw[fill=color3,draw opacity=0,fill opacity=0.8] (axis cs:2.53125,0) rectangle (axis cs:2.74375,4.6715207383614);
\draw[fill=color3,draw opacity=0,fill opacity=0.8] (axis cs:3.53125,0) rectangle (axis cs:3.74375,4.67174776179401);
\draw[fill=color3,draw opacity=0,fill opacity=0.8] (axis cs:4.53125,0) rectangle (axis cs:4.74375,4.67190426185832);
\draw[fill=color3,draw opacity=0,fill opacity=0.8] (axis cs:5.53125,0) rectangle (axis cs:5.74375,4.67233414533839);
\draw[fill=color3,draw opacity=0,fill opacity=0.8] (axis cs:6.53125,0) rectangle (axis cs:6.74375,4.67300425782202);
\end{axis}

\end{tikzpicture}
    }
  }
  \hfill

  \hfill
  \subfloat[\emph{amazon-2008}]{
    \scalebox{0.7}{
\begin{tikzpicture}

\definecolor{color0}{rgb}{0.12156862745098,0.466666666666667,0.705882352941177}
\definecolor{color1}{rgb}{1,0.498039215686275,0.0549019607843137}
\definecolor{color2}{rgb}{0.172549019607843,0.627450980392157,0.172549019607843}
\definecolor{color3}{rgb}{0.83921568627451,0.152941176470588,0.156862745098039}

\begin{axis}[
legend cell align={left},
legend style={fill opacity=0.8, draw opacity=1, text opacity=1, at={(0.03,0.97)}, anchor=north west, draw=white!80.0!black},
tick align=outside,
tick pos=both,
x grid style={white!69.01960784313725!black},
xlabel={\(\displaystyle \delta\)},
xmin=-0.44875, xmax=7.08625,
xtick style={color=black},
xtick={0.31875,1.31875,2.31875,3.31875,4.31875,5.31875,6.31875},
xticklabels={2,5,10,20,50,100,200},
y grid style={white!69.01960784313725!black},
ylabel={bits per link},
ymin=0, ymax=95.158542707528,
ytick style={color=black}
]
\draw[fill=color0,draw opacity=0,fill opacity=0.8] (axis cs:-0.10625,0) rectangle (axis cs:0.10625,8.32822657000598);
\addlegendimage{ybar,ybar legend,fill=color0,draw opacity=0,fill opacity=0.8};
\addlegendentry{h=1}

\draw[fill=color0,draw opacity=0,fill opacity=0.8] (axis cs:0.89375,0) rectangle (axis cs:1.10625,8.43387818054788);
\draw[fill=color0,draw opacity=0,fill opacity=0.8] (axis cs:1.89375,0) rectangle (axis cs:2.10625,9.32445097189277);
\draw[fill=color0,draw opacity=0,fill opacity=0.8] (axis cs:2.89375,0) rectangle (axis cs:3.10625,11.9046678923726);
\draw[fill=color0,draw opacity=0,fill opacity=0.8] (axis cs:3.89375,0) rectangle (axis cs:4.10625,12.3375907357105);
\draw[fill=color0,draw opacity=0,fill opacity=0.8] (axis cs:4.89375,0) rectangle (axis cs:5.10625,12.2470011949469);
\draw[fill=color0,draw opacity=0,fill opacity=0.8] (axis cs:5.89375,0) rectangle (axis cs:6.10625,12.2178075786467);
\draw[fill=color1,draw opacity=0,fill opacity=0.8] (axis cs:0.10625,0) rectangle (axis cs:0.31875,8.32979605256526);
\addlegendimage{ybar,ybar legend,fill=color1,draw opacity=0,fill opacity=0.8};
\addlegendentry{h=2}

\draw[fill=color1,draw opacity=0,fill opacity=0.8] (axis cs:1.10625,0) rectangle (axis cs:1.31875,8.48300980849056);
\draw[fill=color1,draw opacity=0,fill opacity=0.8] (axis cs:2.10625,0) rectangle (axis cs:2.31875,10.2112520423047);
\draw[fill=color1,draw opacity=0,fill opacity=0.8] (axis cs:3.10625,0) rectangle (axis cs:3.31875,17.7490425303409);
\draw[fill=color1,draw opacity=0,fill opacity=0.8] (axis cs:4.10625,0) rectangle (axis cs:4.31875,21.5402796377473);
\draw[fill=color1,draw opacity=0,fill opacity=0.8] (axis cs:5.10625,0) rectangle (axis cs:5.31875,21.9507768706038);
\draw[fill=color1,draw opacity=0,fill opacity=0.8] (axis cs:6.10625,0) rectangle (axis cs:6.31875,22.0252807660067);
\draw[fill=color2,draw opacity=0,fill opacity=0.8] (axis cs:0.31875,0) rectangle (axis cs:0.53125,8.3110677211563);
\addlegendimage{ybar,ybar legend,fill=color2,draw opacity=0,fill opacity=0.8};
\addlegendentry{h=3}

\draw[fill=color2,draw opacity=0,fill opacity=0.8] (axis cs:1.31875,0) rectangle (axis cs:1.53125,8.33112359907785);
\draw[fill=color2,draw opacity=0,fill opacity=0.8] (axis cs:2.31875,0) rectangle (axis cs:2.53125,8.59449890159484);
\draw[fill=color2,draw opacity=0,fill opacity=0.8] (axis cs:3.31875,0) rectangle (axis cs:3.53125,22.5858651966467);
\draw[fill=color2,draw opacity=0,fill opacity=0.8] (axis cs:4.31875,0) rectangle (axis cs:4.53125,73.7030622744935);
\draw[fill=color2,draw opacity=0,fill opacity=0.8] (axis cs:5.31875,0) rectangle (axis cs:5.53125,87.2500106622456);
\draw[fill=color2,draw opacity=0,fill opacity=0.8] (axis cs:6.31875,0) rectangle (axis cs:6.53125,90.6271835309791);
\draw[fill=color3,draw opacity=0,fill opacity=0.8] (axis cs:0.53125,0) rectangle (axis cs:0.74375,8.31015580836494);
\addlegendimage{ybar,ybar legend,fill=color3,draw opacity=0,fill opacity=0.8};
\addlegendentry{h=4}

\draw[fill=color3,draw opacity=0,fill opacity=0.8] (axis cs:1.53125,0) rectangle (axis cs:1.74375,8.31423770371674);
\draw[fill=color3,draw opacity=0,fill opacity=0.8] (axis cs:2.53125,0) rectangle (axis cs:2.74375,8.34512487234384);
\draw[fill=color3,draw opacity=0,fill opacity=0.8] (axis cs:3.53125,0) rectangle (axis cs:3.74375,12.4993715090838);
\draw[fill=color3,draw opacity=0,fill opacity=0.8] (axis cs:4.53125,0) rectangle (axis cs:4.74375,41.5192529138948);
\draw[fill=color3,draw opacity=0,fill opacity=0.8] (axis cs:5.53125,0) rectangle (axis cs:5.74375,67.3688601943088);
\draw[fill=color3,draw opacity=0,fill opacity=0.8] (axis cs:6.53125,0) rectangle (axis cs:6.74375,82.6358560077295);
\end{axis}

\end{tikzpicture}
    }
  }
  \hfill\hfill\hfill
  \subfloat[\emph{ljournal-2008}]{
    \scalebox{0.7}{
\begin{tikzpicture}

\definecolor{color0}{rgb}{0.12156862745098,0.466666666666667,0.705882352941177}
\definecolor{color1}{rgb}{1,0.498039215686275,0.0549019607843137}
\definecolor{color2}{rgb}{0.172549019607843,0.627450980392157,0.172549019607843}
\definecolor{color3}{rgb}{0.83921568627451,0.152941176470588,0.156862745098039}

\begin{axis}[
legend cell align={left},
legend style={fill opacity=0.8, draw opacity=1, text opacity=1, at={(0.03,0.97)}, anchor=north west, draw=white!80.0!black},
tick align=outside,
tick pos=both,
x grid style={white!69.01960784313725!black},
xlabel={\(\displaystyle \delta\)},
xmin=-0.44875, xmax=7.08625,
xtick style={color=black},
xtick={0.31875,1.31875,2.31875,3.31875,4.31875,5.31875,6.31875},
xticklabels={2,5,10,20,50,100,200},
y grid style={white!69.01960784313725!black},
ylabel={bits per link},
ymin=0, ymax=47.3715621785831,
ytick style={color=black}
]
\draw[fill=color0,draw opacity=0,fill opacity=0.8] (axis cs:-0.10625,0) rectangle (axis cs:0.10625,8.75813113075155);
\addlegendimage{ybar,ybar legend,fill=color0,draw opacity=0,fill opacity=0.8};
\addlegendentry{h=1}

\draw[fill=color0,draw opacity=0,fill opacity=0.8] (axis cs:0.89375,0) rectangle (axis cs:1.10625,8.85796861886357);
\draw[fill=color0,draw opacity=0,fill opacity=0.8] (axis cs:1.89375,0) rectangle (axis cs:2.10625,9.04178474705549);
\draw[fill=color0,draw opacity=0,fill opacity=0.8] (axis cs:2.89375,0) rectangle (axis cs:3.10625,9.40403569374652);
\draw[fill=color0,draw opacity=0,fill opacity=0.8] (axis cs:3.89375,0) rectangle (axis cs:4.10625,10.1690940104609);
\draw[fill=color0,draw opacity=0,fill opacity=0.8] (axis cs:4.89375,0) rectangle (axis cs:5.10625,10.7692697918794);
\draw[fill=color0,draw opacity=0,fill opacity=0.8] (axis cs:5.89375,0) rectangle (axis cs:6.10625,11.0543382848533);
\draw[fill=color1,draw opacity=0,fill opacity=0.8] (axis cs:0.10625,0) rectangle (axis cs:0.31875,8.76255150674723);
\addlegendimage{ybar,ybar legend,fill=color1,draw opacity=0,fill opacity=0.8};
\addlegendentry{h=2}

\draw[fill=color1,draw opacity=0,fill opacity=0.8] (axis cs:1.10625,0) rectangle (axis cs:1.31875,8.96301005090382);
\draw[fill=color1,draw opacity=0,fill opacity=0.8] (axis cs:2.10625,0) rectangle (axis cs:2.31875,9.76038538179107);
\draw[fill=color1,draw opacity=0,fill opacity=0.8] (axis cs:3.10625,0) rectangle (axis cs:3.31875,12.6082822674907);
\draw[fill=color1,draw opacity=0,fill opacity=0.8] (axis cs:4.10625,0) rectangle (axis cs:4.31875,23.3746810016742);
\draw[fill=color1,draw opacity=0,fill opacity=0.8] (axis cs:5.10625,0) rectangle (axis cs:5.31875,35.0297567262005);
\draw[fill=color1,draw opacity=0,fill opacity=0.8] (axis cs:6.10625,0) rectangle (axis cs:6.31875,45.1157735034125);
\draw[fill=color2,draw opacity=0,fill opacity=0.8] (axis cs:0.31875,0) rectangle (axis cs:0.53125,8.735676948912);
\addlegendimage{ybar,ybar legend,fill=color2,draw opacity=0,fill opacity=0.8};
\addlegendentry{h=3}

\draw[fill=color2,draw opacity=0,fill opacity=0.8] (axis cs:1.31875,0) rectangle (axis cs:1.53125,8.7703499311632);
\draw[fill=color2,draw opacity=0,fill opacity=0.8] (axis cs:2.31875,0) rectangle (axis cs:2.53125,8.87928895563277);
\draw[fill=color2,draw opacity=0,fill opacity=0.8] (axis cs:3.31875,0) rectangle (axis cs:3.53125,9.19161670387644);
\draw[fill=color2,draw opacity=0,fill opacity=0.8] (axis cs:4.31875,0) rectangle (axis cs:4.53125,10.727933040172);
\draw[fill=color2,draw opacity=0,fill opacity=0.8] (axis cs:5.31875,0) rectangle (axis cs:5.53125,14.7600671205911);
\draw[fill=color2,draw opacity=0,fill opacity=0.8] (axis cs:6.31875,0) rectangle (axis cs:6.53125,22.8931726354287);
\draw[fill=color3,draw opacity=0,fill opacity=0.8] (axis cs:0.53125,0) rectangle (axis cs:0.74375,8.73383485561736);
\addlegendimage{ybar,ybar legend,fill=color3,draw opacity=0,fill opacity=0.8};
\addlegendentry{h=4}

\draw[fill=color3,draw opacity=0,fill opacity=0.8] (axis cs:1.53125,0) rectangle (axis cs:1.74375,8.7401641407779);
\draw[fill=color3,draw opacity=0,fill opacity=0.8] (axis cs:2.53125,0) rectangle (axis cs:2.74375,8.75492396898114);
\draw[fill=color3,draw opacity=0,fill opacity=0.8] (axis cs:3.53125,0) rectangle (axis cs:3.74375,8.78966295721322);
\draw[fill=color3,draw opacity=0,fill opacity=0.8] (axis cs:4.53125,0) rectangle (axis cs:4.74375,8.92762396109231);
\draw[fill=color3,draw opacity=0,fill opacity=0.8] (axis cs:5.53125,0) rectangle (axis cs:5.74375,9.2647926350486);
\draw[fill=color3,draw opacity=0,fill opacity=0.8] (axis cs:6.53125,0) rectangle (axis cs:6.74375,9.9910495586217);
\end{axis}

\end{tikzpicture}
    }
  }
  \hfill

  \caption{\label{fig:social-bpl} Compression ratios of our algorithm for
    different values of the parameters $h$ and $\delta$ for the datasets $(a)$
    \emph{dblp-2010},  $(b)$ \emph{hollywood-2009}, $(c)$ \emph{amazon-2008},
    and $(d)$ \emph{ljournal-2008}.}
\end{figure}

\begin{figure}
  \centering
  \hfill
  \subfloat[\emph{dblp-2010}]{
    \scalebox{0.7}{
\begin{tikzpicture}

\definecolor{color0}{rgb}{0.12156862745098,0.466666666666667,0.705882352941177}
\definecolor{color1}{rgb}{1,0.498039215686275,0.0549019607843137}
\definecolor{color2}{rgb}{0.172549019607843,0.627450980392157,0.172549019607843}
\definecolor{color3}{rgb}{0.83921568627451,0.152941176470588,0.156862745098039}

\begin{axis}[
legend cell align={left},
legend style={fill opacity=0.8, draw opacity=1, text opacity=1, at={(0.03,0.97)}, anchor=north west, draw=white!80.0!black},
tick align=outside,
tick pos=both,
x grid style={white!69.01960784313725!black},
xlabel={\(\displaystyle \delta\)},
xmin=-0.44875, xmax=7.08625,
xtick style={color=black},
xtick={0.31875,1.31875,2.31875,3.31875,4.31875,5.31875,6.31875},
xticklabels={2,5,10,20,50,100,200},
y grid style={white!69.01960784313725!black},
ylabel={Encode time (s)},
ymin=0, ymax=10.09645245,
ytick style={color=black}
]
\draw[fill=color0,draw opacity=0,fill opacity=0.8] (axis cs:-0.10625,0) rectangle (axis cs:0.10625,1.965534);
\addlegendimage{ybar,ybar legend,fill=color0,draw opacity=0,fill opacity=0.8};
\addlegendentry{h=1}

\draw[fill=color0,draw opacity=0,fill opacity=0.8] (axis cs:0.89375,0) rectangle (axis cs:1.10625,2.19981);
\draw[fill=color0,draw opacity=0,fill opacity=0.8] (axis cs:1.89375,0) rectangle (axis cs:2.10625,2.583364);
\draw[fill=color0,draw opacity=0,fill opacity=0.8] (axis cs:2.89375,0) rectangle (axis cs:3.10625,2.985231);
\draw[fill=color0,draw opacity=0,fill opacity=0.8] (axis cs:3.89375,0) rectangle (axis cs:4.10625,3.533922);
\draw[fill=color0,draw opacity=0,fill opacity=0.8] (axis cs:4.89375,0) rectangle (axis cs:5.10625,3.84356);
\draw[fill=color0,draw opacity=0,fill opacity=0.8] (axis cs:5.89375,0) rectangle (axis cs:6.10625,3.955763);
\draw[fill=color1,draw opacity=0,fill opacity=0.8] (axis cs:0.10625,0) rectangle (axis cs:0.31875,2.139204);
\addlegendimage{ybar,ybar legend,fill=color1,draw opacity=0,fill opacity=0.8};
\addlegendentry{h=2}

\draw[fill=color1,draw opacity=0,fill opacity=0.8] (axis cs:1.10625,0) rectangle (axis cs:1.31875,2.471301);
\draw[fill=color1,draw opacity=0,fill opacity=0.8] (axis cs:2.10625,0) rectangle (axis cs:2.31875,2.884702);
\draw[fill=color1,draw opacity=0,fill opacity=0.8] (axis cs:3.10625,0) rectangle (axis cs:3.31875,3.494993);
\draw[fill=color1,draw opacity=0,fill opacity=0.8] (axis cs:4.10625,0) rectangle (axis cs:4.31875,4.159016);
\draw[fill=color1,draw opacity=0,fill opacity=0.8] (axis cs:5.10625,0) rectangle (axis cs:5.31875,4.4303);
\draw[fill=color1,draw opacity=0,fill opacity=0.8] (axis cs:6.10625,0) rectangle (axis cs:6.31875,4.545951);
\draw[fill=color2,draw opacity=0,fill opacity=0.8] (axis cs:0.31875,0) rectangle (axis cs:0.53125,2.265874);
\addlegendimage{ybar,ybar legend,fill=color2,draw opacity=0,fill opacity=0.8};
\addlegendentry{h=3}

\draw[fill=color2,draw opacity=0,fill opacity=0.8] (axis cs:1.31875,0) rectangle (axis cs:1.53125,2.356486);
\draw[fill=color2,draw opacity=0,fill opacity=0.8] (axis cs:2.31875,0) rectangle (axis cs:2.53125,2.701516);
\draw[fill=color2,draw opacity=0,fill opacity=0.8] (axis cs:3.31875,0) rectangle (axis cs:3.53125,3.468285);
\draw[fill=color2,draw opacity=0,fill opacity=0.8] (axis cs:4.31875,0) rectangle (axis cs:4.53125,5.650814);
\draw[fill=color2,draw opacity=0,fill opacity=0.8] (axis cs:5.31875,0) rectangle (axis cs:5.53125,7.979529);
\draw[fill=color2,draw opacity=0,fill opacity=0.8] (axis cs:6.31875,0) rectangle (axis cs:6.53125,9.615669);
\draw[fill=color3,draw opacity=0,fill opacity=0.8] (axis cs:0.53125,0) rectangle (axis cs:0.74375,2.250078);
\addlegendimage{ybar,ybar legend,fill=color3,draw opacity=0,fill opacity=0.8};
\addlegendentry{h=4}

\draw[fill=color3,draw opacity=0,fill opacity=0.8] (axis cs:1.53125,0) rectangle (axis cs:1.74375,2.396126);
\draw[fill=color3,draw opacity=0,fill opacity=0.8] (axis cs:2.53125,0) rectangle (axis cs:2.74375,2.762621);
\draw[fill=color3,draw opacity=0,fill opacity=0.8] (axis cs:3.53125,0) rectangle (axis cs:3.74375,3.037032);
\draw[fill=color3,draw opacity=0,fill opacity=0.8] (axis cs:4.53125,0) rectangle (axis cs:4.74375,4.152027);
\draw[fill=color3,draw opacity=0,fill opacity=0.8] (axis cs:5.53125,0) rectangle (axis cs:5.74375,5.957357);
\draw[fill=color3,draw opacity=0,fill opacity=0.8] (axis cs:6.53125,0) rectangle (axis cs:6.74375,9.447404);
\end{axis}

\end{tikzpicture}
    }
  }
  \hfill\hfill\hfill
  \subfloat[\emph{hollywood-2009}]{
    \scalebox{0.7}{
\begin{tikzpicture}

\definecolor{color0}{rgb}{0.12156862745098,0.466666666666667,0.705882352941177}
\definecolor{color1}{rgb}{1,0.498039215686275,0.0549019607843137}
\definecolor{color2}{rgb}{0.172549019607843,0.627450980392157,0.172549019607843}
\definecolor{color3}{rgb}{0.83921568627451,0.152941176470588,0.156862745098039}

\begin{axis}[
legend cell align={left},
legend style={fill opacity=0.8, draw opacity=1, text opacity=1, at={(0.03,0.97)}, anchor=north west, draw=white!80.0!black},
tick align=outside,
tick pos=both,
x grid style={white!69.01960784313725!black},
xlabel={\(\displaystyle \delta\)},
xmin=-0.44875, xmax=7.08625,
xtick style={color=black},
xtick={0.31875,1.31875,2.31875,3.31875,4.31875,5.31875,6.31875},
xticklabels={2,5,10,20,50,100,200},
y grid style={white!69.01960784313725!black},
ylabel={Encode time (s)},
ymin=0, ymax=74.61427155,
ytick style={color=black}
]
\draw[fill=color0,draw opacity=0,fill opacity=0.8] (axis cs:-0.10625,0) rectangle (axis cs:0.10625,24.138763);
\addlegendimage{ybar,ybar legend,fill=color0,draw opacity=0,fill opacity=0.8};
\addlegendentry{h=1}

\draw[fill=color0,draw opacity=0,fill opacity=0.8] (axis cs:0.89375,0) rectangle (axis cs:1.10625,21.914867);
\draw[fill=color0,draw opacity=0,fill opacity=0.8] (axis cs:1.89375,0) rectangle (axis cs:2.10625,21.524487);
\draw[fill=color0,draw opacity=0,fill opacity=0.8] (axis cs:2.89375,0) rectangle (axis cs:3.10625,22.169771);
\draw[fill=color0,draw opacity=0,fill opacity=0.8] (axis cs:3.89375,0) rectangle (axis cs:4.10625,29.46246);
\draw[fill=color0,draw opacity=0,fill opacity=0.8] (axis cs:4.89375,0) rectangle (axis cs:5.10625,44.935295);
\draw[fill=color0,draw opacity=0,fill opacity=0.8] (axis cs:5.89375,0) rectangle (axis cs:6.10625,68.786446);
\draw[fill=color1,draw opacity=0,fill opacity=0.8] (axis cs:0.10625,0) rectangle (axis cs:0.31875,23.997644);
\addlegendimage{ybar,ybar legend,fill=color1,draw opacity=0,fill opacity=0.8};
\addlegendentry{h=2}

\draw[fill=color1,draw opacity=0,fill opacity=0.8] (axis cs:1.10625,0) rectangle (axis cs:1.31875,21.128117);
\draw[fill=color1,draw opacity=0,fill opacity=0.8] (axis cs:2.10625,0) rectangle (axis cs:2.31875,22.264208);
\draw[fill=color1,draw opacity=0,fill opacity=0.8] (axis cs:3.10625,0) rectangle (axis cs:3.31875,23.414709);
\draw[fill=color1,draw opacity=0,fill opacity=0.8] (axis cs:4.10625,0) rectangle (axis cs:4.31875,30.493858);
\draw[fill=color1,draw opacity=0,fill opacity=0.8] (axis cs:5.10625,0) rectangle (axis cs:5.31875,44.265831);
\draw[fill=color1,draw opacity=0,fill opacity=0.8] (axis cs:6.10625,0) rectangle (axis cs:6.31875,71.061211);
\draw[fill=color2,draw opacity=0,fill opacity=0.8] (axis cs:0.31875,0) rectangle (axis cs:0.53125,28.075876);
\addlegendimage{ybar,ybar legend,fill=color2,draw opacity=0,fill opacity=0.8};
\addlegendentry{h=3}

\draw[fill=color2,draw opacity=0,fill opacity=0.8] (axis cs:1.31875,0) rectangle (axis cs:1.53125,24.635887);
\draw[fill=color2,draw opacity=0,fill opacity=0.8] (axis cs:2.31875,0) rectangle (axis cs:2.53125,27.815092);
\draw[fill=color2,draw opacity=0,fill opacity=0.8] (axis cs:3.31875,0) rectangle (axis cs:3.53125,28.686848);
\draw[fill=color2,draw opacity=0,fill opacity=0.8] (axis cs:4.31875,0) rectangle (axis cs:4.53125,30.401709);
\draw[fill=color2,draw opacity=0,fill opacity=0.8] (axis cs:5.31875,0) rectangle (axis cs:5.53125,38.918716);
\draw[fill=color2,draw opacity=0,fill opacity=0.8] (axis cs:6.31875,0) rectangle (axis cs:6.53125,58.367096);
\draw[fill=color3,draw opacity=0,fill opacity=0.8] (axis cs:0.53125,0) rectangle (axis cs:0.74375,33.67099);
\addlegendimage{ybar,ybar legend,fill=color3,draw opacity=0,fill opacity=0.8};
\addlegendentry{h=4}

\draw[fill=color3,draw opacity=0,fill opacity=0.8] (axis cs:1.53125,0) rectangle (axis cs:1.74375,34.794197);
\draw[fill=color3,draw opacity=0,fill opacity=0.8] (axis cs:2.53125,0) rectangle (axis cs:2.74375,28.852343);
\draw[fill=color3,draw opacity=0,fill opacity=0.8] (axis cs:3.53125,0) rectangle (axis cs:3.74375,30.616114);
\draw[fill=color3,draw opacity=0,fill opacity=0.8] (axis cs:4.53125,0) rectangle (axis cs:4.74375,32.265781);
\draw[fill=color3,draw opacity=0,fill opacity=0.8] (axis cs:5.53125,0) rectangle (axis cs:5.74375,38.817451);
\draw[fill=color3,draw opacity=0,fill opacity=0.8] (axis cs:6.53125,0) rectangle (axis cs:6.74375,48.37492);
\end{axis}

\end{tikzpicture}
    }
  }
  \hfill

  \hfill
  \subfloat[\emph{amazon-2008}]{
    \scalebox{0.7}{
\begin{tikzpicture}

\definecolor{color0}{rgb}{0.12156862745098,0.466666666666667,0.705882352941177}
\definecolor{color1}{rgb}{1,0.498039215686275,0.0549019607843137}
\definecolor{color2}{rgb}{0.172549019607843,0.627450980392157,0.172549019607843}
\definecolor{color3}{rgb}{0.83921568627451,0.152941176470588,0.156862745098039}

\begin{axis}[
legend cell align={left},
legend style={fill opacity=0.8, draw opacity=1, text opacity=1, at={(0.03,0.97)}, anchor=north west, draw=white!80.0!black},
tick align=outside,
tick pos=both,
x grid style={white!69.01960784313725!black},
xlabel={\(\displaystyle \delta\)},
xmin=-0.44875, xmax=7.08625,
xtick style={color=black},
xtick={0.31875,1.31875,2.31875,3.31875,4.31875,5.31875,6.31875},
xticklabels={2,5,10,20,50,100,200},
y grid style={white!69.01960784313725!black},
ylabel={Encode time (s)},
ymin=0, ymax=57.49042005,
ytick style={color=black}
]
\draw[fill=color0,draw opacity=0,fill opacity=0.8] (axis cs:-0.10625,0) rectangle (axis cs:0.10625,5.861033);
\addlegendimage{ybar,ybar legend,fill=color0,draw opacity=0,fill opacity=0.8};
\addlegendentry{h=1}

\draw[fill=color0,draw opacity=0,fill opacity=0.8] (axis cs:0.89375,0) rectangle (axis cs:1.10625,6.488851);
\draw[fill=color0,draw opacity=0,fill opacity=0.8] (axis cs:1.89375,0) rectangle (axis cs:2.10625,7.8356);
\draw[fill=color0,draw opacity=0,fill opacity=0.8] (axis cs:2.89375,0) rectangle (axis cs:3.10625,13.096467);
\draw[fill=color0,draw opacity=0,fill opacity=0.8] (axis cs:3.89375,0) rectangle (axis cs:4.10625,15.299745);
\draw[fill=color0,draw opacity=0,fill opacity=0.8] (axis cs:4.89375,0) rectangle (axis cs:5.10625,15.872508);
\draw[fill=color0,draw opacity=0,fill opacity=0.8] (axis cs:5.89375,0) rectangle (axis cs:6.10625,15.276147);
\draw[fill=color1,draw opacity=0,fill opacity=0.8] (axis cs:0.10625,0) rectangle (axis cs:0.31875,6.497275);
\addlegendimage{ybar,ybar legend,fill=color1,draw opacity=0,fill opacity=0.8};
\addlegendentry{h=2}

\draw[fill=color1,draw opacity=0,fill opacity=0.8] (axis cs:1.10625,0) rectangle (axis cs:1.31875,6.240958);
\draw[fill=color1,draw opacity=0,fill opacity=0.8] (axis cs:2.10625,0) rectangle (axis cs:2.31875,8.457186);
\draw[fill=color1,draw opacity=0,fill opacity=0.8] (axis cs:3.10625,0) rectangle (axis cs:3.31875,17.631184);
\draw[fill=color1,draw opacity=0,fill opacity=0.8] (axis cs:4.10625,0) rectangle (axis cs:4.31875,23.727114);
\draw[fill=color1,draw opacity=0,fill opacity=0.8] (axis cs:5.10625,0) rectangle (axis cs:5.31875,24.616974);
\draw[fill=color1,draw opacity=0,fill opacity=0.8] (axis cs:6.10625,0) rectangle (axis cs:6.31875,24.824755);
\draw[fill=color2,draw opacity=0,fill opacity=0.8] (axis cs:0.31875,0) rectangle (axis cs:0.53125,5.980128);
\addlegendimage{ybar,ybar legend,fill=color2,draw opacity=0,fill opacity=0.8};
\addlegendentry{h=3}

\draw[fill=color2,draw opacity=0,fill opacity=0.8] (axis cs:1.31875,0) rectangle (axis cs:1.53125,6.167818);
\draw[fill=color2,draw opacity=0,fill opacity=0.8] (axis cs:2.31875,0) rectangle (axis cs:2.53125,6.786859);
\draw[fill=color2,draw opacity=0,fill opacity=0.8] (axis cs:3.31875,0) rectangle (axis cs:3.53125,15.521864);
\draw[fill=color2,draw opacity=0,fill opacity=0.8] (axis cs:4.31875,0) rectangle (axis cs:4.53125,43.244881);
\draw[fill=color2,draw opacity=0,fill opacity=0.8] (axis cs:5.31875,0) rectangle (axis cs:5.53125,50.709923);
\draw[fill=color2,draw opacity=0,fill opacity=0.8] (axis cs:6.31875,0) rectangle (axis cs:6.53125,52.475979);
\draw[fill=color3,draw opacity=0,fill opacity=0.8] (axis cs:0.53125,0) rectangle (axis cs:0.74375,5.811893);
\addlegendimage{ybar,ybar legend,fill=color3,draw opacity=0,fill opacity=0.8};
\addlegendentry{h=4}

\draw[fill=color3,draw opacity=0,fill opacity=0.8] (axis cs:1.53125,0) rectangle (axis cs:1.74375,6.42582);
\draw[fill=color3,draw opacity=0,fill opacity=0.8] (axis cs:2.53125,0) rectangle (axis cs:2.74375,6.87492);
\draw[fill=color3,draw opacity=0,fill opacity=0.8] (axis cs:3.53125,0) rectangle (axis cs:3.74375,12.016849);
\draw[fill=color3,draw opacity=0,fill opacity=0.8] (axis cs:4.53125,0) rectangle (axis cs:4.74375,32.578854);
\draw[fill=color3,draw opacity=0,fill opacity=0.8] (axis cs:5.53125,0) rectangle (axis cs:5.74375,46.695576);
\draw[fill=color3,draw opacity=0,fill opacity=0.8] (axis cs:6.53125,0) rectangle (axis cs:6.74375,54.752781);
\end{axis}

\end{tikzpicture}
    }
  }
  \hfill\hfill\hfill
  \subfloat[\emph{ljournal-2008}]{
    \scalebox{0.7}{
\begin{tikzpicture}

\definecolor{color0}{rgb}{0.12156862745098,0.466666666666667,0.705882352941177}
\definecolor{color1}{rgb}{1,0.498039215686275,0.0549019607843137}
\definecolor{color2}{rgb}{0.172549019607843,0.627450980392157,0.172549019607843}
\definecolor{color3}{rgb}{0.83921568627451,0.152941176470588,0.156862745098039}

\begin{axis}[
legend cell align={left},
legend style={fill opacity=0.8, draw opacity=1, text opacity=1, at={(0.03,0.97)}, anchor=north west, draw=white!80.0!black},
tick align=outside,
tick pos=both,
x grid style={white!69.01960784313725!black},
xlabel={\(\displaystyle \delta\)},
xmin=-0.44875, xmax=7.08625,
xtick style={color=black},
xtick={0.31875,1.31875,2.31875,3.31875,4.31875,5.31875,6.31875},
xticklabels={2,5,10,20,50,100,200},
y grid style={white!69.01960784313725!black},
ylabel={Encode time (s)},
ymin=0, ymax=961.0176912,
ytick style={color=black}
]
\draw[fill=color0,draw opacity=0,fill opacity=0.8] (axis cs:-0.10625,0) rectangle (axis cs:0.10625,74.683411);
\addlegendimage{ybar,ybar legend,fill=color0,draw opacity=0,fill opacity=0.8};
\addlegendentry{h=1}

\draw[fill=color0,draw opacity=0,fill opacity=0.8] (axis cs:0.89375,0) rectangle (axis cs:1.10625,76.816887);
\draw[fill=color0,draw opacity=0,fill opacity=0.8] (axis cs:1.89375,0) rectangle (axis cs:2.10625,75.631561);
\draw[fill=color0,draw opacity=0,fill opacity=0.8] (axis cs:2.89375,0) rectangle (axis cs:3.10625,86.064064);
\draw[fill=color0,draw opacity=0,fill opacity=0.8] (axis cs:3.89375,0) rectangle (axis cs:4.10625,128.108902);
\draw[fill=color0,draw opacity=0,fill opacity=0.8] (axis cs:4.89375,0) rectangle (axis cs:5.10625,166.593903);
\draw[fill=color0,draw opacity=0,fill opacity=0.8] (axis cs:5.89375,0) rectangle (axis cs:6.10625,208.43515);
\draw[fill=color1,draw opacity=0,fill opacity=0.8] (axis cs:0.10625,0) rectangle (axis cs:0.31875,74.269981);
\addlegendimage{ybar,ybar legend,fill=color1,draw opacity=0,fill opacity=0.8};
\addlegendentry{h=2}

\draw[fill=color1,draw opacity=0,fill opacity=0.8] (axis cs:1.10625,0) rectangle (axis cs:1.31875,76.266144);
\draw[fill=color1,draw opacity=0,fill opacity=0.8] (axis cs:2.10625,0) rectangle (axis cs:2.31875,95.014442);
\draw[fill=color1,draw opacity=0,fill opacity=0.8] (axis cs:3.10625,0) rectangle (axis cs:3.31875,151.438019);
\draw[fill=color1,draw opacity=0,fill opacity=0.8] (axis cs:4.10625,0) rectangle (axis cs:4.31875,336.171875);
\draw[fill=color1,draw opacity=0,fill opacity=0.8] (axis cs:5.10625,0) rectangle (axis cs:5.31875,553.985413);
\draw[fill=color1,draw opacity=0,fill opacity=0.8] (axis cs:6.10625,0) rectangle (axis cs:6.31875,915.254944);
\draw[fill=color2,draw opacity=0,fill opacity=0.8] (axis cs:0.31875,0) rectangle (axis cs:0.53125,71.416336);
\addlegendimage{ybar,ybar legend,fill=color2,draw opacity=0,fill opacity=0.8};
\addlegendentry{h=3}

\draw[fill=color2,draw opacity=0,fill opacity=0.8] (axis cs:1.31875,0) rectangle (axis cs:1.53125,73.41378);
\draw[fill=color2,draw opacity=0,fill opacity=0.8] (axis cs:2.31875,0) rectangle (axis cs:2.53125,76.311234);
\draw[fill=color2,draw opacity=0,fill opacity=0.8] (axis cs:3.31875,0) rectangle (axis cs:3.53125,83.86451);
\draw[fill=color2,draw opacity=0,fill opacity=0.8] (axis cs:4.31875,0) rectangle (axis cs:4.53125,108.737968);
\draw[fill=color2,draw opacity=0,fill opacity=0.8] (axis cs:5.31875,0) rectangle (axis cs:5.53125,160.344208);
\draw[fill=color2,draw opacity=0,fill opacity=0.8] (axis cs:6.31875,0) rectangle (axis cs:6.53125,258.258759);
\draw[fill=color3,draw opacity=0,fill opacity=0.8] (axis cs:0.53125,0) rectangle (axis cs:0.74375,74.763321);
\addlegendimage{ybar,ybar legend,fill=color3,draw opacity=0,fill opacity=0.8};
\addlegendentry{h=4}

\draw[fill=color3,draw opacity=0,fill opacity=0.8] (axis cs:1.53125,0) rectangle (axis cs:1.74375,75.9571);
\draw[fill=color3,draw opacity=0,fill opacity=0.8] (axis cs:2.53125,0) rectangle (axis cs:2.74375,78.437149);
\draw[fill=color3,draw opacity=0,fill opacity=0.8] (axis cs:3.53125,0) rectangle (axis cs:3.74375,83.919449);
\draw[fill=color3,draw opacity=0,fill opacity=0.8] (axis cs:4.53125,0) rectangle (axis cs:4.74375,101.308708);
\draw[fill=color3,draw opacity=0,fill opacity=0.8] (axis cs:5.53125,0) rectangle (axis cs:5.74375,129.916977);
\draw[fill=color3,draw opacity=0,fill opacity=0.8] (axis cs:6.53125,0) rectangle (axis cs:6.74375,187.669846);
\end{axis}

\end{tikzpicture}
    }
  }
  \hfill

  \caption{\label{fig:social-enc-time} Compression times of our algorithm in seconds for
    different values of the parameters $h$ and $\delta$ for the datasets $(a)$
    \emph{dblp-2010},  $(b)$ \emph{hollywood-2009}, $(c)$ \emph{amazon-2008},
    and $(d)$ \emph{ljournal-2008}.}
\end{figure}

\begin{figure}
  \centering
  \hfill
  \subfloat[\emph{dblp-2010}]{
    \scalebox{0.7}{
\begin{tikzpicture}

\definecolor{color0}{rgb}{0.12156862745098,0.466666666666667,0.705882352941177}
\definecolor{color1}{rgb}{1,0.498039215686275,0.0549019607843137}
\definecolor{color2}{rgb}{0.172549019607843,0.627450980392157,0.172549019607843}
\definecolor{color3}{rgb}{0.83921568627451,0.152941176470588,0.156862745098039}

\begin{axis}[
legend cell align={left},
legend style={fill opacity=0.8, draw opacity=1, text opacity=1, at={(0.03,0.97)}, anchor=north west, draw=white!80.0!black},
tick align=outside,
tick pos=both,
x grid style={white!69.01960784313725!black},
xlabel={\(\displaystyle \delta\)},
xmin=-0.44875, xmax=7.08625,
xtick style={color=black},
xtick={0.31875,1.31875,2.31875,3.31875,4.31875,5.31875,6.31875},
xticklabels={2,5,10,20,50,100,200},
y grid style={white!69.01960784313725!black},
ylabel={Decode time (s)},
ymin=0, ymax=32.49299865,
ytick style={color=black}
]
\draw[fill=color0,draw opacity=0,fill opacity=0.8] (axis cs:-0.10625,0) rectangle (axis cs:0.10625,7.433669);
\addlegendimage{ybar,ybar legend,fill=color0,draw opacity=0,fill opacity=0.8};
\addlegendentry{h=1}

\draw[fill=color0,draw opacity=0,fill opacity=0.8] (axis cs:0.89375,0) rectangle (axis cs:1.10625,8.907521);
\draw[fill=color0,draw opacity=0,fill opacity=0.8] (axis cs:1.89375,0) rectangle (axis cs:2.10625,12.08515);
\draw[fill=color0,draw opacity=0,fill opacity=0.8] (axis cs:2.89375,0) rectangle (axis cs:3.10625,17.031805);
\draw[fill=color0,draw opacity=0,fill opacity=0.8] (axis cs:3.89375,0) rectangle (axis cs:4.10625,24.852262);
\draw[fill=color0,draw opacity=0,fill opacity=0.8] (axis cs:4.89375,0) rectangle (axis cs:5.10625,30.945713);
\draw[fill=color0,draw opacity=0,fill opacity=0.8] (axis cs:5.89375,0) rectangle (axis cs:6.10625,29.856785);
\draw[fill=color1,draw opacity=0,fill opacity=0.8] (axis cs:0.10625,0) rectangle (axis cs:0.31875,7.84613);
\addlegendimage{ybar,ybar legend,fill=color1,draw opacity=0,fill opacity=0.8};
\addlegendentry{h=2}

\draw[fill=color1,draw opacity=0,fill opacity=0.8] (axis cs:1.10625,0) rectangle (axis cs:1.31875,9.744107);
\draw[fill=color1,draw opacity=0,fill opacity=0.8] (axis cs:2.10625,0) rectangle (axis cs:2.31875,13.344392);
\draw[fill=color1,draw opacity=0,fill opacity=0.8] (axis cs:3.10625,0) rectangle (axis cs:3.31875,13.965195);
\draw[fill=color1,draw opacity=0,fill opacity=0.8] (axis cs:4.10625,0) rectangle (axis cs:4.31875,15.785624);
\draw[fill=color1,draw opacity=0,fill opacity=0.8] (axis cs:5.10625,0) rectangle (axis cs:5.31875,16.211861);
\draw[fill=color1,draw opacity=0,fill opacity=0.8] (axis cs:6.10625,0) rectangle (axis cs:6.31875,16.033735);
\draw[fill=color2,draw opacity=0,fill opacity=0.8] (axis cs:0.31875,0) rectangle (axis cs:0.53125,7.696007);
\addlegendimage{ybar,ybar legend,fill=color2,draw opacity=0,fill opacity=0.8};
\addlegendentry{h=3}

\draw[fill=color2,draw opacity=0,fill opacity=0.8] (axis cs:1.31875,0) rectangle (axis cs:1.53125,8.990904);
\draw[fill=color2,draw opacity=0,fill opacity=0.8] (axis cs:2.31875,0) rectangle (axis cs:2.53125,9.801561);
\draw[fill=color2,draw opacity=0,fill opacity=0.8] (axis cs:3.31875,0) rectangle (axis cs:3.53125,10.57585);
\draw[fill=color2,draw opacity=0,fill opacity=0.8] (axis cs:4.31875,0) rectangle (axis cs:4.53125,11.771028);
\draw[fill=color2,draw opacity=0,fill opacity=0.8] (axis cs:5.31875,0) rectangle (axis cs:5.53125,13.072452);
\draw[fill=color2,draw opacity=0,fill opacity=0.8] (axis cs:6.31875,0) rectangle (axis cs:6.53125,14.121787);
\draw[fill=color3,draw opacity=0,fill opacity=0.8] (axis cs:0.53125,0) rectangle (axis cs:0.74375,7.624701);
\addlegendimage{ybar,ybar legend,fill=color3,draw opacity=0,fill opacity=0.8};
\addlegendentry{h=4}

\draw[fill=color3,draw opacity=0,fill opacity=0.8] (axis cs:1.53125,0) rectangle (axis cs:1.74375,8.970397);
\draw[fill=color3,draw opacity=0,fill opacity=0.8] (axis cs:2.53125,0) rectangle (axis cs:2.74375,9.586815);
\draw[fill=color3,draw opacity=0,fill opacity=0.8] (axis cs:3.53125,0) rectangle (axis cs:3.74375,9.730128);
\draw[fill=color3,draw opacity=0,fill opacity=0.8] (axis cs:4.53125,0) rectangle (axis cs:4.74375,10.242282);
\draw[fill=color3,draw opacity=0,fill opacity=0.8] (axis cs:5.53125,0) rectangle (axis cs:5.74375,10.973615);
\draw[fill=color3,draw opacity=0,fill opacity=0.8] (axis cs:6.53125,0) rectangle (axis cs:6.74375,12.913285);
\end{axis}

\end{tikzpicture}
    }
  }
  \hfill\hfill\hfill
  \subfloat[\emph{hollywood-2009}]{
    \scalebox{0.7}{
\begin{tikzpicture}

\definecolor{color0}{rgb}{0.12156862745098,0.466666666666667,0.705882352941177}
\definecolor{color1}{rgb}{1,0.498039215686275,0.0549019607843137}
\definecolor{color2}{rgb}{0.172549019607843,0.627450980392157,0.172549019607843}
\definecolor{color3}{rgb}{0.83921568627451,0.152941176470588,0.156862745098039}

\begin{axis}[
legend cell align={left},
legend style={fill opacity=0.8, draw opacity=1, text opacity=1, at={(0.03,0.97)}, anchor=north west, draw=white!80.0!black},
tick align=outside,
tick pos=both,
x grid style={white!69.01960784313725!black},
xlabel={\(\displaystyle \delta\)},
xmin=-0.44875, xmax=7.08625,
xtick style={color=black},
xtick={0.31875,1.31875,2.31875,3.31875,4.31875,5.31875,6.31875},
xticklabels={2,5,10,20,50,100,200},
y grid style={white!69.01960784313725!black},
ylabel={Decode time (s)},
ymin=0, ymax=607.15795035,
ytick style={color=black}
]
\draw[fill=color0,draw opacity=0,fill opacity=0.8] (axis cs:-0.10625,0) rectangle (axis cs:0.10625,42.781376);
\addlegendimage{ybar,ybar legend,fill=color0,draw opacity=0,fill opacity=0.8};
\addlegendentry{h=1}

\draw[fill=color0,draw opacity=0,fill opacity=0.8] (axis cs:0.89375,0) rectangle (axis cs:1.10625,44.116344);
\draw[fill=color0,draw opacity=0,fill opacity=0.8] (axis cs:1.89375,0) rectangle (axis cs:2.10625,47.979172);
\draw[fill=color0,draw opacity=0,fill opacity=0.8] (axis cs:2.89375,0) rectangle (axis cs:3.10625,58.87125);
\draw[fill=color0,draw opacity=0,fill opacity=0.8] (axis cs:3.89375,0) rectangle (axis cs:4.10625,155.11821);
\draw[fill=color0,draw opacity=0,fill opacity=0.8] (axis cs:4.89375,0) rectangle (axis cs:5.10625,272.870575);
\draw[fill=color0,draw opacity=0,fill opacity=0.8] (axis cs:5.89375,0) rectangle (axis cs:6.10625,578.245667);
\draw[fill=color1,draw opacity=0,fill opacity=0.8] (axis cs:0.10625,0) rectangle (axis cs:0.31875,42.383118);
\addlegendimage{ybar,ybar legend,fill=color1,draw opacity=0,fill opacity=0.8};
\addlegendentry{h=2}

\draw[fill=color1,draw opacity=0,fill opacity=0.8] (axis cs:1.10625,0) rectangle (axis cs:1.31875,45.569881);
\draw[fill=color1,draw opacity=0,fill opacity=0.8] (axis cs:2.10625,0) rectangle (axis cs:2.31875,49.491745);
\draw[fill=color1,draw opacity=0,fill opacity=0.8] (axis cs:3.10625,0) rectangle (axis cs:3.31875,56.254223);
\draw[fill=color1,draw opacity=0,fill opacity=0.8] (axis cs:4.10625,0) rectangle (axis cs:4.31875,82.653252);
\draw[fill=color1,draw opacity=0,fill opacity=0.8] (axis cs:5.10625,0) rectangle (axis cs:5.31875,114.347313);
\draw[fill=color1,draw opacity=0,fill opacity=0.8] (axis cs:6.10625,0) rectangle (axis cs:6.31875,157.577057);
\draw[fill=color2,draw opacity=0,fill opacity=0.8] (axis cs:0.31875,0) rectangle (axis cs:0.53125,41.53009);
\addlegendimage{ybar,ybar legend,fill=color2,draw opacity=0,fill opacity=0.8};
\addlegendentry{h=3}

\draw[fill=color2,draw opacity=0,fill opacity=0.8] (axis cs:1.31875,0) rectangle (axis cs:1.53125,50.2813);
\draw[fill=color2,draw opacity=0,fill opacity=0.8] (axis cs:2.31875,0) rectangle (axis cs:2.53125,51.199657);
\draw[fill=color2,draw opacity=0,fill opacity=0.8] (axis cs:3.31875,0) rectangle (axis cs:3.53125,48.60548);
\draw[fill=color2,draw opacity=0,fill opacity=0.8] (axis cs:4.31875,0) rectangle (axis cs:4.53125,51.917053);
\draw[fill=color2,draw opacity=0,fill opacity=0.8] (axis cs:5.31875,0) rectangle (axis cs:5.53125,56.295811);
\draw[fill=color2,draw opacity=0,fill opacity=0.8] (axis cs:6.31875,0) rectangle (axis cs:6.53125,52.757732);
\draw[fill=color3,draw opacity=0,fill opacity=0.8] (axis cs:0.53125,0) rectangle (axis cs:0.74375,41.887871);
\addlegendimage{ybar,ybar legend,fill=color3,draw opacity=0,fill opacity=0.8};
\addlegendentry{h=4}

\draw[fill=color3,draw opacity=0,fill opacity=0.8] (axis cs:1.53125,0) rectangle (axis cs:1.74375,47.08884);
\draw[fill=color3,draw opacity=0,fill opacity=0.8] (axis cs:2.53125,0) rectangle (axis cs:2.74375,44.359695);
\draw[fill=color3,draw opacity=0,fill opacity=0.8] (axis cs:3.53125,0) rectangle (axis cs:3.74375,44.892685);
\draw[fill=color3,draw opacity=0,fill opacity=0.8] (axis cs:4.53125,0) rectangle (axis cs:4.74375,46.001659);
\draw[fill=color3,draw opacity=0,fill opacity=0.8] (axis cs:5.53125,0) rectangle (axis cs:5.74375,45.886158);
\draw[fill=color3,draw opacity=0,fill opacity=0.8] (axis cs:6.53125,0) rectangle (axis cs:6.74375,45.711681);
\end{axis}

\end{tikzpicture}
    }
  }
  \hfill

  \hfill
  \subfloat[\emph{amazon-2008}]{
    \scalebox{0.7}{
\begin{tikzpicture}

\definecolor{color0}{rgb}{0.12156862745098,0.466666666666667,0.705882352941177}
\definecolor{color1}{rgb}{1,0.498039215686275,0.0549019607843137}
\definecolor{color2}{rgb}{0.172549019607843,0.627450980392157,0.172549019607843}
\definecolor{color3}{rgb}{0.83921568627451,0.152941176470588,0.156862745098039}

\begin{axis}[
legend cell align={left},
legend style={fill opacity=0.8, draw opacity=1, text opacity=1, at={(0.03,0.97)}, anchor=north west, draw=white!80.0!black},
tick align=outside,
tick pos=both,
x grid style={white!69.01960784313725!black},
xlabel={\(\displaystyle \delta\)},
xmin=-0.44875, xmax=7.08625,
xtick style={color=black},
xtick={0.31875,1.31875,2.31875,3.31875,4.31875,5.31875,6.31875},
xticklabels={2,5,10,20,50,100,200},
y grid style={white!69.01960784313725!black},
ylabel={Decode time (s)},
ymin=0, ymax=121.16742225,
ytick style={color=black}
]
\draw[fill=color0,draw opacity=0,fill opacity=0.8] (axis cs:-0.10625,0) rectangle (axis cs:0.10625,21.373745);
\addlegendimage{ybar,ybar legend,fill=color0,draw opacity=0,fill opacity=0.8};
\addlegendentry{h=1}

\draw[fill=color0,draw opacity=0,fill opacity=0.8] (axis cs:0.89375,0) rectangle (axis cs:1.10625,22.516367);
\draw[fill=color0,draw opacity=0,fill opacity=0.8] (axis cs:1.89375,0) rectangle (axis cs:2.10625,28.402121);
\draw[fill=color0,draw opacity=0,fill opacity=0.8] (axis cs:2.89375,0) rectangle (axis cs:3.10625,76.210373);
\draw[fill=color0,draw opacity=0,fill opacity=0.8] (axis cs:3.89375,0) rectangle (axis cs:4.10625,115.397545);
\draw[fill=color0,draw opacity=0,fill opacity=0.8] (axis cs:4.89375,0) rectangle (axis cs:5.10625,107.696991);
\draw[fill=color0,draw opacity=0,fill opacity=0.8] (axis cs:5.89375,0) rectangle (axis cs:6.10625,112.115616);
\draw[fill=color1,draw opacity=0,fill opacity=0.8] (axis cs:0.10625,0) rectangle (axis cs:0.31875,21.217279);
\addlegendimage{ybar,ybar legend,fill=color1,draw opacity=0,fill opacity=0.8};
\addlegendentry{h=2}

\draw[fill=color1,draw opacity=0,fill opacity=0.8] (axis cs:1.10625,0) rectangle (axis cs:1.31875,22.700354);
\draw[fill=color1,draw opacity=0,fill opacity=0.8] (axis cs:2.10625,0) rectangle (axis cs:2.31875,28.226927);
\draw[fill=color1,draw opacity=0,fill opacity=0.8] (axis cs:3.10625,0) rectangle (axis cs:3.31875,46.453236);
\draw[fill=color1,draw opacity=0,fill opacity=0.8] (axis cs:4.10625,0) rectangle (axis cs:4.31875,56.135361);
\draw[fill=color1,draw opacity=0,fill opacity=0.8] (axis cs:5.10625,0) rectangle (axis cs:5.31875,57.634087);
\draw[fill=color1,draw opacity=0,fill opacity=0.8] (axis cs:6.10625,0) rectangle (axis cs:6.31875,56.719547);
\draw[fill=color2,draw opacity=0,fill opacity=0.8] (axis cs:0.31875,0) rectangle (axis cs:0.53125,19.569799);
\addlegendimage{ybar,ybar legend,fill=color2,draw opacity=0,fill opacity=0.8};
\addlegendentry{h=3}

\draw[fill=color2,draw opacity=0,fill opacity=0.8] (axis cs:1.31875,0) rectangle (axis cs:1.53125,21.764645);
\draw[fill=color2,draw opacity=0,fill opacity=0.8] (axis cs:2.31875,0) rectangle (axis cs:2.53125,22.852215);
\draw[fill=color2,draw opacity=0,fill opacity=0.8] (axis cs:3.31875,0) rectangle (axis cs:3.53125,27.226738);
\draw[fill=color2,draw opacity=0,fill opacity=0.8] (axis cs:4.31875,0) rectangle (axis cs:4.53125,39.62495);
\draw[fill=color2,draw opacity=0,fill opacity=0.8] (axis cs:5.31875,0) rectangle (axis cs:5.53125,43.050999);
\draw[fill=color2,draw opacity=0,fill opacity=0.8] (axis cs:6.31875,0) rectangle (axis cs:6.53125,44.508667);
\draw[fill=color3,draw opacity=0,fill opacity=0.8] (axis cs:0.53125,0) rectangle (axis cs:0.74375,17.842281);
\addlegendimage{ybar,ybar legend,fill=color3,draw opacity=0,fill opacity=0.8};
\addlegendentry{h=4}

\draw[fill=color3,draw opacity=0,fill opacity=0.8] (axis cs:1.53125,0) rectangle (axis cs:1.74375,20.573689);
\draw[fill=color3,draw opacity=0,fill opacity=0.8] (axis cs:2.53125,0) rectangle (axis cs:2.74375,21.909494);
\draw[fill=color3,draw opacity=0,fill opacity=0.8] (axis cs:3.53125,0) rectangle (axis cs:3.74375,24.45171);
\draw[fill=color3,draw opacity=0,fill opacity=0.8] (axis cs:4.53125,0) rectangle (axis cs:4.74375,30.604904);
\draw[fill=color3,draw opacity=0,fill opacity=0.8] (axis cs:5.53125,0) rectangle (axis cs:5.74375,36.751518);
\draw[fill=color3,draw opacity=0,fill opacity=0.8] (axis cs:6.53125,0) rectangle (axis cs:6.74375,40.594299);
\end{axis}

\end{tikzpicture}
    }
  }
  \hfill\hfill\hfill
  \subfloat[\emph{ljournal-2008}]{
    \scalebox{0.7}{
\begin{tikzpicture}

\definecolor{color0}{rgb}{0.12156862745098,0.466666666666667,0.705882352941177}
\definecolor{color1}{rgb}{1,0.498039215686275,0.0549019607843137}
\definecolor{color2}{rgb}{0.172549019607843,0.627450980392157,0.172549019607843}
\definecolor{color3}{rgb}{0.83921568627451,0.152941176470588,0.156862745098039}

\begin{axis}[
legend cell align={left},
legend style={fill opacity=0.8, draw opacity=1, text opacity=1, at={(0.03,0.97)}, anchor=north west, draw=white!80.0!black},
tick align=outside,
tick pos=both,
x grid style={white!69.01960784313725!black},
xlabel={\(\displaystyle \delta\)},
xmin=-0.44875, xmax=7.08625,
xtick style={color=black},
xtick={0.31875,1.31875,2.31875,3.31875,4.31875,5.31875,6.31875},
xticklabels={2,5,10,20,50,100,200},
y grid style={white!69.01960784313725!black},
ylabel={Decode time (s)},
ymin=0, ymax=1431.99479325,
ytick style={color=black}
]
\draw[fill=color0,draw opacity=0,fill opacity=0.8] (axis cs:-0.10625,0) rectangle (axis cs:0.10625,207.581879);
\addlegendimage{ybar,ybar legend,fill=color0,draw opacity=0,fill opacity=0.8};
\addlegendentry{h=1}

\draw[fill=color0,draw opacity=0,fill opacity=0.8] (axis cs:0.89375,0) rectangle (axis cs:1.10625,250.790039);
\draw[fill=color0,draw opacity=0,fill opacity=0.8] (axis cs:1.89375,0) rectangle (axis cs:2.10625,247.990738);
\draw[fill=color0,draw opacity=0,fill opacity=0.8] (axis cs:2.89375,0) rectangle (axis cs:3.10625,305.192963);
\draw[fill=color0,draw opacity=0,fill opacity=0.8] (axis cs:3.89375,0) rectangle (axis cs:4.10625,568.950134);
\draw[fill=color0,draw opacity=0,fill opacity=0.8] (axis cs:4.89375,0) rectangle (axis cs:5.10625,935.869934);
\draw[fill=color0,draw opacity=0,fill opacity=0.8] (axis cs:5.89375,0) rectangle (axis cs:6.10625,1363.804565);
\draw[fill=color1,draw opacity=0,fill opacity=0.8] (axis cs:0.10625,0) rectangle (axis cs:0.31875,224.292297);
\addlegendimage{ybar,ybar legend,fill=color1,draw opacity=0,fill opacity=0.8};
\addlegendentry{h=2}

\draw[fill=color1,draw opacity=0,fill opacity=0.8] (axis cs:1.10625,0) rectangle (axis cs:1.31875,240.494141);
\draw[fill=color1,draw opacity=0,fill opacity=0.8] (axis cs:2.10625,0) rectangle (axis cs:2.31875,255.49295);
\draw[fill=color1,draw opacity=0,fill opacity=0.8] (axis cs:3.10625,0) rectangle (axis cs:3.31875,288.264038);
\draw[fill=color1,draw opacity=0,fill opacity=0.8] (axis cs:4.10625,0) rectangle (axis cs:4.31875,381.096283);
\draw[fill=color1,draw opacity=0,fill opacity=0.8] (axis cs:5.10625,0) rectangle (axis cs:5.31875,487.833435);
\draw[fill=color1,draw opacity=0,fill opacity=0.8] (axis cs:6.10625,0) rectangle (axis cs:6.31875,619.589844);
\draw[fill=color2,draw opacity=0,fill opacity=0.8] (axis cs:0.31875,0) rectangle (axis cs:0.53125,215.539276);
\addlegendimage{ybar,ybar legend,fill=color2,draw opacity=0,fill opacity=0.8};
\addlegendentry{h=3}

\draw[fill=color2,draw opacity=0,fill opacity=0.8] (axis cs:1.31875,0) rectangle (axis cs:1.53125,225.992447);
\draw[fill=color2,draw opacity=0,fill opacity=0.8] (axis cs:2.31875,0) rectangle (axis cs:2.53125,231.350159);
\draw[fill=color2,draw opacity=0,fill opacity=0.8] (axis cs:3.31875,0) rectangle (axis cs:3.53125,235.59584);
\draw[fill=color2,draw opacity=0,fill opacity=0.8] (axis cs:4.31875,0) rectangle (axis cs:4.53125,241.780502);
\draw[fill=color2,draw opacity=0,fill opacity=0.8] (axis cs:5.31875,0) rectangle (axis cs:5.53125,255.355698);
\draw[fill=color2,draw opacity=0,fill opacity=0.8] (axis cs:6.31875,0) rectangle (axis cs:6.53125,283.189606);
\draw[fill=color3,draw opacity=0,fill opacity=0.8] (axis cs:0.53125,0) rectangle (axis cs:0.74375,210.541168);
\addlegendimage{ybar,ybar legend,fill=color3,draw opacity=0,fill opacity=0.8};
\addlegendentry{h=4}

\draw[fill=color3,draw opacity=0,fill opacity=0.8] (axis cs:1.53125,0) rectangle (axis cs:1.74375,221.943756);
\draw[fill=color3,draw opacity=0,fill opacity=0.8] (axis cs:2.53125,0) rectangle (axis cs:2.74375,224.610794);
\draw[fill=color3,draw opacity=0,fill opacity=0.8] (axis cs:3.53125,0) rectangle (axis cs:3.74375,228.682236);
\draw[fill=color3,draw opacity=0,fill opacity=0.8] (axis cs:4.53125,0) rectangle (axis cs:4.74375,244.226639);
\draw[fill=color3,draw opacity=0,fill opacity=0.8] (axis cs:5.53125,0) rectangle (axis cs:5.74375,235.265518);
\draw[fill=color3,draw opacity=0,fill opacity=0.8] (axis cs:6.53125,0) rectangle (axis cs:6.74375,236.529663);
\end{axis}

\end{tikzpicture}
    }
  }
  \hfill

  \caption{\label{fig:social-dec-time} Decompression times of our algorithm in seconds for
    different values of the parameters $h$ and $\delta$ for the datasets $(a)$
    \emph{dblp-2010},  $(b)$ \emph{hollywood-2009}, $(c)$ \emph{amazon-2008},
    and $(d)$ \emph{ljournal-2008}.}
\end{figure}

Figures~\ref{fig:social-bpl}, \ref{fig:social-enc-time}, and
\ref{fig:social-dec-time} illustrate the compression ratio, compression time,
and decompression time of our algorithm for these datasets, respectively, for
different values of $h$ and $\delta$. 
 In Table~\ref{tab:social-first}, we report the best ratios
over the values of $h$ and $\delta$, and compare them  with
the ones in \cite{boldi2011layered}, which are the best in
the literature to the best of our knowledge. We have also reported  in
Table~\ref{tab:social-first} the time it takes to
encode and decode each  graph for the corresponding value of $h$ and $\delta$.

\begin{table}
  \centering
  \begin{tabular}{@{}lcccr@{}}
    \toprule
    Dataset & \cite{boldi2011layered} & best 40\% BPL [relative\%] &  value of & encode/decode \\
    & (BPL) & & $(h, \delta)$  & time (sec) \\
    \midrule
    \emph{dblp-2010} & 6.78 & 7.13 [-5.16\%] & (1,10) & 2.53/12.1\\
    \emph{amazon-2008} & 9.12 &  11.1 [-21.7\%] & (1,15) & 10.96/57.1 \\
    \emph{hollywood-2009} & 5.14 &  5.22 [-1.6\%] & (2,200) & 71.1/157.57\\
    \emph{ljournal-2008} & 10.90 &  10.77 [1.2\%] & (1,100) & 611.7/515\\
    \bottomrule
  \end{tabular}
  \caption{\label{tab:social-40p}Best compression rates of our algorithm subject
    to choosing $(h, \delta)$ so that at most $40\%$ of the edges in the graph
    are star edges. For comparison, the values from \cite{boldi2011layered} are
    also presented. Moreover, the values of the $(h, \delta)$ parameters achieving
    the reported results together with encoding/decoding times are presented.}
\end{table}

Similar to our observation from road
networks in Section~\ref{sec:exp-tree}, we see that the value of $\delta$ for the $(h, \delta)$
pair that optimizes the compression ratio is small. Similar to our discussion in
Section~\ref{sec:exp-tree}, one possible explanation of this can be that we are not yet in the
asymptotic regime. In order to address this, similar to our approach in Section~\ref{sec:exp-tree}, we look for the values of $(h, \delta)$ which result in at most 40\% of the
edges in the graph 
becoming star edges, and find the best compression ratio
among such $(h, \delta)$ pairs. This is reported in Table~\ref{tab:social-40p}. As we can see, in most cases,
our results are comparable with those from \cite{boldi2011layered}, and in some
cases they are even better. 

Note that
\begin{enumerate}
\item 
As we discussed
  earlier, 
  from  Table~\ref{tab:social-graphs}, we see that all the social graphs we consider
  here have relatively large average clustering coefficient, and hence they are
  not locally tree--like. Consequently, the assumption of
  Theorem~\ref{thm:optimality-complexity-main} requiring the
  limit to be supported on $\mTb_*$ does not hold. This means that the
  theoretical optimality guarantee of
  Theorem~\ref{thm:optimality-complexity-main} does not hold for such datasets. 
\item The method in \cite{boldi2011layered} is tailored for social graphs,
  whereas our algorithm is universal.
\end{enumerate}

The two observations above imply that we should not expect our compression
ratios to outperform those of \cite{boldi2011layered}. However, as
Tables~\ref{tab:social-first} and \ref{tab:social-40p} suggest, we see, surprisingly,
that they are comparable in most cases, and 
our
compression ratios even outperform those of \cite{boldi2011layered} in some cases. 


\editstart

\section{Details of the Algorithm}
\label{sec:alg-details}

In this section, we discuss the details of our compression and decompression
algorithms which were sketched in Section~\ref{sec:algorithm-general}. More
precisely, in
Section~\ref{sec:enc-alg}, we discuss the details of the compression algorithm,
and in Section~\ref{sec:dec-alg}, we discuss the details of the decompression
algorithm. Furthermore, we analyze the complexities of the compression and the
decompression algorithms in Sections~\ref{sec:enc-complexity} and
\ref{sec:main-decode-complexity}, respectively.


\subsection{The Compression Algorithm}
\label{sec:enc-alg}
  
Here, we introduce our compression algorithm in detail. As in
Section~\ref{sec:algorithm-general}, we assume that a simple marked graph $\Gn$
is given in the edge list representation, and positive integers $h$ and $\delta$ are
fixed. 
\ifreplace
Our compression algorithm is explained in Algorithm~\ref{alg:graph-encode-new}
below.
\color{black}
\fi

\begin{myalg}[Encoding a simple marked graph \label{alg:graph-encode-new}]
  \begin{algorithmic}[1]
    \INPUT
    \Statex $n$: number of vertices
    \Statex $\Gn$: A simple marked graph $\Gn$ on the vertex set $[n]$, vertex mark set
    $\vermark = \{1,    \dots, |\vermark|\}$ and edge mark set $\edgemark = \{1,
    \dots, |\edgemark|\}$ given as follows:
    \begin{itemize}
\item its vertex mark sequence $\vthetan = (\thetan_v: v \in [n])$,
      where $\thetan_v \in \vermark$ is the mark of vertex $v$ in $\Gn$
      \item $\mathsf{EdgeList} = (\mathsf{EdgeList}_i: 1 \leq i \leq \mn)$: the
list of edges in $\Gn$ where  $\mathsf{EdgeList}_i = (v_i, w_i, x_i, x'_i)$ for $1 \leq i \leq
\mn$, where $\mn$ denotes the total number of edges in $\Gn$, and for $1 \leq i
\leq \mn$, the tuple $(v_i, w_i, x_i, x'_i)$ represents an edge between the
vertices $v_i$ and $w_i$ with mark $x_i$ towards $v_i$ and mark $x'_i$ towards
$w_i$, i.e.\ $\xi_{\Gn}(w_i, v_i) = x_i$ and $\xi_{\Gn}(v_i, w_i) = x'_i$
    \end{itemize}
    \Statex $\delta$: degree threshold hyperparameter, $\delta \geq 1$
    \Statex $h$: depth hyperparameter, $h \geq 1$ 
    \OUTPUT
    \Statex $\mathsf{Output}$: A bit sequence in $\{0,1\}^* - \emptyset$
    representing $\Gn$ in compressed form. 
    \Function{MarkedGraphEncode}{$n, \Gn, \delta, h$}
    \State $\mathsf{Output} \gets $ empty bit sequence \Comment{initialize the
      output with empty bit sequence}
    \State $(\vthetan, \vdn, \vgamman, \vtgamman, \vec{x}^{(n)}, \vec{x'}^{(n)}) \gets
    \textproc{Preprocess}(n, \vthetan, \mathsf{EdgeList})$ \label{l:enc-prep}
    \Statex \Comment{Algorithm~\ref{alg:preprocess} in
      Section~\ref{sec:detail-preprocessing}, this finds the equivalent neighbor list representation}
    \State $(\vec{c}, \mathsf{TCount}, \mathsf{TIsStar}, \mathsf{TMark}) \gets
    \textproc{ExtractTypes}(n, \Gn, \delta, h)$ \label{l:encode-extract-types}
    \Comment{Algorithm~\ref{alg:type-extract-message-passing-new} in Section~\ref{sec:MP}}
    \State $\mathsf{Output} \gets \mathsf{Output} \concat
    \edelta(\mathsf{1+TCount})$ \label{l:enc-tcount-elias} \Comment{use the Elias
      delta code to represent $\mathsf{TCount}$}
    \For{$1 \leq i \leq \mathsf{TCount}$} \label{l:enc-tiistar-tmark-for}
    \State $\mathsf{Output} \gets \mathsf{Output} \concat \mathsf{TIsStar}(i)
    \concat \mathsf{TMark}(i)$\label{l:enc-tiisstar-tmark-write} \Comment{use $1+ \lfloor   \log_2 |\edgemark|
      \rfloor$ bits to encode $\mathsf{TMark}(i)$}
    \EndFor \label{l:enc-tisstar-tmark-encode-for-end}
    \State $\textproc{EncodeStarVertices}$ \label{l:enc-star-vertices}
    \Comment{Algorithm~\ref{alg:encode-star-vertices} in Section~\ref{sec:enc-star-vertices}}
    \State $\textproc{EncodeStarEdges}$ \label{l:enc-star-edges}
    \Comment{Algorithm~\ref{alg:encode-star-edges} in Section~\ref{sec:encode-star-edges}}
    \State $\mathsf{Deg} = (\mathsf{Deg}_v: 1 \leq v \leq n) \gets \Array \text{
      of }
    \Dictionary(\nats \times \nats \rightarrow \nats)$ \label{l:encode-deg-def}
    \State $\textproc{FindDeg}$\label{l:enc-find-deg-call} \Comment{Algorithm~\ref{alg:encode-find-deg} in Section~\ref{sec:encode-find-Deg}}
    \State $\textproc{EncodeVertexTypes}$ \label{l:enc-encode-vertex-types-call}
    \Comment{Algorithm~\ref{alg:encode-vertex-types} in Section~\ref{sec:encode-vertex-types}}
    \State $\mathsf{PartitionAdjList} \gets \Dictionary(\nats \times \nats
    \rightarrow \Array \text{ of } \Array \text{ of integers})$ \label{l:enc-partition-adj-list-def}
    \State $\mathsf{PartitionDeg} \gets \Dictionary(\nats \times \nats
    \rightarrow \Array \text{ of integers})$ \label{l:enc-partition-deg-def}
    \State $\mathsf{PartitionIndex} = (\mathsf{PartitionIndex}_v: 1 \leq v \leq
    n) \gets \Array \text{ of } \Dictionary(\nats \times \nats \rightarrow
    \nats)$ \label{l:enc-partition-index-def}
    \State $\textproc{FindPartitionGraphs}$ \label{l:enc-find-partition-graphs-call}
    \Comment{Algorithm~\ref{alg:encode-find-partition-graphs} in Section~\ref{sec:encode-find-partition-graphs}}
    \State $k \gets $ number of keys in
    $\mathsf{PartitionAdjList}$\label{l:enc-nu-partition-graphs} \Comment{number
      of partitions graphs to be encoded}
    \State $\mathsf{Ouptut} \gets \mathsf{Output} \concat \edelta(k+1)$ \label{l:enc-k-output}
    \For{$(i,i') \in \mathsf{PartitionAdjList}.\textproc{Keys}$} \label{l:enc-part-graph-for}
    \If{$i < i'$}
    \State $\mathsf{Output} \gets \mathsf{Output} \concat i \concat i'$ \label{l:enc-part-neq-i-write}
    \Comment{use $1+\lfloor  \log_2 \mathsf{TCount} \rfloor$ bits to encode $i$ and $i'$}
    \State $\va \gets \mathsf{PartitionDeg}(i,i'), \vb \gets
    \mathsf{PartitionDeg}(i',i)$ \label{l:enc-part-neq-va-vb} \Comment{left and
      right degree sequences}
    \State $f \gets \textproc{BEncodeGraph}(\textproc{Size}(\va),
    \textproc{Size}(\vb), \va, \vb,
    \mathsf{PartitionAdjList}(i,i'))$ \label{l:enc-part-neq-call-bencodegraph}
    \Statex \Comment{Algorithm~\ref{alg:bip-comp} in
   Section~\ref{sec:bipartite-compression}}
    \State $\mathsf{Output} \gets \mathsf{Output} \concat
    \edelta( 1 + f)$ \label{l:enc-part-neq-write-f-output}
    \EndIf
    \If{$i = i'$}
        \State $\mathsf{Output} \gets \mathsf{Output} \concat i \concat i'$ \label{l:enc-part-eq-i-write}
    \Comment{use $1+\lfloor  \log_2 \mathsf{TCount} \rfloor$ bits to encode $i$ and $i'$}
    \State $\va \gets \mathsf{PartitionDeg}(i,i)$ \label{l:enc-part-eq-va-def}
    \Comment{degree sequence}
    \State $(f, \vec{\tilde{f}}) \gets \textproc{EncodeGraph}(\textproc{Size}(\va), \va,
    \mathsf{PartitionAdjList}(i,i))$ \label{l:enc-part-eq-call-encodegraph}
    \Statex \Comment{Algorithm~\ref{alg:self-find-f} in 
   Section~\ref{sec:simple-graph-compression}}
    \State $\mathsf{Output} \gets \mathsf{Output} \concat \edelta(1+f) $ \label{l:enc-part-eq-f-write}
    \State $\mathsf{Output} \gets \mathsf{Output} \concat \edelta(1 +
    \textproc{Size}(\vec{\tilde{f}}))$ \label{l:enc-part-eq-size-tvf-write}
    \For{$1 \leq j \leq \textproc{Size}(\vec{\tilde{f}})$} \label{l:enc-part-eq-ftilde-for}
    \State $\mathsf{Output} \gets \mathsf{Output} \concat \edelta(1+\tilde{f}_j)$ \label{l:enc-part-eq-ftilde-write}
    \EndFor \label{l:enc-part-eq-tilde-f-write-end-for}
    \EndIf
    \EndFor \label{l:enc-part-graph-for-end}
    \EndFunction
  \end{algorithmic}
\end{myalg}

Recall that the vertex mark set  and the edge mark set  are of the form $\vermark =
\{1, \dots, |\vermark|\}$ and $\edgemark = \{1, \dots, |\edgemark|\}$
respectively. We assume that the input graph $\Gn$ is given via its
edge list representation. More specifically, the array $\vthetan = (\thetan_v: v
\in [n])$ of integers is given such that $\thetan_v$ is the mark of vertex $v$
in $\Gn$. Moreover, an array $\mathsf{EdgeList} = (\mathsf{EdgeList}_i: 1 \leq i
\leq \mn)$  is given where $\mn$ is the total number of edges in $\Gn$, and for
$1 \leq i \leq \mn$, $\mathsf{EdgeList}_i = (v_i, w_i, x_i, x'_i)$ is an edge in
$\Gn$ between vertices $v_i$ and $w_i$, with mark $x_i$ towards $v_i$, and mark
$x'_i$ towards $w_i$. Also, positive integers $h$ and $\delta$ are given as
hyperparameters of the algorithm. 
\ifreplace
We now discuss the steps of our compression algorithm.
\color{black}
\fi

\begin{enumerate}
\ifreplace
\item First, using Algorithm~\ref{alg:preprocess} which will be explained in
  Section~\ref{sec:detail-preprocessing},
  we find the neighbor list
  representation 
  of $\Gn$,
  i.e. 
  $(\vthetan, \vdn, \vgamman, \vtgamman, \vec{x}^{(n)}, \vec{x'}^{(n)})$, as was
   discussed in Section~\ref{sec:alg-high-level-preprocessing}.
\color{black}
\fi
\ifreplace
 \item Next, 
 using
   Algorithm~\ref{alg:type-extract-message-passing-new}, which will be discussed
   in Section~\ref{sec:MP}, 
 we find integer representations of the edge types as we discussed in
   Section~\ref{sec:alg-high-level-edge-type-def}. More precisely, we 
   find $\vec{c}, \mathsf{TCount},
   \mathsf{TIsStar}$, and $\mathsf{TMark}$ as was explained in
   Section~\ref{sec:alg-high-level-edge-type-def}.
   Here $\vec{c}=
   (c_{v,i}: v \in [n], i \in [\dn_v])$ is an array of array of integers, where for $v \in [n]$ and
   $1 \leq i \leq d_v$, $c_{v,i}$ is a pair of integers representing
   $\type{v}{\gamman_{v,i}}$, i.e.\ the type of
   the edge between $v$ and $\gamman_{v,i}$. Moreover, each component of the
   pair $c_{v,i}$ is in the range $\{1, \dots, \mathsf{TCount}\}$. On the other
   hand, $\mathsf{TIsStar} = (\mathsf{TIsStar}(i): 1 \leq i \leq
   \mathsf{TCount})$ is an array of zeros and ones. Also, $\mathsf{TMark} =
   (\mathsf{TMark}(i): 1 \leq i \leq \mathsf{TCount})$ is an array of integers.
   Moreover, recalling our discussion in
   Section~\ref{sec:alg-high-level-edge-type-def}, for an integer $i$ in the range $\{1, \dots, \mathsf{TCount}\}$ which
   corresponds to some $t = J_n^{-1}(i) \in \mFbardeltah$, $\mathsf{TIsStar}(i)$ is 1 if $t
   \in \mFbardeltah \setminus \mFdeltah$, i.e.\ $t$ is of the form $\star_x$ for
   some $x \in \edgemark$, and zero otherwise. Also, $\mathsf{TMark}(i)$ is equal to $t[m]$, i.e.\
   the mark component of $t$, if $t \in \mFdeltah$, and $\mathsf{TMark}(i) = x$
   if $t = \star_x$.
\color{black}
\fi
\ifreplace
\item We then write $\mathsf{TCount}, \mathsf{TIsStar}$, and $\mathsf{TMark}$
   to the output. In order to do so,
   we first
   use the Elias delta code to append $\edelta(\mathsf{1+TCount})$ to the output.
   Here, we encode
   $1+\mathsf{TCount}$ to make sure that the input to the Elias delta encoder is
   positive. Subsequently, in
   the loop of line~\ref{l:enc-tiistar-tmark-for}, for each $1 \leq i \leq
   \mathsf{TCount}$, we use 1 bit to represent $\mathsf{TIsStar}(i)$, and $1 +
   \lfloor \log_2|\edgemark| \rfloor$ bits to represent $\mathsf{TMark}(i)$.
\color{black}
\fi
 \item In line~\ref{l:enc-star-vertices}, we use
   Algorithm~\ref{alg:encode-star-vertices} which will be explained in
   Section~\ref{sec:enc-star-vertices} to encode star vertices, as we discussed
   in Section~\ref{sec:main-star-vertices}.
 \item In line~\ref{l:enc-star-edges}, we use
   Algorithm~\ref{alg:encode-star-edges} which will be discussed in Section~\ref{sec:encode-star-edges}
   to encode star edges, as we discussed in Section~\ref{sec:main-star-edges}.
 \item The next step is to encode vertex types, as we discussed in
   Section~\ref{sec:main-encode-vertex-types}. In order to do so, we must first
   find $\Dn(v)$ for each vertex $v \in [n]$ in $\Gn$,  as was defined in~\eqref{eq:Dn-def}.
   In order to have an efficient representation of $\vDn$, for a vertex $1 \leq
   v \leq n$, we only store $\Dn_{t,t'}(v)$ which are nonzero.
   In order to do so, we employ the dictionary data structure to store $\Dn(v)$. 
More precisely, motivated by the
   fact that we have an integer representation for the elements in $\mFdeltah$,
for each $v \in [n]$, we
   store $\Dn(v)$ as a dictionary, where each key is a pair of integers $(i,\ip)$
   in the
   range $\{1, \dots, \mathsf{TCount}\}$, and the corresponding value is
   $\Dn_{J_n^{-1}(i), J_n^{-1}(\ip)}(v)$. A pair of integers $(i,\pp{i})$ such that $i, \pp{i} \in \{1,
   \dots, \mathsf{TCount}\}$ is in the set of keys
   of the dictionary at vertex $v$ 
   iff $\Dn_{J_n^{-1}(i), J_n^{-1}(\ip)}(v) > 0$.  We denote
   the dictionary corresponding to vertex $v$ by $\mathsf{Deg}_v$, and the array
   of $\mathsf{Deg}_v$ for $1 \leq v \leq n$ is denoted by $\mathsf{Deg}$. 
   This
   is defined in line~\ref{l:encode-deg-def}. Then, in
   line~\ref{l:enc-find-deg-call}, we call Algorithm~\ref{alg:encode-find-deg},
   which will be explained in Section~\ref{sec:encode-find-Deg}, to construct
   $\mathsf{Deg}$ as was explained above.
Then, in line~\ref{l:enc-encode-vertex-types-call}, we call
   Algorithm~\ref{alg:encode-vertex-types}, which will be explained in
   Section~\ref{sec:encode-vertex-types}, to encode the types of vertices in
   $\Gn$. 
   Recall that the type of a vertex $v$ is the pair $(\thetan(v),
   \Dn(v))$. See Figure~\ref{fig:Deg-2-7-example} for an illustration of
   $\mathsf{Deg}$. 

   \begin{figure}
   \begin{center}
     \begin{tabular}[t]{l@{\qquad\qquad\qquad}r}
       $\mathsf{Deg}_2$ =
       \begin{tabular}{@{}lc@{}}
         \toprule
         Key & Value \\ \midrule
         $(3,4)$ & $1$ \\
         \bottomrule
       \end{tabular}%
       &%
       $\mathsf{Deg}_7$ =
       \begin{tabular}{@{}lc@{}}
         \toprule
         Key & Value \\ \midrule
         $(4,3)$ & $1$ \\
         $(5,5)$ & $1$ \\
         \bottomrule
       \end{tabular}
     \end{tabular}
   \end{center}
   \caption{$\mathsf{Deg}_2$ and $\mathsf{Deg}_7$ for the graph of
     Figure~\ref{fig:sample-graph-for-compression} with parameters $h = 2$ and
     $\delta = 4$. See Figure~\ref{fig:message-passing-example-summary} for the
     type of edges in this example, and Figure~\ref{fig:Dn-example} for
     $\Dn(2)$ and $\Dn(7)$. Note that since all edges connected to vertex $1$
     in this example are star edges, $\mathsf{Deg}_1$ is an empty dictionary.
     Moreover, due to the symmetry in the graph,  $\mathsf{Deg}_v$ for $v \in
     \{2, \dots, 6\}$, and also $\mathsf{Deg}_w$ for $w \in \{7, \dots, 16\}$
     have the same values. \label{fig:Deg-2-7-example}}
 \end{figure}
 
 \item At this point, we are ready to find and encode the partition graphs
   $(\Gn_{t,\pp{t}}: (t,\pp{t}) \in \mEn_{\leq})$, as was
   explained in Section~\ref{sec:main-encode-partition}. 
   In order to do so, we define the following
   variables. In what follows, we describe what we expect to be the content of
   these variables. In 
   Line~\ref{l:enc-find-partition-graphs-call}, we call
   Algorithm~\ref{alg:encode-find-partition-graphs} which will be explained in
   Section~\ref{sec:encode-find-partition-graphs} to find these variables as we
   expect. 
   \begin{itemize}
   \item $\mathsf{PartitionAdjList}$:
which is defined in
     line~\ref{l:enc-partition-adj-list-def}, stores the adjacency lists of
     partition graphs. 
More precisely, $\mathsf{PartitionAdjList}$ is a dictionary, where each key is of
     the form $(i,\pp{i})$, $i \leq \pp{i} \in \{1, \dots,
     \mathsf{TCount}\}$, such that with $t = J_n^{-1}(i)$ and $\pp{t} = J_n^{-1}(\ip)$ being the  elements in $\mFdeltah$ corresponding to $i$ and
     $\pp{i}$ respectively, we have $(t, \pp{t}) \in
     \mEn_{\leq}$. Recalling the definition of $\mEn_{\leq}$, this is equivalent
     to $(t,\tp) \in \mEn$ and $i \leq \ip$. For such $(i,\pp{i})$,
     and the corresponding $(t,\pp{t})$, the value corresponding to the key
     $(i,\pp{i})$ in the dictionary $\mathsf{PartitionAdjList}$, which we denote
     by $\mathsf{PartitionAdjList}(i,\pp{i})$, is an array of
     array of integers with the following property. If $i < \pp{i}$, then for $1
     \leq p \leq \Nn_{t,\pp{t}}$, $(\mathsf{PartitionAdjList}(i,\pp{i}))_p$ is
     an array of integers which stores the adjacency list of the left node $p$
     in $\Gn_{t,\pp{t}}$ in an increasing order. Likewise, recalling the notion
     \emph{forward adjacency list} from Section~\ref{sec:main-self-comp}, if $i = \pp{i}$, for
     $1 \leq p \leq \Nn_{t, t}$, $(\mathsf{PartitionAdjList}(i,\pp{i}))_p$
     is an array of integers, which stores the forward adjacency
     list of the node
     $p$ in $\Gn_{t,t}$ in an increasing order.
See Figure~\ref{fig:partition-adj-list-example} for an example.

\begin{figure}
\begin{center}

  $\mathsf{PartitionAdjList}: \qquad$
  \begin{tabular}{@{}lc@{}}
    \toprule
    Key & Value \\
    \midrule
    $(3,4)$ & $((1,2), (3,4), (5,6), (7,8), (9,10))$ \\
    $(5,5)$ & $((2), (), (4), (), (6), (), (8), (), (10), ())$ \\
    \bottomrule
  \end{tabular}
\end{center}
\caption[The content of $\mathsf{PartitionAdjList}$]{The content of $\mathsf{PartitionAdjList}$ for the graph of
  Figure~\ref{fig:sample-graph-for-compression} with parameters $h = 2$ and
  $\delta = 4$. See Figure~\ref{fig:partition-graphs-example} for the
  partition graphs in this example.  For instance, the value corresponding to
  the key $(3,4)$ represents the adjacency lists of the left vertices in
  $\Gn_{\tthree{0.5}\,,\,\tfour{0.5}}$, while the value corresponding to the
  key $(5,5)$ represents the forward adjacency list of the vertices in $\Gn_{\tfive{0.5}\,,\,\tfive{0.5}}$.\label{fig:partition-adj-list-example}}
\end{figure}

   \item $\mathsf{PartitionDeg}$: 
   which is defined in
     line~\ref{l:enc-partition-deg-def}, stores the degree of vertices in
     partition graphs. 
     More precisely, $\mathsf{PartitionDeg}$ is a dictionary,
     where each key is of the form $(i,\pp{i})$, $i,\pp{i} \in \{1, \dots,
     \mathsf{TCount}\}$, such that with $t = J_n^{-1}(i)$ and $\pp{t} = J_n^{-1}(\ip)$ being the  elements in $\mFdeltah$ corresponding to $i$ and
     $\pp{i}$ respectively, we have $(t,\pp{t}) \in \mEn$. For such $(i,\pp{i})$,
     and the corresponding $(t,\pp{t})$, the value corresponding to the key
     $(i,\pp{i})$ in the dictionary $\mathsf{PartitionDeg}$, which we denote by
     $\mathsf{PartitionDeg}(i,\pp{i})$, is an array of integers of size
     $\Nn_{t,\pp{t}}$ with the following property. If $i < \pp{i}$, for $1 \leq
     p \leq \Nn_{t,\pp{t}}$, $(\mathsf{PartitionDeg}(i,\pp{i}))_p$ is the degree
     of the left node $p$ in $\Gn_{t,\pp{t}}$. Likewise, if $i > \pp{i}$, for $1 \leq
     p \leq \Nn_{t,\pp{t}}$, $(\mathsf{PartitionDeg}(i,\pp{i}))_p$ is the degree
     of the right node $p$ in $\Gn_{\pp{t}, t}$. Moreover, if $i = \pp{i}$, then
     for $1 \leq p \leq \Nn_{t,t}$, $(\mathsf{PartitionDeg}(i,i))_p$ is the
     degree of the node $p$ in $\Gn_{t,t}$.
See Figure~\ref{fig:partition-deg-example} for an example.

\begin{figure}
\begin{center}
  
  $\mathsf{PartitionDeg}: \qquad $
  \begin{tabular}{@{}lc@{}}
    \toprule
    Key & Value \\
    \midrule
    $(3,4)$ & $(2,2,2,2,2)$ \\
    $(4,3)$ & $(1,1,1,1,1,1,1,1,1,1)$ \\
    $(5,5)$ & $(1,1,1,1,1,1,1,1,1,1)$ \\
              \bottomrule
  \end{tabular}
\end{center}
\caption[Example of $\mathsf{PartitionDeg}$]{The content of $\mathsf{PartitionDeg}$ for the graph of
  Figure~\ref{fig:sample-graph-for-compression} with parameters $h = 2$ and
  $\delta = 4$. See Figure~\ref{fig:partition-graphs-example} for partition
  graphs in this example. For instance, the value corresponding to the key
  $(3,4)$ is the degree sequence of the left nodes in $\Gn_{\tthree{0.5}\,,\,
    \tfour{0.5}}$. But from Figure~\ref{fig:partition-graphs-example}, $\Gn_{\tthree{0.5}\,,\,
    \tfour{0.5}}$ has five left nodes, each with degree 2. Moreover, the value
  corresponding to the key $(5,5)$ is the degree sequence of the vertices in
  $\Gn_{\tfive{0.5}\,,\,\tfive{0.5}}$. But
  $\Gn_{\tfive{0.5}\,,\,\tfive{0.5}}$ has 10 nodes, each with degree 1.   \label{fig:partition-deg-example}}
\end{figure}

   \item $\mathsf{PartitionIndex}$: 
   which is defined in
     line~\ref{l:enc-partition-index-def}, stores the mappings $(\In_{t,\pp{t}}:
     (t,\pp{t}) \in \mEn)$. 
     Recall that for $(t,\pp{t}) \in \mEn$, $\In_{t,\pp{t}}$  maps the index of vertices in $\mVn_{t,\pp{t}}$ to those in
     the partition graph $\Gn_{t,\pp{t}}$. More precisely, $\mathsf{PartitionIndex}$ is a array
     of size $n$, where $\mathsf{PartitionIndex}_v$ for $1 \leq v \leq n$ is a
     dictionary, with each key being a pair of integers $(i,\pp{i})$, 
     $i,\pp{i} \in \{1, \dots, \mathsf{TCount}\}$, such that with $t =
     J_n^{-1}(i)$ and $ \pp{t}  = J_n^{-1}(\pp{i})$  being the elements in $\mFdeltah$ corresponding to $i$ and
     $\pp{i}$ respectively, we have $(t,\pp{t}) \in \mEn$ and $v \in
     \mVn_{t,\pp{t}}$. For such $(i,\pp{i})$, and the corresponding
     $(t,\pp{t})$, the value corresponding to the key $(i,\pp{i})$ in
     $\mathsf{ParitionIndex}_v$, which we denote by
     $\mathsf{PartitionIndex}_v(i,\pp{i})$, is precisely $\In_{t,\pp{t}}(v)$.
     See Figure~\ref{fig:partition-index-example} for an example. 

\begin{figure}
     \begin{center}
       $\mathsf{PartitionIndex}_7:\qquad$
       \begin{tabular}{@{}lc@{}}
         \toprule
         Key & Value \\
         \midrule
         $(4,3)$ & 1 \\
         $(5,5)$ & 1 \\
         \bottomrule
       \end{tabular}%
       \qquad \qquad
       $\mathsf{PartitionIndex}_3:\qquad $
       \begin{tabular}{@{}lc@{}}
         \toprule
         Key & Value \\
         \midrule
         $(3,4)$ & 2 \\
         \bottomrule
       \end{tabular}
     \end{center}
     \caption[Content of $\mathsf{PartitionIndex}$]{The content of $\mathsf{PartitionIndex}_7$ and
       $\mathsf{PartitionIndex}_3$ for the graph of
       Figure~\ref{fig:sample-graph-for-compression} with parameters $h=2$ and
       $\delta = 4$. From Figure~\ref{fig:mEn-mVn-In}, we know that vertex $7$
       appears in $\mVn_{\tfour{0.5}\,,\,\tthree{0.5}}$ and
       $\mVn_{\tfive{0.5}\,,\,\tfive{0.5}}$. Therefore, recalling the values of
       $J_n$ from Figure~\ref{fig:message-passing-example-summary}, we realize
       that the keys $(4,3)$ and $(5,5)$ must exist in
       $\mathsf{PartitionIndex}_7$. Moreover, since
       $\In_{\tfour{0.5}\,,\,\tthree{0.5}}(7) = 1$ and
       $\In_{\tfive{0.5}\,,\,\tfive{0.5}}(7) = 1$, we realize that the values
       corresponding to both of these keys must be $1$, as we see above. \label{fig:partition-index-example}}
   \end{figure}
   
   \end{itemize}
   As was mentioned above, in line~\ref{l:enc-find-partition-graphs-call}, we call
   Algorithm~\ref{alg:encode-find-partition-graphs} which will be explained in
   Section~\ref{sec:encode-find-partition-graphs} to find these variables as
   was explained above. 
   Subsequently, in line~\ref{l:enc-nu-partition-graphs},  we find the number of partition graphs of the form
   $\Gn_{t,t'}, (t,t') \in \mEn_{\leq}$, which is
   precisely $|\mEn_\leq|$. 
  Note that due to
   the definition, a pair of integers $(i,i')$ is in the set of keys in
   $\mathsf{PartitionAdjList}$ iff $(t, t') \in \mEn_{\leq}$, where
   $t = J_n^{-1}(i)$ and $t' = J_n^{-1}(\ip)$ are the elements in $\mFdeltah$ corresponding
   to $i$ and $i'$ respectively. With this, the number of partition graphs,
   which is precisely $|\mEn_{\leq}|$, is the number of keys in
   $\mathsf{PartitionAdjList}$.  
   \ifreplace
   We then write this number plus $1$ using the Elias delta code, so that the decoder knows how many partition graphs it must
   expect to read and decode.
   Note
     that it can be the case that all the edges in $\Gn$ are star edges, in
     which case the number of partition graphs is zero, hence it is important to add $1$ to this number
     since the Elias delta code expects to
   receive a positive integer as its input.
   \color{black}
   \fi

   Afterwards, in the loop of line~\ref{l:enc-part-graph-for}, we encode
   each partition graph by going through the set of keys in
   $\mathsf{PartitionAdjList}$ of the form $(i,i')$. First
   assume that $i < i'$ and let $t = J_n^{-1}(i)$ and $t' = J_n^{-1}(\ip)$ be the members in
   $\mFdeltah$ corresponding to $i$ and $i'$ respectively. In this case, 
   we
   first write $i$ and $i'$ to the output in line~\ref{l:enc-part-neq-i-write}
   using $1 + \lfloor \log_2 \mathsf{TCount} \rfloor$ bits each. 
   Note that by
   definition, the left and right degree sequences of the partition graph
   $\Gn_{t, t'}$ are $\mathsf{PartitionDeg}(i,i')$ and
   $\mathsf{PartitionDeg}(i',i)$ respectively. 
   These are stored in
   line~\ref{l:enc-part-neq-va-vb} in arrays $\va$
   and $\vb$ respectively. 
   Then,  in
   line~\ref{l:enc-part-neq-call-bencodegraph} we call
   Algorithm~\ref{alg:bip-comp}, which will be explained in
   Section~\ref{sec:bipartite-compression}, to encode $\Gn_{t,
     t'}$. This algorithm is designed to compress a simple unmarked
   bipartite graph given its degree sequence, as we discussed in
   Section~\ref{sec:main-bip-comp}. Therefore, it receives as input the number
   of left and right nodes, left and right degree sequences, and the adjacency
   list of the left nodes in the bipartite graph $\Gn_{t,\pp{t}}$. Note that the number of left and right nodes in
   $\Gn_{t,t'}$ are precisely the size of $\va$ and $\vb$
   respectively. Moreover, the adjacency list of the left nodes in $\Gn_{t,t'}$ is stored in
   $\mathsf{PartitionAdjList}(i,i')$. Then, motivated by
   Proposition~\ref{prop:bipartite-color-graph-compress}, the algorithm returns an integer
   representing $\Gn_{t,t'}$, 
   which we store in variable $f$ in
   line~\ref{l:enc-part-neq-call-bencodegraph}. 
   Finally, 
   we use the Elias delta code
   in line~\ref{l:enc-part-neq-write-f-output} 
   to write $f$ to the output. Note
   that since the input to the Elias
   delta code must be a positive integer, we encode $1+f$ instead of $f$.

   Now, consider the case $i = i'$ and let $t = J_n^{-1}(i)$ be the element in
   $\mFdeltah$ corresponding to $i$. Similar to the previous case, 
   we write $i$
   to the output in line~\ref{l:enc-part-eq-i-write}. 
   Note that since the
   decoder does not know whether the incoming partition graph is bipartite or
   simple, we write both $i$ and $i'$, even though we know they are equal in
   this case. Also, 
   we set $\va$ to be $\mathsf{PartitionDeg}(i,i)$ in
   line~\ref{l:enc-part-eq-va-def}, 
   which is precisely the degree sequence of
   $\Gn_{t, t}$ by the definition of  $\mathsf{PartitionDeg}$.
   Then, we call Algorithm~\ref{alg:self-find-f}, which will be explained in
   Section~\ref{sec:simple-graph-compression}, to encode $\Gn_{t,
     t}$. This algorithm is designed to compress a simple unmarked graph
   given its degree sequence as we discussed in
   Section~\ref{sec:main-self-comp}. Therefore, it receives as input the number
   of nodes, the degree sequence, and the forward
   adjacency list of a simple unmarked graph. Note that the number of nodes in
   $\Gn_{t, t}$ is precisely the size of $\va$.
   Additionally, the forward adjacency list of
   $\Gn_{t, t}$ is stored in $\mathsf{PartitionAdjList}(i,i)$. Then, motivated
   by Proposition~\ref{prop:equal-color-graph-compress}, Algorithm~\ref{alg:self-find-f} returns an integer which
   we store in variable $f$, together with an array of integers, 
   which we store
   in array $\vec{\tilde{f}}$ in line~\ref{l:enc-part-eq-call-encodegraph}.
   Finally, in lines~\ref{l:enc-part-eq-f-write}
   through~\ref{l:enc-part-eq-tilde-f-write-end-for}, we write $f$ and
   $\vec{\tilde{f}}$ to the output 
   using the Elias delta code. 
\end{enumerate}


\editfinish

\editstart


\subsubsection{Preprocessing the graph}
\label{sec:detail-preprocessing}

In this section, we discuss how to find the neighbor list representation
$(\vthetan, \vdn, \vgamman, \vtgamman, \vec{x}^{(n)}, \vec{x'}^{(n)})$ of a simple marked graph $\Gn \in
\mGb_n$ given in its edge list representation. 
As we
discussed in Section~\ref{sec:enc-alg}, we assume that the vertex mark sequence
$\vthetan$ and the edge list $\mathsf{EdgeList} = (\mathsf{EdgeList}_i: 1 \leq i
\leq \mn)$ are given such that $\mathsf{EdgeList}_i = (v_i, w_i, x_i, x'_i)$. Algorithm~\ref{alg:preprocess} below illustrates how to
obtain the neighbor list representation.
\ifreplace
We first go through $\mathsf{EdgeList}$
and modify each index to ensure that for $1 \leq i \leq \mn$, we have $v_i <
w_i$. This is done in the loop of
line~\ref{l:preprocess-init-for}. 
\color{black}
\fi
Note that the orientation in representing an
edge does not matter, 
hence
we are allowed to simultaneously swap $v_i$ with $w_i$ and $x_i$ with $x'_i$ in line
\ref{l:preprocess-swap} if necessary. 
Next, in 
line~\ref{l:preprocess-sort}, we sort $\mathsf{EdgeList}$  with respect to the
lexicographic order of the pair $(v_i,w_i)$. 
That is, we change the ordering of
edges to ensure that for $1 \leq
i < j \leq \mn$, we either have $v_i < v_j$ or $v_i = v_j$ and $w_i < w_j$. 
Then, it is easy to see that the neighbors of a vertex $v$ appear in the correct
order in this list. More precisely, for $1 \leq v \leq n$, there are integers $1 \leq i_1
< i_2 < \dots < i_{\dn_v} \leq \mn$ such that for $1 \leq j \leq \dn_v$, either
$v_{i_j} = \gamman_{v, j}$ or $w_{i_j} = \gamman_{v, j}$.  
Motivated by this, in
the loop of line~\ref{l:preprocess-update-for}, 
we go over the edges in the list
$\mathsf{EdgeList}$
and update the variables $\vdn, \vgamman$, $\vtgamman$, $\vec{x}^{(n)}$, and
$\vec{x'}^{(n)}$.

\begin{myalg}[Preprocess a simple marked graph to find its equivalent
  neighbor list representation\label{alg:preprocess}]
  \begin{algorithmic}[1]
    \INPUT
    \Statex $n$: number of vertices 
    \Statex A simple marked graph $\Gn$ represented by
    \begin{itemize}
    \item its vertex mark sequence $\vthetan = (\thetan_v: v \in [n])$,
      where $\thetan_v \in \vermark$ is the mark of vertex $v$ in $\Gn$.
      \item $\mathsf{EdgeList} = (\mathsf{EdgeList}_i: 1 \leq i \leq \mn)$: the
list of edges in $\Gn$ where  $\mathsf{EdgeList}_i = (v_i, w_i, x_i, x'_i)$ for $1 \leq i \leq
\mn$. Here, $\mn$ denotes the total number of edges in $\Gn$, and for $1 \leq i
\leq \mn$, the tuple $(v_i, w_i, x_i, x'_i)$ represents an edge between the
vertices $v_i$ and $w_i$ with mark $x_i$ towards $v_i$ and mark $x'_i$ towards
$w_i$, i.e.\ $\xi_{\Gn}(w_i, v_i) = x_i$ and $\xi_{\Gn}(v_i, w_i) = x'_i$.
\end{itemize}
\OUTPUT
\Statex The equivalent representation of $\Gn$ of the form
\begin{itemize}
\item $\vthetan = (\thetan_v: v \in [n])$ where $\thetan_v$
  denotes the vertex mark of $v$. 
\item $\vdn = (\dnn_v: v \in [n])$  such that $\dnn_v$ for $1 \leq v
  \leq n$ is the degree of vertex $v$ in $\Gn$. 
\item $\vgamman$: $\Array$ of $\Array$ of integers, such that for $1 \leq v \leq
  n$, the neighbors of vertex $v$ in $\Gn$ is stored in 
  an increasing order as $1 \leq \gamman_{v,1} < \gamman_{v,2} < \dots <
  \gamman_{v,\dnn_v} \leq n$.
\item $\vec{\tilde{\gamma}}^{(n)}$: $\Array$ of $\Array$ of integers, such that for $1 \leq v \leq n$ and $1 \leq i \leq \dn_v$, $\tgamman_{v,i}$ denotes
  the index of $v$ among the neighbors of $\gamman_{v,i}$, so that
  $\gamman_{\gamman_{v,i}, \tgamman_{v,i}} = v$. 
\item $\vec{x}^{(n)}$ and ${\vec{x'}}^{(n)}$: $\Array$ of $\Array$ of integers, such that 
  for $1 \leq v \leq n$ and $1 \leq i \leq
  \dn_v$, $\xn_{v,i}$ and $\xnp_{v,i}$ denote the two edge marks corresponding to the
  edge connecting $v$ to $\gamman_{v,i}$, so that $\xn_{v,i} =
  \xi_{\Gn}(\gamman_{v,i}, v)$ and $\xnp_{v,i} = \xi_{\Gn}(v, \gamman_{v,i})$.
\end{itemize}
\Function{Preprocess}{$n, \vthetan, \mathsf{EdgeList}$}
\State $\vdn \gets \Array $ of integers of size $n$ \Comment{initialize $\vdn$}
\State $\vgamman, \vec{\tilde{\gamma}}^{n}, \vec{x}^{(n)}, {\vec{x'}}^{(n)}
\gets \Array \text{ of } \Array$ of integers of size $n$ \Comment{initialize
  $\vgamman, \vec{\tilde{\gamma}}^{n}, \vec{x}^{(n)}$, and ${\vec{x'}}^{(n)}$}
\For{$1 \leq i \leq n$}
\State $\dnn_i \gets 0$ \Comment{initialize degree sequence with zero}
\EndFor
\For{$1 \leq i \leq \mn$} \label{l:preprocess-init-for}

\If{$v_i > w_i$}
\State $\textproc{Swap}(v_i, w_i)$, $\textproc{Swap}(x_i,
x'_i)$ \label{l:preprocess-swap} \Comment{to
  make sure that for all $1 \leq i \leq \mn$, we have $v_i < w_i$}
\EndIf
\EndFor
\State $\mathsf{EdgeList} \gets \textproc{Sort}(\mathsf{EdgeList})$ \label{l:preprocess-sort} \Comment{sort $\mathsf{EdgeList}$
  with respect to the lexicographic order of the pair $(v_i, w_i)$}
\For{$1 \leq i \leq \mn$} \label{l:preprocess-update-for}
\State append $w_i$ to $\gamman_{v_i} $ \Comment{add $w_i$ to the neighbor
  list of $v_i$}
\State append $x_i$ to $\xn_{v_i}$, append $x'_i$ to $\xnp_{v_i}$
\Comment{append the mark pair}
\State append $1+\dnn_{w_i}$ to $\tgamman_{v_i}$
\Statex \Comment{$v_i$ is the newly
  added neighbor of $w_i$, and its index among the neighbors of $w_i$ should  be one plus the number of existing
  neighbors of $w_i$, i.e.\ $1+\dn_{w_i}$}
\State append $v_i$ to $\gamman_{w_i}$ \Comment{add $v_i$ to the
  neighbor list of $w_i$}
\State append $x'_i$ to $\xn_{w_i} $, append $x_i$ to $\xnp_{w_i}$ \Comment{append the mark pair}
\State append $1+\dnn_{v_i}$ to $\tgamman_{w_i}$
\Statex \Comment{$w_i$ is the newly
  added neighbor of $v_i$, and its index among the neighbors of $v_i$ should be one plus the number of existing
  neighbors of $v_i$, i.e.\ $1+\dn_{v_i}$}
\State $\dnn_{v_i} \gets \dnn_{v_i} + 1 , \dnn_{w_i} \gets \dnn_{w_i} + 1$
\Comment{add one to the number of existing neighbors}
\EndFor
\State \textbf{return} $(\vthetan, \vdn, \vgamman, \vtgamman, \vec{x}^{(n)}, \vec{x'}^{(n)})$
\EndFunction
  \end{algorithmic}
\end{myalg}

Note that the dominant part of the calculation in Algorithm~\ref{alg:preprocess}
is sorting $\mathsf{EdgeList}$ in line~\ref{l:preprocess-sort}, 
which is
performed in $O(\mn \log \mn (\log n + \log |\edgemark|))$ time. Moreover, since there are $\mn$ edges in
the graph, the memory complexity of Algorithm~\ref{alg:preprocess} is
$O(\mn(\log n + \log(|\edgemark| + |\vermark|)))$.
To sum up, we have 
\begin{lem}
  \label{lem:preprocess-complexity}
  The time and memory complexities of Algorithm~\ref{alg:preprocess} are
  $O(\mn \log \mn(\log n + \log |\edgemark|))$ and $O(\mn(\log n +
  \log(|\edgemark| + |\vermark|)))$ respectively. 
\end{lem}

\editfinish

\editstart
\subsubsection{Compressing a Sequence of Nonnegative Integers}
\label{sec:compress-sequence}


In this section, we prove the first part of Proposition~\ref{prop:encode-sequence} by
presenting an encoding algorithm that produces a prefix free bit string given a
sequence $\vy = (y_i, i \in [n])$ of nonnegative integers. The second part of
this proposition is proved in Section~\ref{sec:decode-sequence} which presents the corresponding
decoding algorithm. 

\begin{figure}
  \centering
  \begin{tikzpicture}
    \begin{scope}[xshift=-4cm]
      \node[scale=1.2] at (0,0) {$\vy = (0,0,1,0,2,0,1,2)$};
      \node at (0,-2.5) {$(a)$};
    \end{scope}
    \begin{scope}[xshift=4cm]
      \begin{scope}[xshift=-1.3cm,scale=0.5]
        \foreach \y/\i in {3.5/1,2.5/2,1.5/3,0.5/4,-0.5/5,-1.5/6,-2.5/7,-3.5/8}{
          \node[fill,draw,circle,inner sep=1pt] at (0,\y) (l\i) {};
          \nodelabel{l\i}{180}{\i}
        }
      \end{scope}
      \begin{scope}[xshift=1.3cm,scale=0.5]
        \foreach \y/\i in {1/1,0/2,-1/3}{
          \node[fill,draw,circle,inner sep=1pt] at (0,\y) (r\i) {};
          \nodelabel{r\i}{0}{\i}
        }
      \end{scope}
      \draw
      (l1) -- (r1)
      (l2) -- (r1)
      (l3) -- (r2)
      (l4) -- (r1)
      (l5) -- (r3)
      (l6) -- (r1)
      (l7) -- (r2)
      (l8) -- (r3);
      \node at (0,-2.5) {$(b)$};
    \end{scope}
  \end{tikzpicture}
  \caption{\label{fig:seq-bip-graph-example} The auxiliary simple unmarked bipartite
    graph $G$ associated to the sequence $\vy$ in $(a)$ is illustrated in $(b)$.
  With $K := 1 + \max\{y_i: i \in [n]\}$, we construct $G$ by connecting a left node $1 \leq i \leq n$ to a right node
  $1 \leq j K$ if $y_i = j-1$.} 
\end{figure}
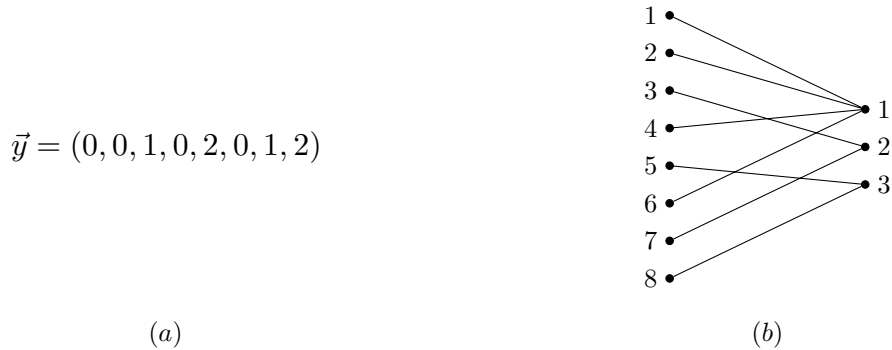

Our compression scheme constructs an auxiliary simple unmarked bipartite graph
$G$ corresponding to $\vy$ and then uses the algorithm introduced in
Proposition~\ref{prop:bipartite-color-graph-compress} (and in
Section~\ref{sec:bipartite-compression} in detail) to encode $G$. More precisely, with $K
:= 1 + \max\{y_i: i \in [n]\}$, we construct a simple unmarked bipartite graph
$G$ with $n$ left nodes and $K$ right nodes, where a left node $1 \leq i \leq n$
is connected to a right node $1 \leq j \leq K$ if $y_i = j-1$.  See
Figure~\ref{fig:seq-bip-graph-example} for an example.

Algorithm~\ref{alg:compress-sequence} below implements the above procedure. More
precisely, given  $\vy$, 
we find the maximum value of the elements in $\vy$ in
the loop of  line~\ref{l:comp-seq-for-K}
and 
add $1$
\color{black}
to it to find $K$.
Moreover, we need to find the left and right
degree sequences of the bipartite graph $G$ corresponding to $\vy$ in order to
use the scheme of Proposition~\ref{prop:bipartite-color-graph-compress}. Note that
with our above construction, the degree of a left node $1 \leq i \leq n$ is
$1$, while the degree of a right node $1 \leq j \leq K$ is the frequency of $j-1$
in $\vy$, which we denote by $\phi_{j-1}$. 
These degree sequences are found inside the loop of
line~\ref{l:comp-seq-for-abgamma} 
and are stored in arrays $\vsfa$ and $\vsfb$
for the left and right nodes, respectively, so that $\sfa_i = 1$ for $1 \leq i
\leq n$ and $\sfb_j = \phi_{j-1}$ for $1 \leq j \leq K$.  Additionally, we store the
adjacency list of the corresponding bipartite graph in the array $\vgamma$ of
size $n$,
where $\vgamma_i$ for $1 \leq i \leq n$ is an array of size 1 containing
$1+y_i$. Recall that $0 \leq y_i \leq K-1$, therefore the left node $i$ in the
corresponding bipartite graph is connected to the right node $1 \leq 1 + y_i
\leq K$. 
With these, using Proposition~\ref{prop:bipartite-color-graph-compress}, more
specifically Algorithm~\ref{alg:bip-comp} in
Section~\ref{sec:bipartite-compression}, we can compress $G$ and convert it into
an integer in the range $[0,\lceil  n! / \prod_{j=0}^{K-1}
\phi_j! \rceil]$. 
This is done in line~\ref{l:comp-seq-bencodegraph} 
and the
result is stored in an integer $f$. In order for the decoder to be able to
reconstruct $\vy$ given $f$, we also need to store  $K$ and the array
$\vsfb$. 
We do so using the Elias delta code in
lines~\ref{l:comp-seq-ed-K} through~\ref{l:comp-seq-ed-f}. 
Recall that the
input to the Elias delta code must be a positive integer, that is why 
we have
added $1$ to $\sfb_j$ and $f$ in lines~\ref{l:comp-seq-ed-bj} and
\ref{l:comp-seq-ed-f}.

\begin{myalg}[Compressing an array $\vy$ consisting of nonnegative integers \label{alg:compress-sequence}]
  \begin{algorithmic}[1]
    \INPUT
    \Statex $n$: size of the array
    \Statex $\vy = (y_1, \dots, y_{n})$: array of nonnegative integers
    \OUTPUT
    \Statex $\mathsf{Output}$: A prefix--free bit sequence in $\{0,1\}^* -
    \emptyset$ representing $\vy$
    \Function{EncodeSequence}{$n, \vy$}
    \State $\mathsf{Output} \gets$ empty bit sequence
    \State $K \gets 0$
    \For{$1 \leq i \leq n$} \label{l:comp-seq-for-K}
    \State $K \gets \max\{K, y_i\}$
    \EndFor
    \State $K \gets K + 1$ \Comment{$y_i$'s are in the range $[0,K-1]$}
    \State $\vsfa \gets \Array$ of integers of size $n$ \Comment{left degree sequence}
    \State $\vsfb \gets \Array $ of integers of size $K$ \Comment{right degree sequence}
    \State $\vgamma \gets \Array \text{ of } \Array$ of integers of size $n$,
    where $\vgamma_i, 1\leq i \leq n$ is of size $1$ \Comment{the adjacency list}
    \For{$1 \leq i \leq n$} \label{l:comp-seq-for-abgamma}
    \State $\sfa_i \gets 1$
    \State $\sfb_{1+y_i} \gets \sfb_{1+y_i} + 1$
    \State $\vgamma_i \gets$ array of size 1 containing $1+y_i$ \label{l:comp-seq-gamma-i-1+yi}
    \EndFor
    \State $f \gets \textproc{BEncodeGraph}(n, K, \vsfa, \vsfb,
    \vgamma)$ \label{l:comp-seq-bencodegraph}
    \Comment{Algorithm~\ref{alg:bip-comp} in Section~\ref{sec:bip-enc-pseudocode}}
    \State $\mathsf{Output} \gets \mathsf{Output} \concat \edelta(K)$ \label{l:comp-seq-ed-K}
    \For{$1 \leq j \leq K$}
    \State $\mathsf{Output} \gets \mathsf{Output} \concat \edelta(1+\sfb_j)$ \label{l:comp-seq-ed-bj}
    \EndFor \label{l:comp-seq-ed-bj-for-end}
    \State $\mathsf{Output} \gets \mathsf{Output} \concat \edelta(1+f)$ \label{l:comp-seq-ed-f}
    \EndFunction
  \end{algorithmic}
\end{myalg}

In order to analyze the complexity of this algorithm, note that the dominant
part of the calculation in terms of complexity 
is 
line~\ref{l:comp-seq-bencodegraph} where we use Algorithm~\ref{alg:bip-comp}.
Therefore, using Proposition~\ref{prop:bipartite-color-graph-compress}, since the degree of each
left node is $1$ and the number of the right nodes is $K \leq n$, the time and memory complexities of
Algorithm~\ref{alg:compress-sequence} above are $O(n \log^4 n \log \log
n)$ and $O(n \log n)$, respectively. 

Let $\ell_{\vy}$ denote the length of the binary sequence generated by this
algorithm to encode $\vy$. In order to find an upper bound on $\ell_{\vy}$,
recall that the length of the encoded version of an integer $N$ using the Elias
delta code is $\lfloor  \log_2 N \rfloor + O(\log \log N)$. Therefore, since
$K \leq n$ and $\phi_j \leq n$ for $0 \leq j \leq K-1$, we have
\begin{align*}
  \ell_{\vy} &\leq \lfloor \log_2 K \rfloor + O(\log \log K) + \sum_{j=0}^{K-1} \lfloor  \log_2(1+\phi_j) \rfloor + O(\log \log (1+\phi_j)) \\
             &\quad + \left\lfloor \log_2 \left(1+\left\lceil   \frac{n!}{\prod_{j=0}^{K-1} \phi_j!} \right\rceil\right)\right \rfloor + O\left( \log \log \left( 1+\left\lceil \frac{n!}{\prod_{j=0}^{K-1} \phi_j!}\right \rceil \right) \right) \\
             &\leq K \lfloor \log_2(1+n) \rfloor + \log_2\left( 1 + \left\lceil \frac{n!}{\prod_{j=0}^{K-1} \phi_j! }\right \rceil \right) + O(\log n) + O(K \log \log n)\\
  &= \log_2 \left( 1+ \left \lceil \frac{n!}{\prod_{j=0}^{K-1} \phi_j!}\right \rceil \right) + O(K \log n),
\end{align*}
which is precisely the desired bound in~\eqref{eq:encode-sequence-length-bound}.
This completes the proof of the first part of
Proposition~\ref{prop:encode-sequence}. We will discuss the corresponding
decoding scheme in Section~\ref{sec:decode-sequence}.

\subsubsection{Encoding $\Vns$}
\label{sec:enc-star-vertices}

In this section, we explain how to encode $\Vns$.
Recall that a vertex $v$ belongs to $\Vns$ if for some $w \sim_{\Gn} v$, we have
$\type{v}{w} = (\star_{\xi_{\Gn}(w,v)}, \star_{\xi_{\Gn}(v,w)})$. Recall that we
have access to $\type{v}{w}$ for all $w \sim_{\Gn} v$ via the array $\vec{c}$
produced 
by $\textproc{ExtractTypes}$ in line~\ref{l:encode-extract-types} 
of
Algorithm~\ref{alg:graph-encode-new}. Therefore, the set $\{c_{v,k}: 1 \leq k \leq
\dn_v\}$ completely identifies $\{\type{v}{w}: w \sim_{\Gn} v \}$. Moreover, for $1 \leq k \leq
\dn_v$, if $c_{v,k} = (i,\ip)$, then $i$ or $\ip$ correspond to $\star_x$ for some
$x \in \edgemark$ iff $\mathsf{TIsStar}(i) = 1$ or
$\mathsf{TIsStar}(\ip) = 1$. Therefore, we can easily determine whether
a vertex $v \in [n]$ belongs to $\Vns$ by examining $c_{v,k} $ for $1 \leq k \leq \dn_v$. This is done
in Algorithm~\ref{alg:encode-star-vertices} below.
This algorithm introduces a procedure which is part of
Algorithm~\ref{alg:graph-encode-new}, hence it has  access to the
variables defined in Algorithm~\ref{alg:graph-encode-new}. In this procedure, $\vs$ is
an array of bits where $s_v$ becomes 1 if $v \in \Vns$, and zero otherwise.
Finally, we encode the zero--one array $\vs$ using
Algorithm~\ref{alg:compress-sequence} which was explained in
Section~\ref{sec:compress-sequence}.

\begin{myalg}[Encoding Star Vertices
(part of
Algorithm~\ref{alg:graph-encode-new})
  \label{alg:encode-star-vertices}]
  \begin{algorithmic}[1]
    \Procedure{EncodeStarVertices}{} \Comment{line~\ref{l:enc-star-vertices} in Algorithm~\ref{alg:graph-encode-new}}
    \State $\vs = (s_v: 1 \leq v \leq n) \gets \Array$ of zero ones, each
    element  initialized with zero
    \For{$1 \leq v \leq n $}
    \For{$1 \leq k \leq \dn_v$}
    \State $(i, \ip) \gets c_{v,k}$ 
    \If{$\mathsf{TIsStar}(i) = 1$ or  $\mathsf{TIsStar}(\ip) = 1$}
    \State $s_v \gets 1$
    \EndIf
    \EndFor
    \EndFor
    \State $\mathsf{Output} \gets \mathsf{Output} \concat
    \textproc{EncodeSequence}(n,\vs)$ \label{l:encode-vs} \Comment{using
      Algorithm~\ref{alg:compress-sequence} in Section~\ref{sec:compress-sequence}}
    \EndProcedure
  \end{algorithmic}
\end{myalg}

Note that as we go through all the vertices and the neighbors of each vertex,
forming the array $\vs$ takes $O(\mn)$ time. Recall that $\mn$ denotes the
number of edges in $\Gn$. Moreover, using Proposition~\ref{prop:encode-sequence},
encoding $\vs$ in line~\ref{l:encode-vs} 
takes $O(n \log^4 n \log \log n)$ time
and $O(n \log n)$ memory. Therefore, we have
\begin{lem}
  \label{lem:encode-star-vertices-complx}
The time and memory complexities of 
  Algorithm~\ref{alg:encode-star-vertices} are $O(\mn + n \log^4 n \log \log n)$
  and $O(n \log n)$ respectively. 
\end{lem}

\editfinish

\editstart

\subsubsection{Encoding Star Edges}
\label{sec:encode-star-edges}
Next, we encode star edges, i.e.\ edges $(v,w)$ in $\Gn$ such that $\type{v}{w}$ is of the form
$(\star_x, \star_{\pp{x}})$ for some $x, \pp{x} \in \edgemark$. 
Similar to the
procedure in Algorithm~\ref{alg:encode-star-vertices} above, this can be done by
going through all the vertices $v \in [n]$ and examining $c_{v,k}$ for $1 \leq k
\leq \dn_v$. Recalling the construction of $\vs$ in
Algorithm~\ref{alg:encode-star-vertices} above, there is an edge adjacent to a vertex $v$
with type $(\star_x, \star_{\pp{x}})$ only if $s_v = 1$. Thereby, we only need
to go through those vertices $v$ with $s_v = 1$. In order to avoid representing
an edge twice, we represent an edge $(v,w)$ only if $w > v$. This is done
in Algorithm~\ref{alg:encode-star-edges} below.
This algorithm  introduces a procedure which is part of
Algorithm~\ref{alg:graph-encode-new}, hence it has  access to
variables defined in Algorithm~\ref{alg:graph-encode-new}. 
In
lines~\ref{l:encode-star-edges-tisstar} and \ref{l:encode-star-edges-xx'}, we
check 
whether, for some $1 \leq k \leq \dn_v$, we have $\type{v}{\gamman_{v,k}} = (\star_x, \star_{\pp{x}})$  and
$\gamman_{v,k}> v$. This is done with the aid of the array $\mathsf{TIsStar}$.
For such $v$ and $k$,
we first output a 1 followed by the binary representation of
$\gamman_{v,k}$. For $v$ with $s_v = 1$, after going over all the neighbors of
$v$, we output a 0 so that the decoder realizes that we are done with vertex
$v$. Assuming that the decoder has access to the sequence $\vs$, it can perform
the reverse procedure and decode for the edges with type $(\star_x,
\star_{x'})$. See Figure~\ref{fig:encode-star-edge-example} for an example of
the output bit sequence generated by  this algorithm. 


\begin{myalg}[Encoding Star Edges (part of Algorithm~\ref{alg:graph-encode-new})\label{alg:encode-star-edges}]
  \begin{algorithmic}[1]
    \Procedure{EncodeStarEdges}{} \Comment{line~\ref{l:enc-star-edges} in Algorithm~\ref{alg:graph-encode-new}}
    \For{$1 \leq x \leq |\edgemark|$}
    \For{$1 \leq \pp{x} \leq |\edgemark|$}
    \For{$1 \leq v \leq n$}
    \If{$s_v = 1$}
    \For{$1 \leq k \leq \dn_v$}
    \State $(i, \ip) \gets c_{v,k}$
    \If{$\mathsf{TIsStar}(i) = 1$ or $\mathsf{TIsStar}(\ip) = 1$} \label{l:encode-star-edges-tisstar}
    \If{$x_{v,k} = x$ and $x'_{v,k} = x'$ and $\gamman_{v,k} > v$} \label{l:encode-star-edges-xx'}
    \State $\mathsf{Output} \gets \mathsf{Output} \concat 1 \concat
    \gamman_{v,k}$ \label{l:encode-star-edges-1-gamma-vi} \Comment{use $1 +\lfloor  
      \log_2 n \rfloor$ bits to represent $\gamman_{v,k}$}
    \EndIf
    \EndIf
    \EndFor
    \State $\mathsf{Output} \gets \mathsf{Output} \concat 0$ \label{l:encode-star-edges-0}
    \EndIf
    \EndFor
    \EndFor
    \EndFor
    \EndProcedure
  \end{algorithmic}
\end{myalg}

In the above procedure, for each pair $(x, \pp{x}) \in \edgemark \times
\edgemark$, we go through
the star vertices in the graph, and check all their neighbors. Therefore, for
each pair $(x,\pp{x})$, we go
over each edge at most twice. For each such edge, we check if it is a star edges
and compare its pair of marks
with $(x, \pp{x})$, which takes $O(\log |\edgemark|)$ time. If the marks match,
we use $O(\log n )$ bits to write that edge to the output. Therefore, the time
complexity of this algorithm is $O(|\edgemark|^2 (n+\mn (\log n + \log
|\edgemark|)))$. On the other hand, we write at most $\mn$ many edges to the
output, and for each edge we use $O(\log n)$ bits. Hence, the memory complexity
is $O(\mn \log n)$.

\begin{lem}
  \label{lem:encode-star-edges-complexity}
  The time and memory complexities  of
  Algorithm~\ref{alg:encode-star-edges} are $O(|\edgemark|^2 (n+\mn (\log n + \log
|\edgemark|)))$ and $O(\mn \log n)$ respectively. 
\end{lem}

\editfinish

\editstart

\subsubsection{Finding Vertex Degree Profiles}
\label{sec:encode-find-Deg}
In this section we explain how to find $\mathsf{Deg} = (\mathsf{Deg}_v: v \in
[n])$ as was defined 
in Section~\ref{sec:enc-alg} and
in Algorithm~\ref{alg:graph-encode-new}.  
Recall that $\mathsf{Deg}_v$
represents $\Dn(v) = (\Dn_{t,\pp{t}}(v) : t, \pp{t} \in \mFdeltah)$. Therefore,
in order to find $\mathsf{Deg}_v$  for $v \in [n]$,  we investigate $c_{v,k} =
(i,\ip)$ for $k \in [\dn_v]$.
We first
check whether $i$ and $\ip$ correspond to elements in $\mFdeltah$. For this, we check
whether both $\mathsf{TIsStar}(i)$ and $\mathsf{TIsStar}(\ip)$ are 0. If
this holds, there are two possibilities: either $(i,\ip)$ is already in the set
of keys in $\mathsf{Deg}_v$, which is denoted by
$\mathsf{Deg}_v.\textproc{Keys}$, or it is a new pair not present in
$\mathsf{Deg}_v.\textproc{Keys}$. If $(i,\ip) \in
\mathsf{Deg}_v.\textproc{Keys}$, we should simply 
add $1$
\color{black}
to the value
corresponding to $(i,\ip)$, which is denoted by $\mathsf{Deg}_v(i,\ip)$.
Otherwise, we should insert the new key--value pair $((i,\ip), 1)$ into
$\mathsf{Deg}_v$. The above procedure is implemented in
Algorithm~\ref{alg:encode-find-deg} below. This procedure is part of
Algorithm~\ref{alg:graph-encode-new}, hence it has  access to
variables defined in Algorithm~\ref{alg:graph-encode-new}.

\begin{myalg}[Finding vertex degree profiles, i.e.\ the variable $\mathsf{Deg}$
  (part of Algorithm~\ref{alg:graph-encode-new}) \label{alg:encode-find-deg}]
  \begin{algorithmic}[1]
    \Procedure{FindDeg}{} \Comment{line~\ref{l:enc-find-deg-call} in Algorithm~\ref{alg:graph-encode-new}}
        \For{$1 \leq v \leq n$}
    \For{$1 \leq k \leq \dn_v$}
    \State $(i,\ip) \gets c_{v,k}$
    \If{$\mathsf{TIsStar}(i) = 0$ and $\mathsf{TIsStar}(\ip) =
      0$}
    \If{$(i,\ip) \in \mathsf{Deg}_v.\textproc{Keys}$}
    \State $\mathsf{Deg}_v(i,\ip) \gets \mathsf{Deg}_v(i,\ip) + 1$
    \Comment{increment the corresponding degree value}
    \Else
    \State $\mathsf{Deg}_v.\textproc{Insert}((i,\ip) , 1)$ \Comment{this is the
      first edge observed with this type}
    \EndIf
    \EndIf
    \EndFor
    \EndFor
    \EndProcedure
  \end{algorithmic}
\end{myalg}

In this algorithm, going through  $c_{v,k}$ for $v \in [n]$ and $1 \leq k \leq \dn_v$
takes $O(\mn)$ time. Moreover, from~\eqref{eq:sum-Dn-ttp-delta}, the number of
key--value pairs in $\mathsf{Deg}_v$ is at most $\delta$ for all $v \in [n]$.
Therefore, the three tasks (1) searching for the key $(i,\ip)$ in $\mathsf{Deg}_v.\textproc{Keys}$,
(2) accessing $\mathsf{Deg}_v(i,\ip)$, and (3) inserting $((i,\ip), 1)$ into
$\mathsf{Deg}_v.\textproc{Keys}$ take $O(\log \delta)$ time. Putting all the
above together, the time complexity of Algorithm~\ref{alg:encode-find-deg} is
$O(\mn + n \delta \log \delta)$. Also, the array $\mathsf{\Deg}$ needs $O(n
\delta(\log |\edgemark| + \log n))$ bits of memory to be stored.
To sum up, we have

\begin{lem}
  \label{lem:enc-find-deg-complexity}
  The time and memory complexities of Algorithm~\ref{alg:encode-find-deg} above
  are $O(\mn + n \delta \log \delta)$ and $O(n \delta(\log |\edgemark| + \log n))$ respectively. 
\end{lem}

\editfinish

\editstart

\subsubsection{Encoding Vertex Types}
\label{sec:encode-vertex-types}

In this section, we discuss how to encode the type of vertices in the graph,
i.e.\ $(\thetan_v, \Dn(v))$ for $1 \leq v \leq n$. In order to do so, we use
$\mathsf{\Deg} = (\mathsf{\Deg}_v: v \in [n])$, which was constructed above in
Section~\ref{sec:encode-find-Deg}. More precisely, for each vertex $v \in [n]$,
we construct an array of integers $\vec{\nu}$ 
\ifreplace
which is in one-to-one
correspondence to $(\thetan_v, \Dn(v))$.
\color{black}
\fi
We define the array $\vec{\nu}$ corresponding to
a vertex $v$ such that $\vec{\nu}$ has at most $1+3\delta$ elements, and each
element is a nonnegative integer whose value is at most $|\edgemark| \vee
\mathsf{TCount} \vee \delta$.
To define $\vec{\nu}$, we first set $\nu_1 = \thetan_v$. Then, we go over
each key--value pair $((i,\ip), l)$ in $\mathsf{Deg}_v$ and append the three
integers $(i,\ip, l)$ at the end of $\vec{\nu}$.
From~\eqref{eq:sum-Dn-ttp-delta}, there are at most $\delta$ such key--value
pairs in $\mathsf{Deg}_v$. Note that since
$\mathsf{Deg}_v$ is a dictionary, the key--value pairs $((i,\ip), l)$ are stored
in an increasing order with respect to the lexicographic order of $(i,\ip)$.
Therefore, as we go through the set of key--value pairs in $\mathsf{Deg}_v$, the
$(i,\ip, l)$ tuples added to $\vec{\nu}$ are lexicographically sorted.
Hence, $\vec{\nu}$ constructed this way is in one-to-one correspondence to
$(\thetan_v, \Dn(v))$. 
For instance, in the graph of Figure~\ref{fig:sample-graph-for-compression},
referring to Figure~\ref{fig:Deg-2-7-example}, the array $\vec{\nu}$
corresponding to a vertex $v$ in the set $\{2, \dots, 6\}$ is of the form $(1,3,4,1)$, while the array
$\vec{\nu}$ corresponding to a vertex $v$  in the set $\{7, \dots, 16\}$ is of
the form $(2,4,3,1,5,5,1)$. Moreover, since all vertices connected to vertex $1$
in this example are star edges, the array $\vec{\nu}$ corresponding to vertex 1
only has the vertex mark component, and is of the form $(2)$. 

In order to encode the arrays $\vec{\nu}$  corresponding
to vertices in the graph, we form an array $\vy = (y_v: 1 \leq v \leq n)$ of
integers, such that for $1 \leq v, w \leq n$, $y_v = y_w$ iff the vector
$\vec{\nu}$ corresponding  to $v$ is the same as that corresponding to $w$. This
way, $y_v = y_w$ iff $v$ and $w$ are of the same type. In
order to construct $\vy$, we use a dictionary called
$\mathsf{VertexTypesDictionary}$ which maps the arrays $\vec{\nu}$ at vertices
in the graph to integers. Hence, the key objects in this dictionary are arrays of integers, and
the corresponding values are integers. For each vertex $1 \leq v \leq n$, after
we construct the array $\vec{\nu}$ at $v$  as above, we search for $\vec{\nu}$ in the set of
keys in $\mathsf{VertexTypesDictionary}$. If such key does not exist, it means
that $\vec{\nu}$ is a new vertex type. Therefore, we 
add $1$
\color{black}
to the number of
distinct vertex types found so far, which we store in a variable $k$, and insert the key value pair $(\vec{\nu}, k)$ into
$\mathsf{VertexTypesDictionary}$. With this, we set $y_v$ to the integer
corresponding to the key $\vec{\nu}$ in $\mathsf{VertexTypesDictionary}$. See
Figure~\ref{fig:vertex-type-dictionary-vy} for the values of
$\mathsf{VertexTypesDictionary}$ and $\vy$ for the graph of
Figure~\ref{fig:sample-graph-for-compression} with parameters $h = 2$ and
$\delta = 4$. 

\begin{figure}
  \centering
  \begin{tabular}[c]{l@{\qquad\qquad}r}
    $\mathsf{VertexTypesDictionary}:\quad $
    \begin{tabular}[c]{@{}lc@{}}
      \toprule
      Key & Value \\\midrule
      $(2)$ & 1 \\
      $(1,3,4,1)$ & 2 \\
      $(2,4,3,1,5,5,1)$ & 3 \\
      \bottomrule
    \end{tabular}%
          &%
            $\vy = (1,2,2,2,2,2,3,3,3,3,3,3,3,3,3,3)$
\end{tabular}
  \caption{$\mathsf{VertexTypeDictionary}$ and $\vy$ for the graph of
    Figure~\ref{fig:sample-graph-for-compression} with parameters $h = 2$ and
    $\delta = 4$. See Figure~\ref{fig:Deg-2-7-example} for the values of
    $\mathsf{Deg}$ in this example.\label{fig:vertex-type-dictionary-vy}}
\end{figure}

The
variables $\mathsf{VertexTypesDictionary}$ together with $\vy$ have all the
information we need to store vertex types. Therefore, we first write
$\mathsf{VertexTypesDictionary}$ to the output. In order to do so, we first
write $k$, which is the number of key--value pairs in
$\mathsf{VertexTypesDictionary}$. We do this using $1 + \lfloor  \log_2 n
\rfloor$ bits. Next, we go through
all the key--value pairs $(\vec{\nu}, i)$ in $\mathsf{VertexTypesDictionary}$.
First, 
we write the size of the array $\vec{\nu}$ to the output. As we discussed above,
the size of $\vec{\nu}$ is at most $1 + 3 \delta$. Hence, we can write
$\textproc{Size}(\vec{\nu})$ to the output using $1 + \lfloor \log_2 (1+3\delta)
\rfloor$ bits. Recall that for $1 \leq j \leq \textproc{Size}(\vec{\nu})$,
$\nu_j$ is an integer bounded by $|\edgemark| \vee \mathsf{TCount} \vee \delta$.
Hence, we may write $\nu_j$ for $1 \leq j \leq \textproc{Size}(\vec{\nu})$ using
$1 + \lfloor  \log_2 (|\edgemark| \vee \mathsf{TCount} \vee \delta) \rfloor$
bits. Finally, we use $1 + \lfloor \log_2 n \rfloor$ bits to encode the integer
$i$ corresponding to $\vec{\nu}$.  Finally, we use
Algorithm~\ref{alg:compress-sequence} in Section~\ref{sec:compress-sequence} to
encode $\vy$. Note that since we have stored $\mathsf{VertexTypesDictionary}$,
as we will see in Section~\ref{sec:dec-ver-deg-profiles}, 
the decoder can map the value of $y_v$ for a vertex $v \in [n]$ to the type
$(\thetan_v, \Dn(v))$ of vertex $v$.

The above procedure is depicted in Algorithm~\ref{alg:encode-vertex-types}
below. Algorithm~\ref{alg:encode-vertex-types} implements a procedure which is
part of Algorithm~\ref{alg:graph-encode-new}, hence  it has access to variables
defined in Algorithm~\ref{alg:graph-encode-new}.

\begin{myalg}[Encoding Vertex Types
  (part of Algorithm~\ref{alg:graph-encode-new}) \label{alg:encode-vertex-types}]
  \begin{algorithmic}[1]
    \Procedure{EncodeVertexTypes}{}
    \Comment{line~\ref{l:enc-encode-vertex-types-call} in Algorithm~\ref{alg:graph-encode-new}}
    \State $\mathsf{VertexTypesDictionary} \gets \Dictionary(\Array \text{ of
      integers } \rightarrow \nats)$
    \State $k \gets 0$ \Comment{number of distinct vertex types found}
    \State $\vy = (y_v : 1 \leq v \leq n) \gets \mathsf{Array} \text{ of
      integers}$
    \State $\vec{\nu} \gets \mathsf{Array } \text{ of integers}$
    \For{$1 \leq v \leq n$}
    \State $\vec{\nu} \gets \emptyset$ \Comment{erasing $\vec{\nu}$ to get a
      fresh array}
    \State $\nu_1 \gets \thetan_v$
    \For{$((i,\ip), l) \in \mathsf{Deg}_v$}
    \State append $(i,\ip, l)$ at the end of $\vec{\nu}$
    \EndFor
    \If{$\vec{\nu} \notin \mathsf{VertexTypesDictionary}.\textproc{Keys}$}  \label{l:find-v-type-dic-search}
    \State $k \gets k + 1$\Comment{a new vertex type is discovered}
    \State $\mathsf{VertexTypesDictionary}.\textproc{Insert}(\vec{\nu}, k)$ \label{l:find-v-type-dic-insert}
    \EndIf
    \State $y_v \gets \mathsf{VertexTypesDictionary}(\vec{\nu})$ \label{l:find-v-type-dic-access}
    \EndFor
    \State $\mathsf{Output} \gets \mathsf{Output} \concat
    k$ \label{l:find-v-types-k-write} \Comment{use $1 + \lfloor  \log_2 n \rfloor$ bits to represent the number of key--value pairs}
    \For{$(\vec{\nu}, i) \in \mathsf{VertexTypesDictionary}$} \label{l:find-v-types-nu-i-for}
    \State $\mathsf{Output} \gets \mathsf{Output} \concat
    \textproc{Size}(\vec{\nu})$ \label{l:find-v-types-write-size-nu} \Comment{use $1 + \lfloor   \log_2 (1+3\delta) \rfloor$
      bits to encode $\textproc{Size}(\vec{\nu})$}
    \For{$1 \leq j \leq \textproc{Size}(\vec{\nu})$}
    \State $\mathsf{Output} \gets \mathsf{Output} \concat
    \nu_j$ \label{l:find-v-types-nu-j-write} \Comment{use
      $1+\lfloor \log_2(|\edgemark| \vee \mathsf{TCount} \vee \delta) \rfloor$ bits to encode $\nu_j$}
    \EndFor
    \State $\mathsf{Output} \gets \mathsf{Output} \concat
    i$ \label{l:find-v-types-write-i} \Comment{use $1 +
      \lfloor \log_2 n \rfloor$ bits to encode $i$}
    \EndFor \label{l:find-v-types-for-nu-VTD-end}
    \State $\mathsf{Output} \gets \mathsf{Output} \concat
    \textproc{EncodeSequence}(n,\vy)$\label{l:find-type-encode-sequence}
    \Comment{using Algorithm~\ref{alg:compress-sequence} in Section~\ref{sec:compress-sequence}}
    \EndProcedure
  \end{algorithmic}
\end{myalg}

Note that finding $\vec{\nu}$ for vertices $1 \leq v
\leq n$ takes $O(n \delta)$ time. On the other hand, the dictionary
$\mathsf{VertexTypesDictionary}$ has at most $n$ key--value pairs. Also,
comparing two key objects in this dictionary takes $O(\delta \log (|\edgemark| \vee
\mathsf{TCount} \vee \delta)) = O(\delta \log (|\edgemark| + \mn + \delta))$ time.
Note that $\mn \leq n^2$ and $\delta \leq n$, therefore $\log(|\edgemark| + \mn
+ \delta) = O(\log|\edgemark| + \log n)$.
Thereby, all the three tasks 
(1) searching in the set of keys in
line~\ref{l:find-v-type-dic-search}, (2) insertion in
line~\ref{l:find-v-type-dic-insert}, and (3) accessing in
line~\ref{l:find-v-type-dic-access} 
take overall $O(n \log n \delta (\log
|\edgemark| + \log n ))$ time. Furthermore, writing $\mathsf{VertexTypesDictionary}$
to the output takes $O(n \delta \log (|\edgemark| + \mn + \delta)) = O(n \delta
(\log n + \log |\edgemark))$ time.
Additionally, using Proposition~\ref{prop:encode-sequence}, 
encoding $\vy$ in
line~\ref{l:find-type-encode-sequence} 
has time and memory complexities $O(n
\log^4 n \log \log n)$ and $O(n \log n)$ respectively. Putting these together,
the overall time complexity of Algorithm~\ref{alg:encode-vertex-types} is $O(n
\log^4 n \log \log n + n \log n \delta (\log |\edgemark| + \log n ))$.

Motivated by the discussion above, storing $\mathsf{VertexTypesDictionary}$
requires $O(n \delta \log (|\edgemark| + \mn + \delta)) = O(n \delta (\log n +
\log |\edgemark|))$ bits. Also, the array
$\vy$ requires $O(n \log n)$ bits to be stored.  Furthermore, we need $O( \delta
\log(|\edgemark| + \mn + \delta)) = O(\delta (\log n + \log |\edgemark|))$ bits to store the array $\vec{\nu}$. Consequently,
the memory complexity of Algorithm~\ref{alg:encode-vertex-types} is $O(n \delta
(\log n + \log |\edgemark|))$ bits. To sum up, we have

\begin{lem}
  \label{lem:alg-find-types-complexity}
  The time and memory complexities of Algorithm~\ref{alg:encode-vertex-types}
  above are $O(n \log^4 n \log \log n + n \log n \delta(\log n + \log |\edgemark|))$ and $O(n \delta
(\log n + \log |\edgemark|))$ respectively. 
\end{lem}

\editfinish

\editstart

\subsubsection{Finding Partition Graphs}
\label{sec:encode-find-partition-graphs}
In this section we discuss how to find the partition graphs, i.e.\ finding the
variables $\mathsf{PartitionIndex}$, $\mathsf{PartitionDeg}$, and
$\mathsf{PartitionAdjList}$, which were introduced in
Section~\ref{sec:enc-alg}. 
Algorithm~\ref{sec:encode-find-partition-graphs} below illustrates a procedure
which performs this task. This procedure is part of
Algorithm~\ref{alg:graph-encode-new}, hence it has access to
variables defined in Algorithm~\ref{alg:graph-encode-new}.

First, we find $\mathsf{PartitionIndex}$ and $\mathsf{PartitionDeg}$ inside the
loop of line~\ref{l:find-part-graphs-part-deg-part-index-for} 
(recall
Figures~\ref{fig:partition-adj-list-example} and \ref{fig:partition-deg-example}
for examples of $\mathsf{PartitionAdjList}$ and $\mathsf{PartitionDeg}$ in the
graph of Figure~\ref{fig:sample-graph-for-compression}).
In order to do
so, we iterate over the vertices $1 \leq v \leq n$ in  an increasing order. For
a vertex $v \in [n]$, we look at the types of non--star edges connected to $v$.
Recalling the definition of $\mathsf{Deg}_v$, we can do so by investigating all
key--value pairs
$((i,\ip), k)$ in $\mathsf{Deg}_v$. 
This is done in the loop of
line~\ref{l:find-part-graphs-Deg-v-for}. 
Take any such pair $((i,\ip), k)$ and let
$t = J_n^{-1}(i)$ and $\tp = J_n^{-1}(\ip)$  be the elements in $\mFdeltah$ corresponding to $i$ and $\ip$
respectively. Note that since $(i,\ip)$ is in the set of keys in
$\mathsf{Deg}_v$, we have $\Dn_{t,\tp}(v) = k > 0$ and hence $v \in \mVn_{t,\tp}$. Now, there are two possible
cases. If $(i,\ip)$ is not in the set of keys in $\mathsf{PartitionDeg}$, it
means that $v$ is the vertex in $\mVn_{t,\tp}$ with the least index. 
In this
case, which  is handled in
line~\ref{l:find-parti-graphs-if-iip-notin-partitiondeg}, 
we have
$I^{(n)}_{t,\tp}(v) = 1$ and $(i,\ip)$ must be added to the set of keys in
$\mathsf{PartitionDeg}$. Furthermore, since $((i,\ip), k) \in \mathsf{Deg}_v$,
it means that $\Dn_{t,\tp}(v) = k$ and hence the degree of $I^{(n)}_{t,\tp}(v)$ in
$\Gn_{t,\tp}$ is $k$. Therefore, 
since $I^{(n)}_{t,\tp}(v) = 1$, in
line~\ref{l:find-part-graphs-if-iip-notin-partitionindex-update}, 
we insert the
key--value pair $((i,\ip), 1)$ into $\mathsf{PartitionIndex}_v$. Additionally,
 since $v$ is
the first vertex we have observed so far which is in $\mVn_{t,\tp}$, the
value corresponding to $(i,\ip)$ in $\mathsf{PartitionDeg}$ should be an array
of length one containing $\Dn_{t,\tp}(v) = k$. 
This is performed in line~\ref{l:find-part-graphs-if-iip-notin-partitiondeg-update}.

Now, consider the second case where $(i,\ip)$ is in the set of keys in $\mathsf{PartitionDeg}$.
This means that there were vertices $1 \leq w < v$ such that $w \in \mVn_{t,
  \tp}$. But the number of such nodes is precisely the size of the array
$\mathsf{PartitionDeg}(i,\ip)$, since each vertex $1 \leq w < v$ such that $w
\in \mVn_{t,\tp}$ adds one element to $\mathsf{PartitionDeg}(i,\ip)$. Therefore,
the value corresponding to $(i,\ip)$ in $\mathsf{PartitionIndex}_v$ must be $1 +
\textproc{Size}(\mathsf{PartitionDeg}(i,\ip))$. 
This is done in line~\ref{l:find-part-graphs-if-iip-notin-partitionindex-insert}. 
On the
other hand, since
$k = \Dn_{t,\tp}(v)$, we should append $k$ at the end of
$\mathsf{PartitionDeg}(i,\ip)$. 
This is done in line~\ref{l:find-part-graphs-if-iip-notin-append-k}.

Next, in the loop of line~\ref{l:find-part-graphs-part-adj-list-for}, we
initialize $\mathsf{PartitionAdjList}$. 
Recall that the keys in
$\mathsf{PartitionAdjList}$ are pairs of integers $(i,\ip)$ such that with $t =
J_n^{-1}(i)$ and 
$\tp = J_n^{-1}(\ip)$ being elements in $\mFdeltah$ corresponding to $i$ and $\ip$ respectively,
we have $(t, \tp) \in \mEn_{\leq}$, or equivalently $(t,\tp) \in \mEn$ and $i \leq \ip$. Note that, by
construction, the set of keys in $\mathsf{PartitionDeg}$ is in one-to-one
correspondence with $\mEn$. Therefore, a pair of integers $(i,\ip)$ is in the
set of keys of 
$\mathsf{PartitionAdjList}$ iff $(i,\ip)$ is in the set of keys of
$\mathsf{PartitionDeg}$ and $i \leq \ip$ (recall
Figures~\ref{fig:partition-adj-list-example} and \ref{fig:partition-deg-example}
for examples of $\mathsf{PartitionAdjList}$ and $\mathsf{PartitionDeg}$ in the
graph of Figure~\ref{fig:sample-graph-for-compression}). Take such $(i,\ip)$ with their
corresponding $t,\tp \in \mFdeltah$ as above.
With this, the value corresponding to
$(i,\ip)$ in $\mathsf{PartitionAdjList}$ represents the adjacency list of the
left nodes in $\Gn_{t,\tp}$ if $t \neq \tp$, or the forward  adjacency list of
$\Gn_{t,t}$ if $t = \tp$. Therefore, we initialize
$\mathsf{PartitionAdjList}(i,\ip)$ as an array of size $\Nn_{t,\tp} = \textproc{Size}(\mathsf{PartitionDeg}(i,\ip))$, where each
element of this array is an empty array, so that later during this algorithm we
can append elements to $(\mathsf{PartitionAdjList}(i,\ip))_p$ for $1 \leq p \leq
\Nn_{t,\tp}$.
This initialization is performed in
line~\ref{l:find-part-graphs-partitionadjlist-init}.


Finally, in the loop of line~\ref{l:find-part-graphs-PAL-FD-update}, we update
$\mathsf{PartitionAdjList}$. 
In order to do so, we
go through each vertex $v \in [n]$ and each of its neighbors $\gamman_{v,k}$ for
$1 \leq k \leq \dn_v$. For such $v$ and $k$, let $(i,\ip) = c_{v,k}$  represent
the type of the edge connecting $v$ to $\gamman_{v,k}$. Recall that only those
edges in $\Gn$ which are not star edges appear in partition graphs. 
In
line~\ref{l:find-part-graphs-if-i-ip-not-star}, we rule out star edges 
by
checking $\mathsf{TIsStar}(i)$ and $\mathsf{TIsStar}(\ip)$. Now,
assume that the edge between $v$ and $w = \gamman_{v,k}$ is not a star edges, 
and let $t = J_n^{-1}(i)$ and $\tp = J_n^{-1}(\ip)$ be the elements in $\mFdeltah$ corresponding to $i$ and
$\ip$ respectively. Recall that the edge between $v$ and $w$ in $\Gn$
corresponds to an edge in the partition graph $\Gn_{t,\tp}$. On the other hand,
recall that, in order to avoid redundancy, we only store  $\Gn_{t,\tp}$ if
$(t,\tp) \in \mEn_{\leq}$, or equivalently $i \leq \ip$. Therefore, we consider
two cases: $i < \ip$, or $i = \ip$. First, assume that $i < \ip$. Then, by
definition, corresponding to the edge between $v$ and $w$ in $\Gn$, an edge exists between the left node $I^{(n)}_{t,\tp}(v)$ and the
right node $I^{(n)}_{\tp,t}(w)$ in the partition graph $\Gn_{t,\tp}$. Note that
by definition, $I^{(n)}_{t, \tp}(v)$ is precisely
$\mathsf{PartitionIndex}_v(i,\ip)$, 
which we denote by $p$ in line~\ref{l:find-part-graphs-p}, 
and $I^{(n)}_{\tp,
  t}(w)$ is precisely $\mathsf{PartitionIndex}_w(\ip, i)$, 
  which we denote by
$q$ in line~\ref{l:find-part-graphs-q}. 
With these, in line~\ref{l:find-part-graphs-PAL-append-bip}, we append $q$ 
at the end of the array
$(\mathsf{PartitionAdjList}(i,\ip))_p$, which stores the adjacency list of the
left node $p$ in $\Gn_{t,\tp}$. Now, consider the case $i = \ip$. Recall that in
this case, $(\mathsf{PartitionAdjList}(i,i))_p$ represents the forward adjacency
list of the node $p = I^{(n)}_{t,t}$ in $\Gn_{t,t}$. Therefore, we append $q$ to the end of
    $(\mathsf{PartitionAdjList}(i,i))_p$    only when $q > p$. 
    This is done in
    line~\ref{l:find-partition-graphs-PAL-append-sim}.
Note that 
since in
the loop of line~\ref{l:find-part-graphs-PAL-FD-update} we go over 
the vertices
$v$ in the graph in an increasing order, and $(\gamman_{v,k}: 1 \leq k \leq \dn_v)$ is an increasing
list of neighbors of $v$, in both cases $i = \ip$ and $i < \ip$ above, the
elements in 
$(\mathsf{PartitionAdjList}(i,\ip))_p$ will be in an increasing order, as we
expect.

\begin{myalg}[Finding Partition Graphs (part of
  Algorithm~\ref{alg:graph-encode-new}) \label{alg:encode-find-partition-graphs}]
  \begin{algorithmic}[1]
    \Procedure{FindPartitionGraphs}{}
    \Comment{line~\ref{l:enc-find-partition-graphs-call} in Algorithm~\ref{alg:graph-encode-new}}
        \For{$1 \leq v \leq
          n$}  \label{l:find-part-graphs-part-deg-part-index-for} \Comment{find
      $\mathsf{PartitionIndex}$ and $\mathsf{PartitionDeg}$}
    \For{$((i, \ip), k) \in
      \mathsf{Deg}_v$} \label{l:find-part-graphs-Deg-v-for} 
    \If{$(i, \ip) \notin \mathsf{PartitionDeg}.\textproc{Keys}$} \label{l:find-parti-graphs-if-iip-notin-partitiondeg}
    \State  $\mathsf{PartitionIndex}_v.\textproc{Insert}((i, \ip), 1)$ \label{l:find-part-graphs-if-iip-notin-partitionindex-update}
    \State $\mathsf{PartitionDeg}.\textproc{Insert}((i, \ip), (k))$ \label{l:find-part-graphs-if-iip-notin-partitiondeg-update}
    \Comment{$\mathsf{PartitionDeg}(i, \ip)$ now becomes an array of length 1
      containing $k$}
    \Else \label{l:find-parti-graphs-if-iip-in-partitiondeg}
    \State $\mathsf{PartitionIndex}_v.\textproc{Insert}((i, \ip),
    \textproc{Size}(\mathsf{PartitionDeg}(i, \ip)) + 1)$ \label{l:find-part-graphs-if-iip-notin-partitionindex-insert}
    \State append $k$ at the end of $\mathsf{PartitionDeg}(i, \ip)$ \label{l:find-part-graphs-if-iip-notin-append-k}
    \EndIf
    \EndFor
    \EndFor
    \For{$(i, \ip) \in
      \mathsf{PartitionDeg}.\textproc{Keys}$} \label{l:find-part-graphs-part-adj-list-for}
    \Comment{initialize $\mathsf{PartitionDeg}$}
    \If{$i \leq \ip$}
    \State Insert key $(i, \ip)$ in $\mathsf{PartitionAdjList}$ with value being
    an array of size $\textproc{Size}(\mathsf{PartitionDeg}(i, \ip))$, such that
    each element of this array is 
    an empty array \label{l:find-part-graphs-partitionadjlist-init}
    \EndIf
    \EndFor
    \For{$1 \leq v \leq n$} \label{l:find-part-graphs-PAL-FD-update}
    \Comment{update $\mathsf{PartitionIndex}$ and $\mathsf{PartitionAdjList}$}
    \For{$1 \leq k \leq \dn_v$}
    \State $w \gets \gamman_{v,k}$
    \State $(i, \ip) \gets c_{v,k}$
    \If{$\mathsf{TIsStar}(i)=0$ and
      $\mathsf{TIsStar}(\ip)=0$} \label{l:find-part-graphs-if-i-ip-not-star}
    \State $p \gets \mathsf{PartitionIndex}_v(i,
    \ip)$ \label{l:find-part-graphs-p} \Comment{index of $v$}
    \State $q \gets \mathsf{PartitionIndex}_w(\ip,
    i)$ \label{l:find-part-graphs-q} \Comment{index of $w$}
    \If{$i < \ip$}
    \State append $q$ at the end of $(\mathsf{PartitionAdjList}(i, \ip))_p$ \label{l:find-part-graphs-PAL-append-bip}
    \EndIf
    \If{$i = \ip$ and $q > p$}
    \State append $q$ at the end of $(\mathsf{PartitionAdjList}(i,i))_p$ \label{l:find-partition-graphs-PAL-append-sim}
    \EndIf
    \EndIf
    \EndFor
    \EndFor
    \EndProcedure
  \end{algorithmic}
\end{myalg}

Now, we analyze the complexity of
Algorithm~\ref{alg:encode-find-partition-graphs} above. Note that due
to~\eqref{eq:sum-Dn-ttp-delta}, the number of key--value pairs in
$\mathsf{Deg}_v$ for $1 \leq v \leq n$ is at most $\delta$. Therefore, for each
$v \in [n]$, the loop in line~\ref{l:find-part-graphs-Deg-v-for} runs for at
most $\delta$ iterations. Furthermore, the number of key--value pairs in
$\mathsf{PartitionDeg}$ is eventually the size of $\mEn$. Observe that in the worst case every
edge $v \sim_{\Gn} w$ in $\Gn$ can contribute two types $\type{v}{w}$ and
$\type{w}{v}$ to $\mEn$, hence $|\mEn| \leq 2\mn$.\footnote{another upper bound
  for the size of $\mEn$, motivated by~\eqref{eq:set-of-types}, is
  $|\mCdeltah|$. However, we choose to use the bound $2\mn$ here since the size
  of $\mCdeltah$ grows exponentially in $h$.} Thereby, at any point in time,
there are at most $2\mn$ key--value pairs in $\mathsf{PartitionDeg}$. 
Additionally, for $(i,\ip)$ in the set of keys in $\mathsf{Deg}_v$, we have
$i,\ip = O(\mn)$. 
Hence, the
complexity of searching $(i,\ip)$ in $\mathsf{PartitionDeg}.\textproc{Keys}$ in
line~\ref{l:find-parti-graphs-if-iip-notin-partitiondeg} is $O(\log^2 \mn)$.
Similarly, the operations in both
lines~\ref{l:find-part-graphs-if-iip-notin-partitiondeg-update} and
\ref{l:find-part-graphs-if-iip-notin-append-k} take $O(\log^2 \mn)$ time. 
Similar to $\mathsf{Deg}_v$, $\mathsf{PartitionIndex}_v$ has at most $\delta$
many key--value pairs. Therefore, since $i, \ip = O(\mn)$, the insertions in both
lines~\ref{l:find-part-graphs-if-iip-notin-partitionindex-update}
and~\ref{l:find-part-graphs-if-iip-notin-partitionindex-insert} take $O( \log
\delta \log \mn)$ time. Overall, the complexity of the loop in
line~\ref{l:find-part-graphs-part-deg-part-index-for} is $O(n \delta \log \mn (\log \mn +
\log \delta))$.

Now, we analyze the loop in line~\ref{l:find-part-graphs-part-adj-list-for}.
Similar to $\mathsf{PartitionDeg}$, $\mathsf{PartitionAdjList}$ has at most
$2\mn$ key--value pairs. Therefore, since $i, \ip = O(\mn)$, inserting the key $(i,\ip)$ in
line~\ref{l:find-part-graphs-partitionadjlist-init} takes $O(\log^2 \mn)$ time.
Moreover, with $t = J_n^{-1}(i)$ and $\tp = J_n^{-1}(\ip)$ being elements in $\mFdeltah$ corresponding to $i$
and $\ip$ respectively, since $\textproc{Size}(\mathsf{PartitionDeg}(i,\ip)) =
\Nn_{t,\tp}$, 
initializing the value in $\mathsf{PartitionAdjList}$ corresponding to $(i,\ip)$
in line~\ref{l:find-part-graphs-partitionadjlist-init}, which is an array of
size $\textproc{Size}(\mathsf{PartitionDeg}(i,\ip))$, takes $O(\Nn_{t,\tp})$
time.
Overall, the time complexity of the loop in
line~\ref{l:find-part-graphs-part-adj-list-for} becomes
\begin{equation}
\label{eq:find-part-complexity-sum-Nn-ttp}
  O(\mn \log^2 \mn) + O\left( \sum_{(t,\tp) \in \mEn_{\leq}} \Nn_{t,\tp} \right).
\end{equation}
Observe that
\begin{equation}
  \label{eq:sum-Nn-ttp-ndelta}
\begin{aligned}
  \sum_{(t,\tp) \in \mEn_{\leq}} \Nn_{t,\tp} & \leq \sum_{(t,\tp) \in \mEn} \Nn_{t,\tp} \\
  & = \sum_{(t,\tp) \in \mEn} |\mVn_{t,\tp}| \\
  &= \sum_{(t,\tp) \in \mEn} \sum_{v=1}^n \one{\Dn_{t,\tp}(v) > 0} \\
  &\leq \sum_{v=1}^n \sum_{(t,\tp) \in \mEn} \Dn_{t,\tp}(v) \\
  &\leq \sum_{v=1}^n \sum_{t, \tp \in \mFdeltah} \Dn_{t,\tp}(v) \\
  &\leq n \delta,
\end{aligned}
\end{equation}
where the last inequality uses~\eqref{eq:sum-Dn-ttp-delta}. Using this back
into~\eqref{eq:find-part-complexity-sum-Nn-ttp}, the overall time complexity of
the loop in line~\ref{l:find-part-graphs-part-adj-list-for} is $O(\mn \log^2 \mn +
n \delta)$.

Now, we study the complexity of the loop in
line~\ref{l:find-part-graphs-PAL-FD-update}. 
Fix $v \in [n]$ and $k \in
[\dn_v]$. Recall from above that $\mathsf{PartitionIndex}_v$ has
at most $\delta$ key--value pairs. Moreover, each key in
$\mathsf{PartitionIndex}_v$ is a pair of integers each being $O(\mn)$, hence
comparing two such pairs takes $O(\log \mn)$ time. 
Thereby, finding $p$ in
line~\ref{l:find-part-graphs-p} takes $O(\log \delta \log \mn)$ time. The same
complexity bound
holds for finding $q$ in line~\ref{l:find-part-graphs-q}. Motivated by our
discussion above, accessing $(\mathsf{PartitionAdjList}(i,\ip))_p$ in
line~\ref{l:find-part-graphs-PAL-append-bip} and in
line~\ref{l:find-partition-graphs-PAL-append-sim}, and appending $q$ at its end take $O(\log^2 \mn)$ time.
Since the
total number of edges connected to a vertex $v$ which are not star edges is at most
$\delta$, the overall time complexity of the loop in line~\ref{l:find-part-graphs-PAL-FD-update}
is $O(n \delta \log \mn(\log \mn + \log \delta))$.

Now, we investigate the memory required to store the variables
$\mathsf{PartitionDeg}$, $\mathsf{PartitionIndex}$, and $\mathsf{PartitionAdjList}$. As we discussed above, the number of key--value pairs
in $\mathsf{PartitionIndex}_v$ for each $v \in [n]$ is at most $\delta$.
Furthermore, each key is a pair of integers bounded by $\mathsf{TCount} \leq 4\mn$, and the
corresponding value is an integer bounded by $n$.  Thereby,
the memory needed to store $\mathsf{PartitionIndex}$ is $O(n \delta(\log \mn +
\log n))$. On the
other hand, for each $(i,\ip) \in \mathsf{PartitionDeg}.\textproc{Keys}$,
$\mathsf{PartitionDeg}(i,\ip)$ is an array of size $\Nn_{t,\tp}$ of integers
bounded by $\delta$, where $t = J_n^{-1}(i)$ and $\tp = J_n^{-1}(\ip)$ are elements in $\mFdeltah$
corresponding to $i$ and $\ip$ respectively. But $(i,\ip) \in
\mathsf{PartitionDeg}.\textproc{Keys}$ iff $(t,\tp) \in \mEn$.
Also, such
$i$ and $\ip$ are bounded by $\mathsf{TCount} \leq 4\mn$. Therefore, the memory required to
store $\mathsf{PartitionDeg}$ is $O(\sum_{(t,\tp) \in \mEn}
\log \mn+ 
\Nn_{t,\tp} \log \delta)$,
which is $O(n \delta \log \delta + \mn \log \mn)$ due
to~\eqref{eq:sum-Nn-ttp-ndelta}.
Additionally, $(i,\ip) \in
\mathsf{PartitionAdjList}.\textproc{Keys}$ iff $(t,\tp) \in \mEn_{\leq}$ where
$t = J_n^{-1}(i)$ and $\tp = J_n^{-1}(\ip)$. Also, for such $(i,\ip)$, $\mathsf{PartitionAdjList}(i,\ip)$ is an
array of size $\Nn_{t,\tp}$, where for $1 \leq p \leq \Nn_{t,\tp}$,
$(\mathsf{PartitionAdjList}(i,\ip))_p$ is an array of size
$\Dn_{t,\tp}((I^{(n)}_{t,\tp})^{-1}(p))$ with each element being bounded by $n$.
Hence, the memory required to store $\mathsf{PartitionAdjList}$ is bounded by
\begin{align*}
  O\left( \sum_{(t,\tp) \in \mEn_{\leq}} \left (\log \mn + \log n\sum_{p=1}^{\Nn_{t,\tp}} \Dn_{t,\tp}((I^{(n)}_{t,\tp})^{-1}(p))\right ) \right) &\leq O(\mn \log \mn) + O\left(\log n \sum_{v=1}^n \sum_{(t,\tp) \in \mEn} \Dn_{t,\tp}(v) \right) \\
  &= O(\mn \log \mn  + n \delta \log n),
\end{align*}
where the last step uses~\eqref{eq:sum-Dn-ttp-delta}. Since $\log \mn = O(\log
n)$,  the overall memory
required to store the above variables is $O((\mn +n \delta) \log n)$.
To summarize, we have

\begin{lem}
  \label{lem:find-partition-graphs-complexity}
  The time complexity of Algorithm~\ref{alg:encode-find-partition-graphs} is
  $O(n \delta \log \mn (\log \mn + \log \delta) + \mn \log^2 \mn)$. Moreover, the memory
  complexity of this algorithm is $O((\mn +n \delta) \log n)$. 
\end{lem}


\editfinish

\editstart

\subsection{Complexity of the Encoding Algorithm}
\label{sec:enc-complexity}
  
In this section, we analyze the complexity of the encoding algorithm, i.e.\
Algorithm~\ref{alg:graph-encode-new} described in Section~\ref{sec:enc-alg}.

First, from Lemma~\ref{lem:preprocess-complexity} in
Section~\ref{sec:detail-preprocessing}, the time and memory complexities of the preprocessing in line~\ref{l:enc-prep} are
$O(\mn \log \mn (\log n + \log |\edgemark|))$ and $O(\mn(\log n +
\log(|\edgemark| + |\vermark|)))$ respectively.
Using Proposition~\ref{prop:MP}, the time and memory complexities of extracting
edge types in line~\ref{l:encode-extract-types} are
$O((\mn+n) h \delta \log n (\log n + \log(|\edgemark| + |\vermark|)))$
 and $O((\mn +\delta)( \log n +  \log (|\edgemark| +
 |\vermark|)))$
respectively. 
Using the bound $\mathsf{TCount}
\leq 4\mn$, both the memory and time complexities of
writing $\mathsf{TCount}$ in line~\ref{l:enc-tcount-elias} are $O(\log \mn)$. Using the same bound, both the time
and memory complexities of the loop in
line~\ref{l:enc-tiistar-tmark-for} are $O(\mn \log |\edgemark|)$. Using
Lemma~\ref{lem:encode-star-vertices-complx} in Section~\ref{sec:enc-star-vertices}, the time and memory complexities of
encoding star vertices in line~\ref{l:enc-star-vertices} are $O(\mn + n \log^4 n
\log \log n)$ and $O(n \log n)$ respectively. Also, from
Lemma~\ref{lem:encode-star-edges-complexity} in
Section~\ref{sec:encode-star-edges}, the time  and memory complexities 
of encoding star edges in line~\ref{l:enc-star-edges} are $O(|\edgemark|^2 (n+\mn
(\log n + \log |\edgemark|)))$ and $O(\mn \log n)$ respectively. From
Lemma~\ref{lem:enc-find-deg-complexity} in Section~\ref{sec:encode-find-Deg},
the time and memory complexities of finding vertex  degree profiles in
line~\ref{l:enc-find-deg-call} are $O(\mn + n \delta \log \delta)$ and $O(n
\delta(\log |\edgemark| + \log n))$ respectively. From Lemma~\ref{lem:alg-find-types-complexity} in
Section~\ref{sec:encode-vertex-types}, the time and memory complexities of
encoding vertex types in line~\ref{l:enc-encode-vertex-types-call} are $O(n
\log^4 n \log \log n + n \log n\delta(\log n + \log |\edgemark|))$ and $O(n
\delta (\log n + \log |\edgemark|))$ respectively. From
Lemma~\ref{lem:find-partition-graphs-complexity} in
Section~\ref{sec:encode-find-partition-graphs}, the time and memory complexities
of finding partition graphs in line~\ref{l:enc-find-partition-graphs-call} are
$O(n \delta \log \mn (\log \mn + \log \delta) + \mn \log^2 \mn)$ and $O((\mn + n \delta) \log n)$ respectively. 

Now, we
turn to finding the complexity of the loop in line~\ref{l:enc-part-graph-for}.
Take $(i,\ip)$ in $\mathsf{PartitionAdjList}.\textproc{Keys}$ and let $t = J_n^{-1}(i)$ and
$\tp = J_n^{-1}(\ip)$ be the elements in $\mFdeltah$ corresponding to $i$ and $\ip$ respectively. 
First, consider the case $i < \ip$. 
Since the number of key--value pairs in $\mathsf{PartitionDeg}$ is $O(\mn)$, and
$i$ and $\ip$ are $O(\mn)$, finding $\va$ and $\vb$ in
line~\ref{l:enc-part-neq-va-vb} takes $O(\log^2 \mn) = O(\log^2 n)$ time. 
Moreover, with $\tNn_{t, \tp} := \max
\{\Nn_{t, \tp}, \Nn_{\tp, t}\}$, using
Proposition~\ref{prop:bipartite-color-graph-compress}, since
from~\eqref{eq:sum-Dn-ttp-delta}, the degree of all
vertices in the partition graphs are bounded by $\delta$, the time and memory
complexities of compressing the bipartite partition graph $\Gn_{t,\tp}$ in
line~\ref{l:enc-part-neq-call-bencodegraph} are $O(\delta \tNn_{t,\tp} \log^4
\tNn_{t,\tp} \log \log \tNn_{t,\tp}) = O(\delta \tNn_{t,\tp} \log^4 n \log \log n)$ and $O(\delta \tNn_{t,\tp} \log
\tNn_{t,\tp}) = O(\delta \tNn_{t,\tp} \log n)$ respectively. Moreover,
Proposition~\ref{prop:bipartite-color-graph-compress} implies that the integer
$f$ returned by Algorithm~\ref{alg:bip-comp} in
line~\ref{l:enc-part-neq-call-bencodegraph} has $O(\delta \Nn_{t,\tp}  \log
(\Nn_{t,\tp} \delta)) = O(\Nn_{t,\tp} \delta \log n)$ bits. Hence, writing $f$ to the output in
line~\ref{l:enc-part-neq-write-f-output} has $O(\delta \Nn_{t, \tp} \log n)$
time and memory complexities. Thereby, the overall time and memory complexity of encoding the partition
bipartite graph $\Gn_{t,\tp}$ are $O(\delta \tNn_{t,\tp} \log^4 n \log \log n)$
and $O(\delta \tNn_{t,\tp} \log n)$ respectively. 

Now, consider the case $i = \ip$. Similar to the previous case, finding $\va$ in
line~\ref{l:enc-part-eq-va-def} takes $O(\log^2 \mn) = O(\log^2 n)$ time. Moreover, using
Proposition~\ref{prop:equal-color-graph-compress}, since
from~\eqref{eq:sum-Dn-ttp-delta}, the degree of all
vertices in the partition graphs are bounded by $\delta$, running
Algorithm~\ref{alg:self-find-f} in line~\ref{l:enc-part-eq-call-encodegraph} has
time and memory complexities $O(\delta \Nn_{t,t} \log^4 \Nn_{t,t} \log \log
\Nn_{t,t}) = O(\delta \Nn_{t,t} \log ^4 n \log \log n)$ and $O(\delta \Nn_{t,t}
\log \Nn_{t,t}) = O(\delta \Nn_{t,t} \log n)$ respectively. Moreover, from
Proposition~\ref{prop:equal-color-graph-compress}, integer $f$ returned in
line~\ref{l:enc-part-eq-call-encodegraph} has $O(\delta \Nn_{t,t} \log (\delta
\Nn_{t,t})) = O(\delta \Nn_{t,t} \log n)$ bits, and the array $\vec{\tilde{f}}$
has size $O(\Nn_{t,t} / \log^2 \Nn_{t,t})$, and each of its elements is bounded
by $\Nn_{t,t} \log \delta$. Therefore, the time and memory complexities
of writing $f$ to the output in line~\ref{l:enc-part-eq-f-write} is $O(\delta
\Nn_{t,t} \log n)$. Additionally, the time and memory complexities of writing the array
$\vec{\tilde{f}}$ in lines~\ref{l:enc-part-eq-size-tvf-write}
through~\ref{l:enc-part-eq-tilde-f-write-end-for} is $O(\frac{\Nn_{t,t}}{\log^2
  \Nn_{t,t}} \log (\Nn_{t,t} \delta)) = O(\Nn_{t,t} \log \delta)$. To sum up,
the overall time and memory complexities of compressing the partition graph
$\Gn_{t,t}$ are $O(\delta \Nn_{t,t} \log^4 n \log \log n)$ and $O(\delta
\Nn_{t,t} \log n)$ respectively.

Putting the two above cases together, with $\mEn_{<}$ denoting the set of
$(t,\tp) \in \mEn_{\leq}$ such that $t \neq \tp$, the overall time complexity of
the loop in line~\ref{l:enc-part-graph-for} is 
\begin{equation}
  \label{eq:enc-overall-part-enc-time-complexity}
  O\left( \delta \left( \sum_{(t,\tp) \in \mEn_<} \tNn_{t,\tp} + \sum_{(t,\tp) \in \mEn: t = \tp} \Nn_{t,t}\right) \log^4 n \log \log n\right).
\end{equation}
Similar to~\eqref{eq:sum-Nn-ttp-ndelta}, we can write
\begin{equation}
  \label{eq:sum-tNn-Nn-ndelta}
\begin{aligned}
  \sum_{(t,\tp) \in \mEn_<} \tNn_{t,\tp} + \sum_{(t,\tp) \in \mEn: t = \tp} \Nn_{t,t} &\leq \sum_{(t,\tp) \in \mEn_{<}} \Nn_{t,\tp} + \Nn_{\tp, t} + \sum_{(t,\tp) \in \mEn: t = \tp} \Nn_{t,t} \\
&= \sum_{(t,\tp) \in \mEn} \Nn_{t,\tp} \\
&= \sum_{(t,\tp) \in \mEn} |\mVn_{t,\tp}| \\
&= \sum_{(t, \tp) \in \mEn} \sum_{v=1}^n \one{\Dn_{t,\tp}(v) > 0} \\
&\leq \sum_{(t,\tp) \in \mEn} \sum_{v=1}^n \Dn_{t,\tp} (v) \\
&= \sum_{v=1}^n \sum_{(t,\tp) \in \mEn} \Dn_{t,\tp}(v) \\
&\leq n \delta
\end{aligned}
\end{equation}
where the last inequality uses~\eqref{eq:sum-Dn-ttp-delta}. Substituting this
into~\eqref{eq:enc-overall-part-enc-time-complexity}, the overall time
complexity of the loop in line~\ref{l:enc-part-graph-for} is  $O(n \delta^2
\log^4n \log \log n)$.  Likewise, the overall memory complexity of this loop is
$O(n \delta^2 \log n)$. 

Putting all the above together and simplifying using $\mn = O(n^2)$ and $\delta
\leq n$, we realize that the time complexity of
Algorithm~\ref{alg:graph-encode-new} is $O((\mn +n ) h \delta \log n (\log n +
\log(|\edgemark| + |\vermark|)) + n \delta^2 \log^4 n \log \log n +
|\edgemark|^2 (n+\mn (\log n + \log |\edgemark|)))$. Also, its
memory complexity is
$O((\mn + \delta)(\log n+ \log(|\edgemark| + |\vermark|)) + n \delta \log
|\edgemark| + n \delta^2 \log n)$. 


\editfinish

\editstart

\subsection{Decoding Algorithm}
\label{sec:dec-alg}


In this section, we discuss our decompression algorithm. 
\ifreplace
We
assume that we are given $\fn(\Gn)$, and we want to reconstruct $\Gn$. Here, $\fn(\Gn)$ refers to the
bits sequence generated by our compression algorithm discussed in
Section~\ref{sec:enc-alg}. 
Algorithm~\ref{alg:graph-decode} carries out the following steps.
\color{black}
\fi

\begin{enumerate}
\item First, we decode for $\mathsf{TCount}$,
  $\mathsf{TIsStar}$, and $\mathsf{TMark}$, which were encoded in
  lines~\ref{l:enc-tcount-elias}
  through~\ref{l:enc-tisstar-tmark-encode-for-end} in Algorithm~\ref{alg:graph-encode-new}. This is done in
  lines~\ref{l:dec-tcount-read} through~\ref{l:dec-tmark-i-read}
  of Algorithm~\ref{alg:graph-decode}.
  \color{black}
  Recall that
  $\edelta^{-1}$ refers to the Elias delta decoding function. 
\item Then, in line~\ref{l:dec-Vns}, we decode for star vertices. Recall from
  Algorithm~\ref{alg:encode-star-vertices} in
  Section~\ref{sec:enc-star-vertices} that the star vertices are encoded as a
  binary sequence $\vs = (s_v: v \in [n])$ where $s_v = 1$ for $v \in [n]$ means
  that vertex $v$ is a star vertex. Then, during the compression phase, 
  we encode this sequence using
  the sequence compression procedure of Algorithm~\ref{alg:compress-sequence}
  which was explained in Section~\ref{sec:compress-sequence}. Motivated by this, we introduce the
  decompression counterpart of Algorithm~\ref{alg:compress-sequence} as
  Algorithm~\ref{alg:sequence-decompress} which is explained in
  Section~\ref{sec:decode-sequence} below, and use it in line~\ref{l:dec-Vns} to decode for the binary
  sequence $\vs = (s_v: v \in [n])$ which represents star vertices.
\item Then, we decode for the star edges in line~\ref{l:dec-star-edges}, which
  calls Algorithm~\ref{alg:decode-star-edges} explained in
  Section~\ref{sec:dec-star-edges} below. These star edges are appended to the
  list of decoded edges which we denote by $\mathsf{EdgeListDec}$ 
  defined in line~\ref{l:dec-edgelistdec-def}. $\mathsf{EdgeListDec}$ is an
  array with each element being an edge represented in the form $(v, w, x, \pp{x})$,
  where $v$ and $w$ are the endpoints of the edge, and $x$ and $\pp{x}$ are  the
  marks towards $v$ and $w$ respectively. 
\item Then, in line~\ref{l:dec-vertex-deg-profiles}, we decode for $(\thetan_v,
  \Dn(v))$ for $1 \leq v \leq n$.
  Recall from Section~\ref{sec:enc-alg} that $\vDn$ is stored via the 
  data structure $\mathsf{Deg} = (\mathsf{Deg}_v: 1 \leq v \leq n)$ where
  $\mathsf{Deg}_v$ is a dictionary, such that a key--value pair $((i,\ip), k)$
  indicates that with $t = J_n^{-1}(i)$ and $\tp = J_n^{-1}(\ip)$ being the elements in $\mFdeltah$ corresponding to
  $i$ and $\ip$, respectively, we have $\Dn_v(t, \tp) = k$. The procedure for
  decoding $\mathsf{Deg}$ and $\vthetan$ is explained in
  Algorithm~\ref{alg:decode-vertex-deg-profiles} in
  Section~\ref{sec:dec-ver-deg-profiles}.
\item Next, we aim to decode for partition graphs. Before that, we find the
  following variables. We will explain in
  Algorithm~\ref{alg:dec-partition-deg-original-index} in
  Section~\ref{sec:dec-find-partition-deg-original-index} how to find these
  variables. 
  \begin{itemize}
  \item $\mathsf{PartitionDeg}$ defined in line~\ref{l:dec-partition-deg-def},
    as was explained in Section~\ref{sec:enc-alg}.
    Recall that if $(i,\ip)$ is a key in $\mathsf{PartitionDeg}$, with $t = J_n^{-1}(i)$ and
    $\tp = J_n^{-1}(\ip)$ being  the elements in $\mFdeltah$ corresponding to $i$ and $\ip$,
    respectively, we have $(t,\tp) \in \mEn$, and $\mathsf{PartitionDeg}(i,\ip)$ is an array of integers with size
    $\Nn_{t,\tp}$ such that: if $i < \ip$, for $1 \leq p \leq \Nn_{t,\tp}$,
    $(\mathsf{PartitionDeg}(i,\ip))_p$ is the degree of the left node $p$ in
    $\Gn_{t,\tp}$; if $i > \ip$, for $1 \leq p \leq \Nn_{t,\tp}$,
    $(\mathsf{PartitionDeg}(i,\ip))_p$ is the degree of the right node $p$ in
    $\Gn_{\tp, t}$; and if $i = \ip$, for $1 \leq p \leq \Nn_{t,t}$,
    $(\mathsf{PartitionDeg}(i,i))_p$ is the degree of the node $p$ in
    $\Gn_{t,t}$. Recall Figure~\ref{fig:partition-deg-example} for an example. 
  \item $\mathsf{OriginalIndex}$ defined in line~\ref{l:dec-originalindex-def},
    is a dictionary with key objects being pairs of integers $(i,\ip)$, 
    $1 \leq i, \ip \leq \mathsf{TCount}$, such that with $t = J_n^{-1}(i)$ and
    $\tp = J_n^{-1}(\ip)$ being
    the elements in $\mFdeltah$ corresponding to $i$ and $\ip$, respectively, we
    have $(t, \tp) \in \mEn$. Moreover, for such $(i,\ip)$,
    the value corresponding to the key $(i,\ip)$, which we denote by $\mathsf{OriginalIndex}(i,\ip)$, is an array of integers of size
    $\Nn_{t,\tp}$ such that for $1 \leq p \leq \Nn_{t,\tp}$,
    $(\mathsf{OriginalIndex}(i,\ip))_p$  is precisely $(\In_{t,\tp})^{-1}(p)$. In
    fact, $\mathsf{OriginalIndex}$ keeps track of the index of the vertex in $\Gn$
    corresponding to each vertex in the partition graphs, hence the name
    $\mathsf{OriginalIndex}$. See Figure~\ref{fig:original-index-example} for an
    example. 

    \begin{figure}
      \centering

      $\mathsf{OriginalIndex}: \qquad \qquad $
      \begin{tabular}{@{}lc@{}}
        \toprule
        Key & Value \\
        \midrule 
        $(3,4)$ & $(2,3,4,5,6)$ \\
        $(4,3)$ & $(7,8,9,10,11,12,13,14,15,16)$ \\
        $(5,5)$ & $(7,8,9,10,11,12,13,14,15,16)$ \\
        \bottomrule
      \end{tabular}
      \caption{The content of $\mathsf{OriginalIndex}$ for the graph of
        Figure~\ref{fig:sample-graph-for-compression} with parameters $h = 2$
        and $\delta = 4$. See Figure~\ref{fig:mEn-mVn-In} for the values of
        $(\In_{t,\tp}: (t, \tp) \in \mEn)$, and
        Figure~\ref{fig:partition-graphs-example} for partition graphs in this
        example. Also, see Figure~\ref{fig:message-passing-example-summary} for
        the function $J_n$. \label{fig:original-index-example}}
    \end{figure}

  \end{itemize}
\item Now, we are ready to decode for the partition graphs. Recalling the
  procedure of encoding partition graphs (lines~\ref{l:enc-k-output} through
  \ref{l:enc-part-graph-for-end} in Algorithm~\ref{alg:graph-encode-new}), we
  first read the number of partition graphs in
  line~\ref{l:dec-nu-part-graphs-read} and store it in an integer variable $K$. Note that since we
  encode 1 plus the number of partition graphs in
  Algorithm~\ref{alg:graph-encode-new}, we subtract 1 here to get the correct
  number of partition graphs.  Then, for each $1
  \leq k \leq K$, we read $i$ and $\ip$ in lines~\ref{l:dec-part-read-i} and
  \ref{l:dec-part-read-ip} respectively. From
  Algorithm~\ref{alg:graph-encode-new} we know that $i$ and $\ip$ are stored
  using $1 + \lfloor \log_2 \mathsf{TCount} \rfloor$ bits each. We know that if
  $t = J_n^{-1}(i)$ and $\tp = J_n^{-1}(\ip)$  are the members in $\mFdeltah$ corresponding to $i$ and $\ip$,
  respectively, then what follows in $\mathsf{Input}$ is the compressed form of
  the partition graph $\Gn_{t,\tp}$. From Algorithm~\ref{alg:graph-encode-new}
  we know that either $i < \ip$ or $i = \ip$.
  \begin{itemize}
  \item If $i < \ip$, we know from line~\ref{l:enc-part-neq-write-f-output} in
    Algorithm~\ref{alg:graph-encode-new} that
    $1$ plus the compressed form of $\Gn_{t,\tp}$, which is an integer generated by
    Algorithm~\ref{alg:bip-comp}, is stored using the Elias delta code. Therefore,
    we decode this integer in line~\ref{l:dec-y-read-neq} using the Elias delta decoder.
    Then, in line~\ref{l:dec-part-b-graph-decode}, we use
    Algorithm~\ref{alg:b-dec} in Section~\ref{sec:bipartite-decode} to decode for the adjacency list of $\Gn_{t,\tp}$,
    which is stored in $\mathsf{AdjList}$.
    In order to do so, we need the degree sequence of left and right nodes in
    $\Gn_{t,\tp}$. But by definition, these are precisely $\mathsf{PartitionDeg}(i,\ip)$ and
    $\mathsf{PartitionDeg}(\ip, i)$, respectively.
  \item If $i = \ip$, from Algorithm~\ref{alg:graph-encode-new} we know that
    $\Gn_{t,t}$ is encoded using an integer $f$ and a array of integers
    $\vec{\tilde{f}}$. Therefore, in addition to the integer $f$, which is  read in
    line~\ref{l:dec-y-read-eq}, we read the array $\vec{\tilde{f}}$ in
    lines~\ref{l:dec-sim-L-read} through \ref{l:dec-sim-for-L-end}. Then, we use
    Algorithm~\ref{alg:s-dec-main} in Section~\ref{sec:self-decompression} to decode for the adjacency list of
    $\Gn_{t,t}$, which is stored in $\mathsf{AdjList}$ in line~\ref{l:dec-sim-graph-decode}. In order to do so, we need the degree sequence of $\Gn_{t,t}$,
    which is precisely $\mathsf{PartitionDeg}(i,i)$. 
  \end{itemize}
  Now that we have decoded the adjacency list of  $\Gn_{t,\tp}$ and stored it in
  $\mathsf{AdjList}$, we use it to extract the edges in $\Gn$ corresponding to the
  partition graph $\Gn_{t,\tp}$. Recall that by definition, if $t \neq \tp$, an
  edge between a left node $v$ and a right node $w$ in $\Gn_{t, \tp}$ mean
  that there exists an edge between nodes $(\In_{t,\tp})^{-1}(v)$ and
  $(\In_{\tp, t})^{-1}(w)$ in $\Gn$ with mark $x := t[m]$ towards $(\In_{t,\tp})^{-1}(v)$ and
  mark $\pp{x} := \tp[m]$ towards $(\In_{\tp, t})^{-1}(w)$. Similarly, if $t = \tp$, an edge
  between vertices $v$ and $w$ in $\Gn_{t,t}$ means that there exists an edge
  between vertices $(\In_{t,t})^{-1}(v)$ and $(\In_{t,t})^{-1}(w)$ in $\Gn$ with mark
  $x = \pp{x} := t[m]$ towards both endpoints. But by definition, we have $x =
  \mathsf{TMark}(i)$ and $\pp{x}= \mathsf{TMark}(\ip)$, as are implemented in
  lines~\ref{l:dec-x-mark-i} and \ref{l:dec-xp-mark-ip}, respectively.
  Then, in
  the loop of line~\ref{l:dec-for-v-partition-deg}, we go over each edge $(v,w)$
  in the
  adjacency list of $\Gn_{t,\tp}$. 
Note that if $i \neq \ip$, this means that an edge exists in $\Gn_{t,\tp}$
between the left node $v$ and the right node $w$. Moreover, if $ i = \ip$, this
indicates an edge between vertices $v$ and $w$ in $\Gn_{t,t}$. However,
motivated by the discussion above, in either case, we should add an edge in the
decoded graph between the nodes $(\In_{t,\tp})^{-1}(v)$ and $(\In_{\tp,
  t})^{-1}(w)$. 
But by the definition of
  $\mathsf{OriginalIndex}$ as above, we have $(\In_{t,\tp})^{-1}(v) =
  (\mathsf{OriginalIndex}(i,\ip))_v =:\pp{v}$ and $(\In_{\tp, t})^{-1}(w) =
  (\mathsf{OriginalIndex}(\ip, i))_w =: \pp{w}$, as are implemented in
lines~\ref{l:dec-vp-def} and \ref{l:dec-wp-def}, respectively. Finally,
motivated by the discussion above, we add an edge between vertices $\pp{v}$ and
$\pp{w}$ with mark $x$ towards $\pp{v}$ and $\pp{x}$ towards $\pp{w}$. This edge
is added to the list of decode edges, which we denote by $\mathsf{EdgeListDec}$,
in line~\ref{l:dec-append-edge-list}. 
\end{enumerate}

{
\begin{myalg}[Decoding a simple marked graph \label{alg:graph-decode}]
  \begin{algorithmic}[1]
    \INPUT
    $\mathsf{Input} = \fn(\Gn)$ for a simple marked graph $\Gn$ on the vertex set $[n]$.
    Here, $\fn(\Gn)$ refers to the
bit sequence generated by our compression procedure discussed in
Algorithm~\ref{alg:graph-encode-new} of 
Section~\ref{sec:enc-alg}.
    \OUTPUT
    $\hGn$ a reconstruction of $\Gn$ represented in the edge list
    form, i.e.\
    \begin{itemize}
    \item $\vec{\theta}^{(n)}$: sequence of vertex marks in $\hGn$.
    \item $\mathsf{EdgeListDec}$: list of edges in $\hGn$. 
    \end{itemize}
    \Function{MarkedGraphDecode}{$\Gn$}
    \State $\mathsf{TCount} \gets \edelta^{-1}(\mathsf{Input})
    -1$ \label{l:dec-tcount-read} \Comment{we encode $1+\mathsf{TCount}$
      in Algorithm~\ref{alg:graph-encode-new}}
    \State $\mathsf{TIsStar} \gets \Array \text{ of bits of size }
    \mathsf{TCount}$ \label{l:dec-tisstar-deg}
    \State $\mathsf{TMark} \gets \Array \text{ of integers of size }
    \mathsf{TCount}$ \label{l:dec-tmark-def}
    \For{$1 \leq i \leq \mathsf{TCount}$} \label{l:dec-for-i-tcount}
    \State $\mathsf{TIsStar}(i) \gets $ read 1 bit from $\mathsf{Input}$ \label{l:dec-tisstar-i-read}
    \State $\mathsf{TMark}(i) \gets $ read $1 + \lfloor   \log_2 |\edgemark|
    \rfloor$ bits from $\mathsf{Input}$ \label{l:dec-tmark-i-read}
    \EndFor
    \State $\vs \gets \textproc{DecodeSequence}(n,
    \mathsf{Input})$ \label{l:dec-Vns} 
    \Statex \Comment{decode for star vertices using Algorithm~\ref{alg:sequence-decompress} in Section~\ref{sec:decode-sequence}}
    \State $\mathsf{EdgeListDec} \gets \Array \text{ of } \nats \times \nats
    \times \nats \times \nats$ \label{l:dec-edgelistdec-def}
\Statex     \Comment{$\mathsf{EdgeListDec}$ is the decoded edge list, each index of the form $(v,w, x, \pp{x})$, where
    $x = \xi_{\hGn}(w,v)$ and $\pp{x} = \xi_{\hGn}(v,w)$}
    \State $\textproc{DecodeStarEdges}$ \label{l:dec-star-edges}
    \Comment{Algorithm~\ref{alg:decode-star-edges} in Section~\ref{sec:dec-star-edges}}
\State $\mathsf{Deg} = (\mathsf{Deg}_v: 1 \leq v \leq n) \gets \Array \text{
      of }
    \Dictionary(\nats \times \nats \rightarrow \nats)$ \label{l:decode-deg-def}
    \State $\vec{\theta}^{(n)} \gets \Array \text{ of integers}$
    \State
    $\textproc{DecodeVertexDegreeProfiles}$ \label{l:dec-vertex-deg-profiles}
    \Comment{Algorithm~\ref{alg:decode-vertex-deg-profiles} in Section~\ref{sec:dec-ver-deg-profiles}}
    \State $\mathsf{PartitionDeg} \gets \Dictionary(\nats \times \nats
    \rightarrow \Array \text{ of integers})$ \label{l:dec-partition-deg-def}
    \State $\mathsf{OriginalIndex} \gets \Dictionary(\nats \times \nats
    \rightarrow \Array \text{ of integers})$ \label{l:dec-originalindex-def}
    \State
    $\textproc{DecodePartitionDegOriginalIndex}$ \label{l:main-dec-part-indx-orig-indx}
    \Comment{Algorithm~\ref{alg:dec-partition-deg-original-index} in Section~\ref{sec:dec-find-partition-deg-original-index}}
    \State $K\gets \edelta^{-1}(\mathsf{Input})-1$ \label{l:dec-nu-part-graphs-read}
    \Comment{number of partition graphs}
    \For{$1 \leq k \leq K$} \label{l:dec-part-graph-for}
    \State $i \gets$ read $1+\lfloor   \log_2 \mathsf{TCount} \rfloor$ bits from
    $\mathsf{Input}$ \label{l:dec-part-read-i}
    \State $i' \gets$ read $1+\lfloor   \log_2 \mathsf{TCount} \rfloor$ bits from
    $\mathsf{Input}$ \label{l:dec-part-read-ip}
    \If{$i < i'$} \label{l:l:dec-if-i-less-ip}
    \State $f \gets \edelta^{-1}(\mathsf{Input}) - 1$ \label{l:dec-y-read-neq}
    \State $\mathsf{AdjList} \gets
    \textproc{BDecodeGraph}(f,\mathsf{PartitionDeg}(i,i'),
    \mathsf{PartitionDeg}(i',i))$ \label{l:dec-part-b-graph-decode}
    \Statex \Comment{Algorithm~\ref{alg:b-dec} in Section~\ref{sec:bipartite-decode}}
    \Else
    \State $f \gets \edelta^{-1}(\mathsf{Input}) - 1$ \label{l:dec-y-read-eq}
    \State $L \gets \edelta^{-1}(\mathsf{Input}) - 1$ \label{l:dec-sim-L-read}
    \State $\vec{\tilde{f}} \gets \Array \text{ of integers with size } L$ \label{l:dec-sim-vy-def}
    \For{$1 \leq l \leq L$} \label{l:dec-sim-for-L}
    \State $\tilde{f}_l \gets \edelta^{-1}(\mathsf{Input})-1$ \label{l:dec-sim-yl-read}
    \EndFor \label{l:dec-sim-for-L-end}
    \State $\mathsf{AdjList} \gets
    \textproc{GraphDecode}(f,\vec{\tilde{f}},\mathsf{PartitionDeg}(i,i))$ \label{l:dec-sim-graph-decode}
    \Comment{Algorithm~\ref{alg:s-dec-main} in Section~\ref{sec:self-decompression}}
    \EndIf
    \State $x \gets \mathsf{TMark}(i)$ \label{l:dec-x-mark-i}
    \State $x' \gets \mathsf{TMark}(i')$ \label{l:dec-xp-mark-ip}
    \State $A \gets \mathsf{OriginalIndex}(i,\ip)$ \label{l:dec-OI-i-ip-A}
    \State $B \gets \mathsf{OriginalIndex}(\ip, i)$ \label{l:dec-IO-ip-i-B}
    \For{$1 \leq v \leq
      \textproc{Size}(\mathsf{AdjList})$} \label{l:dec-for-v-partition-deg}
    \State $v' \gets A_v$ \label{l:dec-vp-def}
    \For{$1 \leq j \leq \textproc{Size}(\mathsf{AdjList}_v)$} \label{l:dec-for-j-partition-deg-iip}
    \State $w \gets \mathsf{AdjList}_{v,j}$ \label{l:dec-w-def}
    \State $w' \gets B_w$ \label{l:dec-wp-def}
    \State append $(v', w', x, x')$ at the end of $\mathsf{EdgeListDec}$ \label{l:dec-append-edge-list}
    \EndFor
    \EndFor
    \EndFor
    \State $\textbf{return } (\vthetan, \mathsf{EdgeListDec})$
    \EndFunction
    \end{algorithmic}
  \end{myalg}

\editfinish

\editstart

\subsubsection{Decompressing a Sequence}
\label{sec:decode-sequence}

  
In this section, we discuss the reverse of
Algorithm~\ref{alg:compress-sequence}, 
which was discussed in Section~\ref{sec:compress-sequence}, to decompress a sequence of nonnegative
integers. 
\ifreplace
We implement this part of our decoding procedure in
Algorithm~\ref{alg:sequence-decompress} below.
\color{black}
\fi
In
line~\ref{l:seq-dec-K-read}, we use the Elias delta decoder to read $K$, an upper
bound on the number of distinct nonnegative integers that show up in the
compressed sequence. More precisely, if $\vy = (y_1, \dots, y_n)$ denotes the
sequence to be decoded, we have $0 \leq y_i \leq K - 1$ for all $1 \leq i \leq
n$. Recall from Section~\ref{sec:compress-sequence} that in order to  encode
$\vy$, we 
first represent  it with a bipartite graph $G$ and then compress $G$ using
the procedure in Section~\ref{sec:bipartite-compression}. This bipartite graph
has $n$ left nodes and $K$ right nodes, where a left node $i$ is connected to
the right node $1+y_i$. We denote the degree sequences of the left and the right
nodes of $G$ as $\sfa$ and $\sfb$ in lines~\ref{l:seq-dec-a-def} and
\ref{l:seq-dec-b-def} respectively. Note that by definition, $\sfa_i = 1$ for $1
\leq i \leq n$, as is set in line~\ref{l:seq-dec-a-def}. 
Moreover, in the loop of
line~\ref{l:seq-dec-b-read-for}, for each $1 \leq j \leq K$, we read $\sfb_j$
using the Elias delta decoder and then subtract 1 from it. This is because in
line~\ref{l:comp-seq-ed-bj} of Algorithm~\ref{alg:compress-sequence}, we encode $1 + \sfb_j$.
Furthermore, in line~\ref{l:seq-dec-f-read}, we use the Elias delta decoder to read $f$ which represents the
bipartite graph $G$. Next, we subtract 1 from $f$. The reason is that again in
line~\ref{l:comp-seq-ed-f} in Algorithm~\ref{alg:compress-sequence}, we encode
$1+f$. Then, in line~\ref{l:seq-dec-call-b-dec-graph}, we
use Algorithm~\ref{alg:b-dec} in Section~\ref{sec:bdec-main} to find the
adjacency list of the bipartite graph $G$, which is denoted by 
$\vgamma_{[1:n]}$. Recall from Section~\ref{sec:compress-sequence} that
$\vgamma_i$ for $1 \leq i \leq n$ is an array of size 1 containing $y_i + 1$.
Therefore, in line~\ref{l:seq-dec-yi-gamma-i-1}, we set $y_i$ to $\gamma_{i,1} -
1$. This completes our decompression procedure.

\begin{myalg}[Decompressing an array consisting of nonnegative integers \label{alg:sequence-decompress}]
  \begin{algorithmic}[1]
    \INPUT
    \Statex $n$: size of the array
    \Statex $\mathsf{Input}$: sequence of bits which contains the compressed
    form of an array generated by Algorithm~\ref{alg:compress-sequence} in
    Section~\ref{sec:compress-sequence}
    \OUTPUT
    \Statex $\vy = (y_1, \dots, y_n)$: the decoded array consisting of nonnegative integers
    \Function{DecodeSequence}{$n, \mathsf{Input}$}
    \State $K \gets \edelta^{-1}(\mathsf{Input})$ \label{l:seq-dec-K-read} \Comment{Symbols in $\vy$ are
      in the range $[0,K-1]$}
    \State $\sfa \gets \Array \text{ of nonnegative integers of size } n \text{ where all elements are 1}$ \label{l:seq-dec-a-def}
    \State $\sfb \gets \Array \text{ of nonnegative integers of size } K$ \label{l:seq-dec-b-def}
    \For{$1 \leq j \leq K$} \label{l:seq-dec-b-read-for}
    \State $\sfb_j \gets \edelta^{-1}(\mathsf{Input})$ \label{l:seq-dec-bj-read}
    \State $\sfb_j \gets \sfb_j - 1$ \label{l:seq-dec-bj--1}\Comment{We encode $1+\sfb_j$ in
      line~\ref{l:comp-seq-ed-bj} of Algorithm~\ref{alg:compress-sequence}}
    \EndFor
    \State $f \gets \edelta^{-1}(\mathsf{Input})$ \label{l:seq-dec-f-read}
    \State $f \gets f - 1$ \label{l:seq-dec-f--1}
    \State $\vgamma_{[1:n]} \gets \textproc{BDecodeGraph}(f, \vsfa, \vsfb)$ \label{l:seq-dec-call-b-dec-graph}
    \Comment{Algorithm~\ref{alg:b-dec} in Section~\ref{sec:bdec-main}}
    \For{$1 \leq i \leq n$} \label{l:seq-dec-yi-for}
    \State $y_i \gets \gamma_{i,1} - 1$ \label{l:seq-dec-yi-gamma-i-1} \Comment{as in
      line~\ref{l:comp-seq-gamma-i-1+yi} of
      Algorithm~\ref{alg:compress-sequence}, $\vgamma_i$ is an array of size 1
      containing $1+y_i$}
    \EndFor
    \State \textbf{return} $\vy$
    \EndFunction
  \end{algorithmic}
\end{myalg}


Similar to the compression algorithm in Section~\ref{sec:compress-sequence}, the
dominant part of this algorithm in terms of complexity is
line~\ref{l:seq-dec-call-b-dec-graph} where we use Algorithm~\ref{alg:b-dec} to
decode for the bipartite graph which represents the sequence $\vy$. Using
Proposition~\ref{prop:bipartite-color-graph-compress}, since the degree of each
left node is $1$ and the number of the right nodes is $K \leq n$, the time and
memory complexities of this part are $O(n \log^4 n \log \log n)$ and $O(n \log
n)$ respectively, as was claimed in Proposition~\ref{prop:encode-sequence}.
This together with our discussion in Section~\ref{sec:compress-sequence}
completes the proof of Proposition~\ref{prop:encode-sequence}. 



\editfinish

\editstart

\subsubsection{Decoding Star Edges}
\label{sec:dec-star-edges}

In this section, we perform the operations of
Section~\ref{sec:encode-star-edges} (and specifically
Algorithm~\ref{alg:encode-star-edges} therein) in reverse in order to decode for star
edges. This is illustrated in Algorithm~\ref{alg:decode-star-edges} below. Recall from line~\ref{l:dec-Vns} of Algorithm~\ref{alg:graph-decode} that
the array $\vs$ encodes the star vertices. Therefore, for each pair of edge
marks $x, \pp{x} \in \edgemark$, we go over the vertices $1 \leq v \leq n$ and, if
$s_v = 1$, we decode the star edges connected to $v$. For $x, \pp{x}$ and $v$
fixed, if $s_v = 1$, recalling the procedure in Algorithm~\ref{alg:encode-star-edges}, we encode star edges connected to $v$ by writing a flag
bit with value 1 for every star edge followed by the binary representation of
the index of the other endpoint, and then a flag bit with value 0 to
indicate that the list of the star edges connected to $v$ is over. Therefore, to decode star edges connected to $v$,
we read one flag bit $b$ from the input and if its value is 1, we find $w$, the index of
the other endpoint of the star edge, by reading $1 + \lfloor
\log_2 n \rfloor$ bits from the input. Then, in line~\ref{l:decode-star-edges-append-edge}, we append $(v, w, x, \pp{x})$ at the
end of the list of the decoded edges. We continue this process until the flag bit
$b$ becomes zero. 
The Algorithm~\ref{alg:decode-star-edges} below introduces a procedure which is part of
Algorithm~\ref{alg:graph-decode}, hence it has  access to the
variables defined in Algorithm~\ref{alg:graph-decode}.

\begin{myalg}[Decoding Star Edges\label{alg:decode-star-edges}]
  \begin{algorithmic}[1]
    \Procedure{DecodeStarEdges}{} \Comment{line~\ref{l:dec-star-edges} in Algorithm~\ref{alg:graph-decode}}
    \For{$1 \leq x \leq |\edgemark|$}
    \For{$1 \leq \pp{x} \leq |\edgemark|$}
    \For{$1 \leq v \leq n$}
    \If{$s_v = 1$}
    \State $b \gets $ read 1 bit from $\mathsf{Input}$
    \While{$b \neq 0$}
    \State $w \gets $ read $1 + \lfloor   \log_2 n \rfloor$ bits from $\mathsf{Input}$
    \State append $(v,w,x, x')$ at the end of $\mathsf{EdgeListDec}$ \label{l:decode-star-edges-append-edge}
    \State $b \gets $ read 1 bit from $\mathsf{Input}$
    \EndWhile
    \EndIf
    \EndFor
    \EndFor
    \EndFor
    \EndProcedure
  \end{algorithmic}
\end{myalg}

Note that we decode for  at most $\mn$  many star edges. For each such edge, we
read $O(\log n)$ bits from the input and write $O( \log |\edgemark| + \log n)$
bits to $\mathsf{EdgeListDec}$. Also, since for each pair $x, \pp{x} \in
\edgemark$, we go through every vertex $v$ in the graph and check $s_v$, the
overall time complexity of Algorithm~\ref{alg:decode-star-edges} is $O(\mn(\log
|\edgemark| + \log n) + |\edgemark|^2 n)$. On the other hand, storing the
decoded edges in $\mathsf{EdgeListDec}$ requires $O(\mn(\log n + \log
|\edgemark|))$ bits of memory. 
To summarize, we have

\begin{lem}
  \label{lem:dec-star-edges-complexity}
  The time and memory complexities of Algorithm~\ref{alg:decode-star-edges} above are $O(\mn(\log
|\edgemark| + \log n) + |\edgemark|^2 n)$ and $O(\mn(\log n + \log
|\edgemark|))$, respectively.
\end{lem}

\editfinish 

\editstart

\subsubsection{Decoding Vertex Types}
\label{sec:dec-ver-deg-profiles}

In this section, we discuss how to decode for the degree profiles of the vertices in
the graph, i.e.\ the pair $(\thetan_v, \Dn(v))$ for vertices $1 \leq v \leq n$. Recall from
Section~\ref{sec:encode-vertex-types} that $\Dn(v)$ is encoded via
$\mathsf{Deg}_v$, which is a dictionary where each  key is a pair of integers
representing $t, \pp{t} \in \mFdeltah$, the corresponding value being
$\Dn_{t,\pp{t}}(v)$. Also, recall from Section~\ref{sec:encode-vertex-types}
that in Algorithm~\ref{alg:encode-vertex-types}, associated to each vertex $v$,
we construct an array of integers $\vnu$ which uniquely determines $(\thetan_v,
\Dn(v))$. Moreover, in  Algorithm~\ref{alg:encode-vertex-types}, we form an
array of integers $(y_v :v \in [n])$ where $y_v$ uniquely identifies the array
$\vnu$ associated to vertex $v$ as above. In order to do this, as we discussed
in Section~\ref{sec:encode-vertex-types}, we form a dictionary
named $\mathsf{VertexTypesDictionary}$ which maps an array $\vnu$ as above to its
corresponding integer.
Then, we write every key--value
pair $(\vnu, i)$ to the output.

Therefore, in order to decode for vertex
types, we perform these actions in Algorithm~\ref{alg:encode-vertex-types} in 
reverse. This is done in
Algorithm~\ref{alg:decode-vertex-deg-profiles} below. 
Algorithm~\ref{alg:decode-vertex-deg-profiles} introduces a procedure which is part of
Algorithm~\ref{alg:graph-decode}, hence it has  access to the
variables defined in Algorithm~\ref{alg:graph-decode}.
More precisely, we first decode the key--value pairs $(\vnu, i)$ in
$\mathsf{VertexTypesDictionary}$. We store the result in an array called
$\mathsf{VertexTypesList}$ defined in
line~\ref{l:dec-deg-vertex-types-list-def} where each index is an array of
integers. In this array, we have $\mathsf{VertexTypesList}(i) = \vnu$ iff the
key--value pair $(\vnu, i)$ is in $\mathsf{VertexTypesDictionary}$.
We read $\mathsf{VertexTypesList}$  between lines~\ref{l:dec-deg-read-K}
and~\ref{l:dec-deg-for-k-end}, motivated by  the actions in 
lines~\ref{l:find-v-types-k-write} through~\ref{l:find-v-types-for-nu-VTD-end}
in Algorithm~\ref{alg:encode-vertex-types}. 
Then, in line \ref{l:dec-deg-decode-sequenece},
we
decode for the sequence $\vy$.
Next, we use this sequence together with $\mathsf{VertexTypesList}$ to decode
for $\mathsf{Deg} = (\mathsf{Deg}_v: 1 \leq v \leq n)$. Recall that
$\mathsf{Deg}_v$ for a vertex $v \in [n]$ encodes $\Dn(v)$. First, we initialize
$\mathsf{Deg}$ in line~\ref{l:dec-deg-def-deg}. Then, in the loop of
line~\ref{l:dec-deg-for-v-n}, for each vertex $v \in [n]$, we use
$\mathsf{VertexTypesList}$ to find the array $\vnu$ corresponding to $v$.
Namely, in line~\ref{l:dec-deg-vec-nnu-VTL-yv}, we set $\vnu$ to be
$\mathsf{VertexTypesList}(y_v)$. Recall from
Algorithm~\ref{alg:encode-vertex-types} that the first index in $\vnu$ is
$\theta_v$, and for each key--value pair $((i,\ip), j)$ in $\mathsf{Deg}_v$, we
include the triple $(i,\ip, j)$ in $\vnu$. Therefore, we extract $\theta_v$ from
the first index of $\vnu$ in line~\ref{l:dec-deg-theta-v}, and then in the loop
of line~\ref{l:dec-deg-for-k-vnu} we extract triples from $\vnu$ and insert
them into $\mathsf{Deg}_v$.


\begin{myalg}[Decoding Vertex Degree Profiles \label{alg:decode-vertex-deg-profiles}]
  \begin{algorithmic}[1]
    \Procedure{DecodeVertexDegreeProfiles}{}
    \Comment{line~\ref{l:dec-vertex-deg-profiles} in Algorithm~\ref{alg:graph-decode}}
    \State $\mathsf{VertexTypesList} \gets \Array \text{ of } \Array $ of
    integers \label{l:dec-deg-vertex-types-list-def}
    \State $\vec{\nu} \gets \mathsf{Array } \text{ of integers}$
    \State $K \gets$ read $1 + \lfloor   \log_2 n \rfloor$ bits from $\mathsf{Input}$
    \label{l:dec-deg-read-K} \Comment{number of distinct vertex types}
    \State resize $\mathsf{VertexTypesList}$ to have $K$ elements, each being an
    empty array \label{l:dec-deg-resize-vertex-types-list}
    \For{$1 \leq k \leq K$} \label{l:dec-deg-for-k}
    \State $\vec{\nu} \gets \emptyset$ \label{l:dec-deg-for-nu-empty}
    \State $l \gets$ read $1 + \lfloor \log_2 (1+3\delta) \rfloor$  bits from
    $\mathsf{Input}$ \label{l:dec-deg-l-read} \Comment{number of elements in $\vec{\nu}$}
    \For{$ 1 \leq j \leq l$} \label{l:dec-deg-for-j-l}
    \State read $1 + \lfloor \log_2(|\edgemark| \vee \mathsf{TCount} \vee \delta) \rfloor$ bits from $\mathsf{Input}$ and append
    to $\vec{\nu}$ \label{l:dec-deg-read-nu-j}
    \EndFor
    \State $i \gets$ read $1 + \lfloor  \log_2 n \rfloor$ bits from
    $\mathsf{Input}$ \label{l:dec-deg-read-i}
    \State $\mathsf{VertexTypesList}(i) \gets \vec{\nu}$ \label{l:dec-deg-vertex-types-list-i-read}
    \EndFor \label{l:dec-deg-for-k-end}
    \State $\vy \gets
    \textproc{DecodeSequence}(n, \mathsf{Input})$ \label{l:dec-deg-decode-sequenece}
    \Comment{Algorithm~\ref{alg:sequence-decompress} in Section~\ref{sec:decode-sequence}}
    \State $\mathsf{Deg} = (\mathsf{Deg}_v: 1 \leq v \leq n) \gets \Array \text{
      of }
    \Dictionary(\nats \times \nats \rightarrow \nats)$ \label{l:dec-deg-def-deg}
    \For{$1 \leq v \leq n$} \label{l:dec-deg-for-v-n}
    \State $\vec{\nu} \gets \mathsf{VertexTypesList}(y_v)$ \label{l:dec-deg-vec-nnu-VTL-yv}
    \State $\theta_v \gets \nu_1$ \label{l:dec-deg-theta-v}\Comment{decode the vertex mark of $v$}
    \For{$1 \leq k \leq (\textproc{Size}(\vec{\nu})-1)/3$} \label{l:dec-deg-for-k-vnu}
    \State $i \gets \nu_{1+(3k-2)}$ \label{l:dec-deg-i-nu}
    \State $\ip \gets \nu_{1+(3k-1)}$ \label{l:dec-deg-ip-nu}
    \State $j \gets \nu_{1+3k}$ \label{l:dec-deg-j-nu}
    \State $\mathsf{Deg}_v.\textproc{Insert}((i,\ip), j)$ \label{l:dec-deg-Deg-v-insert}
    \EndFor
    \EndFor
    \EndProcedure
  \end{algorithmic}
\end{myalg}

Now, we analyze the complexity of
Algorithm~\ref{alg:decode-vertex-deg-profiles}. Note that there are at most $n$
possible vertex types in the graph. Thereby, the variable $K$ in
line~\ref{l:dec-deg-read-K} is at most $n$ and the loop of
line~\ref{l:dec-deg-for-k} runs at most $n$ times. Recalling the discussion in
Section~\ref{sec:encode-vertex-types}, the array $\vnu$ has $O(\delta)$ elements,
each having  $O(\log_2(|\edgemark| \vee \mathsf{TCount} \vee \delta)) = O(\log |\edgemark| +
\log n)$ bits. Thereby, the overall time complexity of the loop in
line~\ref{l:dec-deg-for-k} is $O(n \delta(\log |\edgemark| + \log n))$ and the
memory required to store $\mathsf{VertexTypesList}$ is $O(n \delta (\log
|\edgemark| + \log n))$ bits. From Proposition~\ref{prop:encode-sequence}, the
time and memory complexities associated to the  decoding for  $\vy$ in
line~\ref{l:dec-deg-decode-sequenece} are $O(n \log^4 n \log \log n)$ and $O(n
\log n)$ respectively. Moreover, for each $1 \leq v \leq n$ inside the loop of
line~\ref{l:dec-deg-for-v-n}, we perform $O(\delta)$ insertions into $\mathsf{Deg}_v$
in line~\ref{l:dec-deg-Deg-v-insert}. Since $\mathsf{Deg}_v$ has at most
$\delta$ key--value pairs, and the integers $i, \ip$ and $j$ have $O(\log
|\edgemark| + \log n)$ bits, the overall time complexity of the loop in
line~\ref{l:dec-deg-for-v-n} is $O(n \delta \log \delta (\log |\edgemark| + \log
n))$. Additionally, similar to the discussion in
Section~\ref{sec:encode-find-Deg}, the memory required to store $\mathsf{Deg}$
is $O(n \delta(\log |\edgemark| + \log n))$.
Putting all the above together, we have

\begin{lem}
  \label{lem:dec-deg-complexity}
  The time and memory complexities of
  Algorithm~\ref{alg:decode-vertex-deg-profiles} are $O(n \delta \log
  \delta(\log |\edgemark| + \log n) + n \log^4 n \log \log n)$ and $O(n \delta
  (\log |\edgemark| + \log n))$, respectively.
\end{lem}

\editfinish

\editstart

\subsubsection{Decoding for $\mathsf{PartitionDeg}$ and $\mathsf{OriginalIndex}$}
\label{sec:dec-find-partition-deg-original-index}

In this section we discuss how the decoder finds $\mathsf{PartitionDeg}$ and
$\mathsf{OriginalIndex}$ (which were defined in Section~\ref{sec:dec-alg}).
Algorithm~\ref{alg:dec-partition-deg-original-index} below performs this.
Algorithm~\ref{alg:dec-partition-deg-original-index} introduces a
procedure which is part of Algorithm~\ref{alg:graph-decode}, hence it has
access to the variables defined in Algorithm~\ref{alg:graph-decode}.

Since we have constructed $\mathsf{Deg}$ in the previous step, we can use it to find
$\mathsf{PartitionDeg}$ in a way identical to
Algorithm~\ref{alg:encode-find-partition-graphs} in
Section~\ref{sec:encode-find-partition-graphs}. In order to find
$\mathsf{OriginalIndex}$,
\color{black}
{\color{peditcolor}
consider  a vertex $v$ and a key--value pair $((i,\ip), k) \in
\mathsf{Deg}_v$. Note that if $(i,\ip)$ is not in the set of key--value pairs of
$\mathsf{PartitionDeg}$, 
then 
$v$ is the first vertex in $\mVn_{t,\pp{t}}$, where $t=
J_n^{-1}(i)$ and $t' = J_n^{-1}(i')$.}
Therefore, in this case, 
$\mathsf{OriginalIndex}(i,\ip)$ must be initialized with an array of size 1
containing $v$. This is done in
line~\ref{l:dec-part-deg-orig-index--notin-orig-index}. On the other hand, if
$(i,\ip)$ is in the set of key--value pairs of $\mathsf{PartitionDeg}$, we
append $v$ at the end of $\mathsf{OriginalIndex}(i,\ip)$ in
line~\ref{l:dec-part-deg-orig-index--in-orig-index}.

\begin{myalg}[Finding Degree Sequences of Partition Graphs and Relative Vertex
  Indexing \label{alg:dec-partition-deg-original-index}]
  \begin{algorithmic}[1]
    \Procedure{DecodePartitionDegOriginalIndex}{}
    \Comment{line~\ref{l:main-dec-part-indx-orig-indx} in Algorithm~\ref{alg:graph-decode}}
    \For{$1 \leq v \leq n$}
    \For{$((i,\ip), k) \in \mathsf{Deg}_v$}
    \If{$(i,\ip) \notin \mathsf{PartitionDeg}.\textproc{Keys}$}
    \State $\mathsf{OriginalIndex}.\textproc{Insert}((i,\ip),(v))$ \label{l:dec-part-deg-orig-index--notin-orig-index}
    \Comment{$\mathsf{OriginalIndex}(i,\ip)$ is now an array with length 1
      containing $v$}
    \State $\mathsf{PartitionDeg}.\textproc{Insert}((i,\ip), (k))$
    \Comment{$\mathsf{PartitionDeg}(i,\ip)$ now becomes an array of length 1
      containing $k$}
    \Else
    \State append $v$ at the end of $\mathsf{OriginalIndex}(i,\ip)$ \label{l:dec-part-deg-orig-index--in-orig-index}
    \State append $k$ at the end of $\mathsf{PartitionDeg}(i,\ip)$
    \EndIf
    \EndFor
    \EndFor
    \EndProcedure
  \end{algorithmic}
\end{myalg}

Now, we analyze the complexity of
Algorithm~\ref{alg:dec-partition-deg-original-index} above. Recall that for $1
\leq v \leq n$, $\Deg_v$ has at most $\delta$ key--value pairs. Moreover, recall from
Section~\ref{sec:encode-find-partition-graphs} that $\mathsf{PartitionDeg}$ has
$O(\mn)$ key--value pairs. Similarly, 
$\mathsf{OriginalIndex}$ has $O(\mn)$ key--value pairs.
Moreover, since we have $i, i' = O(\mn)$, the complexity of searching and
inserting $(i,i')$ in $\mathsf{PartitionDeg}$ or $\mathsf{OriginalIndex}$ is
$O(\log^2 \mn)$. 
 Therefore, the time
complexity of the algorithm is $O(n \delta \log^2 \mn)$.
Additionally, recalling our discussion from
Section~\ref{sec:encode-find-partition-graphs}, the memory required to store
$\mathsf{PartitionDeg}$ is $O(n \delta \log \delta + \mn \log \mn)$. On the
other hand, each key in $\mathsf{OriginalIndex}$ corresponds to a pair $(t,\tp)
\in \mEn$, with the corresponding value being an array of size $\Nn_{t,t'}$ with
elements bounded by $n$. Consequently, using the bound $|\mEn| \leq 2 \mn$, the memory required to store
$\mathsf{OriginalIndex}$ is $O(\sum_{(t,\pp{t}) \in \mEn} \log \mn +
\Nn_{t,\pp{t}} \log n)$. Using ~\eqref{eq:sum-Nn-ttp-ndelta}
and  $|\mEn| \leq 2 \mn$, this is  $O(\mn \log \mn + n \delta \log n)$.  To summarize, we have

\begin{lem}
  \label{lem:dec-find-part-deg-orig-index-complexity}
  The time and memory complexities of
  Algorithm~\ref{alg:dec-partition-deg-original-index} are $O(n \delta \log^2
  \mn)$ and $O(\mn \log \mn + n \delta \log n)$, respectively.
\end{lem}

\editfinish

\editstart

\subsection{Complexity of the Decoding Algorithm}
\label{sec:main-decode-complexity}

In this section, we analyze the complexity of the decoding algorithm, i.e.\
Algorithm~\ref{alg:graph-decode} described in Section~\ref{sec:dec-alg}.

First note that, using the bound $\mathsf{TCount} \leq 4 \mn$, the time and memory complexities of decoding the variables $\mathsf{TCount}$,
$\mathsf{TIsStar}$, and $\mathsf{TMark}$ are each $O(\mn \log |\edgemark|)$.
Furthermore, from Proposition~\ref{prop:encode-sequence}, the time and memory
complexities of decoding star vertices in line~\ref{l:dec-Vns} are $O(n \log^4 n
\log \log n)$ and $O(n \log n)$, respectively. Moreover,
Lemma~\ref{lem:dec-star-edges-complexity} in Section~\ref{sec:dec-star-edges}
implies that the time and memory complexities of decoding star edges in
line~\ref{l:dec-star-edges} are $O(\mn(\log |\edgemark| + \log n) + |\edgemark|^2
n)$ and $O(\mn(\log n + \log |\edgemark|))$, respectively. From
Lemma~\ref{lem:dec-deg-complexity} in Section~\ref{sec:dec-ver-deg-profiles},
the time and memory complexities of decoding vertex degree profiles in
line~\ref{l:dec-vertex-deg-profiles} are 
$O(n \delta \log
\delta(\log |\edgemark| + \log n) + n \log^4 n \log \log n)$ and $O(n \delta
(\log |\edgemark| + \log n))$, respectively. From
Lemma~\ref{lem:dec-find-part-deg-orig-index-complexity} in
Section~\ref{sec:dec-find-partition-deg-original-index}, finding
$\mathsf{PartitionDeg}$ and $\mathsf{OriginalIndex}$ in
line~\ref{l:main-dec-part-indx-orig-indx} has time and memory complexities  $O(n \delta \log^2
\mn)$ and $O(\mn \log \mn + n \delta \log n)$, respectively.


Now, we turn to finding the complexity of the loop in
line~\ref{l:dec-part-graph-for}. For $1 \leq k \leq K$, reading $i$ and $\ip$
is performed in $O(\log \mathsf{TCount}) = O( \log \mn)$ time. Let $t =
J_n^{-1}(i)$ and $\tp = J_n^{-1}(i')$ be elements in
$\mFdeltah$ corresponding to $i$ and $\ip$, respectively. First, consider the
case that $i < \ip$. Similar to the discussion in
Section~\ref{sec:enc-complexity},
Proposition~\ref{prop:bipartite-color-graph-compress} implies that the integer
$f$ which we read in line~\ref{l:dec-y-read-neq} has $O(\delta \Nn_{t,\tp} \log
(\Nn_{t,\tp} \delta)) = O(\Nn_{t,\tp} \delta \log n)$ bits.
Therefore, reading
it takes $O(\Nn_{t,\tp} \delta \log n)$ time. Since all degrees in $\Gn_{t,\tp}$
are bounded by $\delta$ by definition, as implied by
Proposition~\ref{prop:bipartite-color-graph-compress}, with $\tNn_{t,\tp} :=
\max \{ \Nn_{t,\tp}, \Nn_{\tp, t}\}$, the time and memory complexities of
decoding the adjacency list of $\Gn_{t,\tp}$ in
line~\ref{l:dec-part-b-graph-decode} are $O(\delta \tNn_{t,\tp} \log^4
\tNn_{t,\tp} \log \log \tNn_{t,\tp}) = O(\delta \tNn_{t,\tp} \log^4 n \log \log n)$ and $O(\delta \tNn_{t,\tp} \log
\tNn_{t,\tp}) = O(\delta \tNn_{t,\tp} \log n)$, respectively.
Note that the number of key--value pairs in $\mathsf{PartitionDeg}$ is at most
$2 \mn$, and for each key $(i,\ip)$ in $\mathsf{PartitionDeg}$, we have $i, \ip
\leq \mathsf{TCount} \leq 4 \mn$. This implies that the time complexity of accessing
$\mathsf{PartitionDeg}(i,\ip)$ in line~\ref{l:dec-part-b-graph-decode} is $O(\log^2 \mn)$. 
Now,
consider the case $i = \ip$. Similar to the discussion in
Section~\ref{sec:enc-complexity}, implied by
Proposition~\ref{prop:equal-color-graph-compress}, reading the integer $f$ and the array
$\vec{\tilde{f}}$ has both time and memory complexities $O(\delta \Nn_{t,t} \log
n)$. Furthermore, Proposition~\ref{prop:equal-color-graph-compress} implies that
the time and memory complexities of finding the forward adjacency list of
$\Gn_{t,t}$ in line~\ref{l:dec-sim-graph-decode} are $O(\delta \Nn_{t,t} \log^5
\Nn_{t,t} \log \log \Nn_{t,t}) = O(\delta \Nn_{t,t} \log^5 n \log \log n)$ and
$O(\delta \Nn_{t,t} \log \Nn_{t,t}) = O(\delta \Nn_{t,t} \log n)$,
respectively.
Similar to the previous case, accessing $\mathsf{PartitionDeg}(i,i)$ in
line~\ref{l:dec-sim-graph-decode} takes
$O(\log^2 \mn)$ time. 
Note that the number of key--value pairs in $\mathsf{OriginalIndex}$ is
$|\mEn_{\leq}| = O(\mn)$, and each key is a pair of integers bounded by
$\mathsf{TCount}$. Therefore, the time complexity of accessing
$\mathsf{OriginalIndex}(i,\ip)$ and $\mathsf{OriginalIndex}(\ip,i)$ in
lines~\ref{l:dec-OI-i-ip-A} and \ref{l:dec-IO-ip-i-B} are $O(\log^2 \mn)$.
Finally, in the loop of line~\ref{l:dec-for-v-partition-deg}, for each edge in
the partition graph $\Gn_{t,\tp}$, we append an edge to
$\mathsf{EdgeListDec}$. Since there are at most $\tNn_{t,\tp} \delta$ many edges
in $\Gn_{t,\tp}$, the overall time complexity of the loop of
line~\ref{l:dec-for-v-partition-deg} is $O(\tNn_{t,\tp} \delta)$. Recall that 
$\tNn_{t,\tp} = \max \{\Nn_{t,\tp}, \Nn_{\tp, t}\}$.
Putting the above discussion together, we realize that the overall time
complexity of the loop in line~\ref{l:dec-part-graph-for} is 
\begin{align*}
  & O\Bigg (\sum_{(t,\tp) \in \mEn_<}  \Nn_{t,\tp} \delta \log n + \delta \tNn_{t,\tp} \log^4 n \log \log n + \log^2 \mn   \\
& \qquad + \sum_{(t,t) \in \mEn}  \Nn_{t,t}  \delta \log n + \delta \Nn_{t,t} \log^5 n \log \log n + \log^2 \mn  \Bigg ).
\end{align*}
Using~\eqref{eq:sum-tNn-Nn-ndelta} together with the bound $|\mEn| \leq 2 \mn$
and simplifying, we realize that the overall time complexity of the loop in
line~\ref{l:dec-part-graph-for} is $O(\mn \log^2 \mn + n \delta^2 \log^5 n \log
\log n)$. Similarly, the overall memory complexity of the loop in line~\ref{l:dec-part-graph-for} is
\begin{equation*}
  O\left(\sum_{(t,\tp) \in \mEn_< } \delta \tNn_{t,\tp} \log n + \sum_{(t,t) \in \mEn} \delta \Nn_{t,t} \log n \right ) = O(n \delta^2 \log n).
\end{equation*}

Putting all the above together, we realize that the overall time and memory
complexities of Algorithm~\ref{alg:graph-decode} are $O(\mn(\log |\edgemark| +
\log n ) + n \delta \log \delta \log |\edgemark| + n \delta^2 \log^5 n \log \log n +
|\edgemark|^2 n + \mn \log^2 \mn)$ and $O((\mn + n \delta)
(\log |\edgemark| + \log n ) + \mn \log \mn + n \delta^2 \log n)$, respectively. 


\editfinish


\editstart

\section{Proof of Optimality}
\label{sec:optimality-proof}

In this section we prove that the compression algorithm  we described in
the previous sections is universally optimal, i.e.\ we prove the first part  of
Theorem~\ref{thm:optimality-complexity-main}.
The structure of this section is as follows. We begin with reviewing and
introducing some
notation. 
We then
\color{black}
find an
upper bound for the codeword length associated to a simple marked graph in
Proposition~\ref{prop:opt-aa12-nat-count}. This bound holds for both cases $h>1$ and
$h=1$. Then, we use this to prove the bound~\eqref{eq:opt-statement-eta-0-bound}
in Theorem~\ref{thm:optimality-complexity-main} 
separately
\color{black}
for the two cases $h>1$ and
$h=1$ in Propositions~\ref{prop:opt-h>1} and \ref{prop:opt-h=1}, respectively.
Afterwards, we use these to prove the first part of Theorem~\ref{thm:optimality-complexity-main}.
The proofs of  Propositions~\ref{prop:opt-h>1} and \ref{prop:opt-h=1} will be given in
Sections~\ref{sec:opt-h>1-proof} and \ref{sec:opt-h=1-proof}, respectively.




  
To start,
\color{black}
recall from Section~\ref{sec:main-star-vertices} that an edge $(v,w)$ in $\Gn$ is called
a star edge if $\type{v}{w} = (\star_x, \star_{\pp{x}})$ where $x =
\xi_{\Gn}(w,v)$ and $\pp{x} = \xi_{\Gn}(v,w)$. 
Let $\mn_\star$ denote 
the number of star edges in $\Gn$. Also,
recall that $\Vns$ denotes the subset of vertices in $\Gn$ which have at least
one star edge connected to them. 
We defined $\mEn$ to be the set of $c \in \mCdeltah = \mFdeltah \times
\mFdeltah$ such that $\type{v}{w} = c$ for some $v \sim_{\Gn} w$. In other
words, $\mEn$ is the set of all edge types that appear in $\Gn$, excluding star
edges. Moreover, we defined $\mEn_\leq$ be the set
of $(t,t') \in \mEn$ such that with $\tilde{t}= J_n(t)$ and $\tilde{t}' = J_n(\pp{t})$
being the integer representations of $t$ and $t'$ respectively,
we have  $\tilde{t} \leq \tilde{t}'$.
  For $[T,o] \in \mTb_*$, define $\theta([T,o]) := \tau_T(o) \in \vermark$
    to be the mark at the root in $[T,o]$. Moreover, with $h \geq 1$, and
    $\delta \geq 1$, let $D([T,o]) =
    (D_{t,t'}([T,o]): t,t' \in \mFdeltah)$ be defined as follows. If $\deg_T(o) \leq
    \delta$, for $t, t' \in \mFdeltah$, we define $D_{t,t'}([T,o])$ to be 
    the number of vertices $v \sim_T o$ such that $\deg_T(v) \leq \delta$, 
    $T[v,o]_{h-1} = t$, and $T[o,v]_{h-1} = t'$. If $\deg_T(o) > \delta$, we
    define $D_{t,t'} ([T,o])$ to be zero for all $t,t' \in \mFdeltah$. 
    In other words, we can write
    \begin{equation}
      \label{eq:D-T-o-def}
      D_{t,\pp{t}}([T,o]) := |\{v \sim_T o: \deg_T(o) \leq \delta, \deg_T(v) \leq \delta, T[v,o]_{h-1} = t, T[o,v]_{h-1} = t'\}|.
    \end{equation}
Comparing this to~\eqref{eq:type-v-w-equivalent} and \eqref{eq:Dn-def}, for $v \in [n]$,
    we have
    \begin{equation}
      \label{eq:thetan-Dn--UC-Gn}
      (\thetan_v , \Dn(v)) = (\theta([\UC_v(\Gn),v]), D([\UC_v(\Gn), v])).
    \end{equation}
Note that $D([T,o])$ is defined for $[T,o] \in \mTb_*$, and $[T,o]$ is not
restricted to be a member of $\mTb_*^h$, i.e.\ its depth can be bigger than $h$. 
For the sake of simplicity, we have not included $h$ and $\delta$ in this
notation, with the understanding that $h$ and $\delta$  are fixed parameters and
can be determined  from
the context. 
Note that by this definition, if $D_{t,t'}([T,o]) \neq 0$, we must have
$\deg_T(o) \leq \delta$. Additionally, in this case, we have $D_{t,t'}([T,o])
\leq \deg_T(o) \leq \delta$. Consequently, we realize that in general, we have
    \begin{equation}
      \label{eq:D-T-o-delta}
      D_{t,t'}([T,o]) \leq \delta \qquad \forall [T,o] \in \mTb_*, t,t' \in \mFdeltah.
    \end{equation}
In addition to this, define 
\begin{equation*}
 \mA^{(n)}_{\delta} := \{(\thetan_v, \Dn(v)): v \in [n]\},
\end{equation*}
which is the set of pairs $(\thetan_v, \Dn(v))$ that appear in $\Gn$.
Furthermore, define
\begin{equation*}
  \mA_\delta := \{(\theta([T,o]), D([T,o])): [T,o] \in \mTb_*\}.
\end{equation*}
From~\eqref{eq:thetan-Dn--UC-Gn}, we realize that $\mA^{(n)}_\delta \subseteq \mA_\delta$.
Recall from~\eqref{eq:deg-star-deg} in Section~\ref{sec:Problem-statement} that
for $h\geq 1$, $\delta \geq 1$,  and $[T,o] \in \mTb_*$, we define $\deg_\star^{(h, \delta)}([T,o])$ to
be 
the number of neighbors $v \sim_T o$ such that $T[v,o]_{h-1} \notin \mFdeltah$ or
$T[o,v]_{h-1} \notin \mFdeltah$ or $\deg_T(o) > \delta$ or $\deg_T(v) > \delta$. 

In the following Proposition~\ref{prop:opt-aa12-nat-count}, we find an upper bound on the codeword length
corresponding to a simple marked graph $\Gn$ which holds for both cases $h>1$
and $h=1$. The proof of Proposition~\ref{prop:opt-aa12-nat-count} is given in
Appendix~\ref{sec:app-nat-count}.


For $c \in \mCdeltah = \mFdeltah \times \mFdeltah$, we define $\Sn_c :=
\sum_{v=1}^n \Dn_c(v)$. 
In particular, for 
$(t,\tp) \in \mEn$, we write $\Sn_{t,\tp} = \sum_{v=1}^n
\Dn_{t,\tp}(v)$.
\color{black}

\begin{prop}
  \label{prop:opt-aa12-nat-count}
  For our compression algorithm explained in Section~\ref{sec:alg-details}, as
  $h \geq 1$ and $\delta \geq 1$ are fixed and $n \rightarrow \infty$, we have
  \begin{equation}
    \label{eq:opt-nat-bound-1}
  \begin{aligned}
    \nat(\fn_{h, \delta}(\Gn)) &\leq o(n) + \log \binom{n}{|\Vns|} + |\edgemark|^2 \times |\Vns| \log 2 + 2 \mn_{\star} \log 2 + \mn_{\star} \log n \\
    &\qquad +\log \left( \frac{n!}{\prod_{(\theta, D) \in \mA_\delta} |\{ v \in [n]: (\thetan_v, \Dn(v)) = (\theta, D)\}|!}  \right) + \sum_{(t,\pp{t}) \in \mEn_{\leq}} \log \eln_{t,\pp{t}},
  \end{aligned}
\end{equation}
where for $(t,\tp) \in \mEn_{\leq}$, with  $\Sn_{t,\tp} = \sum_{v=1}^n
\Dn_{t,\tp}(v)$, when $t \neq \tp$, we define
\begin{equation*}
  \eln_{t,t'} :=\frac{\Sn_{t,t'}!}{\prod_{v=1}^n \Dn_{t,t'}(v)! \prod_{v=1}^n \Dn_{t',t}(v)!},
\end{equation*}
and when $t = \tp$, we define
\begin{equation*}
  \eln_{t,t} := \frac{(\Sn_{t,t} - 1)!!}{\prod_{v=1}^n \Dn_{t,t}(v)!}.
\end{equation*}
Moreover, for $(t,\pp{t}) \in \mEn_{\leq}$, we have $\eln_{t,\pp{t}} \geq 1$. 
\end{prop}

The proof this proposition, which is provided in Appendix~\ref{sec:app-nat-count}, is based on
investigating the details of the compression algorithm provided in
Section~\ref{sec:alg-details}. Here we give a rough explanation on  how
each term in~\eqref{eq:opt-nat-bound-1} correspond to the steps of our
algorithm, as we discussed in Section~\ref{sec:algorithm-general}:
\begin{itemize}
\item The term $\log
\binom{n}{|\Vns|}$ corresponds to encoding the set $\Vns$ as in
Section~\ref{sec:main-star-vertices}. 
\item The 
sum of
\color{black}
terms $|\edgemark|^2 \times |\Vns| \log 2 + 2 \mn_{\star} \log 2 +
  \mn_{\star} \log n$ corresponds to encoding star edges as in
  Section~\ref{sec:main-star-edges}. Note that~\eqref{eq:opt-nat-bound-1}
  studies the number of nats instead of the number of bits, that is why  $\log 2$
  appears for conversion.
\item The term
  \begin{equation*}
    \log \left( \frac{n!}{\prod_{(\theta, D) \in \mA_\delta} |\{ v \in [n]: (\thetan_v, \Dn(v)) = (\theta, D)\}|!}  \right),
  \end{equation*}
  corresponds to encoding vertex types as in
  Section~\ref{sec:main-encode-vertex-types}.
\item The summation $\sum_{(t,\pp{t}) \in \mEn_{\leq}} \log \eln_{t,\pp{t}}$
  corresponds to encoding partition graphs as in
  Section~\ref{sec:main-encode-partition}. 
  Note
  \color{black}
  the similarity between
  $\eln_{t,t'}$ for $t \neq \pp{t}$ and the upper bound in
  Proposition~\ref{prop:bipartite-color-graph-compress}, and also between
  $\eln_{t,t}$ and the upper bound in Proposition~\ref{prop:equal-color-graph-compress}.
\end{itemize}

As we discussed above, we split the proof of the
bound~\eqref{eq:opt-statement-eta-0-bound} of
Theorem~\ref{thm:optimality-complexity-main} into two cases. First, in
Proposition~\ref{prop:opt-h>1}, which is proved in
Section~\ref{sec:opt-h>1-proof}, we address the case $h>1$.  Then, in
Proposition~\ref{prop:opt-h=1}, which is proved in
Section~\ref{sec:opt-h=1-proof}, we address the case $h=1$. 

\begin{prop}
  \label{prop:opt-h>1}
  With the assumptions of Theorem~\ref{thm:optimality-complexity-main}, 
  if $h > 1$ and $\delta \geq 1$, we have
  \begin{equation*}
      \limsup_{n \rightarrow \infty} \frac{\nat(\fn_{h, \delta}(\Gn)) - \mn \log n}{n} \leq J_h(\mu_h) + \eta_0(\mu; h, \delta).
  \end{equation*}
\end{prop}

\begin{prop}
  \label{prop:opt-h=1}
  With the assumptions of Theorem~\ref{thm:optimality-complexity-main}, 
  if $h = 1$ and $\delta \geq 1$, we have
  \begin{equation*}
      \limsup_{n \rightarrow \infty} \frac{\nat(\fn_{h, \delta}(\Gn)) - \mn \log n}{n} \leq J_h(\mu_h) + \eta_0(\mu; h, \delta).
  \end{equation*}
\end{prop}

Proposition~\ref{prop:opt-h>1} and Proposition~\ref{prop:opt-h=1} depend respectively on 
Lemma~\ref{lem:opt-h>1-simplifications} and 
Lemma~\ref{lem:opt-h-1-properties}, which deal with the cases $h>1$ and $h=1$
and are also proved in Section~\ref{sec:opt-h>1-proof} and 
Section~\ref{sec:opt-h=1-proof} respectively. We first state Lemma~\ref{lem:opt-h>1-simplifications},
which applies when $h > 1$.

\begin{lem}
  \label{lem:opt-h>1-simplifications}
  If $h > 1$, with $\delta \geq 1$ and for all $[T,o] \in \mTb_*$, we have
  \begin{equation}
    \label{eq:h>1-deg-star-simplification}
    \deg_\star^{(h, \delta)}([T,o]) = |\{ v \sim_T o: T[o,v]_{h-1} \notin \mFdeltah \text{ or } T[v,o]_{h-1} \notin \mFdeltah\}|.
  \end{equation}
 In particular, since the right hand side is a function of $[T,o]_h$, we have
  \begin{equation}
    \label{eq:deg-star-function-To-h}
    \deg_\star^{(h, \delta)}([T,o]) =     \deg_\star^{(h, \delta)}([T,o]_h).
  \end{equation}
  Furthermore, for all $[T,o] \in \mTb_*$ and $t,t' \in \mFdeltah$, we have
\begin{equation}
  \label{eq:D-T-o-is-Eh-for-h>1}
  D_{t,t'}([T,o]) = E_h(t,t')([T,o]),
\end{equation}
where $E_h(t,t')$ is defined in~\eqref{eq:Eh-g-g'}.
Moreover, since the right hand side is a function of $[T,o]_h$,  we have
\begin{equation}
  \label{eq:h>1-D-is-funct-of-T-o-h}
  D([T,o]) = D([T,o]_h).
\end{equation}
\end{lem}

In
Figure~\ref{fig:h-1-simplifications-do-not-hold} we illustrate via an example why 
these claims do
not necessarily hold when $h=1$. In fact, a crucial difference is that when
$h=1$, the degree of an offspring of the root $v \sim_T o$ in $[T,o] \in
\mTb_*$, i.e.\ $\deg_T(v)$, 
is not necessarily the same as the degree of $v$ in $[T,o]_1$. Indeed, since
such a $v$ is a leaf node in $[T,o]_1$, its degree is 1 in $[T,o]_1$, while
$\deg_T(v)$ can be bigger than one. Note that this phenomenon does not occur when
$h>1$. 

\begin{figure}
  \centering
  \newcommand{\tonerr}[2][0]{\scalebox{#2}{\tikz[baseline=(p.base)]{
      \node (p) at (0,#1-0.5) {};
      \node[nodeR] (r) at (0,0) {};
      \draw[edgeO] (r) -- ($(r) + (0,0.3)$);
    }}}

  \newcommand{\tonerb}[2][0]{\scalebox{#2}{\tikz[baseline=(p.base)]{
      \node (p) at (0,#1-0.5) {};
      \node[nodeB] (r) at (0,0) {};
      \draw[edgeO] (r) -- ($(r) + (0,0.3)$);
    }}}

  \newcommand{\tonebr}[2][0]{\scalebox{#2}{\tikz[baseline=(p.base)]{
      \node (p) at (0,#1-0.5) {};
      \node[nodeR] (r) at (0,0) {};
      \draw[edgeB] (r) -- ($(r) + (0,0.3)$);
    }}}

  \newcommand{\tonebb}[2][0]{\scalebox{#2}{\tikz[baseline=(p.base)]{
      \node (p) at (0,#1-0.5) {};
      \node[nodeB] (r) at (0,0) {};
      \draw[edgeB] (r) -- ($(r) + (0,0.3)$);
    }}}

  \begin{tikzpicture}
    \begin{scope}[xshift=-3cm]
      \node[nodeB] (r) at (0,0) {};
      \node[nodeR] (a) at (-1,-1) {};
      \node[nodeR] (b) at (0,-1) {};
      \node[nodeR] (c) at (1,-1) {};
      \node[nodeB] (aa) at (-1.5,-2) {};
      \node[nodeB] (ab) at (-1,-2) {};
      \node[nodeB] (ac) at (-0.5,-2) {};

      \drawedge{r}{a}{B}{O}
      \drawedge{r}{b}{B}{O}
      \drawedge{r}{c}{B}{O}
      \drawedge{a}{aa}{B}{B}
      \drawedge{a}{ab}{B}{B}
      \drawedge{a}{ac}{B}{B}

      \node at (0,-2.5) {$(a)$};

    \end{scope}

    \begin{scope}[xshift=3cm]
      \node[nodeB] (r) at (0,0) {};
      \node[nodeR] (a) at (-1,-1) {};
      \node[nodeR] (b) at (0,-1) {};
      \node[nodeR] (c) at (1,-1) {};

      \drawedge{r}{a}{B}{O}
      \drawedge{r}{b}{B}{O}
      \drawedge{r}{c}{B}{O}

      \node at (0,-2.5) {$(b)$};

    \end{scope}

  \end{tikzpicture}
  \caption[What happens when
  $h=1$]{\label{fig:h-1-simplifications-do-not-hold}The results of
    Lemma~\ref{lem:opt-h>1-simplifications} do not necessarily hold when $h=1$.
    To see this, we consider $[T,o] \in \mTb_*$ as shown in $(a)$, and set $h=1$
    and $\delta = 3$. We also set $\edgemark =
  \{\text{\color{blueedgecolor} Blue (solid)}, \text{\color{orangeedgecolor}
    Orange (wavy)} \}$ and $\vermark = \{\tikz{\node[nodeB] at (0,0) {};}, \tikz{\node[nodeR] at (0,0)
  {};}\}$. 
Observe that since $h-1 = 0$, we have
    $\mFdeltah = \{\tonebb[0.5]{1}, \tonebr[0.5]{1},\tonerb[0.5]{1},
    \tonerr[0.5]{1}\} \equiv \edgemark \times \vermark$. This means that for all
    $[T,o] \in \mTb_*$ and $v \sim_T o$, we have $T[v,o]_0 \in \mFdeltah$ and
    $T[o,v]_0 \in \mFdeltah$. Thereby, the right hand side
    of~\eqref{eq:h>1-deg-star-simplification} is zero. 
 On the other hand, following the definition, we have $\deg_\star^{(h, \delta)}([T,o]) =
1$, the reason being that the degree of the leftmost offspring of the root is $4
> \delta = 3$. This means that \eqref{eq:h>1-deg-star-simplification} does not
hold. Furthermore, with $[T,o]_1$ illustrated in $(b)$, we have $\deg_\star^{(h,
  \delta)}([T,o]_1) = 0 \neq \deg_\star^{(h, \delta)}([T,o])$
and~\eqref{eq:deg-star-function-To-h} also does not hold. In addition to this,
we have $D_{\tonebb[0.5]{0.8}, \tonerr[0.5]{0.8}}([T,o]) = 2$ while
$E_1(\tonebb[0.5]{0.8}, \tonerr[0.5]{0.8})([T,o]) = 3$. Thereby,
\eqref{eq:D-T-o-is-Eh-for-h>1} does not hold. Furthermore,
$D_{\tonebb[0.5]{0.8}, \tonerr[0.5]{0.8}}([T,o]_1) = 3 \neq 2 =
D_{\tonebb[0.5]{0.8}, \tonerr[0.5]{0.8}}([T,o])$ which means that
\eqref{eq:h>1-D-is-funct-of-T-o-h} also does not hold. Note that this happens
because the degree of the leftmost offspring of the root in $[T,o]_1$ is 1,
while its degree in $[T,o]$ is 4.} 
\end{figure}


We next state Lemma~\ref{lem:opt-h-1-properties}, which applies when $h=1$.

\begin{lem}
  \label{lem:opt-h-1-properties}
  Assume that $h=1$ and $\delta \geq 1$. Then, the following hold:
  \begin{enumerate}
  \item \label{it:loh1-1} $\mFdeltah = \edgemark \times \mTb_*^0$. In particular, $|\mFdeltah| =
    |\edgemark| \times |\vermark|$.
  \item \label{it:loh1-2} For all $[T,o] \in \mTb_*$, and all $v \sim_T o$, we have $T[o,v]_{h-1}
    \in \mFdeltah$ and $T[v,o]_{h-1} \in \mFdeltah$.
  \item \label{it:loh1-3} For all $[T,o] \in \mTb_*$, we have $\deg_\star^{(h, \delta)}([T,o]) =
    |\{v \sim_T o : \deg_T(o) >  \delta \text{ or } \deg_T(v) > \delta\}|$.
  \item \label{it:loh1-4} For $[T,o] \in \mTb_*$, if $\deg_\star^{(h, \delta)}([T,o]) = 0$, then
    for all $t, \pp{t} \in \mFdeltah $, we have $D_{t,\pp{t}}([T,o]) =
    E_1(t,\pp{t})([T,o])$.
  \item\label{it:loh1-5}  For $[T,o] \in \mTb_*$, $D([T,o])$ and
    $\deg_\star^{(h, \delta)}([T,o])$ are functions of $[T,o]_2$.
  \end{enumerate}
\end{lem}
\color{black}

With the aid of the discussion above, we are now ready to give the proof of the
first part of Theorem~\ref{thm:optimality-complexity-main}.

\begin{proof}[Proof of Theorem~\ref{thm:optimality-complexity-main}, part 1]
 The  bound~\eqref{eq:opt-statement-eta-0-bound} follows from
 Propositions~\ref{prop:opt-h>1} and \ref{prop:opt-h=1}. In order to
 show~\eqref{eq:optimality-thm-statement-delta-n}, we 
 show that $\eta_0(\mu; h, \delta) \rightarrow 0$ as $\delta
\rightarrow \infty$. In order to do this, since $|\vermark|$ and $|\edgemark|$
are constants, it suffices to  show that for $1 \leq i \leq 4$, $\eta_i(\mu; h, \delta) \rightarrow 0$
as $\delta \rightarrow \infty$. Note that as $\delta \rightarrow \infty$,
$\mFdeltah \uparrow \edgemark \times \mTb_*^{h-1}$. Thereby, for $h \geq 1$, we
have
\begin{equation}
  \label{eq:lim-deg*-delta-infty}
  \lim_{\delta \rightarrow \infty} \deg_\star^{(h, \delta)} ([T,o]) = 0 \qquad \forall [T,o] \in \mTb_*.
\end{equation}
This implies that $\eta_1(\mu; h, \delta) \rightarrow 0$ as $\delta \rightarrow
\infty$. On the other hand, as $\deg_*^{(h, \delta)}([T,o]) \leq \deg_T(o)$ and
$\evwrt{\mu}{\deg_T(o)} = \deg(\mu) < \infty$, \eqref{eq:lim-deg*-delta-infty}
together with dominated convergence theorem implies that $\eta_2(\mu; h, \delta)
\rightarrow 0$ as $\delta \rightarrow \infty$. A similar argument together with
the assumption that $\evwrt{\mu}{\deg_T(o) \log \deg_T(o)} < \infty$ implies
that $\eta_4(\mu; h, \delta) \rightarrow 0$ as $\delta \rightarrow \infty$. Now,
we claim that for $k \geq 1$, 
\begin{equation}
  \label{eq:opt-ev-deg-log-muh-claim}
  \evwrt{\mu}{\deg_T(o) \log \frac{1}{\mu_k([T,o]_k)}} < \infty.
\end{equation}
Note that with $P := \mu_k$, we have
\begin{equation*}
  \evwrt{\mu}{\deg_T(o) \log \frac{1}{\mu_k([T,o]_k)}} = - \sum_{[T,o] \in \mTb_*^h} P([T,o]) \deg_T(o) \log P([T,o]).
\end{equation*}
On the other hand, we have assumed that $\evwrt{\mu}{\deg_T(o) \log \deg_T(o)} <
\infty$. Therefore, since $\mu \in \mP_u(\mTb_*)$, and $0 < \deg(\mu) < \infty$, part 2 of Theorem~\ref{thm:Jh} implies that $H(P) = H(\mu_k)
< \infty$. Moreover, we have
\begin{equation*}
  \evwrt{\mu}{\deg_T(o) \log \deg_T(o)} = - \sum_{[T,o] \in \mTb_*^h} P([T,o]) \deg_T(o) \log \deg_T(o) < \infty.
\end{equation*}
Consequently, 
Lemma~5.4 in~\cite{bordenave2015large}  implies that $- \sum_{[T,o] \in
  \mTb_*^h} P([T,o]) \deg_T(o) \log P([T,o]) < \infty$
and~\eqref{eq:opt-ev-deg-log-muh-claim} follows. Thereby,
 \eqref{eq:lim-deg*-delta-infty} and \eqref{eq:opt-ev-deg-log-muh-claim}
 together with dominated convergence theorem imply that $\eta_3(\mu; h,
 \delta) \rightarrow 0$ as $\delta \rightarrow \infty$. 
Putting all the above
 together, since $|\edgemark|$ and $|\vermark|$ are finite, we realize that $\eta_0(\mu; h,
 \delta) \rightarrow 0$ as $\delta \rightarrow \infty$. Hence, by sending
 $\delta \rightarrow \infty$ in \eqref{eq:opt-statement-eta-0-bound}, we get
 \begin{equation*}
\limsup_{\delta \rightarrow \infty} \limsup_{n \rightarrow \infty} \frac{\nat(\fn_{h, \delta}(\Gn)) - \mn \log n}{n} \leq J_h(\mu_h),
\end{equation*}
which is precisely \eqref{eq:optimality-thm-statement-delta-n}.
Finally, note that  we have assumed that $\evwrt{\mu}{\deg_T(o) \log \deg_T(o)}< \infty$,
$\mu \in \mP_u(\mTb_*)$, and $0 < \deg(\mu) < \infty$.
Therefore, part 2 of 
Theorem~\ref{thm:Jh} implies that $J_h(\mu_h)
\rightarrow \bch(\mu)$ as $h \rightarrow \infty$, which
implies~\eqref{eq:optimality-thm-statement}. This completes the proof.
\end{proof}


Finally, before moving on the proofs in the individual cases $h > 1$ and $h=1$,
in Sections~\ref{sec:opt-h>1-proof} and \ref{sec:opt-h=1-proof} respectively,
we introduce 
\color{black}
Lemmas~\ref{lem:sum-Snc-mn-mn-star}
through~\ref{lem:opt-ee16-general-bound-for-ln-ttp}, which hold for both cases
$h>1$ and $h=1$, and are useful in our further
analysis. The proofs of these lemmas are given in Appendix~\ref{sec:opt-lemmas}. 

\begin{lem}
  \label{lem:sum-Snc-mn-mn-star}
  For $h \geq 1$ and $\delta \geq 1$, we have
  \begin{equation*}
    \frac{1}{2} \sum_{c \in \mCdeltah} \Sn_c = \mn - \mn_\star.
  \end{equation*}
\end{lem}

\begin{lem}
  \label{lem:opt-bb13-Vns}
  With the assumptions in the first part of Theorem~\ref{thm:optimality-complexity-main},
  and with $h \geq 1$ and $\delta \geq 1$ fixed, we have 
  \begin{equation}
  \label{eq:Vns-n-eta1-delta}
  \lim_{n \rightarrow \infty} \frac{|\Vns|}{n} 
  = \eta_1(\mu; h , \delta),
\end{equation}
and
\begin{equation}
  \label{eq:opt-limsup-n-choose-Vns-eta1}
  \limsup_{n \rightarrow \infty} \frac{1}{n} \log \binom{n}{|\Vns|} \le \eta_1(\mu;h, \delta)(1 - \log \eta_1(\mu; h, \delta)).
  \color{black}
\end{equation}
Recall that $\eta_1(\mu; h, \delta) = \prwrt{\mu}{\deg_\star^{(h, \delta)}([T,o]) > 0}$.
\end{lem}



\begin{lem}
  \label{lem:opt-cc14-general-bounds-for-Dn}
  With the assumptions in the first part of Theorem~\ref{thm:optimality-complexity-main},
  and with $h \geq 1$ and $\delta \geq 1$ fixed,  for $c \in \mCdeltah =
  \mFdeltah \times \mFdeltah$, we have
  \begin{equation}
  \label{eq:lim-sum-Dn-c-E-D-To}
  \lim_{n \rightarrow \infty} \frac{1}{n} \sum_{v=1}^n \Dn_c(v) = \evwrt{\mu}{D_c([T,o])},
\end{equation}
and
\begin{equation}
  \label{eq:lim-sum-Dnc-factorial-Emu-log-Dc-factorial}
  \lim_{n \rightarrow \infty} \frac{1}{n} \sum_{v=1}^n \log \Dn_c(v)! = \evwrt{\mu}{\log D_c([T,o])!}.
\end{equation}
\end{lem}





\begin{lem}
  \label{lem:opt-dd15-mnstar-bound}
  With the assumptions in the first part of Theorem~\ref{thm:optimality-complexity-main},
and  with $h \geq 1$ and $\delta \geq 1$ fixed, we have
  \begin{equation}
    \label{eq:opt-lim-mn-star}
    \lim_{n \rightarrow \infty} \frac{\mn_\star}{n} = \eta_2(\mu; h, \delta).
  \end{equation}
  Recall that $\eta_2(\mu; h, \delta) =  \frac{1}{2} \evwrt{\mu}{\deg_\star^{(h, \delta)}([T,o])} $.
\end{lem}



\begin{lem}
  \label{lem:theta-D-fraction-limit}
  With the assumptions in the first part of Theorem~\ref{thm:optimality-complexity-main},
  and with $h \geq 1$ and $\delta \geq 1$ fixed, for $(\theta, D) \in \mA_\delta =
  \{(\theta([T,o]), D([T,o])): [T,o] \in \mTb_*\}$, we have 
  \begin{equation*}
    \lim_{n \rightarrow \infty} \frac{1}{n} \sum_{v=1}^n \one{\thetan_v = \theta, \Dn(v) = D} = \prwrt{\mu}{\theta([T,o]) = \theta, D([T,o]) = D}.
  \end{equation*}
\end{lem}


\begin{lem}
  \label{lem:opt-ee16-general-bound-for-ln-ttp}
  With the assumptions in the first part of Theorem~\ref{thm:optimality-complexity-main},
  and with $h \geq 1$ and $\delta \geq 1$ fixed, we have 
  \begin{equation}
  \label{eq:sum-log-eln-ttp-bound-1}
  \sum_{(t,\pp{t}) \in \mEn_{\leq}} \log \eln_{t,\pp{t}} \leq \frac{1}{2} \sum_{c \in \mCdeltah} \left( \Sn_c \log n + \Sn_c \log \frac{\Sn_c}{n} - \Sn_c - 2 \sum_{v=1}^n \log \Dn_c(v)! \right) + O( \log n).
\end{equation}
\end{lem}





\subsection{Proof of Proposition~\ref{prop:opt-h>1}: $h>1$}
\label{sec:opt-h>1-proof}

In this section, we 
complete the discussion of 
\color{black}
the case $h>1$, i.e.\ 
we prove
\color{black}
Proposition~\ref{prop:opt-h>1}.
In order to do so, we first give the proof of
Lemma~\ref{lem:opt-h>1-simplifications}.


\begin{proof}[Proof of Lemma~\ref{lem:opt-h>1-simplifications}]
Let $[T,o]  \in \mTb_*$. If $\deg_T(o) = 0$ then both sides of~\eqref{eq:h>1-deg-star-simplification}
equal $0$. If $\deg_T(o) > 0$ we can write
\color{black}
\begin{align*}
  \deg_\star^{(h, \delta)}([T,o]) &= |\{ v\sim_To : T[o,v]_{h-1} \notin \mFdeltah \text{ or } T[v,o]_{h-1} \notin \mFdeltah \text{ or } \deg_T(o) > \delta \text{ or } \deg_T(v) > \delta\}| \\
                                  &= \deg_T(o) \\
  &\quad - |\{ v\sim_To : T[o,v]_{h-1} \in \mFdeltah \text{ and } T[v,o]_{h-1} \in \mFdeltah \text{ and } \deg_T(o) \leq \delta \text{ and } \deg_T(v) \leq \delta\}|.
\end{align*}
Observe that  following the definition of
$\mFdeltah$, for $v \sim_T o$, $T[o,v]_{h-1} \in \mFdeltah$ implies that the degree at the root in
the subgraph component of $T[o,v]_{h-1}$ is strictly less than $\delta$. But
since $h- 1 \geq 1$, the degree at the root in the subgraph component of
$T[o,v]_{h-1}$ is precisely $\deg_T(v) - 1$. Thereby, $T[o,v]_{h-1} \in
\mFdeltah$ implies that $\deg_T(v) \leq \delta$. Similarly, $T[v,o]_{h-1} \in
\mFdeltah$ implies that $\deg_T(o) \leq \delta$. 
 Thereby, we may relax the conditions $\det_T(o) \leq \delta$ and $\deg_T(v)
 \leq \delta$ in the above and simply write
 \begin{align*}
   \deg_\star^{(h, \delta)}([T,o]) &= \deg_T(o) - |\{ v\sim_To : T[o,v]_{h-1} \in \mFdeltah \text{ and } T[v,o]_{h-1} \in \mFdeltah \}| \\
   &= |\{ v\sim_To : T[o,v]_{h-1} \notin \mFdeltah \text{ or } T[v,o]_{h-1} \notin \mFdeltah \}|,
 \end{align*}
 which is precisely~\eqref{eq:h>1-deg-star-simplification}. 
 
 Next, 
 \color{black}
 note that both
 $T[o,v]_{h-1}$ and $T[v,o]_{h-1}$ are functions of $[T,o]_h$. This implies~\eqref{eq:deg-star-function-To-h}.

 Using the same argument, for $[T,o] \in \mTb_*$, $v \sim_T o$, and $t,t' \in
 \mFdeltah$, if we have $T[v,o]_{h-1} = t$ and $T[o,v]_{h-1} = t'$, we automatically
 have $\deg_T(o) \leq \delta$ and $\deg_T(v) \leq \delta$.
 Thereby, we have
 \begin{align*}
   D_{t,\pp{t}}([T,o]) &:= |\{v \sim_T o: \deg_T(o) \leq \delta, \deg_T(v) \leq \delta, T[v,o]_{h-1} = t, T[o,v]_{h-1} = t'\}| \\
                       &= |\{v \sim_T o:  T[v,o]_{h-1} = t, T[o,v]_{h-1} = t'\}| \\
   &= E_h(t,\pp{t})([T,o]),
 \end{align*}
which is precisely~\eqref{eq:D-T-o-is-Eh-for-h>1}. Finally, note that
$E_h(t,t')([T,o])$ is a function of $[T,o]_h$.
Consequently, $D_{t,\pp{t}}([T,o]_h) = E_h(t,\pp{t})([T,o]_h) =
E_h(t,\pp{t})([T,o]) = D_{t,\pp{t}}([T,o])$ which 
implies~\eqref{eq:h>1-D-is-funct-of-T-o-h}  and completes the proof.
\end{proof}


Now, we are ready to prove Proposition~\ref{prop:opt-h>1}.

\begin{proof}[Proof of Proposition~\ref{prop:opt-h>1}]
 Recall from~\eqref{eq:D-T-o-is-Eh-for-h>1} in
 Lemma~\ref{lem:opt-h>1-simplifications} that when
$h > 1$, for all $[T,o] \in \mTb_* $ and $t, t' \in \mFdeltah$ we have
$D_{t,t'}([T,o]) = E_h(t,t')([T,o])$. But $E_h(t,t')([T,o]) =
E_h(t,t')([T,o]_h)$. Therefore, if $P := \mu_h$ denotes the law of $[T,o]_h$
when $[T,o] \sim \mu$, using~\eqref{eq:lim-sum-Dn-c-E-D-To} in Lemma~\ref{lem:opt-cc14-general-bounds-for-Dn}, we have
\begin{equation}
  \label{eq:opt-lim-Dn-c-ep-c}
  \lim_{n \rightarrow \infty} \frac{1}{n} \sum_{v=1}^n \Dn_c(v) = \lim_{n \rightarrow \infty} \frac{1}{n} \Sn_c = \evwrt{\mu}{E_h(t,\pp{t})([T,o]_h)} = \evwrt{P}{E_h(t,t')([T,o])} = e_P(c) \qquad 
  \forall c = (t,\pp{t}) \in \mCdeltah,
  \color{black}
\end{equation}
where $e_P(c)$ is defined in~\eqref{eq:esubPdefinition}.
\color{black}
Recall that $\Sn_c = \sum_{v=1}^n \Dn_c(v)$ for $c \in \mCdeltah = \mFdeltah
\times \mFdeltah$.
Similarly, using~\eqref{eq:lim-sum-Dnc-factorial-Emu-log-Dc-factorial} in Lemma~\ref{lem:opt-cc14-general-bounds-for-Dn}, we get 
\begin{equation}
  \label{eq:lim-Dnc-factorial-Ep-log-fact}
  \lim_{n \rightarrow \infty} \frac{1}{n} \sum_{v=1}^n \log \Dn_c(v)! 
  = \evwrt{P}{\log E_h(c)([T,o])!} 
  \color{black}
  \qquad \forall c \in \mCdeltah.
\end{equation}

We proceed by bounding the last two terms in the
bound~\eqref{eq:opt-nat-bound-1} from Proposition~\ref{prop:opt-aa12-nat-count}.  
Recalling $\mA_\delta = \{(\theta([T,o]), D([T,o])): [T,o] \in
\mTb_*\}$, for $(\theta, D) \in \mA_{\delta}$, we have
\begin{equation}
  \label{eq:lim-average-theta-D-ev-theta-D}
\begin{aligned}
  \lim_{n \rightarrow \infty} \frac{1}{n} \sum_{v=1}^n \one{\thetan_v = \theta, \Dn(v) = D} &\stackrel{(a)}{=}  \prwrt{\mu}{\theta([T,o]) = \theta, D([T,o]) = D} \\
                                                                                            &\stackrel{(b)}{=} \prwrt{\mu}{\theta([T,o]_h) = \theta, D([T,o]_h) = D} \\
                                                                                            &= \prwrt{P}{\theta([T,o]) = \theta, D([T,o]) = D},
                                                                                          \end{aligned}
\end{equation}
where $(a)$ employs Lemma~\ref{lem:theta-D-fraction-limit}, and in $(b)$ we have
used~\eqref{eq:h>1-D-is-funct-of-T-o-h} in Lemma~\ref{lem:opt-h>1-simplifications}.
Note that since $\mA_\delta$ is finite, with $R$ being the law of
$(\theta([T,o]), D([T,o])) = (\theta([T,o]_h), D([T,o]_h))$ when $[T,o] \sim P =
\mu_h$, we conclude that using~\eqref{eq:lim-average-theta-D-ev-theta-D}, 
\begin{equation}
  \label{eq:opt-lim-nfactorial-HP}
  \lim_{n \rightarrow \infty} \frac{1}{n} \log \left( \frac{n!}{\prod_{(\theta, D) \in \mA_\delta} |\{ v \in [n]: (\thetan_v, \Dn(v)) = (\theta, D)\}|!}  \right) = H(R) \leq H(P),
\end{equation}
where the last inequality employs the data processing inequality.

Next we move to bounding $\sum_{(t,\pp{t}) \in \mEn_\leq} \log \eln_{t,\pp{t}}$.
From Lemma~\ref{lem:sum-Snc-mn-mn-star}, we have $\sum_{c \in \mCdeltah} \Sn_c/2 = \mn - \mn_\star$. Using
this together with~\eqref{eq:opt-lim-Dn-c-ep-c}
and~\eqref{eq:lim-Dnc-factorial-Ep-log-fact} and substituting in the bound
\eqref{eq:sum-log-eln-ttp-bound-1} of Lemma~\ref{lem:opt-ee16-general-bound-for-ln-ttp}, we get
\begin{equation}
  \label{eq:limsup-eln}
  \sum_{(t,\pp{t}) \in \mEn_{\leq}} \log \eln_{t,\pp{t}} \leq (\mn - \mn_\star) \log n + \frac{n}{2} 
  \left( \sum_{c \in \mCdeltah} e_P(c) \log e_P(c) -  e_P(c) 
  - 2\evwrt{P}{\log E_h(c)([T,o])!} \right) 
  \color{black}
  + o(n).
\end{equation}
Note that $\mCdeltah$ is a finite set, hence we are allowed to employ \eqref{eq:opt-lim-Dn-c-ep-c}
and~\eqref{eq:lim-Dnc-factorial-Ep-log-fact} in the above calculation.
Using the bounds \eqref{eq:Vns-n-eta1-delta} and
\eqref{eq:opt-limsup-n-choose-Vns-eta1} in Lemma~\ref{lem:opt-bb13-Vns},
\eqref{eq:opt-lim-mn-star} in Lemma~\ref{lem:opt-dd15-mnstar-bound},
\eqref{eq:opt-lim-nfactorial-HP}, and \eqref{eq:limsup-eln} back
into the bound~\eqref{eq:opt-nat-bound-1} in Proposition~\ref{prop:opt-aa12-nat-count}, we get
\begin{equation}
  \label{eq:opt-bound-2}
  \begin{aligned}
    \limsup_{n \rightarrow \infty} \frac{\nat(\fn_{h, \delta}(\Gn)) - \mn \log n}{n} &\leq H(P) + \frac{1}{2} \left( \sum_{c \in \mCdeltah} e_P(c) \log e_P(c) - e_P(c) \right) \\
    &\quad - \sum_{c \in \mCdeltah} 
    \evwrt{P}{\log E_h(c)([T,o])!} 
    \color{black}
    \\
    &\quad + \eta_1(\mu; h, \delta) (1 + |\edgemark|^2 \log 2 - \log \eta_1(\mu; h, \delta)) + 2\eta_2(\mu; h, \delta) \log 2.
    \end{aligned}
\end{equation}
Now we address the two summations over $\mCdeltah$ in the right hand side one by one.

 Note that with $\mC:= (\edgemark \times \mTb_*^{h-1})
\times (\edgemark \times \mTb_*^{h-1})$ and $d := \deg(\mu) = \sum_{c \in \mC} e_P(c)$, we have
\begin{equation}
\label{eq:opt-sum-ep-log-ep-abs-summable}
    \begin{aligned}
      \sum_{c \in \mC} |e_P(c) \log e_P(c) | &\leq \sum_{c \in \mC} |e_P(c) \log e_P(c) - e_P(c) \log d| + |e_P(c)\log d| \\
      &= \sum_{c \in \mC} e_P(c) \log \frac{d}{e_P(c)} + \sum_{c \in \mC} e_P(c) |\log d| \\
      &= d H(\pi_P) + d |\log d|.
    \end{aligned}
\end{equation}
On the other hand, since we have assumed that $\evwrt{\mu}{\deg_T(o) \log
  \deg_T(o)} < \infty$, the second part of Theorem~\ref{thm:Jh} implies that the
sequence $(J_k(\mu_k): k \geq 1)$ is nonincreasing
and converges to $\bch(\mu)$ as $k \rightarrow \infty$. Thereby, since we have
assumed that $\bch(\mu) > -\infty$, we have $J_h(\mu_h) > -\infty$.  In
particular, recalling the definition of $J_h(\mu_h)$ in~\eqref{eq:Jh-def}, this implies that $H(\pi_P) < \infty$. Therefore,
\eqref{eq:opt-sum-ep-log-ep-abs-summable} above implies that
\begin{equation}
  \label{eq:opt-h>1-ep-log-ep-abs-summable}
  \sum_{c \in \mC}|e_P(c) \log e_P(c) | < \infty.
\end{equation}
Additionally,
\begin{equation}
  \label{eq:opt-h>1-ep-abs-summable}
 \sum_{c \in \mC} |e_P(c)| = \sum_{c \in \mC} e_P(c) = \deg(\mu) < \infty.
\end{equation}
Consequently, we may write
\begin{equation}
  \label{eq:sum-ep-c-mC-mCdelta}
  \sum_{c \in \mCdeltah} \left( e_P(c) \log e_P(c) - e_P(c) \right) = \sum_{c \in \mC}  \left( e_P(c) \log e_P(c) - e_P(c)  \right) - \sum_{c \in \mC \setminus \mCdeltah} e_P(c) \log e_P(c) + \sum_{c \in \mC \setminus \mCdeltah} e_P(c).
\end{equation}
Moreover,
\begin{equation}
  \label{eq:sum-mC-mCdelta-ep-log-ep}
\begin{aligned}
  \sum_{c \in \mC \setminus \mCdeltah} e_P(c) \log e_P(c) &= \sum_{c \in \mC \setminus \mCdeltah} \left( \sum_{s \in \mTb_*^h} P(s) E_h(c)(s) \right) \log \left( \sum_{s \in \mTb_*^h} P(s) E_h(c)(s)\right) \\
                                                         &\stackrel{(a)}{\geq} \sum_{c \in \mC \setminus \mCdeltah}  \sum_{s \in \mTb_*^h} P(s) E_h(c)(s) \log\left(  P(s) E_h(c)(s) \right) \\
                                                         &\stackrel{(b)}{\geq}  \sum_{c \in \mC \setminus \mCdeltah}  \sum_{s \in \mTb_*^h} P(s) E_h(c)(s) \log P(s) \\
                                                         &\stackrel{(c)}{=} \sum_{s \in \mTb_*^h} P(s) \deg_\star^{(h, \delta)}(s) \log P(s) \\
                                                         &= - \evwrt{\mu}{\deg_\star^{(h, \delta)}([T,o]_h) \log \frac{1}{\mu_h([T,o]_h)}} \\
                                                         &\stackrel{(d)}{=} - \evwrt{\mu}{\deg_\star^{(h, \delta)}([T,o]) \log\frac{1}{\mu_h([T,o]_h)}} = -2\eta_3(\mu; h, \delta),
\end{aligned}
\end{equation}
where in $(a)$, we have used the fact that $\log(.)$ is increasing, in
$(b)$ we have again used the fact that $\log$ is increasing as well as that
$E_h(c)(s)$ is an integer, and in $(c)$ we have employed
\eqref{eq:h>1-deg-star-simplification} in
Lemma~\ref{lem:opt-h>1-simplifications}, and in $(d)$ we have
used~\eqref{eq:deg-star-function-To-h} in Lemma~\ref{lem:opt-h>1-simplifications}.
 On the other hand, we have 
 \begin{equation}
   \label{eq:sum-mC-mCdelta-eP-c}
\begin{aligned}
  \sum_{c \in \mC\setminus \mCdeltah} e_P(c) &= \sum_{c \in \mC \setminus \mCdeltah} \sum_{[T,o] \in \mTb_*^h} P([T,o]) E_h(c)([T,o]) \\
                                            &= \sum_{[T,o] \in \mTb_*^h} P([T,o]) \sum_{c \in \mC \setminus \mCdeltah} E_h(c)([T,o]) \\
                                            &\stackrel{(a)}{=}  \sum_{[T,o] \in \mTb_*^h} P([T,o]) \deg_\star^{(h, \delta)}([T,o]) \\
                                            &= \evwrt{\mu}{\deg_\star^{(h, \delta)}([T,o]_h)} \\
                                            &\stackrel{(b)}{=} \evwrt{\mu}{\deg_\star^{(h, \delta)}([T,o])} \\
                                            &= 2 \eta_2(\mu; h, \delta),
\end{aligned}
\end{equation}
where in $(a)$,   we have used \eqref{eq:h>1-deg-star-simplification} in
Lemma~\ref{lem:opt-h>1-simplifications}
and in $(b)$ we have
used~\eqref{eq:deg-star-function-To-h} in Lemma~\ref{lem:opt-h>1-simplifications}.
Additionally, \eqref{eq:opt-h>1-ep-log-ep-abs-summable}
and \eqref{eq:opt-h>1-ep-abs-summable} allow us to write
\begin{equation}
  \label{eq:sum-C-ep-log-ep-d}
\begin{aligned}
  \sum_{c \in \mC} \left( e_P(c) \log e_P(c) - e_P(c)  \right) &= \sum_{c \in \mC} e_P(c) \log \frac{e_P(c)}{d} + \sum_{c \in \mC} e_P(c) \log d - \sum_{c \in \mC} e_P(c) \\
                                                               &= d \sum_{c \in \mC} \frac{e_P(c)}{d} \log \frac{e_P(c)}{d} + d \log d - d \\
                                                               &= d \sum_{c \in \mC} \pi_P(c) \log \pi_P(c) - 2 s(d) \\
                                                               &= - d H(\pi_P) - 2 s(d).
\end{aligned}
\end{equation}
Using~\eqref{eq:sum-mC-mCdelta-ep-log-ep}, \eqref{eq:sum-mC-mCdelta-eP-c}, and
\eqref{eq:sum-C-ep-log-ep-d} back in~\eqref{eq:sum-ep-c-mC-mCdelta}, we get
\begin{equation}
  \label{eq:sum-mCdelta-eP-log-eP-bound}
  \frac{1}{2} \left(  \sum_{c \in \mCdeltah} e_P(c) \log e_P(c) - e_P(c)  \right) \leq -s(d) - \frac{d}{2} H(\pi_P) + \eta_2(\mu;h, \delta) + \eta_3(\mu; h, \delta).
\end{equation}

We now turn to the second summation over $\mCdeltah$ in the right hand side of~\eqref{eq:opt-bound-2}.
Note that for all $c \in \mC$ and $[T,o] \in \mTb_*$, $E_h(c)([T,o])$ is an
integer, whereby $\log E_h(c)([T,o])! \geq 0$. This allows us to write 
\begin{equation}
  \label{eq:opt-log-ep-factorial-C-Cdelta}
    - \sum_{c \in \mCdeltah} \evwrt{P}{\log E_h(c)([T,o])!} = - \sum_{c \in \mC} \evwrt{P}{\log E_h(c)([T,o])!} + \sum_{c \in \mC \setminus \mCdeltah} \evwrt{P}{\log E_h(c)([T,o])!}.
    \color{black}
\end{equation}
Note that
\begin{align*}
\sum_{c \in \mC \setminus \mCdeltah} \evwrt{P}{\log E_h(c)([T,o])!} &\leq   \sum_{c \in \mC \setminus \mCdeltah} \evwrt{P}{E_h(c)([T,o]) \log E_h(c)([T,o])} \\
                                                              &= \evwrt{P}{  \sum_{c \in \mC \setminus \mCdeltah} E_h(c)([T,o]) \log E_h(c)([T,o])} \\
                                                              &\leq \evwrt{P}{\left(   \sum_{c \in \mC \setminus \mCdeltah} E_h(c)([T,o]) \right) \log \left(   \sum_{c \in \mC \setminus \mCdeltah} E_h(c)([T,o]) \right)} \\
                                                              &\stackrel{(a)}{=} \evwrt{P}{\deg_\star^{(h, \delta)}([T,o]) \log \deg_\star^{(h, \delta)}([T,o])} \\
                                                              &\leq \evwrt{P}{\deg_\star^{(h, \delta)}([T,o]) \log \deg_T(o)} \\
                                                               & = \evwrt{\mu}{\deg_\star^{(h, \delta)}([T,o]_h) \log \deg_T(o)} \\
                                                               &\stackrel{(b)}{=} \evwrt{\mu}{\deg_\star^{(h, \delta)}([T,o]) \log \deg_T(o)} \\
  &= \eta_4(\mu; h, \delta),
\end{align*}
\color{black}
where in $(a)$, we have used~\eqref{eq:h>1-deg-star-simplification} in
Lemma~\ref{lem:opt-h>1-simplifications}, and in $(b)$, we have
used~\eqref{eq:deg-star-function-To-h} in Lemma~\ref{lem:opt-h>1-simplifications}.
Putting this back into~\eqref{eq:opt-log-ep-factorial-C-Cdelta}, we get
\begin{equation}
  \label{eq:opt-sum-log-Ep-factorial-eta4-bound}
  - \sum_{c \in \mCdeltah} \evwrt{P}{\log E_h(c)([T,o])!} \leq -   \sum_{c \in \mC} \evwrt{P}{\log E_h(c)([T,o])!} + \eta_4(\mu; h, \delta).
\end{equation}
\color{black}
Substituting~\eqref{eq:sum-mCdelta-eP-log-eP-bound} and
\eqref{eq:opt-sum-log-Ep-factorial-eta4-bound} back into~\eqref{eq:opt-bound-2},
we realize that
\begin{align*}
  \limsup_{n \rightarrow \infty} \frac{\nat(\fn_{h, \delta}(\Gn)) - \mn \log n}{n} &\leq 
  -s(d) + H(P) - \frac{d}{2} H(\pi_P) - \sum_{c \in \mC} \evwrt{P}{\log E_h(c)([T,o])!}   \\
  \color{black}
                                                                                   &\qquad + \eta_2(\mu; h, \delta) + \eta_3(\mu; h , \delta) + \eta_4(\mu; h, \delta)  \\
  &\qquad + \eta_1(\mu; h , \delta)(1 + |\edgemark|^2 \log 2 - \log \eta_1(\mu; h , \delta))  + 2\eta_2(\mu; h, \delta) \log 2 \\
  &=  J_h(\mu_h) + \eta_1(\mu; h, \delta) (1 +|\edgemark|^2 \log 2 - \log \eta_1(\mu; h, \delta)) \\
  &\qquad + \eta_2(\mu; h, \delta)(1 + \log 4) + \eta_3(\mu; h, \delta) + \eta_4(\mu; h, \delta) \\
  &\leq J_h(\mu_h) + \eta_0(\mu; h, \delta),
\end{align*}
where in the last step, we have used the facts $\eta_1(\mu; h , \delta) =
\prwrt{\mu}{\deg_\star^{(h, \delta)}([T,o])> 0} \in [0,1]$, and $\eta_4(\mu; h ,
\delta) \geq 0$. 
This completes the proof.
\end{proof}




\subsection{Proof of Proposition~\ref{prop:opt-h=1}: $h=1$}
\label{sec:opt-h=1-proof}

In this section, 
we complete the discussion of the case $h=1$, i.e.
\color{black}
we prove Proposition~\ref{prop:opt-h=1}.
In order to do this, we first prove Lemma~\ref{lem:opt-h-1-properties}.
\color{black}

\begin{proof}[Proof of Lemma~\ref{lem:opt-h-1-properties}]
  Recall that by definition, $\mFdeltah$ consists of $t \in \edgemark
  \times \mTb_*^{h-1}$ such that the degree at the root in  the subgraph
  component $t[s]$ is strictly
  less than $\delta$, while the degree of every other vertex in $t[s]$ is no more
  than $\delta$. But since $h-1 = 0$, $t[s] \in \mTb_*^0$ is an isolated root,
  with the degree of the root being zero. Since $\delta \geq 1$, and there is no
  vertex other than the root in $t[s]$, 
  the above
  degree condition automatically holds for all $t \in \edgemark \times
  \mTb_*^{h-1}$, hence $\mFdeltah = \edgemark \times \mTb_*^{h-1}$. In particular,
  since $\mTb_*^0 \equiv \vermark$, we have $|\mFdeltah| = |\edgemark| \times
  |\vermark|$. To see part~\ref{it:loh1-2}, note that  $T[o,v]_{h-1}$ and $T[v,o]_{h-1}$ are
  members in $\edgemark \times \mTb_*^{h-1} $, which is the same as $\mFdeltah$
  from part~\ref{it:loh1-1}. For part~\ref{it:loh1-3}, note that from
  part~\ref{it:loh1-2}, for $[T,o] \in \mTb_*$, we have
  \begin{align*}
    \deg_\star^{(h, \delta)}([T,o]) &= |\{ v \sim_T o: \deg_T(o) > \delta \text{ or } \deg_T(v) > \delta \text{ or } T[o,v]_{h-1} \notin \mFdeltah \text{ or } T[v,o]_{h-1} \notin \mFdeltah \}| \\
    &= |\{v \sim_T o : \deg_T(o) >  \delta \text{ or } \deg_T(v) > \delta\}|.
  \end{align*}
  To see part~\ref{it:loh1-4}, note that from part~\ref{it:loh1-3}, if
  $\deg_\star([T,o]) = 0$, we have $\deg_T(o) \leq \delta$ and $\deg_T(v) \leq
  \delta$ for all $v\sim_T o$. Therefore, for $t, \pp{t} \in \mFdeltah$,
  recalling the definition of $D_{t,\pp{t}}([T,o])$ from~\eqref{eq:D-T-o-def},
  we have
  \begin{align*}
    D_{t,\pp{t}}([T,o])  &=|\{v \sim_T o: \deg_T(o) \leq \delta, \deg_T(v) \leq \delta, T[v,o]_{h-1} = t, T[o,v]_{h-1} = \pp{t}\}| \\
    &= |\{v \sim_T o: T[v,o]_{h-1} = t, T[o,v]_{h-1} = \pp{t}\}| \\
    &= E_1(t,\pp{t})([T,o]).
  \end{align*}
  Finally, to see part~\ref{it:loh1-5}, note that for $[T,o] \in \mTb_*$,
  $\deg_T(o)$ as well as $\deg_T(v)$, $T[o,v]_{h-1}$, and $T[v,o]_{h-1}$ for $v
  \sim_T o$ can be determined given $[T,o]_2$. Therefore, recalling the
  definition of $D([T,o])$ from~\eqref{eq:D-T-o-def}, we conclude that
  $D([T,o])$ is a function of $[T,o]_2$. Likewise, 
  since $\deg_T(o)$ and $\deg_T(v)$ for $v \sim_T o$ are functions of $[T,o]_2$,
  using part~\ref{it:loh1-3}, $\deg_\star^{(h, \delta)}([T,o])$ is also a
  function of $[T,o]_2$.
This completes the proof.
\end{proof}

We are now ready to discuss the proof of Proposition~\ref{prop:opt-h=1}.

\begin{proof}[Proof of Proposition~\ref{prop:opt-h=1}]
First, we find upper bounds for the last two terms in the
bound~\eqref{eq:opt-nat-bound-1} from Proposition~\ref{prop:opt-aa12-nat-count}.
From Lemma~\ref{lem:theta-D-fraction-limit}, for $(\theta, D) \in \mA_\delta$,  we have
\begin{equation*}
  \lim_{n \rightarrow \infty} \frac{1}{n} \sum_{v=1}^n \one{\thetan_v = \theta, \Dn(v) = D} = \prwrt{\mu}{\theta([T,o]), D([T,o])}.
\end{equation*}
Thereby, since $\mA_\delta$ is finite, we have
\begin{equation}
  \label{eq:opt-h-1-lim-nfactorial-HR}
  \lim_{n \rightarrow \infty} \frac{1}{n} \log \left( \frac{n!}{\prod_{(\theta, D) \in \mA_\delta} |\{ v \in [n]: (\thetan_v, \Dn(v)) = (\theta, D)\}|!}  \right) = H(R),
\end{equation}
with  $R$ being the distribution of $(\theta([T,o]), D([T,o]))$ when $[T,o] \sim \mu$.
Note that unlike the  case of $h>1$ as in
Lemma~\ref{lem:opt-h>1-simplifications}, when $h= 1$,  $D([T,o])$ is not necessarily a function of
$[T,o]_1$ (also see Figure~\ref{fig:h-1-simplifications-do-not-hold} for an
example).  Instead, as we saw in part~\ref{it:loh1-5} of
Lemma~\ref{lem:opt-h-1-properties}, it is a function of $[T,o]_2$. Therefore,
unlike Proposition~\ref{prop:opt-h>1} where we used the data
processing inequality to bound $H(R)$ by $H(\mu_1)$,  here we use a different
technique to bound $H(R)$. With $[T,o] \sim \mu$,
define the random variable $X$ as follows
\begin{equation}
  \label{eq:opt-h-1-X-def}
  X =
  \begin{cases}
    1 & \deg_\star^{(h,\delta)}([T,o]) = 0 \\
    2 & \deg_\star^{(h,\delta)}([T,o]) > 0 \\
  \end{cases}
\end{equation}
With this, we may write
\begin{equation}
  \label{eq:opt-h-1-H-R-with-X}
  \begin{aligned}
    H(R) &= H(\theta([T,o]), D([T,o])) \leq H(\theta([T,o]), D([T,o]), X) \\
    &= H(X) + H(\theta([T,o]), D([T,o]) | X) \\
    &= H(X) + H(\theta([T,o]), D([T,o]) | X = 1) \pr{X=1} \\
    &\qquad + H(\theta([T,o]), D([T,o]) | X = 2) \pr{X=2} \\
  \end{aligned}
\end{equation}
Now, we bound each term separately. Note that $\pr{X = 2} = \eta_1(\mu; h,
\delta)$. Hence
\begin{equation}
  \label{eq:opt-h-1-HX-bound}
  H(X) = - \eta_1(\mu; h, \delta) \log \eta_1(\mu; h, \delta) - (1 - \eta_1(\mu; h, \delta)) \log (1 - \eta_1(\mu; h, \delta)) \leq \eta_1(\mu; h, \delta)(1 - \log \eta_1(\mu; h, \delta)),
\end{equation}
where the last inequality uses $-(1-x) \log (1-x) \leq x$ which holds for $0
\leq x \leq 1$.
Moreover, conditioned on $X=1$, since $\deg_\star^{(h, \delta)}([T,o]) = 0$,
using part~\ref{it:loh1-4} of Lemma~\ref{lem:opt-h-1-properties}, for all $t,
\pp{t} \in \mFdeltah$, we have $D_{t,\pp{t}}([T,o]) = E_1(t,\pp{t})([T,o])$,
which  is a
function of $[T,o]_1$.
Consequently, with $P := \mu_1$ being the law of
$[T,o]_1$ when $[T,o] \sim \mu$, using the data processing inequality, we have 
\begin{equation}
\label{eq:opt-h-1-ent-X=1-bound}
  H(\theta([T,o]), D([T,o]) | X = 1) \pr{X=1} \leq H(P| X = 1) \pr{X=1} \leq H(P|X) \leq H(P).
\end{equation}
Now we focus on the case $X = 2$. From part~\ref{it:loh1-1} of
Lemma~\ref{lem:opt-h-1-properties}, we have  $|\mFdeltah| =
|\edgemark| \times |\vermark|$. Therefore, using~\eqref{eq:D-T-o-delta},
$(D_{t,t'}([T,o]): t, \tp \in \mF^{(\delta,h)})$ has at most $(\delta +
1)^{(|\vermark| \times |\edgemark|)^2}$ many possible values. Thereby, we have
\begin{equation*}
  H(\theta([T,o]), D([T,o])| X = 2) \leq \log (|\vermark| \times (\delta + 1)^{(|\vermark| \times |\edgemark|)^2}) = \log |\vermark| + (|\edgemark| \times |\vermark|)^2 \log (1+\delta).
\end{equation*}
Furthermore, since $\pr{X = 2} = \prwrt{\mu}{\deg_\star^{(h, \delta)}([T,o]) > 0} =
\eta_1(\mu; h, \delta)$, we have
\begin{equation}
\label{eq:opt-h-1-ent-X=2-bound-1}
  H(\theta([T,o]), D([T,o])| X = 2) \pr{X = 2} \leq \eta_1(\mu; h, \delta) \left( \log |\vermark| + (|\edgemark| \times |\vermark|)^2 \log (1+\delta) \right).
\end{equation}
Using part~\ref{it:loh1-3} of Lemma~\ref{lem:opt-h-1-properties}, we have 
\begin{equation}
  \label{eq:h-1-deg-star-equal-and-bound}
\begin{aligned}
  \deg_\star^{(h, \delta)}([T,o]) &= |\{ v \sim_T o : \deg_T(v) > \delta \text{ or } \deg_T(o) > \delta\}| \\
  &\leq \deg_T(o) \one{\deg_T(o) > \delta} + \sum_{v \sim_T o} \one{\deg_T(v) > \delta}.
\end{aligned}
\end{equation}
Using this, we can write
\begin{align*}
  \eta_1(\mu; h, \delta) \log(1+\delta) &= \log (1+\delta) \pr{\deg_\star^{(h, \delta)}([T,o])>  0} \\
                                        &\leq \log(1+\delta) \evwrt{\mu}{\deg_\star^{(h, \delta)}([T,o])} \\
                                        &\leq \log(1+\delta) \left( \evwrt{\mu}{\deg_T(o) \one{\deg_T(o)>\delta}}  + \evwrt{\mu}{\sum_{v \sim_T o} \one{\deg_T(v) > \delta}} \right) \\
                                        &\stackrel{(a)}{=} \log(1+\delta) \left( \evwrt{\mu}{\deg_T(o) \one{\deg_T(o)>\delta}}  + \evwrt{\mu}{\sum_{v \sim_T o} \one{\deg_T(o) > \delta}} \right) \\
                                        &= 2\log(1+\delta) \evwrt{\mu}{\deg_T(o) \one{\deg_T(o) > \delta}} \\
                                        &\leq 2 \evwrt{\mu}{\deg_T(o) \log \deg_T(o) \one{\deg_T(o) > \delta}} \\
                                        &\stackrel{(b)}{=} 2 \evwrt{\mu}{\deg_\star^{(h, \delta)}([T,o]) \log \deg_T(o) \one{\deg_T(o) > \delta}}  \\
                                        &\leq 2 \evwrt{\mu}{\deg_\star^{(h, \delta)}([T,o]) \log \deg_T(o)} \\
                                        &= 2 \eta_4(\mu; h, \delta),
\end{align*}
where $(a)$ uses the fact that $\mu$ is unimodular, and in $(b)$, we have
employed the fact that when $\deg_T(o) > \delta$,
from part~\ref{it:loh1-3} in Lemma~\ref{lem:opt-h-1-properties},
we have $\deg_T(o) = \deg_\star^{(h,\delta)}([T,o])$.
Using this in~\eqref{eq:opt-h-1-ent-X=2-bound-1}, we get
\begin{equation}
  \label{eq:opt-h-1-ent-X=2-bound-2}
  H(\theta([T,o]), D([T,o])| X = 2) \pr{X = 2} \leq \eta_1(\mu; h, \delta) \log |\vermark| + 2(|\vermark| \times |\edgemark|)^2 \eta_4(\mu; h, \delta).
\end{equation}
By substituting~\eqref{eq:opt-h-1-HX-bound}, \eqref{eq:opt-h-1-ent-X=1-bound},
and~\eqref{eq:opt-h-1-ent-X=2-bound-2} back into \eqref{eq:opt-h-1-H-R-with-X},
and comparing with~\eqref{eq:opt-h-1-lim-nfactorial-HR},
we get 
\begin{equation}
  \label{eq:opt-h-1-HR-upper-bound}
\begin{aligned}
  &\lim_{n \rightarrow \infty} \frac{1}{n} \log \left( \frac{n!}{\prod_{(\theta, D) \in \mA_\delta} |\{ v \in [n]: (\thetan_v, \Dn(v)) = (\theta, D)\}|!}  \right) \leq H(P) \\
  &\qquad \qquad + \eta_1(\mu; h, \delta)\left(1 + \log |\vermark| -  \log \eta_1(\mu; h, \delta) \right) + 2(|\edgemark| \times |\vermark|)^2 \eta_4(\mu; h, \delta).
  \end{aligned}
\end{equation}

Now we turn to bounding $\sum_{(t,\pp{t}) \in \mEn_{\leq}} \log
\eln_{t,\pp{t}}$. From Lemma~\ref{lem:sum-Snc-mn-mn-star}, we have  $\sum_{c \in \mCdeltah} \Sn_c / 2 =
\mn - \mn_\star$.
Therefore, simplifying the bound of
Lemma~\ref{lem:opt-ee16-general-bound-for-ln-ttp} using
Lemma~\ref{lem:opt-cc14-general-bounds-for-Dn}, we get
\begin{equation}
  \label{eq:opt-h-1-sum-log-elln-tt'-bound-1}
  \begin{aligned}
    \sum_{(t,t') \in \mEn_{\leq}} \log \eln_{t,t'} &\leq (\mn - \mn_\star) \log n + \frac{n}{2} \Bigg ( \sum_{c \in \mCdeltah} \evwrt{\mu}{D_c([T,o])} \log \evwrt{\mu}{D_c([T,o])} \\
    &\qquad - \evwrt{\mu}{D_c([T,o])} - 2 \evwrt{\mu}{\log D_c([T,o])!} \Bigg ) + o(n).
\end{aligned}
\end{equation}
Note that $\mCdeltah$ is a finite set, hence we are allowed to employ
Lemma~\ref{lem:opt-cc14-general-bounds-for-Dn} in the above calculation.
Using~\eqref{eq:Vns-n-eta1-delta} and 
\eqref{eq:opt-limsup-n-choose-Vns-eta1} in Lemma~\ref{lem:opt-bb13-Vns},
\eqref{eq:opt-lim-mn-star} in Lemma~\ref{lem:opt-dd15-mnstar-bound},
\eqref{eq:opt-h-1-HR-upper-bound},
and~\eqref{eq:opt-h-1-sum-log-elln-tt'-bound-1} in the bound~\eqref{eq:opt-nat-bound-1} in
Proposition~\ref{prop:opt-aa12-nat-count},  we get
\begin{equation}
  \label{eq:opt-h-1-bound-2}
\begin{aligned}
  \limsup_{n \rightarrow \infty} \frac{\nat(\fn_{h, \delta}(\Gn)) - \mn \log n}{n} &\leq H(P) + \frac{1}{2} \left( \sum_{c \in \mCdeltah} \evwrt{\mu}{D_c([T,o])} \log \evwrt{\mu}{D_c([T,o])} - \evwrt{\mu}{D_c([T,o])} \right) \\
  &\quad -  \sum_{c \in \mCdeltah} \evwrt{\mu}{\log D_c([T,o])!} \\
  &\quad + \eta_1(\mu; h, \delta) \left ( |\edgemark|^2 \log 2 + \log |\vermark|  + 2 - 2 \log \eta_1(\mu; h, \delta) \right )\\
  &\qquad \qquad +\eta_2(\mu; h, \delta ) \log 4 +  2(|\edgemark|\times|\vermark|)^2\eta_4(\mu; h, \delta).
\end{aligned}
\end{equation}
Now, we focus on finding upper bounds for the summations over $\mCdeltah$ in the
above expression. For $t,
t' \in \mFdeltah$ and $[T,o] \in \mTb_*$, define
\begin{equation*}
  \tD_{t,t'}([T,o]) := |\left\{ v \sim_T o: T[v,o]_0 = t \text{ and } T[o,v]_0 = t' \text{ and } ( \deg_T(o) > \delta \text{ or } \deg_T(v) > \delta )\right\}|.
\end{equation*}
Note that in the above, $\tD_{t,\pp{t}}([T,o])$ is defined for all $[T,o] \in
\mTb_*$, and $[T,o]$ is not restricted to be a member of $\mTb_*^1$.
For all $[T,o] \in \mTb_*$ and $t,t' \in \mFdeltah$, we may write
\begin{equation}
\label{eq:opt-h-1-E1-sum-D-tD}
\begin{aligned}
  E_1(t,t')([T,o]) &= |\{v \sim_T o: T[v,o]_0 = t, T[o,v]_0 = t'\} \\
  &= |\{v \sim_T o: T[v,o]_0 = t, T[o,v]_0 = t', \deg_T(o) \leq \delta, \deg_T(v) \leq \delta\} \\ 
&\quad + |\{v \sim_T o: T[v,o]_0 = t \text{ and } T[o,v]_0 = t' \text{ and } (\deg_T(o) > \delta \text{ or } \deg_T(v) >  \delta)\} \\
&= D_{t,t'}([T,o]) + \tD_{t,t'}([T,o]).
\end{aligned}
\end{equation}
In particular, using part~\ref{it:loh1-1} in Lemma~\ref{lem:opt-h-1-properties},
we have
\begin{equation}
\label{eq:h-1-deg-sum-D-tD}
  \deg_T(o) = \sum_{t, \pp{t} \in \edgemark \times \mTb_*^0} E_1(t,\pp{t})([T,o]) = \sum_{t, t' \in \mFdeltah} E_1(t,t')([T,o]) = \sum_{t,t' \in \mFdeltah} D_{t,t'}([T,o]) + \tD_{t,t'}([T,o]).
\end{equation}
Additionally, for $[T,o] \in \mTb_*$, using parts~\ref{it:loh1-1} and \ref{it:loh1-3} in
Lemma~\ref{lem:opt-h-1-properties}, we have 
\begin{equation}
  \label{eq:h-1-deg-star-sum-of-tD}
  \begin{aligned}
    \deg_\star^{(h, \delta)}([T,o]) &= |\{v\sim_T o:  \deg_T(o) > \delta \text{ or } \deg_T(v) > \delta\}|\\
    &= \sum_{t, \pp{t} \in \edgemark \times \mTb_*^0} |\{v \sim_T o: T[v,o]_0 =t \text{ and } T[o,v]_0 = \pp{t} \text{ and } (\deg_T(o) > \delta \text{ or } \deg_T(v) > \delta) \}| \\
    &= \sum_{t, \pp{t} \in \mFdeltah} |\{v \sim_T o: T[v,o]_0 =t \text{ and } T[o,v]_0 = \pp{t} \text{ and } (\deg_T(o) > \delta \text{ or } \deg_T(v) > \delta) \}| \\
    &= \sum_{t, \pp{t} \in \mFdeltah} \tD_{t,\pp{t}}([T,o]).
  \end{aligned}
\end{equation}
To simplify the notation, we may use $\evwrt{\mu}{D_c}$ and $\evwrt{\mu}{\tD_c}$
instead of $\evwrt{\mu}{D_c([T,o])}$ and $\evwrt{\mu}{\tD_c([T,o])}$
respectively. With these, we may write
\begin{equation}
  \label{eq:opt-h-1-sum-ED-log-ED-pm-tD}
  \begin{aligned}
    \sum_{c \in \mCdeltah } \evwrt{\mu}{D_c} \log \evwrt{\mu}{D_c} &= \sum_{c \in \mCdeltah} \evwrt{\mu}{D_c} \log \evwrt{\mu}{D_c} + \evwrt{\mu}{\tD_c} \log \evwrt{\mu}{\tD_c} \\
    &\quad - \sum_{c \in \mCdeltah} \evwrt{\mu}{\tD_c} \log \evwrt{\mu}{\tD_c}.
\end{aligned}
\end{equation}
We define $\tilde{\pi}$ to be a probability distribution on the set $\{0,1\}
\times (\edgemark \times \mTb_*^0) \times (\edgemark \times \mTb_*^0) = \{0,1\}
\times \mFdeltah \times \mFdeltah = \{0,1\} \times \mCdeltah$ (where we have
used part~\ref{it:loh1-1} of Lemma~\ref{lem:opt-h-1-properties} to write
$\mFdeltah = \edgemark \times \mTb_*^0$) such that
with $d := \deg(\mu)$ and $t, t' \in \mFdeltah$, we have 
\begin{equation*}
  \tilde{\pi}(0,t,t') := \frac{1}{d} \evwrt{\mu}{D_{t,t'}} \qquad   \tilde{\pi}(1,t,t') := \frac{1}{d} \evwrt{\mu}{\tD_{t,t'}}.
\end{equation*}
To verify that $\tilde{\pi}$ is indeed a probability distribution,
note that using~\eqref{eq:h-1-deg-sum-D-tD}, we have
\begin{equation*}
  d := \deg(\mu) = \evwrt{\mu}{\deg_T(o)} = \sum_{c \in \mCdeltah} \evwrt{\mu}{D_c} + \evwrt{\mu}{\tD_c},
\end{equation*}
which implies that $\sum_{t, \pp{t} \in \mFdeltah} \tilde{\pi}(0,t,\pp{t}) +
\tilde{\pi}(1,t, \pp{t}) = 1$.
Moreover, using this, we can write
\begin{equation}
  \label{eq:opt-h-1-sum-D-tD-H-tpi}
\begin{aligned}
  \sum_{c \in \mCdeltah} \evwrt{\mu}{D_c} \log \evwrt{\mu}{D_c} + \evwrt{\mu}{\tD_c} \log \evwrt{\mu}{\tD_c} &= d  \sum_{c \in \mCdeltah}  \Bigg( \frac{\evwrt{\mu}{D_c}}{d} \log \frac{\evwrt{\mu}{D_c}}{d}\\
  &\qquad \qquad + \frac{\evwrt{\mu}{\tD_c}}{d} \log \frac{\evwrt{\mu}{\tD_c}}{d} \Bigg) \\
  &\qquad \qquad + \sum_{c \in \mCdeltah} \left(\evwrt{\mu}{D_c} + \evwrt{\mu}{\tD_c}\right) \log d \\
  &= - d H(\tilde{\pi}) + d \log d.
\end{aligned}
\end{equation}
We know from part~\ref{it:loh1-5} of Lemma~\ref{lem:opt-h-1-properties} that for all $[T,o] \in \mTb_*$ and $t,\pp{t}
\in \mFdeltah$, $D_{t,\pp{t}}([T,o])$ is
a  function of $[T,o]_2$. It is easy to see that the same argument implies that $\tD_{t,\pp{t}}([T,o])$ is  also
a function of $[T,o]_2$.
 Therefore,
with $\tP := \mu_2$ being the law of $[T,o]_2$ when $[T,o] \sim \mu$, we may
write
\begin{align*}
  \sum_{c \in \mCdeltah} \evwrt{\mu}{\tD_c} \log \evwrt{\mu}{\tD_c} &= \sum_{c \in \mCdeltah} \left( \sum_{[T,o] \in \mTb_*^2} \tP([T,o]) \tD_c([T,o]) \right) \times \\
                                                                    &\quad  \log \left( \sum_{[T,o] \in \mTb_*^2} \tP([T,o]) \tD_c([T,o]) \right) \\
                                                                    &\geq \sum_{c \in \mCdeltah} \sum_{[T,o] \in \mTb_*^2} \tP([T,o]) \tD_c([T,o]) \log (\tP([T,o]) \tD_c([T,o])) \\
                                                                    &\geq \sum_{c \in \mCdeltah} \sum_{[T,o] \in \mTb_*^2} \tP([T,o]) \tD_c([T,o]) \log \tP([T,o]) \\
                                                                    &\stackrel{(a)}{=} \sum_{[T,o] \in \mTb_*^2} \tP([T,o])  \deg_\star^{(h, \delta)}([T,o]) \log \tP([T,o]) \\
                                                                    &= -\evwrt{\mu}{\deg_*^{(h,\delta)}([T,o]_2) \frac{1}{\mu_2([T,o])_2}} \\
                                                                    &\stackrel{(b)}{=} -\evwrt{\mu}{\deg_*^{(h,\delta)}([T,o]) \frac{1}{\mu_2([T,o])_2}} \\
                                                                    &= -2\eta_3(\mu; h, \delta),
\end{align*}
where $(a)$ employs~\eqref{eq:h-1-deg-star-sum-of-tD} and $(b)$ uses
part~\ref{it:loh1-5} in Lemma~\ref{lem:opt-h-1-properties}.
Substituting this together with~\eqref{eq:opt-h-1-sum-D-tD-H-tpi} back in
\eqref{eq:opt-h-1-sum-ED-log-ED-pm-tD}, we get
\begin{equation}
  \label{eq:sum-Dc-leq-H-tilde-P}
  \sum_{c \in \mCdeltah} \evwrt{\mu}{D_c} \log \evwrt{\mu}{D_c} \leq -d H(\tilde{\pi}) + d \log d + 2 \eta_3(\mu; h , \delta).
\end{equation}
Recall from Section~\ref{sec:marked-BC-entropy} that with $P = \mu_1$, the probability distribution $\pi_P$ is defined on
the set $(\edgemark \times \mTb_*^0) \times (\edgemark \times \mTb_*^0)$ as
$\pi_P(t,t') = e_P(t,t') / d$ where $e_P(t,t') = \evwrt{P}{E_1(t,t')[T,o]}$. But
using~\eqref{eq:opt-h-1-E1-sum-D-tD}, for $t, \pp{t} \in \edgemark \times \mTb_*
= \mFdeltah$, we have
\begin{equation*}
  \pi_P(t,t') = \frac{1}{d} \evwrt{P}{E_1(t,t')[T,o]} = \frac{1}{d} \evwrt{\mu}{E_1(t,t')[T,o]} = \frac{1}{d} \evwrt{\mu}{D_{t,t'}} + \frac{1}{d} \evwrt{\mu}{\tD_{t,t'}} = \tilde{\pi}(0,t,t') + \tilde{\pi}(1,t,t').
\end{equation*}
Thereby, we have $H(\tilde{\pi}) \geq H(\pi_P)$. Using this
in~\eqref{eq:sum-Dc-leq-H-tilde-P}, we get
\begin{equation}
  \label{eq:opt-h-1-sum-Dc-leq-H-piP}
    \sum_{c \in \mCdeltah} \evwrt{\mu}{D_c} \log \evwrt{\mu}{D_c} \leq -d H(\pi_P) + d \log d + 2 \eta_3(\mu; h , \delta).
\end{equation}
Additionally, from~\eqref{eq:h-1-deg-sum-D-tD}, we have
\begin{equation}
  \label{eq:opt-h-1-sum-EDc-d-2eta2}
  \begin{aligned}
    \sum_{c \in \mCdeltah} \evwrt{\mu}{D_c([T,o])} &= d - \sum_{c \in \mCdeltah} \evwrt{\mu}{\tD_c([T,o])} \\
    &= d - \evwrt{\mu}{\sum_{c \in \mCdeltah} \tD_c([T,o])} \\
    &\stackrel{(a)}{=} d - \evwrt{\mu}{\deg_\star^{(h, \delta)}([T,o])} \\
    &= d - 2 \eta_2(\mu; h, \delta),
  \end{aligned}
\end{equation}
where in $(a)$, we have used~\eqref{eq:h-1-deg-star-sum-of-tD}.

Now we find a lower bound for the term $\sum_{c \in \mCdeltah} \evwrt{\mu}{\log
  D_c([T,o])!}$ in~\eqref{eq:opt-h-1-bound-2}. Similar to the above, to simplify the notation, for $c = (t,t')
\in \mCdeltah$, we may use $\tD_c$ and $E_1(c)$ instead of $\tD_c([T,o])$ and
$E_1(t,t')([T,o])$, respectively, inside the expectations with respect to $\mu$,
with the understanding that $[T,o] \sim \mu$.
Using~\eqref{eq:opt-h-1-E1-sum-D-tD}, we may write 
\begin{equation}
  \label{eq:opt-h-1-sum-log-Dc-fact-1}
\begin{aligned}
  \sum_{c \in \mCdeltah} \evwrt{\mu}{\log D_c!} &= \sum_{c \in \mCdeltah} \evwrt{\mu}{\log(E_1(c) - \tD_c)!} \\
  &= \sum_{c \in \mCdeltah} \evwrt{\mu}{\log E_1(c)!} - \sum_{c \in \mCdeltah} \evwrt{\mu}{\log \frac{E_1(c)!}{(E_1(c) - \tD_c)!}}.
\end{aligned}
\end{equation}
Using the inequality $\log \frac{a!}{(a-b)!} \leq b \log a$ which holds for $a \geq
b \geq 0$ (by identifying $0 \log 0 = 0$ as usual), we have
\begin{equation}
  \label{eq:opt-h-1-sum-ev-E1-E1-tDc}
  \begin{aligned}
    \sum_{c \in \mCdeltah} \evwrt{\mu}{\log \frac{E_1(c)!}{(E_1(c) - \tD_c)!}} &\leq \sum_{c \in \mCdeltah} \evwrt{\mu}{\tD_c \log E_1(c)} \\
    &\leq \sum_{c \in \mCdeltah} \evwrt{\mu}{\tD_c \log \deg_T(o)} \\
    &\stackrel{(a)}{=} \evwrt{\mu}{\deg_\star^{(h, \delta)}([T,o]) \log \deg_T(o)}\\
    &=  \eta_4(\mu; h, \delta),
  \end{aligned}
\end{equation}
where $(a)$ uses~\eqref{eq:h-1-deg-star-sum-of-tD}.
Note that for all $[T,o] \in \mTb_*$ and $c \in \mCdeltah$, $E_1(c)([T,o]) =
E_1(c)([T,o]_1)$. Therefore, for all $c \in \mCdeltah$, we have
$\evwrt{\mu}{\log E_1(c)!} = \evwrt{P}{\log E_1(c)!}$. Using these 
back in~\eqref{eq:opt-h-1-sum-log-Dc-fact-1}, and using part~\ref{it:loh1-1} of
Lemma~\ref{lem:opt-h-1-properties} to write $\mFdeltah = \edgemark \times
\mTb_*^0$, we get
\begin{equation}
\label{eq:opt-h-1-sum-D-factorial-final-bound}
  \sum_{c \in \mCdeltah} \evwrt{\mu}{\log D_c!}  \geq \sum_{t, t' \in \edgemark \times \mTb_*^0} \evwrt{P}{\log E_1(t,t')([T,o])!} -  \eta_4(\mu; h, \delta).
\end{equation}
Substituting~\eqref{eq:opt-h-1-sum-Dc-leq-H-piP},
\eqref{eq:opt-h-1-sum-EDc-d-2eta2}, and \eqref{eq:opt-h-1-sum-D-factorial-final-bound}
back in~\eqref{eq:opt-h-1-bound-2} and simplifying, we get
\begin{align*}
  \limsup_{n \rightarrow \infty} \frac{\nat(\fn_{h, \delta}(\Gn)) - \mn \log n}{n} &\leq -s(d) + H(P) - \frac{d}{2} H(\pi_P) - \sum_{t,\pp{t} \in \edgemark \times \mTb_*^0} \evwrt{P}{\log E_1(t,\pp{t})!} \\
&\quad  + \eta_1(\mu; h, \delta)(|\edgemark|^2 \log 2 + \log |\vermark| + 2 - 2 \log \eta_1(\mu; h, \delta)) \\
                                                                                             &\quad + (1 + \log 4) \eta_2(\mu; h, \delta) + \eta_3(\mu; h, \delta) + (1 + 2(|\edgemark| \times |\vermark|)^2) \eta_4(\mu; h, \delta) \\
  &= J_h(\mu_h) + \eta_0(\mu; h, \delta),
\end{align*}
which completes the proof.
\end{proof}

\editfinish



\editstart

\section{Finding Edge Types}
\label{sec:MP}

In this section, we prove Proposition~\ref{prop:MP}. More precisely, we  present an algorithm which given a
  simple marked graph $\Gn$ and parameters $h \ge 1$ and $\delta \ge 1$, extracts the edge
  types as we discussed in Section~\ref{sec:main-find-edge-type}. This algorithm employs a message passing
  framework to achieve this goal. 
  Before presenting the algorithm, we introduce a family of mappings
  $(\Lambda_k: k \geq 0)$ such that $\Lambda_k$  maps
  elements in 
  $\edgemark \times \mTb_*^k$ to sequences of nonnegative integers. Later in
  Section~\ref{sec:mp-alg-detials}, we will use these functions to design our
  algorithm.
  Finally, we analyze the complexity of the
  algorithm in Section~\ref{sec:MP-complexity}, specifically 
  Lemma~\ref{lem:message-passing-complexity} therein.
  Throughout this section, we will use the sample marked graph in
  Figure~\ref{fig:sample-graph-for-compression} to demonstrate the definitions
  and the steps of our algorithm. We have redrawn this graph in
  Figure~\ref{fig:sample-graph-for-compression-repeat} here. 


\begin{figure}
  \centering
  \begin{tikzpicture}
    \node[nodeR, label={[label distance=1mm]54:1}] (n1) at (0,0) {};
    \node[nodeB, label={[label distance=1mm]90:2}] (n2) at (90:2) {};
    \node[nodeB, label={[label distance=1mm]90:3}] (n3) at (162:2) {};
    \node[nodeB, label={[label distance=1mm]90:4}] (n4) at (234:2) {};
    \node[nodeB, label={[label distance=1mm]90:5}] (n5) at (306:2) {};
    \node[nodeB, label={[label distance=1mm]90:6}] (n6) at (18:2) {};
    \foreach \i in {2,...,6}
    {
      \drawedge{n1}{n\i}{O}{B};
      \begin{scope}[shift={(n\i)}]
        \begin{scope}[rotate={90+72*(\i-2)}]
          \node[nodeR] (n1\i) at (45:1) {};
          \node at (45:1.4) {\pgfmathparse{2*\i+4} \pgfmathprintnumber{\pgfmathresult}};
          \node[nodeR] (n2\i) at (-45:1) {};
          \node at (-45:1.4) {\pgfmathparse{2*\i+3} \pgfmathprintnumber{\pgfmathresult}};
          \drawedge{n\i}{n1\i}{B}{B};
          \drawedge{n\i}{n2\i}{B}{B};
          \draw[edgeO] (n1\i) -- (n2\i);
        \end{scope}
      \end{scope}
    }
  \end{tikzpicture}
  \caption[Sample graph for demonstrating our compression algorithm, redrawn]{The simple
    marked graph  of Figure~\ref{fig:sample-graph-for-compression}. Here, $\edgemark =
  \{\text{\color{blueedgecolor} Blue (solid)}, \text{\color{orangeedgecolor}
    Orange (wavy)} \}$ and $\vermark = \{\tikz{\node[nodeB] at (0,0) {};}, \tikz{\node[nodeR] at (0,0)
  {};}\}$. In order to represent elements in $\edgemark$ and $\vermark$ with
positive integers, we
write $\edgemark = \{\xone, \xtwo\}$, and $\vermark = \{\tone, \ttwo\}$.}
  \label{fig:sample-graph-for-compression-repeat}
\end{figure}

  
Let $\integers_+^{*+} := \cup_{n \ge 1} \integers_+^n$ denote the set of finite and nonempty sequences of nonnegative
integers.
For two sequences $a, b \in \integers_+^{*+}$, let $(a,b) \in \integers_+^{*+}$ denote the
  concatenation of $a$ and $b$. Also, for nonnegative integers $i, j$ and a sequence $a \in
  \integers_+^{*+}$, let $(i,a,j) \in \integers_+^{*+}$ denote the sequence which
  starts with $i$, follows with $a$, and ends with $j$. Also, $(i,a)$ and
  $(a,j)$ are defined similarly.
Recall that  we  assume that the edge and vertex mark
  sets are of the form $\edgemark = \{1, 2, \dots, |\edgemark|\}$ and $\vermark
  = \{ 1, 2, \dots, |\vermark|\}$.
  For integer $k \geq 0$, let $\Lambda_k : \edgemark \times \mTb_*^k
  \rightarrow \integers_+^{*+}$ be defined as follows.
For $k = 0$ and $(x, [T,o]) \in \edgemark \times \mTb_*^0$, we define
$\Lambda_0(x, [T,o]) := (\tau_T(o), 0, x)$.
Given $k \geq 1$, assuming that $\Lambda_{k-1}(.)$ is
    defined, for $(x, [T,o]) \in \edgemark \times \mTb_*^k$, we define
    $\Lambda_k(x, [T,o])$ as follows. If $\deg_T(o) \geq \delta$, we let
    $\Lambda_k(x,[T,o]) := (0,x)$. Otherwise, we form the list
    $(s_v: v \sim_T o)$ where $s_v$ for $v\sim_T o$ is the concatenation of
    $\Lambda_{k-1}(T[o,v])$ and $\xi_T(v,o)$. Then, we sort $s_v, v \sim_T o,$
    in an increasing lexicographic order. If any of the
    sequences $s_v$ start with a zero, we set $\Lambda_k(x,[T,o]) = (0,x)$. Otherwise,
    we define
    $s$ to be the concatenation of the sequences $s_v$ for $ v \sim_T o$ after
    sorting, and finally set
    $\Lambda_k(x,[T,o]) := (\tau_T(o), \deg_T(o), s, x)$. 
Note that as we sort the list $(s_v: v \sim_T o)$ in the process of defining
$\Lambda_k(.)$, the result does not depend on any ordering of the neighbors $v
\sim_T o$. This implies that $\Lambda_k(.)$ is well defined.
See    Figure~\ref{fig:Lambda-example-figure} for an example.

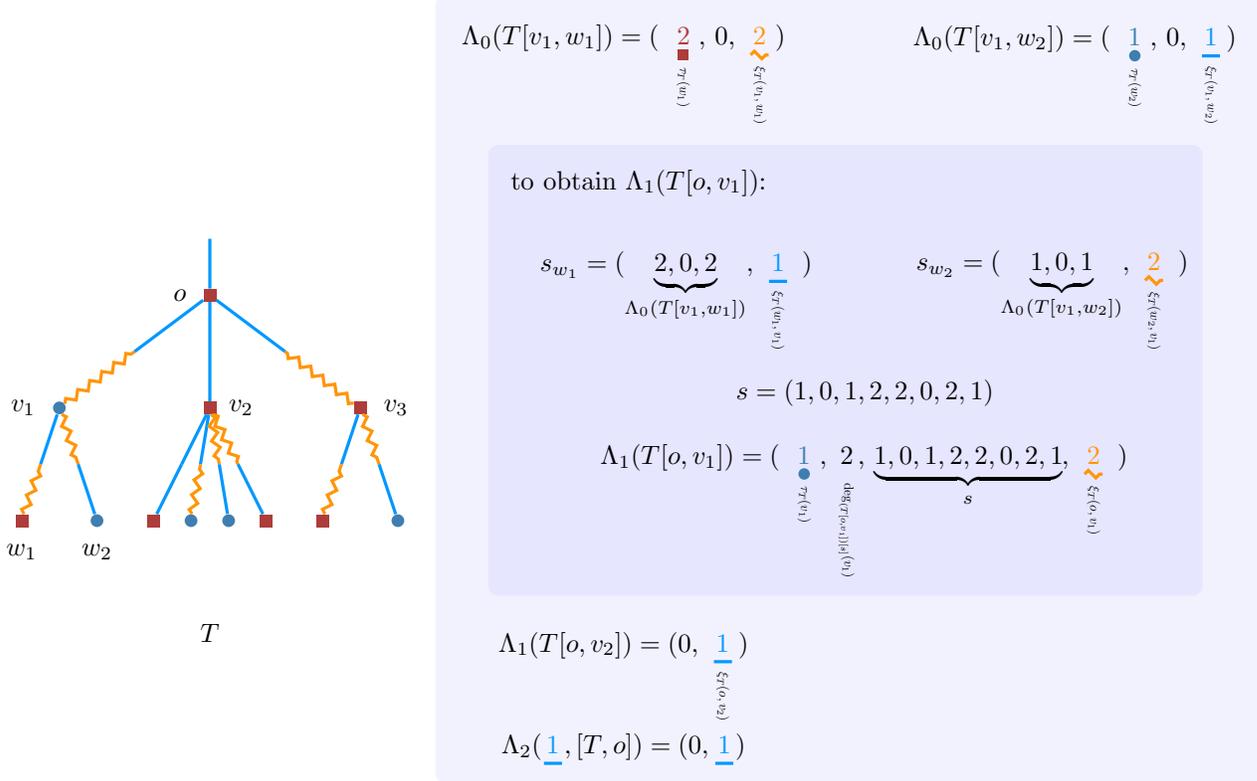
\begin{figure}
\centering
  \begin{tikzpicture}
    \begin{scope}[yshift=-3cm,yscale=1.5] 
      \node[nodeR,label={[label distance=1mm]180:$o$}] (o) at (0,0) {};

      \node[nodeB,label={[label distance=1mm]180:$v_1$}] (vone) at (-2,-1) {};
      \node[nodeR,label={[label distance=0.4mm]0:$v_2$}] (vtwo) at (0,-1) {};
      \node[nodeR,label={[label distance=1mm]0:$v_3$}] (vthree) at (2,-1) {};

      \drawedge{o}{vone}{B}{O}
      \drawedge{o}{vtwo}{B}{B}
      \drawedge{o}{vthree}{B}{O}

      \draw[edgeB] (o) -- ($(o)+(0,0.5)$);

      \begin{scope}[shift=(vone)]
        \node[nodeR,label={[label distance=1mm]270:$w_1$}] (wone) at ($(-0.5,-1)$) {};
        \node[nodeB,label={[label distance=1mm]270:$w_2$}] (wtwo) at ($(0.5,-1)$) {};
        \drawedge{vone}{wone}{B}{O};
        \drawedge{vone}{wtwo}{O}{B};
      \end{scope}

      \begin{scope}[shift=(vtwo)]
      \node[nodeR] (vtwoone) at ($(-0.75,-1)$) {};
      \node[nodeB] (vtwotwo) at ($(-0.25,-1)$) {};
      \node[nodeB] (vtwothree) at ($(0.25,-1)$) {};
      \node[nodeR] (vtwofour) at ($(0.75,-1)$) {};
      \drawedge{vtwo}{vtwoone}{B}{B}
      \drawedge{vtwo}{vtwotwo}{B}{O}
      \drawedge{vtwo}{vtwothree}{O}{B}
      \drawedge{vtwo}{vtwofour}{O}{B}
      
      \end{scope}

      \begin{scope}[shift=(vthree)]
        \node[nodeR] (wone) at ($(-0.5,-1)$) {};
        \node[nodeB] (wtwo) at ($(0.5,-1)$) {};
        \drawedge{vthree}{wone}{B}{O};
        \drawedge{vthree}{wtwo}{O}{B};
        
      \end{scope}

      \node at (0,-3) {$T$};
    \end{scope} 

    \begin{scope}[xshift=5.5cm] 
      \fill[blue!5,rounded corners] (-2.5,1) rectangle (8.5,-9.5);
      \begin{scope}
        \node at (0,0) {$\Lambda_0(T[v_1,w_1])=($ \stack{\ttwo}{$\tau_T(w_1)$}, 0, \stack{\xtwo}{$\xi_T(v_1,w_1)$})};
      \end{scope}
      \begin{scope}[xshift=6cm]
        \node at (0,0) {$\Lambda_0(T[v_1,w_2])=($ \stack{\tone}{$\tau_T(w_2)$}, 0, \stack{\xone}{$\xi_T(v_1,w_2)$})};
      \end{scope}

      \begin{scope}[xshift=0.7cm,yshift=-1.5cm] 
        \fill[blue!10,rounded corners] (-2.5,0.5) rectangle (7,-5.5);
        \node at (-0.5,0) {to obtain $\Lambda_1(T[o,v_1])$:};
        \begin{scope}[yshift=-1.5cm] 
          \node at (0,0) {$s_{w_1} = (\underbrace{2,0,2}_{\Lambda_0(T[v_1,w_1])},$ \stack{\xone}{$\xi_T(w_1,v_1)$} $)$};
          \node at (5,0) {$s_{w_2} = (\underbrace{1,0,1}_{\Lambda_0(T[v_1,w_2])},$ \stack{\xtwo}{$\xi_T(w_2,v_1)$} $)$};
          \begin{scope}[yshift=-1.3cm] 
            \node at (2.5,0) {$s = (1,0,1,2,2,0,2,1)$};
            \begin{scope}[yshift=-1.5cm] 
              \node at (2.5,0) {$\Lambda_1(T[o,v_1]) = ($ \stack{\tone}{$\tau_T(v_1)$}, \stack{2}{$\deg_{(T[o,v_1])[s]}(v_1)$}, $\underbrace{1,0,1,2,2,0,2,1}_{s}$, \stack{\xtwo}{$\xi_T(o,v_1)$} $)$};
            \end{scope} 
          \end{scope} 
        \end{scope} 

      \end{scope} 

      \begin{scope}[yshift=-8cm] 
        \node at (0,0) {$\Lambda_1(T[o,v_2]) = (0,$ \stack{\xone}{$\xi_T(o,v_2)$}$)$};
      \end{scope} 
      \begin{scope}[yshift=-9cm] 
        \node at (0,0) {$\Lambda_2(\xone,[T,o])=(0,\xone)$};
      \end{scope} 
    \end{scope} 
  \end{tikzpicture}
  \caption[How to find Lambda functions]{For $(\xone,[T,o]) \in \edgemark \times \mTb_*^2$ drawn on the left,
    we explain on the right how to find $\Lambda_2(\xone,[T,o])$ step by step.
    Here, we assume that $\delta = 4$. 
    First, following the definition of $\Lambda_0$, we can find
    $\Lambda_0(T[v_1, w_1])$ and $\Lambda_0(T[v_1, w_2])$. In order to obtain
    $\Lambda_1(T[o,v_1])$, we first find $s_{w_i} = (\Lambda_0(T[v_1, w_i]) ,
    \xi_T(w_i, v_i))$ for
    $i \in \{1, 2\}$. Then, we form the sequence $s$ by first sorting $s_{w_1}$
    and $s_{w_2}$ in an increasing lexicographical order, and then concatenate
    them. Note that since $s_{w_2}$ starts with a 1, while $s_{w_1}$ starts with
    a 2, $s_{w_2}$ is lexicographically smaller and appears first in $s$. 
Since none of the sequences $s_{w_i}$, $i \in \{1, 2\}$  start
    with a zero, and the degree at the root in the subgraph component of
    $T[o,v_1]$ is $\deg_{(T[o,v_1])[s]}(v_1) = 2 < \delta$, $\Lambda_1(T[o,v_1])$ is obtained by concatenating
    $\tau_T(v_1)$, the degree at the root in the subgraph component of
    $T[o,v_1]$, the sequence $s$, and the mark component of $T[o,v_1]$ which is
    $\xi_T(o,v_1)$. On the other hand, since the degree at the root in the
    subgraph component of $T[o,v_2]$  is $4 \geq \delta$, $\Lambda_1(T[o,v_2]) =
    (0,\xone)$. Finally, in order to find $\Lambda_2(\xone, [T,o])$, note that
    although the degree at the root in $[T,o]$ is $3 < \delta$, since
    $\Lambda_1(T[o,v_2])$ starts with a zero, $s_{v_2}$ starts with a zero and
    $\Lambda_2(\xone, [T,o]) = (0,\xone)$.
    \label{fig:Lambda-example-figure}}
\end{figure}

In the following lemma, we show that for $k \geq 0$, the function $\Lambda_k(.)$
introduces a prefix--free code.
{\color{peditcolor}
Therefore, an object $(x, [T,o]) \in \edgemark \times \mTb_*^{k}$ is uniquely
determined by knowing 
$\Lambda_k(x,[T,o])$.}

    \begin{lem}
      \label{lem:Lambda-prefix-free}
      For $k \geq 0$, $\Lambda_k(.)$ is {\color{peditcolor}a prefix--free
        function on $\edgemark \times \mTb_*^k$}
    \end{lem}

    \begin{proof}
      We prove this by induction. For $k = 0$, $\Lambda_0(x, [T,o]) = (\tau_T(o), 0,x)$ and hence is
      prefix--free. For $k \geq 1$, we either have $\Lambda_k(x,[T,o]) = (0,x)$
      or $\Lambda_k(x,[T,o]) = (\tau_T(o), \deg_T(o),s,x)$. But $\tau_T(o) \geq
      1$. Hence, by looking at the first index, we can determine which of these
      two cases hold. If the first index is $0$, we read the second index, which
      is $x$, and we are done. Otherwise, we read $\tau_T(o)$ and read
      $\deg_T(o)$ in the second index. Then, we know that $\deg_T(o)$ many
      sequences $(\Lambda_{k-1}(T[o,v]), \xi_T(v,o))$ for $v \sim_T o$ follow. But by
      induction, $\Lambda_{k-1}(.)$ is prefix--free. Hence, we can read
      $s$ unambiguously. After this, we read the last index, which is $x$, and we
      are done. 
    \end{proof}

The following lemma shows that for $k
\geq 0$ and $t \in \edgemark \times \mTb_*^k$, $\Lambda_k(t)$ determines whether $t$ falls in $\mF^{(\delta, k+1)}$, and this
can be realized  by looking at the first index in $\Lambda_k(t)$.

        \begin{lem}
      \label{lem:Lambda-unique}
      For $k \geq 0$, $\Lambda_k(x,[T,o]) = (0,x)$ iff $(x,[T,o]) \notin
      \mFdeltakp$. Moreover, if $(x,[T,o]) \in \mFdeltakp$, then
      $\Lambda_k(x,[T,o])$ does not start with a zero and $(x,[T,o])$ is
      uniquely determined by $\Lambda_k(x,[T,o])$. 
    \end{lem}
    \begin{proof}
      We prove this by induction. For $k = 0$, $\mF^{(\delta, 1)} = \edgemark \times
      \mTb_*^0$ and $\Lambda_0(x,[T,o]) = (\tau_T(o), 0, x)$. Note that
      $\tau_T(o) > 0$ by definition. Therefore, $[T,o] \in \mTb_*^0$ is uniquely determined by $\tau_T(o)$,
      as $o$ has no offspring.

      For $k \geq 1$, by definition of $\mFdeltakp$, we have $(x,[T,o]) \in
      \mFdeltakp$ iff $\deg_T(o) < \delta$ and for all $v \sim_T o$, $T[o,v] \in
      \mF^{(\delta, k)}$. On the other hand, by definition of $\Lambda_k(.)$,
      $\Lambda_k(x,[T,o]) = (0,x)$ iff $\deg_T(o) \geq \delta$ or
      $\Lambda_{k-1}(T[o,v]) = (0,\xi_T(o,v))$ for some $v \sim_T o$. By
      induction, $\Lambda_{k-1}(T[o,v]) = (0,\xi_T(o,v))$ for some $v \sim_T o$
      iff $T[o,v] \notin \mFdeltak$. Hence, $\Lambda_k(x,[T,o]) = (0,x)$
      iff $\deg_T(o) \geq \delta$ or for some $v \sim_T o $, $T[o,v] \notin
      \mFdeltak$. But this is precisely  equivalent
      to $(x,[T,o]) \notin \mFdeltakp$.

      Now, we take $(x,[T,o]) \in \mFdeltakp$ and show that it is uniquely
      determined by $\Lambda_k(x,[T,o])$. The above discussion implies that
      $\Lambda_k(x,[T,o]) = (\tau_T(o), \deg_T(o), s, x)$ where $s$ is the
      concatenation of $(\Lambda_{k-1}(T[o,v]), \xi_T(v,o))$ for $v \sim_T o$
      after sorting lexicographically. Observe that $\tau_T(o) > 0$ and hence
      $\Lambda_k(x,[T,o])$ does not start with a zero.
Also note that given $\Lambda_k(x,[T,o])$, we
      can extract $\deg_T(o)$ from its second index. Moreover, 
      Lemma~\ref{lem:Lambda-prefix-free} implies that $\Lambda_{k-1}(.)$ is
      prefix free. Therefore, we
      can recover $(\Lambda_{k-1}(T[o,v]), \xi_T(v,o))$ for all $v \sim_T o$. 
      Using the induction hypothesis, this suffices to  unambiguously find
      $(T[o,v], \xi_T(v,o))$ for
      all $v \sim_T o$. Since $[T,o]$ is a marked rooted tree, we can put these
      together with $\tau_T(o)$ to unambiguously reconstruct $[T,o]$. Note that
      the order of vertices $v \sim_T o$  in the list does not matter, since we
      are only interested in the equivalence class $[T,o]$.  Finally, we have
      access to $x$
      from the last index of $\Lambda_k(x,[T,o])$. Therefore, we can unambiguously
      reconstruct $(x,[T,o])$, which completes the proof. 
    \end{proof}


Motivated by the above discussion, we can introduce the following message
passing scheme to find $\Lambda_{h-1}(\UC_v(\Gn)[w,v]_{h-1})$ for adjacent
vertices $v \sim_{\Gn} w$ in the input graph $\Gn$. As we saw in
Lemma~\ref{lem:Lambda-unique}, this is helpful to identify
those $\T{v}{w} = \UC_v(\Gn)[w,v]_{h-1}$ which are in $\mFdeltah$. We will
later use this to identify $\TT{v}{w}$ and $\type{v}{w}$
using~\eqref{eq:tilde-t-v-w} and \eqref{eq:type-v-w}.

For $0 \leq k \leq h-1$,  we define the messages $\{M_k(v,w): v
\sim_{\Gn} w\}$ for a given
 simple marked graph $\Gn$ such that $M_k(v,w) \in \mathbb{Z}_+^*$ are defined as follows. We first initialize the messages as
 $M_0(v,w) = (\tau_{\Gn}(v), 0, \xi_{\Gn}(w,v))$ for $v \sim_{\Gn} w$. Then, we define
 $M_k(.,.)$ for $k \geq 1$ inductively. For $v \sim_{\Gn} w$, if $\deg_\Gn(v) >
 \delta$, we let $M_k(v, w) = (0,\xi_{\Gn}(w,v))$. Otherwise, we form the list
 $(s_u: u \sim_{\Gn} v, u \neq w)$, where for $u \sim_{\Gn} v, u \neq w$, we set
 $s_u = (M_{k-1}(u,v), \xi_{\Gn}(u,v))$. If for some $u \sim_{\Gn} v, u \neq w$,
 the sequence $s_u$ starts with a zero, we set $M_k(v,w) = (0,\xi_{\Gn}(w,v))$.
 Otherwise, we sort the sequences $s_u$ increasingly  with
 respect to the lexicographic order in $\integers_+^{*+}$, and set $s$ to be
 the concatenation of the sorted list. Finally, we set $M_k(v,w) =
 (\tau_{\Gn}(v), \deg_{\Gn}(v) -1, s, \xi_{\Gn}(w,v))$.

In the following proposition, we show that $M_k(v,w)$ for $v \sim_{\Gn} w$ represents
$\UC_v(\Gn)[w,v]_k$. Later in this section, we use a modified version of $M_k$ to
design our message passing algorithm.

      \begin{prop}
        \label{prop:message-pass-M-Lambda}
        Given a simple marked graph $\Gn$, adjacent vertices $v \sim_{\Gn} w$,
        and integer $0 \leq k \leq h-1$, we have
        \begin{equation*}
          M_k(v,w) = \Lambda_k(\UC_v({\Gn})[w,v]_k).
        \end{equation*}
      \end{prop}
      \begin{proof}
        We prove this by induction on $k$. If $k = 0$, we have $M_0(v,w) =
        (\tau_{\Gn}(v), 0, \xi_{\Gn}(w,v))$ while $\Lambda_0(\UC_v({\Gn})[w,v]_0) =
        (\tau_{\UC_v({\Gn})}(v), 0, \xi_{\UC_v(\Gn)}(w,v))$. But $\tau_{\UC_v({\Gn})}(v) =
        \tau_{\Gn}(v)$ and $\xi_{\UC_v({\Gn})}(w,v) = \xi_{\Gn}(w,v)$. Thereby, $M_0(v,w) =
        \Lambda_0(\UC_v({\Gn})[w,v]_0)$.

        Now, assume that $k > 1$ and the statement holds for $k-1$. To simplify
        the notation, we set $x$ and $[T,v]$ to be the mark and the subgraph
        components of $\UC_v({\Gn})[w,v]_k$, respectively. Recalling the definition
        of $\Lambda_k(.)$, if $\deg_T(v) \geq \delta$, we set
        $\Lambda_k(x,[T,v]) = (0,x)$. But $\deg_T(v) = \deg_{\Gn}(v) - 1$, which
        means that $\deg_T(v) \geq \delta$ is equivalent to $\deg_{\Gn}(v) >
        \delta$. However, in this case, we have $M_k(v,w) = (0,\xi_{\Gn}(w,v))$. But
        $x = \xi_{\Gn}(w,v)$. Hence, if $\deg_{\Gn}(v) > \delta$, $M_k(v,w) =
        \Lambda_k(x,[T,v])$. 

        Now, assume that  $\deg_{\Gn}(v) \leq \delta$.  In this case, recall that in order to
        define $M_k(v,w)$, we first consider the list of sequences $\mS = (s_u: u
        \sim_{\Gn} v: u \neq w)$ where $s_u = (M_{k-1}(u,v), \xi_{\Gn}(u,v))$ for $u
        \sim_{\Gn} v, u \neq w$. On the other hand, in order to define
        $\Lambda_k(x,[T,v])$, we form the list $\tmS = ( \ts_{\tu}: \tu \sim_T v)$
        where $\ts_{\tu} = (\Lambda_{k-1}(T[v,\tu]), \xi_T(\tu,v))$ for $\tu
        \sim_T v$. Note that
        due to the definition of $\UC_v({\Gn})$ and $\UC_v({\Gn})[w,v]_k$, there is a
        one to one correspondence  between the vertices $u \sim_{\Gn} v, u \neq w$
        and $\tu \sim_T v$. If we denote the vertex in $T$ corresponding to $u
        \sim_{\Gn} v, u \neq w$ by $\tu$, we have $\xi_T(\tu, v) =
        \xi_{\UC_v({\Gn})}(u,v) = \xi_{\Gn}(u,v)$. On the other hand, using the
        induction hypothesis, for all $u \sim_{\Gn} v, u \neq w$, we have 
        \begin{equation*}
          M_{k-1}(u,v) = \Lambda_{k-1}(\UC_u({\Gn})[v,u]_{k-1}).
        \end{equation*}
        However, from Lemma~\ref{lem:UCv-UCu}, we have $\UC_u({\Gn})[v,u]_{k-1} =
        \UC_v({\Gn})[v,u]_{k-1}$. Moreover,
        we have $\UC_v({\Gn})[v,u]_{k-1} = T[v,\tilde{u}]$.
        Consequently, for all $u \sim_{\Gn} v, u \neq w$, we have
        \begin{equation*}
          M_{k-1}(u,v) = \Lambda_{k-1}(T[v,\tu]).
        \end{equation*}
        Also, as we showed above, $\xi_T(\tu,v) = \xi_{\Gn}(u,v)$. Therefore, we
        have $s_u = \ts_{\tu}$ for all $u \sim_{\Gn} v, u \neq w$. This means that the lists $\mS$ and $\tmS$ are
        the same. Hence, there is a sequence in $\mS$ starting with zero iff
        there is such a sequence in $\tmS$. In this case, by definition, we set
        $M_k(v,w) = (0,\xi_{\Gn}(w,v)) = (0,x) = \Lambda_k(x,[T,v])$. Otherwise, we
        sort the sequences in $\mS$ in an increasing order and set $s$ to be
        the concatenation of these sequences. Finally, we have $M_k(v,w) =
        (\tau_{\Gn}(v), \deg_{\Gn}(v) - 1, s, \xi_{\Gn}(w,v))$. On the other hand, since
        $\mS$ and $\tmS$ are the same, we have $\Lambda_k(x,[T,v]) = (\tau_T(v),
        \deg_T(v), s, x)$. But $\tau_T(v) = \tau_{\Gn}(v)$, $\deg_T(v) = \deg_{\Gn}(v)
        -1$, and $x = \xi_{\Gn}(w,v)$. Thus, we have $M_k(v,w) = \Lambda_k(x,[T,v])$
        and the proof is complete. 
      \end{proof}

\subsection{A Message Passing Algorithm to Find Edge Types}
\label{sec:mp-alg-detials}

Now, motivated by the messages $M_k(.,.)$, we design our main message passing
algorithm. Observe that in principle, as $k$ grows, the length of $M_k(.,.)$ can
grow exponentially in $\delta$. In order to avoid dealing with large sequences,
in the following we introduce a scheme that at each step $0 \leq k \leq h-1$,
the messages which are sequences of integers are converted to integers.

For $0 \leq k \leq h-1$, we inductively define the integer $T_k(v,w)$ for $v \sim_{\Gn} w$
as follows. At each step $k$, we store $(T_k(v,w): v \sim_{\Gn} w)$ in a two
dimensional array $(\sfT_{v,i}: v \in [n], i \in [\dn_v])$, where $\sfT_{v,i} =
T(v,\gamman_{v,i})$ for $v \in [n]$ and $i \in [\dn_v]$. For 
$k = 0$, we go over vertices $1 \leq v \leq n$ and its neighbors $\gamman_{v,1} <
\gamman_{v,2} < \dots < \gamman_{v,\dn_v}$ in an increasing order. On the other
hand, let $\mathsf{TDictionary}$ be a $\Dictionary(\Array \text{ of integers}
\rightarrow \nats)$ and $\mathsf{TCount}$ be an integer initialized with zero.
Furthermore, let $\mathsf{TMark}$ and $\mathsf{TIsStar}$ be initially empty
arrays of integers and bits, respectively. With these, for $1 \leq v \leq n$ and
$1 \leq i \leq \dn_v$, we search for $M_0(v, \gamman_{v,i})$ in the set of keys in $\mathsf{TDictionary}$. If we find $M_0(v, \gamman_{v,i})$, we set
$T_0(v, \gamman_{v,i})$ to be the corresponding value. Otherwise, we detect
$M_0(v, \gamman_{v,i})$ as a new message, and  we insert the key--value
pair $(M_0(v, \gamman_{v,i}),1 + \mathsf{TCount} )$ into $\mathsf{TDictionary}$,
set $T_0(v, \gamman_{v,i}) = 1 + \mathsf{TCount}$, and add $1$ to
$\mathsf{TCount}$. Moreover, we append $0$ to the $\mathsf{TIsStar}$ array. Also, we append the last index of $M_0(v, \gamman_{v,i})$ to the 
$\mathsf{TMark}$ array.
Figure~\ref{fig:mp-run-example-k0} illustrates an example of this procedure for the
graph of Figure~\ref{fig:sample-graph-for-compression-repeat}.

For $k > 0$, we inductively define $T_k$ based on $T_{k-1}$ as follows. Again, we
go through $1 \leq v \leq n$ and $1 \leq i \leq \dn_v$ in an increasing order.
Similar to the above, let $\mathsf{TCount}$ be an integer initialized with $0$,
and $\mathsf{TDictionary}$ be an empty $\Dictionary(\Array \text{ of integers}
\rightarrow \nats)$. Let $\mathsf{TMarkOld}$ and $\mathsf{TIsStarOld}$ be copies
of the $\mathsf{TMark}$ and $\mathsf{TIsStar}$ arrays, respectively,  from step
$k-1$.  Furthermore, let $\mathsf{TMark}$ and $\mathsf{TIsStar}$ be
initially empty arrays of integers and bits, respectively. For $1 \leq v \leq n$
and $1 \leq i \leq \dn_v$, if $\deg_{\Gn}(v) > \delta$, define $\tM_{k}(v,
\gamman_{v,i}) = (0,\xi_{\Gn}(\gamman_{v,i}, v))$. Also, if
$\mathsf{TIsStarOld}(T_{k-1}(u,v))= 1$ for some $u \sim_{\Gn} v, u \neq
\gamman_{v,i}$, we set $\tM_k(v, \gamman_{v,i}) = (0,\xi_{\Gn}(\gamman_{v,i},
v))$. Otherwise, for $u \sim_{\Gn} v, u \neq \gamman_{v,i}$, let $s_u =
(T_{k-1}(u,v), \xi_{\Gn}(u,v))$ and sort the array $\vs = (s_{u}: u \sim_{\Gn} v, u
\neq \gamman_{v,i})$ increasingly with respect to the lexicographic order. Then,
let $\ts$ to be the concatenation of the elements in the sorted array $\vs$ and
define $\tM_k(v, \gamman_{v,i}) = (\tau_{\Gn}(v), \deg_{\Gn}(v) - 1, s, \xi_{\Gn}(
\gamman_{v,i},v))$. Finally, we search for $\tM_k(v, \gamman_{v,i})$ in the
set of keys in $\mathsf{TDictionary}$. If we find $\tM_k(v,
\gamman_{v,i})$, we set $T_k(v, \gamman_{v,i})$ to be the corresponding value.
Otherwise, if we do not find $\tM_k(v, \gamman_{v,i})$ in the
set of keys in $\mathsf{TDictionary}$, we do the following:
\begin{enumerate}
\item We insert the key--value pair $(\tM_k(v, \gamman_{v,i}), 1 +
  \mathsf{TCount})$ into $\mathsf{TDictionary}$.
\item We set $T_k(v,
\gamman_{v,i}) = 1 + \mathsf{TCount}$, and add $1$ to $\mathsf{TCount}$.
\item If the first index in $\tM_k(v, \gamman_{v,i})$ is 0, we append 1
  to $\mathsf{TIsStar}$; otherwise, we append 0 to $\mathsf{TIsStar}$.
  \item  We append the last
index of $\tM_k(v, \gamman_{v,i})$, which is precisely $\xi_{\Gn}(\gamma^{(n)}_{v,i},v)$, to $\mathsf{TMark}$. 
\end{enumerate}
As we mentioned above, we keep the values of $(T_k(v,w): v \sim_{\Gn} w)$ in an
array $(\sfT_{v,i}: v \in [n], i \in [\dn_{v}])$. Since we need $T_{k-1}(.,.)$ to
find $T_k(.,.)$, at step $k > 0$, we first copy the values of $(\sfT_{v,i}:v \in [n], i \in [\dn_v])$ in
an array called $(\tilde{\sfT}_{v,i}: v \in [n], i \in [\dn_v])$ so that
$\tilde{\sfT}_{v,i} = T_{k-1}(v,\gamman_{v,i})$. Then, we use this array to
update $(\sfT_{v,i}: v \in [n], i \in [\dn_v])$ so that $\sfT_{v,i} = T_k(v,\gamman_{v,i})$.

Figure~\ref{fig:mp-run-example-k1} illustrates an example of the above procedure for
finding $T_k(.,.)$ for
$k=1$ in the graph of Figure~\ref{fig:sample-graph-for-compression-repeat}. 

In Lemma~\ref{lem:MP-T-M-equivalent} below, we study the connection between the
quantities $T_k(v,w), \tM_k(v,w)$, and $M_k(v,w)$ for $v \sim_{\Gn} w$ which were defined
above. In order to do so, we first make some definitions. Motivated
by~\eqref{eq:t-v-w}, for $v \sim_{\Gn} w$
and $k \geq 1$, let
\begin{equation}
  \label{eq:tnk}
  t^{(n)}_k(v,w) := (\UC_v(\Gn))[w,v]_{k-1}.
\end{equation}
Moreover, for $v \sim_{\Gn} w$
and $k \geq 1$, let
\begin{equation}
  \label{eq:that-def}
  \hat{t}^{(n)}_{k, \delta}(v,w) :=
  \begin{cases}
    t^{(n)}_k(v,w) & t^{(n)}_k(v,w) \in \mF^{(\delta, k)} \\
    \star_{\xi_{\Gn}(w,v)} & \text{otherwise}
  \end{cases}
\end{equation}
For $k \geq 0$, let $\mathsf{TIsStar}^{(k)}$  be the value of the array
$\mathsf{TIsStar}$ at step $k$ of the above procedure. Therefore, at step $k
>0$,
$\mathsf{TIsStarOld}$ is precisely $\mathsf{TIsStar}^{(k-1)}$.

\begin{lem}
  \label{lem:MP-T-M-equivalent}
  For $0 \leq k \leq h-1$, $v \sim_{\Gn} w$, and $\pp{v} \sim_{\Gn} \pp{w}$, the
  following are equivalent:
  \begin{enumerate}[(1)]
  \item $T_k(v,w) = T_k(\pp{v}, \pp{w})$
  \item $\tM_k(v,w) = \tM_k(\pp{v}, \pp{w})$
  \item $M_k(v,w) = M_k(\pp{v}, \pp{w})$
  \item $\hat{t}^{(n)}_{k+1, \delta}(v,w) = \hat{t}^{(n)}_{k+1, \delta}(\pp{v},
    \pp{w})$
  \end{enumerate}
  Here, for $k=0$, we define $\tM_0(v,w) := M_0(v,w)$ for all $v \sim_{\Gn} w$. 
  Furthermore, 
  \begin{equation}
    \label{eq:tisstar-1-iff-t-notin}
    \mathsf{TIsStar}^{(k)}(T_k(v,w)) = 1 \qquad \text{iff} \qquad t^{(n)}_{k+1}(v,w)
  \notin \mF^{(\delta, k+1)}.
  \end{equation}
\end{lem}

In fact, this lemma implies that the value of $T_k(v,w)$ uniquely determines
$\hat{t}^{(n)}_{k+1, \delta}(v,w)$, which is relevant to the type of the edge
$(v,w)$ motivated by~\eqref{eq:tilde-t-v-w}.

\begin{proof}[Proof of Lemma~\ref{lem:MP-T-M-equivalent}]

  We show this by induction on $k$. First, assume that $k = 0$. Note that by
  definition, $T_0(v,w)$ is defined by searching  $M_0(v,w) = \tM_0(v,w)$ in
  $\mathsf{TDictionary}$. On the other hand, from
  Proposition~\ref{prop:message-pass-M-Lambda}, $M_0(v,w) = M_0(\pp{v}, \pp{w})$
  iff $\Lambda_0(t^{(n)}_1(v,w)) = \Lambda_0(t^{(n)}_1(\pp{v}, \pp{w}))$. Note that $\mF^{(\delta, 1)}
  = \edgemark \times \mTb_*^0$. Therefore, from Lemma~\ref{lem:Lambda-unique},
  $\Lambda_0(t^{(n)}_1(v,w)) = \Lambda_0(t^{(n)}_1(\pp{v}, \pp{w}))$ is
  equivalent to $t^{(n)}_1(v,w) = t^{(n)}_1(\pp{v}, \pp{w})$. On the other hand,
  since $\mF^{(\delta, 1)}
  = \edgemark \times \mTb_*^0$, we have $\hat{t}^{(n)}_{1, \delta}(v,w) =
  t^{(n)}_1(v,w)$ for all $v \sim_{\Gn} w$. Consequently, all of the 4
  conditions are equivalent for $k = 0$. Additionally, by construction, all
  values in 
  $\mathsf{TIsStar}^{(0)}$ are zero, and since $\mF^{(\delta, 1)}
  = \edgemark \times \mTb_*^0$, $t^{(n)}_1(v,w) \in \mF^{(\delta, 1)}$ for all $
  v \sim_{\Gn} w$. Hence, \eqref{eq:tisstar-1-iff-t-notin} is verified, and the
  statement of the lemma holds for $k =0$.

  Now, take $k > 0$. Note that  $T_k(v,w)$ and $T_k(\pp{v}, \pp{w})$ are defined by
  searching for $\tM_k(v,w)$ and $\tM_k(\pp{v}, \pp{w})$ in
  $\mathsf{TDictionary}$, respectively. Hence, $T_k(v,w) = T_k(\pp{v}, \pp{w})$ implies
  $\tM_k(v,w) = \tM_k(\pp{v}, \pp{w})$, i.e.\ $(1) \Rightarrow (2)$.
  To show $(2) \Rightarrow (3)$, note that if $\tM_k(v,w) =
  \tM_k(\pp{v}, \pp{w})$, one of the following two cases must hold:

  \underline{Case 1:} $\tM_k(v,w) = \tM_k(\pp{v}, \pp{w}) =
  (0,\xi_{\Gn}(w,v)) = (0,\xi_{\Gn}(\pp{w}, \pp{v}))$. Therefore, either $\deg_{\Gn}(v) > \delta$, or
  $\mathsf{TIsStar}^{(k-1)}(T_{k-1}(u,v)) = 1$ for some $u \sim_{\Gn} v$, $u
  \neq w$. By induction hypothesis, for  $u \sim_{\Gn} v$, $u  \neq w$, we know
  that $\mathsf{TIsStar}^{(k-1)}(T_{k-1}(u,v)) = 1$
  is equivalent to $t^{(n)}_k(u,v) \notin
  \mF^{(\delta, k)}$.
  Recalling the definition of $\mF^{(\delta, k)}$ and $\mF^{(\delta, k+1)}$, 
  the condition ``$\deg_{\Gn}(v) > \delta$ or $t^{(n)}_k(u,v) \notin
  \mF^{(\delta,k )}$ for some $u \sim_{\Gn}v, u \neq w$'' implies that
  $t^{(n)}_{k+1}(v,w) \notin \mF^{(\delta, k+1)}$. This together with
  Lemma~\ref{lem:Lambda-unique} and Proposition~\ref{prop:message-pass-M-Lambda}
  imply that $M_k(v,w) = (0,\xi_{\Gn}(w,v))$. Similarly, we have $M_k(\pp{v},
  \pp{w}) = (0,\xi_{\Gn}(\pp{w}, \pp{v}))$. Since $\tM_k(v,w) =
  \tM_k(\pp{v}, \pp{w})$, we have $\xi_{\Gn}(w,v) =
  \xi_{\Gn}(\pp{w}, \pp{v})$. Thereby, $(3)$ holds.

  \underline{Case 2:} The first index in $\tM_k(v,w) = \tM_k(\pp{v},
  \pp{w})$ is not zero. Therefore, motivated by the definition of
  $\tM_k(v,w)$ and $\tM_k(\pp{v}, \pp{w})$, we realize that the
  followings must  hold:
  \begin{enumerate}[$(i)$]
  \item $\tau_{\Gn}(v) = \tau_{\Gn}(\pp{v})$.
  \item $\deg_{\Gn}(v) = \deg_{\Gn}(\pp{v}) \leq \delta $.
  \item $\xi_{\Gn}(w,v) = \xi_{\Gn}(\pp{w}, \pp{v})$.
  \item $\mathsf{TIsStar}^{(k-1)}(T_{k-1}(u,v)) = 0$ for all $u \sim_{\Gn} v, u
    \neq w$. Similarly $\mathsf{TIsStar}^{(k-1)}(T_{k-1}(\pp{u}, \pp{v})) = 0$ for
    all $\pp{u} \sim \pp{v}, \pp{u} \neq \pp{w}$. 
  \item There is a bijection $\pi: \{ u \sim_{\Gn} v: u \neq w\} \rightarrow
    \{\pp{u} \sim_{\Gn} \pp{v}: \pp{u} \neq \pp{w}\}$, such that for all $u
    \sim_{\Gn} v, u \neq w$, we have $T_{k-1}(u,v) = T_{k-1}(\pi(u), \pp{v})$,
    and $\xi_{\Gn}(u,v) = \xi_{\Gn}(\pi(u), \pp{v})$. 
  \end{enumerate}
Now, using induction hypothesis together with condition $(v)$, we realize that
$M_{k-1}(u,v) = M_{k-1}(\pi(u), \pp{v})$ for all $u \sim_{\Gn} v, u \neq w$.
Moreover,  from condition $(iv)$ and \eqref{eq:tisstar-1-iff-t-notin} for $k-1$, we
realize that $t^{(n)}_k(u,v) \in \mF^{(\delta, k)}$ for all $u \sim_{\Gn} v, u
\neq w$. Thereby, Lemma~\ref{lem:Lambda-unique} together with 
Proposition~\ref{prop:message-pass-M-Lambda} imply that for all $u
\sim_{\Gn} v, u \neq w$, $M_{k-1}(u,v)$ does not
start with a zero. Similarly, for all $\pp{u} \sim_{\Gn} \pp{v}, \pp{u} \neq
\pp{w}$, $M_{k-1}(\pp{u}, \pp{v})$ does not start
with a zero. Putting these
together with $(i), (ii)$, and $(iii)$ above, and recalling the definition of
$M_k(v,w)$ and $M_k(\pp{v}, \pp{w})$, we realize that $M_k(v,w) = M_k(\pp{v},
\pp{w})$.

Now, we show $(3) \Rightarrow (4)$.
From Proposition~\ref{prop:message-pass-M-Lambda}, we have $M_k(v,w) =
\Lambda_k(t^{(n)}_{k+1}(v,w)) = M_k(\pp{v}, \pp{w}) =
\Lambda_k(t^{(n)}_{k+1}(\pp{v}, \pp{w}))$.
%
Therefore, if $t^{(n)}_{k+1}(v,w) \in \mF^{(\delta, k+1)}$,
Lemma~\ref{lem:Lambda-unique} implies that 
$\Lambda_k(t^{(n)}_{k+1}(v,w)) = \Lambda_k(t^{(n)}_{k+1}(v',w'))$ do not start
with a zero, and another usage of 
Lemma~\ref{lem:Lambda-unique} implies that  $t^{(n)}_{k+1}(v,w) =
t^{(n)}_{k+1}(v',w') \in \mF^{(\delta, k+1)}$. This means that $\hat{t}^{(n)}_{k+1,
\delta}(v,w) = \hat{t}^{(n)}_{k+1, \delta}(v',w')$. On the other hand, if
$t^{(n)}_{k+1}(v,w) \notin \mF^{(\delta, k+1)}$, Lemma~\ref{lem:Lambda-unique}
implies that $\Lambda_k(t^{(n)}_{k+1}(v,w)) = (0,\xi_{\Gn}(w,v)) =
\Lambda_k(t^{(n)}_{k+1}(v',w'))$. Hence, Lemma~\ref{lem:Lambda-unique} implies
that we also have $t^{(n)}_{k+1}(v',w') \notin \mF^{(\delta, k+1)}$ and
$\xi_{\Gn}(w',v') = \xi_{\Gn}(w,v)$. Therefore, $\hat{t}^{(n)}_{k+1,
\delta}(v,w) = \star_{\xi_{\Gn}(w,v)} = \star_{\xi_{\Gn}(w',v')} =
\hat{t}^{(n)}_{k+1, \delta}(v',w')$. Consequently, $(4)$ is established.

In order to show $(4) \Rightarrow (1)$, we take two cases:

\underline{Case 1:} $\hat{t}^{(n)}_{k+1, \delta}(v,w) = \star_{\xi_{\Gn}(w,v)} =
\hat{t}^{(n)}_{k+1,\delta}(\pp{v}, \pp{w}) = \star_{\xi_{\Gn}(\pp{w}, \pp{v})}$,
which in particular implies that $\xi_{\Gn}(w,v) = \xi_{\Gn}(\pp{w}, \pp{v})$,
$t^{(n)}_{k+1}(v,w) \notin \mF^{(\delta, k+1)}$, and
$t^{(n)}_{k+1}(\pp{v},\pp{w}) \notin \mF^{(\delta, k+1)}$. Note that
$t^{(n)}_{k+1}(v,w) \notin \mF^{(\delta, k+1)}$ implies that either
$\deg_{\Gn}(v) > \delta$, or $t^{(n)}_k(u,v) \notin \mF^{(\delta, k)}$ for some
$u \sim_{\Gn} v, u \neq w$. If $\deg_{\Gn}(v) > \delta$, by definition, we have
$\tM_k(v,w) = (0,\xi_{\Gn}(w,v))$. If $t^{(n)}_k(u,v) \notin \mF^{(\delta,
  k)}$ for some $u \sim_{\Gn} v, u \neq w$, by \eqref{eq:tisstar-1-iff-t-notin}
for $k-1$, we have
$\mathsf{TIsStar}^{(k-1)}(T_{k-1}(u,v)) = 1$. Thereby, again by definition, we
have $\tM_k(v,w) = (0,\xi_{\Gn}(w,v))$. Thereby, in either case we have 
$\tM_k(v,w) = (0,\xi_{\Gn}(w,v))$. Similarly, since
$t^{(n)}_{k+1}(\pp{v}, \pp{w}) \notin \mF^{(\delta, k+1)}$, we have
$\tM_k(\pp{v}, \pp{w}) = (0,\xi_{\Gn}(\pp{w}, \pp{v}))$. Since
$\xi_{\Gn}(w,v) = \xi_{\Gn}(\pp{w}, \pp{v})$, we realize that $\tM_k(v,w)
= \tM_k(\pp{v},\pp{w})$, which by definition implies that $T_k(v,w) =
T_k(\pp{v}, \pp{w})$.

\underline{Case 2:} $t^{(n)}_{k+1}(v,w) \in \mF^{(\delta, k+1)}$ and
$t^{(n)}_{k+1}(\pp{v}, \pp{w}) \in \mF^{(\delta, k+1)}$, which implies that
$\hat{t}^{(n)}_{k+1, \delta}(v,w) = t^{(n)}_{k+1}(v,w)$, and $\hat{t}^{(n)}_{k+1,
  \delta}(\pp{v}, \pp{w}) = t^{(n)}_{k+1}(\pp{v}, \pp{w})$. Thereby, we have
$t^{(n)}_{k+1}(v,w) = t^{(n)}_{k+1}(\pp{v}, \pp{w}) \in \mF^{(\delta, k+1)}$.
This implies that the following must hold:
\begin{enumerate}[$(i)$]
\item $\tau_{\Gn}(v) = \tau_{\Gn}(\pp{v})$.
\item $\deg_{\Gn}(v) = \deg_{\Gn}(\pp{v}) \leq \delta$.
\item $\xi_{\Gn}(w,v) = \xi_{\Gn}(\pp{w}, \pp{v})$.
\item $t^{(n)}_k(u,v) \in \mF^{(\delta, k)}$ for all $u \sim_{\Gn} v, u \neq w$.
  Similarly, $t^{(n)}_k(\pp{u}, \pp{v}) \in \mF^{(\delta, k)}$ for all $\pp{u}
  \sim_{\Gn} \pp{v}, \pp{u} \neq \pp{w}$.
  \item There exists a bijection $\pi: \{u \sim_{\Gn} v: u \neq w\} \rightarrow
    \{\pp{u} \sim_{\Gn} \pp{v}: \pp{u} \neq \pp{w}\}$, such that for all $u
    \sim_{\Gn} v, u \neq w$, we have $t^{(n)}_k(u,v) = t^{(n)}_k(\pi(u),
    \pp{v})$, and $\xi_{\Gn}(u,v) = \xi_{\Gn}(\pi(u), \pp{v})$.
\end{enumerate}
Now,  $(iv)$ together with $(v)$ implies that for all $u \sim_{\Gn} v,
u \neq w$, we have $\hat{t}^{(n)}_{k,\delta}(u,v) = t^{(n)}_k(u,v) =
t^{(n)}_k(\pi(u), \pp{v}) = \hat{t}^{(n)}_{k, \delta}(\pi(u), \pp{v})$.
Therefore, induction hypothesis implies that $T_{k-1}(u,v) =
T_{k-1}(\pi(u), \pp{v})$ for all $u \sim_{\Gn} v, u \neq w$. Moreover, $(iv)$
together with the induction hypothesis~\eqref{eq:tisstar-1-iff-t-notin} implies
that $\mathsf{TIsStar}^{(k-1)}(T_{k-1}(u,v)) = \mathsf{TIsStar}^{(k-1)}(T_{k-1}(\pi(u), \pp{v})) =0$ for all $u \sim_{\Gn} v, u \neq w$. Putting these together with conditions
$(i)$, $(ii)$, and $(iii)$ above, we realize that $\tM_k(v,w) =
\tM_k(\pp{v}, \pp{w})$, and hence $T_k(v,w) = T_k(\pp{v}, \pp{w})$.

To complete the inductive proof, we now show \eqref{eq:tisstar-1-iff-t-notin},
i.e.\  $\mathsf{TIsStar}^{(k)}(v,w) =
1$ iff $t^{(n)}_{k+1}(v,w) \notin \mF^{(\delta, k+1)}$. Note that, by
definition, we have $\mathsf{TIsStar}^{(k+1)}(v,w) = 1$ iff $\tM_k(v,w)$
starts with zero, which happens precisely when either $\deg_{\Gn}(v) > \delta$,
or for some $u \sim_{\Gn} v, u \neq w$, we have $\mathsf{TIsStar}^{(k-1)}(T_{k-1}(u,v)) =
1$. But by induction, the latter is equivalent to $t^{(n)}_{k, \delta}(u,v)
\notin \mF^{(\delta, k)}$ for some $u \sim_{\Gn} v, u \neq w$. To sum up, we
showed that $\mathsf{TIsStar}^{(k)}(v,w) =
1$ is equivalent to the condition that $\deg_{\Gn}(v) > \delta$, or for some $u
\sim_{\Gn} v, u \neq w$, $t^{(n)}_{k, \delta}(u,v)
\notin \mF^{(\delta, k)}$. But this is equivalent to $t^{(n)}_{k+1}(v,w) \notin
\mF^{(\delta, k+1)}$. This completes the proof.
\end{proof}

Algorithm~\ref{alg:type-extract-message-passing-new} below performs the above
procedure. At each step $0 \leq k \leq h-1$, the array $\sfT = (\sfT_{v,i}: v \in [n], i \in [\dn_v])$ keeps the
integer messages $T_k(.,.)$ discussed above, such that $\sfT_{v,i} = T_k(v,
\gamman_{v,i})$ for $v \in [n]$ and $i \in [\dn_v]$.   Note that
at each step $1 \leq k \leq h-1$, in order to update $T_k(.,.)$, we only need
the messages $T_{k-1}(.,.)$ form step $k-1$.  Therefore, at each step $0 \leq k
\leq h-1$, it suffices to store the messages corresponding to the
current step in array $\sfT$, and the messages corresponding to step $k-1$  in
another array  called $\tilde{\sfT} = (\tilde{\sfT}_{v,i}: v \in [n], i \in [\dn_v])$,
so that $\tilde{\sfT}_{v,i} = T_{k-1}(v,\gamman_{v,i})$ for $v \in [n]$ and $i \in
[\dn_v]$. Also, as we discussed above, since we need $\mathsf{TIsStar}^{(k-1)}$ in order to
update messages at step $k$,  we make a copy of the
$\mathsf{TIsStar}$ array from step $k-1$ and store it in another array named
$\mathsf{TIsStarOld}$. 

Now, we explain Algorithm~\ref{alg:type-extract-message-passing-new} in more
details. In the loop of line~\ref{l:message-init-zero}, we initialize the messages
for step $0$. In order to do so, we use the procedure $\textproc{SendMessage}$
described in Algorithm~\ref{alg:send-message}.
This Algorithm, which will be discussed later, implements the part of our above
discussion for converting messages $\tM_k(.,.)$ into integer messages $T_k(.,.)$
via the dictionary $\mathsf{TDictionary}$. More precisely, as we will see later,
at step $k$, 
$\textproc{SendMessage}(v,i,t)$ converts the message $t =
\tM_k(v,\gamman_{v,i})$ to an integer via $\mathsf{TDictionary}$ and updates the
arrays $\sfT$, $\mathsf{TIsStar}$, and $\mathsf{TMark}$. For instance, $\textproc{SendMessage}(v, i, (\thetan_v,
0,\xn_{v,i}))$ in line~\ref{l:mp-send-message-0} sends a message at step $0$ which is the
array $(\thetan_v, 0, \xn_{v,i})$ of size 3 from vertex $v$ towards its $i$th
neighbors which is $\gamman_{v,i}$. 
Figure~\ref{fig:mp-run-example-k0} illustrates the phase $k=0$ of the algorithm for
the graph of Figure~\ref{fig:sample-graph-for-compression-repeat}. 

\begin{figure}
  \centering
\begin{tabular}[t]{c@{\qquad}c}
  \begin{tabular}{@{}lcccr@{}}
\toprule
 $v$& $i$   &$\gamman_{v,i}$  & $\tM_0(v, \gamman_{v,i})$  &  $\sfT_{v,i} = T_0(v, \gamman_{v,i})$ \\ \midrule
 1 &  1 &  2 &  (\stack{\ttwo}{$\thetan_1$}, 0, \stack{\xtwo}{$\xn_{1,1}$})&  1\\
    2 &  1 &  1 &  (\stack{\tone}{$\thetan_2$}, 0, \stack{\xone}{$\xn_{2,1}$})&  2\\
    2 & 2 & 7 &  (\stack{\tone}{$\thetan_2$}, 0, \stack{\xone}{$\xn_{2,2}$}) & 2 \\
    7 & 1 & 2 & (\stack{\ttwo}{$\thetan_7$}, 0, \stack{\xone}{$\xn_{7,1}$}) & 3 \\
    7 & 2 & 8 & (\stack{\ttwo}{$\thetan_7$}, 0, \stack{\xtwo}{$\xn_{7,2}$}) & 1\\
                                                                               \bottomrule
  \end{tabular}%
    &%
      \begin{tikzpicture}[baseline=(pivot.center)]
        \node (pivot) at (0,-4) {};
        \begin{scope}
          \node at (0,0) {$\mathsf{TDictionary}$:};
          \begin{scope}[yshift=-1cm]
            \node at (0,0) {Key};
            \node at (1.5,0) {Value};
            \node[scale=0.9,text width=3.5cm,align=center] at (4,0) {corresponding element in $\bar{\mF}^{(4,1)}$};
            \draw (-0.8,-0.5) -- (5.5,-0.5);
            \begin{scope}[yshift=-1cm]
              \node at (0,0) {(\ttwo, 0, \xtwo)};
              \node at (1.5,0) {1};
              \begin{scope}[xshift=4cm]
                \node[nodeR] (r) at (0,0) {};
                \draw[edgeO] ($(r)+(0,0.3)$) -- (r);
              \end{scope}
            \end{scope}
            \begin{scope}[yshift=-2cm]
              \node at (0,0) {(\tone, 0, \xone)};
              \node at (1.5,0) {2};
              \begin{scope}[xshift=4cm]
                \node[nodeB] (r) at (0,0) {};
                \draw[edgeB] ($(r)+(0,0.3)$) -- (r);
              \end{scope}
            \end{scope}
            \begin{scope}[yshift=-3cm]
              \node at (0,0) {(\ttwo, 0, \xone)};
              \node at (1.5,0) {3};
              \begin{scope}[xshift=4cm]
                \node[nodeR] (r) at (0,0) {};
                \draw[edgeB] ($(r)+(0,0.3)$) -- (r);
              \end{scope}
            \end{scope}
          \end{scope}
        \end{scope} 
        \begin{scope}[yshift=-5.5cm]
          \node at (0,0) {$\mathsf{TMark}$:};
          \begin{scope}[xshift=1.5cm,scale=0.9]
            \draw[thick] (-0.5,-0.5) rectangle (2.5,0.5);
            \draw (0.5,-0.5) -- (0.5,0.5) (1.5,-0.5) -- (1.5,0.5);
            \node at (0,0) {\xtwo};
            \node at (0,-0.9) {1};
            \node at (1,0) {\xone};
            \node at (1,-0.9) {2};
            \node at (2,0) {\xone};
            \node at (2,-0.9) {3};
          \end{scope}
        \end{scope}
        \begin{scope}[yshift=-7.5cm]
          \node at (0,0) {$\mathsf{TIsStar}$:};
          \begin{scope}[xshift=1.5cm,scale=0.9]
            \draw[thick] (-0.5,-0.5) rectangle (2.5,0.5);
            \draw (0.5,-0.5) -- (0.5,0.5) (1.5,-0.5) -- (1.5,0.5);
            \node at (0,0) {0};
            \node at (0,-0.9) {1};
            \node at (1,0) {0};
            \node at (1,-0.9) {2};
            \node at (2,0) {0};
            \node at (2,-0.9) {3};
          \end{scope}
        \end{scope}
      \end{tikzpicture}
\end{tabular}
\caption{Example of running step $k=0$ of the message passing algorithm for finding edge
    types (Algorithm~\ref{alg:type-extract-message-passing-new}, loop of
    line~\ref{l:message-init-zero}) for the graph
    of Figure~\ref{fig:sample-graph-for-compression-repeat} with parameters
    $\delta=4$ and $h = 2$. Here, $\tM_0(v,w)$ is defined to be $M_0(v,w)$ for
    $v \sim_{\Gn} w$. Due to the symmetry in the graph, we have only
    presented a subset of $(v,i)$ pairs, e.g.\ $v=1, i \in \{2,\dots, 5\}$ is
    identical to $v=1,i=1$.}
\label{fig:mp-run-example-k0}
\end{figure}


Next, for each $1 \leq k \leq h-1$, we try to update messages for
depth $k$ in the loop of line~\ref{l:mp-for-k}. In order to do so, we first make
copies of $\mathsf{TIsStar}$ and $\sfT$ and store them in $\mathsf{TIsStarOld}$
and $\tilde{\sfT}$ respectively. Then, we  iterate
over all the vertices $1 \leq v \leq n$ in the loop of line~\ref{l:mp-for-v}. If
$\dn_v > \delta$,  by definition, the message sent from $v$ towards all its
neighbors must be of the form $(0,x)$ for some $x \in \edgemark$. This is done in the loop of
line~\ref{l:mp-dv-star-messages} using the $\textproc{SendMessage}$ procedure.
If $\dn_v \leq \delta$, we go over all the neighbors of $v$ in the loop of
line~\ref{l:mp-si-aggregate-for} to collect messages from the previous step. For
the $i$th neighbors of $v$, $1 \leq i \leq \dn_v$, we store the message sent from
$\gamman_{v,i}$ from the previous step, which is precisely
$\tilde{\sfT}_{\gamman_{v,i}, \tilde{\gamma}^{(n)}_{v,i}}$, together with the mark of
the corresponding edge $\xn_{v,i} =
\xi_{\Gn}(\gamman_{v,i}, v)$ in the $i$th index of an array $\vs$. This is done in
line~\ref{l:mp-si-set}. Note that the array $\vs$, which is initialized in line~\ref{l:mp-s-init}, needs to store at most $\delta$ many entries. Also, we check
whether the received message $\tilde{\sfT}_{\gamman_{v,i},
  \tilde{\gamma}^{(n)}_{v,i}}$ is a star message. This is  verified by evaluating
$\mathsf{TIsStarOld}(\tilde{\sfT}_{\gamman_{v,i}, \tilde{\gamma}^{(n)}_{v,i}})$ in
line~\ref{l:mp-tisstar-check}. If the message is star, we add one to the
variable $n_\star$, and keep the index $i$ in $i_\star$. Therefore, when the
loop is over, $n_\star$ will be the number of neighbors of $v$ which have sent a
star message towards $v$ at step $k-1$. Dependent on the value of $n_\star$, we
have  the
following three cases:
\begin{enumerate}
\item If $n_\star \geq 2$, following the
definition, indeed all the
neighbors of $v$ will receive a star message from $v$ at step $k$.
\item If $n_\star = 1$, all the  neighbor of $v$ expect for the one neighbor which had sent a star message
towards $v$ at step $k-1$, i.e.\ $\gamman_{v,i_\star}$, will receive a star
message from $v$ at step $k$.
\item If $n_\star = 0$, all the neighbors
of $v$ will receive non--star messages from $v$ at step $k$.
\end{enumerate}
 In order to figure
out the non--star messages at step $k$, we sort the array $s_{[1:\dn_v]}$ in
line~\ref{l:mp-sort-s} lexicographically and store the sorted array in a new
array called $\ts$. We also keep the ordering of original indices in another
array called $\pi$, so that $\ts_i = s_{\pi_i}, 1 \leq i \leq \dn_v$. Having the sorted array $\ts$,
in order to send a non--star message towards the $i$th neighbor of  $v$, we only need to
concatenate indices $1 \leq j \leq \dn_v$ of $\ts$ for which $\pi_j \neq i$.
This is done in lines~\ref{l:mp-append-t-ns-1} and \ref{l:mp-append-t-ns-0} for
$n_\star = 1$ and $n_\star = 0$, respectively. The result is stored in the array
$\vec{t}$ which basically stores the value of  vector messages $\tM_k(.,.)$
discussed above. Notice that since
$\dn_v \leq \delta$, the size of $\vec{t}$ is no more than $3 + 2 \delta$,
hence we need to only reserve this much memory for $\vec{t}$, as is done in line~\ref{l:mp-t-def}.
Figure~\ref{fig:mp-run-example-k1} illustrates the result of performing this
phase of the algorithm for $k=1$  for the graph of
Figure~\ref{fig:sample-graph-for-compression-repeat}.

Finally, after the last step $k = h-1$, we can form the array $\vec{c}$. In
order to do this, recall from~\eqref{eq:type-v-w-equivalent} that for $v \sim_{\Gn} w$, if $\UC_v(\Gn)[w,v]_{h-1}
\notin \mFdeltah$, or $\UC_v(\Gn)[v,w]_{h-1} \notin \mFdeltah$, or
$\deg_{\Gn}(v) > \delta$, or $\deg_{\Gn}(w) > \delta$, then $\type{v}{w}$
 must be
$(\star_{\xi_{\Gn}(w,v)}, \star_{\xi_{\Gn}(v,w)})$. In order to make sure this condition
is satisfied, we check all the messages after step $k = h-1$ in the loop of
line~\ref{l:mp-star-sym-for}. More precisely, for a vertex $v \in [n]$ and $i
\in [\dn_v]$, if $\dn_v = \deg_{\Gn}(v) > \delta$ or $\dn_{\gamman_{v,i}} =
\deg_{\Gn}(\gamman_{v,i}) > \delta$ or $\sfT_{\gamman_{v,i}, \tilde{\gamma}^{(n)}_{v,i}}$, i.e.\ the message sent
towards $v$ from its $i$th neighbor, is a star message, then the message that
$v$ sends towards its $i$th neighbor must also be a star message. This is
ensured in line~\ref{l:mp-star-sym}. Note that if $\mathsf{TIsStar}(\sfT_{v,i})
= 1$, then $\sfT_{v,i}$ has already been set to be a star message in the previously
discussed steps and no modification is required.
Finally, in the loop of line~\ref{l:mp-c-for}, for $1
\leq v \leq n$ and $1 \leq i \leq \dn_v$, we set $c_{v,i} = (\sfT_{v,i},
\sfT_{\gamman_{v,i}, \tilde{\gamma}^{(n)}_{v,i}})$.
Figure~\ref{fig:mp-run-example-after-sym} illustrates the result of performing
this phase for the graph of
Figure~\ref{fig:sample-graph-for-compression-repeat}.

Observe that at each point of time, using a loose bound, $\mathsf{TDictionary}$ has at most $4\mn$ many
key--value pairs. The reason is that there are $2\mn$ many neighbor pairs $v
\sim_{\Gn} w$  in the graph, and each time the $\textproc{SendMessage}$
procedure is called, at most one key--value pair is added to
$\mathsf{TDictionary}$. Moreover, in the loop of line~\ref{l:mp-star-sym-for},
using a loose bound, at most $2\mn$ extra key--value pairs can be added to
$\mathsf{TDictionary}$, yielding the $4\mn$ bound mentioned above. 
Therefore, for $v
\in [n]$ and $i \in [\dn_v]$, $\sfT_{v,i}$ is an integer with $O(\log \mn)$ bits.
Consequently, when collecting messages from the previous step, all the elements
of array $\vec{t}$ defined in line~\ref{l:mp-t-def} are bounded by $\max\{4m, n,
|\edgemark|, |\vermark|\}$. Likewise, every key in $\mathsf{TDictionary}$ is an
array of integers of size $O(\delta)$ with each of its indices being bounded by $\max\{4m, n,
|\edgemark|, |\vermark|\}$.

\begin{myalg}[Extracting edge types for a simple marked graph \label{alg:type-extract-message-passing-new}]
  \begin{algorithmic}[1]
    \INPUT
    \Statex $n$: number of vertices 
    \Statex $\Gn$: A simple marked graph on the vertex set $[n]$, vertex mark set
    $\vermark = \{1,    \dots, |\vermark|\}$ and edge mark set $\edgemark = \{1,
    \dots, |\edgemark|\}$ given in its neighbor list representation. More
    precisely, for a vertex $1 \leq v \leq n$, the
    following are given 
    \begin{itemize}
    \item $\dn_v$: the degree of vertex $v$
    \item $\thetan_v$: the vertex mark of $v$
    \item for $1 \leq i \leq \dn_v$, the tuple $(\gamman_{v,i}, \xn_{v, i}, \xnp_{v, i})$
      where $\gamman_{v,1} < \gamman_{v,2} < \dots < \gamman_{v,\dn_v}$ are the neighbors of vertex
      $v$ and for $1 \leq i \leq \dn_v$, $\xn_{v,i} = \xi_{\Gn}(\gamman_{v,i}, v)$ and
      $\xnp_{v,i} = \xi_{\Gn}(v, \gamman_{v,i})$.
      \item for $1 \leq i \leq \dn_v$, $\tilde{\gamma}^{(n)}_{v,i}$ which is the index
        of $v$ among the neighbors of $\gamman_{v,i}$, i.e.\
        $\gamman_{\gamman_{v,i}, \tilde{\gamma}^{(n)}_{v,i}} = v$.
    \end{itemize}
    \Statex $h$: the depth parameter
    \Statex $\delta$: the degree parameter
    \OUTPUT
    \Statex $\vec{c} = (c_{v,i}: v \in [n], i \in [\dn_v])$: $\Array \text{ of }
    \Array \text{ of integers}$, where for a vertex $v \in
    [n]$ and $1 \leq i \leq \dn_v$, $c_{v,i}$ represents $\type{v}{\gamman_{v,i}}$, the
    type of the edge between vertex $v$ and its $i$th neighbor. The object $c_{v,i}$ is a
    pair of integers where the first component corresponds to $\TT{v}{\gamman_{v,i}}$  and the second corresponds to $\TT{\gamman_{v,i}}{v}$. More precisely,
    with the mapping $J_n$ in Proposition~\ref{prop:MP}, we have $c_{v,i} =
    (J_n(\TT{v}{\gamman_{v,i}}), J_n(\TT{\gamman_{v,i}}{v}))$ for all $v \in
    [n]$ and $i \in [\dn_v]$. 
    \Statex $\mathsf{TCount}$: the number of explored messages at step $h-1$.
    \Statex $\mathsf{TIsStar}$: an $\Array$ of bits with size $\mathsf{TCount}$, where
    for $1 \leq i \leq \mathsf{TCount}$, $\mathsf{TIsStar}(i)$ is 1 if the
    member of $\mFbardeltah$ corresponding to integer $i$, i.e. $J_n^{-1}(i)$,
    is of the form $\star_x$, and $\mathsf{TIsStar}(i) = 0$  otherwise. In other words,
    $\mathsf{TIsStar}(i) = \one{J_n^{-1}(i) \notin \mFdeltah}$.
    \Statex $\mathsf{TMark}$: an $\Array$ of integers of size $\mathsf{TCount}$,
    where for $1 \leq i \leq \mathsf{TCount}$, $\mathsf{TMark}(i)$ is the mark
    component associated to the member of $\mFbardeltah$ corresponding to integer $i$,
    i.e. $J_n^{-1}(i)$.  In other words, if
    $\mathsf{TIsStar}(i) = 1$, with $J_n^{-1}(i) = \star_x$, we have
    $\mathsf{TMark}(i) = x$; otherwise, if $\mathsf{TIsStar}(i) = 0$, we have
    $\mathsf{TMark}(i) = (J_n^{-1}(i))[m]$, i.e.\ the mark component of
    $J_n^{-1}(i) \in \mFdeltah$.
    \Function{ExtractTypes}{$n, \Gn, \delta, h$}
    \State $\mathsf{TDictionary} \gets \Dictionary(\Array \text{ of integers }
    \rightarrow \nats)$
    \State $\mathsf{TMark} \gets \Array \text{ of integers }$
    \State $\mathsf{TIsStar} \gets \Array \text{ of bits}$
    \State $\sfT = (\sfT_{v,i}: v\in [n], i \in [\dn_v]) \gets \Array \text{ of }
    \Array \text{ of integers}$ \Comment{array of messages}
    \State $\mathsf{TCount} \gets 0$ \Comment{Number of elements in $\mathsf{TDictionary}$}
    \For{$1 \leq v \leq n$} \label{l:message-init-zero}\Comment{initialize messages at step $0$} 
    \For{$1 \leq i \leq \dn_v$}
    \State $\textproc{SendMessage}(v,i,(\thetan_v, 0,
    \xn_{v,i}))$ \label{l:mp-send-message-0} \Comment{Algorithm~\ref{alg:send-message}}
    \EndFor
    \EndFor
    \State $\vs \gets \Array \text{ of } \nats \times \nats$ with size $\delta$ \label{l:mp-s-init}
    \For{$1 \leq k \leq h-1$} \label{l:mp-for-k}
    \State $\mathsf{TCount} \gets 0$
    \State $\mathsf{TIsStarOld} \gets \mathsf{TIsStar}$ \label{l:mp-tisstarold}\Comment{corresponding
      to the previous step}
    \State $\tilde{\sfT} \gets \sfT$ \label{l:mp:Ttile-copy} \Comment{messages from  the previous step}
    \State $\mathsf{TDictionary}, \mathsf{TIsStar}, \mathsf{TMark} \gets
    \emptyset$  \Comment{erase for the current step}
    \For{$1 \leq v \leq n$} \label{l:mp-for-v}
    \If{$\dn_v > \delta$} \Comment{in this case, all the neighbors will receive star messages}
    \For{$1 \leq i \leq \dn_v$} \label{l:mp-dv-star-messages}
    \State $\textproc{SendMessage}(v,i,(0,\xn_{v,i}))$ \Comment{Algorithm~\ref{alg:send-message}}
    \EndFor
    \Else
    \State $n_\star \gets 0$ \Comment{number of neighbors which have sent a star
      message in the previous step}
    \For{$1 \leq i \leq \dn_v$} \label{l:mp-si-aggregate-for}
    \State $s_i \gets (\tilde{\sfT}_{\gamman_{v,i},
      \tilde{\gamma}^{(n)}_{v,i}}, \xn_{v,i})$ \label{l:mp-si-set}
    \If{$\mathsf{TIsStarOld}(\tilde{\sfT}_{\gamman_{v,i}, \tilde{\gamma}^{(n)}_{v,i}}) =
      1$} \label{l:mp-tisstar-check}
    \State $n_\star \gets n_\star + 1$
    \State $i_\star \gets i$ \Comment{index of the neighbor who has sent a star
      message}
    \EndIf
    \EndFor
    \If{$n_\star \geq 2$} \Comment{all the neighbors will receive a star message}
    \For{$1 \leq i \leq \dn_v$} \label{l:mp-nstar-2-send}
    \State $\textproc{SendMessage}(v,i,(0,\xn_{v,i}))$ \Comment{Algorithm~\ref{alg:send-message}}
    \EndFor
    \Else
    \State $(\pi, \tilde{s}) \gets \textproc{Sort}(s_{[1:\dn_v]})$ \label{l:mp-sort-s}
    \Comment{$\tilde{s}$ is the sorted array such that $\tilde{s}_i =
      s_{\pi_i}$} \label{l:mp-t-def}
    \State $\vec{t} \gets \Array \text{ of integers with maximum size } 3+2 \delta$ \label{l:mp-t-def}
    \If{$n_\star = 1$}
    \State $t_1 \gets \thetan_v, t_2 \gets \dn_v - 1$ \Comment{preparing
      $\mathsf{T}_{v,i_\star}$, the message towards
      $i_\star$}
    \For{$ 1 \leq i \leq \dn_v$}
    \If{$\pi_i \neq i_\star$}
    \State append the first and the second component of $\tilde{s}_i$ to $\vec{t}$ \label{l:mp-append-t-ns-1}
    \State $\textproc{SendMessage}(v,i,(0,\xn_{v,i}))$ \Comment{Algorithm~\ref{alg:send-message}}
    \EndIf
    \EndFor
    \State append $\xn_{v,i_\star}$ to $\vec{t}$
    \State $\textproc{SendMessage}(v,i_\star, \vec{t})$ \Comment{Algorithm~\ref{alg:send-message}}
    \EndIf
    \If{$n_\star = 0$}
    \For{$1 \leq i \leq \dn_v$} 
    \State $\vec{t} \gets \Array $ of size $2$
    \State $t_1 \gets \thetan_v, t_2 \gets \dn_v - 1$
    \For{$1 \leq j \leq \dn_v$}
    \If{$\pi_j \neq i$}
    \State append the first and the second components of $\tilde{s}_j$ to $\vec{t}$ \label{l:mp-append-t-ns-0}
    \EndIf
    \EndFor
    \State append $\xn_{v,i}$ to $\vec{t}$
    \State $\textproc{SendMessage}(v,i,\vec{t})$ \Comment{Algorithm~\ref{alg:send-message}}
    \EndFor
    \EndIf
    \EndIf
    \EndIf
    \EndFor
    \EndFor
    \For{$1 \leq v \leq n$} \label{l:mp-star-sym-for}
    \For{$1 \leq i \leq \dn_v$}
    \If{$\mathsf{TIsStar}(\mathsf{T}_{v,i}) = 0$ and
      ($\mathsf{TIsStar}(\mathsf{T}_{\gamman_{v,i}, \tilde{\gamma}^{(n)}_{v,i}})
      = 1$ or $\dn_v > \delta$ or $\dn_{\gamman_{v,i}} > \delta$) }
    \State $\textproc{SendMessage}(v,i,(0,\xn_{v,i}))$  \label{l:mp-star-sym} \Comment{Algorithm~\ref{alg:send-message}}
    \EndIf
    \EndFor
    \EndFor
    \State $\vec{c} = (c_{v,i}: v \in [n], i \in [\dn_v]) \gets \Array \text{ of }
    \nats \times \nats$ \Comment{type of edges}
    \For{$1 \leq v \leq n$} \label{l:mp-c-for}
    \For{$1 \leq i \leq \dn_v$}
    \State $c_{v,i} \gets (\mathsf{T}_{v,i}, \mathsf{T}_{\gamman_{v,i}, \tilde{\gamma}^{(n)}_{v,i}})$
    \EndFor
    \EndFor
    \State \textbf{return} $(\vec{c}, \mathsf{TCount}, \mathsf{TIsStar}, \mathsf{TMark})$
    \EndFunction
  \end{algorithmic}
\end{myalg}

The procedure $\textproc{SendMessage}$ is explained in
Algorithm~\ref{alg:send-message} below. In order to send a message $t$ from a
vertex $v \in [n]$ towards its $i$th neighbor, we first search for $t$ among the
keys in $\mathsf{TDictionary}$. If we find $t$, we set $\sfT_{v,i}$ to be
the corresponding value. Otherwise, $t$ should be recognized as a new object, and
its corresponding value should be $1 + \mathsf{TCount}$. Note that by
definition, $t$ is a star message iff its first index is zero. Also, the mark
component of $t$ is its last index, whether $t$ is a star message or not. These
facts  are used to update
$\mathsf{TIsStar}$ and $\mathsf{TMark}$ arrays. Note that
Algorithm~\ref{alg:send-message} introduces a process which is part of
Algorithm~\ref{alg:type-extract-message-passing-new}, hence it has access to the
variables defined in Algorithm~\ref{alg:type-extract-message-passing-new}.

\begin{myalg}[Sending a  message from a node to one of its neighbors \label{alg:send-message}]
  \begin{algorithmic}[1]
    \INPUT
    \Statex $v$: the vertex from which the message is originated
    \Statex $i$: the index of the neighbor of $v$ to whom the message is being sent, so
    the message is from $v$ towards $\gamman_{v,i}$
    \Statex $t$: the message, which is an $\Array$ of integers
    \OUTPUT
    \Statex updates $\mathsf{TDictionary}, \mathsf{TMark}, \mathsf{TIsStar}$ and
    $\mathsf{T}$
    \Procedure{SendMessage}{$v, i, t$}
    \If{$t \in \mathsf{TDictionary}.\textproc{Keys}$}
    \State $\mathsf{T}_{v,i} \gets \mathsf{TDictionary}(t)$
    \Else
    \State $\mathsf{TDictionary}.\textproc{Insert}(t,
    1+\mathsf{TCount})$
    \State $\mathsf{T}_{v,i} \gets 1 +
    \mathsf{TCount}$ \label{l:send-message-Tvi-1+Tcount}
    \State $\mathsf{TCount} \gets 1 + \mathsf{TCount}$
    \State append $t_{\textproc{Size}(t)}$ at the end of $\mathsf{TMark}$
    \Comment{the mark component is the last index in array $t$}
    \If{$t_1 = 0$} \Comment{$t$ is a star message iff its first component is zero}
    \State append 1 at the end of $\mathsf{TIsStar}$
    \Else
    \State append 0 at the end of $\mathsf{TIsStar}$
    \EndIf \label{l:send-message-TIsStar-end-if}
    \EndIf 
    \EndProcedure
  \end{algorithmic}
\end{myalg}

Now, we are ready to  state and prove the following
Lemma~\ref{lem:message-passing-c-J-Tn-relation} to address the existence of the
mapping $J_n$ as was  discussed previously in Section~\ref{sec:main-find-edge-type},
which maps the set $\mTn$ defined in~\eqref{eq:mTn-deg} to the set $\{1, \dots,
\mathsf{TCount}\}$. Figure~\ref{fig:mp-Jn-example} illustrates the mapping $J_n$
for the graph of Figure~\ref{fig:sample-graph-for-compression-repeat} with
parameters $h = 2$ and $ \delta = 4$.

\begin{lem}
  \label{lem:message-passing-c-J-Tn-relation}
  After the termination of Algorithm~\ref{alg:type-extract-message-passing-new},
  with $\mTn$ defined in~\eqref{eq:mTn-deg}, i.e.\
  \begin{align*}
    \mTn &:= \{\TT{v}{w}: v \sim_{\Gn} w \} \cup \{\T{v}{w}: (v,w) \in \mBn\},\\ 
    \mBn &:= \{(v,w): v \sim_{\Gn} w, \T{v}{w} \in \mFdeltah \text{ and } (\T{w}{v} \notin \mFdeltah \text{ or } \deg_{\Gn}(v) > \delta \text{ or } \deg_{\Gn}(w) > \delta)\},
  \end{align*}
  there exists a one to one mapping $J_n: \mTn \rightarrow \{1, \dots,
  \mathsf{TCount}\}$, such that for all $v \in [n]$ and $i \in [\dn_v]$, we have
  \begin{equation}
    \label{eq:mp-Jn-lemma-cvi}
    c_{v,i} = (J_n(\TT{v}{\gamman_{v,i}}), J_n(\TT{\gamman_{v,i}}{v})).
  \end{equation}
  Moreover, we have $\mathsf{TCount} \leq 4\mn$.
\end{lem}

\begin{proof}
  Motivated by our above discussion, in
  Algorithm~\ref{alg:type-extract-message-passing-new},
  prior to the final modification phase in the loop in
  line~\ref{l:mp-star-sym-for}, we have $\sfT_{v,i} = T_{h-1}(v, \gamman_{v,i})$
  for all $v \in [n]$ and $i \in [\dn_v]$. Moreover, prior to the loop in line~\ref{l:mp-star-sym-for}, using
  Lemma~\ref{lem:MP-T-M-equivalent} for $k = h-1$, we realize that there is a
  one to one mapping from the set $\{\hat{t}^{(n)}_{h,\delta}(v,w) : v
  \sim_{\Gn} w\}$ to the set of integers that appear in $\{\sfT_{v,i}: v \in
  [n], i \in [\dn_v]\}$. Additionally, in the loop of
  line~\ref{l:mp-star-sym-for}, we modify $\sfT_{v,i}$ for those $v \in [n]$ and
  $i \in
  [\dn_v]$ where at least one of the following properties hold:
  \begin{enumerate}
  \item   $\mathsf{TIsStar}(\sfT_{v,i}) = 0$ and   $\mathsf{TIsStar}(\sfT_{\gamman_{v,i}, \tilde{\gamma}^{(n)}_{v,i}})= 1$.
  \item $\mathsf{TIsStar}(\sfT_{v,i}) = 0$ and ($\deg_{\Gn}(v) > \delta$ or
    $\deg_{\Gn}(w) > \delta$).
  \end{enumerate}
  Using~\eqref{eq:tisstar-1-iff-t-notin} in Lemma~\ref{lem:MP-T-M-equivalent},
  the first scenario takes place precisely for $v \in [n], i \in [\dn_v]$ such
  that $t^{(n)}_h(v,\gamman_{v,i}) \in \mFdeltah$ but $t^{(n)}_h(\gamman_{v,i},
  v) \notin \mFdeltah$. In this case, $\sfT_{v,i}$ will become an integer
  equivalent to $\star_{\xn_{v,i}}$. Similarly, the second condition takes place
  iff $t^{(n)}_h(v,\gamman_{v,i}) \in \mFdeltah$ but $\deg_{\Gn}(v) > \delta$ or
  $\deg_{\Gn}(w) > \delta$. 
Therefore, after the loop in
  line~\ref{l:mp-star-sym-for} terminates, there will be a one to one mapping $J_n$
  from the set
  \begin{align*}
    \tilde{\mT}^{(n)} &:= \{\hat{t}^{(n)}_h(v,w): v \sim_{\Gn} w\} \cup \{ \star_{\xi_{\Gn}(w,v)}: v \sim_{\Gn} w, t^{(n)}_h(v,w) \in \mFdeltah \text{ and } t^{(n)}_h(w,v) \notin \mFdeltah\}\\
    & \quad \cup \{  \star_{\xi_{\Gn}(w,v)} : v \sim_{\Gn}w, \T{v}{w} \in \mFdeltah \text{ and } (\deg_{\Gn}(v) > \delta \text{ or } \deg_{\Gn}(w) > \delta ) \}
  \end{align*}
  to the set of integers that appear as values in $\mathsf{TDictionary}$, i.e.\
  $\{1, \dots, \mathsf{TCount}\}$. Moreover, with this mapping, we have
  $\sfT_{v,i} = J_n(\tilde{t}^{(n)}_{h,\delta}(v, \gamman_{v,i}))$ for all $v
  \in [n]$ and $i \in [\dn_v]$. To complete the argument, it remains to show that
  $\tilde{\mT}^{(n)} = \mTn$. In order for this, let $\mAn := \{(v,w): v
  \sim_{\Gn} w\} - \mBn$, and note that for
  $(v,w) \in \mAn$,  $\tilde{t}^{(n)}_{h,\delta}(v,w) =
  \hat{t}^{(n)}_{h,\delta}(v,w)$. Moreover, for $(v,w)
  \in \mBn$, we have $\hat{t}^{(n)}_{h,\delta}(v,w) = t^{(n)}_h(v,w)$, while
  $\tilde{t}^{(n)}_{h, \delta}(v,w) = \star_{\xi_{\Gn}(w,v)}$. Therefore,
  we may write
  \begin{align*}
    \tilde{\mT}^{(n)} &= \{\hat{t}^{(n)}_{h,\delta}(v,w): (v,w) \in \mAn\} \cup \{\hat{t}^{(n)}_{h,\delta}(v,w): (v,w) \in \mBn\} \cup \{\star_{\xi_{\Gn}(w,v)}: (v,w) \in \mBn\} \\
                      &= \{\tilde{t}^{(n)}_{h,\delta}(v,w): (v,w) \in \mAn\} \cup \{t^{(n)}_h(v,w): (v,w) \in \mBn\} \cup \{\tilde{t}^{(n)}_{h,\delta}(v,w): (v,w) \in \mBn\} \\
                      &= \{ \tilde{t}^{(n)}_{h,\delta}(v,w): v \sim_{\Gn} w\} \cup \{ t^{(n)}_h(v,w): (v,w) \in \mBn\} \\
                      &= \mTn.
  \end{align*}
  Finally, since there are $2\mn$ many neighbor pairs $v \sim_{\Gn} w$, we have
  $|\mBn| \leq 2 \mn$ and 
  $|\mTn| \leq 4\mn$. Consequently, we have  $\mathsf{TCount} \leq 4\mn$ and the
  proof is complete.
\end{proof}






\subsection{Complexity of the Message Passing Algorithm}
\label{sec:MP-complexity}

In this section, we analyze the complexity of extracting edge types in
Algorithm~\ref{alg:type-extract-message-passing-new}. Prior to that, we analyze
the complexity of the $\textproc{SendMessage}$ procedure in
Algorithm~\ref{alg:send-message}. 

\begin{lem}
  \label{lem:send-message-complexity}
  The time complexity of the $\textproc{SendMessage}$ procedure in
  Algorithm~\ref{alg:send-message} is $O(\delta \log {\mn} (\log n + \log(|\edgemark|
  + |\vermark|)))$, where $n$ and ${\mn}$ are the number of vertices and edges in the
  input graph, respectively. 
\end{lem}

\begin{proof}
   As was discussed earlier, each key in $\mathsf{TDictionary}$ is an array 
  of integers of length $O(\delta)$, with each coordinate being an integer
  bounded  by $\max\{4\mn, n, |\edgemark|, |\vermark|\} = O({\mn} + n + |\edgemark| +
  |\vermark|)$. Therefore, the lexicographic comparison of two such arrays takes
  $O(\delta \log ({\mn} + n + |\edgemark| + |\vermark|))$ time.
  Using ${\mn} \leq n^2 $, we have $\log({\mn} + n + |\edgemark| + |\vermark|) =
  O(\log({\mn}+n) + \log (|\edgemark| + |\vermark|))= O(\log n + \log
  (|\edgemark|+|\vermark|))$.
  Also, as we discussed before, $\mathsf{TDictionary}$ has at most $4\mn$ many
  key--value pairs.
Thereby, the time
  complexity of searching for the array $t$ in
  $\mathsf{TDictionary}.\textproc{Keys}$ is $O(\delta \log {\mn} (\log n + \log (|\edgemark| + |\vermark|)))$. If we find $t$, we set $\sfT_{v,i}$ to be the
  corresponding value. Otherwise, we need to insert the key--value pair $(t,
  1+\mathsf{TCount})$ in $\mathsf{TDictionary}$, which similar to the above has a time complexity of
  $O(\delta \log {\mn} (\log n + \log(|\edgemark| + |\vermark|)))$.
  Consequently, the overall time complexity
  of the $\textproc{SendMessage}$ procedure is $O(\delta \log {\mn} (\log n +\log(
  |\edgemark| + |\vermark|)))$. 
\end{proof}

\begin{lem}
  \label{lem:message-passing-complexity}
  Given a simple marked graph ${\Gn}$ on the vertex set $[n]$ with ${\mn}$ edges,
  extracting types with parameters $\delta$ and $h$ in
  Algorithm~\ref{alg:type-extract-message-passing-new} has time complexity $O(({\mn}+n) h \delta \log n (\log n + \log(|\edgemark| + |\vermark|)))$  and requires storing
  $O(({\mn}+\delta) (\log n + \log (|\edgemark| + |\vermark|)))$ bits.
\end{lem}

\begin{proof}
  Using Lemma~\ref{lem:send-message-complexity} above, the time complexity of
  initializing messages in the loop of  line~\ref{l:message-init-zero} is $O(\delta {\mn} \log {\mn}
  (\log n + \log( |\edgemark| + |\vermark|)))$. Now, fix $1 \leq k \leq h-1$ and
  note that copying $\mathsf{TIsStar}$ and $\sfT$ in lines~\ref{l:mp-tisstarold}
  and \ref{l:mp:Ttile-copy} takes $O({\mn} \log {\mn})$ time as each have $O({\mn})$
  elements with each element having $O(\log {\mn})$ bits. Furthermore, for a fixed $1 \leq
  v \leq n$, if $\dn_v > \delta$, using Lemma~\ref{lem:send-message-complexity},
  the time complexity of sending messages to the neighbors of $v$ in the loop of
  line~\ref{l:mp-dv-star-messages} is $O(\delta \dn_v \log {\mn} (\log n + \log( |\edgemark| + |\vermark|)))$. Now, assume that $\dn_v \leq \delta$. In this case,
  going through the neighbors of $v$ in the loop of
  line~\ref{l:mp-si-aggregate-for} and filling the array $\vs$ takes
  $O(\delta(\log {\mn} + \log |\edgemark|))$ time.
  If $n_\star \geq 2$,
  similar to the above discussion, sending messages to the neighbors of $v$ in the
  loop of line~\ref{l:mp-nstar-2-send} takes $O(\delta \dn_v \log {\mn} (\log n + \log(|\edgemark| + |\vermark|)))$ time. If $n_\star < 2$, sorting $s_{[1:\dn_v]}$ in
  line~\ref{l:mp-sort-s} takes $O(\delta \log \delta (\log {\mn} + \log
  |\edgemark|))$ time, as every index in $\vs$ is a pair with first and the
  second coordinates being bounded by $4\mn$ and $|\edgemark|$, respectively.
  On the other hand, if $n_\star = 1$, sending messages to all neighbors of $v$,
including $i_\star$, takes $O(\delta \dn_v \log {\mn} (\log n + \log (|\edgemark| +
|\vermark|)))$ time. If $n_\star = 0$, preparing the messages for the $\dn_v
\leq \delta$ neighbors of $v$ takes $O({(\dn_v)}^2\log ({\mn} + n + |\edgemark| +
|\vermark|) )= O({(\dn_v)}^2 (\log n + \log (|\edgemark| + |\vermark|))) =
O(\delta \dn_v (\log n + \log (|\edgemark| + |\vermark|)) )$ time. 
Moreover, sending these
messages to the neighbors takes $O(\delta \dn_v \log {\mn} (\log n + \log (|\edgemark| +
|\vermark|)))$ time.

 To sum up, for a vertex $1 \leq v \leq n$, the time complexity of the loop of
 line~\ref{l:mp-for-v} is 
 \begin{equation*}
 O(\delta \dn_v \log {\mn} (\log n + \log (|\edgemark| +
 |\vermark|)) + \delta \log \delta (\log {\mn} + \log |\edgemark|)).
 \end{equation*}
Taking the summation over  $1 \leq v \leq n$, using $\sum_{v=1}^n \dn_v = 2\mn$, and then
taking the summation over $1 \leq k \leq h-1$, we realize that the time complexity of the
loop in line~\ref{l:mp-for-k} is
\begin{equation*}
  O\left(mh \delta \log {\mn} (\log n + \log(|\edgemark| + |\vermark|)) + n h \delta \log \delta (\log {\mn} + \log |\edgemark|)\right).
\end{equation*}
A similar analysis shows that the time complexity of the loop in
line~\ref{l:mp-star-sym-for} is $O({\mn} \delta \log {\mn} (\log n + \log (|\edgemark| +
|\vermark|)))$. Finally, forming the array $\vec{c}$ in the loop of
line~\ref{l:mp-c-for} takes $O({\mn} \log {\mn})$ time. 
Putting the above together, we
realize that the overall time complexity of the algorithm is 
\begin{equation*}
  O\left(mh \delta \log {\mn} (\log n + \log(|\edgemark| + |\vermark|)) + n h \delta \log \delta (\log {\mn} + \log |\edgemark|)\right) = O(({\mn}+n) h \delta \log n (\log n + \log(|\edgemark| + |\vermark|))),
\end{equation*}
where we have used the bounds $\log \mn =O(\log n)$ and $\delta \leq n$ to simplify the
terms.

Now, we analyze the memory complexity of the algorithm. Note that as we
discussed previously, each integer $\sfT_{v,i}, v \in [n], i \in [\dn_v]$ is
bounded by $4\mn$. Therefore, the memory required to store the array $\sfT$ and its copy
$\tilde{\sfT}$ is $O({\mn} \log {\mn})$. On the other hand, since
$\mathsf{TCount} \leq 4 \mn$,  the memory required to store the arrays $\mathsf{TMark}$ and
$\mathsf{TIsStar}$ is $O( {\mn} \log |\edgemark|)$. Furthermore, each key in
$\mathsf{TDictionary}$  is an array of length $O(\delta)$, where each element in
this array is bounded by $\max \{4\mn , n , |\edgemark|, |\vermark|\}$. Moreover,
the value corresponding to this key is $O({\mn})$. Also, there are $O({\mn})$ many
key--value pairs in the dictionary at any given time. Thereby, the memory
required to store $\mathsf{TDictionary}$ is $O({\mn} \log ({\mn} + n + |\edgemark| +
|\vermark|)) = O({\mn} (\log n + \log(|\edgemark| + |\vermark|)))$. In addition,
since the size of the array $\vec{s}$ defined in line~\ref{l:mp-s-init} is  $\delta$
and each of its elements is a pair of integers bounded by $\max\{4\mn, |\edgemark|\}$,
$\vec{s}$ requires $O(\delta \log ({\mn} + |\edgemark|))$ bits to be stored. The
same bound holds for the array $\tilde{s}$ defined in line~\ref{l:mp-sort-s} which is
the sorted version of $\vec{s}$. Also, the array $\pi$ defined in
line~\ref{l:mp-sort-s} requires $O(\delta \log \delta)$ bits to be stored. 
Additionally, the array $\vec{t}$ defined in line~\ref{l:mp-t-def} has $O(\delta)$
elements, each bounded by $\max \{4\mn, n, |\edgemark|, |\vermark|\}$, hence
requires $O( \delta \log ({\mn} + n + |\edgemark| + |\vermark|)) = O(\delta (\log n
+ \log(|\edgemark| + |\vermark|)))$ bits to be stored.
Finally, the memory required to store the array $\vec{c}$ is similar to that of
storing $\sfT$ and is $O({\mn} \log {\mn})$. Putting all the above together, the memory
complexity of the algorithm is $O(({\mn}+\delta) (\log n + \log (|\edgemark| +
|\vermark|)))$. Similar to the discussion on the time complexity, we have used
the bounds $\log \mn = O( \log n)$ and $\delta \leq n$ to simplify the terms. 
This completes the proof.
\end{proof}

 \begin{figure}
  \centering

\begin{tabular}[t]{c@{\qquad}c}
  \begin{tabular}{@{}lcccr@{}}
\toprule
 $v$& $i$   &$\gamman_{v,i}$  & $\tM_1(v, \gamman_{v,i})$  &  \begin{tabular}{@{}c@{}}$\sfT_{v,i}$ \\ {\tiny($=T_1(v,\gamman_{v,i})$)}\end{tabular} \\ \midrule
 1 &  1 &  2 &  (0, \stack{\xtwo}{$\xn_{1,1}$})&  1\\
    2 &  1 &  1 &  (\stack{\tone}{$\thetan_2$}, \stack{2}{$\dn_2-1$}, \stack{3}{$T_0(7,2)$}, \stack{\xone}{$\xn_{2,2}$}, \stack{3}{$T_0(8,2)$}, \stack{\xone}{$\xn_{2,3}$}, \stack{\xone}{$\xn_{2,1}$})&  2\\
    2 & 2 & 7 &  (\stack{\tone}{$\thetan_2$}, \stack{2}{$\dn_2-1$}, \stack{1}{$T_0(1,2)$}, \stack{\xone}{$\xn_{2,1}$}, \stack{3}{$T_0(8,2)$}, \stack{\xone}{$\xn_{2,3}$}, \stack{\xone}{$\xn_{2,2}$}) & 3 \\
    7 & 1 & 2 & (\stack{\ttwo}{$\thetan_7$}, \stack{1}{$\dn_7-1$}, \stack{1}{$T_0(8,7)$}, \stack{\xtwo}{$\xn_{7,2}$}, \stack{\xone}{$\xn_{7,1}$}) & 4 \\
    7 & 2 & 8 & (\stack{\ttwo}{$\thetan_7$}, \stack{1}{$\dn_7-1$}, \stack{2}{$T_0(2,7)$}, \stack{\xone}{$\xn_{7,1}$}, \stack{\xtwo}{$\xn_{7,2}$}) & 5\\
                                                                               \bottomrule
  \end{tabular}%
    &%
      \begin{tikzpicture}[baseline=(pivot.center)]
        \node (pivot) at (0,-5) {};
        \begin{scope}
          \node at (0,0) {$\mathsf{TDictionary}$:};
          \begin{scope}[yshift=-1cm]
            \node at (0,0) {Key};
            \node at (1.8,0) {Value};
            \node[scale=0.9,text width=3.5cm,align=center] at (4,0) {corresponding element in $\bar{\mF}^{(4,2)}$};
            \draw (-0.8,-0.5) -- (5.5,-0.5);
            \begin{scope}[yshift=-1cm]
              \node at (0,0) {(0,\xtwo)};
              \node at (1.8,0) {1};
              \begin{scope}[xshift=4cm]
                \node at (0,0) {$\star_{\xtwo}$};
              \end{scope}
            \end{scope}
            \begin{scope}[yshift=-2cm]
              \node at (0,0) {(\tone,2,3,\xone,3,\xone,\xone)};
              \node at (1.8,0) {2};
              \begin{scope}[xshift=4cm]
                \node[nodeB] (r) at (0,0) {};
                \draw[edgeB] ($(r)+(0,0.3)$) -- (r);
                \node[nodeR] (r1) at (-0.4,-0.4) {};
                \node[nodeR] (r2) at (0.4,-0.4) {};
                \drawedge{r}{r1}{B}{B}
                \drawedge{r}{r2}{B}{B}
              \end{scope}
            \end{scope}
            \begin{scope}[yshift=-3cm]
              \node at (0,0) {(\tone,2,1,\xone,3,\xone,\xone)};
              \node at (1.8,0) {3};
              \begin{scope}[xshift=4cm]
                \node[nodeB] (r) at (0,0) {};
                \node[nodeR] (r1) at (-0.4,-0.4) {};
                \node[nodeR] (r2) at (0.4,-0.4) {};
                \draw[edgeB] ($(r)+(0,0.3)$) -- (r);
                \drawedge{r}{r1}{B}{B}
                \drawedge{r}{r2}{B}{O}
              \end{scope}
            \end{scope}
            \begin{scope}[yshift=-4cm]
              \node at (0,0) {(\ttwo,1,1,\xtwo,\xone)};
              \node at (1.8,0) {4};
              \begin{scope}[xshift=4cm]
                \node[nodeR] (r) at (0,0) {};
                \draw[edgeB] ($(r)+(0,0.3)$) -- (r);
                \node[nodeR] (r1) at (0,-0.4) {};
                \draw[edgeO] (r) -- (r1);
              \end{scope}
            \end{scope}
            \begin{scope}[yshift=-5cm]
              \node at (0,0) {(\ttwo,1,2,\xone,\xtwo)};
              \node at (1.8,0) {5};
              \begin{scope}[xshift=4cm]
                \node[nodeR] (r) at (0,0) {};
                \draw[edgeO] ($(r)+(0,0.3)$) -- (r);
                \node[nodeB] (r1) at (0,-0.4) {};
                \drawedge{r}{r1}{B}{B}
              \end{scope}
            \end{scope}
          \end{scope}
        \end{scope} 
        \begin{scope}[xshift=-1cm,yshift=-7.5cm]
          \node at (0,0) {$\mathsf{TMark}$:};
          \begin{scope}[xshift=1.5cm,scale=0.9]
            \draw[thick] (-0.5,-0.5) rectangle (4.5,0.5);
            \draw (0.5,-0.5) -- (0.5,0.5) (1.5,-0.5) -- (1.5,0.5) (2.5,-0.5) -- (2.5,0.5) (3.5,-0.5) -- (3.5,0.5);
            \node at (0,0) {\xtwo};
            \node at (0,-0.9) {1};
            \node at (1,0) {\xone};
            \node at (1,-0.9) {2};
            \node at (2,0) {\xone};
            \node at (2,-0.9) {3};
            \node at (3,0) {\xone};
            \node at (3,-0.9) {4};
            \node at (4,0) {\xtwo};
            \node at (4,-0.9) {5};
          \end{scope}
        \end{scope}
        \begin{scope}[xshift=-1cm,yshift=-9.5cm]
          \node at (0,0) {$\mathsf{TIsStar}$:};
          \begin{scope}[xshift=1.5cm,scale=0.9]
            \draw[thick] (-0.5,-0.5) rectangle (4.5,0.5);
            \draw (0.5,-0.5) -- (0.5,0.5) (1.5,-0.5) -- (1.5,0.5) (2.5,-0.5) -- (2.5,0.5) (3.5,-0.5) -- (3.5,0.5);
            \node at (0,0) {1};
            \node at (0,-0.9) {1};
            \node at (1,0) {0};
            \node at (1,-0.9) {2};
            \node at (2,0) {0};
            \node at (2,-0.9) {3};
            \node at (3,0) {0};
            \node at (3,-0.9) {4};
            \node at (4,0) {0};
            \node at (4,-0.9) {5};
          \end{scope}
        \end{scope}
      \end{tikzpicture}
\end{tabular}
  \caption{Example of running step $k=1$ of the message passing algorithm for finding edge
    types (Algorithm~\ref{alg:type-extract-message-passing-new}, loop of
    line~\ref{l:mp-for-k} with $k=1$) for the graph
    of Figure~\ref{fig:sample-graph-for-compression-repeat} with parameters
    $\delta=4$ and $h = 2$. Due to the symmetry in the graph, we have only
    presented a subset of $(v,i)$ pairs, e.g.\ $v=1, i \in \{2,\dots, 5\}$ is
    identical to $v=1,i=1$. The values of $\tM_1(.,.)$ are derived with
    reference to the values corresponding to step 0 which were presented in
    Figure~\ref{fig:mp-run-example-k0}.}
  \label{fig:mp-run-example-k1}
\end{figure}


 \begin{figure}
  \centering
  \begin{tabular}[t]{c@{\qquad}c}
  \begin{tabular}{@{}lcccr@{}}
\toprule
 $v$& $i$   &$\gamman_{v,i}$  & $\sfT_{v,i}$ &  $c_{v,i}$ \\ \midrule
 1 &  1 &  2 &    1 &(1,6)\\
    2 &  1 &  1 &  6 & (6,1)\\
    2 & 2 & 7 &  3 & (3,4)\\
    7 & 1 & 2 & 4 & (4,3)\\
    7 & 2 & 8 &  5& (5,5)\\
\bottomrule
  \end{tabular}%
    &%
      \begin{tikzpicture}[baseline=(pivot.center)]
        \node (pivot) at (0,-5) {};
        \begin{scope}
          \node at (0,0) {$\mathsf{TDictionary}$:};
          \begin{scope}[yshift=-1cm]
            \node at (0,0) {Key};
            \node at (1.8,0) {Value};
            \node[scale=0.9,text width=3.5cm,align=center] at (4,0) {corresponding element in $\bar{\mF}^{(4,2)}$};
            \draw (-0.8,-0.5) -- (5.5,-0.5);
            \begin{scope}[yshift=-1cm]
              \node at (0,0) {(0,\xtwo)};
              \node at (1.8,0) {1};
              \begin{scope}[xshift=4cm]
                \node at (0,0) {$\star_{\xtwo}$};
              \end{scope}
            \end{scope}
            \begin{scope}[yshift=-2cm]
              \node at (0,0) {(\tone,2,3,\xone,3,\xone,\xone)};
              \node at (1.8,0) {2};
              \begin{scope}[xshift=4cm]
                \node[nodeB] (r) at (0,0) {};
                \draw[edgeB] ($(r)+(0,0.3)$) -- (r);
                \node[nodeR] (r1) at (-0.4,-0.4) {};
                \node[nodeR] (r2) at (0.4,-0.4) {};
                \drawedge{r}{r1}{B}{B}
                \drawedge{r}{r2}{B}{B}
              \end{scope}
            \end{scope}
            \begin{scope}[yshift=-3cm]
              \node at (0,0) {(\tone,2,1,\xone,3,\xone,\xone)};
              \node at (1.8,0) {3};
              \begin{scope}[xshift=4cm]
                \node[nodeB] (r) at (0,0) {};
                \node[nodeR] (r1) at (-0.4,-0.4) {};
                \node[nodeR] (r2) at (0.4,-0.4) {};
                \draw[edgeB] ($(r)+(0,0.3)$) -- (r);
                \drawedge{r}{r1}{B}{B}
                \drawedge{r}{r2}{B}{O}
              \end{scope}
            \end{scope}
            \begin{scope}[yshift=-4cm]
              \node at (0,0) {(\ttwo,1,1,\xtwo,\xone)};
              \node at (1.8,0) {4};
              \begin{scope}[xshift=4cm]
                \node[nodeR] (r) at (0,0) {};
                \draw[edgeB] ($(r)+(0,0.3)$) -- (r);
                \node[nodeR] (r1) at (0,-0.4) {};
                \draw[edgeO] (r) -- (r1);
              \end{scope}
            \end{scope}
            \begin{scope}[yshift=-5cm]
              \node at (0,0) {(\ttwo,1,2,\xone,\xtwo)};
              \node at (1.8,0) {5};
              \begin{scope}[xshift=4cm]
                \node[nodeR] (r) at (0,0) {};
                \draw[edgeO] ($(r)+(0,0.3)$) -- (r);
                \node[nodeB] (r1) at (0,-0.4) {};
                \drawedge{r}{r1}{B}{B}
              \end{scope}
            \end{scope}
            \begin{scope}[yshift=-6cm]
              \node at (0,0) {(0,\xone)};
              \node at (1.8,0) {6};
              \begin{scope}[xshift=4cm]
                \node at (0,0) {$\star_{\xone}$};
              \end{scope}
            \end{scope}
          \end{scope}
        \end{scope} 
        \begin{scope}[yshift=-8.5cm]
          \node at (0,0) {$\mathsf{TMark}$:};
          \begin{scope}[xshift=1.5cm,scale=0.9]
            \draw[thick] (-0.5,-0.5) rectangle (5.5,0.5);
            \draw (0.5,-0.5) -- (0.5,0.5) (1.5,-0.5) -- (1.5,0.5) (2.5,-0.5) -- (2.5,0.5) (3.5,-0.5) -- (3.5,0.5) (4.5,-0.5) -- (4.5,0.5);
            \node at (0,0) {\xtwo};
            \node at (0,-0.9) {1};
            \node at (1,0) {\xone};
            \node at (1,-0.9) {2};
            \node at (2,0) {\xone};
            \node at (2,-0.9) {3};
            \node at (3,0) {\xone};
            \node at (3,-0.9) {4};
            \node at (4,0) {\xtwo};
            \node at (4,-0.9) {5};
            \node at (5,0) {\xone};
            \node at (5,-0.9) {6};
          \end{scope}
        \end{scope}
        \begin{scope}[yshift=-10.5cm]
          \node at (0,0) {$\mathsf{TIsStar}$:};
          \begin{scope}[xshift=1.5cm,scale=0.9]
            \draw[thick] (-0.5,-0.5) rectangle (5.5,0.5);
            \draw (0.5,-0.5) -- (0.5,0.5) (1.5,-0.5) -- (1.5,0.5) (2.5,-0.5) -- (2.5,0.5) (3.5,-0.5) -- (3.5,0.5) (4.5,-0.5) -- (4.5,0.5);
            \node at (0,0) {1};
            \node at (0,-0.9) {1};
            \node at (1,0) {0};
            \node at (1,-0.9) {2};
            \node at (2,0) {0};
            \node at (2,-0.9) {3};
            \node at (3,0) {0};
            \node at (3,-0.9) {4};
            \node at (4,0) {0};
            \node at (4,-0.9) {5};
            \node at (5,0) {1};
            \node at (5,-0.9) {6};
          \end{scope}
        \end{scope}

      \end{tikzpicture}
\end{tabular}
  \caption[MP example]{Example of running
    Algorithm~\ref{alg:type-extract-message-passing-new} on the graph of
    Figure~\ref{fig:sample-graph-for-compression-repeat} with parameters $\delta
    =4$ and $h = 2$. The values of the arrays $\sfT$ and $\vec{c}$ are presented
    after the algorithm termination.
    See Figures~\ref{fig:mp-run-example-k0} and \ref{fig:mp-run-example-k1} for
    the running procedure for steps $k=0$ and $k=1$ respectively. Due to the symmetry in the graph, we have only
    presented a subset of $(v,i)$ pairs, e.g.\ $v=1, i \in \{2,\dots, 5\}$ is
    identical to $v=1,i=1$.
    The values in this figure are  different from those in 
  Figure~\ref{fig:mp-run-example-k1} for step $k=1$ due to the final
  modification of the loop in line~\ref{l:mp-star-sym-for} of
  Algorithm~\ref{alg:type-extract-message-passing-new}. More precisely, since in
  Figure~\ref{fig:mp-run-example-k1}, the message that vertex 2 sends towards
  vertex 1 is not a star message, i.e.\ $\mathsf{TIsStar}(\sfT_{2,1}) = 0$,
  while the message that vertex 1 sends towards vertex 2 is a star message,
  i.e.\ 
  $\mathsf{TIsStar}(\sfT_{1,1}) = 1$, we should update the message $\sfT_{2,1}$
   so that it  becomes a star message representing $\star_{\xi_{\Gn}(1,2)}= \star_{\xone}$. This is done
   by adding the key--value pair $((0,\xone), 6)$ to $\mathsf{TDictionary}$ and
   setting $\sfT_{2,1} = 5$. 
}
  \label{fig:mp-run-example-after-sym}
\end{figure}
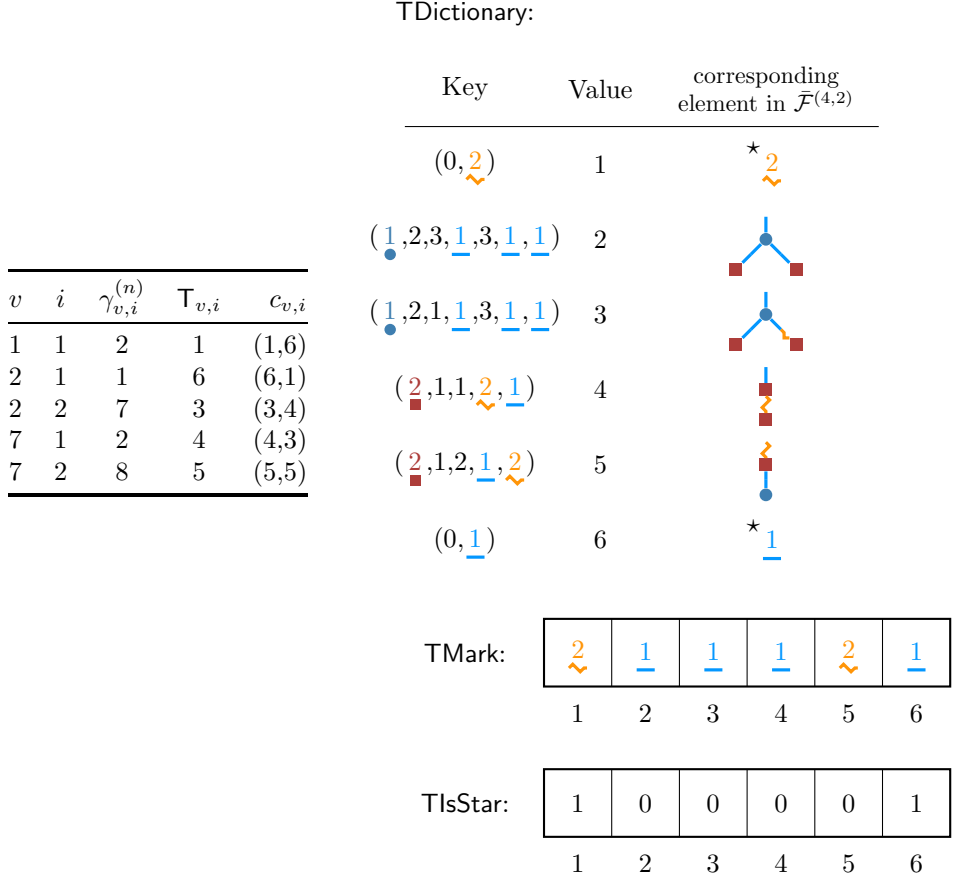


 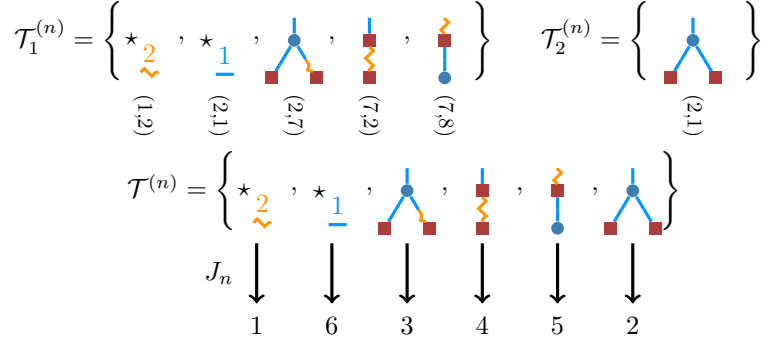
\begin{figure}
  \centering
  \begin{tikzpicture}
    \begin{scope}[xshift=-5cm]
      \node at (0,0) {$\mTn_1 =\Bigg \{$};
      \begin{scope}[xshift=1cm]
        \node at (0,-0.2) {$\star_{\xtwo}$};
      \end{scope}
      \begin{scope}[xshift=2cm]
        \node at (0,-0.2) {$\star_{\xone}$};
      \end{scope}
      \begin{scope}[xshift=3cm]
        \node[nodeB] (r) at (0,0) {};
        \node[nodeR] (r1) at (-0.3,-0.5) {};
        \node[nodeR] (r2) at (0.3,-0.5) {};
        \draw[edgeB] (r) -- ($(r)+(0,0.3)$);
        \drawedge{r}{r1}{B}{B}
        \drawedge{r}{r2}{B}{O}
      \end{scope}
      \begin{scope}[xshift=4cm]
        \node[nodeR] (r) at (0,0) {};
        \node[nodeR] (r1) at (0,-0.5) {};
        \draw[edgeB] (r) -- ($(r)+(0,0.3)$);
        \draw[edgeO] (r) -- (r1);
      \end{scope}
      \begin{scope}[xshift=5cm]
        \node[nodeR] (r) at (0,0) {};
        \node[nodeB] (r1) at (0,-0.5) {};
        \draw[edgeO] (r) -- ($(r)+(0,0.3)$);
        \drawedge{r}{r1}{B}{B}
      \end{scope}
      \foreach \x in {1.5,2.5,...,4.5}
      \node at (\x,0) {,};
      \node[rotate=-90,scale=0.8] at (1,-1) {(1,2)};
      \node[rotate=-90,scale=0.8] at (2,-1) {(2,1)};
      \node[rotate=-90,scale=0.8] at (3,-1) {(2,7)};
      \node[rotate=-90,scale=0.8] at (4,-1) {(7,2)};
      \node[rotate=-90,scale=0.8] at (5,-1) {(7,8)};
      
      \node at (5.5,0) {$\Bigg \}$};

    \end{scope}

    \begin{scope}[xshift=2cm]
      \node at (0,0) {$\mTn_2 =\Bigg \{$};
      \begin{scope}[xshift=1.3cm]
        \node[nodeB] (r) at (0,0) {};
        \node[nodeR] (r1) at (-0.3,-0.5) {};
        \node[nodeR] (r2) at (0.3,-0.5) {};
        \draw[edgeB] (r) -- ($(r)+(0,0.3)$);
        \drawedge{r}{r1}{B}{B}
        \drawedge{r}{r2}{B}{B}
      \end{scope}
      \node[rotate=-90,scale=0.8] at (1.3,-1) {(2,1)};
      
      \node at (2.1,0) {$\Bigg \}$};

    \end{scope}

    \begin{scope}[xshift=-3.5cm,yshift=-2cm]
      \node at (0,0) {$\mTn =\Bigg \{$};
      \begin{scope}[xshift=1cm]
        \node at (0,-0.2) {$\star_{\xtwo}$};
      \end{scope}
      \begin{scope}[xshift=2cm]
        \node at (0,-0.2) {$\star_{\xone}$};
      \end{scope}
      \begin{scope}[xshift=3cm]
        \node[nodeB] (r) at (0,0) {};
        \node[nodeR] (r1) at (-0.3,-0.5) {};
        \node[nodeR] (r2) at (0.3,-0.5) {};
        \draw[edgeB] (r) -- ($(r)+(0,0.3)$);
        \drawedge{r}{r1}{B}{B}
        \drawedge{r}{r2}{B}{O}
      \end{scope}
      \begin{scope}[xshift=4cm]
        \node[nodeR] (r) at (0,0) {};
        \node[nodeR] (r1) at (0,-0.5) {};
        \draw[edgeB] (r) -- ($(r)+(0,0.3)$);
        \draw[edgeO] (r) -- (r1);
      \end{scope}
      \begin{scope}[xshift=5cm]
        \node[nodeR] (r) at (0,0) {};
        \node[nodeB] (r1) at (0,-0.5) {};
        \draw[edgeO] (r) -- ($(r)+(0,0.3)$);
        \drawedge{r}{r1}{B}{B}
      \end{scope}
      \begin{scope}[xshift=6cm]
        \node[nodeB] (r) at (0,0) {};
        \node[nodeR] (r1) at (-0.3,-0.5) {};
        \node[nodeR] (r2) at (0.3,-0.5) {};
        \draw[edgeB] (r) -- ($(r)+(0,0.3)$);
        \drawedge{r}{r1}{B}{B}
        \drawedge{r}{r2}{B}{B}
      \end{scope}

      \foreach \x in {1.5,2.5,...,5.5}
      \node at (\x,0) {,};
      \node at (6.5,0) {$\Bigg \}$};

      \foreach \x in {1,...,6}
      \draw[->,very thick] (\x,-0.7) -- (\x,-1.5);
      \node at (0.5,-1.1) {$J_n$};
      \node at (1,-1.8) {1};
      \node at (2,-1.8) {6};
      \node at (3,-1.8) {3};
      \node at (4,-1.8) {4};
      \node at (5,-1.8) {5};
      \node at (6,-1.8) {2};
    \end{scope}
  \end{tikzpicture}
  \caption{The mapping $J_n$ for the graph of
    Figure~\ref{fig:sample-graph-for-compression-repeat} with parameters $h = 2$
  and $\delta =4$. Here, $\mTn_1$ denotes the set
  $\{\tilde{t}^{(n)}_{h,\delta}(v,w): v \sim_{\Gn} w\}$, while $\mTn_2$ denotes the
set $\{t^{(n)}_h(v,w): (v,w) \in \mBn\}$, so that $\mTn = \mTn_1 \cup \mTn_2$. For each element in $\mTn_1$ or
$\mTn_2$, we write one  corresponding pair $(v,w)$ below that element (note that
there might be more than one pair resulting in that element). In this
example, we have $\mBn = \{(v,1): 2 \leq v \leq 6\}$. Also, $\mathsf{TCount} = 6$, and the mapping $J_n$ can be found by
looking at the key--value mapping of $\mathsf{TDictionary}$ in
Figure~\ref{fig:mp-run-example-after-sym}. Comparing this with
Figure~\ref{fig:mp-run-example-after-sym}, it is easy to verify that the
relation~\eqref{eq:mp-Jn-lemma-cvi} holds in this example.}\label{fig:mp-Jn-example}
\end{figure}


\editfinish

\editstart

\section{A Compression Algorithm for Simple  Bipartite Unmarked Graphs with Given
  Degree Sequence}
\label{sec:bipartite-compression}

\ifreplace
In this section, we prove Proposition~\ref{prop:bipartite-color-graph-compress} in Section~\ref{sec:main-bip-comp}.
\color{black}
\fi
Recall that $\mGnlr_{\va, \vb}$ denotes
the set of simple bipartite  unmarked graphs $G$ with $n_l$ left vertices 
and  $n_r$
right vertices  
such that for $1 \leq i \leq n_i$ and $1
\leq j \leq n_r$, the degree of the left vertex $i$ is $a_i$, and the degree of
the right vertex $j$ is
$b_j$. We use $\pn$ for $\max\{n_l, n_r\}$. We assume that the degree sequence pair  $(\va, \vb)$  is graphical,
i.e.\ $\mGnlr_{\va, \vb} \neq \emptyset$.
\ifreplace
Our goal is to define
an encoding function $\fnlr_{\va, \vb} : \mGnlr_{\va, \vb} \rightarrow
\{0,1\}^* - \emptyset$, and a decoding function $\gnlr_{\va, \vb}$, such that $\fnlr_{\va,
  \vb}$ is prefix--free, and for all $G \in \mGnlr_{\va, \vb}$ we have
$\gnlr_{\va, \vb} \circ \fnlr_{\va, \vb}(G) = G$.
\color{black}
\fi


\ifreplace
A
graph $G  \in \mGnlr_{\va, \vb}$ can be represented via its adjacency list
$\vgamma^G = (\vgamma^G_v: v \in [n_l])$, where for $1 \leq v \leq n_l$,
$\vgamma^G_v = (\gamma^G_{v,1} <  \dots < \gamma^G_{v,a_v})$ are the right nodes
adjacent to the left node $v$.
\color{black}
\fi
\ifreplace
For $G \in \mGnlr_{\va, \vb}$, 
let
$A^G_{i,j} = 1$ if there is an edge between the left vertex $i$ and
right vertex $j$ in $G$, and zero otherwise.
\color{black}
\fi
\ifreplace
We 
write
    \begin{equation}
      \label{eq:bip-vA-vector}
      \vA^G = (A^G_{1,1}, \dots, A^G_{1,n_r}, \dots, A^G_{n_l,1}, \dots, A^G_{n_l,n_r}).
    \end{equation}
\color{black}
\fi
For $1 \leq i \leq n_l$, let $\vA^G_i$ denote the subvector $(A^G_{i,1},
    \dots, A^G_{i,n_r})$. Moreover, for $1 \leq i \leq n_l$ and $ 1 \leq j \leq n_r$, let $\vA^G_{\leq
      (i,j)}$ denote the subvector of $\vA$ up to the index $(i,j)$, i.e.\
    \begin{equation*}
      \vA^G_{\leq (i,j)} := (A^G_{1,1}, \dots A^G_{1,n_r}, \dots, A^G_{i,1}, \dots, A^G_{i,j}).
    \end{equation*}
We similarly define $\vA^G_{< (i,j)}$  with the difference that it excludes the
last index $(i,j)$.

\ifreplace
We now define
a set
$\mCnlr_{\va, \vb}$ of configurations. For each right node $1 \leq j \leq n_r$, we attach a set of
    distinct half--edges $R_{j} =
    \{r_{j,1}, \dots, r_{j,b_{j}}\}$  to $j$. 
\color{black}
\fi
With this, a
    configuration $C \in \mCnlr_{\va, \vb}$ is defined to be a partition  $(C_i: 1
    \leq i \leq n_l)$ of the set $R := \bigcup_{j=1}^{n_r} R_{j}$ such that $|C_i| =
    a_i$ for $1 \leq i \leq n_l$. Note that we have 
    \begin{equation*}
      |\mCnlr_{\va, \vb}| = \frac{S!}{\prod_{i=1}^{n_l} a_i!}.
    \end{equation*}   
where $S = \sum_{i=1}^{n_l} a_i = \sum_{j=1}^{n_r} b_j$.
\color{black}
    Given a configuration $C = (C_1, \dots, C_{n_l}) \in \mCnlr_{\va, \vb}$ as above, we construct  a
    bipartite multigraph with $n_l$ left nodes and $n_r$ right nodes, such that 
    the number of edges between a left node $i$ and a right node $j$
    is the number of half--edges belonging to node $j$ that appear in $C_i$. With this,
    define $\vA^C = (A^C_{i,j}: 1 \leq i \leq n_l, 1 \leq j \leq n_r)$ such that
    $A^C_{i,j}$ is the
    number of edges between the left node $i$ and the right node $j $ in the
    bipartite multigraph corresponding to $C$. 
    Similar to \eqref{eq:bip-vA-vector}, we represent $\vA^C$ as a row vector.
    We also define $\vA^C_i$, $\vA^C_{\leq (i,j)}$, and $\vA^C_{< (i,j)}$ for $1
    \leq i \leq n_l$ and $ 1 \leq j \leq n_r$ similar to the above.

    With this, for $G \in \mGnlr_{\va, \vb}$, we define
    \begin{equation*}
      N^{(n_l, n_r)}_{\va, \vb}(G) := |\{ C \in \mCnlr_{\va, \vb}: \vA^C \prec \vA^G\}|,
    \end{equation*}
    where $\prec$ denotes the lexicographic order when $\vA^C$ and $\vA^G$ are
    viewed as row vectors as in~\eqref{eq:bip-vA-vector}. To simplify the
    notation, we may remove $n_l, n_r, \va$ and $\vb$ from the notation whenever
    it is clear from the context and
    simply write $N(G)$. Finally, for $G \in \mGnlr_{\va,\vb}$, we define
    \begin{equation}
      \label{eq:bip-fn-va-vb-def} 
      \fnlr_{\va, \vb}(G) := \left\lceil  \frac{N(G)}{\prod_{j=1}^{n_r} b_j!} \right\rceil,
    \end{equation}
    to be the integer that encodes $G$. 
\ifreplace
The following lemma proves
    why $\fnlr_{\va, \vb}(G)$ is enough to uniquely determine $G$. 
\color{black}
\fi

     \begin{lem}
       \label{lem:bip-NG-uniquely-determines-G}
       Assume $G, \pp{G} \in \mGnlr_{\va, \vb}$ are distinct. Then, we have
       $\fnlr_{\va, \vb}(G) \neq \fnlr_{\va, \vb}(\pp{G})$. 
     \end{lem}
     \begin{proof}
       Without loss of generality, assume that $\vA^G \prec \vA^{\pp{G}}$. Then,
       \begin{equation}
         \label{eq:bip-NGp-NG}
         \begin{aligned}
           N(\pp{G}) &= \left| \left\{ C \in \mCnlr_{\va, \vb}: \vA^C \prec \vA^{\pp{G}} \right\} \right| \\
           &\geq \left| \left\{ C \in \mCnlr_{\va, \vb}: \vA^C \prec \vA^G \right\} \right| + \left| \left\{ C \in \mCnlr_{\va, \vb}: \vA^C = \vA^G \right\} \right|.
         \end{aligned}
       \end{equation}
       Note that $\vA^C = \vA^G$ implies that for all right nodes $1 \leq j \leq
       n_r$, $C$ assigns the half--edges of $j$ to the neighbors of $j$ in $G$.
       Since $G$ is a simple bipartite graph, the neighbors of a right node $j$  are $b_j$ 
       distinct left nodes. Hence there are $b_j! $ many ways of assigning half--edges of node $j$ to these left nodes. Consequently, we have
       \begin{equation*}
         \left| \left\{ C \in \mCnlr_{\va, \vb}: \vA^C = \vA^G \right\} \right| = \prod_{j=1}^{n_r} b_j!.
       \end{equation*}
       Substituting this into \eqref{eq:bip-NGp-NG}, we have
       \begin{equation*}
         N(\pp{G}) \geq N(G) + \prod_{j=1}^{n_r} b_j!.
       \end{equation*}
       Dividing both sides by $ \prod_{j=1}^{\pp{n}} b_j!$ and taking the
       ceiling, we realize that $\fnlr_{\va, \vb}(\pp{G}) \geq \fnlr_{\va, \vb}(G) +
       1$, which completes the proof.
     \end{proof}

    \editfinish
    


     \subsection{Compression Algorithm}
\label{sec:bip-encode}

\editstart

In this section, we introduce an algorithm to efficiently compute $\fnlr_{\va,\vb}(G)$
given $G \in \mGnlr_{\va, \vb}$. 
Recall that for $G \in \mGnlr_{\va, \vb}$ and a left node $1 \leq v \leq n_l$, we
denote by
$\gamma^G_{v,1} < \dots < \gamma^G_{v, a_v}$ the right nodes adjacent to
$v$ in $G$.
\ifreplace
For  $1 \leq i \leq n_l$ and $1 \leq j \leq n_r$, let
\color{black}
\fi
    \begin{equation*}
      b^G_j(i) := b_j - \sum_{k=1}^{i-1} \one{k \sim_G j}.
    \end{equation*}
Effectively, $b^G_j(i)$ denotes the
    number of free half--edges at the right node $j$ in a configuration $C$,
    assuming that the connections for the left vertices $1, \dots, i-1$ are made
    according to  $G$, i.e.\ $\vA^C_k = \vA^G_k$ for $1 \leq k \leq
    i-1$. 
    As the summation over an empty set is defined to be zero,
    we have $b^G_j(1) = b_j$.
Moreover, for $1 \leq i \leq n_l$ and $1 \leq j \leq n_r$ define
    \begin{equation*}
      U^G_j(i) := \sum_{k=j}^{n_r} b^G_k(i).
    \end{equation*}
    Also, for $j > n_r$, we define $U^G_j(i)$ to be zero.

    The following lemma gives an expression for $N(G)$ based on the above
    quantities.  Later on, we will introduce an efficient procedure to compute this
    expression.

    \begin{lem}
      \label{lem:NG-counting_bipartite}
      For $G \in \mGnlr_{\va, \vb}$, we have 
      \begin{equation*}
        N(G) = \sum_{i=1}^{n_l} \sum_{j=1}^{a_i} \left( \prod_{k=1}^{i-1} l^G_k \right) \left( \prod_{j'=1}^{j-1} b^G_{\gamma^G_{i,\pp{j}}}(i)  \right) \binom{U^G_{1+\gamma^G_{i,j}}(i)}{a_i - j + 1} \frac{S_{i+1}!}{\prod_{i'=i+1}^{n_l} a_{i'}!}.
      \end{equation*}
      Here, for $1 \leq k \leq n_l$,
      we define
      \begin{align*}
        l^G_k &:= \prod_{\pp{j}=1}^{a_k} b^G_{\gamma^G_{k,\pp{j}}}(k),  \\
        S_k &:= \sum_{\pp{k}=k}^{n_l} a_{\pp{k}},
      \end{align*}
      and $S_{n_l+1} := 0$.
    \end{lem}

        \begin{proof}
      Note that for $C \in \mCnlr_{\va, \vb}$, we have $\vA^C \prec \vA^G$ iff for
      some $1 \leq i \leq n_l$ and $1 \leq j \leq a_i$, we have
      \begin{subequations}
        \begin{gather}
          \vA^{C}_{<(i,\gamma^G_{i,j})} = \vA^G_{<(i,\gamma^G_{i,j})}, \label{eq:Bij-bip-1} \\
          A^C_{i,\gamma^G_{i,j}} = 0. \label{eq:Bij-bip-2}
        \end{gather}
      \end{subequations}
      Now, we fix $1 \leq i \leq n_l$ and $1 \leq j \leq a_i$ and count the
      number of configurations $C$ satisfying~\eqref{eq:Bij-bip-1} and \eqref{eq:Bij-bip-2}.
Observe that for these two
      conditions to
      hold for given $i$ and $j$, the following conditions must be satisfied: $(1)$ for $1 \leq k
      \leq i-1$, $C_k$ contains half--edges corresponding to the right
      nodes $\gamma^G_{k,1}, \dots, \gamma^G_{k,a_k}$,  $(2)$ $C_i$ contains half--edges
      corresponding to the right nodes $\gamma^G_{i,1}, \dots, \gamma^G_{i,j-1}$, and $(3)$ the rest
      of the $a_i - j + 1 $ members of $C_i$  must be chosen among the half--edges
      corresponding to the right nodes in the
      range $\gamma^G_{i,j} + 1, \dots, n_r$. Recall that for $1 \leq k < i$ and $1 \leq
      j' \leq a_k$, $b^G_{\gamma^G_{k,j'}}(k)$ is the number of free half--edges at the
      right node $\gamma^G_{k,j'}$ assuming that $\vA^C_{\pp{k}} = \vA^G_{\pp{k}}$ for $1
      \leq \pp{k} \leq k-1$. Therefore, to satisfy condition $(1)$ above, there
      are
      \begin{equation*}
        \prod_{k=1}^{i-1} \prod_{\pp{j} = 1}^{a_k} b^G_{\gamma^G_{k,\pp{j}}}(k) = \prod_{k=1}^{i-1} l^G_k,
      \end{equation*}
      many ways to choose $C_1, \dots, C_{i-1}$. Given this, to satisfy
      condition $(2)$, there are precisely
      \begin{equation*}
        \prod_{\pp{j} = 1}^{j-1} b^G_{\gamma^G_{i,\pp{j}}}(i),
      \end{equation*}
      many ways to pick free half--edges attached to the right nodes $\gamma^G_{i,1}, \dots,
      \gamma^G_{i,j-1}$. At this point, the total number of free half--edges corresponding to the
      right nodes in the range $\gamma^G_{i,j} + 1, \dots, n_r$ is precisely
      $U^G_{1+\gamma^G_{i,j}}(i)$. Therefore, to pick the remaining $a_i - j + 1$
      elements of $C_i$ to satisfy the condition $(3)$, we have
      \begin{equation*}
        \binom{U^G_{1+\gamma^G_{i,j}}(i)}{a_i - j + 1},
      \end{equation*}
      many options. Recall that we define $\binom{r}{s} = 0$ if $r < s$.
      Thereby, the above quantity evaluates to zero if
      $U^G_{1+\gamma^G_{i,j}}(i) < a_i - j +1$, i.e.\ if there are no enough free
      half--edges to satisfy condition (3).
Finally, to choose $C_{i+1}, \dots, C_{n_l}$, there are precisely
      \begin{equation*}
        \frac{S_{i+1}!}{\prod_{\pp{i} = i+1}^{n_l} a_{i'}!},
      \end{equation*}
      many ways. Putting the above together and summing over $i$ and $j$ in the
      specified range, we get the desired form for $N(G)$.
    \end{proof}

\subsubsection{A Recursive Form for $N(G)$}
\label{sec:bip-NG-recursive}

Now, we introduce a recursive form for $N(G)$ which is helpful for efficiently
computing it. 
for $1 \leq i \leq j \leq n_l$, we define
    \begin{subequations}
    \begin{align}
      l^G_{i,j} &:= \prod_{k=i}^j l^G_k \label{eq:benc-lij}\\
      z^G_i &:= \sum_{k=1}^{a_i} \left( \prod_{\pp{k}=1}^{k-1} b^G_{\gamma^G_{i,\pp{k}}}(i) \right)  \binom{U^G_{1+\gamma^G_{i,k}}(i)}{a_i - k + 1} \label{eq:benc-zi}\\
      r_i &:= \frac{S_i!}{a_i! S_{i+1}!} = \binom{S_i}{a_i}. \label{eq:benc-ri}
    \end{align}
  \end{subequations}
    Recall that $S_{n_l+1} = 0$ by definition, hence $r_{n_l} = 1$.
    Moreover, for $1 \leq i \leq j \leq n_l$, we define
    \begin{equation*}
      r_{i,j} := \prod_{k=i}^j r_k.
    \end{equation*}
    Also, following the convention that a product over an empty set evaluates
    to one, we let
    \begin{equation*}
      l^G_{i,j} = r_{i,j} = 1 \qquad \text{ for } i > j.
    \end{equation*}
Now, with the above definitions, we can rewrite the result of
    Lemma~\ref{lem:NG-counting_bipartite} as follows:
    \begin{equation*}
      N(G) = \sum_{i=1}^{n_l} l^G_{1,i-1} z^G_i r_{i+1,n_l}.
    \end{equation*}
Motivated by this form, for $1 \leq i \leq j \leq n_l$, we define 
    \begin{equation*}
      N_{i,j}(G) := \sum_{k=i}^j l^G_{i,k-1} z^G_k r_{k+1,j}.
    \end{equation*}
Also, following the convention that a summation over an empty set evaluates to
zero, we define
\begin{equation*}
  N_{i,j}(G) := 0 \qquad \text{ for } i > j.
\end{equation*}
Note that we have $N(G) = N_{1,n_l}(G)$ and $z^G_i = N_{i,i}(G)$  for $1 \leq i
\leq n_l$. 
Furthermore, for $1 \leq i \leq k \leq j \leq n_l$, we have the following
    recursive relation
    \begin{equation}
      \label{eq:bip-Nij-recurvie}
      N_{i,j}(G) = N_{i,k}(G) r_{k+1, j} + l^G_{i,k} N_{k+1, j}(G).
    \end{equation}

\subsubsection{The Encoding Algorithm}
\label{sec:bip-enc-pseudocode}

Motivated by the recursive form in \eqref{eq:bip-Nij-recurvie}, we introduce a
recursive algorithm that, given $G \in \mGnlr_{\va, \vb}$ and an interval $[i,j]$ for $1 \leq i \leq j \leq
n_l$, computes $N_{i,j}(G)$. In
order to compute $N(G)$, we use this algorithm for $i = 1$ and $j = n_l$. When
$i = j$, we use the formula in~\eqref{eq:benc-zi} to compute $N_{i,i}(G) =
z^G_i$. On the other hand,
in order to compute $N_{i,j}(G)$ for $j > i$, we recursively compute $N_{i,k}(G)$ and
$N_{k+1, j}(G)$ with $k
= \lfloor (i+j)/2 \rfloor$, and use \eqref{eq:bip-Nij-recurvie} to find
$N_{i,j}(G)$.
Note that when $i=j$, in order to compute $N_{i,i}(G)$,  from~\eqref{eq:benc-zi}, we need  $b^G_k(i)$ and
$U^G_k(i)$ for $1 \leq k \leq n_r$. For this, we assume that when the algorithm
is called for the interval $[i,j]$, an array $\vsfb =
(\sfb_k: 1 \leq k \leq n_r)$ is given as an input such that $\sfb_k =
b^G_k(i)$ for $1 \leq k \leq n_r$. Recall that $U^G_k(i) = \sum_{\pp{k} =
  k}^{n_r} b^G_{\pp{k}}(i)$ involves partial summations of the list
$(b^G_{\pp{k}}: 1 \leq \pp{k} \leq n_r)$. In order to efficiently find such
quantities, we employ a  Fenwick tree (see Section~\ref{sec:ds-fenwick} in
Appendix~\ref{sec:data-structures} for a review of this data structure).
More specifically, we assume that when the algorithm is called for the interval $[i,j]$,
a Fenwick tree $\sfU$ is given as an input where
$\sfU.\textproc{Sum}(k) = U^G_k(i)$ for $1 \leq k \leq n_r$.  The outputs of our
algorithm when called for the interval $[i,j]$ are: $(i)$ $N_{i,j}(G)$, $(ii)$
$l^G_{i,j}$, $(iii)$ the array $\vsfb$ updated so that $\sfb_k = b^G_k(j+1)$ for $1 \leq k \leq
 n_r$, and $(iv)$ the Fenwick tree $\sfU$ updated so that $\sfU.\textproc{Sum}(k) =
 U^G_k(j+1)$ for $1 \leq k \leq n_r$. In the following, we discuss the
 procedure for computing $N_{i,j}(G)$ for the two cases where $i=j$ and $i < j$
separately.

\underline{Case I: $i=j$} As we discussed above, in this case, we need to find
$N_{i,i}(G) = z^G_i$ defined in~\eqref{eq:benc-zi}. In order to compute the binomial coefficient
$\binom{U^G_{1+\gamma^G_{i,k}}(i)}{a_i - k + 1}$ in the definition of $z^G_i$ for $1 \leq k \leq a_i$, we use the function
$\textproc{ComputeProduct}$ of Algorithm~\ref{alg:compute-product}, which will
be  explained in
Section~\ref{sec:aux-algs-bip} below. For nonnegative integers $p, k$, and $s$, this
function returns
\begin{equation}
  \label{eq:compute-product-return}
  \textproc{ComputeProduct}(p,k,s) =
  \begin{cases}
    1, & \text{ if } k = 0; \\
    0, & \text{ if } k > 0, p-(k-1)s \leq 0; \\
    \prod_{\pp{k}=0}^{k-1} (p-\pp{k} s), & \text{ if } k > 0, p-(k-1)s > 0.
  \end{cases}
\end{equation}
In other words, we have
\begin{equation}
  \label{eq:compute-product-prod-plus-part}
  \textproc{ComputeProduct}(p,k,s) = \prod_{\pp{k}=0}^{k-1} (p-\pp{k} s)^+,
\end{equation}
where $(n)^+ := \max\{n, 0\}$ for an integer $n$, and the product over the empty set
evaluates to 1 by convention. With this, it is easy to see that for an integer
$k\geq 0$ we have $k! = \textproc{ComputeProduct}(k,k,1)$, and for integers $n
\geq 0$ and $ m \geq 0$, we have
\begin{equation*}
  \binom{n}{m} = \textproc{ComputeProduct}(n,m, 1) / \textproc{ComputeProduct}(m,m, 1).
\end{equation*}
Recall that $\binom{n}{m}$ is defined to be 
$\frac{n!}{m!(n-m)!}$ 
\color{black}
if $n \geq m$,
and zero otherwise. Employing this, for $1 \leq k \leq a_i$, we compute
$\binom{U^G_{1+\gamma^G_{i,k}}(i)}{a_i - k + 1}$ in line~\ref{l:bNij=y} of
Algorithm~\ref{alg:bip-Nij} below.  Furthermore, we use this to update $z_i$ and
$l_i$ in lines~\ref{l:bNij=z} and~\ref{l:bNij=l} so that $z_i$ becomes 
\begin{equation*}
  \sum_{k'=1}^{k} \left( \prod_{k''=1}^{k'-1} b^G_{\gamma^G_{i,\ppp{k}}}(i) \right)  \binom{U^G_{1+\gamma^G_{i,k'}}(i)}{a_i - k' + 1},
\end{equation*}
and $l_i$ becomes $\prod_{\pp{k} = 1}^k b^G_{\gamma^G_{i,\pp{k}}}(i)$. Therefore,
when the loop over $k$ is finished and $k$ reaches the upper limit $a_i$, $z_i$ becomes $z_i^G$ and $l_i$ becomes
$l_i^G$. Also, we update $\sfU$ and $\vsfb$ in lines~\ref{l:bNij=U} and
\ref{l:bNij=beta}, respectively.

\underline{Case II: $i<j$} In this case, with $k = \lfloor   (i+j) / 2 \rfloor$, we
compute $N_{i,k}(G)$ and $N_{k+1, j}(G)$ in lines~\ref{l:bNij-Nik} and
\ref{l:bNij-Nkj}, respectively. This is done in a recursive fashion by calling
Algorithm~\ref{alg:bip-Nij} for the intervals $[i,k]$ and $[k+1, j]$. These two
calls return $N_{i,k} = N_{i,k}(G)$, $N_{k+1,j} = N_{k+1, j}(G)$, as well as $l_{i,k} = l_{i,k}^G$ and $l_{k+1, j} = l_{k+1, j}^G$.
Comparing with the recursive form of~\eqref{eq:bip-Nij-recurvie}, the only
missing term for computing $N_{i,j}(G)$ is $r_{k+1, j}$. Recalling the
definition, we have
\begin{equation}
  \label{eq:bip-compression-r-k+1-j}
  r_{k+1, j} = \frac{S_{k+1}(S_{k+1}-1) \dots (S_{j+1}+1)}{a_{k+1}! \dots a_j!} = \frac{\prod_{\pp{k}=0}^{S_{k+1} - S_{j+1} - 1} (S_{k+1} - \pp{k})}{\prod_{\pp{k}= k+1}^j a_{\pp{k}} !}.
\end{equation}
If we compute $S_{k+1}$ and $S_{j+1}$, motivated
by~\eqref{eq:compute-product-return}, the numerator is precisely
$\textproc{ComputeProduct}(S_{k+1}, S_{k+1} - S_{j+1}, 1)$. In order to compute
$S_{k+1}$, note that after calling Algorithm~\ref{alg:bip-Nij} for the interval
$[i, k]$ in line~\ref{l:bNij-Nik}, the Fenwick tree $\sfU$ is updated so that
for $1 \leq s \leq n_r$, we have $\sfU.\textproc{Sum}(s) = U^G_s(k+1)$. In
particular, we have
\begin{equation}
  \label{eq:bip-U-1-is-S}
\begin{aligned}
  \sfU.\textproc{Sum}(1) = U^G_1(k+1) &= \sum_{s=1}^{n_r} b^G_s(k+1) \\
                                      &= \sum_{s=1}^{n_r} \left( b_s - \sum_{\pp{s}=1}^k \one{\pp{s} \sim_G s} \right )\\
                                      &= S - \sum_{\pp{s} = 1}^k \sum_{s=1}^{n_r} \one{\pp{s} \sim_G s} \\
                                      &= S - \sum_{\pp{s} = 1}^k a_{\pp{s}} \\
  &= \sum_{\pp{s} = k+1}^{n_l} a_{\pp{s}} = S_{k+1}.
\end{aligned}
\end{equation}
Hence, the calculation in line~\ref{l:bNij-Sk+1} correctly finds $S_{k+1}$.
Similarly, we compute $S_{j+1}$ in line~\ref{l:bNij-Sj+1}.  On the other hand,
in order to compute the denominator of~\eqref{eq:bip-compression-r-k+1-j}, which
is 
$\prod_{\pp{k} = k+1}^j a_{\pp{k}}!$, we use the function
$\textproc{ProdFactorial}$ of Algorithm~\ref{alg:comp-ProdAFactorial}, which will
be explained
in Section~\ref{sec:aux-algs-bip} below. This algorithm takes an array of integers
$\va$ and a pair of integers $p \leq q$, and returns $\prod_{v = p}^q a_v!$.
This is done in line~\ref{l:bNij-y} and the result is stored in variable $y$. Finally, we
divide the numerator by the denominator of~\eqref{eq:bip-compression-r-k+1-j} in
line~\ref{l:bNij-r} to find $r_{k+1, j}$. This is further used in
line~\ref{l:bNij-N} to compute $N_{i,j}$. Moreover, $l_{i,j}$ is calculated in
line~\ref{l:bNij-l} using the identity $l^G_{i,j} = l^G_{i,k} \times l^G_{k+1, j}$.


\begin{myalg}[Computing $N_{i,j}(G)$ for a simple unmarked bipartite graph $G \in
  \mGnlr_{\va, \vb}$ \label{alg:bip-Nij}]
\begin{algorithmic}[1]
  \INPUT
  \Statex $n_l, n_r$: the number of left and right nodes, respectively
  \Statex $\va$: left  degree sequence. 
  \Statex $\vsfb = (\sfb_v: v \in [n_r])$: $\sfb_v = b_v^G(i)$ for $v \in [n_r]$
  \Statex $i,j$: endpoints of the interval, such that  $1 \leq i \leq j \leq n_l$
\Statex $\vgamma^G_{[i:j]} = \vgamma^G_i,\dots, \vgamma^G_j$: adjacency list of vertices $i \leq v \leq j$,
where $\vgamma^G_v = (\gamma^G_{v,1}, \dots, \gamma^G_{v,a_v})$ such that $1 \leq
\gamma^G_{v,1} < \gamma^G_{v,2} < \dots < \gamma^G_{v,a_v} \leq n_r$ are the right nodes
adjacent to the left node $v$ in $G$
\Statex $\sfU$: Fenwick tree, where for $v \in [n_r]$, $\sfU.\textproc{Sum}(v) =
U^G_v(i) = \sum_{k=v}^{n_r} b^G_k(i)$
\OUTPUT
\Statex $N_{i,j} = N_{i,j}(G)$
\Statex $\vsfb = (\sfb_v: v \in [n_r])$: vector $\vsfb$ updated, so
that $\sfb_v = b^G_v(j+1)$ for $v \in [n_r]$
\Statex $\sfU$: Fenwick tree $\sfU$ updated, so that for $v \in [n_r]$, we have
$\sfU.\textproc{Sum}(v) = U^G_v(j+1) = \sum_{k=v}^{n_r} b^G_k(j+1)$
\Statex $l_{i,j} = l_{i,j}^G$
\Function{BComputeN}{$n_l, n_r, \va, \vsfb, i,j,\vgamma^G_{[i:j]}, \sfU$}
\Comment{B stands for Bipartite}
\If{$i=j$}
\State $z_i \gets 0, l_i \gets 1$
\For{$1 \leq k \leq a_i$}
\State $c \gets a_i - k + 1$
\State $y \gets \textproc{ComputeProduct}(\sfU.\textproc{Sum}(1+\gamma^G_{i,k}), c,1) \div
\textproc{ComputeProduct}(c, c, 1)$ 
\label{l:bNij=y}
\Statex \Comment{Algorithm~\ref{alg:compute-product} in Section~\ref{sec:aux-algs-bip}}
\State $z_i \gets z_i + l_i \times y$  \label{l:bNij=z}
\State $l_i \gets l_i \times \sfb_{\gamma^G_{i,k}}$ \label{l:bNij=l}
\State $\sfU.\textproc{Add}(\gamma^G_{i,k}, -1)$ \label{l:bNij=U}
\State $\sfb_{\gamma^G_{i,k}} \gets \sfb_{\gamma^G_{i,k}} - 1$ \label{l:bNij=beta}
\EndFor
\State $N_{i,i} \gets z_i$
\State \textbf{return} $(N_{i,i}, \vsfb, \sfU, l_i)$
\Else
\State $k \gets (i + j ) \div 2$ \Comment{$k = \lfloor (i+j)/2 \rfloor$}
\State $(N_{i,k}, \vsfb, \sfU, l_{i,k}) \gets \textproc{BComputeN}(n_l, n_r,
\va, \vsfb, i,k,
\vgamma^G_{[i:k]}, \sfU)$ \label{l:bNij-Nik}
\Comment{Algorithm~\ref{alg:bip-Nij} (recursive)}
\State $S_{k+1} \gets \sfU.\textproc{Sum}(1)$
\label{l:bNij-Sk+1}
\State $(N_{k+1, j}, \vsfb, \sfU, l_{k+1, j}) \gets \textproc{BComputeN}(n_l,
n_r, \va, \vsfb, k+1, j,
\vgamma^G_{[k+1:j]},  \sfU)$ \label{l:bNij-Nkj} 
\Statex \Comment{Algorithm~\ref{alg:bip-Nij} (recursive)}
\State $S_{j+1} \gets \sfU.\textproc{Sum}(1)$
\label{l:bNij-Sj+1}
\State $y \gets \textproc{ProdFactorial}(\va, k+1,j)$  \label{l:bNij-y} \Comment{$y=
  \prod_{v=k+1}^j a_v!$, Algorithm~\ref{alg:comp-ProdAFactorial} in Sec.~\ref{sec:aux-algs-bip}}
\State $r_{k+1,j} \gets \textproc{ComputeProduct}(S_{k+1}, S_{k+1} - S_{j+1}, 1) 
\div y$ \label{l:bNij-r} \Comment{Algorithm~\ref{alg:compute-product} in Sec.~\ref{sec:aux-algs-bip}}
\State $N_{i,j} \gets N_{i,k} \times r_{k+1, j} + l_{i,k} \times N_{k+1, j}$ \label{l:bNij-N}
\State $l_{i,j} \gets l_{i,k} \times l_{k+1, j}$ \label{l:bNij-l}
\State \textbf{return} $(N_{i,j}, \vsfb, \sfU, l_{i,j})$
\EndIf
\EndFunction
\end{algorithmic}
\end{myalg}

\begin{rem}
  \label{rem:b-enc-U-b-pointer}
Note that due to the order in which we recursively call subintervals $[i,k]$ and $[k+1, j]$,
we always move from left to right in the process of  encoding. Therefore, instead of passing
the whole data structures $\sfU$ and $\sfb$ to the subprocesses, we may keep
them in one location of the memory and send the pointer to that location to
subprocesses.
In fact, the  inductive argument in this section implies that if we modify this
shared location of the memory, when the procedure is called for the interval
$[i,j]$, we have $\sfU.\textproc{Sum}(v) = U^G_v(i)$ and $\sfb_v = b^G_v(i)$ for
$1 \leq v \leq n_r$. Moreover, when the procedure is finished for this interval, we have
$\sfU.\textproc{Sum}(v) = U^G_v(j+1)$ and $\sfb_v = b^G_v(j+1)$ for $1 \leq v
\leq n_r$. 
 With this, we can  keep updating the same $\sfU$ and $\sfb$ and
avoid storing unnecessary copies. Note that since all
the partial sums of the form $\sum_{k=v}^{n_r} b_v^G(i)$ for $1 \leq i \leq n_l$
and $1 \leq v \leq n_r$, are bounded by $\pn \delta$, the memory required to
store $\sfU$ is $O(\pn \log \pn)$, where we recall that $\pn = \max\{n_l, n_r\}$. Likewise, the array $\vsfb$ can be stored
using $O(\pn \log \pn)$ bits. Similarly, since the array $\va$ is not changed
throughout the process, we can also store it in one location of the memory, and
pass the pointer to that location. Since $a_i \leq \delta$ for $1 \leq i \leq
n_l$, the memory required to store $\va$ is $O(\pn \log \delta) = O(\pn \log \pn)$.
\end{rem}

Finally, after we compute $N(G)$ using the above, we use~\eqref{eq:bip-fn-va-vb-def} to find
$\fnlr_{\va, \vb}(G)$ for $G \in \mGnlr_{\va, \vb}$.
  It is easy to see that
  \begin{equation}
    \label{eq:prod-b!-l1n}
    \prod_{j=1}^{n_r} b_j! = l^G_{1,n_l} = \prod_{i=1}^{n_l} \prod_{k=1}^{a_i} b^G_{\gamma^G_{i,k}}(i).
  \end{equation}
  To verify this, note that for $1 \leq j \leq n_r$, if  
  $y^G_{j,1} < \dots < y^G_{j,b_j}$
  \color{black}
  denote the left neighbors of the right node $j$ in $G$, we have
  $b^G_j(y^G_{j,i}) = (b_j - i + 1)$ 
  \color{black}
  for $1 \leq i \leq b_j$. Therefore,
  \begin{align*}
    l^G_{1,n_l} = \prod_{i=1}^{n_l} \prod_{k=1}^{a_i} b^G_{\gamma^G_{i,k}}(i) = \prod_{j=1}^{n_r} 
    \prod_{i=1}^{b_j} b^G_j(y^G_{j,i})
    \color{black}
    = \prod_{j=1}^{n_r} \prod_{i=1}^{b_j} (b_j - i + 1) = \prod_{j=1}^{n_r} b_j!,
  \end{align*}
  which establishes~\eqref{eq:prod-b!-l1n}. Recall that $l^G_{1,n_l}$ was
  computed and returned by Algorithm~\ref{alg:bip-Nij}. 
Algorithm~\ref{alg:bip-comp}
below illustrates
the process of computing $\fnlr_{\va, vb}(G)$ for $G \in \mGnlr_{\va, \vb}$.

\begin{myalg}[Finding $\fnlr_{\va, \vb}(G)$ for $G \in \mGnlr_{\va, \vb}$ \label{alg:bip-comp}]
  \begin{algorithmic}[1]
    \INPUT
    \Statex $n_l, n_r$: the number of left and right vertices, respectively
    \Statex $\va = (a_i: i \in [n_l])$: left degree vector
    \Statex $\vb = (b_i: i \in [n_r])$: right degree vector
    \Statex $(\vgamma^G_v: v \in [n_l])$: adjacency list of left nodes, where
    $\vgamma^G_v = (\gamma^G_{v,1}, \dots, \gamma^G_{v,a_v})$ for $v \in [n_l]$, such
    that $1 \leq \gamma^G_{v,1} < \dots < \gamma^G_{v,a_v} \leq n_r$ are the right nodes adjacent to
    the left node $v$
    \OUTPUT
    \Statex $\fnlr_{\va, \vb}(G)$: an integer representing $G$ in the compressed
    form
    \Function{BEncodeGraph}{$n_l, n_r, \va, \vb, (\vgamma^G_v: v \in [n_l])$}
    \State $\sfU \gets$ Fenwick tree initialized with array $\vb$
    \Comment{$\sfU.\textproc{Sum}(v) = \sum_{k=v}^{n_r} b_k = U^G_v(1)$ for $v \in [n_r]$}
    \State $\vsfb \gets \vb$ \Comment{$\sfb_v = b^G_v(1) = b_v$ for $v \in [n_r]$}
    \State  $(N_{1,n_l}, \vsfb, \sfU, l_{1,n_l}) \gets \textproc{BComputeN}(n_l,
    n_r, \va, \vsfb, 1, n_l,
    \vgamma^G_{[1:n_l]}, \sfU)$\label{l:bip-comp-N} \Comment{Algorithm~\ref{alg:bip-Nij} above}
    \State $f \gets N \div
    l_{1,n_l}$ \label{l:bip-enc-f-N-div-l}\Comment{$l_{1,n_l} = l^G_{1,n_l} =
      \prod_{j=1}^{n_r} b_j!$ (see~\eqref{eq:prod-b!-l1n})}
    \If{$f \times l_{1,n_l} < N$} \label{l:bip-enc-f-times-l-N} \Comment{this means $N \mod l_{1,n_l} \neq 0$}
    \State $f \gets f+1$ \Comment{so that $f = \fnlr_{\va,\vb}(G) = \lceil N / l_{1,n_l} \rceil$}
    \EndIf
    \State \textbf{return} $f$ 
    \EndFunction
  \end{algorithmic}
\end{myalg}

\subsubsection{Auxiliary Algorithms}
\label{sec:aux-algs-bip}

In this section, we present the two functions $\textproc{ComputeProduct}$ and
$\textproc{ProdFactorial}$ used in the above encoding procedure. 
Algorithm~\ref{alg:compute-product} below computes 
$\textproc{ComputeProduct}(p,k,s)$ as defined in~\eqref{eq:compute-product-return}. 
When $k > 1$ and $p-(k-1)s > 0$, in order to find $\prod_{i = 0}^{k-1}
(p-is)$, we split the terms into two halves, compute each of them
recursively, and return their product. More precisely, when $k>1$ and $p-(k-1)s
> 0$, we define  $k':= \lfloor k/2 \rfloor$ and find
$L := \prod_{i = 0}^{k'-1} (p-is)$ and $R := \prod_{i=0}^{k-k'-1}
(p-\pp{k}s - i s)$. Note that $R = \prod_{i = \pp{k}}^{k-1} (p- is)$.
Therefore, $\prod_{i = 0}^{k-1} (p - i s) = L \times R$.

\begin{myalg}[Computing $\prod_{\pp{k}=0}^{k-1} (p - \pp{k}s)^+$ \label{alg:compute-product}]
  \begin{algorithmic}[1]
    \INPUT
    \Statex Integer $p \geq 0$: the first term in the product
    \Statex Integer $k\geq 0$: the number of terms in the product
    \Statex Integer $s \geq 0$: the difference between successive terms
    \OUTPUT
    \Statex If $k=0$, return $1$. If $k>0$ and $p - (k-1)s \leq 0$, return $0$.
    If $k > 0 $ and $p - (k-1)s > 0$, returns $\prod_{i=0}^{k-1}(p - i s)$.
    \Function{ComputeProduct}{$p, k, s$}
    \If{$k=0$} \Comment{product over empty set is $1$}
    \State \textbf{return} $1$
    \EndIf
    \If{$p - (k-1) s \leq 0$}
    \State \textbf{return} $0$
    \EndIf
    \If{$k = 1$} \Comment{there is only one term}
    \State \textbf{return} $p$
    \EndIf
    \State $k' \gets k \div 2$ \Comment{we compute the product by
      dividing the terms into two halves}
    \State $L \gets \textproc{ComputeProduct}(p, k', s)$ \Comment{product
      of the first half}
    \State $R \gets \textproc{ComputeProduct}(p-k's, k-k',s)$
    \Comment{product of the second half}
    \State \textbf{return} $L \times R$ \Comment{aggregate the two
      pieces to get the final result} \label{line:alg:compute-product_LtimesR}
    \EndFunction
  \end{algorithmic}
\end{myalg}

\begin{lem}
  \label{lem:compute-product-complexity}
  If the input $p$ in Algorithm~\ref{alg:compute-product} has at most $q$ bits,
  this algorithm runs in $O(qk \log k \log (qk) \log \log (qk))$ time and its memory
  complexity is $O(qk)$. Moreover, the result has $O(q k)$ bits.
\end{lem}

\begin{proof}
  Let $f(q,k,s)$ be the time complexity of this algorithm given inputs $k$ and
  $s$, and when the input $p$ has
  at most $q$ bits. The dominant part of the calculation is
  line~\ref{line:alg:compute-product_LtimesR} when we multiply $L$ by $R$. But  each
  of these two variables has at most $O(qk)$ bits. The reason is that each is
  the product of $\max\{\pp{k}, k - \pp{k}\} = O(k)$ integers, each having at
  most $q$ bits. 
Hence, the cost of this multiplication is
  \begin{equation*}
    \Mul(qk) = O(qk \log( qk) \log \log (qk)). 
  \end{equation*}
  Therefore, we have
  \begin{equation*}
    f(q,k,s) = f\left( q, \left\lfloor \frac{k}{2} \right \rfloor ,s\right) + f\left(q, \left\lceil \frac{k}{2}  \right\rceil,s \right) + O(qk \log (qk) \log \log (qk)). 
  \end{equation*}
  On the other hand, we have $f(q,1,s) = O(q)$. Solving for this recursion, we get
  \begin{equation*}
    f(q,k,s) = O(qk \log k \log (qk) \log \log (qk)). 
  \end{equation*}
  Furthermore, let $g(q, k,s)$ denote the  memory complexity given $k$ and $s$
  and when the input $p$ has at most
  $q$ bits. Since we need to store $L$ and $R$, which have $O(qk)$ bits each, we
  have
  \begin{equation*}
    g(q,k,s) = \max \left \{g\left( q, \left\lfloor \frac{k}{2} \right \rfloor ,s\right),  g\left(q, \left\lceil \frac{k}{2}  \right\rceil,s\right) \right \} + O(qk).
  \end{equation*}
  Solving for this, we get $g(q,k,s) = O(qk)$, as was claimed.
  Finally, note that the final result is the product of $k$ integers, each
  having $O(q)$ bits. Therefore, the result has $O(qk)$ bits. This completes the proof.
\end{proof}

In Algorithm~\ref{alg:comp-ProdAFactorial} below, we  compute $\prod_{\pp{i}=i}^j
v_{\pp{i}}!$ given an array $\vv = (v_1, \dots, v_k)$ of integers and $1 \leq i \leq j
\leq k$. Note that when $i =j$, this reduces to
$v_i! = \textproc{ComputeProduct}(v_i, v_i, 1)$. 
In order to do this, similar to the idea in Algorithm~\ref{alg:compute-product}
above, we divide the interval $[i,j]$ into two halves, compute the product for
each half recursively, and then multiply the results.

\begin{myalg}[Computing $\prod_{\pp{i}=i}^j v_{\pp{i}}!$ for an array $\vv$ of nonnegative integers
   \label{alg:comp-ProdAFactorial}]
  \begin{algorithmic}[1]
    \INPUT
    \Statex $\vv$: array of nonnegative integers
    \Statex $i,j$: endpoints of the interval for which the product is being computed
    \OUTPUT
    \Statex $\prod_{\pp{i}=i}^j v_{\pp{i}}!$
    \Function{ProdFactorial}{$\vv, i, j$}
    \If{$i = j$}
    \State \textbf{return} $\textproc{ComputeProduct}(v_i, v_i, 1)$ \label{l:prod-factorial-v1-fact}
    \Comment{return $v_i!$ using Algorithm~\ref{alg:compute-product} above}
    \Else 
    \State $m \gets (i+j) \div 2$ \Comment{split the interval into two halves}
    \State $L \gets \textproc{ProdFactorial}(\vv, i, m)$ \label{l:paf-x}
    \State $R \gets \textproc{ProdFactorial}(\vv, m+1, j)$ \label{l:paf-y}
    \State \textbf{return} $L \times R$
    \EndIf
    \EndFunction
  \end{algorithmic}
\end{myalg}

Now, we analyze the complexity of Algorithm~\ref{alg:comp-ProdAFactorial} above.
We do this in two regimes. First, in
Lemma~\ref{lem:prod-a-factorial-complexity-each-delta} below, we do this with the
assumption that each element in the array is bounded by an integer $\delta$.
Then, in Lemma~\ref{lem:prod-factorial-complexity-sum-bound}, we study a less
restrictive regime where instead of an elementwise bound, the average of the
elements in the array is bounded by $\delta$, and we prove a looser  complexity bound
compared to that in Lemma~\ref{lem:prod-a-factorial-complexity-each-delta}. These
two different regimes will be useful for our future analysis. 

\begin{lem}
  \label{lem:prod-a-factorial-complexity-each-delta}
  Assume that the array $\vv$ in Algorithm~\ref{alg:comp-ProdAFactorial}  is
  such that $v_p \leq\delta$ for $i \leq p \leq j$. Then, with $k := j - i + 1$
  and  $C:= \max\{k, \delta\}$, the algorithm runs in $O(k \delta
  \log \delta \log^2 C\log \log C)$ time and its
  memory complexity is $O(k \delta \log \delta)$.  Furthermore, the result has
  $O(k \delta \log \delta)$ bits.
\end{lem}

\begin{proof}
  For $i \leq \pp{i} \leq \pp{j} \leq j$, let $f_{\pp{i}, \pp{j}}$ denote the time complexity of
  computing $\prod_{p = \pp{i}}^{\pp{j}} v_p!$.
  If $\pp{i} = \pp{j}$, using Lemma~\ref{lem:compute-product-complexity}, the cost of
  computing $v_{\pp{i}}!$ in line~\ref{l:prod-factorial-v1-fact} is $O(\delta \log^3 \delta \log \log \delta)$ and requires
  $O(\delta \log \delta)$ bits of memory. Furthermore, the result $v_{\pp{i}}!$ has $O(\delta
  \log \delta)$ bits. On the other hand, if $\pp{i} < \pp{j}$,  $L$ and $R$
  defined in lines~\ref{l:paf-x} and \ref{l:paf-y} have $O((\pp{j} - \pp{i} + 1)\delta \log \delta)$ bits each,
  hence the time complexity of computing $L \times R$ is
  \begin{align*}
   \Mul((\pp{j} - \pp{i} + 1)\delta \log \delta) &= O((\pp{j} - \pp{i} + 1)
  \delta \log \delta\log ((\pp{j} - \pp{i} + 1)\delta \log \delta) \log \log  ((\pp{j} - \pp{i} + 1) \delta \log
                                                   \delta)) \\
    &= O((\pp{j} - \pp{i} + 1) \delta \log \delta \log C \log \log C).
  \end{align*}
   Therefore,  with $m = \lfloor  (\pp{i} + \pp{j}) /2 \rfloor$, we have 
  \begin{align*}
    f_{\pp{i}, \pp{j}} &= f_{\pp{i}, m} + f_{m+1, \pp{j}} + O((\pp{j} - \pp{i} + 1\}) \delta \log \delta \log C \log \log C), \qquad \qquad \text{ for } i \leq \pp{i} < \pp{j} \leq j;\\
    f_{\pp{i}, \pp{i}} &= O(\delta \log^3 \delta \log \log \delta).
  \end{align*}
  Solving for this, we get
  \begin{equation*}
    f_{i,j} = O(k \delta \log \delta \log k \log C \log \log C + k \delta \log^3 \delta + \log \log \delta)= O(k \delta \log \delta \log^2 C \log \log C).
  \end{equation*}
  In order to analyze the memory complexity, note that the memory needed to store the
  variables $L$ and $R$ when the algorithm is called for the interval $[\pp{i}
  ,\pp{j}]$ is  $O((\pp{j} - \pp{i} + 1)\delta \log \delta)$. Therefore,
  if $g_{\pp{i}, \pp{j}}$ denotes the memory complexity when the algorithm is
  called for the interval $[\pp{i}, \pp{j}]$, with $m = \lfloor (\pp{i} + \pp{j}
  ) / 2 \rfloor$ we 
  have 
  \begin{align*}
    g_{\pp{i}, \pp{j}} &= \max \left\{ g_{\pp{i}, m} , g_{m + 1, \pp{j}}\right\} + O((\pp{j} - \pp{i} + 1) \delta \log \delta), \qquad \qquad \text{ for } i \leq \pp{i} < \pp{j} \leq j;\\
    g_{\pp{i}, \pp{i}} &= O(\delta \log \delta).
  \end{align*}
  Solving for this, we get $g_{i,j} = O(k \delta \log \delta)$. Finally, note
  that the final result is the multiplication of $k$ integers, each having
  $O(\delta \log \delta)$ bits. Hence, the result has $O(k \delta \log \delta)$ bits.
  This completes the proof.
\end{proof}

\begin{lem}
  \label{lem:prod-factorial-complexity-sum-bound}
  Assume that  the array $\vv$ in Algorithm~\ref{alg:comp-ProdAFactorial}
  is such that $\sum_{p=i}^j v_p \leq k \delta$  where $k := j - i + 1$ and
  $\delta$ is an integer. Then, with $C := \max \{k, \delta, (v_p: i \leq p \leq j)\}$, the
  time complexity of computing $\prod_{p=i}^j v_p!$ in this algorithm  is $O(k
  \delta \log^3 C \log \log C)$ and its memory
  complexity is $O(\delta k \log k \log C)$.
\end{lem}

\begin{proof}
  For $i \leq \pp{i} \leq \pp{j} \leq j$, let $f_{\pp{i}, \pp{j}}$ denote the time complexity of
  computing $\prod_{p = \pp{i}}^{\pp{j}} v_p!$. Using
  Lemma~\ref{lem:compute-product-complexity}, we have $f_{\pp{i}, \pp{i}} =
  O(v_{\pp{i}} \log^3 v_{\pp{i}} \log \log v_{\pp{i}}) = O(v_{\pp{i}} \log^3 C
  \log \log C)$. To simplify the notation, for $i \leq \pp{i} \leq \pp{j} \leq  j $, let $S_{\pp{i},\pp{j}} :=
  \sum_{p=\pp{i}}^{\pp{j}} v_p$. Note that for $\pp{i} < \pp{j}$ and $m := \lfloor (\pp{i}+\pp{j})/ 2 \rfloor$,
  both $\prod_{p= \pp{i}}^m v_{p}!$  and $\prod_{p = m+1}^{\pp{j}}
  v_{p}!$ are bounded by $\prod_{p = \pp{i}}^{\pp{j}} v_{p}!$, which has
  $O(\sum_{p = \pp{i}}^{\pp{j}} v_{p} \log v_{p}) = O(S_{\pp{i}, \pp{j}} \log C)$  bits.
  Therefore, if the algorithm is called for such $\pp{i} < \pp{j}$, both variables $L$ and $R$ defined in lines~\ref{l:paf-x} and
  \ref{l:paf-y} have $O(S_{\pp{i}, \pp{j}} \log C)$ bits, and the time complexity of
  multiplying $L$ by $R$ is
  \begin{align*}
    \Mul(S_{\pp{i}, \pp{j}} \log C) &= O(S_{\pp{i}, \pp{j}} \log C \log (S_{\pp{i}, \pp{j}} \log C) \log \log (S_{\pp{i}, \pp{j}} \log C)) \\
                                    &\stackrel{(a)}{=} O(S_{\pp{i}, \pp{j}} \log C \log (k \delta \log C) \log  \log (k \delta \log C)) \\
    &= O(S_{\pp{i}, \pp{j}} \log^2 C \log \log C)
  \end{align*}
  where in $(a)$, we have used the fact that $S_{\pp{i}, \pp{j}} \leq S_{i,j}
  \leq k \delta$. Consequently, for $i \leq \pp{i} < \pp{j} \leq j$ and $m =
  \lfloor  (\pp{i}+\pp{j}) / 2 \rfloor$, we have
  \begin{equation*}
    f_{\pp{i}, \pp{j}} = f_{\pp{i}, m} + f_{m+1, \pp{j}} + O(S_{\pp{i}, \pp{j}} \log^2 C \log \log C),
  \end{equation*}
  and for $i \leq \pp{i} \leq j$, we have
  \begin{equation*}
    f_{\pp{i}, \pp{i}} = O(v_{\pp{i}} \log^3 C \log \log C).
  \end{equation*}
  Solving for this, we get
  \begin{equation*}
   f_{i,j} = O(k \delta \log k \log^2 C \log \log C + k \delta \log^3 C \log \log C) = O(k \delta \log^3 C \log \log C),
  \end{equation*}
  as was desired.

  Now, for $i \leq \pp{i} \leq \pp{j} \leq j$, let
  $g_{\pp{i}, \pp{j}}$ denote the memory complexity of computing $\prod_{p =
    \pp{i}}^{\pp{j}} v_{p}!$. Recall from the above that both $L$ and $R$ have
  $O(S_{\pp{i}, \pp{j}} \log C) = O(k \delta \log C)$ bits. Hence, for $\pp{i} <
  \pp{j}$ and $m = \lfloor   (\pp{i} + \pp{j}) / 2 \rfloor$, we have 
  \begin{equation*}
    g_{\pp{i}, \pp{j}} = \max \{g_{\pp{i}, m} , g_{m+1, \pp{j}} \} + O(k \delta \log C),
  \end{equation*}
  and $g_{\pp{i}, \pp{i}} = O(v_{\pp{i}} \log v_{\pp{i}}) = O(k \delta \log C)$.
  Therefore, we have $g_{i,j} = O(\delta k \log k \log C)$, as was claimed. 
\end{proof}

\subsubsection{Complexity of Encoding}
\label{sec:bip-compression-complexity}

Now, we analyze the complexity of the encoding algorithm described in
Section~\ref{sec:bip-enc-pseudocode} above. Before
doing this, we first introduce Lemmas~\ref{lem:combinatorial-bound-pqs}, \ref{lem:bip-N+l<r}, and
\ref{lem:bip-r-l-N-bit-count} below, whose proofs are postponed to the end
of this section. Afterwards, we analyze the complexity of
Algorithms~\ref{alg:bip-Nij} and \ref{alg:bip-comp} in
Lemmas~\ref{lem:bip-encode-computeN-complexity} and
\ref{lem:bip-encode-overall-complexity}, respectively.

\begin{lem}
  \label{lem:combinatorial-bound-pqs}
  For integers $q \geq 1$ and $p, s \geq 0$, we have
  \begin{equation}
    \label{eq:comb-bound-pqs-bound}
    \binom{p+s}{q} \geq \binom{p}{q} + s \binom{p}{q-1}.
  \end{equation}
  Recall that for integers $u, v \geq 0$, if $u < v$,  $\binom{u}{v}$ is defined
  to be zero.
\end{lem}

\begin{lem}
  \label{lem:bip-N+l<r}
  Given $G \in \mGnlr_{\va, \vb}$ and $1 \leq i \leq j \leq n_l$, we have
  \begin{equation*}
    N_{i,j}(G) + l^G_{i,j} \leq r_{i,j}. 
  \end{equation*}
\end{lem}

 Here, we give an
informal  explanation for Lemma~\ref{lem:bip-N+l<r}. Roughly speaking, among the
configurations which are consistent with $G$ for the left nodes $1, \dots, i-1$,
the quantity $N_{i,j}(G)$ is  the number of extensions  to the left nodes in the interval $[i,j]$, such that the
resulting extended configuration is lexicographically smaller than $G$. Also,
$l^G_{i,j}$ is the number of such extensions that remain consistent with $G$ in
the interval $[i,j]$, and $r_{i,j}$ is the total number of such extensions. With this
interpretation, the above inequality is apparent. A formal proof is given at the end of this section.


\begin{lem}
  \label{lem:bip-r-l-N-bit-count}
Assume that we have $a_k \leq \delta$ for all $1 \leq k \leq n_l$. Then, given $G \in
\mGnlr_{\va, \vb}$ and  $1 \leq i \leq j \leq n_l$, the variables $r_{i,j}$, $l_{i,j}^G$ and
  $N_{i,j}(G)$ have $O((j-i+1) \delta \log \pn)$ bits where $\pn = \max \{ n_l,
  n_r\}$. 
\end{lem}

Now we are ready to analyze the complexity of Algorithm~\ref{alg:bip-Nij}.

  \begin{lem}
    \label{lem:bip-encode-computeN-complexity}
    In Algorithm~\ref{alg:bip-Nij}, if we have $a_i \leq \delta$ for $ 1 \leq i \leq
    n_l$,  computing $N_{1,n_l}(G)$ for
    $G \in \mGnlr_{\va, \vb}$ takes $O(\pn \delta  \log^4 \pn \log
    \log \pn)$ time and requires
    $O(\delta \pn \log \pn)$ bits of memory, where $\tilde{n} = \max\{n_l, n_r\}$.
  \end{lem}

  \begin{proof}[Proof of Lemma~\ref{lem:bip-encode-computeN-complexity}]
    {
      Note that for $1 \leq i \leq n_l$, $a_i$ is the degree of the $i$th
  left node  in a simple bipartite graph with $n_r$ right nodes. Therefore, we indeed have
  $a_i \leq n_r$. Moreover, with $\delta' := \min\{\delta, n_r\}$, we have $a_i
  \leq \delta' \leq n_r$ for $1 \leq i \leq n_l$. Additionally, note that the
  desired complexity bounds $O(\pn \delta  \log^4 \pn \log \log \pn)$ and
  $O(\delta \pn \log \pn)$ are monotone in $\delta$. Hence, without loss of
  generality, we may assume that $\delta \leq n_r \leq \pn$ for the rest of the proof.}

    We first discuss the complexity when the algorithm is called for the
    interval $[i,j]$ for $1 \leq i \leq j \leq n_l$.  We do this separately for the two
    cases $i = j$ and $i < j$.

    \underline{Case I: $i = j$:} Note that for $1 \leq v \leq n_r$, we have 
    \begin{equation*}
      \sfU.\textproc{Sum}(v) = U^G_v(i) = \sum_{k=v}^{n_r} b^G_k(i) \leq \sum_{k=v}^{n_r} b_k \leq \sum_{k=1}^{n_r} b_k  = \sum_{k=1}^{n_l} a_k \leq \pn \delta.
    \end{equation*}
Therefore, each partial sum in $\sfU$ has $O(\log \pn \delta) = O(\log \pn)$
bits. Hence, for $1 \leq k \leq a_i$, computing $\sfU.\textproc{Sum}(1+\gamma^G_{i,k})$ in 
    line~\ref{l:bNij=y} of Algorithm~\ref{alg:bip-Nij} has time complexity $O(\log^2 \pn)$ and the result has
    $O(\log \pn)$ bits.
Moreover, since in  line~\ref{l:bNij=y} we have $c \leq a_i \leq \delta$, using
    Lemma~\ref{lem:compute-product-complexity}, computing
    $\textproc{ComputeProduct}(\sfU.\textproc{Sum}(1+\gamma^G_{i,k}), c,1)$ takes
    $O(\delta (\log \pn)  (\log \delta) \log (\delta \log \pn) \log \log (\delta \log
    \pn)) = O(\delta \log \delta \log^2 \pn \log \log \pn)$ time and requires
    $O(\delta \log \pn)$ bits of memory. Also, the result has $O(\delta \log
    \pn)$ bits. On the other
    hand, using
    Lemma~\ref{lem:compute-product-complexity}, computing $\textproc{ComputeProduct}(c,c,1)$ takes $O(\delta \log
    \delta \log (\delta \log \delta) \log \log (\delta \log \delta)) = O(\delta
    \log^3 \delta \log \log \delta)$ time and $O(\delta \log \delta )$ bits of
    memory. Also, the result has $O(\delta \log \delta ) = O(\delta \log \pn)$
    bits. Thereby, performing the division in this line takes $\Div(\delta
    \log \pn) = O(\delta \log \pn \log (\delta \log \pn) \log \log (\delta \log
    \pn)) = O(\delta \log^2 \pn \log \log \pn)$, and the result, which is stored in
    $y$, has $O(\delta \log \pn)$ bits. Thereby, the overall time complexity of
    computing $y$ in line~\ref{l:bNij=y} is
    \begin{equation*}
      O(\delta \log \delta \log^2 \pn \log \log \pn + \delta \log^3 \delta \log \log \delta) = O(\delta \log \delta \log^2 \pn \log \log \pn),
    \end{equation*}
    and the overall memory complexity of this line is $O(\delta \log \pn)$.
    In line~\ref{l:bNij=z}, note that $l_i$ is initialized with 1 and will
    eventually become $l^G_i$. Hence
    \begin{equation*}
      l_i \leq l^G_i = \prod_{\pp{i} = 1}^{a_i} b^G_{\gamma^G_{i,\pp{i}}}(i) \leq \prod_{\pp{i} = 1}^{a_i} b_{\gamma^G_{i,\pp{i}}} \leq \pn^{a_i} \leq \pn^\delta,
    \end{equation*}
    which implies that $l_i$ has $O(\delta \log \pn)$
    bits. 
Note that the variable $z_i$ is initialized with zero and eventually becomes
$z^G_i$. Therefore, we have $z_i \leq z^G_i = N_{i,i}(G)$ which has $O(\delta
\log \pn)$ bits from Lemma~\ref{lem:bip-r-l-N-bit-count}.
Since  $y$ has also $O(\delta \log \pn)$ bits, the time complexity of multiplication and addition in line~\ref{l:bNij=z} is
    $\Mul(\delta \log \pn) = O(\delta \log^2 \pn \log \log \pn)$. 
    In line~\ref{l:bNij=l}, $l_i$ has $O(\delta \log \pn)$ bits, and
    $\sfb_{\gamma^G_{i,k}} = b^G_{\gamma^G_{i,k}}(i) \leq \pn$ has $O(\log \pn)$ bits. Hence, the time complexity of
    the multiplication in this line is $\Mul(\delta \log \pn) = O(\delta \log^2
    \pn \log \log \pn)$. In line~\ref{l:bNij=U}, since each partial sum in the
    Fenwick tree $\sfU$ has
    $O(\log \pn)$ bits as we discussed above, updating $\sfU$ in this line takes $O(\log^2\pn)$ time. Finally,
    as $\sfb_{\gamma^G_{i,k}}$ has $O(\log \pn)$ bits as we discussed above, the
    time complexity of updating
    $\sfb_{\gamma^G_{i,k}}$ in line~\ref{l:bNij=beta} is $O(\log \pn)$.
    To sum up,
    if $f(k)$  denotes the time complexity of
    the algorithm when $j - i + 1 = k$, we have
    \begin{equation*}
        f(1) = O(\delta \log \delta \log^2 \pn \log \log \pn + \delta \log^3 \delta \log \log \delta) = O(\delta \log \delta \log^2 \pn \log \log \pn).
    \end{equation*}
    Moreover, if $g(k)$ denote the memory complexity of the algorithm when
    $j-i+1 = k$, excluding the space required to store $\sfU$ and $\sfb$, the
    above discussion implies that
    \begin{equation*}
      g(1) = O(\delta \log \pn).
    \end{equation*}

    \underline{Case II: $i<j$:} Similar to the above, the time complexities of
    finding $\sfU.\textproc{Sum}(1)$ in lines~\ref{l:bNij-Sk+1} and
    \ref{l:bNij-Sj+1} are $O(\log^2 \pn)$. Furthermore, using
    Lemma~\ref{lem:prod-a-factorial-complexity-each-delta}, the time complexity of
    computing $y$ in line~\ref{l:bNij-y} is $O((j-i+1) \delta  \log \delta \log^2
    (\max\{j-1+1, \delta\}) \log \log(\max\{j-i+1, \delta\})) = O((j-i+1) \delta  \log \delta \log^2
    \pn \log \log \pn)$, its memory complexity is  $O((j-i+1) \delta \log \delta)$ bits,
     and the result $y$ has $O((j-i+1) \delta \log \delta)$ bits. 
    In line~\ref{l:bNij-r}, note that
    \begin{equation*}
      S_{k+1} - S_{j+1} = \sum_{v=k+1}^j a_v \leq (j-i+1) \delta.
    \end{equation*}
    Moreover, $S_{k+1} \leq S \leq \pn \delta$, which implies hat $S_{k+1}$ has
    $O(\log (\pn \delta)) = O(\log \pn)$ bits. 
    Therefore, using Lemma~\ref{lem:compute-product-complexity},
        the time complexity of computing $\textproc{ComputeProduct}(S_{k+1}, S_{k+1} -
        S_{j+1}, 1)$ in line~\ref{l:bNij-r} is $O((j-i+1) \delta \log^3 \pn \log
        \log \pn)$, its time complexity is   $O((j-i+1) \delta \log \pn)$ bits,  and
        the result has $O((j-i+1) \delta \log \pn)$ bits. 
        Since $y$ has $O((j-i+1)
        \delta \log \delta)$ bits as we discussed above, the time complexity of
        the division in line~\ref{l:bNij-r} is
        $\Div((j-i+1) \delta \log \pn) = O((j-i+1) \delta \log^2 \pn \log \log
        \pn)$. Therefore, the overall time and memory complexity of
        computing $r_{k+1, j}$ in line~\ref{l:bNij-r} are $O((j-i+1) \delta \log^3 \pn \log
        \log \pn)$ and $O((j-i+1) \delta \log \pn)$, respectively, and the result
        $r_{k+1,j}$ has $O((j-i+1) \delta \log \pn)$ bits. 
        In line~\ref{l:bNij-N}, using Lemma~\ref{lem:bip-r-l-N-bit-count}, all
        terms in the right hand side have $O((j-i+1) \delta \log \pn)$ bits.
        Hence, the time complexity of the multiplications and the addition in this line is $\Mul((j-i+1)
        \delta \log \pn) = O((j-i+1) \delta \log^2 \pn \log \log \pn)$. Also
        note that from Lemma~\ref{lem:bip-r-l-N-bit-count}, the result $N_{i,j}
        = N_{i,j}(G)$ has
        $O((j-i+1) \delta \log \pn)$ bits. 
        Similarly, the multiplication in line~\ref{l:bNij-l} has  time
        complexity $O((j-i+1) \delta \log^2 \pn \log \log \pn)$.

        Putting the above together, recalling that $f(k)$  denotes the
        time complexity of the algorithm when $j-i+1 = k$,  we have
      \begin{align*}
        f(k) &= f\left( \left\lfloor  \frac{k}{2} \right \rfloor\right) + f\left( \left\lceil \frac{k}{2}\right \rceil \right) +  O(k \delta  \log^3 \pn \log \log \pn),\\
        f(1) &= O(\delta \log \delta \log^2 \pn \log \log \pn).
      \end{align*}
      Solving for this, we have $f(k) = O(k \delta \log k  \log^3 \pn
      \log \log \pn)$. In particular, the complexity of finding $N_{1,n_l}(G)$
      is $f(\pn) = O(\pn \delta \log^4 \pn \log \log \pn)$.

      Now, we turn to bounding $g(k)$, which was defined as the memory
      complexity when $j-i+1 = k$, excluding the space required to store $\sfU$
      and $\sfb$. The above discussion implies that
      \begin{align*}
        g(k) &= \max \left\{ g\left( \left\lfloor \frac{k}{2}\right \rfloor \right) , g\left( \left\lceil \frac{k}{2}\right \rceil \right)\right\} + O(k \delta \log \pn), \\
        g(1) &= O(\delta \log \pn).
      \end{align*}
      Solving for this, we have $g(k) = O(k \delta \log \pn)$. Specifically,
      $g(n_l) = O(\delta \pn \log \pn)$. Recall from
      Remark~\ref{rem:b-enc-U-b-pointer} that the space required to store
      $\sfU$ and $\sfb$ is $O(\pn \log \pn)$, and we store these data structures
      in one location of the memory, and use the same location throughout the
      process of finding $N_{i,j}(G)$ in Algorithm~\ref{alg:bip-Nij}. Therefore, the overall memory
      complexity of the algorithm is $O(\delta \pn \log \pn + \pn \log \pn) =
      O(\delta \pn \log \pn)$.  
  \end{proof}

\begin{lem}
  \label{lem:bip-encode-overall-complexity}
    In Algorithm~\ref{alg:bip-comp}, if we have $a_i \leq \delta$ for $ 1 \leq i \leq
    n_l$,  computing $\fnlr_{\va,\vb}(G)$ for
    $G \in \mGnlr_{\va, \vb}$ takes $O(\pn \delta \log^4 \pn \log
    \log \pn)$ time and requires
    $O(\delta \pn \log \pn)$ bits of memory, where $\tilde{n} = \max\{n_l, n_r\}$.
  \end{lem}

  \begin{proof}[Proof of Lemma~\ref{lem:bip-encode-overall-complexity}]
Note that each partial sum in the Fenwick tree $\sfU$, which is initialized with
array $\vb$,  has $O(\log \pn)$ bits. To
see this, note that
    \begin{equation*}
      \sfU.\textproc{Sum}(v) = \sum_{k=v}^{n_r} b_k \leq \sum_{k=1}^{n_r} b_k  = \sum_{k=1}^{n_l} a_k \leq \pn \delta.
    \end{equation*}
Therefore, initializing
    the Fenwick tree $\sfU$ with array $\vb$ takes $O(\pn \log \pn)$ time and
    the space required to store $\sfU$ is $O(\pn \log \pn)$ bits.
Furthermore, since $b_i$ has $O(\log \pn)$ bits, initializing $\vsfb$ with $\vb$
takes $O(\pn \log \pn)$ time, and requires $O(\pn \log \pn)$ bits to be stored.
Also, from Lemma~\ref{lem:bip-r-l-N-bit-count}, the variable $N_{1, n_l} = N_{1,n_l}(G)$
which is computed in line~\ref{l:bip-comp-N} of Algorithm~\ref{alg:bip-comp} has $O(\delta \pn \log \pn)$ bits.
Similarly, from Lemma~\ref{lem:bip-r-l-N-bit-count}, $l_{1,n_l} = l^G_{1, n_l} =
\prod_{i=1}^{n_r} b_r!$ has $O(\delta \pn \log \pn)$ bits. Therefore, the time
complexity of division and multiplication in lines~\ref{l:bip-enc-f-N-div-l} and
\ref{l:bip-enc-f-times-l-N} is $O(\pn \delta  \log^2 \pn \log \log \pn)$.
This together with  Lemma~\ref{lem:bip-encode-computeN-complexity} completes the
proof. 
  \end{proof}

We now turn to the proofs of Lemmas~\ref{lem:combinatorial-bound-pqs}, \ref{lem:bip-N+l<r}, and
\ref{lem:bip-r-l-N-bit-count}.
\color{black}

  \begin{proof}[Proof of Lemma~\ref{lem:combinatorial-bound-pqs}]
  We prove this in three cases. First, assume that $p \geq q$. In this case, the
  left hand side can be interpreted as the number of ways we can pick $q$ balls
  out of $p$ red balls and $s$ blue balls. For this, we can either pick all $q$
  balls among the red balls, or pick one blue ball and $q-1$ red balls. This
  establishes the inequality. Now, assume that $p = q-1$. Then, the inequality
  of interest becomes
  \begin{equation}
    \label{eq:pqs-lemma-case}
   \binom{q-1+s}{q} \geq s. 
  \end{equation}
  If $s = 0$, this becomes $0
  \geq 0$ which is true. If $s \geq 1$, the left hand side
  of~\eqref{eq:pqs-lemma-case} is the number of ways we
  can pick $q$ balls out of $q-1$ red balls and $s$ blue balls. A special
  way of doing this would be to pick one blue ball, and $q-1$ red balls which
  establishes~\eqref{eq:pqs-lemma-case}. Finally, if $p < q-1$, the right hand
  side of~\eqref{eq:comb-bound-pqs-bound} becomes zero, and it automatically holds.
\end{proof}

  \begin{proof}[Proof of Lemma~\ref{lem:bip-N+l<r}]
    First, we show this for $i = j$. In order do so, for $1 \leq k \leq
    a_i$, we define
    \begin{equation*}
      z^G_{i,k} := \sum_{\pp{k} = k}^{a_i} \left[  \prod_{\ppp{k} = k}^{\pp{k} - 1} b^G_{\gamma^G_{i,\ppp{k}}}(i) \right] \binom{U^G_{1+\gamma^G_{i,\pp{k}}}(i)}{a_i - \pp{k} + 1},
    \end{equation*}
    and we 
    claim that
    \begin{equation}
      \label{eq:zik-prod-b<=U-choose-ai-k+1}
      z^G_{i,k} + \prod_{\pp{k} = k}^{a_i} b^G_{\gamma^G_{i,\pp{k}}}(i) \leq \binom{U^G_{\gamma^G_{i,k}}(i)}{a_i - k + 1}.
    \end{equation}
    We do this by backward induction on $k$. For $k = a_i$, we have
    \begin{equation*}
      z^G_{i,a_i} + b^G_{\gamma^G_{i,a_i}}(i) = \binom{U^G_{1+\gamma^G_{i,a_i}}(i)}{1} + b^G_{\gamma^G_{i,a_i}}(i) = b^G_{\gamma^G_{i,a_i}}(i) + \sum_{s=1+\gamma^G_{i,a_i}}^{n_r} b^G_{s}(i) =  U^G_{\gamma^G_{i,a_i}}(i),
    \end{equation*}
    which yields \eqref{eq:zik-prod-b<=U-choose-ai-k+1} for $k = a_i$. Now,
    assuming that~\eqref{eq:zik-prod-b<=U-choose-ai-k+1} holds for $k +1$, we have
    \begin{align*}
      z^G_{i,k} + \prod_{\pp{k} = k}^{a_i} b^G_{\gamma^G_{i,\pp{k}}}(i) &= \binom{U^G_{1+\gamma^G_{i,k}}(i)}{a_i - k + 1} + b^G_{\gamma^G_{i,k}}(i) z^G_{i,k+1} + b^G_{\gamma^G_{i,k}}(i)\prod_{\pp{k} = k+1}^{a_i} b^G_{\gamma^G_{i,\pp{k}}}(i) \\
                                                             &\stackrel{(a)}{\leq} \binom{U^G_{1+\gamma^G_{i,k}}(i) }{a_i - k + 1} + b^G_{\gamma^G_{i,k}}(i) \binom{U^G_{\gamma^G_{i,k+1}}(i)}{a_i - k } \\
                                                             &\stackrel{(b)}{\leq} \binom{U^G_{1+\gamma^G_{i,k}}(i)}{a_i - k + 1} + b^G_{\gamma^G_{i,k}}(i) \binom{U^G_{1+\gamma^G_{i,k}}(i)}{a_i - k} \\
                                                             &\stackrel{(c)}{\leq} \binom{U^G_{1+\gamma^G_{i,k}}(i) + b^G_{\gamma^G_{i,k}}(i)}{a_i - k + 1} \\
      &= \binom{U^G_{\gamma^G_{i,k}}(i)}{a_i - k + 1},
    \end{align*}
    where $(a)$ uses the induction hypothesis for $k+1$, $(b)$ uses
    $U^G_{\gamma^G_{i,k+1}}(i) \leq U^G_{1+\gamma^G_{i,k}}(i)$ which holds since
    $\gamma^G_{i,k+1} \geq 1+\gamma^G_{i,k}$, and $(c)$  employs
    Lemma~\ref{lem:combinatorial-bound-pqs}.
    This completes the proof of
    \eqref{eq:zik-prod-b<=U-choose-ai-k+1}. In particular, using
    \eqref{eq:zik-prod-b<=U-choose-ai-k+1} for $k = 1$, we get
    \begin{equation}
      \label{eq:Nii-li-UG-1-ai}
      N_{i,i}(G) + l^G_{i,i} = z^G_{i,1} + l^G_i \leq \binom{U^G_{\gamma^G_{i,1}}(i)}{a_i} \leq \binom{U^G_{1}(i)}{a_i}. 
    \end{equation}
    On the other hand, we may write 
    \begin{align*}
      S_i = \sum_{k=i}^{n_l} a_k &= S - \sum_{k=1}^{i-1} a_k \\
                                 &= S - \sum_{k=1}^{i-1} \sum_{j=1}^{n_r} \one{k \sim_G j} \\
                                 &= \sum_{j=1}^{n_r} b_j - \sum_{j=1}^{n_r} \sum_{k=1}^{i-1} \one{k \sim_G j} \\
                                 &= \sum_{j=1}^{n_r} \left( b_j - \sum_{k=1}^{i-1} \one{k \sim_G j}\right) \\
                                 &= \sum_{j=1}^{n_r} b^G_j(i) \\
      &= U^G_1(i).
    \end{align*}
    Substituting into~\eqref{eq:Nii-li-UG-1-ai}, we realize that
    \begin{equation*}
      N_{i,i}(G) + l^G_i \leq \binom{S_i}{a_i} = r_i.
    \end{equation*}
    This completes the proof of the  desired result for $i = j$.

    We prove the general case for $i \leq j$ using induction on $j-i$. The base
    case $j - i = 0$ was proved above. For $i < j$, with $k := \lfloor  (i+j) /
    2 \rfloor$, using~\eqref{eq:bip-Nij-recurvie} and the identity $l_{i,j}^G = l_{i,k}^G
    l_{k+1, j}^G$, we have
    \begin{align*}
      N_{i,j}(G) + l^G_{i,j} &= N_{i,k}(G) r_{k+1, j} + l^G_{i,k} N_{k+1, j}(G)  + l^G_{i,k}l^G_{k+1, j} \\
                           &= N_{i,k}(G) r_{k+1, j} + l^G_{i,k}(N_{k+1, j}(G) + l^G_{k+1,j}) \\
                           &\stackrel{(a)}{\leq} N_{i,k}(G) r_{k+1, j} + l^G_{i,k} r_{k+1, j} \\
                           &\stackrel{(b)}{\leq} r_{i,k} r_{k+1, j} \\
      &= r_{i,j}
    \end{align*}
    where in $(a)$ we have used the induction hypothesis for the interval $[k+1,
    j]$ with  $j - (k+1) < j- i$, and in $(b)$, we have used the induction
    hypothesis for the interval $[i,k]$ with $k - i < j - i $. This
    completes the proof. 
  \end{proof}

  \begin{proof}[Proof of Lemma~\ref{lem:bip-r-l-N-bit-count}]
  Following the definition, we have
  \begin{equation*}
    r_{i,j} = \frac{S_i!}{\left( \prod_{k=i}^j a_k! \right) S_{j+1}!} \leq \frac{S_i!}{S_{j+1}!} \leq S^{S_i - S_{j+1}} = S^{\sum_{k=i}^j a_k}.
  \end{equation*}
  Using $\sum_{k=i}^j a_k \leq (j-i+1) \delta$ and $S \leq \pn \delta$, we
  realize that $r_{i,j}$ has $O((j-i+1)\delta \log (\pn \delta)) = O((j-i+1)
  \delta \log \pn)$ bits. 
  Lemma~\ref{lem:bip-N+l<r} above implies that this is also an upper bound for
  the number of bits in $N_{i,j}(G)$ and $l_{i,j}^G$. This completes the proof.
\end{proof}

\subsection{Decompression Algorithm}
\label{sec:bipartite-decode}

At the decoder, given the sequences $\va$, $\vb$, and  $\fnlr_{\va,
    \vb}(G)$, we want to decode for the simple bipartite graph $G$. Recall that
  $\fnlr_{\va,\vb}(G) = \lceil  N(G) / \prod_{j=1}^{n_r} b_j! \rceil$.
  Therefore, we introduce a proxy for $N(G) = N_{1,n_l}(G)$ defined as
  \begin{equation*}
    \tN_{1,n_l} := \fnlr_{\va, \vb}(G) \left( \prod_{j=1}^{n_r} b_j! \right).
  \end{equation*}
  Indeed, we have
  \begin{equation}
    \label{eq:tN-N-N+proj-b!}
    N_{1,n_l}(G) \leq \tN_{1,n_l} < N_{1,n_l}(G) + \prod_{j=1}^{n_r} b_j!.
  \end{equation}
  Recall from~\eqref{eq:prod-b!-l1n} that $\prod_{j=1}^{n_r} b_j! =
  l^G_{1,n_l}$. Substituting this in the above, we get 
  \begin{equation}
    \label{eq:tN-N-l1n}
    N_{1,n_l}(G) \leq \tN_{1,n_l} < N_{1,n_l}(G) + l^G_{1,n_l}.
  \end{equation}
We employ a recursive 
approach
to decode for the graph $G$ by using $\tN_{1, n_l}$. More precisely, we
define a procedure that, given $1 \leq i \leq j \leq n_l$, decodes for the
adjacency list of the left vertices in the interval $[i,j]$. This
procedure is initially called for $i = 1, j = n_l$. Once called for $i \leq
j$, if $i \neq j$, it calls sub--procedures for the intervals $[i,k]$ and
$[k+1, j]$, where $k = \lfloor  (i+j) / 2 \rfloor$. On the other hand, if $i =
j$, it decodes for the adjacency list of the left vertex $i$ and finds
$\gamma^G_{i,1}, \dots, \gamma^G_{i,a_i}$. We describe this procedure below in more
detail.
When the procedure is called for the interval $[i,j]$, we inductively assume that
the following conditions hold
\begin{enumerate}[label=\textbf{A\arabic*}]
\item \label{a:bdec-1-i-1} The adjacency lists of the left vertices $1, \dots, i-1$ are
  correctly decoded. 
\item \label{a:bdec-tN-N} Integer $\tN_{i,j}$ is given such that $N_{i,j}(G)
  \leq \tN_{i,j} < N_{i,j}(G) + l^G_{i,j}$.
\end{enumerate}
In particular, condition~\ref{a:bdec-1-i-1} implies that $b^G_k(i)$ and $U^G_k(i)$ for $1 \leq
k \leq n_r$ are known. Given these, we design an algorithm that: 
\begin{enumerate}
\item Decodes for the adjacency lists of the left vertices in the range
  $[i,j]$;
\item Finds $N_{i,j}(G)$ and $l^G_{i,j}$. 
\end{enumerate}
We will prove that our algorithm satisfies these properties
inductively.
Note that for $i=1$ and 
\color{red}
$j= n_l$, 
\color{black}
the condition~\ref{a:bdec-1-i-1} automatically
holds since the interval $[1,i-1]$ is empty. Condition~\ref{a:bdec-tN-N}
also holds from~\eqref{eq:tN-N-l1n}.
The inductive structure of our argument is as follows: define the \emph{process
tree} to be a binary rooted tree where each node corresponds to an interval
of the form $[i,j], 1 \leq i \leq j \leq n_l$. The root corresponds to the interval
$[1,n_l]$, and each node $[i,j]$ with $i < j$ has two children, where the left child
corresponds to the interval $[i,k]$ and the right node corresponds to the
interval $[k+1, j]$, where $k = \lfloor (i+j)/2 \rfloor$. Moreover, the leaf
nodes correspond to the intervals $[i,i]$ for $1 \leq i \leq n_l$.
The running of the algorithm can be visualized on this tree, where each node
first calls its left node, waits for its result, then calls its right node, and
combines the results received from the children to return its result to the
parent. We will prove that the conditions~\ref{a:bdec-1-i-1} and
\ref{a:bdec-tN-N} hold when we traverse this tree from top to bottom by
calling sub-processes, and will show that each node delivers the two items
mentioned above from bottom to top when returning to its parent. 

Now, we present the decoding procedure for the interval $[i,j]$. First,
we introduce the scheme when $i = j$, and later we study the case $i < j$.

\subsubsection{Decoding the Adjacency List of One Node }
\label{sec:bip-dec-one-node}

In this section we consider the case $i = j$. 
Define $\tz_i := \tN_{i,i}$ and note that due to
the condition~\ref{a:bdec-tN-N} we can think of  $\tz_i$ as a proxy
for $z^G_i = N_{i,i}(G)$. We claim that
  \begin{equation}
    \label{eq:bip-dec-xi1-claim}
    \gamma^G_{i,1} = \min \left\{ 1 \leq v \leq n_r : \binom{U^G_{1+v}(i)}{a_i} \leq \tz_i \right\}.
  \end{equation}
  Note that, as we discussed above, the decoder has access to $U^G_v(i)$ for $1
  \leq v \leq n_r$. Thereby, assuming  \eqref{eq:bip-dec-xi1-claim} holds, it can be used at
  the decoder to find $\gamma^G_{i,1}$.
  Before proving \eqref{eq:bip-dec-xi1-claim}, 
considering the definition of  $z^G_i$ in~\eqref{eq:benc-zi} and the fact
  that we use $\tz_i$ as a proxy for $z^G_i$, let $\tz_{i,2}$ be obtained from $\tz_i$ by removing the contribution of
  $\gamma^G_{i,1}$, i.e.\
  \begin{equation*}
    \tz_{i,2} := \left\lfloor \frac{\tz_i - \binom{U^G_{1+\gamma^G_{i,1}}(i)}{a_i}}{b^G_{\gamma^G_{i,1}}(i)} \right \rfloor.
  \end{equation*}
  Again, note that, assuming $\gamma^G_{i,1}$ is found, $\tz_{i,2}$ can be computed at the decoder. We claim that
    \begin{equation*}
    \gamma^G_{i,2} = \min\left\{ \gamma^G_{i,1} < v \leq n_r : \binom{U^G_{1+v}(i)}{a_i - 1} \leq \tz_{i,2}  \right\}.
  \end{equation*}
  Assuming this holds, it can be used to find $\gamma^G_{i,2}$ at the decoder.
In general, for $2 \leq k \leq a_i$, we claim that
  \begin{equation}
    \label{eq:bip-dec-xik-claim}
    \gamma^G_{i,k} = \min \left\{  \gamma^G_{i,k-1} < v \leq n_r : \binom{U^G_{1+v}(i)}{a_i - k + 1} \leq \tz_{i,k} \right\},
  \end{equation}
where $\tz_{i,k}$ is defined inductively as 
\begin{equation}
  \label{eq:bip-decomp--tzk+1-tz-def}
  \tz_{i,k} := \left\lfloor \frac{\tz_{i,k-1} - \binom{U^G_{1+\gamma^G_{i,{k-1}}}(i)}{a_i-k+2}}{b^G_{\gamma^G_{i,k-1}}(i)} \right \rfloor.
\end{equation}
Before proving these claims, we need to define some notation and introduce some
lemmas. For $1 \leq k \leq a_i$, we define
\begin{equation}
  \label{eq:bip-decomp-zik}
  z^G_{i,k} := \sum_{\pp{k} = k}^{a_i} \left[  \prod_{\ppp{k} = k}^{\pp{k} - 1} b^G_{\gamma^G_{i,\ppp{k}}}(i) \right] \binom{U^G_{1+\gamma^G_{i,\pp{k}}}(i)}{a_i - \pp{k} + 1}.
\end{equation}
In particular, note that $z^G_{i,1} = z^G_i$. Also, we define $\tz_{i,1} := \tz$.
In order to prove the above claims, we state and prove two  lemmas.

\begin{lem}
  \label{lem:bip-decomp-tzik-zik-bound}
  For $1 \leq i \leq n_l$ and $1 \leq k \leq a_i$, we have
  \begin{equation*}
    z^G_{i,k} \leq \tz_{i,k} < z^G_{i,k} + \prod_{\pp{k} = k}^{a_i} b^G_{\gamma^G_{i,\pp{k}}}(i).
  \end{equation*}
\end{lem}

\begin{proof}
      We prove this by induction on $k$. For $k = 1$, this reduces to $z^G_i \leq
      \tN_{i,i} < z^G_i + l^G_i$ which is known by assumption. Now we assume it
      holds for $k$ and show it also holds for $k+1$. It is easy to see that
      \begin{equation}
        \label{eq:bip-dec-zik-recursive}
        z^G_{i,k} = \binom{U^G_{1+\gamma^G_{i,k}}(i)}{a_i - k + 1} + b^G_{\gamma^G_{i,k}}(i) z^G_{i,k+1}.
      \end{equation}
      Using this together with the assumption for $k$, we get
      \begin{equation*}
        z^G_{i,k+1} \leq \frac{\tz_{i,k} - \binom{U^G_{1+\gamma^G_{i,k}}(i)}{a_i - k + 1}}{b^G_{\gamma^G_{i,k}}(i)} < z^G_{i,k+1} + \prod_{\pp{k} = k + 1}^{a_i} b^G_{\gamma^G_{i,\pp{k}}}(i).
      \end{equation*}
      Since both the lower and upper bounds are integers, we get the desired result by
      taking the floor and comparing with the definition of $\tz_{i,k+1}$ in \eqref{eq:bip-decomp--tzk+1-tz-def}.
    \end{proof}

        \begin{lem}
      \label{lem:bip-deconp-v-xik}
      For $1 \leq k \leq a_i$, if $v < \gamma^G_{i,k}$, then we have
      \begin{equation}
        \label{eq:bip-decomp-U1+v-zik}
        \binom{U^G_{1+v}(i)}{a_i - k + 1} \geq z^G_{i,k} + \prod_{\pp{k} = k}^{a_i} b^G_{\gamma^G_{i,\pp{k}}}(i).
      \end{equation}
    \end{lem}
    \begin{proof}
      We prove this by backward induction on $k$. Note that if $v <
      \gamma^G_{i,k}$, recalling the definition of $U^G_{1+v}(i)$, we have
      \begin{equation*}
        U^G_{1+v}(i) = \sum_{s=1+v}^{n_r} b^G_s(i) \geq b^G_{\gamma^G_{i,k}}(i) + \sum_{s=1+\gamma^G_{i,k}}^{n_r} b^G_s(i) = b^G_{\gamma^G_{i,k}}(i) + U^G_{1+\gamma^G_{i,k}}(i).
      \end{equation*}
      Thereby, for $k = a_i$, if $v < \gamma^G_{i,a_i}$, we have 
      \begin{equation*}
        \binom{U^G_{1+v}(i)}{1} \geq \binom{U^G_{1+\gamma^G_{i,a_i}}(i) + b^G_{\gamma^G_{i,a_i}}(i)}{1} = \binom{U^G_{1+\gamma^G_{i,a_i}}(i)}{1} + b^G_{\gamma^G_{i,a_i}}(i) = z^G_{i,a_i} + b^G_{\gamma^G_{i,a_i}}(i),
      \end{equation*}
      which is precisely \eqref{eq:bip-decomp-U1+v-zik} for $k = a_i$. Now,
      assume $k < a_i$, $v < \gamma^G_{i,k}$ and the statement holds for $k+1$.
      We may write 
      \begin{align*}
        \binom{U^G_{1+v}(i)}{a_i - k + 1} &\geq \binom{U^G_{1+\gamma^G_{i,k}}(i) + b^G_{\gamma^G_{i,k}}(i)}{a_i - k + 1} \\
                                        &\stackrel{(a)}{\geq} \binom{U^G_{1+\gamma^G_{i,k}}(i)}{a_i - k + 1} + b^G_{\gamma^G_{i,k}}(i) \binom{U^G_{1+\gamma^G_{i,k}}(i)}{a_i - (k+1) + 1} \\
                                        &\stackrel{(b)}{\geq}  \binom{U^G_{1+\gamma^G_{i,k}}(i)}{a_i - k + 1} + b^G_{\gamma^G_{i,k}}(i) \left( z^G_{i,k+1} + \prod_{\pp{k}= k+1}^{a_i} b^G_{\gamma^G_{i,\pp{k}}}(i) \right) \\
                                        &=  \binom{U^G_{1+\gamma^G_{i,k}}(i)}{a_i - k + 1} + b^G_{\gamma^G_{i,k}}(i) z^G_{i,k+1} + \prod_{\pp{k}= k}^{a_i} b^G_{\gamma^G_{i,\pp{k}}}(i)  \\
                                        &\stackrel{(c)}{=} z^G_{i,k} + \prod_{\pp{k}= k}^{a_i} b^G_{\gamma^G_{i,\pp{k}}}(i),
      \end{align*}
      where $(a)$ uses Lemma~\ref{lem:combinatorial-bound-pqs},
      $(b)$ uses the induction hypothesis for $k+1$ with $\gamma^G_{i,k} < \gamma^G_{i,k+1}$, and $(c)$
      employs~\eqref{eq:bip-dec-zik-recursive}. Note that this is
      precisely~\eqref{eq:bip-decomp-U1+v-zik}, and completes the proof.
    \end{proof}

We are now ready to show \eqref{eq:bip-dec-xi1-claim} and
\eqref{eq:bip-dec-xik-claim}. In order to show \eqref{eq:bip-dec-xi1-claim},
note that if $v \geq \gamma^G_{i,1}$, we have $U^G_{1+v}(i) \leq U^G_{1+\gamma^G_{i,1}}(i)$.
Thereby,
\begin{equation}
  \label{eq:v>=xi1}
  \binom{U^G_{1+v}(i)}{a_i} \leq \binom{U^G_{1+\gamma^G_{i,1}}(i)}{a_i} \leq z^G_{i,1} \leq \tz_{i,1},
\end{equation}
where the last inequality uses Lemma~\ref{lem:bip-decomp-tzik-zik-bound}
with $k=1$.
On the other hand, if $v < \gamma^G_{i,1}$, using
Lemma~\ref{lem:bip-deconp-v-xik} and then
Lemma~\ref{lem:bip-decomp-tzik-zik-bound} for $k=1$, we get
\begin{equation}
  \label{eq:v<xi1}
  \binom{U^G_{1+v}(i)}{a_i} \geq z^G_{i,1} + \prod_{\pp{k} = 1}^{a_i} b^G_{\gamma^G_{i,\pp{k}}}(i) > \tz_{i,1}.
\end{equation}
Also, observe that $\binom{U^G_{1+v}(i)}{a_i}$ is
nonincreasing for $1 \leq v \leq n_r$. This together with 
the bounds in \eqref{eq:v>=xi1} and~\eqref{eq:v<xi1}   imply~\eqref{eq:bip-dec-xi1-claim}. The proof of \eqref{eq:bip-dec-xik-claim} is similar.

Motivated by~ \eqref{eq:bip-dec-xi1-claim} and
\eqref{eq:bip-dec-xik-claim},  we employ a binary
search scheme to decode for $\gamma^G_{i,1}, \dots, \gamma^G_{i,a_i}$. 
Note that once we finish decoding for $\gamma^G_{i,1}, \dots, \gamma^G_{i,a_i}$, we can
compute $N_{i,i}(G)$ and $l^G_i$. Hence, for $i =j$, the algorithm satisfies the
two conditions promised earlier, i.e. it decodes for the adjacency list of the left
vertex $i$ and finds $N_{i,i}(G)$ and $l^G_{i}$.

Algorithm~\ref{alg:bip-dec-i} below performs the above decompression scheme for
a left node $i$. 
Similar to the compression algorithm, we assume that the algorithm is provided
with a Fenwick tree $\sfU$ such that $\sfU.\textproc{Sum}(v) = U^G_v(i)$ for $1
\leq v \leq n_r$, and an array $\vsfb = (\sfb_v: v \in [n_r])$ such that $\sfb_v
= b^G_v(i)$ for $v \in [n_r]$. For each $1 \leq k \leq a_i$, in the loop of
line~\ref{l:bdec-k-for}, the variable $\tz$
keeps track of the value $\tz_{i,k}$ defined above and uses it to perform the binary
search in the loop of line~\ref{l:bdec-i-while} to find $\gamma_{i,k} =
\gamma^G_{i,k}$. Then, in line~\ref{l:bdec-i-tz}, we update $\tz$ 
to $\tz_{i,k+1}$. Furthermore, $z_i$ and $l_i$ are updated in
lines~\ref{l:bdec-i-z} and \ref{l:bdec-i-l}, respectively, so that once the loop
over $k$ is over, we have $z_i = z^G_i$ and $l_i = l^G_i$. Finally, the Fenwick
tree $\sfU$ and the array
$\vsfb$ are updated in lines~\ref{i:bdec-i-U} and~\ref{i:bdec-i-beta},
respectively.



\begin{myalg}[Decoding the adjacency list of a left vertex $1 \leq i \leq n_l$
    given $\tN_{i,i}$ for a simple unmarked bipartite graph $G \in \mGnlr_{\va,
    \vb}$. \label{alg:bip-dec-i}]
  \begin{algorithmic}[1]
    \INPUT
    \Statex $n_l, n_r$: the number of left and right nodes, respectively
  \Statex $i$: the index of the left node to be decoded, $1 \leq i \leq n_l$
    \Statex $\va$: left degree sequence
  \Statex $\vsfb = (\sfb_v: v \in [n_r])$: $\Array$ of integers where $\sfb_v = b^G_v(i)$ for
  $1 \leq v \leq n_r $
  \Statex $\tN_{i,i}$: integer satisfying $N_{i,i}(G) \leq \tN_{i,i}  < N_{i,i}(G) + l^G_i$
  \Statex $\sfU$: Fenwick tree, where for $1 \leq v \leq n_r $, $\sfU.\textproc{Sum}(v) =
  U^G_v(i) =   \sum_{k=v}^{n_r} b^G_k(i)$
  \OUTPUT
  \Statex  $N_{i,i} = N_{i,i}(G)$
  \Statex $\vgamma_i$: the decoded adjacency list of  vertex $i$, so
  that $\vgamma_i = \vgamma^G_i = (\gamma^G_{i,k}: 1 \leq k \leq a_i)$ 
  \Statex $\vsfb = (\sfb_v: 1 \leq v \leq n_r )$: array $\vsfb$ updated so
  that $\sfb_v = b^G_v(i+1)$ for $1 \leq v \leq n_r$
  \Statex $\sfU$: Fenwick tree $\sfU$ updated, so that for $1 \leq v \leq n_r$, we
  have $\sfU.\textproc{Sum}(v) = U^G_v(i+1) = \sum_{k=v}^{n_r} b^G_k(i+1)$
  \Statex $l_i$: integer such that $l_i = l_i^G$
  \Function{BDecodeNode}{$n_l, n_r, i, \va, \vsfb,\tN_{i,i}, \sfU$}
  \State $\tz \gets \tN_{i,i}$ \label{l:bdec-i-tz-init}
  \State $z_i \gets 0, l_i \gets 1$ 
  \For{$1 \leq k \leq a_i$} \label{l:bdec-k-for}
  \State $q \gets a_i - k + 1$
  \State $L \gets 1, R \gets n_r$ \Comment{$L$ and $R$ are the endpoints of the
    binary search interval}
  \If{$k > 1$}
  \State $L \gets 1 + \gamma_{i,k-1}$ \Comment{If $k > 1$, $1+\gamma_{i,k-1} \leq
    \gamma_{i,k}$, so limit the search}
  \EndIf
  \While{$R > L$} \label{l:bdec-i-while} \Comment{binary search on the interval $[f,g]$ to find
    $\gamma^G_{i,k}$} 
  \State $v \gets (L+R)\div 2$ \Comment{$v = \lfloor  (L+R)/2 \rfloor$ is the midpoint}
  \State $y \gets \textproc{ComputeProduct}(\sfU.\textproc{Sum}(1+v), q, 1) \div
  \textproc{ComputeProduct}(q,q,1)$ \label{l:bdec-i-ywhile}
  \Statex \Comment{$y = \binom{U^G_{1+v}(i)}{a_i - k + 1}$, Algorithm~\ref{alg:compute-product} in Section~\ref{sec:aux-algs-bip}}
  \If{$y \leq \tz$}
  \State $R \gets v$ \Comment{switch to interval $[L,v]$}
  \Else
  \State $L \gets v+1$ \Comment{switch to interval $[v+1,R]$}
  \EndIf
  \EndWhile
  \State $\gamma_{i,k} \gets L$
  \State $y \gets \textproc{ComputeProduct}(\sfU.\textproc{Sum}(1+\gamma_{i,k}), q, 1) \div
  \textproc{ComputeProduct}(q,q,1)$ \label{l:bdec-i-y} 
  \Statex \Comment{$y = \binom{U^G_{1+\gamma^G_{i,k}}(i)}{a_i - k + 1}$, Algorithm~\ref{alg:compute-product} in Section~\ref{sec:aux-algs-bip}}
  \State $\tz \gets (\tz - y) \div \sfb_{\gamma_{i,k}}$ \label{l:bdec-i-tz}
  \State $z_i \gets z_i + l_i \times y$ \label{l:bdec-i-z} \Comment{here, $l_i =
    \prod_{\pp{k} = 1}^{k-1} b^G_{\gamma^G_{i,\pp{k}}}(i)$}
  \State $l_i \gets l_i \times \sfb_{\gamma_{i,k}}$ \label{l:bdec-i-l}
  \Comment{$l_i$ becomes $\prod_{\pp{k} = 1}^{k} b^G_{\gamma^G_{i,\pp{k}}}(i)$}
  \State $\sfU.\textproc{Add}(\gamma_{i,k}, -1)$ \label{i:bdec-i-U}
  \State $\sfb_{\gamma_{i,k}} \gets \sfb_{\gamma_{i,k}} - 1$ \label{i:bdec-i-beta}
  \EndFor
  \State $N_{i,i} \gets z_i$
  \State \textbf{return} $(N_{i,i}, \vgamma_i, \sfb, \sfU, l_i)$
  \EndFunction
\end{algorithmic}
\end{myalg}

\subsubsection{Decoding the Adjacency List of an Interval of Nodes}
\label{sec:bip-dec-interval}
Now we introduce our procedure for decoding the interval $[i,j]$ when $i
  < j$. Assuming that \ref{a:bdec-1-i-1} and \ref{a:bdec-tN-N} hold for the
  interval $[i,j]$,  we will break the decoding task  into two subproblems, one for the
  interval $[i,k]$ and one for the interval $[k+1, j]$ where $k = \lfloor (i+j) / 2 \rfloor$.
Motivated by \eqref{eq:bip-Nij-recurvie}, and the fact that from condition~\ref{a:bdec-tN-N},
  $\tN_{i,j}$ is a proxy for $N_{i,j}(G)$, we define
  \begin{equation*}
    \tN_{i,k} := \left\lfloor \frac{\tN_{i,j}}{r_{k+1,j}} \right \rfloor.
  \end{equation*}  
Recall that, by definition, we have
  \begin{equation*}
    r_{k+1, j} = \frac{S_{k+1} (S_{k+1} - 1) \dots (1 + S_{j+1})}{\prod_{v=k+1}^j a_v!}.
  \end{equation*}
  Note that all the terms on the right hand side 
 depend   only on the sequence $\va$ which is known to the decoder. Hence, we
  can compute $r_{k+1,j}$ and hence $\tN_{i,k}$ at the decoder.
  Therefore, we can call the decoding procedure for $[i,k]$ with $\tN_{i,k}$.

  Now, we check that the conditions~\ref{a:bdec-1-i-1} and \ref{a:bdec-tN-N}
    hold for the subinterval $[i,k]$.
Note that since \ref{a:bdec-1-i-1} holds for the interval $[i,j]$, it
automatically holds for $[i,k]$. To show~\ref{a:bdec-tN-N}, we first state the
      following lemma, whose proof is postponed until the end of this section. 

\begin{lem}
  \label{lem:N-N1-N2}
  Assume that nonnegative integers $N, N_1, N_2$ and positive integers $l_1, l_2, r$ are
  given such that $N = N_1 r + l_1 N_2$ and  $N_2 + l_2 \leq r$. Then, if
  an integer $\tN$ is given such
  that $N \leq \tN < N + l_1 l_2$, we have
  \begin{equation}
    \label{eq:lem-NN1N2l1l2r--1}
    N_1 \leq \left\lfloor  \frac{\tN}{r} \right\rfloor < N_1 + l_1,
  \end{equation}
  and
  \begin{equation}
    \label{eq:lem-NN1N2l1l2r--2}
    N_2 \leq \left\lfloor  \frac{\tN - N_1 r }{l_1} \right\rfloor < N_2 + l_2.
  \end{equation}
\end{lem}

Now we want to use Lemma~\ref{lem:N-N1-N2} with $N \equiv N_{i,j}(G)$, $N_1 \equiv
  N_{i,k}(G)$, $N_2 \equiv N_{k+1, j}(G)$, $l_1 \equiv l^G_{i,k}$, $l_2 \equiv
  l^G_{k+1, j}$, $\tN \equiv \tN_{i,j}$, and $r \equiv r_{k+1,j}$. The assumptions of the lemma hold
  because of Lemma~\ref{lem:bip-N+l<r}, \eqref{eq:bip-Nij-recurvie}, the
  Condition~\ref{a:bdec-tN-N} for the interval $[i,j]$, and the fact that
  $l^G_{i,j} = l^G_{i,k} l^G_{k+1, j}$.
  Therefore, 
  from~\eqref{eq:lem-NN1N2l1l2r--1}, we realize that the condition \ref{a:bdec-tN-N} also
  holds for the interval $[i,k]$. Thereby, by induction, the decoding procedure
  for the interval $[i,k]$ successfully decodes for the adjacency list of the
  left vertices in $[i,k]$, and it also finds $N_{i,k}(G)$ and $l^G_{i,k}$.
Next,   again motivated by \eqref{eq:bip-Nij-recurvie}, we
    define
    \begin{equation*}
      \tN_{k+1, j} := \left\lfloor \frac{\tN_{i,j} - N_{i,k}(G) r_{k+1, j}}{l^G_{i,k}} \right\rfloor.
    \end{equation*}
Note that all the terms are known to the decoder at this point, hence the
decoder can
compute $\tN_{k+1, j}$ unambiguously.
Then, we call the procedure for decoding the interval $[k+1, j]$ given $\tN_{k+1,
      j}$. Here, we  check that both \ref{a:bdec-1-i-1} and
    \ref{a:bdec-tN-N} hold for this interval. Condition~\ref{a:bdec-1-i-1} for
    the interval $[k+1, j]$ holds
    since the condition \ref{a:bdec-1-i-1} for the interval $[i,j]$ implies
    that prior to calling the procedure for the interval $[i,j]$, all the left
    ndoes in the
    interval $[1,i-1]$ were decoded correctly. In addition, the decoder has
    just decoded for the interval $[i,k]$. This means that when the decoder
    starts decoding for the interval $[k+1, j]$, all the left nodes in the
    interval $[1,k]$ are successfully decoded. 
    On the other hand, from \eqref{eq:lem-NN1N2l1l2r--2} in 
    Lemma~\ref{lem:N-N1-N2}, we realize that the condition \ref{a:bdec-tN-N} also holds
    for  the interval $[k+1, j]$. Hence, the procedure for decoding $[k+1, j]$ successfully decodes the adjacency list for
      the left nodes in this  interval  and returns $N_{k+1, j}(G)$ and $l^G_{k+1, j}$. At
      this point, we put the results of the two intervals $[i,k]$ and $[k+1, j]$
      together to obtain the adjacency list of the interval $[i,j]$. Also, we
       compute $N_{i,j}(G)$ and $l^G_{i,j}$ using
      \eqref{eq:bip-Nij-recurvie} and the identity $l^G_{i,j} = l^G_{i,k} l^G_{k+1,
        j}$.

Algorithm~\ref{alg:bip-dec-ij} below illustrates  the above procedure for
decoding the interval $[i,j]$ for $1 \leq i \leq j \leq n$. Similar to
Algorithm~\ref{alg:bip-dec-i}, we assume that an array $\vsfb = (\sfb_v: v \in
[n_r])$ is given such that $\sfb_v = b^G_v(i)$ for $1 \leq v \leq n_r$, and a
Fenwick tree $\sfU$ is given so that $\sfU.\textproc{Sum}(v) = U^G_v(i) =
\sum_{\pp{k} = v}^{n_r} b^G_{\pp{k}}(i)$ for $1
\leq v \leq n_r$. Moreover, in order to compute $r_{k+1, j}$, we assume that
another Fenwick tree $\sfW$ is given so that for $1 \leq v \leq n_l$, we have
$\sfW.\textproc{Sum}(v) = \sum_{\pp{k}=v}^{n_l} a_{\pp{k}}$. With this, we have
$S_{k+1} = \sfW.\textproc{Sum}(k+1)$ and $S_{j+1} = \sfW.\textproc{Sum}(j+1)$.
We use this together with $\textproc{ComputeProduct}$ of
Algorithm~\ref{alg:compute-product} and $\textproc{ProdFactorial}$ of
Algorithm~\ref{alg:comp-ProdAFactorial} in line~\ref{l:bdec-ij-r} to compute
$r_{k+1, j}$.


\begin{myalg}[Decoding the adjacency list of  the left vertices $i \leq v \leq j$ for $1 \leq i \leq j \leq n_l$ given $\tN_{i,j}$ for a simple unmarked
  bipartite graph $G \in \mGnlr_{\va,\vb}$. \label{alg:bip-dec-ij}]
  \begin{algorithmic}[1]
  \INPUT
  \Statex $n_l, n_r$: the number of left and right nodes, respectively
  \Statex $1 \leq i \leq j  \leq n_l$: the endpoints of the interval to be decoded
  \Statex $\va$: left degree sequence
  \Statex $\vsfb = (\sfb_v: 1 \leq v \leq n_r )$ where $\sfb_v = b^G_v(i)$ for
  $1 \leq v \leq n_r $
  \Statex $\tN_{i,j}$: integer satisfying $N_{i,j}(G) \leq \tN_{i,j} <
  N_{i,j}(G) + l^G_{i,j}$
  \Statex $\sfU$: Fenwick tree, where for $1 \leq v \leq n_r $,
  $\sfU.\textproc{Sum}(v) = U^G_v(i) =
  \sum_{\pp{k}=v}^{n_r} b^G_{\pp{k}}(i)$
  \Statex $\sfW$: Fenwick tree, where for $1 \leq v \leq n_l$, we have $\sfW.\textproc{Sum}(v) =
  \sum_{\pp{k}=v}^{n_l} a_{\pp{k}}$.
  \OUTPUT
  \Statex  $N_{i,j} = N_{i,j}(G)$
  \Statex $\vgamma_{[i:j]}$: the decoded adjacency list of the vertices $i \leq v
  \leq j$, so that $\vgamma_v = \vgamma_v^G$ for $i \leq v \leq j$
  \Statex $\vsfb = (\sfb_v: 1 \leq v \leq n_r )$: array $\vsfb$ updated so
  that $\sfb_v = b^G_v(j+1)$ for $1 \leq v \leq n_r$
  \Statex $\sfU$: Fenwick tree $\sfU$ updated, so that for $1 \leq v \leq n_r$, we
  have $\sfU.\textproc{Sum}(v) = U^G_v(j+1)= \sum_{\pp{k}=v}^{n_r} b^G_{\pp{k}}(j+1)$
  \Statex $l_{i,j} = l_{i,j}^G$
  \Function{BDecodeInterval}{$n_l, n_r, i,j, \va, \vsfb, \tN_{i,j},  \sfU, \sfW$}
  \If{$i=j$}
  \State \textbf{return} $\textproc{BDecodeNode}(n_l, n_r, i, \va, \vsfb, 
  \tN_{i,i},\sfU)$ \label{l:bdec-ij-call-ii} \Comment{Algorithm~\ref{alg:bip-dec-i} in Section~\ref{sec:bip-dec-one-node}}
  \Else
  \State $k \gets (i+j) \div 2$ \label{l:bdec-ij-k-ij-2-def}
  \State $S_{k+1} \gets \sfW.\textproc{Sum}(k+1)$ \label{l:bdec-Sk+1-Wk+1}
  \State $S_{j+1} \gets \sfW.\textproc{Sum}(j+1)$ \label{l:bdec-Sj+1-Wj+1}
  \State $r_{k+1, j} \gets \textproc{ComputeProduct}(S_{k+1}, S_{k+1} - S_{j+1},
  1) / \textproc{ProdFactorial}(\va, k+1,j)$ \label{l:bdec-ij-r}
  \Statex \Comment{Algorithms~\ref{alg:compute-product} and
    \ref{alg:comp-ProdAFactorial} in Section~\ref{sec:aux-algs-bip}}
  \State $\tN_{i,k} \gets \tN_{i,j} \div r_{k+1, j}$ \label{l:bdec-ij-tNik}
  \State $(N_{i,k}, \vgamma_{[i,k]}, \vsfb, \sfU, l_{i,k}) \gets
  \textproc{BDecodeInterval}(n_l, n_r,i,k, \va, \vsfb, \tN_{i,k}, \sfU,
  \sfW)$ \label{l:bdec-ij-ikcall} 
  \Statex \Comment{Algorithm~\ref{alg:bip-dec-ij} (recursive)}
  \State $\tN_{k+1, j} \gets (\tN_{i,j} - N_{i,k} \times r_{k+1, j}) \div
  l_{i,k}$ \label{l:bdec-ij-tNkj}
  \State $(N_{k+1, j}, \vgamma_{[k+1, j]}, \vsfb, \sfU, l_{k+1, j}) \gets
  \textproc{BDecodeInterval}(n_l, n_r, k+1, j, \va, \vsfb,  \tN_{k+1, j}, \sfU,
  \sfW)$ \label{l:bdec-ij-kjcall}
  \Statex \Comment{Algorithm~\ref{alg:bip-dec-ij} (recursive)}
  \State $N_{i,j} \gets N_{i,k} \times r_{k+1, j} + l_{i,k} \times N_{k+1, j}$ \label{l:bdec-ij-Nij}
  \State $l_{i,j} \gets l_{i,k} \times l_{k+1, j}$ \label{l:bdec-ij-lij}
  \State \textbf{return} $(N_{i,j}, \vgamma_{[i:j]}, \vsfb, \sfU, l_{i,j})$  \label{l:bdec-ij-return}
  \EndIf
  \EndFunction
\end{algorithmic}
\end{myalg}

Finally, we present the proof of Lemma~\ref{lem:N-N1-N2}.

\begin{proof}[Proof of Lemma~\ref{lem:N-N1-N2}]
  We have
  \begin{align*}
    N_1 = \frac{N_1 r}{r} &\leq \frac{N}{r} \\
          &\leq \frac{\tN}{r}  \\
    &<\frac{N+l_1 l_2}{r} 
        = \frac{N_1 r + l_1 N_2 + l_1 l_2}{r} 
        \leq \frac{N_1 r + l_1 r}{r} 
    = N_1 + l_1,
  \end{align*}
  which implies that $N_1 \leq \tN / r < N_1 + l_1$. Since $N_1$ and $N_1
  + l_1$ are integers, we arrive at \eqref{eq:lem-NN1N2l1l2r--1} by taking the floor
  from both sides. On the other hand,
  \begin{align*}
    N_2 = \frac{N_1 r + l_1 N_2 - N_1 r}{l_1} &= \frac{N- N_1 r}{l_1}\\
        &\leq \frac{\tN - N_1 r}{l_1}\\
        &< \frac{N+l_1 l_2 - N_1 r}{l_1} = \frac{N_1 r + l_1 N_2 + l_1 l_2 - N_1 r}{l_1} = N_2 + l_2,
  \end{align*}
  which implies $N_2 \leq \frac{\tN - N_1 r}{l_1} < N_2 + l_2$. Again,
  since $N_2$ and $N_2 + l_2$ are integers, we arrive at
  \eqref{eq:lem-NN1N2l1l2r--2} by taking the floor from both sides. 
\end{proof}

\subsubsection{The Main Decoding Algorithm}
\label{sec:bdec-main}

With this, we are ready to present the main decoding algorithm as follows.

\begin{myalg}[Decoding for a simple unmarked bipartite graph $G \in \mGnlr_{\va,
  \vb}$ given $\fnlr_{\va, \vb}(G)$ \label{alg:b-dec}]
\begin{algorithmic}[1]
  \INPUT
  \Statex $f$: integer, which is $\fnlr_{\va, \vb}(G)$ for the graph $G$ that
  was given to the encoder during the compression phase
  \Statex $\va$: array of left degrees
  \Statex $\vb$: array of right degrees
  \OUTPUT
  \Statex $\vgamma_{[1:n_l]}$: adjacency list of left nodes such that $\vgamma_v
  = \vgamma^G_v = (\gamma^G_{v,1}< \dots < \gamma^G_{v,a_v})$ for $1 \leq v \leq
  n_l$
  \Function{BDecodeGraph}{$f, \va, \vb$}
  \State $n_l \gets \textproc{Size}(\va)$
  \State $n_r \gets \textproc{Size}(\vb)$
  \State $c \gets \textproc{ProdFactorial}(\vb, 1,
  n_r)$ \label{l:bdec-prod-b-factorial} \Comment{$c = \prod_{i=1}^{n_r}
    b_i!$ using Algorithm~\ref{alg:comp-ProdAFactorial} in Sec.~\ref{sec:aux-algs-bip}}
  \State $\tN_{1, n_l} \gets f \times c$ \label{l:bdec-tN-1-nl-y-c}
  \State $\sfU \gets$ Fenwick tree initialized with array $\vb$ \label{l:bdec-U-init}
  \State $\sfW \gets$ Fenwick tree initialized with array $\va$ \label{l:bdec-W-init}
  \State $\vsfb \gets \vb$
  \State $(N_{1, n_l}, \vgamma_{[1:n_l]}, \vb, \sfU, l_{1,n_l}) \gets
  \textproc{BDecodeInterval}(n_l, n_r, 1, n_l, \va, \vsfb,  \tN_{1, n_l}, \sfU,
  \sfW)$ \label{l:bdec-bdecode-interval-call}
  \Statex \Comment{Algorithm~\ref{alg:bip-dec-ij} in Sec.~\ref{sec:bip-dec-interval}}
  
  \State \textbf{return} $\vgamma_{[1:n_l]}$
  \EndFunction
\end{algorithmic}
\end{myalg}

\begin{rem}
  \label{rem:b-dec-U-b-pointer}
  Similar to the discussion in Remark~\ref{rem:b-enc-U-b-pointer}, we  can store
  the Fenwick tree $\sfU$ and the array $\vsfb$ in one location of the memory,
  send the pointer of this location to the decoding algorithm and modify the
  same location for each interval $[i,j]$. This results in an improvement in the
  memory required to perform the algorithm. Also, since the Fenwick tree $\sfW$
  is not modified after initialization, we may keep it in one location and
  only pass the pointer to this location. The same property holds for the array $\va$.
Similar to
  Remark~\ref{rem:b-enc-U-b-pointer}, the memory required to store $\sfU$,
  $\vsfb$, $\sfW$, and $\va$ is $O(\pn \log \pn)$. 
\end{rem}

\subsubsection{Complexity of Decoding}
\label{sec:bdec-complexity}

In this section, we analyze the complexity of Algorithms~\ref{alg:bip-dec-i},
\ref{alg:bip-dec-ij} and \ref{alg:b-dec} discussed above, in
Lemmas~\ref{lem:bdec-i-complexity}, \ref{lem:bdec-ij-complexity}, and
\ref{lem:bdec-overall-complexity} below.

\begin{lem}
  \label{lem:bdec-i-complexity}
Assume that we have $a_v \leq \delta$ for all $1 \leq v \leq  n_l$. Also, assume
that the input $\tN_{i,i}$ to Algorithm~\ref{alg:bip-dec-i} satisfies
$N_{i,i}(G) \leq \tN_{i,i} < N_{i,i}(G) + l^G_i$ for the graph $G \in
\mGnlr_{\va, \vb}$ that was given to the encoder during the compression phase. Then, the time complexity of
decoding for the adjacency list of a left vertex $1 \leq i \leq n_l$ is
$O(\delta \log \delta \log^3 \pn \log
\log \pn)$. Moreover, the memory required to perform this
  algorithm, excluding the memory required to store the Fenwick tree $\sfU$ and the
  arrays $\vsfb$ and $\va$ is $O(\delta \log \pn)$.  Here, $\pn = \max\{n_l, n_r\}$.
\end{lem}

\begin{proof}
  Fix $1 \leq k \leq a_i$. The loop in line~\ref{l:bdec-i-while} of Algorithm~\ref{alg:bip-dec-i} corresponds to the binary search for
  finding $\gamma_{i,k}$, therefore it runs for  $O(\log \pn)$ many iterations.
The dominant computation inside this loop is line~\ref{l:bdec-i-ywhile}
  where we compute $y$. Similar to our analysis for the complexity of line~\ref{l:bNij=y} in
  Algorithm~\ref{alg:bip-Nij}, as was done in the proof of
  Lemma~\ref{lem:bip-encode-computeN-complexity}, the time complexity of
  computing $y$ here is $O(\delta \log \delta \log^2 \pn \log \log \pn)$, its
  memory complexity is $O(\delta \log \pn)$, and the result $y$ has $O(\delta
  \log \pn)$ bits. Hence, the overall time and memory complexity of the loop starting in
  line~\ref{l:bdec-i-while} are $O(\delta \log \delta \log^3 \pn \log \log \pn)$
  and $O(\delta \log \pn)$, respectively. Similarly, the time and memory
  complexity of computing $y$ in line~\ref{l:bdec-i-y} are $O(\delta \log \delta
  \log^2 \pn \log \log \pn)$ and $O(\delta
  \log \pn)$, respectively, and the result has $O(\delta \log \pn)$ bits. 
Also, using Lemma~\ref{lem:bip-r-l-N-bit-count} and the assumption $\tN_{i,i} \leq
  N_{i,i}(G) + l^G_i$, the variable $\tz$ initialized in
  line~\ref{l:bdec-i-tz-init} has $O(\delta \log \pn)$ bits.  Moreover, as we update this
  variable in line~\ref{l:bdec-i-tz} in a nonincreasing fashion, it remains to
  have $O(\delta \log \pn)$ bits.
On the other hand,  we have $\sfb_{\gamma_{i,k}} \leq n_l$ and hence
$\sfb_{\gamma_{i,k}}$ has $O(\log \pn)$
  bits. The above discussion also implies that $\tz - y$ has $O(\delta \log \pn)$ bits.
  Consequently, the time complexity of performing the division in line~\ref{l:bdec-i-tz} is
  $\Div(\delta \log \pn) = O(\delta \log^2 \pn \log \log \pn)$.
  Note that lines~\ref{l:bdec-i-z} through \ref{i:bdec-i-beta} perform
  similar calculations as in lines~\ref{l:bNij=z} through \ref{l:bNij=beta} of
  Algorithm~\ref{alg:bip-Nij}. As a result, following the proof of
  Lemma~\ref{lem:bip-encode-computeN-complexity}, the overall time complexity
  associated to  these
  lines is $O(\delta \log^2 \pn \log \log \pn)$. Putting the above together, we
  realize that the overall time complexity of Algorithm~\ref{alg:bip-dec-i} is
  $O(\delta \log \delta \log^3 \pn \log \log \pn)$. Furthermore,  its memory complexity, excluding the
  memory required to store $\sfU$ and the arrays  $\vsfb$ and $\va$, is $O(\delta \log \pn)$.
\end{proof}

\begin{lem}
  \label{lem:bdec-ij-complexity}
  Assume that we have $a_v \leq \delta$ for all $1 \leq v \leq n_l$. Also,
  assume that the input $\tN_{1,n_l}$ to Algorithm~\ref{alg:bip-dec-ij}
  satisfies $N_{1,n_l}(G) \leq \tN_{1,n_l} < N_{1,n_l}(G) + l^G_{1,n_l}$ for the
  graph $G \in \mGnlr_{\va,\vb}$ that was given to the encoder during the
  compression phase. Then, the time complexity of decoding for the adjacency
  lists of left vertices $1, \dots, n_l$ is $O(\pn \delta \log^4 \pn \log \log \pn)$ and the memory required to
  perform this is $O(\pn \delta \log \pn)$ bits. Here, $\pn = \max \{n_l, n_r\}$.
\end{lem}

\begin{proof}
Recall that our inductive argument implies that the
conditions~\ref{a:bdec-1-i-1} and \ref{a:bdec-tN-N} hold when 
Algorithm~\ref{alg:bip-dec-ij} is called for any interval $[i,j]$, $1 \leq i
\leq j \leq n_l$.
When Algorithm~\ref{alg:bip-dec-ij} is
called for an interval $[i,j]$ such that $i = j$, Algorithm~\ref{alg:bip-dec-i}
is called  in line~\ref{l:bdec-ij-call-ii}, whose complexity was analyzed in
Lemma~\ref{lem:bdec-i-complexity} above.

Now, assume that
Algorithm~\ref{alg:bip-dec-ij} is called for an interval $[i,j]$ such that $j >
i$. In this case, lines~\ref{l:bdec-ij-k-ij-2-def} through
\ref{l:bdec-ij-return} in Algorithm~\ref{alg:bip-dec-ij} are executed.
Since we have assumed $a_v \leq \delta$ for $1 \leq v \leq n_l$, each partial
sum in the Fenwick tree $\sfW$, which is  of the form $\sum_{k=v}^{n_l} a_k$  for some $1
\leq k \leq n_l$, has $O(\log (\pn \delta)) = O(\log \pn)$ bits.
Hence, the time
complexities of computing $S_{k+1}$ and $S_{j+1}$ in lines~\ref{l:bdec-Sk+1-Wk+1}
and \ref{l:bdec-Sj+1-Wj+1} are both $O(\log^2 \pn)$. Furthermore, both variables
$S_{k+1}$ and $S_{j+1}$ have $O(\log
\pn)$ bits. Hence, using Lemma~\ref{lem:compute-product-complexity}, the time
complexity of computing $\textproc{ComputeProduct}(S_{k+1}, S_{k+1} - S_{j+1},
1)$ in line~\ref{l:bdec-ij-r} is 
\begin{equation*}
  O((j-k) \delta \log \pn \log ((j-k) \delta) \log ((j-k)\delta \log \pn) \log \log ((j-k)\delta \log \pn)) = O((j-i+1) \delta \log^3 \pn \log \log \pn),
\end{equation*}
and  its memory complexity is  $O((j-1+1) \delta \log \pn)$.
Also, the
result has $O((j-i+1) \delta \log \pn)$ bits.
Furthermore, using Lemma~\ref{lem:prod-a-factorial-complexity-each-delta},
the time and memory complexity of computing $\textproc{ProdFactorial}(\va, k+1,
j)$ are $O((j-i+1) \delta \log \delta \log^2 \pn \log \log \pn)$ and $O((j-i+1)
\delta \log \delta)$, respectively, and the result has $O((j-i+1) \delta \log
\delta)$ bits.  This implies that the division in  line~\ref{l:bdec-ij-r} has
time complexity $\Div((j-i+1) \delta \log \pn) = O((j-i+1) \delta \log^2 \pn
\log \log \pn)$.
To sum up, the time and memory complexity of computing $r_{k+1, j}$ in
line~\ref{l:bdec-ij-r} are $O((j-i+1) \delta \log^3 \pn \log \log \pn)$ and
$O((j-i+1) \delta \log \pn)$, respectively.

In line~\ref{l:bdec-ij-tNik}, using the assumption $\tN_{i,j} < N_{i,j}(G) +
l^G_{i,j}$ and Lemma~\ref{lem:bip-N+l<r}, we have $\tN_{i,j} \leq r_{i,j}$.
Thereby, from Lemma~\ref{lem:bip-r-l-N-bit-count}, $\tN_{i,j}$ has $O((j-i+1)
\delta \log \pn)$ bits. Lemma~\ref{lem:bip-r-l-N-bit-count} also implies that
$r_{k+1, j}$ has $O((j-i+1) \delta \log \pn)$ bits. Hence, the time complexity of
the division in line~\ref{l:bdec-ij-tNik} is $\Div((j-i+1) \delta \log \pn) =
O((j-i+1) \delta \log^2 \pn \log \log \pn)$. 

In line~\ref{l:bdec-ij-tNkj}, from Lemma~\ref{lem:bip-r-l-N-bit-count},
the variables $N_{i,k}= N_{i,k}(G)$, $r_{k+1, j}$  and $l_{i,k} = l^G_{i,k}$ have $O((j-i+1) \delta \log \pn)$ bits
each. Hence, the time complexity of computing $N_{i,k} \times r_{k+1, j}$ is
$O((j-i+1) \delta \log^2 \pn \log \log \pn)$, and the result has $O((j-i+1)
\delta \log \pn)$ bits. Since $\tN_{i,j}$ has also
$O((j-i+1)\delta \log \pn)$ bits, the complexity of the subtraction is $O((j-i+1)
\delta \log \pn)$, and the result has $O((j-i+1) \delta \log \pn)$ bits.
Therefore, the time complexity of the division in this line is $O((j-i+1) \delta
\log^2 \pn \log \log \pn)$. Consequently, the overall time complexity of
line~\ref{l:bdec-ij-tNkj} is $O((j-i+1) \delta \log^2 \pn \log \log \pn)$. 

Note that the computations in lines~\ref{l:bdec-ij-Nij} and \ref{l:bdec-ij-lij}
are similar to those in lines~\ref{l:bNij-N} and \ref{l:bNij-l} of
Algorithm~\ref{alg:bip-Nij}. Thus, recalling the proof of
Lemma~\ref{lem:bip-encode-computeN-complexity}, the time complexity associated
to  these two lines is $O((j-i+1) \delta \log^2 \pn \log \log \pn)$. 

Putting the above together, if $f(m)$ denotes the time complexity of
Algorithm~\ref{alg:bip-dec-ij} when $j-i+1 = m$, for $m > 1$, we have
\begin{equation*}
  f(m) = f\left( \left\lfloor \frac{m}{2} \right \rfloor \right) + f\left( \left\lceil \frac{m}{2} \right \rceil \right) + O (m \delta \log^3 \pn \log \log \pn).
\end{equation*}
Also, from Lemma~\ref{lem:bdec-i-complexity} above, we have
\begin{equation*}
  f(1) = O(\delta \log \delta \log^3 \pn \log \log \pn).
\end{equation*}
Solving for this recursive equation, we get
\begin{equation*}
  f(n_l) = O(\pn \delta \log^4 \pn \log \log \pn),
\end{equation*}
as was claimed.

In order to study the memory complexity of the algorithm, let $g(m)$ denote the
required memory for running the algorithm when $j - i + 1 = m$. Recall from
Remark~\ref{rem:b-dec-U-b-pointer} that we store the Fenwick trees $\sfU$,
$\sfW$ and the arrays $\vsfb$ and $\va$ in one location in the memory using $O(\pn \log
\pn)$ bits. Motivated by this, let $\tilde{g}(m)$ denote the memory complexity
of the algorithm when $j-i + 1 = m$,
excluding the memory required to store $\sfU, \sfW$, $\vsfb$, and $\va$. With this, we
have $g(n_l) = \tilde{g}(n_l) + O(\pn \log \pn)$. Also, the discussion above
implies that  for $m > 1$, we have 
\begin{equation*}
  \tilde{g}(m) = \max \left\{ \tilde{g}\left( \left\lfloor  \frac{m}{2}\right \rfloor \right) , \tilde{g}\left( \left\lceil \frac{m}{2} \right \rceil \right)\right\} + O(m \delta \log \pn).
\end{equation*}
On the other hand, from Lemma~\ref{lem:bdec-i-complexity}, we have
\begin{equation*}
  \tilde{g}(1) = O(\delta \log \pn).
\end{equation*}
Solving for this recursive equation, we get $\tilde{g}(n_l) = O(\pn \delta \log
\pn)$. Consequently, we have 
$g(n_l) = O(\pn \delta \log \pn)$. This completes the proof.
\end{proof}

\begin{lem}
  \label{lem:bdec-overall-complexity}
  Assume that $a_v \leq \delta$ for $1 \leq v \leq n_l$. Then, the time and
  memory complexities associated to  decoding for the adjacency list of $G \in \mGnlr_{\va,
  \vb}$ given $\fnlr_{\va,\vb}(G)$ in Algorithm~\ref{alg:b-dec} are $O(\pn \delta
\log^4 \pn \log \log \pn)$ and
$O(\pn \delta \log^2 \pn)$, respectively. Here, $\pn = \max\{n_l, n_r\}$.
\end{lem}

\begin{proof}
Note that we have $\sum_{i=1}^{n_r} b_i = \sum_{j=1}^{n_l} a_j \leq \pn \delta$.
Therefore, using Lemma~\ref{lem:prod-factorial-complexity-sum-bound}, the time
and memory complexity of computing $c = \prod_{i=1}^{n_r} b_i!$ in
line~\ref{l:bdec-prod-b-factorial} are $O(\pn \delta \log^3 \pn \log \log \pn)$
and $O(\pn \delta \log^2 \pn)$, respectively. Recall from~\eqref{eq:prod-b!-l1n}
that $c = l_{1,n_l}^G$. Thereby, from Lemma~\ref{lem:bip-r-l-N-bit-count}, $c$
has $O(\delta \pn \log \pn)$ bits. On the other hand, since $f = \fnlr_{\va,
  \vb}(G) < N_{1,n_l}(G) +l^G_{1, n_l}$ for some $G \in \mGnlr_{\va, \vb}$, another usage
of Lemma~\ref{lem:bip-r-l-N-bit-count} implies that $f$ also has $O(\delta \pn
\log \pn)$ bits. Hence, the time complexity of performing the multiplication in
line~\ref{l:bdec-tN-1-nl-y-c} is $\Mul(\delta \pn \log \pn) = O(\delta \pn
\log^2 \pn \log \log \pn)$. Note that partial sums in the array $\vb$ are
bounded by $\sum_{k=1}^{n_r} b_k = \sum_{k=1}^{n_l} a_k \leq \pn \delta$, which
has $O(\log \pn)$ bits. Hence, the time complexity of initializing $\sfU$ is $O(\pn
\log \pn)$. Similarly, the time complexity of initializing $\sfW$ in
line~\ref{l:bdec-W-init} is $O(\pn \log \pn)$. On the other hand, since $b_k
\leq \pn$ for $1 \leq k \leq n_r$, initializing $\vsfb$ with $\vb$
takes $O(\pn \log \pn)$ time. 
Also, from Remark~\ref{rem:b-dec-U-b-pointer}, the memory required to store
$\sfU$, $\sfW$, $\vsfb$, and $\va$ is $O(\pn \log \pn)$.
The proof is complete by combining the
above with Lemma~\ref{lem:bdec-ij-complexity}.  
\end{proof}

\editfinish


\section{A Compression Algorithm for Simple Unmarked Graphs with Given Degree
  Sequence}
\label{sec:simple-graph-compression}

\editstart

\ifreplace
In this section, we prove Proposition~\ref{prop:equal-color-graph-compress} in
Section~\ref{sec:main-self-comp}.
\color{black}
\fi
Recall that $\mGnt_{\va}$ denotes the set of simple
unmarked graphs on the vertex set $\{1, \dots, \pn\}$ such that the degree of the vertex $1
\leq i \leq \pn$ is $a_i$. 
 We assume that $\pn \geq 2$ and that the degree sequence $\va$ is graphical, 
 i.e.
 $\mGnt_{\va}
 \neq \emptyset$.
\ifreplace
Here, to avoid confusion,  we have used the symbol $\pn$ rather than $n$ to
denote the number of vertices.
\color{black}
\fi

Given $\pn$ and $\va$ as above, we introduce an encoding
mechanism which, given a graph $G \in \mGnt_{\va}$, generates a
nonnegative integer
$\fnt_{\va}(G) \in \mathbb{Z}_+$ 
\color{black}
together with an array
$\vec{\tilde{f}}^{(\pn)}_{\va}(G) = (\tilde{f}^{(\pn)}_{\va, i}(G): 1 \leq i \leq \lfloor 16 \pn / \log^2 \pn
    \rfloor) $
of length $\lfloor  16 \pn / \log^2 \pn \rfloor$ consisting of nonnegative integers.
In fact, $\fnt_{\va}(G)$ is sufficient
to unambiguously decode for $G$ (see
Lemma~\ref{lem:self-NG-uniquely-determines-G} below). However, as we will see
later in Section~\ref{sec:self-decompression}, the role of
$\vec{\tilde{f}}^{\pn}_{\va}(G)$ is to help the decoder
find $G$ more efficiently. 

Given $G \in \mGnt_{\va}$, let $\vA^G = (A^G_{i,j}: 1 \leq i \leq j \leq \pn)$ be the upper triangular
  part of the adjacency matrix of $G$, i.e.\ for $1 \leq i \leq j \leq \pn$,
  $A^G_{i,j} = 1$ if $i$ and $j$ are adjacent in $G$ and zero otherwise.
We represent $\vA^G$ as a row vector of the form 
\begin{equation}
  \label{eq:self-vAg}
    \vA^G = (A^G_{i,j}: 1 \leq i \leq j \leq \pn) = (A^G_{1,1}, \dots, A^G_{1,\pn}, A^G_{2,2}, \dots, A^G_{\pn-1,\pn-1}, A^G_{\pn-1,\pn},
  A^G_{\pn,\pn}).
  \end{equation}
Since $G$ is simple, we have $A^G_{i,i} = 0$ for $1 \leq i \leq \pn$.
For $1 \leq i \leq \pn$, let $\vA^G_i$ denote  $(A^G_{i,i}, \dots,
  A^G_{i,\pn})$.
Also, for $1 \leq i \leq j \leq \pn$, define  
  \begin{equation*}
    \vA^G_{\leq (i,j)} := (A^G_{1,1}, \dots, A^G_{1,\pn}, \dots, A^G_{i,i}, \dots, A^G_{i,j}).
  \end{equation*}
Let $\mCnt_{\va}$ be the set of configurations defined as follows. For
each node $1 \leq i \leq \pn$, we attach a set of distinct half--edges $R_i =
\{r_{i, 1}, \dots, r_{i, a_i}\}$ to $i$. Then, a configuration $C \in
\mCnt_{\va}$ is a matching in the set of all half--edges $R :=
\bigcup_{i=1}^{\pn} R_i$. 
Note that we have $  |\mCnt_{\va}| = (S-1)!!$, where 
$  S := \sum_{i=1}^{\pn} a_i$
\color{black}
is twice the number of edges in the graph. 
Given a configuration $C \in \mCnt_{\va}$, we construct a multigraph with $\pn$
vertices, such that the number of edges between nodes $1 \leq i , j \leq \pn$
is the number of half--edges at node $i$ that are matched with  half--edges at
node $j$. With this, define $\vA^C = (A^C_{i,j}: 1 \leq i \leq j \leq \pn)$ such
that $\vA^C_{i,j}$ for $1 \leq i \leq j \leq \pn$ is the number of edges between
$i$ and $j$ in the multigraph associated to $C$. Similar to \eqref{eq:self-vAg},
we represent $\vA^C$ as a row vector. We also define $\vA^C_i$ and $\vA^C_{\leq
  (i,j)}$ for $1 \leq i \leq j \leq \pn$ in a way that is similar to what was done in defining $\vA^G_i$ and $\vA^G_{\leq(i,j)}$. 


With this, for $G \in \mGnt_{\va}$, we define
\begin{equation*}
    N^{\pn}_{\va}(G) := |\{C \in \mCnt_{\va}: \vA^C \prec \vA^G\}|,
\end{equation*}
where $\prec$ denotes the lexicographic order when $\vA^C$ and $\vA^G$ are
viewed as row vectors as in~\eqref{eq:self-vAg}. To simplify the notation, we
may remove $\pn$ and $\va$ from the notation whenever
they are clear from the context and simply write $N(G)$.
Finally, for $G \in \mGnt_{\va}$, we define
  \begin{equation}
    \label{eq:self-fna-def}
    \fnt_{\va}(G) := \left\lceil \frac{N(G)}{\prod_{i=1}^{\pn} a_i!} \right\rceil.
  \end{equation}
\ifreplace
We now argue why $\fnt_{\va}(G)$ uniquely determines the graph $G \in \mGnt_{\va}$.
\color{black}
\fi
\begin{lem}
  \label{lem:self-NG-uniquely-determines-G}
  Assume $G, G' \in \mGnt_{\va}$ are distinct. Then, we have $\fnt_{\va}(G) \neq
  \fnt_{\va}(G')$. 
\end{lem}

\begin{proof}
  Without loss of generality, we assume $\vA^G \prec \vA^{G'}$. Then,
  we have
  \begin{equation}
    \label{eq:self-N-Gp-N-G-inequality}
    N(G') = |\{C \in \mCnt_{\va}: \vA^C \prec \vA^{G'}\}| \geq \left| \left\{ C \in \mCnt_{\va}: \vA^C \prec \vA^G \right\} \right| + \left| \left\{ C \in \mCnt_{\va}: \vA^C = \vA^G \right\} \right|.
    \color{black}
  \end{equation}
  Note that $\vA^C = \vA^G$ implies that for all the vertices $1 \leq v \leq \pn$,
  $C$ connects all of the half--edges of $v$ precisely to the neighbors of $v$ in $G$. As
  $G$ is simple, all of these neighbors are distinct, and  there are $a_v!$
  many ways to assign half--edges of $v$ to its neighbors for all $1 \leq v \leq \pn$. Consequently, we have
  \begin{equation*}
    \left| \left\{ C \in \mCnt_{\va}: \vA^C = \vA^G \right\} \right| = \prod_{i=1}^{\pn} a_i!.
  \end{equation*}
  Using this in~\eqref{eq:self-N-Gp-N-G-inequality}, we get 
  $N(G') \geq N(G) + \prod_{i=1}^{\pn} a_i!$. Dividing both sides by
  $\prod_{i=1}^{\pn} a_i!$ and taking the ceiling completes the proof. 
\end{proof}




\subsection{Compression Algorithm}
\label{sec:self-encode}

In this section, we discuss an efficient way for computing $\fnt_{\va}(G)$ given
$G \in \mGnt_{\va}$. 
\ifreplace
We also define $\vec{\tilde{f}}^{(n)}_{\va}(G)$.
\color{black}
\fi
Recall from
Section~\ref{sec:main-self-comp} that for 
a vertex $1 \leq v \leq \pn$, we denote by $\hat{a}^G_v$ the
number of vertices $w$ adjacent to $v$ in $G$ such that $w > v$. Also, recall
that  
$\gamma^G_{v,1}< \dots< \gamma^G_{v,\hat{a}^G_v}$
\color{black}
are
  these neighbors in an increasing order
  (here $v < \gamma^G_{v,1}$), 
  \color{black}
  and  $\vgamma^G_v = (\vgamma^G_{v,
    1}, \dots, \vgamma^G_{v,\hat{a}^G_v})$ is called the  ``forward adjacency list'' of
  vertex $v$. Note that $\hat{a}^G_v \leq a_v$ for
  all $1 \leq v \leq \pn$. Furthermore, for $1 \leq i \leq j \leq \pn$, we define
  \begin{equation}
    \label{eq:self-di(v)-def}
    a^G_j(i) := a_j - \sum_{k=1}^{i-1} \one{k \sim_G j}.
  \end{equation}
Effectively, $a^G_j(i)$ counts the number of free half--edges at node $j$ in a
configuration $C$ assuming that the connections for the nodes $1, \dots, i-1$
are made according to $G$, i.e.\ $\vA^C_k = \vA^G_k$ for $1 \leq k \leq i-1$.
In particular, observe that $a^G_i(i) = \hat{a}^G_i$. 
Moreover, for $1 \leq i \leq v \leq \pn$, we define
  \begin{equation}
    \label{eq:self-U-def}
    U^G_v(i) := \sum_{w=v}^{\pn} a^G_w(i).
  \end{equation}
In fact, $U^G_v(i)$ denotes the total number of free half--edges connected to the
  vertices in the range $[v,\pn]$ in a configuration $C$ assuming that the
  connections for the nodes $1, \dots, i-1$ are made according to $G$.
\ifreplace
Furthermore, for $1 \leq i \leq \pn$, define
\color{black}
\fi
  \begin{equation}
    \label{eq:self-Si-def}
    S^G_i := \sum_{v=i}^{\pn} a^G_v(i).
  \end{equation}
  In fact, $S^G_i$ is the total number of free half--edges in a configuration assuming that the
  connections for vertices $1, \dots, i-1$ are made according to $G$.  
  We let $S^G_{\pn+1}:= 0$. Notice that for $1 \leq i < \pn$, we have 
  \begin{equation}
    \label{eq:self-Si+1-Si}
    \begin{aligned}
      S^G_{i+1} &= \sum_{v=i+1}^{\pn} a^G_v(i+1) = \sum_{v=i+1}^{\pn} \left( a^G_v(i) - \one{i \sim_G v} \right) \\
      &= \left( \sum_{v=i+1}^{\pn} a^G_v(i)  \right) - \left( \sum_{v=1}^{\pn} \one{i \sim_G v} - \sum_{v=1}^{i} \one{i \sim_G v} \right) \\ 
      &= S^G_i - a^G_i(i) - \left( a_i - \sum_{v=1}^{i} \one{i \sim_G v} \right) \\
      &\stackrel{(*)}{=} S^G_i - a^G_i(i) - \left( a_i - \sum_{v=1}^{i-1} \one{i \sim_G v} \right) \\
      &= S^G_i - 2a^G_i(i) = S^G_i - 2 \hat{a}^G_i,
    \end{aligned}
  \end{equation}
  where in $(*)$, we have used the fact that since $G$ is simple, $\one{i \sim_G
    i} = 0$.
  Note that for $1 \leq v \leq \pn$, we have $a^G_v(1) = a_v$. Therefore, 
  $S^G_1 =S$,
  \color{black}
  where we recall that $S:= \sum_{i=1}^{\pn} a_i$,
  \color{black}
  and \eqref{eq:self-Si+1-Si}
  in particular implies that for $1 \leq i \leq \pn$, we have 
  \begin{equation}
    \label{eq:self-SGi-S-ahat}
    S^G_i = S - \sum_{j=1}^{i-1} 2\hat{a}^G_i.
  \end{equation}
  
With this, we are ready to give a closed form expression for $N(G)$ for $G
\in \mGnt_{\va}$. 

  \begin{lem}
    \label{lem:self-N-count}
    For a simple unmarked graph $G \in \mGnt_{\va}$, we have 
    \begin{equation*}
      N(G) = \sum_{i=1}^{\pn} \sum_{j=1}^{\hat{a}^G_i} \left( \prod_{k=1}^{i-1} l^G_k \right) \left( \prod_{\pp{j}=1}^{j-1} (\hat{a}^G_i - \pp{j} + 1)a^G_{\gamma^G_{i,\pp{j}}}(i) \right) (U^G_{1+\gamma^G_{i,j}}(i))_{\hat{a}^G_i - j + 1}   (S^G_{i+1}-1)!!,
    \end{equation*}
    where for nonnegative integers $r$ and $s$,  $(r)_s := r(r-1)(r-2) \dots (r-(s-1))$
    denotes the falling factorial,
    with $(r)_s := 0$ if $r < s$.
    \color{black}
    Here, for $1 \leq i \leq \pn$, we define
    \begin{equation*}
      l^G_i := \prod_{j=1}^{\hat{a}^G_i} (\hat{a}^G_i - j + 1) a^G_{\gamma^G_{i,j}}(i).
    \end{equation*}
  \end{lem}

    \begin{proof}
Note that for $C \in \mCnt_{\va}$, we have $\vA^C \prec \vA^G$ iff for some
  $1 \leq i \leq \pn$ and $1 \leq j \leq \hat{a}^G_i$, we have
  \begin{subequations}
    \begin{gather}
      \vA^C_{\leq (i,\gamma^G_{i,j} - 1)} = \vA^G_{\leq (i,\gamma^G_{i,j} - 1)}, \label{eq:Bij-1}\\
      A^C_{i,\gamma^G_{i,j}} = 0. \label{eq:Bij-2}
    \end{gather}
  \end{subequations}
Therefore, if for  $1 \leq i \leq \pn$ and $1 \leq j \leq \hat{a}^G_i$, $B_{i,j}$ denotes the
  set of configurations $C \in \mCnt_{\va}$ satisfying \eqref{eq:Bij-1} and
  \eqref{eq:Bij-2}, we have 
  \begin{equation}
    \label{eq:self-N-count-sum-Bij}
    N(G) = \sum_{i=1}^{\pn} \sum_{j=1}^{\hat{a}^G_i} |B_{i,j}|.
  \end{equation}
Hence, it suffices to count $|B_{i,j}|$ for $1 \leq i \leq \pn$ and $1 \leq j
\leq \hat{a}^G_i$. We start by counting $|B_{1,1}|$ when $\hat{a}^G_1 > 0$.
Note that if  $C \in B_{1,1}$, all of the $\hat{a}^G_1 = a_1$ half--edges at vertex $1$ must
connect to half--edges that belong to vertices in the range $\gamma^G_{1,1} + 1, \dots, \pn$.
  Notice that since $\gamma^G_{1,1} > 1$, a configuration $C \in B_{1,1}$ can not construct a
  self loop at vertex $1$. Thereby, all the half--edges connected to vertices in
  the range $\gamma^G_{1,1} + 1, \dots, \pn$ are distinct from those that are 
  connected to vertex $1$. Motivated by this, there are precisely $(U^G_{1+\gamma^G_{1,1}}(1))_{\hat{a}^G_1}$
  many ways to connect the half--edges at vertex $1$ so that all of
such half--edges 
  are connected to half--edges connected to vertices in
  the range $\gamma^G_{1,1} + 1, \dots, \pn$. 
Note that if $U^G_{1+\gamma^G_{1,1}}(1) < \hat{a}^G_1$, there is no way to make
the connections of the half--edges at vertex 1 to satisfy the above condition.
However, if this is the case,  then by definition
$(U^G_{1+\gamma^G_{1,1}}(1))_{\hat{a}^G_1} $ is automatically zero.
Once we make connections for these $\hat{a}^G_1$ many half--edges at vertex $1$ in the above fashion, there are
    precisely $(S^G_2 - 1)!!$ many ways to match the  remaining $S^G_2 = S -
    2a_1$ half--edges, all of them resulting in configurations in $B_{1,1}$.
To sum up, if $\hat{a}^G_1 > 0$, we have 
    \begin{equation*}
      |B_{1,1}| = (U^G_{1+\gamma^G_{1,1}}(1))_{\hat{a}^G_1} (S^G_2 - 1)!!.
    \end{equation*}

Now, we count $|B_{1,j}|$ for $1 < j \leq \hat{a}^G_1$ when $\hat{a}^G_1 > 1$.
Note that in this case,  a
configuration $C \in B_{1,j}$ must
    connect $j-1$ half--edges at vertex $1$ to half--edges associated to vertices $\gamma^G_{1,1}, \dots,
    \gamma^G_{1,j-1}$. Since at this stage there are $ a_{\gamma^G_{1,\pp{j}}} =
    a^G_{\gamma^G_{1,\pp{j}}}(1)$ many free half--edges connected to  vertex
    $\gamma^G_{1,\pp{j}}$ for $1 \leq \pp{j} \leq j-1$, connecting $j-1$ half--edges at vertex $1$ to half--edges associated to vertices $\gamma^G_{1,1}, \dots, \gamma^G_{1,j-1}$ can be done in $\prod_{\pp{j}=1}^{j-1} (\hat{a}^G_1 - \pp{j} + 1) a^G_{\gamma^G_{1,\pp{j}}}(1)$ many ways.
On the other hand, since our configuration is in $B_{1,j}$, the remaining $\hat{a}^G_1 - j +1$ 
half--edges at vertex 1 must
    connect to vertices in the range $1+\gamma^G_{1,j}, \dots, \pn$. But this can be done in $(U^G_{1+\gamma^G_{1,j}}(1))_{\hat{a}^G_1 - j +1}$ many ways.
Note that if $U^G_{1+\gamma^G_{1,j}}(1) < \hat{a}^G_1 - j +1$, the number
of ways we can do this is zero, and by definition, we
have $(U^G_{1+\gamma^G_{1,j}}(1))_{\hat{a}^G_1 - j +1} = 0$.
    Similar to the above, there are $(S^G_2 - 1)!!$ many ways to connect
    half--edges at vertices $2, \dots, \pn$.
    Thus far, we have shown that for $1 \leq j \leq \hat{a}^G_1$, we have 
    \begin{equation*}
        |B_{1,j}| = \left( \prod_{\pp{j}=1}^{j-1} (\hat{a}^G_1 - \pp{j} + 1) a^G_{\gamma^G_{1,\pp{j}}}(1) \right) (U^G_{1+\gamma^G_{1,j}}(1))_{\hat{a}^G_1 - j +1} (S^G_2 - 1) !!
      \end{equation*}

      Now, we count $|B_{i,j}|$ for  $i > 1$ and $1 \leq j \leq \hat{a}^G_i$.
Note that if $C \in B_{i,j}$,
by definition we have  $\vA^C_k = \vA^G_k$ for $1 \leq k < i$.
This means that $C$ connects half--edges at vertices $1 \leq k < i$ in
      a way consistent with $G$.
Motivated by our earlier discussion, there are precisely $
\prod_{j=1}^{\hat{a}^G_1} (\hat{a}^G_1 - j + 1) a^G_{\gamma^G_{i,j}}(i) = l^G_1$
      many possibilities to connect half--edges at vertex $1$ so that $\vA^C_1 =
      \vA^G_1$. 
Furthermore, for  $1 < k \leq i$, if we maintain connections of vertices  $1,
      \dots, k-1$ so that $A^C_{\pp{k}} = A^G_{\pp{k}}$ for $1 \leq \pp{k} \leq
      k-1$, then connected to a vertex
      $v \geq k$, there are precisely $ a^G_v(k)$
      many free half--edges. Therefore, 
in order to satisfy $\vA^C_k = \vA^G_k$, there are $ \prod_{p=1}^{\hat{a}^G_k}
(\hat{a}^G_k - p + 1) a^G_{\gamma^G_{k,p}}(k) = l^G_k$
      many choices to connect the $\hat{a}^G_k$ half--edges at vertex $k$.
Similarly, given $\vA^C_k = \vA^G_k$ for $1 \leq k \leq i-1$, to maintain
      $(A^C_{i,i}, \dots, A^C_{i,\gamma^G_{i,j}-1}) = (A^G_{i,i}, \dots,
      A^G_{i,\gamma^G_{i,j}-1})$, $C$ must connect $j-1$ half--edges at vertex
      $i$ to half--edges at vertices $\gamma^G_{i,1}, \dots, \gamma^G_{i,j-1}$.
      To achieve this, there are precisely 
      \begin{equation*}
        \prod_{\pp{j}=1}^{j-1} (\hat{a}^G_i - \pp{j} + 1) a^G_{\gamma^G_{i,\pp{j}}}(i),
      \end{equation*}
      many possibilities.
 Furthermore, to satisfy $C \in B_{i,j}$, the remaining $\hat{a}^G_i - j +
        1$ half--edges at vertex $i$ must connect to half--edges associated to vertices in the range
        $1+\gamma^G_{i,j}, \dots, \pn$.
Observe that since $\gamma^G_{i,j} > i$, $C \in B_{i,j}$ can not make any self loops at
        vertex $i$ and the half--edges associated to vertices in the range
        $1+\gamma^G_{i,j}, \dots, \pn$ are distinct from those connected to
        vertex $i$. Thus, there are precisely $      (U^G_{1+\gamma^G_{i,j}}(i))_{\hat{a}^G_i - j + 1}$
        many ways to connect the remaining $\hat{a}^G_i - j + 1$ half--edges at
        vertex $i$ in the way described above. Finally, to connect the remaining
        $S^G_{i+1}$ half--edges at vertices $i+1, \dots, \pn$, there are $(S^G_{i+1}-1)!!$ many
        choices.
Putting the above together, we conclude that
        \begin{equation*}
          |B_{i,j}| = \left(\prod_{k=1}^{i-1} l^G_k \right)  \left(  \prod_{\pp{j}=1}^{j-1} (\hat{a}^G_i - \pp{j} + 1) a^G_{\gamma^G_{i,\pp{j}}}(i) \right) (U^G_{1+\gamma^G_{i,j}}(i))_{\hat{a}^G_i - j + 1} (S^G_{i+1}-1)!!.
        \end{equation*}
        The proof is complete by putting the above together and comparing with \eqref{eq:self-N-count-sum-Bij}.
    \end{proof}

\subsubsection{A Recursive Form for $N(G)$}
\label{sec:self-NG-recursive}

In order to efficiently compute $N(G)$, we introduce a recursive formula for it. First, we define some notation. 
Assume that $G \in \mGnt_{\va}$ is given. For $1 \leq i \leq \pn$, let
    \begin{subequations}
    \begin{align}
      z^G_i &:= \sum_{j=1}^{\hat{a}^G_i} \left( \prod_{\pp{j} = 1}^{j-1} (\hat{a}^G_i - \pp{j} + 1) a^G_{\gamma^G_{i,\pp{j}}}(i) \right) (U^G_{1+\gamma^G_{i,j}}(i))_{\hat{a}^G_i - j + 1} \label{eq:senc-zi}\\
      r^G_i &:= \frac{(S^G_i-1)!!}{(S^G_{i+1}-1)!!}. \label{eq:senc-ri}
    \end{align}
    \end{subequations}
    Observe that $S^G_{\pn} = a^G_{\pn}(\pn) \hat{a}^G_{\pn} = 0$. Also, recall
    that by definition $
    S^G_{\pn+1} = 0$ and $(-1)!! := 1$. Therefore,
    $r^G_{\pn} = 1$.
With these, for $1 \leq i \leq j \leq \pn$, we define
    \begin{align*}
      l^G_{i,j} &:= \prod_{k=i}^j l^G_k \\
      r^G_{i,j} &:= \prod_{k=i}^j r^G_k.
    \end{align*}
    Moreover, for $i > j$, we define $l^G_{i,j} = r^G_{i,j} = 1$.
Employing the above notation, from Lemma~\ref{lem:self-N-count}, we have
    \begin{equation}
      \label{eq:self-NG-lzr}
      N(G) = \sum_{i=1}^{\pn} l^G_{1,i-1} z^G_i r^G_{i+1, \pn}.
    \end{equation}    
Moreover, for $1 \leq i \leq j \leq \pn$, we define
    \begin{equation}
      \label{eq:self-Nij-def}
      N_{i,j}(G) = \sum_{k=i}^j l^G_{i,k-1} z^G_k r^G_{k+1, j}.
    \end{equation}
Also, following the convention that a summation over an empty set evaluates to
zero, we define $N_{i,j}(G) = 0$ for $i > j$. 
Notice that in particular, $N(G) = N_{1,\pn}(G)$ and for $1
\leq i \leq \pn$, $z_i^G = N_{i,i}(G)$. Furthermore, it is easy to see that 
for $1 \leq i \leq k \leq j \leq \pn$, we have 
    \begin{equation}
      \label{eq:self-Nij-recursive}
      N_{i,j}(G) = N_{i,k}(G) r^G_{k+1, j} + l^G_{i,k} N_{k+1, j}(G).
    \end{equation}

\subsubsection{The Encoding Algorithm}
\label{sec:self-encode-alg}

In this section, motivated by~\eqref{eq:self-Nij-recursive},  we introduce a
recursive  algorithm that, given  $G \in \mGnt_{\va}$ and an interval $[i,j]$ for $1 \leq i \leq j \leq
\pn$, computes $N_{i,j}(G)$. Our procedure
is similar in nature to that in Section~\ref{sec:bip-enc-pseudocode} for
bipartite unmarked graphs. 
More precisely, when $i = j$, we use~\eqref{eq:senc-zi} to compute $N_{i,i}(G) =
z^G_i$, and when $j > i$, we recursively compute $N_{i,k}(G)$ and $N_{k+1,
  j}(G)$ with $k = \lfloor   (i+j) / 2 \rfloor$ and
use~\eqref{eq:self-Nij-recursive} to compute $N_{i,j}(G)$.
When the algorithm is called for the interval $[i,j]$, we assume that an array
$\vsfa = (\sfa_k: 1 \leq k \leq \pn)$ is given such that $\sfa_k = a^G_k(i)$ for
$i \leq k \leq \pn$. Furthermore, we assume that a Fenwick tree $\sfU$ is given
so that for $1 \leq k \leq \pn$, we have $\sfU.\textproc{Sum}(k) = U^G_k(i)$.
After the algorithm concludes its calculations for the interval $[i,j]$, it is
guaranteed to return the following: $(1)$ $N_{i,j}(G)$, $(2)$
$l^G_{i,j}$, $(3)$ the array $\vsfa$ updated so that $\sfa_k = a^G_k(j+1)$ for
$j+1 \leq k \leq \pn$, and $(4)$ the Fenwick tree $\sfU$ updated so that
$\sfU.\textproc{Sum}(k) = U^G_k(j+1)$ for $j+1 \leq k \leq \pn$.

In addition to the above, in order to achieve an efficient decompression, as we
will see later during our discussion of the 
  decompression phase in Section~\ref{sec:self-decompression}, we need to store the values of $S^G_k$ for certain
  values of $k$ as discussed below. When the algorithm is called for the
  interval $[i,j]$, we check whether $j - i + 1 > \lfloor  \log_2 \pn  \rfloor^2$.
  If this is the case, we store  $S^G_{k+1}$ for $k = \lfloor  (i+j)/2 \rfloor$.
In order to do this, we assign a unique integer to each of the intervals
  called during the encoding procedure. 
To start with, we  assign index $1$ to the initial interval $[1,\pn]$. 
Furthermore, if index $I$ is assigned to an interval $[i,j]$, and if $j > i$,
we assign indices $2I$
  and $2I+1$ to the subintervals $[i,k]$ and $[k+1, j]$, respectively, where $k = \lfloor  (i+j)
 / 2\rfloor$. 
It is easy to see that the index assigned to each interval called during the
encoding procedure is unique. 
If an interval $[i,j]$ is called with index $I$ and  $j - i+1 > \lfloor  \log_2 \pn \rfloor^2$, we store
  $S^G_{k+1}$ in index $I$ of the array $\vec{\tilde{f}}^{\pn}_{\va}(G)$. In
  order to simplify the notation, we may use $\vec{\tilde{f}}$ to denote
  $\vec{\tilde{f}}^{\pn}_{\va}(G)$. 

  We claim that the
  index corresponding to an interval with length more than 
$\lfloor  \log_2 \pn \rfloor^2$ 
is bounded by $\lfloor 16 \pn / \log^2 \pn\rfloor$. 
 This means that it is sufficient to
 set the size of the array $\vec{\tilde{f}}$ to  $ \lfloor 16 \pn / \log^2
 \pn \rfloor$. 
In order to prove the above claim, consider a binary tree where
the root corresponds to the interval $[1,\pn]$. Furthermore, each node corresponds to an
interval $[i,j]$ for some $1 \leq i \leq j \leq \pn$. If $i \neq j$,  such a node has two
children corresponding to intervals $[i,k]$ and $[k+1, j]$ with $k = \lfloor
(i+j)/2 \rfloor$. It
    is easy to see that the length of an interval in depth $k$ of this tree is no
    more than $1+\frac{\pn}{2^k}$.
    Therefore, if $k_1$ is the minimum among  integers $k$ such that
    \begin{equation}
      \label{eq:n-2k+1-log2n}
      \frac{\pn}{2^{k+1}} + 1 \leq \lfloor  \log_2 \pn \rfloor^2,
    \end{equation}
    then the intervals with length more than $ \lfloor  \log_2 \pn \rfloor^2$ are completely contained
    in depths $0$ through $k_1$ of the above binary tree. 
    First, assume that  $\pn \geq 4$. 
In this case, $ \lfloor  \log_2 \pn \rfloor^2 \geq 2$ and for
    \eqref{eq:n-2k+1-log2n} to hold, it suffices that 
    \begin{equation}
      \label{eq:tS-size-k2-sufficient}
      2^{k+1} \geq 2\pn / \lfloor  \log_2 \pn \rfloor^2.
    \end{equation}
Let  $k_2$ be the smallest integer satisfying
~\eqref{eq:tS-size-k2-sufficient}. 
 The above discussion implies that $k_2 \geq
     k_1$. Also, since $k_2$ is the smallest integer satisfying
     \eqref{eq:tS-size-k2-sufficient}, we have $2^{k_2} < 2 \pn / \lfloor \log_2
     \pn\rfloor^2$.
     Observe that the index corresponding to each interval in depth $k$ of the above
     binary tree is bounded by $2^{k+1}-1$.
Therefore, since each interval with length more than $\lfloor \log_2 \pn
\rfloor^2$ has depth at most $k_1$ in the binary tree, its corresponding index
is bounded by $2^{k_1+1} - 1 \leq 2^{k_2+1} \leq 4 \pn / \lfloor \log_2 \pn \rfloor^2$.
    It is easy to directly check that 
    when $\pn < 4$, the index corresponding to each interval  with length more than $\lfloor
    \log_2 \pn \rfloor^2$ is no more than $2 \pn / \lfloor  \log_2 \pn \rfloor^2$.
    Also, it is easy to show that for $\pn \geq 2$, we have $\lfloor  \log_2 \pn
    \rfloor \geq \frac{1}{2} \log \pn$.
    Putting the above together, we realize
    that it suffices to set the length of the array $\vec{\tilde{f}}$ to be
    $ \lfloor 16 \pn / \log^2 \pn \rfloor$.


In the following, we discuss our procedure for computing $N_{i,j}(G)$ in more
detail.

\underline{Case I: ($i = j$)} In this case, we need to compute $z^G_i = N_{i,i}(G)$ defined
in~\eqref{eq:senc-zi}. In order to compute
$(U^G_{1+\gamma^G_{i,k}}(i))_{\hat{a}^G_i - k + 1}$ for $1 \leq k \leq \hat{a}^G_i$, we use
Algorithm~\ref{alg:compute-product} from Section~\ref{sec:aux-algs-bip}. More specifically, motivated by our earlier
discussion, we have
$U^G_{1+\gamma^G_{i,k}}(i) = \sfU.\textproc{Sum}(1+\gamma^G_{i,k})$ and
$\hat{a}^G_i = \sfa_{i}$. Therefore,
using~\eqref{eq:compute-product-prod-plus-part}, we have 
\begin{equation*}
(U^G_{1+\gamma^G_{i,k}}(i))_{\hat{a}^G_i - k + 1} = \prod_{\pp{k}=0}^{\hat{a}^G_i - k} (U^G_{1+\gamma^G_{i,k}}(i) - \pp{k})^+ = \textproc{ComputeProduct}(\sfU.\textproc{Sum}(1+\gamma^G_{i,k}), \sfa_i - k+1, 1).
\end{equation*}
This is computed in line~\ref{line:sNij-i-y} of
Algorithm~\ref{alg:ComputeNij_graph} below. The variable $l_i$ is initialized
with 1 and is updated in the loop of line~\ref{l:sNij-i-for} so that when the
algorithm arrives at line~\ref{l:sNij-i-z}, the value of $l_i$ is $\prod_{\pp{k}= 1}^{k-1} (\hat{a}^G_i - \pp{k} + 1)
a^G_{\gamma^G_{i,\pp{k}}}(i)$. We use this in line~\ref{l:sNij-i-z} to update
 $z_i$. 
The variable $z_i$ is initialized with
$0$ and is updated in line~\ref{l:sNij-i-z} so that it becomes
\begin{equation*}
  \sum_{\pp{k}= 1}^k \left( \prod_{k'' = 1}^{\pp{k}-1} (\hat{a}^G_i - k'' + 1) a^G_{\gamma^G_{i,k''}}(i) \right) (U^G_{1+\gamma^G_{i,\pp{k}}}(i))_{\hat{a}^G_i - \pp{k} + 1}.
\end{equation*}
Furthermore, we update $l_i$
in line~\ref{l:sNij-i-l} inside the loop of
line~\ref{l:sNij-i-for} so that it becomes 
$\prod_{\pp{k}= 1}^{k} (\hat{a}^G_i - \pp{k} + 1)
a^G_{\gamma^G_{i,\pp{k}}}(i)$.
Therefore, when the loop is over, we have  $l_i =
l^G_i$ and $z_i = z^G_i$.

\editfinish

\underline{Case II: ($i < j$)} In this case, with $k = \lfloor (i+j)/2 \rfloor$, we recursively compute $N_{i,k}(G)$
and $N_{k+1, j}(G)$  in
lines~\ref{l:sNij-call-ik} and \ref{l:sNij-call-kj}, respectively. Then, we have
access to the returned variables $N_{i,k} = N_{i,k}(G)$, $l_{i,k} = l^G_{i,k}$,
$N_{k+1, j} = N_{k+1, j}(G)$ and $l_{k+1, j} = l^G_{k+1, j}$. In order to
use the recursive form~\eqref{eq:self-Nij-recursive} to compute $N_{i,j}(G)$, we need to first compute $r^G_{k+1, j}$. 
For this, we need $S^G_{k+1}$ and $S^G_{j+1}$, which are computed in
lines~\ref{l:sNij-Sk} and \ref{l:sNij-Sj}, respectively. Note that in
line~\ref{l:sNij-Sk}, after calling the procedure for the interval $[i,k]$,
$\sfU$ is updated so that $\sfU.\textproc{Sum}(v) = U^G_v(k+1)$ for $1 \leq v
\leq \pn$. Therefore, $\sfU.\textproc{Sum}(k+1) = U^G_{k+1}(k+1)  =
\sum_{v=k+1}^{\pn} a^G_v(k+1) = S^G_{k+1}$. Therefore, we have $S_{k+1} =
S^G_{k+1}$. Similarly, $S_{j+1} = S^G_{j+1}$. 
Moreover, recall that
\begin{equation*}
  r^G_{k+1, j} = \prod_{p=k+1}^j r^G_p = \prod_{p=k+1}^j \frac{(S^G_p - 1)!!}{(S^G_{p+1}-1)!!} = \frac{(S^G_{k+1}-1)!!}{(S^G_{j+1}-1)!!}.
\end{equation*}
By definition, we have $(S^G_{k+1}-1)!! = \prod_{p=0}^{\frac{S^G_{k+1}}{2} - 1}
(S^G_{k+1} - 1 - 2p)$. Likewise, we have
\begin{align*}
  (S^G_{j+1}-1) !! &= \prod_{p=0}^{\frac{S^G_{j+1}}{2} - 1} (S^G_{j+1} - 1 - 2p) \\
                   &=\prod_{p=0}^{\frac{S^G_{j+1}}{2} - 1} \left( S^G_{k+1} - 2 \left( \frac{S^G_{k+1} - S^G_{j+1}}{2} \right) - 1 - 2p  \right)  \\
  &= \prod_{p=\frac{S^G_{k+1} - S^G_{j+1}}{2}}^{\frac{S^G_{k+1}}{2} - 1} \left( S^G_{k+1} - 1 - 2p \right).
\end{align*}
Thereby, we have 
\begin{equation*}
  r^G_{k+1, j} = \frac{(S^G_{k+1} - 1)!!}{(S^G_{j+1}-1)!!} =\frac{ \prod_{p=0}^{ \frac{S^G_{k+1}}{2} - 1 } (S^G_{k+1} - 1 - 2p) }{\prod_{p = \frac{S^G_{k+1} - S^G_{j+1}}{2}}^{\frac{S^G_{k+1}}{2} - 1} (S^G_{k+1} - 1 - 2p)} 
= \prod_{p=0}^{\frac{S^G_{k+1} - S^G_{j+1}}{2} - 1} (S^G_{k+1} - 1 - 2p).
\end{equation*}
Comparing this with~\eqref{eq:compute-product-return}, since $S^G_{k+1} - 1 - 2
\left( \frac{S^G_{k+1} - S^G_{j+1}}{2} - 1 \right) = 1 + S^G_{j+1} > 0$, we
realize that if $S^G_{k+1} > S^G_{j+1}$, we have $r^G_{k+1, j} =
\textproc{ComputeProduct}(S^G_{k+1} - 1, \frac{S^G_{k+1} - S^G_{j+1}}{2}, 2)$.
Moreover, if $S^G_{k+1} = S^G_{j+1}$, again
using~\eqref{eq:compute-product-return}, we have
$\textproc{ComputeProduct}(S^G_{k+1} - 1, \frac{S^G_{k+1} - S^G_{j+1}}{2}, 2) =
r^G_{k+1, j} = 1$. Combining these with $S_{k+1} = S^G_{k+1}$ and $S_{j+1} =
S^G_{j+1}$, we have 
\begin{equation}
  \label{eq:s-enc-rk+1-j-CP}
 r^G_{k+1, j} =
 \textproc{ComputeProduct}(S_{k+1} - 1, \frac{S_{k+1} - S_{j+1}}{2}, 2) 
\end{equation}
Consequently, the variable $r_{k+1, j}$ computed in line~\ref{l:sNij-rkj} is
precisely $r^G_{k+1, j}$. Therefore, using the recursive
form~\eqref{eq:self-Nij-recursive}, $N_{i,j}$ computed in line~\ref{l:sNij-Nij}
coincides with $N_{i,j}(G)$. Finally, we can compute $l^G_{i,j}$ using the
identity $l^G_{i,j} = l^G_{i,k} \times l^G_{k+1, j}$ in line~\ref{l:sNij-lij}.



\begin{myalg}[Computing $N_{i,j}(G)$ for a simple unmarked graph $G \in
   \mGnt_{\va}$ \label{alg:ComputeNij_graph}]
 \begin{algorithmic}[1]
   \INPUT
   \Statex $\pn$: number of nodes
   \Statex $i,j$: endpoints of the interval, such that $1 \leq i \leq j \leq \pn$
   \Statex $\vsfa = (\sfa_v: 1 \leq v \leq \pn)$ where $\sfa_v = a^G_v(i)$ for $i \leq v
   \leq \pn$
   \Statex $\vgamma^G_i, \dots, \vgamma^G_j$: forward adjacency list of vertices $i
   \leq v \leq j$, where $\vgamma^G_v = (\vgamma^G_{v, 1}, \dots, 
   \vgamma^G_{v,\hat{a}^G_v})$ such that $v < \gamma^G_{v,1} < \dots <
   \gamma^G_{v, \hat{a}^G_v} \leq \pn$ are the neighbors of $v$ in $G$ with index
   greater than $v$
   \Statex $\sfU$: Fenwick tree, where for $i \leq v \leq \pn$,
   $\sfU.\textproc{Sum}(v) = U^G_v(i) = \sum_{k = v}^{\pn} a^G_k(i)$
   \Statex $I$: an integer specifying  the interval $[i,j]$
   \Statex $\vec{\tilde{f}} = (\tilde{f}_p: 0 \leq p \leq \lfloor 16\pn / \log^2 \pn \rfloor)$: array of integers
   where if $j - i + 1 > \lfloor  \log_2 \pn
   \rfloor^2$,  $\tilde{f}_I = S^G_{k+1}$  with  $k = \lfloor (i+j)/2 \rfloor$
   and $I$ being the index corresponding to the  interval $[i,j]$ as above
   \OUTPUT
   \Statex $N_{i,j} = N_{i,j}(G)$
   \Statex $\vsfa = (\sfa_v: 1 \leq v \leq \pn)$: vector $\vsfa$ updated, so that
   $\sfa_v = a^G_v(j+1)$ for $j+1 \leq v \leq \pn$
   \Statex $\sfU$: Fenwick tree $U$ updated, so that for $j+1 \leq v \leq \pn$, we have $\sfU.\textproc{Sum}(v) =
   U^G_v(j+1) = \sum_{k=v}^{\pn} a^G_k(j+1)$
   \Statex $l_{i,j} = l^G_{i,j}$
   \Statex $\vec{\tilde{f}}$: array $\vec{\tilde{f}}$ updated, so that if $j - i +1> \lfloor \log_2\pn
   \rfloor^2$, $\tilde{f}_I = S^G_{k+1}$ with $k = \lfloor (i+j) / 2 \rfloor$
   and $I$ being the index corresponding to the  interval $[i,j]$ as above
   \Function{ComputeN}{$\pn, \vsfa, i,j, \vgamma^G_{[i:j]}, \sfU, I, \vec{\tilde{f}}$}
   \If {$i=j$}
   \State $z_i \gets 0, l_i \gets 1$
   \For{$1 \leq k \leq \sfa_i$} \label{l:sNij-i-for}
   \State $y \gets \textproc{ComputeProduct}(\sfU.\textproc{Sum}(1+\gamma^G_{i,k}), \sfa_i - k + 1,
   1)$ \label{line:sNij-i-y} \Comment{Algorithm~\ref{alg:compute-product} in Section~\ref{sec:aux-algs-bip}}
   \State $z_i \gets z_i + l_i \times y$ \label{l:sNij-i-z}
   \State $c \gets (\sfa_i - k + 1) \times \sfa_{\gamma^G_{i,k}}$ \label{l:sNij-i-c}
   \State $l_i \gets l_i \times c$ \label{l:sNij-i-l}
   \State $\sfa_{\gamma^G_{i,k}} \gets \sfa_{\gamma^G_{i,k}} - 1$ \label{l:sNij-i-beta}
   \State $\sfU.\textproc{Add}( \gamma^G_{i,k}, -1)$ \label{l:sNij-i-U}
   \EndFor
   \State $N_{i,i} \gets z_i$
   \State \textbf{return} $(N_{i,i}, \vsfa, \sfU, l_i, \vec{\tilde{f}})$
   \Else
   \State $k \gets (i+j) \div 2$ \label{l:sNij-ij-k}
   \State $(N_{i,k}, \vsfa, \sfU, l_{i,k}, \vec{\tilde{f}}) \gets 
   \textproc{ComputeN}(\pn, \vsfa, i,k,\vgamma^G_{[i:k]}, \sfU, 2I, \vec{\tilde{f}})$ \label{l:sNij-call-ik}
   \State $S_{k+1} \gets \sfU.\textproc{Sum}(k+1)$ \label{l:sNij-Sk}
   \If {$j - i+1 > \lfloor  \log_2 \pn \rfloor^2 $} \label{l:sNij-if-ij-log2n}
   \State $\tilde{f}_I \gets S_{k+1}$ \label{l:sNij-tSI}
   \EndIf
   \State $(N_{k+1, j}, \vsfa, \sfU, l_{k+1, j},\vec{\tilde{f}}) \gets
   \textproc{ComputeN}(\pn, \vsfa,  k+1,
   j, \vgamma^G_{[k+1:j]}, \sfU,2I+1, \vec{\tilde{f}})$ \label{l:sNij-call-kj}
   \State $S_{j+1} \gets \sfU.\textproc{Sum}(j+1)$ \label{l:sNij-Sj}
   \State $r_{k+1,j} \gets \textproc{ComputeProduct}(S_{k+1}-1, (S_{k+1} -
     S_{j+1}) / 2, 2)$ \label{l:sNij-rkj}
     \Comment{Algorithm~\ref{alg:compute-product} in Section~\ref{sec:aux-algs-bip}}
   \State $N_{i,j} \gets N_{i,k} \times r_{k+1, j} + l_{i,k} \times
   N_{k+1,j}$ \label{l:sNij-Nij}
   \State $l_{i,j} \gets l_{i,k} \times l_{k+1, j}$ \label{l:sNij-lij}
   \State \textbf{return} $(N_{i,j}, \vsfa, \sfU, l_{i,j}, \vec{\tilde{f}})$ \label{l:sNij-ij-ret}
   \EndIf
   \EndFunction
 \end{algorithmic}
\end{myalg}

\begin{rem}
  \label{rem:senc-pointer}
  Similar to our discussion in Remark~\ref{rem:b-enc-U-b-pointer}, we can store
  the Fenwick tree $\sfU$ as well as the arrays $\vsfa$ and $\vec{\tilde{f}}$ in one place
  of the memory and only pass the pointer of these locations to
  Algorithm~\ref{alg:ComputeNij_graph}. This way, the algorithm modifies the
  given locations in the memory and avoids copying
  unnecessary data. Observe that since each partial sum in $\sfU$ is at most
  $\pn \delta$, the memory required to store $\sfU$ is $O(\pn \log \pn)$. On the
  other hand, since each element in $\vsfa$ is bounded by $\delta$, the memory
  required to store this array is $O(\pn \log \delta)$. Furthermore, as we
  discussed above, the length of the array $\vec{\tilde{f}}$ is $O(\pn / \log^2 \pn)$, and
  each of its elements is bounded by $\pn \delta$. Therefore, the memory
  required to store $\vec{\tilde{f}}$ is $O(\pn / \log \pn)$ bits. 
\end{rem}

Finally, we use the above together with~\eqref{eq:self-fna-def} to find
$\fnt_{\va}(G)$ for $G \in \mGnlr_{\va}$. It is easy to see that 
\begin{equation}
  \label{eq:self-l-is-prod-a-factorial-claim}
  l^G_{1, \pn} = \prod_{i=1}^{\pn} a_i!.
\end{equation}
The reason is that by definition, $l^G_{1, \pn} = \prod_{i=1}^{\pn}
\prod_{k=1}^{\hat{a}^G_i} (\hat{a}^G_i-k+1)a^G_{\gamma^G_{i,k}}(i)$ is the total
number of configurations $C \in \mCnt_{\va}$ such that $\vA^C = \vA^G$. On the
other hand, since $G$ is simple and each vertex $1 \leq i \leq \pn$ has $a_i$
many half--edges, there are precisely $\prod_{i=1}^{\pn} a_i!$ many
configurations $C$ with $\vA^C = \vA^G$. This establishes
\eqref{eq:self-l-is-prod-a-factorial-claim}. Therefore, we can call
Algorithm~\ref{alg:ComputeNij_graph} in order to find $N_{1,\pn}(G)$ as well as
$l^G_{1, \pn}$ in order to find $\fnt_{\va}(G)$. This is done in Algorithm
\ref{alg:self-find-f} below. 
Note that, as $a^G_i(1) = a_i$ for $1 \leq i \leq \pn$, we initialize the array
$\vsfa$ with $\va$.

\begin{myalg}[Finding $\fnt_{\va}(G)$ and $\vec{\tilde{f}}^{(\pn)}_{\va}(G)$ for $G \in \mGnlr_{\va}$ \label{alg:self-find-f}]
  \begin{algorithmic}[1]
    \INPUT
    \Statex $\pn$: number of nodes
    \Statex $\va = (a_v: 1 \leq v \leq \pn)$: degree vector
    \Statex $\vgamma^G = (\vgamma^G_v: 1 \leq v \leq \pn)$: forward adjacency
    list where $\vgamma^G_v = (\gamma^G_{v, 1} < \dots <
    \gamma^G_{v,\hat{a}^G_v})$ is the forward adjacency list of node $v$
    \OUTPUT
    \Statex  $f = f^{(\pn)}_{\va}(G)$
    \Statex $\vec{\tilde{f}} = (\tilde{f}_i: 1 \leq i \leq  \lfloor 16 \pn / \log^2 \pn
    \rfloor)$: $\Array$ of integers such that $\tilde{f}_i =
    \tilde{f}^{(n)}_{\va, i}(G)$
    \Function{EncodeGraph}{$\pn, \va, \vgamma^G$}
    \State $\sfU \gets$ Fenwick tree initialized with array $\va$
    \State $\vec{\tilde{f}} \gets \Array $ of integers with length $\lfloor
    16\pn / \log^2 \pn \rfloor$
    \State $\vsfa \gets \va$ \Comment{$a^G_i(1) = a_i$ for $1 \leq i \leq \pn$}
    \State $(N_{1, \pn}, \vsfa, \sfU, l_{1, \pn}, \vec{\tilde{f}}) \gets
    \textproc{ComputeN}(\pn, \va, 1, \pn, \vgamma^G_{[1:\pn]},
    \sfU, 1, \vec{\tilde{f}})$ \Comment{Algorithm~\ref{alg:ComputeNij_graph} above}
    \State $f \gets N \div l_{1, \pn}$
    \If{$f \times l_{1, \pn} < N$} \label{l:self-enc-N-div-l}\Comment{this means $N \mod l_{1, \pn} \neq 0$}
    \State $f \gets f + 1$ \label{l:self-enc-f-mul-l} \Comment{so that $f = \lceil N / l_{1,\pn} \rceil$}
    \EndIf
    \State \textbf{return} $(f, \vec{\tilde{f}})$
    \EndFunction
  \end{algorithmic}
\end{myalg}

\subsubsection{Complexity of Encoding}
\label{sec:self-compression-complexity}

In this section, we analyze the complexity of the encoding algorithm described
above. Before doing this, we first state Lemmas~\ref{lem:self-N+l<r} and
\ref{lem:self-r-l-N-bit-count}, whose proofs are postponed until the end of this
section. 

\begin{lem}
  \label{lem:self-N+l<r}
  For a simple unmarked graph $G \in \mGnt_{\va}$ and integers $1 \leq i \leq j
  \leq \pn$, we have
  \begin{equation*}
    N_{i,j}(G) + l^G_{i,j} \leq r^G_{i,j}.
  \end{equation*}
\end{lem}

\begin{lem}
  \label{lem:self-r-l-N-bit-count}
  Assume that we have $a_k \leq \delta$ for all $1 \leq k \leq \pn$. Then, given
  $G \in \mGnt_{\va}$ and $1 \leq i \leq j \leq \pn$, the variables $r^G_{i,j},
  l^G_{i,j}$ and $N_{i,j}(G)$ have $O((j-i+1) \delta \log \pn)$ bits.
\end{lem}

Now we are ready to study the complexity of Algorithm~\ref{alg:ComputeNij_graph}.

\begin{lem}
  \label{lem:s-compute-N-complexity}
  In Algorithm~\ref{alg:ComputeNij_graph}, if we have $a_v \leq \delta$, $1 \leq
  v \leq \pn$,  then computing
  $N_{1,\pn}(G)$ for $G \in \mGnt_{\va}$ takes $O(\pn \delta \log^4 \pn \log \log \pn)$ time and
  requires $O(\pn \delta \log \pn)$ bits of  memory.
\end{lem}

\begin{proof}
  First, we analyze the performance of the algorithm when $i =j$. In
  line~\ref{line:sNij-i-y}, computing $\sfU.\textproc{Sum}(1+\gamma^G_{i,k})$
  for each $1 \leq k \leq \sfa_i$ takes $O(\log^2 \pn)$ time,
  the reason being that $\sfU$ keeps track of partial sums each bounded by $\pn
  \delta$. On the other hand, since $\sfa_i - k +
  1 \leq \delta$, using Lemma~\ref{lem:compute-product-complexity}, computing
  $\textproc{ComputeProduct}(\sfU.\textproc{Sum}(1+\gamma^G_{i,k}), \sfa_i - k + 1, 1)$ takes
  \begin{equation*}
    O(\delta \log \delta \log \pn  \log (\delta \log \pn) \log \log (\delta \log \pn)) = O(\delta \log \delta \log^2 \pn \log \log \pn),
  \end{equation*}
  time and $O(\delta \log \pn)$ memory. Also, the result $y$ has $O(\delta \log
  \pn)$ bits. Note that the variable $l_i$ always remains bounded by $l^G_i$
  inside the loop of line~\ref{l:sNij-i-for}. Thereby,
  Lemma~\ref{lem:self-r-l-N-bit-count} implies that it has $O(\delta \log \pn)$
  bits. Consequently,  computing $l_i  \times y$ in line~\ref{l:sNij-i-z} takes
  $\Mul(\delta \log \pn) = O( \delta \log \pn \log (\delta \log
  \pn) \log \log (\delta \log \pn)) = O(\delta \log^2 \pn \log \log \pn)$  time.
  On the other hand, since the variable $z_i$ remains bounded by $z^G_i$ inside
  the loop of line~\ref{l:sNij-i-for}, Lemma~\ref{lem:self-r-l-N-bit-count}
  implies that it has $O(\delta \log \pn)$ bits.
  Consequently, the addition in
  line~\ref{l:sNij-i-z} takes $O(\delta \log \pn)$ time.
  Putting these together,
  the overall time complexity of line~\ref{l:sNij-i-z} is $O(\delta \log^2 \pn
  \log \log \pn)$.
In line~\ref{l:sNij-i-c}, both $\sfa_i - k + 1$ and $\sfa_{\gamma^G_{i,k}}$ are
bounded by $\delta$. On the other hand, as we discussed above, $l_i$ remains to
have $O(\delta \log \pn)$ bits inside the loop.
As a result, the products in lines~\ref{l:sNij-i-c} and \ref{l:sNij-i-l} take
$\Mul(\delta \log \pn) = O(\delta \log^2 \pn \log \log \pn)$ time.
Also, since $\sfa_{\gamma^G_{i,k}} \leq \delta$, the update in
line~\ref{l:sNij-i-beta} takes $O(\log \delta)$ time.  Moreover,
updating $\sfU$ in line~\ref{l:sNij-i-U} takes $O(\log^2 \pn)$ time, the reason being
that as was discussed above, all the partial terms have $O(\log \pn)$ bits. 
Putting the above discussion together, the time and memory complexities of the algorithm when
$i = j$ are $O(\delta \log \delta \log^2 \pn \log \log \pn)$ and $O(\delta \log \pn)$,
respectively.

Now, assume that $i < j$. Similar to the above, finding
$\sfU.\textproc{Sum}(k+1)$ in line~\ref{l:sNij-Sk} takes $O(\log^2 \pn)$ time.  Note that
we may compute $\lfloor  \log_2 \pn \rfloor^2$ once and store it in the memory.
In order to do so, since the variable $\pn$ is stored in binary format, we can compute $\lfloor
\log_2 \pn  \rfloor$ by simply looking at the length of this binary representation
in $O(\log \pn)$ time. Therefore, the one--time computation of  $\lfloor  \log_2 \pn
\rfloor^2$ takes $O(\log^2 \pn )$ time, and the result has $O(\log \pn)$ bits.
Thereby, the comparison in line~\ref{l:sNij-if-ij-log2n} takes $O(\log \pn)$ time.
Retrieving $\sfU.\textproc{Sum}(j+1)$ in line~\ref{l:sNij-Sj}, similar to line~\ref{l:sNij-Sk},
takes $O(\log^2 \pn)$ time.
In line~\ref{l:sNij-rkj}, from~\eqref{eq:self-SGi-S-ahat}, we have
\begin{equation*}
  \frac{S^G_{k+1} - S^G_{j+1}}{2} = \sum_{v=k+1}^j \hat{a}^G_v \leq (j-k) \delta.
\end{equation*}
Also, $S^G_{k+1}- 1$ has $O(\log \pn)$ bits. Therefore, using
Lemma~\ref{lem:compute-product-complexity}, computing
$\textproc{ComputeProduct}(S^G_{k+1}-1, (S^G_{k+1} - S^G_{j+1})/2, 2)$ takes
\begin{equation*}
  O((j-k) \delta \log \pn \log ((j-k) \delta) \log ((j-k) \delta \log \pn) \log \log ((j-k) \delta \log \pn)) = O((j-k) \delta \log^3 \pn \log \log \pn),
\end{equation*}
time and requires storing $O((j-k) \delta \log \pn)$ memory.
Furthermore, using Lemma~\ref{lem:self-r-l-N-bit-count}, all the variables on
the right hand side of line~\ref{l:sNij-Nij} have $O((j-i+1) \delta \log \pn)$
bits. As a result, the time complexity of performing the arithmetic
in this line is $\Mul((j-i+1)\delta \log \pn) = O((j-i+1) \delta \log^2 \pn \log
\log \pn)$. The time complexity of performing the product in
line~\ref{l:sNij-lij} is similarly $O((j-i+1) \delta \log^2 \pn \log \log \pn)$.

Putting the above discussion together, we realize that if $f(k)$ denotes the
time complexity of the algorithm when $j-i +1 = k$, for $k > 1$, we have
\begin{equation*}
  f(k) = f\left( \left \lfloor \frac{k}{2}\right \rfloor \right) + f\left( \left\lceil \frac{k}{2} \right \rceil \right) + O(k \delta \log^3 \pn \log \log \pn),
\end{equation*}
and
\begin{equation*}
  f(1) = O(\delta \log \delta \log^2 \pn \log \log \pn).
\end{equation*}
Solving for this recursive form, we get
\begin{equation*}
  f(\pn) = O(\pn \delta \log^4 \pn \log \log \pn),
\end{equation*}
as was claimed.

Now, we analyze the memory complexity of the algorithm. Motivated by
Remark~\ref{rem:senc-pointer}, let $\tilde{g}(k)$ be the memory complexity of
the algorithm when $j - i + 1 = k$, excluding the memory required to store
$\sfU$, $\vec{\tilde{f}}$ and $\vsfa$. As we discussed above, when $i < j$, all the variables defined between
lines~\ref{l:sNij-ij-k} and \ref{l:sNij-ij-ret} have $O((j-i+1) \delta \log
\pn)$ bits. Hence, for $k > 1$, we have 
\begin{equation*}
  \tilde{g}(k) = \max \left\{ \tilde{g} \left( \left\lfloor \frac{k}{2} \right \rfloor \right) , \tilde{g} \left( \left \lceil  \frac{k}{2} \right \rceil \right)\right\} + O(k \delta \log \pn).
\end{equation*}
Also, as was discussed above,  we have
\begin{equation*}
  \tilde{g}(1) = O(\delta \log \pn). 
\end{equation*}
Solving for this, we get $\tilde{g}(\pn) = O(\pn \delta \log \pn)$. Furthermore, as we
discussed in Remark~\ref{rem:senc-pointer}, the memory required to store $\vec{\tilde{f}}$,
$\sfU$ and $\vsfa$ is $O(\pn \log \pn)$. Putting all these together, we realize
that the memory complexity of computing $N_{1, \pn}(G)$ is $O(\pn \delta \log \pn)$.
This completes the proof. 
\end{proof}

\begin{lem}
\label{lem:self-find-f-complexity}
  In Algorithm~\ref{alg:self-find-f}, if we have $a_i \leq \delta$ for $1 \leq i
  \leq \pn$, running Algorithm~\ref{alg:self-find-f} for $G \in \mGnt_{\va}$
  takes $O(\pn \delta \log^4 \pn \log \log \pn)$ time and requires $O(\pn \delta
  \log \pn)$ bits of memory. 
\end{lem}

\begin{proof}
  Since each $a_i$ for $1 \leq i \leq \pn$ has $O( \log \pn)$ bits, initializing
  the Fenwick tree $\sfU$ with array $\va$ takes $O( \pn \log^2 \pn)$ time and
  the space required to store $\sfU$ is $O( \pn \log \pn)$. 
  Likewise, initializing $\vsfa$ with $\va$ takes $O(\pn \log \pn)$ time. 
Also, from
  Lemma~\ref{lem:self-r-l-N-bit-count}, the variables $N_{1,\pn} = N_{1, \pn}(G)
 $ and $l_{1,\pn} = l^G_{1, \pn}$ have  $O(\pn \delta \log \pn)$ bits.
 Therefore, the arithmetic in lines~\ref{l:self-enc-N-div-l} and
 \ref{l:self-enc-f-mul-l} take $\Mul(\pn \delta \log \pn) = O(\pn \delta
 \log^2 \pn \log \log \pn)$. This together with
 Lemma~\ref{lem:s-compute-N-complexity} completes the proof. 
\end{proof}

\begin{proof}[Proof of Lemma~\ref{lem:self-N+l<r}]
  We first prove the claim for $i = j$. In this case, the claim reduces to
  showing $z^G_i + l^G_i \leq r^G_i$. Note that if $\hat{a}^G_i = 0$, we have $z^G_i = 0$,
  $l^G_i = 1$ and $r^G_i = 1$, which implies that the inequality holds. Therefore,
  assume $\hat{a}^G_i > 0$. For $1 \leq k \leq \hat{a}^G_i$, define
  \begin{equation}
    \label{eq:self-zik-def}
    z^G_{i,k} := \sum_{\pp{k} = k}^{\hat{a}^G_i} \left[ \prod_{\ppp{k} = k}^{\pp{k} - 1}  (\hat{a}^G_i - \ppp{k} + 1) a^G_{\gamma^G_{i,\ppp{k}}}(i) \right] (U^G_{1+\gamma^G_{i,\pp{k}}}(i))_{\hat{a}^G_i - \pp{k} + 1}.
  \end{equation}
  In particular, note that $z^G_{i,1} = z^G_i$. With this, we claim that for $1 \leq
  k \leq \hat{a}^G_i$,
  \begin{equation}
    \label{eq:s-claim-zik+<U}
    z^G_{i,k} + \prod_{\pp{k}= k }^{\hat{a}^G_i} (\hat{a}^G_i - \pp{k} + 1) a^G_{\gamma^G_{i,\pp{k}}}(i) \leq (U^G_{\gamma^G_{i,k}}(i))_{\hat{a}^G_i - k + 1}.
  \end{equation}
  We prove this by backward induction on $k$. If $k = \hat{a}^G_i$, the left hand
  side becomes $U^G_{1+\gamma^G_{i,\hat{a}^G_i}}(i) + a^G_{\gamma^G_{i,\hat{a}^G_i}}(i) =
  U^G_{\gamma^G_{i,\hat{a}^G_i}}(i)$, which is equal to the right hand side for $k =
  \hat{a}^G_i$. Hence, for $k = \hat{a}^G_i$, \eqref{eq:s-claim-zik+<U} becomes an
  equality. Moreover, for $1 \leq k < \hat{a}^G_i$, it is easy to see from the
  definition of $z^G_{i,k}$ that
  \begin{equation*}
    z^G_{i,k} = (U^G_{1+\gamma^G_{i,k}}(i))_{\hat{a}^G_i - k + 1} + (\hat{a}^G_i - k + 1) a^G_{\gamma^G_{i,k}}(i) z^G_{i,k+1}.
  \end{equation*}
  Using this on the left hand side of \eqref{eq:s-claim-zik+<U} and using the
  induction hypothesis, we get
  \begin{align*}
    z^G_{i,k} +  \prod_{\pp{k}= k }^{\hat{a}^G_i} (\hat{a}^G_i - \pp{k} + 1) a^G_{\gamma^G_{i,\pp{k}}}(i) &= (U^G_{1+\gamma^G_{i,k}}(i))_{\hat{a}^G_i - k + 1} \\
                                                                                         &\quad + (\hat{a}^G_i - k + 1) a^G_{\gamma^G_{i,k}}(i) \left[ z^G_{i,k+1} +  \prod_{\pp{k}= k +1}^{\hat{a}^G_i} (\hat{a}^G_i - \pp{k} + 1) a^G_{\gamma^G_{i,\pp{k}}}(i)\right] \\
                                                                                         &\stackrel{(a)}{\leq} (U^G_{1+\gamma^G_{i,k}}(i))_{\hat{a}^G_i - k + 1} + (\hat{a}^G_i - k + 1) a^G_{\gamma^G_{i,k}}(i) (U^G_{\gamma^G_{i,k+1}}(i))_{\hat{a}^G_i - k} \\
                                                                                         &\stackrel{(b)}{\leq} (U^G_{1+\gamma^G_{i,k}}(i))_{\hat{a}^G_i - k + 1} + (\hat{a}^G_i - k + 1) a^G_{\gamma^G_{i,k}}(i) (U^G_{1+\gamma^G_{i,k}}(i))_{\hat{a}^G_i - k} \\
                                                                                         &\stackrel{(c)}{\leq} (a^G_{\gamma^G_{i,k}}(i) + U^G_{1+\gamma^G_{i,k}}(i))_{\hat{a}^G_i - k + 1} \\
    &= (U^G_{\gamma^G_{i,k}}(i))_{\hat{a}^G_i - k + 1},
  \end{align*}
  where in $(a)$ we have used the induction hypothesis for $k + 1$,  $(b)$
  exploits the fact that $1 + \gamma^G_{i,k} \leq \gamma^G_{i,k+1}$, and $(c)$ uses the classical
  inequality $(p)_q + qs (p)_{q-1} \leq (p+s)_q$ which holds for $ s \geq 0$
  and $p \geq q \geq 1$\footnote{one way to see this is to divide both sides by $q!$
    which reduces the inequality to the more familiar form $\binom{p}{q} + s
    \binom{p}{q-1} \leq \binom{p+s}{q}$}. This establishes
  \eqref{eq:s-claim-zik+<U} for $k$. Now, using this for $k = 1$, we get
  \begin{align*}
    z^G_i + l^G_i &= z^G_{i,1} + \prod_{k=1}^{\hat{a}^G_i} (\hat{a}^G_i - k + 1) a^G_{\gamma^G_{i,k}}(i) \\
              &\leq (U^G_{\gamma^G_{i,1}}(i))_{\hat{a}^G_i} \\
              &\leq (U^G_{i+1}(i))_{\hat{a}^G_i} \\
              &= (S^G_i - \hat{a}^G_i)_{\hat{a}^G_i} \\
              &= (S^G_i - \hat{a}^G_i)(S^G_i - \hat{a}^G_i - 1) \dots (S^G_i - 2 \hat{a}^G_i + 1) \\
              &= \prod_{k=0}^{\hat{a}^G_i - 1} (S^G_i - (\hat{a}^G_i + k)) \\
              &\stackrel{(a)}{\leq} \prod_{k=0}^{\hat{a}^G_i - 1} (S^G_i - (2k + 1)) \\
              &= r^G_i,
  \end{align*}
  where in $(a)$, since $k < \hat{a}^G_i$, $k + \hat{a}^G_i \geq 2k+1$. This is our
  desired result for the case $i = j$.

  Now, we prove the lemma for the general case $i \leq j$ by induction on $j-i$.
  The base case is $j -i =0$ which we just showed. For $j > i$, with $k :=
  \lfloor  (i+j) / 2 \rfloor$, using~\eqref{eq:self-Nij-recursive} and the fact
  that by definition, $l^G_{i,j} = l^G_{i,k} l^G_{k+1, j}$, we get
  \begin{align*}
    N_{i,j}(G) + l^G_{i,j} &= N_{i,k}(G) r^G_{k+1, j} + l^G_{i,k} N_{k+1, j}(G) + l^G_{i,k} l^G_{k+1, j} \\
                         &= N_{i,k} r^G_{k+1, j} + l^G_{i,k} (N_{k+1, j}(G) + l^G_{k+1, j}) \\
                         &\stackrel{(a)}{\leq} N_{i,k}(G) r^G_{k+1, j} + l^G_{i,k} r^G_{k+1, j} \\
                         &\stackrel{(b)}{\leq} r^G_{i,k} r^G_{k+1, j} \\
                         &= r^G_{i,j},
  \end{align*}
  where $(a)$ and $(b)$ use the induction hypothesis for the intervals $[k+1,
  j]$ and $[i,k]$, respectively. This completes the proof.
\end{proof}

\begin{proof}[Proof of Lemma~\ref{lem:self-r-l-N-bit-count}]
  Recalling the definition of $r^G_{i,j}$, we have
  \begin{equation}
    \label{eq:self-r-bit-count-bound}
    r^G_{i,j} = \frac{(S^G_i-1)!!}{(S^G_{j+1}-1)!!} \leq {S^G_i}^{\frac{S^G_i - S^G_{j+1}}{2}}.
  \end{equation}
  But from~\eqref{eq:self-SGi-S-ahat}, we have
  \begin{equation*}
    S^G_i - S^G_{j+1} = \sum_{k=i}^j 2 \hat{a}^G_k \leq 2 \sum_{k=i}^j a_k \leq 2(j-i+1)\delta.
  \end{equation*}
  Using this together with $S^G_i \leq S \leq \pn \delta$
  in~\eqref{eq:self-r-bit-count-bound}, we realize that $r^G_{i,j}$ has
  $O((j-i+1)\delta \log(\pn \delta)) = O((j-i+1) \delta \log \pn)$ bits.
  Lemma~\ref{lem:self-N+l<r} implies that the same bound also holds for the
  number of bits in $N_{i,j}(G)$ as well as $l^G_{i,j}$. This completes the
  proof. 
\end{proof}

\subsection{Decompression Algorithm}
\label{sec:self-decompression}

In this section, we discuss how to decode for $G \in \mGnt_{\va}$ given
$\fnt_{\va}(G)$ and the array $\vec{\tilde{f}}^{\pn}_{\va}(G)$. Recall that
$\fnt_{\va}(G) = \lceil N(G) / \prod_{i=1}^{\pn} a_i! \rceil$. Therefore, we
introduce a proxy for $N(G)$ defined as 
\begin{equation}
  \label{eq:s-dec-tN-1-pn-def}
  \tN_{1,\pn} := \fnt_{\va}(G) \left( \prod_{i=1}^{\pn} b_i! \right).
\end{equation}
Indeed, we have
\begin{equation*}
  N_{1,\pn}(G) \leq \tN_{1,\pn} < N_{1,\pn}(G) + \prod_{i=1}^{\pn} a_i!.
\end{equation*}
Recall from~\eqref{eq:self-l-is-prod-a-factorial-claim} that $\prod_{i=1}^{\pn}
a_i! = l^G_{1,\pn}$. Therefore, we have
\begin{equation}
  \label{eq:self-decode-tN-N-N+l}
  N_{1,\pn}(G) \leq \tN_{1,\pn} < N_{1,\pn}(G) + l^G_{1,\pn}.
\end{equation}
Similar to the decoding procedure in Section~\ref{sec:bipartite-decode}, we
employ a recursive fashion for decoding the graph $G$. More precisely, we
introduce an algorithm that given an interval $[i,j]$ for $1 \leq i \leq j \leq
\pn$, decodes the forward adjacency list of vertices in the interval $[i,j]$. In
order to do this, if $j - i + 1 > \lfloor \log_2 n  \rfloor^2$, we recursively decode
the intervals $[i,k]$ and $[k+1, j]$ with $k = \lfloor (i+j)/2 \rfloor$.
Otherwise, if $1 < j - i + 1 \leq \lfloor  \log_2n \rfloor^2$, we recursively
decode the intervals $[i,i]$ and $[i+1, j]$. When the procedure is called for
the interval $[i,j]$, we assume that the following properties hold:
  \begin{enumerate}[label=\textbf{B\arabic*}]
  \item \label{a:sdec-1-i-1} The forward adjacency list of vertices $1, \dots,
    i-1$ are decoded. In particular, the quantities $a^G_v(i)$ and $U^G_v(i)$ for vertices in the range $i \leq
    v \leq \pn$ are known.
  \item \label{a:sdec-N} Integer $\tN_{i,j}$ is given which satisfies $N_{i,j}(G) \leq \tN_{i,j} < N_{i,j}(G) +
    l^G_{i,j}$.
  \item \label{a:sdec-Sj} $S^G_{j+1}$ is known.
  \item \label{a:sdec-Sk} If $j - i +1 > \lfloor \log_2 \pn \rfloor^2$, $S^G_{k+1}$ is known where $k = \lfloor  (i+j) / 2
    \rfloor$. 
  \end{enumerate}
  Given these, our algorithm is able to do the following: 
  \begin{enumerate}
  \item Decode for the forward adjacency list of the vertices in the range
    $[i,j]$.
  \item Find $N_{i,j}(G)$ and $l^G_{i,j}$.
  \end{enumerate}
First, note that all these properties hold for $i = 1, j = \pn$.
Specifically, \ref{a:sdec-1-i-1} holds since $i=1$ and $a^G_v(i) = a_v$ for $1
  \leq v \leq \pn$ is known, \ref{a:sdec-N} holds from \eqref{eq:self-decode-tN-N-N+l},
  \ref{a:sdec-Sj} holds since $S^G_{\pn+1} = 0$ by definition,
  and~\ref{a:sdec-Sk} holds since we have access to $\vec{\tilde{f}}^{(\pn)}_{\va}(G)$,  and by
  construction, if $\pn > \lfloor \log_2 \pn \rfloor^2$, with $k = \lfloor
  (1+\pn)/2 \rfloor$, we have $S^G_{k+1} = \tilde{f}^{(\pn)}_{\va, 0}(G)$. 
We prove that these properties inductively, using an argument similar in nature
to that in Section~\ref{sec:bipartite-decode}. More specifically, consider a
binary tree which we call the \emph{process tree} in which every node
represents an interval $[i,j]$ for some $1 \leq i \leq j \leq \pn$. The root in
this tree corresponds to the interval $[1, \pn]$. Moreover, each node
corresponds to an interval $[i,j]$ with $j > i$ has two children as follows. If
$j - i + 1 > \lfloor \log_2 \pn \rfloor^2$, with $k = \lfloor (i+j)/2 \rfloor$, the left child  represents the
interval $[i,k]$, while the right child represents the interval $[k+1, j]$. Otherwise,
the left child represents the intervals $[i,i]$ and the right child represents
the interval $[i+1,  j]$. The running
of the algorithm can be visualized on this tree, where each node first calls its
left node, waits for the results, then calls its right node, and combines the
results of the two children before returning to its parent. We will prove that the
properties~\ref{a:sdec-1-i-1}--~\ref{a:sdec-Sk} hold while we traverse on this
process tree from top to bottom by induction.
Moreover, we show that each node delivers the two items
mentioned above while returning to its parent. 

Next, we first introduce the decoding procedure for the interval $[i,j]$ when
$i=j$, and then we study the case where $i < j$.

\subsubsection{Decoding the Forward Adjacency List of One Node}
\label{sec:self-decode-one-node}

In this section, we study the decoding procedure for the interval $[i,j]$ when
$i =j $.Define $\tz_i = \tN_{i,i}$, and note that due to
property~\ref{a:sdec-N}, we can think of $\tz_i$ as a proxy for $z^G_i = N_{i,i}(G)$.
We claim that
    \begin{equation}
      \label{eq:sdec-xi1}
      \gamma^G_{i,1}= \min \left\{ i < v \leq \pn: (U^G_{1+v}(i))_{\hat{a}^G_i}  \leq \tz_i \right\}.
    \end{equation}
Note that as we discussed above, due to~\ref{a:sdec-1-i-1}, the decoder has access to $\hat{a}^G_i =
a^G_i(i)$ and $U^G_{1+v}(i)$ for $i < v \leq \pn$. Therefore, assuming
that~\eqref{eq:sdec-xi1} holds, it can be used at the decoder to find
$\gamma^G_{i,1}$. Before proving~\eqref{eq:sdec-xi1}, assuming that it holds, 
treating $\tz_i$ as a proxy for $z^G_i$ and recalling the definition of
    $z^G_i$ in \eqref{eq:senc-zi}, let $\tz_{i,2}$ be obtained from $\tz_{i}$ by removing the
    contribution of $\gamma^G_{i,1}$, i.e.\
    \begin{equation}
      \label{eq:sdec-tzi2}
      \tz_{i,2} := \left\lfloor \frac{\tz_i - (U^G_{1+\gamma^G_{i,1}}(i))_{\hat{a}^G_i}}{\hat{a}^G_i a^G_{\gamma^G_{i,1}}(i)} \right \rfloor.
    \end{equation}
With this, we claim that
    \begin{equation*}
      \gamma^G_{i,2} = \min \left\{ \gamma^G_{i,1} < v \leq \pn : (U^G_{1+v}(i))_{\hat{a}^G_i-1} \leq \tz_{i,2}\right\}.
    \end{equation*}
    In general, we claim that for $2 \leq k \leq \hat{a}^G_i$, we have
    \begin{equation}
      \label{eq:sdec-xik}
      \gamma^G_{i,k} = \min \left\{ \gamma^G_{i,k-1} < v \leq \pn: (U^G_{1+v}(i))_{\hat{a}^G_i - k + 1} \leq \tz_{i,k} \right\},
    \end{equation}
    where  for $1 \leq k \leq \hat{a}^G_i$, we define $\tz_{i,k}$
    inductively as $\tz_{i,1} := \tz_i$ and for $1 \leq k \leq \hat{a}^G_i-1$, 
    \begin{equation}
      \label{eq:sdec-tz-k+1}
      \tz_{i,k+1} := \left\lfloor \frac{\tz_{i,k} - (U^G_{1+\gamma^G_{i,k}}(i))_{\hat{a}^G_i - k + 1}}{(\hat{a}^G_i - k + 1) a^G_{\gamma^G_{i,k}}(i)}\right \rfloor.
    \end{equation}
    Before proving these claims, for $1 \leq k \leq \hat{a}^G_i$, we define
    \begin{equation}
      \label{eq:sdec-zik-def}
      z^G_{i,k} := \sum_{\pp{k} = k}^{\hat{a}^G_i} \left( \prod_{\ppp{k} = k}^{\pp{k}- 1} (\hat{a}^G_i - \ppp{k} + 1) a^G_{\gamma^G_{i,\pp{k}}}(i) \right) (U^G_{1+\gamma^G_{i,\pp{k}}}(i))_{\hat{a}^G_i - \pp{k} + 1}.
    \end{equation}
    In particular, note that $z^G_{i,1} = z^G_i$.  In order to prove the above claims, consider the following two
    lemmas.

    \begin{lem}
      \label{lem:s-dec-tzik-zik}
      With the above setup, for $1 \leq k \leq \hat{a}^G_i$, we have
      \begin{equation*}
        z^G_{i,k} \leq \tz_{i,k} < z^G_{i,k} + \prod_{\pp{k} = k}^{\hat{a}^G_i} (\hat{a}^G_i - \pp{k} + 1) a^G_{\gamma^G_{i,\pp{k}}}(i).
      \end{equation*}
    \end{lem}
    \begin{proof}
      We prove this by induction on $k$. For $k = 1$, the claim reduces to
      $N_{i,i}(G) \leq \tN_{i,i} < N_{i,i}(G) + l^G_i$, which holds due to 
      \ref{a:sdec-N}. Now, assume that the claim holds
      for $k $
      and we show it for $k+1$. It is easy to verify that
      \begin{equation}
        \label{s-zik-interms-zik+1}
        z^G_{i,k} = (U^G_{1+\gamma^G_{i,k}}(i))_{\hat{a}^G_i - k + 1} + (\hat{a}^G_i - k + 1) a^G_{\gamma^G_{i,k}}(i) z^G_{i,k+1}.
      \end{equation}
      Using this together with the induction hypothesis, we get
      \begin{equation*}
        z^G_{i,k+1} \leq \frac{\tz_{i,k} - (U^G_{1+\gamma^G_{i,k}}(i))_{\hat{a}^G_i - k + 1}}{(\hat{a}^G_i - k + 1) a^G_{\gamma^G_{i,k}}(i)} < z^G_{i,k+1} + \prod_{\pp{k} = k + 1}^{\hat{a}^G_i} (\hat{a}^G_i - \pp{k} + 1) a^G_{x_{i,\pp{k}}}(i).
      \end{equation*}
      Since both the lower and the upper bounds are integers, we get the desired
      result by taking the floor and comparing with the definition of $\tz_{k+1}$ in~\eqref{eq:sdec-tz-k+1}.
    \end{proof}

        \begin{lem}
      \label{lem:sdec-v-xik}
      For $1 \leq k \leq \hat{a}^G_i$, if $v < \gamma^G_{i,k}$, then
      \begin{equation}
        \label{eq:s-v<xik-claim}
        (U^G_{1+v}(i))_{\hat{a}^G_i - k + 1} \geq z^G_{i,k} + \prod_{\pp{k} = k }^{\hat{a}^G_i } (\hat{a}^G_i - \pp{k} + 1 ) a^G_{\gamma^G_{i,\pp{k}}}(i).
      \end{equation}
    \end{lem}

    \begin{proof}
      We prove this by backward induction on $k$. For $k = \hat{a}^G_i$, on the
      left hand side we have
      \begin{equation}
        \label{eq:s-v<xik-for-k-di}
        (U^G_{1+v}(i))_{\hat{a}^G_i - k + 1} = U^G_{1+v}(i) \geq U^G_{\gamma^G_{i,\hat{a}^G_i}}(i) = U^G_{1+\gamma^G_{i,\hat{a}^G_i}}(i) + a^G_{\gamma^G_{i,\hat{a}^G_i}}(i).
      \end{equation}
      On the other hand, the right hand side of~\eqref{eq:s-v<xik-claim} for $k
      = \hat{a}^G_i$ is precisely $U^G_{1+\gamma^G_{i,\hat{a}^G_i}}(i) +
      a^G_{\gamma^G_{i,\hat{a}^G_i}}(i)$. Comparing this to~\eqref{eq:s-v<xik-for-k-di}
      establishes the claim for $k = \hat{a}^G_i$. Now, assume that $k < \hat{a}^G_i$
      and the claim holds for $k+1$. For $v < \gamma^G_{i,k}$, we have 
      \begin{align*}
        (U^G_{1+v}(i))_{\hat{a}^G_i - k + 1} &\geq (a^G_{\gamma^G_{i,k}}(i) + U^G_{1+\gamma^G_{i,k}}(i))_{\hat{a}^G_i - k + 1} \\
                                        &\stackrel{(a)}{\geq} (U^G_{1+\gamma^G_{i,k}}(i))_{\hat{a}^G_i - k + 1} + (\hat{a}^G_i - k + 1) a^G_{\gamma^G_{i,k}}(i) (U^G_{1+\gamma^G_{i,k}}(i))_{\hat{a}^G_i - k} \\
                                        &\stackrel{(b)}{\geq} (U^G_{1+\gamma^G_{i,k}}(i))_{\hat{a}^G_i - k + 1} + (\hat{a}^G_i - k + 1) a^G_{\gamma^G_{i,k}}(i) \left(
                                          z^G_{i,k+1} + \prod_{\pp{k} = k+1}^{\hat{a}^G_i} (\hat{a}^G_i - \pp{k} + 1) a^G_{\gamma^G_{i,\pp{k}}}(i)
                                          \right)\\
                                        &= (U^G_{1+\gamma^G_{i,k}}(i))_{\hat{a}^G_i - k + 1} + (\hat{a}^G_i - k + 1) a^G_{\gamma^G_{i,k}}(i) z^G_{i,k+1} + \prod_{\pp{k} = k}^{\hat{a}^G_i} (\hat{a}^G_i - \pp{k} + 1) a^G_{\gamma^G_{i,\pp{k}}}(i) \\
        &\stackrel{(c)}{=} z^G_{i,k} + \prod_{\pp{k} = k}^{\hat{a}^G_i} (\hat{a}^G_i - \pp{k} + 1) a^G_{\gamma^G_{i,\pp{k}}}(i).
      \end{align*}
      where in $(a)$, we have used the classical
  inequality $(p)_q + qs (p)_{q-1} \leq (p+s)_q$ which holds for $ s \geq 0$
  and $p \geq q \geq 1$\footnote{one way to see this is to divide both sides by $q!$
    which reduces the inequality to the more familiar form $\binom{p}{q} + s
    \binom{p}{q-1} \leq \binom{p+s}{q}$}, $(b)$ uses the induction hypothesis
and the fact that $\gamma^G_{i,k} < \gamma^G_{i,k+1}$,
  and in $(c)$, we have used \eqref{s-zik-interms-zik+1}. This shows the claim
  for $k$ and hence completes the proof.
\end{proof}

Now, we turn to showing the claims~\eqref{eq:sdec-xi1} and \eqref{eq:sdec-xik}.
To show~\eqref{eq:sdec-xi1}, note that if $v \geq \gamma^G_{i,1}$, we have
$U^G_{1+v}(i) \leq U^G_{1+\gamma^G_{i,1}}(i)$ and using the definition of
$z^G_{i,1}$, we get 
  \begin{equation*}
    (U^G_{1+v}(i))_{\hat{a}^G_i} \leq (U^G_{1+\gamma^G_{i,1}}(i))_{\hat{a}^G_i} \leq z^G_{i,1}.
  \end{equation*}
  But from Lemma~\ref{lem:s-dec-tzik-zik}, we have $z^G_{i,1} \leq \tz_{i,1}$.
  Therefore, if $v \geq \gamma^G_{i,1}$, we have
  \begin{equation}
    \label{eq:sdec-i-v>xi1}
   (U^G_{1+v}(i))_{\hat{a}^G_i} \leq \tz_{i,1} =   \tz_i.
 \end{equation}
 On the other hand, if $v < \gamma^G_{i,1}$, using Lemmas~\ref{lem:sdec-v-xik} and
 \ref{lem:s-dec-tzik-zik}, we have
 \begin{equation}
   \label{eq:sdec-i-v<xi1}
   (U^G_{1+v}(i))_{\hat{a}^G_i} \geq z^G_{i,1} + \prod_{k = 1}^{\hat{a}^G_i} (\hat{a}^G_i - k + 1) a^G_{\gamma^G_{i,k}}(i) > \tz_{i,1} = \tz_i.
 \end{equation}
Also, note that for $1 \leq k \leq \hat{a}^G_i$, the sequence
$(U^G_{1+v}(i))_{\hat{a}^G_i - k + 1}$ is nonincreasing for $1 \leq v \leq \pn$.
Therefore, \eqref{eq:sdec-i-v>xi1} and \eqref{eq:sdec-i-v<xi1} together
establish~\eqref{eq:sdec-xi1}. The proof of~\eqref{eq:sdec-xik} for $2 \leq k
\leq \hat{a}^G_i$ is similar.

Motivated by~\eqref{eq:sdec-xi1}, we can employ a binary search to find
$\gamma^G_{i,1}$. In general, once we find $\gamma^G_{i,k}$ for $1 \leq k <
\hat{a}^G_i$, we can compute $\tz_{i,k+1}$ using~\eqref{eq:sdec-tz-k+1} and
inductively find $\gamma^G_{i,k+1}$. Continuing this, we can find the forward
adjacency list of vertex $i$. Algorithm~\ref{alg:sdec-i} below illustrates this
procedure.

Inside the loop of line~\ref{l:s-dec-k-for}, for each $1 \leq k \leq \sfa_i = a^G_i(i)=
\hat{a}^G_i$, we employ a binary search to find $\gamma^G_{i,k}$. More
precisely, if $k = 1$, since $i +1 \leq \gamma^G_{i,1} \leq \pn$, we start with
the interval $[L, R]$ with $L = i+1$ and $R=\pn$. Likewise, when $k > 1$, since
$1 + \gamma^G_{i,k-1} \leq \gamma^G_{i,k}$, we
start with $L = 1 +\gamma^G_{i,k-1}$ and $R = \pn$. In each iteration of the
loop of line~\ref{l:s-dec-while-g-f}, we keep shrinking the interval $[R,L]$ so
that eventually we have $L = R = \gamma^G_{i,k}$. In order to do so, we use
\eqref{eq:sdec-xi1} for $k = 1$ and \eqref{eq:sdec-xik} for $2 \leq k$. More
precisely, with  $v =
\lfloor (L+R)/2 \rfloor$, we 
compare $\textproc{ComputeProduct}(\sfU.\textproc{Sum}(1+v), \sfa_i - k + 1, 1)=
(U^G_{1+v}(i))_{\hat{a}^G_i - k + 1}$ with $\tz = \tz_{i,k}$ in
line~\ref{l:s-dec-CP-Uv} in order to
determine whether we should switch to the interval $[L, v]$ or $[v+1, R]$. After
we find $\gamma_{i,k} = \gamma^G_{i,k}$, we update the variables
in the
following way using $y = (U^G_{1+\gamma^G_{i,k}}(i))_{\hat{a}^G_i - k + 1}$ computed in
line~\ref{l:s-dec-CP-Ux} and $c =
(\hat{a}^G_i - k + 1) a^G_{\gamma^G_{i,k}}(i)$ computed in line~\ref{l:sdec-c}.
 The variable $z_i$ is initialized with zero and is updated in
line~\ref{l:s-dec-zi-update} so that it becomes
\begin{equation*}
  \sum_{j=1}^k \left( \prod_{\pp{j} = 1}^{j-1} (\hat{a}^G_i - \pp{j} + 1) a^G_{\gamma^G_{i,\pp{j}}}(i) \right) (U^G_{1+\gamma^G_{i,j}}(i))_{\hat{a}^G_i - j + 1}.
\end{equation*}
On the other hand, the variable $l_i$ is initialized with $1$ and is updated in
line~\ref{l:s-dec-li-update} so that it becomes $\prod_{j=1}^k (\hat{a}^G_i - j
+ 1) a^G_{\gamma^G_{i,j}}(i)$. Additionally, motivated
by~\eqref{eq:sdec-tz-k+1}, 
we update $\tz$ in
lines~\ref{l:s-dec-tz-y} and \ref{l:s-dec-tz-c} so that it becomes $\tz_{i,
  k+1}$. Furthermore, we update $\vsfa$ and $\sfU$ in
lines~\ref{l:s-dec-beta-update} and \ref{l:s-dec-U-update}, respectively.
Therefore, when the loop over $k$ is over, $z_i$
becomes $z^G_i = N_{i,i}(G)$ and $l_i$ becomes $l^G_i$. 
Also, $\vsfa$ and $\sfU$ are updated as expected.

\begin{myalg}[Decoding the forward adjacency list of a vertex $1 \leq i \leq \pn$
 given $\tN_{i,i}$ for a simple unmarked graph $G \in \mGnt_{\va}$ \label{alg:sdec-i}]
 \begin{algorithmic}[1]
   \INPUT
   \Statex $\pn$: number of nodes in the graph
   \Statex $1 \leq i  \leq \pn$: the index of the node to be decoded
   \Statex $\tN_{i,i}$: integer satisfying $N_{i,i}(G) \leq \tN_{i,i} <
   N_{i,i}(G) + l^G_i$
   \Statex $\vsfa = (\sfa_v: 1 \leq v \leq \pn)$, where $\sfa_v = a^G_v(i)$ for
   $i \leq v \leq \pn$
   \Statex $\sfU$: Fenwick tree, where for $i \leq v \leq \pn$,
   $\sfU.\textproc{Sum}(v) = U^G_v(i) = \sum_{k = v}^{\pn} a^G_k(i)$
   \OUTPUT
   \Statex $N_{i,i} = N_{i,i}(G)$.
   \Statex $\vgamma_i = (\gamma_{i,k} : 1 \leq k \leq \hat{a}^G_i)$: forward adjacency list of the graph for vertex $i$, such
   that $\gamma_{i,k} = \gamma^G_{i,k}$ for $1 \leq k \leq \hat{a}^G_i$
   \Statex $\vsfa = (\sfa_v: 1  \leq v \leq \pn)$: vector $\vsfa$ updated, so that
   $\sfa_v = a^G_v(i+1)$ for $i < v \leq \pn$.
   \Statex $\sfU$: Fenwick tree $\sfU$ updated, so that for $i+1 \leq v \leq \pn$, we have $\sfU.\textproc{Sum}(v) =
   \sum_{k=v}^{\pn} a^G_k(i+1)$
   \Statex $l_{i} = l_{i}^G$.
   \Function{DecodeNode}{$\pn, i,\tN_{i,i}, \vsfa, \sfU$}
      \State $\tz \gets \tN_{i,i}$
      \State $l_i \gets 1$ 
      \State $z_i \gets 0$ 
      \For{$1 \leq k \leq \sfa_i$} \label{l:s-dec-k-for}
         \If{$k=1$}
            \State $L \gets i+1$ \Comment{we do a binary search on the
              interval $[L,R]$}
         \Else
            \State $L \gets \gamma_{i,k-1} + 1$ \Comment{since $\gamma^G_{i,k-1}
              + 1 \leq \gamma^G_{i,k}$}
         \EndIf
         \State $R \gets \pn$  \Comment{since $\gamma^G_{i,k} \leq \pn$}
         \While{$R > L$} \Comment{binary search to find $\gamma_{i,k}$} \label{l:s-dec-while-g-f}
            \State $v \gets   (L+R)\div 2   $
            \If{$\textproc{ComputeProduct}(\sfU.\textproc{Sum}(1+v), \sfa_i - k + 1, 1) \leq
              \tz$} \label{l:s-dec-CP-Uv}
            \Statex \Comment{Algorithm~\ref{alg:compute-product} in Section~\ref{sec:aux-algs-bip}}
               \State $R \gets v$ \Comment{switch to $[L, v]$}
            \Else
               \State $L \gets v+1$ \Comment{switch to $[v+1, R]$}
            \EndIf
         \EndWhile \Comment{when the loop is over, we have $L = R = \gamma^G_{i,k}$}
         \State $\gamma_{i,k} \gets L$ 
         \State $y \gets \textproc{ComputeProduct}(\sfU.\textproc{Sum}(1+\gamma_{i,k}), \sfa_i
         - k + 1, 1)$  \label{l:s-dec-CP-Ux} \Comment{$y =
           (U^G_{1+\gamma^G_{i,k}}(i))_{\hat{a}^G_i - k + 1}$}
         \Statex \Comment{Algorithm~\ref{alg:compute-product} in Section~\ref{sec:aux-algs-bip}}
         \State $z_i \gets z_i + l_i \times y$ \label{l:s-dec-zi-update}
         \State $c \gets (\sfa_i - k + 1) \times
         \sfa_{\gamma_{i,k}}$ \label{l:sdec-c} \Comment{$c = (\hat{a}^G_i - k +
           1) a^G_{\gamma^G_{i,k}}(i)$}
         \State $l_i \gets l_i \times c$ \label{l:s-dec-li-update} \Comment{updating $l_i$} 
         \State $\sfa_{\gamma_{i,k}} \gets \sfa_{\gamma_{i,k}} -
         1$ \label{l:s-dec-beta-update} \Comment{updating
           $\vsfa$} 
         \State $\sfU.\textproc{Add}(\gamma_{i,k}, -1)$ \label{l:s-dec-U-update}
         \Comment{updating the
         Fenwick tree} 
         \State $\tz \gets \tz - y$ \label{l:s-dec-tz-y} \Comment{subtracting the
           contribution of $\gamma_{i,k}$} 
         \State $\tz \gets   \tz \div c $ \label{l:s-dec-tz-c} \Comment{$\tz$ is
         updated so that it becomes $\tz_{i,k+1}$}
      \EndFor
      \State $N_{i,i} \gets z_i$
      \State \textbf{return} $(N_{i,i}, \vgamma_i, \vsfa, \sfU, l_i)$
      \EndFunction
\end{algorithmic}
\end{myalg}

\subsubsection{Decoding the Adjacency List of an Interval of Nodes}
\label{sec:s-dec-interval}

Now, we present our decompression method for an interval $[i,j]$ when $i < j$.
As we discussed above, we assume that properties~\ref{a:sdec-1-i-1} through
\ref{a:sdec-Sk} holds for the interval $[i,j]$ and our scheme is guaranteed to
find the forward adjacency list of vertices $i \leq v \leq j$ as well as
$N_{i,j}(G)$ and $l^G_{i,j}$. Similar to what we did in
Section~\ref{sec:bipartite-decode}, we propose a recursive scheme which first
tries to decode the subinterval $[i,k]$, and the subinterval $[k+1, j]$, where
\begin{equation}
  \label{eq:s-dec-k-def}
  k =
  \begin{cases}
    \lfloor  (i+j)/2 \rfloor & \text{if } j -i + 1 > \lfloor  \log_2 \pn \rfloor^2 \\
     i   & \text{o.t.w.}
  \end{cases}
\end{equation}
The reason for this bifurcation is that we need to guarantee that \ref{a:sdec-Sj} holds for both the intervals $[i,k]$ and $[k+1, j]$.
This assumption automatically holds for the interval $[k+1,j]$ since $S^G_{j+1}$
is known due the \ref{a:sdec-Sj} for $[i,j]$. However, due to
\ref{a:sdec-Sk}, with $k = \lfloor (i+j)/2 \rfloor$, $S^G_{k+1}$ is
known only when $j -i + 1 > \lfloor  \log_2 \pn  \rfloor^2$. On the other hand, if $j - i + 1 \leq \lfloor \log_2 \pn
\rfloor^2$, we can compute $S^G_{k+1}$ with $k = i$ using $S^G_{i+1} = S^G_i -
2\hat{a}^G_i = U^G_i(i) - 2 \hat{a}^G_i$. Note that due to~\ref{a:sdec-1-i-1},
the decoder has access to both $U^G_i(i)$ and $\hat{a}^G_i = a^G_i(i)$, hence
can compute $S^G_{i+1}$.

Note that we initially start decoding the interval $[1,\pn]$.
Therefore, due to the above choice
of $k$, the intervals $[i,j]$ that show up in the decompression procedure, as
long as $j -i + 1 > \lfloor  \log_2 \pn \rfloor^2$, are in one to one
correspondence with those intervals in the compression procedure of Algorithm~\ref{alg:ComputeNij_graph} in
Section~\ref{sec:self-encode-alg}.

With the above setup, here are the steps we take to decode the interval $[i,j]$:
\begin{enumerate}
\item With $k$ as in~\eqref{eq:s-dec-k-def}, we define $\tN_{i,k}$
  appropriately and verify that
  properties~\ref{a:sdec-1-i-1}--\ref{a:sdec-Sk} hold for the interval
  $[i,k]$. \label{step:sdec-tNik}
\item We assume that our procedure correctly decodes for the interval $[i,k]$
  and returns $N_{i,k}(G)$ together with $l^G_{i,k}$. \label{step:sdec-ik}
\item We define $\tN_{k+1, j}$ appropriately and verify that the
  properties~\ref{a:sdec-1-i-1}--\ref{a:sdec-Sk} hold for the interval $[k+1,
  j]$. \label{step:sdec-tNkj}
\item We assume that our procedure correctly decodes for the interval $[k+1,j]$
  and returns $N_{k+1,j}(G)$ together with $l^G_{k+1,j}$. \label{step:sdec-kj}
\item We put the above together to find the forward adjacency list of vertices
  in the interval $[i,j]$ as well as $N_{i,j}(G)$ and $l^G_{i,j}$.  \label{step:sdec-final}
\end{enumerate}

This way, our claim on the ability of the algorithm to decode the forward
adjacency list of vertices in $[i,j]$ as well as finding $N_{i,j}(G)$ and
$l^G_{i,j}$ is established using an induction on the size of the interval, with
the base case being  $i=j$, as was shown in Algorithm~\ref{alg:sdec-i}
above. On the other hand, the claim that the
properties~\ref{a:sdec-1-i-1}--\ref{a:sdec-Sk} hold throughout the decoding
procedure is proved inductively, with the base case being $i=1, j= \pn$ as was
discussed above.

\underline{\textbf{Step~\ref{step:sdec-tNik}:}} For an interval $[i,j]$ and $k$ as
in~\eqref{eq:s-dec-k-def}, in order to appropriately define $\tN_{i,k}$, due
to~\ref{a:sdec-N}, we may treat $\tN_{i,j}$ as a proxy for $N_{i,j}(G)$. Then,
motivated by~\eqref{eq:self-Nij-recursive}, we define
\begin{equation}
  \label{eq:sdec-tNik-def}
  \tN_{i,k} := \left\lfloor \frac{\tN_{i,j}}{r^G_{k+1, j}}\right \rfloor.
\end{equation}
However, in order to find $\tN_{i,k}$ at the decoder, we first need to find
$r^G_{k+1, j} = \frac{(S^G_{k+1}-1)!!}{(S^G_{j+1}-1)!!}$. Note that we know $S^G_{j+1}$
from~\ref{a:sdec-Sj}. Moreover, if $j-i +1> \lfloor \log_2 \pn  \rfloor^2$, we
also know $S^G_{k+1}$ due to \ref{a:sdec-Sk}. On the other hand, if $j -i +1 \leq
\lfloor \log_2n \rfloor^2$, we have $k=i$ and from~\eqref{eq:self-Si+1-Si}, we
have 
\begin{equation}
  \label{eq:s-dec-k-i+1-Sk+1}
  S^G_{k+1} = S^G_i - 2\hat{a}^G_i =  \left( \sum_{v=i}^{\pn} a^G_v(i) \right) - 2a^G_i(i) = U^G_i(i) - 2a^G_i(i).
\end{equation}
But both $U^G_v(i)$ and $a^G_v(i)$ are known due to \ref{a:sdec-1-i-1}. Therefore, we
can compute $r^G_{k+1, j}$ at the decoder and find $\tN_{i,k}$. Now, we verify
that properties~\ref{a:sdec-1-i-1}--\ref{a:sdec-Sk} holds for the interval
$[i,k]$. Property~\ref{a:sdec-1-i-1} holds since $i$ is fixed. In order to
check~\ref{a:sdec-N}, we use Lemma~\ref{lem:N-N1-N2} with $N \equiv N_{i,j}(G)$,
$N_1 \equiv N_{i,k}(G)$, $N_2 \equiv N_{k+1, j}(G)$, $l_1 \equiv l^G_{i,k}$, $l_2
\equiv l^G_{k+1, j}$, $\tN \equiv \tN_{i,j}$, and $r \equiv r^G_{k+1, j}$. The assumptions of the lemma
hold due to~\eqref{eq:self-Nij-recursive}, Lemma~\ref{lem:self-N+l<r} for the
interval $[k+1, j]$, and \ref{a:sdec-N} for the interval $[i,j]$. Thus,
from~\eqref{eq:lem-NN1N2l1l2r--1}, we have
\begin{equation*}
  N_{i,k}(G) \leq \tN_{i,k} < N_{i,k}(G) + l^G_{i,k},
\end{equation*}
which is precisely \ref{a:sdec-N} for the interval $[i,k]$. Property~\ref{a:sdec-Sj} also holds, since as we discussed above, we can compute
$S^G_{k+1}$ at the decoder. In order to verify~\ref{a:sdec-Sk}, note that as we
discussed above, due to
our choice of $k$ in~\eqref{eq:s-dec-k-def}, since we initially start with the
interval $[1,\pn]$, the intervals $[i,j]$ that appear in the decoding procedure,
as long as $j - i + 1 > \lfloor  \log_2n \rfloor^2$, are in a one to one
correspondence with the intervals that show up in the compression scheme of
Section~\ref{sec:self-encode}. Hence, if we use the same indexing scheme for the
intervals as the one we used for compression, when $j - i + 1 > \lfloor \log_2n
\rfloor^2$, $S^G_{k+1}$ is precisely $\tilde{f}_I$, where $I$ is the index corresponding
to the interval $[i,j]$ and $\tilde{f}_I = \tilde{f}^{\pn}_{\va, I}(G)$ is the
$I$--th coordinate of the $\vec{\tilde{f}}^{\pn}_{\va}(G)$ array given to the
decoder.
This argument in fact shows that  property~\ref{a:sdec-Sk}
holds for all the intervals that show up in the decompression procedure,
including the subintervals $[i,k]$ and $[k+1, j]$ here. Note that if $j-i+1 \leq
\lfloor  \log_2\pn \rfloor^2$, then $k - i + 1 \leq \lfloor  \log_2\pn \rfloor^2$
and $j - k \leq  \lfloor  \log_2n \rfloor^2$ and \ref{a:sdec-Sk}
automatically holds for both $[i,k]$ and $[k+1, j]$.

\underline{\textbf{Step~\ref{step:sdec-ik}:}} We call the decoding procedure for
the interval $[i,k]$ and assume that it finds the forward adjacency list of
vertices in the interval $[i,k]$ as well as $N_{i,k}(G)$ and $l^G_{i,k}$.

\underline{\textbf{Step~\ref{step:sdec-tNkj}:}} In order to define $\tN_{k+1,
  j}$, similar to step~\ref{step:sdec-tNik}, treating $\tN_{i,j}$ as a proxy for
$N_{i,j}$ and recalling~\eqref{eq:self-Nij-recursive}, we define
\begin{equation*}
  \tN_{k+1, j} := \left\lfloor \frac{\tN_{i,j} - N_{i,k}(G) r^G_{k+1, j}}{l^G_{i,k}}\right \rfloor.
\end{equation*}
Recall that $N_{i,k}(G)$ and $l^G_{i,k}$ are known from Step~\ref{step:sdec-ik}.
Also, $r^G_{k+1, j}$ was computed in step~\ref{step:sdec-tNik}. Thereby,
$\tN_{k+1, j}$ can be computed at the decoder. Now, we verify that
properties~\ref{a:sdec-1-i-1}--\ref{a:sdec-Sk} hold for the interval $[k+1,
j]$. Note that property~\ref{a:sdec-1-i-1} holds for $[k+1, j]$ due to the
facts that 
property~\ref{a:sdec-1-i-1} holds for the interval $[i,j]$ and  the
forward adjacency list of vertices in $[i,k]$ were decoded in
step~\ref{step:sdec-ik}. Moreover, if we use Lemma~\ref{lem:N-N1-N2} as in
step~\ref{step:sdec-tNik}, we conclude from~\eqref{eq:lem-NN1N2l1l2r--2} that
\begin{equation*}
  N_{k+1, j}(G) \leq \tN_{k+1, j} < N_{k+1, j}(G) + l^G_{k+1, j},
\end{equation*}
which establishes \ref{a:sdec-N} for the interval $[k+1, j]$. Property~\ref{a:sdec-Sj} also holds since $S^G_{j+1}$ is known due to
property~\ref{a:sdec-Sj} for the interval $[i,j]$. Finally, as we discussed in
step~\ref{step:sdec-tNik}, property~\ref{a:sdec-Sk} holds for all the
interval in the decoding process.

\underline{\textbf{Step~\ref{step:sdec-kj}:}} We call the decoding procedure for
the interval $[k+1,j]$ and assume that it find the forward adjacency list of
vertices in the interval $[k+1,j]$ as well as $N_{k+1,j}(G)$ and $l^G_{k+1,j}$.

\underline{\textbf{Step~\ref{step:sdec-final}:}} Putting the decoded forward
adjacency lists of vertices in the intervals $[i,k]$ and $[k+1, j]$ from
steps~\ref{step:sdec-ik} and \ref{step:sdec-kj} above, respectively, we can form
the forward adjacency list of vertices in the interval $[i,j]$ as desired.
Moreover, using~\eqref{eq:self-Nij-recursive}, we can compute
$N_{i,j}(G)$. Note that at this point, all the values on the right hand side of~\eqref{eq:self-Nij-recursive}
are known to the decoder, i.e.\ $N_{i,k}(G)$ and $l^G_{i,k}$ from
step~\ref{step:sdec-ik}, $N_{k+1, j}(G)$ from step~\ref{step:sdec-kj} and
$r^G_{k+1, j}$ from step~\ref{step:sdec-tNik}. Finally, we find $l^G_{i,j} = l^G_{i,k}
l^G_{k+1, j}$ using the two known values $l^G_{i,k}$ and $l^G_{k+1, j}$ obtained from
steps~\ref{step:sdec-ik} and \ref{step:sdec-kj}, respectively.

Algorithm~\ref{alg:sdec-ij} below illustrates the above steps. In this
algorithm, we use a Fenwick tree $\sfU$ to keep track of the partial sums $U^G_v(i)$.
More specifically, we assume that when the algorithm is called for an interval
$[i,j]$, we have
\begin{equation*}
  \sfU.\textproc{Sum}(v) = U^G_v(i) = \sum_{w=v}^{\pn} a^G_w(i) \qquad \qquad \text{for } i \leq v \leq \pn.
\end{equation*}
On the other hand, we assume that an array $\vsfa = (\sfa_v: 1 \leq v \leq \pn)$
of integers is given so that when the
algorithm is called for the interval $[i,j]$, we have
\begin{equation*}
  \sfa_v = a^G_v(i) \qquad \qquad \text{for } i \leq v \leq \pn.
\end{equation*}
We update $\sfU$ and $\vsfa$ when we call the decoding procedure for the two
subintervals $[i,k]$ and $[k+1, j]$ in lines~\ref{l:s-dec-ij-call-ik}
and~\ref{l:s-dec-ij-call-kj}, respectively. 

As we discussed above, an indexing similar to that of the compression scheme is
used, where the variable $I$ keeps track of the index of the current interval.
Also, we pass indices $2I$ and $2I+1$ to the processes corresponding to
intervals $[i,k]$ and $[k+1, j]$ in lines~\ref{l:s-dec-ij-call-ik}
and~\ref{l:s-dec-ij-call-kj}, respectively. With this, in case $j-i+1 > \lfloor
\log_2 \pn \rfloor^2$, as was discussed above, $S^G_{k+1}$ is set to $\tilde{f}_I$ in
line~\ref{l:s-dec-ij-tS-I}. Otherwise, we have $k = i + 1$ and using \eqref{eq:s-dec-k-i+1-Sk+1},
$S^G_{k+1}$ is set to $U^G_i(i) - 2 a^G_i(i) = \sfU.\textproc{Sum}(i) - 2\sfa_i$
in line~\ref{l:s-dec-ij-Ui-betai}. In
line~\ref{l:s-dec-ij-rkj}, we compute $r_{k+1,j} = r^G_{k+1, j}$
using~\eqref{eq:s-enc-rk+1-j-CP} and $\textproc{ComputeProduct}$ of 
Algorithm~\ref{alg:compute-product}.


\begin{myalg}[Decoding the forward adjacency list of vertices $i \leq v \leq j$
 given $\tN_{i,j}$ for a simple unmarked graph $G \in \mGnt_{\va}$ \label{alg:sdec-ij}]
\begin{algorithmic} [1]                   
  \INPUT
  \Statex $\pn$: number of nodes in the graph 
   \Statex $1 \leq i \leq j \leq \pn$: the interval to be decoded
   \Statex $\tN_{i,j}$: integer satisfying $N_{i,j}(G) \leq \tN_{i,j} <
   N_{i,j}(G) + l^G_{i,j}$
   \Statex $\vsfa = (\sfa_v: 1 \leq v \leq \pn)$ where $\sfa_v = a^G_v(i)$ for $i \leq v
   \leq \pn$
   \Statex $\sfU$: Fenwick tree, where for $i \leq v \leq \pn$,
   $\sfU.\textproc{Sum}(v) =U^G_v(i) =  \sum_{k = v}^{\pn} a^G_k(i)$
   \Statex $I$: an integer specifying the interval $[i,j]$
   \Statex $S_{j+1}$: which is $S^G_{j+1} = \sum_{k=j+1}^{\pn} a^G_k(j+1)$
   \Statex $\vec{\tilde{f}} = (\tilde{f}_p: 0 \leq p \leq \lfloor 16\pn / \log^2 \pn \rfloor)$: array of integers
   where if $j - i + 1 > \lfloor  \log_2 \pn
   \rfloor^2$,  $\tilde{f}_I = S^G_{k+1}$  with  $k = \lfloor (i+j)/2 \rfloor$    and $I$ being the index corresponding to the  interval $[i,j]$ as above
   \OUTPUT
   \Statex $N_{i,j} = N_{i,j}(G)$.
   \Statex $\vgamma_i, \dots, \vgamma_j$: forward adjacency list of vertices in
   the interval $[i,j]$, such that $\vgamma_v = (\gamma_{v,k}: 1 \leq k \leq
   \hat{a}^G_v)$ where $\gamma_{v,k} = \gamma^G_{v,k}$ for $i \leq v \leq j$
   and $1 \leq k \leq \hat{a}^G_v$
    \Statex $\vsfa = (\sfa_v: 1 \leq v \leq \pn)$: array $\vsfa$ updated, so that
   $\sfa_v = a^G_v(j+1)$ for $j+1 \leq v \leq \pn$.
   \Statex $\sfU$: Fenwick tree $U$ updated, so that for $j +1 \leq  v\leq \pn$, we
   have $\sfU.\textproc{Sum}(v) = U^G_v(j+1) =
   \sum_{k=v}^{\pn} a^G_k(j+1)$.
   \Statex $l_{i,j} = l^G_{i,j}$.
   \Function{DecodeInterval}{$\pn, i,j,\tN_{i,j}, \vsfa, \sfU, I, S_{j+1}, \vec{\tilde{f}}$}
   \If {$i=j$}
   \State $(N_{i,i} , \vgamma_i, \vsfa, \sfU, l_i) \gets
   \textproc{DecodeNode}(\pn, i,\tN_{i,i}, \vsfa, U)$
   \Comment{Algorithm~\ref{alg:sdec-i} in Section~\ref{sec:self-decode-one-node}}
   \State \textbf{return} $(N_{i,i} , \vgamma_i, \vsfa, \sfU, l_i)$
   \EndIf
      \If{$j - i+1 > \lfloor  \log_2 \pn \rfloor^2$} \Comment{specifying the midpoint $k$ using~\eqref{eq:s-dec-k-def}}
         \State $k \gets  (i+j) \div 2 $ 
         \State $S_{k+1} \gets \tilde{f}_I$ \label{l:s-dec-ij-tS-I}
      \Else
         \State $k \gets i$
         \State $S_{k+1} \gets \sfU.\textproc{Sum}(i)  - 2 \sfa_i$  \label{l:s-dec-ij-Ui-betai}
      \EndIf
      \State $r_{k+1,j} \gets \textproc{ComputeProduct}(S_{k+1}-1, (S_{k+1} -
      S_{j+1}) / 2, 2)$ \label{l:s-dec-ij-rkj}
      \Comment{Algorithm~\ref{alg:compute-product} in Section~\ref{sec:aux-algs-bip}}
      \Statex \Comment{finding $\tN_{i,k}$ for the left interval $[i,k]$ and decoding $[i,k]$:}
      \State $\tN_{i,k} \gets \tN_{i,j} \div r_{k+1,j}$    \label{l:s-dec-ij-tNik}
      \State $(N_{i,k}, \vgamma_{[i:k]}, \vsfa, \sfU, l_{i,k}) \gets
      \textproc{DecodeInterval}(\pn, i,k,\tN_{i,k}, \vsfa, \sfU, 2I, S_{k+1},
      \vec{\tilde{f}})$ \label{l:s-dec-ij-call-ik}
      \Statex \Comment{finding $\tN_{k+1, j}$ for the right interval $[k+1, j]$
        and decoding $[k+1, j]$:}
      \State $\tN_{k+1, j} \gets (\tN_{i,j} -  N_{i,k} \times r_{k+1,j}) \div
      l_{i,k}$  \label{l:s-dec-ij-tNkj}
      \State $(N_{k+1, j}, \vgamma_{[k+1:j]}, \vsfa, \sfU, l_{k+1,j}) \gets
      \textproc{DecodeInterval}(\pn, k+1, j, \tN_{k+1, j}, \vsfa, \sfU, 2I+1, S_{j+1},
      \vec{\tilde{f}})$ \label{l:s-dec-ij-call-kj}
      \State $N_{i,j} \gets N_{i,k} \times r_{k+1,j} + l_{i,k} \times N_{k+1,
        j}$ \label{l:s-dec-ij-Nij}
      \State $l_{i,j} \gets l_{i,k} \times l_{k+1, j}$ \label{l:s-dec-ij-lij}
      \State \textbf{return} $(N_{i,j}, \vgamma_{[i:j]}, \vsfa, \sfU, l_{i,j})$
   \EndFunction
 \end{algorithmic}
\end{myalg}

\subsubsection{The Main Decoding Algorithm}
\label{sec:self-main-dec}

In this section, we present the main decoding algorithm which decodes for $G \in
\mGnt_{\va}$ given $\fnt_{\va}(G)$ and $\vec{\tilde{f}}^{\pn}_{\va}(G)$. In
order to do so, we first compute $\tN_{1, \pn}$
using~\eqref{eq:s-dec-tN-1-pn-def} and then call Algorithm~\ref{alg:sdec-ij}
above. This is illustrated in Algorithm~\ref{alg:s-dec-main} below. In order to
do so, we initialize the Fenwick tree $\sfU$ with the array $\va$, since we have
$U^G_v(i) = \sum_{k=v}^{\pn} a_v$ for $1 \leq v \leq \pn$. Moreover, since
$a^G_v(1) = a_v$ for $1 \leq v \leq \pn$, we initialize $\vsfa$ with $\va$.
Also, in order to compute $\tN_{1, \pn} = \fn_{\va}(G) \times \prod_{v=1}^{\pn}
a_v!$, we use $\textproc{ProdFactorial}$ of
Algorithm~\ref{alg:comp-ProdAFactorial} to compute $\prod_{v=1}^{\pn} a_v!$.

\begin{rem}
  \label{rem:s-dec-U-a-pinter}
  Similar to the discussion in Remark~\ref{rem:senc-pointer}, we can store the
  Fenwick tree $\sfU$ as well as the arrays $\vsfa$ and $\vec{\tilde{f}}$ in one location of the memory and
  pass the pointer of this location to the decoding algorithm. This way, we can
  modify this location of the memory and avoid copying unnecessary data. Similar
  to Remark~\ref{rem:senc-pointer}, the memory required to store $\sfU$,
  $\vec{\tilde{f}}$ and $\vsfa$ is $O(\pn \log \pn)$ bits. 
\end{rem}

\begin{myalg}[Decoding for a simple unmarked graph $G \in \mGnt_{\va}$ given
  $\fnt_{\va}(G)$ and $\vec{\tilde{f}}^{(\pn)}_{\va}(G)$ \label{alg:s-dec-main}]
  \begin{algorithmic}[1]
  \INPUT
  \Statex $f$: integer, which is $\fnt_{\va}(G)$ for the target graph $G \in
  \mGnt_{\va}$ which was given to encoder during the compression phase
  \Statex $\vec{\tilde{f}} = (\tilde{f}_i: 1 \leq i \leq \lfloor  16 \pn /
  \log^2 \pn 
  \rfloor)$: $\Array$ of integers, where $\tilde{f}_i = \tilde{f}^{(\pn)}_{\va,
    i}(G)$ for $1 \leq i \leq \lfloor 16 \pn / \log^2 \pn \rfloor$
  \Statex $\va$: array of vertex degrees
  \OUTPUT
  \Statex $\vgamma_{[1:\pn]}$: the decoded forward adjacency list such that $\vgamma_v =
  \vgamma^G_v = (v < \gamma^G_{v,1} < \dots < \gamma^G_{v, \hat{a}^G_v})$ for $1
  \leq v \leq \pn$
  \Function{GraphDecode}{$f, \vec{\tilde{f}}, \va$}
  \State $\pn \gets \textproc{Size}(\va)$
  \State $c \gets \textproc{ProdFactorial}(\va, 1, \pn)$
  \Comment{Algorithm~\ref{alg:comp-ProdAFactorial} in
    Section~\ref{sec:aux-algs-bip}} \label{l:s-dec-main-c}
  \State $\tN_{1, \pn} \gets f \times c$ \label{l:s-dec-main-fc}
  \State $\sfU \gets$ Fenwick tree initialized with array $\va$ \label{l:s-dec-main-U}
  \State $\vsfa \gets \va$ \Comment{$a^G_v(1) = a_v$ for $1 \leq v \leq \pn$} \label{l:s-dec-main-va}
  \State $(N_{1, \pn}, \vgamma_{[1:\pn]}, \vsfa, \sfU, l_{1, \pn}) \gets
  \textproc{DecodeInterval}(\pn, 1, \pn, \tN_{1, \pn}, \vsfa, \sfU, 0,
  \vec{\tilde{f}})$ \label{l:s-dec-main-call}
  \Statex \Comment{Algorithm~\ref{alg:sdec-ij} in Section~\ref{sec:s-dec-interval}}
  \State \textbf{return} $\vgamma_{[1:\pn]}$
    \EndFunction
  \end{algorithmic}
\end{myalg}

\subsection{Complexity of Decompression}
\label{sec:sdec-complexity}

In this section, we analyze the complexity of decoding for a simple unmarked
graph $G \in \mGnt_{\va}$ given $\fnt_{\va}(G)$ and
$\vec{\tilde{f}}^{(\pn)}_{\va}(G)$. First, we start with the complexity of
decoding the forward adjacency list of one node in Algorithm~\ref{alg:sdec-i}.

\begin{lem}
  \label{lem:sdec-i-complexity}
  Assume that we have $a_v \leq \delta$ for $1 \leq v \leq
  \pn$. Also, assume that the input $\tN_{i,i}$ given to
  Algorithm~\ref{alg:sdec-i} satisfies $N_{i,i}(G) \leq \tN_{i,i} < N_{i,i}(G) +
  l^G_i$ for the graph $G \in \mGnt_{\va}$ that was given to the encoder during
  the compression phase. Then, the time complexity of decoding for the forward
  adjacency list of vertex $i$ is $O(\delta \log \delta \log^3 \pn \log \log \pn)$. 
Moreover, the memory required to perform this
  algorithm, excluding the memory required to store the Fenwick tree $\sfU$ and the
  array $\vsfa$ is $O(\delta \log \pn)$. 
\end{lem}

\begin{proof}
  Note that the binary search of line~\ref{l:s-dec-while-g-f} has $O(\log \pn)$
  iterations. Moreover, similar to our calculations in
  Lemma~\ref{lem:s-compute-N-complexity} for line~\ref{line:sNij-i-y} of
  Algorithm~\ref{alg:ComputeNij_graph}, finding
  $\textproc{ComputeProduct}(\sfU.\textproc{Sum}(1+v), \sfa_i - k + 1, 1)$ in
  line~\ref{l:s-dec-CP-Uv} takes $O(\delta \log \delta \log^2 \pn \log \log \pn)$
  time, requires $O(\delta \log \pn)$ bits of memory, and the result has
  $O(\delta \log \pn)$ bits.
    On the other
  hand, the variable $\tz$ is initialized with $\tN_{i,i}$ and keeps decreasing
  in lines~\ref{l:s-dec-tz-y} and \ref{l:s-dec-tz-c}. Thereby, from
  property~\ref{a:sdec-N} and Lemma~\ref{lem:self-N+l<r}, the variable $\tz$
  remains bounded by $r^G_i$ throughout the loop.
  Therefore, using Lemma~\ref{lem:self-r-l-N-bit-count}, $r^G_i$ has $O(\delta
  \log \pn)$ bits, and hence $\tz$ has $O(\delta \log \pn)$ bits inside  the
  loop. This implies
  that the comparison in line~\ref{l:s-dec-CP-Uv} takes $O(\delta \log \pn)$
  time.
  Thus,
  the entire loop of line~\ref{l:s-dec-while-g-f} runs in $O(\delta \log \delta
  \log^3 \pn \log \log \pn)$ time and
  requires $O(\delta \log \pn)$ bits of memory.

  Note that the calculations in lines~\ref{l:s-dec-CP-Ux} through
  \ref{l:s-dec-U-update} are identical to those in lines~\ref{line:sNij-i-y}
  through~\ref{l:sNij-i-U} of Algorithm~\ref{alg:ComputeNij_graph}.
  Consequently, following the calculations in the proof of
  Lemma~\ref{lem:s-compute-N-complexity}, running lines~\ref{l:s-dec-CP-Ux} through
  \ref{l:s-dec-U-update} takes $O(\delta \log \delta \log^2 \pn \log \log \pn)$ time
  and requires $O(\delta \log \pn)$ bits of memory.
  Also, the variable $y$ and $c$ have
  $O(\delta \log \pn)$ and $O(\log \delta)$ bits, respectively. As a result, the subtraction in
  line~\ref{l:s-dec-tz-y} takes $O(\delta \log \pn)$ time. Moreover, the division in
  line~\ref{l:s-dec-tz-c} takes $\Div(\delta \log \pn) = O(\delta \log^2 \pn
  \log \log \pn)$ time.

  Putting all the above together, the overall time complexity of the algorithm
  is $O(\delta \log \delta \log^3 \pn \log \log \pn)$. Furthermore, its memory complexity, excluding
  the space required to store $\sfU$ and $\vsfa$, is $O(\delta \log \pn)$.
\end{proof}

\begin{lem}
  \label{lem:s-dec-ij-complexity}
  Assume that we have $a_v \leq \delta$ for $1 \leq v \leq \pn$. Also, assume
  that the input $\tN_{1, \pn}$ to Algorithm~\ref{alg:sdec-ij} satisfies
  $N_{1,\pn}(G) \leq \tN_{1, \pn} < N_{1, \pn}(G) + l^G_{1, \pn}$ for the graph
  $G \in \mGnt_{\va}$ that was given to the encoder during the compression
  phase. Moreover, assume that the array $\vec{\tilde{f}}$ given to
  Algorithm~\ref{alg:sdec-ij} is precisely  $\vec{\tilde{f}}^{(\pn)}_{\va}(G)$
  with the same $G$.
  Then, the time complexity of decoding for the forward adjacency list of
  vertices $1, \dots, \pn$ is $O(\delta \pn \log^5 \pn \log \log \pn)$ and the memory required to perform this is
  $O(\pn \delta  \log \pn)$ bits. 
\end{lem}

\begin{proof}
  Recall from our above inductive argument that when Algorithm~\ref{alg:sdec-ij}
  is called for an interval $[i,j]$, $1 \leq i \leq j \leq \pn$,
  properties~\ref{a:sdec-1-i-1}--~\ref{a:sdec-Sk} hold. In particular, if $i =
  j$, we have $N_{i,i}(G) \leq \tN_{i,i} < N_{i,i}(G) + l^G_i$, and from
  Lemma~\ref{lem:sdec-i-complexity} above, the time and memory complexities of decoding for the
  forward adjacency list of vertex $i$ are $O(\delta \log \delta \log^3 \pn \log
  \log n)$ and $O(\delta \log \pn)$, respectively.

  Now, assume that Algorithm~\ref{alg:sdec-ij} is called for an interval $[i,j]$
  with $i < j$. when $j - i + 1 \leq
\lfloor  \log_2 \pn \rfloor^2$, the time complexity of finding
$\sfU.\textproc{Sum}(i)$ in line~\ref{l:s-dec-ij-Ui-betai} is $O(\log^2 \pn)$, since $\sfU$
keeps track of partial sums each 
with $O(\log \pn)$ bits.

Note that the computation in line~\ref{l:s-dec-ij-rkj} is similar to that of
line~\ref{l:sNij-rkj} of Algorithm~\ref{alg:ComputeNij_graph}. More
specifically, $S^G_{k+1} - 1$ has $O(\log \pn)$ bits in both cases $k= i $ or $k =
\lfloor (i+j)/2 \rfloor$. Thereby, from the discussion in
Lemma~\ref{lem:s-compute-N-complexity}, the calculation in
line~\ref{l:s-dec-ij-rkj} has time complexity 
takes $O((j-i+1) \delta \log^3 \pn \log \log \pn)$ and memory complexity
$O((j-i+1) \delta \log \pn)$.
Moreover, from Lemma~\ref{lem:self-r-l-N-bit-count}, the result $r_{k+1, j} =
r^G_{k+1, j}$ has $O((j-i+1)\delta \log \pn)$ bits. Likewise, $r_{i,k} = r^G_{i,k}$
and $r_{i,j} = r^G_{i,j}$ have $O((j-i+1) \delta \log \pn)$ bits.
Furthermore, using $\tN_{i,j} \leq N_{i,j}(G) + l^G_{i,j}$ together with
Lemma~\ref{lem:self-N+l<r} and Lemma~\ref{lem:self-r-l-N-bit-count}, $\tN_{i,j}$
has $O((j-i+1) \delta \log \pn)$ bits. Since $r_{k+1, j}$ also has $O((j-i+1)\delta \log \pn)$ bits, the time
complexity of division in line~\ref{l:s-dec-ij-tNik} is $\Div((j-i+1)\delta \log \pn ) =
O((j-i+1)\delta \log^2 \pn \log \log \pn)$.

Note that implied by Lemma~\ref{lem:self-r-l-N-bit-count}, the variables
$N_{i,k} = N_{i,k}(G)$ and $l_{i,k} = l^G_{i,k}$ returned in
line~\ref{l:s-dec-ij-call-ik} have $O((j-i+1) \delta \log \pn)$ bits. This
together with the fact that $r_{k+1, j}$ also has $O((j-i+1)\delta \log \pn)$
bits implies that the time complexity of the arithmetic in
line~\ref{l:s-dec-ij-tNkj} is $\Mul((j-i+1)\delta \log \pn) + \Div((j-i+1)\delta
\log \pn) = O((j-i+1) \delta \log^2 \pn \log \log \pn)$. 
Another usage of Lemma~\ref{lem:self-r-l-N-bit-count} implies that the variables
$N_{k+1, j} = N_{k+1, j}(G)$ and $l_{k+1, j} = l^G_{k+1, j}$ both have
$O((j-i+1) \delta \log \pn)$ bits, and hence the time complexity of the arithmetic in lines
\ref{l:s-dec-ij-Nij} and \ref{l:s-dec-ij-lij} is $\Mul((j-i+1) \delta \log \pn)
= O((j-i+1) \delta \log^2 \pn \log \log \pn)$.

Putting all the above together, we realize that if $f(m)$ denotes the time
complexity of the algorithm when $j-i + 1 = m$, for $m > 1$, we have
\begin{equation}
  \label{eq:s-dec-ij-complexity-fm-recursive}
  f(m) =
  \begin{cases}
    f\left( \left\lfloor \frac{m}{2} \right \rfloor \right) + f\left( \left\lceil \frac{m}{2}\right \rceil \right) + O(m \delta \log^3\pn \log \log \pn) & \text{if } m > \lfloor \log_2 \pn \rfloor^2 \\
f(1) + f(m-1) + O(m \delta \log^3 \pn \log \log \pn) & \text{otherwise}.
  \end{cases}
\end{equation}
Moreover, from Lemma~\ref{lem:sdec-i-complexity}, we have 
\begin{equation*}
  f(1) = O(\delta \log \delta \log^3 \pn \log \log \pn).
\end{equation*}
Solving for this, we realize that if $m \leq \lfloor  \log_2 \pn \rfloor^2$, we have 
\begin{equation}
  \label{eq:s-dec-ij-complexity-fm<logn}
  f(m) = mf(1) + O(m^2 \delta \log^3 \pn \log \log \pn) = O(\delta \log^7 \pn \log \log \pn). 
\end{equation}
Moreover, we get
\begin{align*}
  f(\pn) &= O(\delta \pn \log^4 \pn \log \log \pn) + O\left(  \frac{\pn}{\log^2 \pn} \delta \log^7 \pn \log \log \pn \right) \\
         &= O(\delta \pn \log^5 \pn \log \log \pn).
\end{align*}

In order to address the memory complexity of the algorithm, recalling
Remark~\ref{rem:s-dec-U-a-pinter}, we store $\sfU$, $\vsfa$ and
$\vec{\tilde{f}}$ using $O(\pn \log \pn)$ bits in one location of the memory.
Therefore, if $g(m)$ denotes the memory complexity of the algorithm when $j - i
+ 1 = m$, excluding the memory required to store $\sfU$, $\vsfa$ and
$\vec{\tilde{f}}$, the overall memory complexity of the algorithm is $g(\pn) +
O(\pn \log \pn)$. On the other hand, motivated by our discussion above, we have
\begin{equation*}
  g(m) =
  \begin{cases}
    \max \left\{  g\left( \left\lfloor \frac{m}{2}\right \rfloor \right) , g\left( \left\lceil  \frac{m}{2}\right \rceil \right) \right\} + O(m \delta \log \pn) & m > \lfloor  \log_2 \pn  \rfloor^2 \\
    \max \left\{ g(1), g(m-1) \right\} + O(m \delta \log \pn) & \text{otherwise}.
  \end{cases}
\end{equation*}
On the other hand, from Lemma~\ref{lem:sdec-i-complexity}, we have $g(1)
= O(\delta \log \pn)$. Solving for this, for $m \leq \lfloor \log_2 \pn
\rfloor^2$, we have $g(m) = O(m^2 \delta \log \pn) =
O(\delta \log^5 \pn)$. Furthermore, we have $g(\pn) =
O(\delta \pn \log \pn) + O(\delta \log^5 \pn) = O(\delta \pn \log \pn)$. This
completes the proof. 
\end{proof}


{
\begin{lem}
  \label{lem:self-dec-main-decode-complexity}
  Assume that we have $a_v \leq \delta$ for $1 \leq v \leq \pn$. Also, assume
  that the input $f$ to Algorithm~\ref{alg:s-dec-main} satisfies
  $f = \fnt_{\va}(G)$ for the graph
  $G \in \mGnt_{\va}$ that was given to the encoder during the compression
  phase. Moreover, assume that the array $\vec{\tilde{f}}$ given to
  Algorithm~\ref{alg:s-dec-main} is precisely  $\vec{\tilde{f}}^{(\pn)}_{\va}(G)$
  with the same $G$. Then, the time complexity of decoding for the forward
  adjacency list of $G$ in Algorithm~\ref{alg:s-dec-main}  is $O(\delta \pn \log^5 \pn \log \log \pn)$ and the memory required to perform this is
  $O(\pn \delta  \log \pn)$ bits. 
\end{lem}

\begin{proof}
Using Lemma~\ref{lem:prod-a-factorial-complexity-each-delta}, the time and
memory complexity of finding $c$ in line~\ref{l:s-dec-main-c} are $O(\pn \delta
\log \delta \log^2 \pn \log \log \pn)$ and $O(\pn \delta \log \delta)$,
respectively, and the result $c$ has $O(\pn \delta \log \delta)$ bits.
  Note that $f = \fnt_{\va}(G) = \lceil N(G) / \prod_{i=1}^{\pn} a_i! \rceil$
  for some graph $G$. Therefore, we have $f \leq N(G) = N_{1,\pn}(G)$. Thereby,
  Lemma~\ref{lem:self-r-l-N-bit-count} implies that $f$ has $O(\pn \delta \log
  \delta)$ bits. Hence, the time complexity of finding $f \times c$ in
  line~\ref{l:s-dec-main-fc} is $\Mul(\pn \delta \log \delta) = O(\pn \delta
  \log \delta \log \pn \log \log \pn)$, and the result which is stored in
  $\tN_{1, \pn}$ has
  $O( \pn \delta \log \delta)$ bits. 
On the other hand, note that by assumption, $f = \fnt_{\va}(G) = \lceil
N_{1,\pn}(G) / \prod_{i=1}^{\pn}a_i! \rceil$ and $c = \prod_{i=1}^{\pn} a_i!$.
Also, recall from~\eqref{eq:self-l-is-prod-a-factorial-claim} that
$\prod_{i=1}^{\pn} a_i! = l^G_{1,\pn}$. Consequently, we have
\begin{equation*}
  N_{1,\pn}(G) \leq f \times c < N_{1,\pn}(G) + \prod_{i=1}^{\pn} a_i! = N_{1,\pn}(G) + l^G_{1,\pn}.
\end{equation*}
  Since each $a_i$ for $1 \leq i \leq \pn$ has $O( \log \pn)$ bits, initializing
  the Fenwick tree $\sfU$ with array $\va$ in line~\ref{l:s-dec-main-U} takes $O( \pn \log^2 \pn)$ time and
  the space required to store $\sfU$ is $O( \pn \log \pn)$. 
  Likewise, initializing $\vsfa$ with $\va$ in line~\ref{l:s-dec-main-va} takes
  $O(\pn \log \pn)$ time.
  Furthermore, Lemma~\ref{lem:s-dec-ij-complexity} above implies that calling
  Algorithm~\ref{alg:sdec-ij} in line~\ref{l:s-dec-main-call} has time and
  memory complexity $O(\delta \pn \log^5 \pn \log \log \pn)$ and $O(\pn \delta
  \log \pn)$, respectively. Putting the above together completes the proof.
\end{proof}
}

\begin{rem}
  \label{rem:why-threshold-log2n2}
Here, we justify our choice of the threshold $\lfloor \log_2 \pn \rfloor^2$ for
the compression/decompression schemes in this section.  Let $T_{\pn}$ be the a candidate threshold for this purpose. recall
from the proof of Lemma~\ref{lem:s-dec-ij-complexity} that there are $O(\pn /
T_{\pn})$ many leaf nodes in the complexity tree in which each node corresponds to
an interval and the leaf nodes correspond to the interval with length no more
than $T_{\pn}$. Moreover, due to \eqref{eq:s-dec-ij-complexity-fm<logn}, the
complexity associated to each of these leaf nodes is $O(T_{\pn}^2 \log^4n \log \log
\pn)$. Consequently, the overall contribution of the leaf nodes in the complexity
is
\begin{equation*}
  O\left( \frac{\pn}{T_{\pn}} T_{\pn}^2 \log^4 \pn \log \log \pn  \right)= O(\pn T_{\pn} \log^4 \pn \log \log \pn).
\end{equation*}
From this, we realize that we had better choose $T_{\pn}$ so that it grows at
most 
logarithmically in $\pn$. On the other hand, as the length of the array $\vec{\tilde{f}}$ is
$O(\pn / T_{\pn})$ and each of its elements has $O(\log \pn)$ bits, we need to use $O(\frac{\pn
\log \pn}{ T_{\pn}})$ bits to transmit the array $\vec{\tilde{f}}$ to the decoder. But from the
proof of Theorem~\ref{thm:optimality-complexity-main}, we can not afford this unless $\frac{\pn \log \pn}{
T_{\pn}} = o(\pn)$. Observe that our choice of $T_{\pn} = \lfloor  \log_2 \pn
\rfloor^2$ satisfies both of the above conditions. 
\end{rem}


\appendix

\section{Algorithms  and their dependencies}
\label{sec:app-algo-dependencies}


In this section, we list all the algorithms in this document as well as their
dependencies. Moreover, for each algorithm, we  point to the part of the document where its description
and complexity analysis is given.

{\footnotesize
  \begin{longtable}{@{}lp{5cm}ccr@{}}
\toprule
  Algorithm & Purpose & Description & Complexity analysis & Uses algorithms \\ \midrule
  \ref{alg:graph-encode-new}& encoding a simple marked graph        &  Sec.~\ref{sec:enc-alg}   & Sec.~\ref{sec:enc-complexity}        &  \ref{alg:preprocess}, \ref{alg:type-extract-message-passing-new}, \ref{alg:encode-star-vertices}, \ref{alg:encode-star-edges}, \ref{alg:encode-find-deg}, \ref{alg:encode-vertex-types}, \ref{alg:encode-find-partition-graphs}, \ref{alg:bip-comp}, \ref{alg:self-find-f}              \\
    \ref{alg:preprocess} & preprocessing a simple marked graph to find its equivalent neighbor list representation & Sec.~\ref{sec:detail-preprocessing} & Lemma~\ref{lem:preprocess-complexity} in Sec.~\ref{sec:detail-preprocessing}& \\
    \ref{alg:compress-sequence} & compressing an array consisting of nonnegative integers & Sec.~\ref{sec:compress-sequence} & Sec.~\ref{sec:compress-sequence} & \ref{alg:bip-comp}\\
    \ref{alg:encode-star-vertices} & encoding star vertices (part of Algorithm~\ref{alg:graph-encode-new}) & Sec.~\ref{sec:enc-star-vertices} & Lemma~\ref{lem:encode-star-vertices-complx} in Sec.~\ref{sec:enc-star-vertices} & \ref{alg:compress-sequence} \\
    \ref{alg:encode-star-edges} & encoding star edges (part of Algorithm~\ref{alg:graph-encode-new}) & Sec.~\ref{sec:encode-star-edges} & Lemma~\ref{lem:encode-star-edges-complexity} in Sec.~\ref{sec:encode-star-edges} & \\
    \ref{alg:encode-find-deg} & finding vertex degree profiles, i.e.\ the variable $\mathsf{Deg}$ (part of Algorithm~\ref{alg:graph-encode-new}) & Sec.~\ref{sec:encode-find-Deg} & Lemma~\ref{lem:enc-find-deg-complexity} in Sec.~\ref{sec:encode-find-Deg} & \\
    \ref{alg:encode-vertex-types} & encoding Vertex Types (part of Algorithm~\ref{alg:graph-encode-new}) & Sec.~\ref{sec:encode-vertex-types} & Lemma~\ref{lem:alg-find-types-complexity} in Sec.~\ref{sec:encode-vertex-types} & \ref{alg:compress-sequence} \\
    \ref{alg:encode-find-partition-graphs} & finding Partition Graphs (part of Algorithm~\ref{alg:graph-encode-new}) & Sec.~\ref{sec:encode-find-partition-graphs} & Lemma~\ref{lem:find-partition-graphs-complexity} in Sec.~\ref{sec:encode-find-partition-graphs} & \\
    \ref{alg:graph-decode} & decoding a simple marked graph & Sec.~\ref{sec:dec-alg} & Sec.~\ref{sec:main-decode-complexity} & \ref{alg:sequence-decompress}, \ref{alg:decode-star-edges}, \ref{alg:decode-vertex-deg-profiles}, \ref{alg:dec-partition-deg-original-index}, \ref{alg:b-dec}, \ref{alg:s-dec-main} \\
    \ref{alg:sequence-decompress} & decompressing an array consisting of nonnegative integers & Sec.~\ref{sec:decode-sequence} & Sec.~\ref{sec:decode-sequence} & \ref{alg:b-dec} \\
    \ref{alg:decode-star-edges} & decoding star edges & Sec.~\ref{sec:dec-star-edges} & Lemma~\ref{lem:dec-star-edges-complexity} in Sec.~\ref{sec:dec-star-edges} & \\
    \ref{alg:decode-vertex-deg-profiles} & decoding vertex degree profiles & Sec.~\ref{sec:dec-ver-deg-profiles} & Lemma~\ref{lem:dec-deg-complexity} in Sec.~\ref{sec:dec-ver-deg-profiles} & \ref{alg:sequence-decompress} \\
    \ref{alg:dec-partition-deg-original-index} & finding degree sequences of partition graphs and relative vertex indexing & Sec.~\ref{sec:dec-find-partition-deg-original-index} & Lemma~\ref{lem:dec-find-part-deg-orig-index-complexity} in Sec.~\ref{sec:dec-find-partition-deg-original-index} & \\
    \ref{alg:type-extract-message-passing-new} & extracting edge types for a simple marked graph & Sec.~\ref{sec:mp-alg-detials} & Lemma~\ref{lem:message-passing-complexity} in Sec.~\ref{sec:MP-complexity} & \ref{alg:send-message} \\
    \ref{alg:send-message} & sending a message from a node to one of its neighbors & Sec.~\ref{sec:mp-alg-detials} & Lemma~\ref{lem:send-message-complexity} in Sec.~\ref{sec:MP-complexity} & \\
    \ref{alg:bip-Nij} & computing $N_{i,j}(G)$ for a simple unmarked bipartite graph $G \in \mGnlr_{\va, \vb}$ & Sec.~\ref{sec:bip-enc-pseudocode} &  Lemma~\ref{lem:bip-encode-computeN-complexity} in Sec.~\ref{sec:bip-compression-complexity} &  \ref{alg:compute-product}, \ref{alg:comp-ProdAFactorial} \\
    \ref{alg:bip-comp} & finding $\fnlr_{\va, \vb}(G)$ for $G \in \mGnlr_{\va, \vb}$ &  Sec.~\ref{sec:bip-enc-pseudocode} & Lemma~\ref{lem:bip-encode-overall-complexity} in Sec.~\ref{sec:bip-compression-complexity} & \ref{alg:bip-Nij} \\
    \ref{alg:compute-product} & computing $\prod_{\pp{k}=0}^{k-1} (p - \pp{k}s)$ & Sec.~\ref{sec:aux-algs-bip} & Lemma~\ref{lem:compute-product-complexity} in Sec.~\ref{sec:aux-algs-bip} & \\
    \ref{alg:comp-ProdAFactorial} & computing $\prod_{\pp{i}=i}^j v_{\pp{i}}!$ for an array $\vv$ of nonnegative integers & Sec.~\ref{sec:aux-algs-bip} & Lemma~\ref{lem:prod-a-factorial-complexity-each-delta} in Sec.~\ref{sec:aux-algs-bip} & \ref{alg:compute-product}\\
    \ref{alg:bip-dec-i} & decoding the adjacency list of a left vertex $1 \leq i \leq n_l$
    given $\tN_{i,i}$ for a simple unmarked bipartite graph $G \in \mGnlr_{\va,
                          \vb}$ & Sec.~\ref{sec:bip-dec-one-node} & Lemma~\ref{lem:bdec-i-complexity} in Sec.~\ref{sec:bdec-complexity} & \ref{alg:compute-product} \\
    \ref{alg:bip-dec-ij} & decoding the adjacency list of the  left vertices $i \leq v \leq j$ for $1 \leq i \leq j \leq n_l$ given $\tN_{i,j}$ for a simple unmarked
  bipartite graph $G \in \mGnlr_{\va,\vb}$ & Sec.~\ref{sec:bip-dec-interval} & Lemma~\ref{lem:bdec-ij-complexity} in Sec.~\ref{sec:bdec-complexity} & \ref{alg:compute-product}, \ref{alg:comp-ProdAFactorial}, \ref{alg:bip-dec-i}\\
\ref{alg:b-dec} & decoding for a simple unmarked bipartite graph $G \in \mGnlr_{\va,
                  \vb}$ given $\fnlr_{\va, \vb}(G)$ & Sec.~\ref{sec:bdec-main} & Lemma~\ref{lem:bdec-overall-complexity} in Sec.~\ref{sec:bdec-complexity} &  \ref{alg:comp-ProdAFactorial}, \ref{alg:bip-dec-ij}\\
    \ref{alg:ComputeNij_graph} & computing $N_{i,j}(G)$ for a simple unmarked graph $G \in
                                 \mGnt_{\va}$ & Sec.~\ref{sec:self-encode-alg} & Lemma~\ref{lem:s-compute-N-complexity} in Sec.~\ref{sec:self-compression-complexity} & \ref{alg:compute-product} \\
    \ref{alg:self-find-f} & finding $\fnt_{\va}(G)$ and $\vec{\tilde{f}}^{(\pn)}_{\va}(G)$ for $G \in \mGnlr_{\va}$ & Sec.~\ref{sec:self-encode-alg} & Lemma~\ref{lem:self-find-f-complexity} in Sec.~\ref{sec:self-compression-complexity} &  \ref{alg:ComputeNij_graph}\\
    \ref{alg:sdec-i} & decoding the forward adjacency list of a vertex $1 \leq i \leq \pn$
                       given $\tN_{i,i}$ for a simple unmarked graph $G \in \mGnt_{\va}$ & Sec.~\ref{sec:self-decode-one-node} & Lemma~\ref{lem:sdec-i-complexity} in Sec.~\ref{sec:sdec-complexity} & \ref{alg:compute-product} \\
    \ref{alg:sdec-ij} & decoding the forward adjacency list of vertices $i \leq v \leq j$
                        given $\tN_{i,j}$ for a simple unmarked graph $G \in \mGnt_{\va}$ & Sec.~\ref{sec:s-dec-interval} & Lemma~\ref{lem:s-dec-ij-complexity} in Sec.~\ref{sec:sdec-complexity} & \ref{alg:compute-product}, \ref{alg:sdec-i} \\
    \ref{alg:s-dec-main} & decoding for a simple unmarked graph $G \in \mGnt_{\va}$ given
  $\fnt_{\va}(G)$ and $\vec{\tilde{f}}^{(\pn)}_{\va}(G)$ &
                                                           Sec.~\ref{sec:self-main-dec}
                                    &
                                      Lemma~\ref{lem:self-dec-main-decode-complexity}
                                      in Sec.\ \ref{sec:sdec-complexity} & \ref{alg:comp-ProdAFactorial}, \ref{alg:sdec-ij} \\
\bottomrule
  \end{longtable}
}


\section{Data Structures and Algorithms}
\label{sec:data-structures}



In this section, we briefly cover the data structures and algorithms used in this document, and
refer the reader to the literature for further reading.



\subsection{Array}
\label{sec:ds-arrays}

An array is a list of data points which allows for adding new elements. We use
the notation $\Array$ for this data structures. We assume that all arrays in
this documents are dynamic, i.e.\ their size does not have to be specified at
initialization, and they allow for resizing and adding new elements. Therefore,
we can either specify the size of an array at initialization, e.g.\ write ``$\va
\gets \Array \text{ of integers of size } n$'', or leave the size unspecified,
e.g.\ write ``$\va \gets \Array \text{ of integers}$''. When the size is
unspecified, we assume that the initial size of the array is zero. The $i$th element of an
array $\va$ is denoted by $a_i$ or $a(i)$. We use both notations interchangeably.
The first element in an array is indexed by $1$. The size of an array $\va$, which is the number of
elements in it,  is denoted by
$\textproc{Size}(\va)$. Therefore, the last element in an array $\va$ is
$a(\textproc{size}(\va))$. We use the term ``append'' to add a new element to an
array. For instance, if $\va$ is an array of integers, after performing ``append
2 to $\va$'', the size of $\va$ is incremented by one, and the last element
becomes 2, i.e.\ $a(\textproc{size}(\va)) = 2$. 

In order to ensure the dynamic property of an array, we always reserve a number
of blocks in the memory for an array which can be more than its size. We call
this the ``capacity'' of an array. In case we want to append an elements to an
array $\va$ whose size is equal to its capacity, we create a new array whose
capacity is twice the size of $\va$, and copy the content of $\va$ to this new
array, and let $\va$ point to this new location in the memory. This way, the
so-called 
\emph{amortized complexity}
of appending a new element to an array is
$\Theta(1)$.
This means that the average complexity of appending
  an element to the array, averaged over several append operations, is
  $\Theta(1)$.
Also, the complexity of accessing the $i$th element in an array is $\Theta(1)$.
Moreover, the memory required to store an array $\va$ is $O(nC)$ where $n = \textproc{Size}(\va)$ is the
number of elements in $a$, and $C$ is the memory required to store each
element. 
The reader is referred to \cite{cormen2009introduction} for more details on
arrays, specifically Section~17.4 therein for dynamic arrays.

We can have nested arrays, i.e.\ an array such that each element is a further
array. For instance, ``$\va \gets \Array \text{ of } \Array \text{ of integers}$''
creates an array such that each of its elements in an array of integers. In this
case, we denote the $j$th element in the $i$th element of $\va$ by $a_{i,j}$. In
fact, $a_{i,j}$ is a shorthand for $(a_i)_j$. In this case, we may use $a(i,j)$ and $a_{i,j}$
interchangeably. 


\subsection{Sort}
\label{sec:app-sort}

Given an array $\va=(a_1, \dots, a_n)$, ``$\vb \gets \textproc{Sort}(\va)$'' stores the values of
$\va$ in an array $\vb$ 
in increasing order,
\color{black}
so that $b_1 \leq b_2 \leq \dots \leq b_n$. Moreover, ``$(\pi, \vb) \gets
\textproc{Sort}(\va)$'' in addition to storing the sorted values in $\vb$,
stores the permutation that needs to be applied to each index in  $\va$ to get
the corresponding index in $\vb$ in an array $\pi = (\pi_1, \dots, \pi_n)$, i.e. $b_i = a_{\pi_i}$
for $1 \leq i \leq n$. The time complexity of sorting an array of size $n$ is
$O(Cn \log n)$, where $C$ is the time complexity of comparing two values in
$\va$ (see, for instance, \cite{cormen2009introduction}).

\subsection{Dictionary}
\label{sec:data-structure-dictionary}

A dictionary is a data structure that maps distinct ``keys'' to ``values''. We
use the notation ``$D \gets \Dictionary(\mA \rightarrow \mB)$'' to denote that $D$
is a dictionary such that key objects come from the set $\mA$ and the
corresponding value
objects come from the set $\mB$. We say that a key--value pair $(a,b)$ for $a \in
\mA$ and $b \in \mB$ is in $D$, and we write $(a,b) \in D$, if $D$ maps $a$ to $b$. Each key  can
appear in at most one key--value pair in a dictionary. The set of keys in a
dictionary $D$ is denoted by $D.\textproc{Keys}$, i.e.\ $D.\textproc{Keys}$ is
the set of $a \in \mA$ such that $(a,b) \in D$ for some $b \in \mB$. If $a \in
D.\textproc{Keys}$, we denote the value object $b \in \mB$ corresponding to $a$
by $D(a)$. If $a \notin D.\textproc{Keys}$ and $b \in \mB$, the operation 
``$D.\textproc{Insert}(a,b)$'' inserts the key--value pair $(a,b)$ into $D$.

A dictionary is an abstract data type and there are several ways to implement it.
However, we assume that all dictionaries in this paper are implemented via a
self--balancing binary search tree, such as a Red--Black tree (see, for
instance, \cite[Chapter 13]{cormen2009introduction}). Therefore, in order to
insert a new key--value pair $(a,b)$ in a dictionary which already contains $n$
key--value pairs, we need to compare $a$ with $O(\log n)$ many existing keys in
the dictionary.  Hence, if each of such comparisons has time
complexity $C$, the overall time complexity of inserting a new key--value
pair is $O(C \log n)$. Likewise, checking whether a key $a$ is in the set of
keys of a dictionary, and if it is, accessing its corresponding value has time
complexity $O(C \log n)$. The memory required to store a dictionary with $n$
elements is $O(M n)$, where $M$ is the memory required to store each key--value
pair. See \cite{cormen2009introduction} for more details. 

\subsection{Fenwick Tree}
\label{sec:ds-fenwick}

A ``Fenwick tree'', or a ``binary indexed tree'', is a data structure that can
efficiently find partial sums in an array \cite{fenwick1994new}. For instance, if $\va = (a_i: i \in
[n])$ is an array of integers,  ``$\sfU \gets \text{Fenwick tree
  initialized with array } \va$'' initializes a Fenwick tree $\sfU$ with the array
$\va$, such that for $1 \leq k \leq n$, $\sfU.\textproc{Sum}(k)$ returns
$\sum_{i = k}^n a_i$. In addition to this, we can dynamically change the values
of the underlying array for a Fenwick tree. More precisely, after performing
``$\sfU.\textproc{Add}(k, c)$'', $\sfU$ calculates partial sums corresponding to
an array $\va' = (a'_i: i \in [n])$ such that $a'_i = a_i$ for $i \neq k$, and
$a'_k = a_k + c$. Note that after performing $\sfU.\textproc{Add}$, the values
of the array $\va$ which was used to initialize $\sfU$ does not change. In fact,
$\sfU$ does not save a copy of $\va$. Instead, it stores enough information to
find partial sums efficiently, and by performing $\sfU.\textproc{Add}$, the
internal content of $\sfU$ is modified in a way which is equivalent to changing
the value of the array.

The Fenwick tree data structure is designed so that the operations of finding
partial sums, i.e.\ calling $\sfU.\textproc{Sum}$ above, and modification, i.e.\
$\sfU.\textproc{Add}$ above, are performed in $O(C \log n)$ time, where $n$ is the
number of elements in the underlying array, and $C$ is an upper bound on the
maximum number of bits among  the binary representations of partial sums. Moreover, the time complexity of
initializing a Fenwick tree with an array of $n$ elements is $O(C n)$. Also, the
memory required to store such a Fenwick tree is $O(C n)$.
Note that compared to the naive approach in which we store the values of an
array and directly compute the partial sums, which has time complexity $O(C n)$
for computing partial sums,
employing a Fenwick tree results in a complexity improvement. This is with the
cost that the complexity of modification is $O(C \log n)$ in a Fenwick tree,
compared to $O(C)$ in the above naive method. In fact, employing a Fenwick tree
can result in an overall performance improvement in scenarios where we need to
compute many partial sums, such as several occasions in the present paper. We refer the reader to
\cite{fenwick1994new} for more details on properties of the Fenwick tree and its
implementation. 




\section{Proof of Lemma~\ref{lem:UCv-UCu}}
\label{sec:uc-lemma-proof}

        Note that the mark component of each object is $\xi_G(v,u)$. Moreover, each vertex in $\UC_u(G)(v,u)$ corresponds to a
        non--backtracking walk in $G$ of the form $v_0 = u, v_1, \dots, v_l$
        for some $l \geq 0$ such that if $l \geq 1$, we have  $v_1 \neq v$. On the other hand, each
        node in $\UC_v(G)(v,u)$ corresponds to a non--backtracking walk of the
        form $\tilde{v}_0 = v, \tilde{v}_1 = u, \dots, \tilde{v}_{\tilde{l}}$
        for some $\tilde{l} \geq 1$. Since every such walk is non--backtracking,
        if $\tilde{l} \geq 2$, we have $\tilde{v}_2 \neq v$. Therefore, by
        setting $l = \tilde{l} -1$ and $v_i = \tilde{v}_{i+1}$ for $0 \leq i
        \leq \tilde{l} - 1$, we find a bijection between the vertices in the
        subgraph components of $\UC_u(G)(v,u)$ and $\UC_v(G)(v,u)$. Such a
        bijection is adjacency preserving by the definition of the universal
        cover. Additionally, since edge
        and vertex marks in both objects come from the same marked graph $G$,
        this bijection also preserves marks. This completes the proof.


\editstart

\section{Proof of Proposition~\ref{prop:opt-aa12-nat-count}}
\label{sec:app-nat-count}

In this section, we give the proof of Proposition~\ref{prop:opt-aa12-nat-count}.
This is done by investigating the compression algorithm as is explained in
Section~\ref{sec:alg-details}. Before giving the proof, we state and prove the
following two lemmas.



\begin{lem}
  \label{lem:bip-bound-greater-1}
  Given integers $n_l, n_r \geq 1$ and sequences $(a_1, \dots, a_{n_l})$ and
  $(b_1, \dots, b_{n_r})$ of nonnegative integers, assume that there exists a simple bipartite graph
  $G$ with $n_l$ left vertices and $n_r$ right vertices such that the degree of each
  left vertex $1 \leq i \leq n_l$ is $a_i$, while the degree of each right vertex
  $1 \leq j \leq n_r$ is $b_j$. Then, with $S := \sum_{i=1}^{n_l} a_i$, we have
  \begin{equation*}
    \frac{S!}{\prod_{i=1}^{n_l} a_i! \prod_{j=1}^{n_r}b_j! } \geq 1.
  \end{equation*}
\end{lem}

\begin{proof}
  First note that the existence of $G$ and the fact that $G$ is simple implies
  that $\sum_{j=1}^{n_r} b_j = \sum_{i=1}^{n_l} a_i = S$. Consider a bipartite
  configuration model with $n_l$ left vertices and $n_r$ right vertices, where we
  attach $a_i$ distinct half--edges to each left vertex $1 \leq i \leq n_l$, and
  $b_j$ distinct half--edges  to each right vertex $1 \leq j \leq n_r$.
  A half--edge connected to a left vertex is only allowed to be connected to a half--edge connected to a right vertex, and multiple edges are allowed to form.
  Therefore, there are $S!$ admissible configurations in total. Now, consider those
  configurations which result in the simple bipartite graph $G$ introduced above. Since
  $G$ is simple, starting with a configuration which results in $G$ and simultaneously applying any of the $a_i!$ permutations on the
  half--edges connected to a left vertex $1 \leq i \leq n_l$, and applying any of the $b_j!$
  permutations on the half--edges connected to a right vertex $1 \leq j \leq n_r$
  yields  a distinct configuration which also results in $G$. Since such
  configurations form a subset of the set of all admissible configurations, we have $S!
  \geq \prod_{i=1}^{n_l} a_i! \prod_{j=1}^{n_r} b_j!$ and the proof is complete.
\end{proof}

\begin{lem}
  \label{lem:self-bound-greater-1}
  Given a graphical degree sequence  $d_1, \dots, d_k$ on $k$ vertices,
  with $S := \sum_{i=1}^k d_i$, we have
  \begin{equation*}
    \frac{(S-1)!!}{\prod_{i=1}^{k} d_i!} \geq 1.
  \end{equation*}
\end{lem}

\begin{proof}
  Consider a configuration model where we attach $d_i$ many distinct half--edges
  to each vertex $1 \leq i \leq k$, so that any two half--edges are allowed to connect to each
  other to form an edge (where multiple edges and self loops are allowed).
  Since there are $S$ half--edges in total, there are overall $(S-1)!!$ possible
  ways of matching them in pairs. On the other hand, since this
  degree sequence is graphical, there is a simple graph $G$ such that the degree
  of vertex $1 \leq i \leq k$ is precisely $d_i$. Given this simple graph $G$, there are $\prod_{i=1}^k d_i!
  $ many possible matchings of the half--edges that result in  $G$. The
  reason is that since $G$ is simple, it does not have self loops or half--edges.
  Therefore, if we start with a configuration that results in $G$, and if we consider any of the $d_i!$ permutations of half--edges
  associated to a
  vertex $1 \leq i \leq k$ along
  the edges connected to $i$ in $G$, we have a valid matching of the total $S$
  half--edges, and each of such configurations are distinct. This means that
  $\prod_{i=1}^k d_i! \leq (S-1)!!$ and completes the proof. 
\end{proof}

\begin{proof}[Proof of Proposition~\ref{prop:opt-aa12-nat-count}]
Following the details of
  the compression algorithm from Section~\ref{sec:alg-details}, specifically
  Algorithm~\ref{alg:graph-encode-new} therein, we realize that
  the codeword $\fn_{h, \delta}(\Gn)$ has the following components:
  \begin{itemize}
  \item We first write $\mathsf{TCount}$ using the Elias delta code (see
    line~\ref{l:enc-tcount-elias} in Algorithm~\ref{alg:graph-encode-new}).
    Recall from Proposition~\ref{prop:MP} that there is a one to one mapping
    $J_n$ from the set $\mTn \subset \mFbardeltah$ to $\{1, \dots,
    \mathsf{TCount}\}$. This implies 
    $\mathsf{TCount} \leq |\mFbardeltah|$. Moreover, since $\delta$ and $h$ are fixed and $n$ is being
    sent to infinity, $\mFbardeltah$ is a finite set, and we have $\len(\edelta(1+\mathsf{TCount})) = O(1)$.
  \item Next, for $1 \leq i \leq \mathsf{TCount}$, we concatenate the output
    with $\mathsf{TIsStar}(i) \concat \mathsf{TMark}(i)$ using $1 + 1 +
    \lfloor \log_2 |\edgemark| \rfloor$ bits
    (see line~\ref{l:enc-tiisstar-tmark-write} in Algorithm~\ref{alg:graph-encode-new}). Again, since $\mathsf{TCount}
    \leq |\mFbardeltah|$ and $|\edgemark|$ is not
    growing with $n$, the contribution of this part is also $O(1)$.
  \item Next, we write the encoded version of  $\Vns$ to the output (see line~\ref{l:enc-star-vertices} in
    Algorithm~\ref{alg:graph-encode-new} and Section~\ref{sec:enc-star-vertices}). 
Recall from Section~\ref{sec:enc-star-vertices} that we encode $\Vns$ by first
forming an array $\vec{s} = (s_v: v \in [n])$ where $s_v = 1$ if $v \in \Vns$
and $s_v = 0$ otherwise. We then encode this array.
Using Proposition~\ref{prop:encode-sequence},
    the number of bits required to do so is at most
    \begin{equation}
      \label{eq:encode-Vns-bit-count}
      \log_2 \left( 1 + \left\lceil  \binom{n}{|\Vns|}\right\rceil \right) + O(\log n) \leq \log_2 \binom{n}{|\Vns|} + O(\log n), 
    \end{equation}
    where we have used the inequality $\log_2(1+\lceil x \rceil) \leq \log_2 x +
    \log_2 3$ which holds for $x \geq 1$.
  \item The next step is to write star edges to the output (see line~\ref{l:enc-star-edges}
    in Algorithm~\ref{alg:graph-encode-new} and
    Section~\ref{sec:encode-star-edges}). Following our procedure in
    Section~\ref{sec:encode-star-edges}, if $\mn_\star$ denotes the total
    number of star edges in $\Gn$, for each star edge, we output a  single bit
    with value one
    followed by $1 + \lfloor \log_2 n \rfloor$ bits. Moreover, for each pair $x,
    x' \in \edgemark$ and $v \in \Vns$, we output a single bit with value zero
    to indicate that we are finished with processing the neighbors of $v$.  Summing up, we realize that we encode
    star edges using
    \begin{equation}
      \label{eq:star-edges-bits-count}
      |\edgemark|^2 \times |\Vns| + \mn_\star(1 + 1 + \lfloor \log_2 n \rfloor) \leq  |\edgemark|^2 \times |\Vns| + 2 \mn_\star + \mn_\star \log_2 n
    \end{equation}
    many bits.
  \item Next, we write  vertex types to the output (see
    line~\ref{l:enc-encode-vertex-types-call} in
    Algorithm~\ref{alg:graph-encode-new} and Section~\ref{sec:encode-vertex-types}).
Recall that $ \mA^{(n)}_{\delta} = \{(\thetan_v, \Dn(v)): v \in [n]\}$ is the
set of distinct vertex types appearing in $\Gn$.
Following
    our procedure in Section~\ref{sec:encode-vertex-types}, we effectively find
    $\mA^{(n)}_{\delta}$ and convert each element in $\mA^{(n)}_{\delta}$ to a
    sequence of size at most $1 + 3\delta$ whose elements are bounded by
    $|\edgemark| \vee \mathsf{TCount} \vee \delta$. Then, we use $1 + \lfloor
    \log_2 n \rfloor$ bits to encode $|\mA^{(n)}_{\delta}|$. After this point,
    we write the sequence representation of the elements in $\mA^{(n)}_{\delta}$
    using at most
    \begin{equation}
      \label{eq:vertex-type-mAndelta-bit-count}
      |\mA^{(n)}_{\delta}| \bigg[ 1 + \lfloor \log_2(1+3\delta) \rfloor + (1+3\delta) \left( 1 + \lfloor \log_2(|\edgemark| \vee \mathsf{TCount} \vee \delta) \rfloor \right) + 1 + \lfloor  \log_2 n \rfloor\bigg]
    \end{equation}
    many bits (see the loop of line~\ref{l:find-v-types-nu-i-for} in
    Algorithm~\ref{alg:encode-vertex-types}). Recall that 
    $\mA^{(n)}_\delta \subset \mA_\delta$. But from~\eqref{eq:D-T-o-delta}, $\mA_\delta$ is a finite set.
    Therefore, $|\mA^{(n)}_\delta|$ remains bounded by a constant independent of
    $n$. Also, $\mathsf{TCount} \leq |\mFbardeltah |$ and $|\mFbardeltah|$ does not grow
    with $n$. Therefore, the quantity
    in~\eqref{eq:vertex-type-mAndelta-bit-count} is $O(\log n)$.
    After encoding $\mA^{(n)}_\delta$, we compress the integer sequence representing the type of
    vertices, where each element in this sequence is bounded by
    $|\mA^{(n)}_{\delta}|$ (this is the sequence $\vy$ in Algoirthm~\ref{alg:encode-vertex-types}). Using
    Proposition~\ref{prop:encode-sequence}, this can be done using at most
    \begin{equation*}
      \log_2 \left( 1 + \left\lceil \frac{n!}{\prod_{(\theta, D) \in \mA^{(n)}_\delta} |\{ v \in [n]: (\thetan_v, \Dn(v)) = (\theta, D)\}|!}\right \rceil \right) + O(|\mA^{(n)}_\delta| \log n)
    \end{equation*}
    many bits. Using the above argument and the inequality $\log_2(1+\lceil x
    \rceil) \leq \log_2 x + \log_2 3$ which holds for $x \geq 1$, we get the
    following upper bound on the number of bits used to encode vertex types
    \begin{equation}
      \label{eq:encode-vertex-types-bit-count}
      \log_2\left( \frac{n!}{\prod_{(\theta, D) \in \mA_\delta} |\{ v \in [n]: (\thetan_v, \Dn(v)) = (\theta, D)\}|!}  \right) + O(\log n).
    \end{equation}
  \item At this point, we start writing the partition graphs to the output. In order to do
    so, we first write $\edelta(1 + |\mEn_{\leq}|)$ to the output
    (line~\ref{l:enc-k-output} in Algorithm~\ref{alg:graph-encode-new}). Since $\mEn_{\leq}
    \subseteq \mCdeltah = \mFdeltah \times \mFdeltah$, and $|\mFdeltah|$ is fixed and does not grow with $n$,
    $\len(\edelta(1 + |\mEn_{\leq}|))$ is $O(1)$. Then, for each $(t,t') \in
    \mEn_{\leq}$,  we use $2(1+\lfloor \log_2 \mathsf{TCount}
    \rfloor)$ bits to encode $i$ and $i'$, where $i = J_n(t)$ and
    $i' = J_n(t')$ are the integer representations of $t$ and $t'$ respectively
    (see lines~\ref{l:enc-part-neq-i-write} and \ref{l:enc-part-eq-i-write} in Algorithm~\ref{alg:graph-encode-new}).
    If $t \neq t'$,
    using Proposition~\ref{prop:bipartite-color-graph-compress}
    and the fact that $\len(\edelta(N)) \leq \log_2 N + O(\log \log N)$ for $N
    \geq 1$, we
    write the integer representation of  $\Gn_{t,t'}$ to the output in
    line~\ref{l:enc-part-neq-write-f-output} in
    Algorithm~\ref{alg:graph-encode-new} with at most $\log_2(1+\lceil \eln_{t,t'}\rceil) + O(\log \log
    \eln_{t,t'})$ bits.
    Note that using~\eqref{eq:Dn-bounded-delta}, the degree of each vertex in
    each partition graph is  bounded by $\delta$, thus we have $\Sn_{t,t'} \leq n
    \delta$. Therefore, using Stirling approximation, we have $\log \log \eln_{t,t'} =
    O(\log n)$. Also note that from Lemma~\ref{lem:bip-bound-greater-1},
    we have
    $\eln_{t,t'} \geq 1$. Thereby, using the
    inequality $\log_2 (1+\lceil x\rceil) \leq \log_2 x + \log_2 3$ which holds
    for $x \geq 1$ and putting the above together, we realize that  the total
    number of bits used to encode $\Gn_{t,t'}$ is at most 
    \begin{equation*}
      \log_2 \eln_{t,t'} + O(\log n). 
    \end{equation*}
    On the other hand, if $t = t'$,
    using Proposition~\ref{prop:equal-color-graph-compress}, the
    total number of bits used to write the encoded version of  $\Gn_{t,t}$ to
    the output in lines~\ref{l:enc-part-eq-f-write}
    through~\ref{l:enc-part-eq-tilde-f-write-end-for} in Algorithm~\ref{alg:graph-encode-new} is at most 
    \begin{equation*}
      \log_2(1+\lceil \eln_{t,t}\rceil) + O(\log \log \eln_{t,t}) + \frac{16n}{\log^2 n} \left( \log_2(1+\Sn_{t,t}) + O(\log \log \Sn_{t,t}) \right) + O(\log n).
    \end{equation*}
    Note that $\Sn_{t,t} \leq n\delta$. Therefore, using Stirling's
    approximation, we have $\log \log \eln_{t,t} = O(\log n)$. Furthermore,
    $\log_2(1+\Sn_{t,t}) = O(\log n)$ and $\log \log \Sn_{t,t} = O(\log \log
    n)$. On the other hand, using Lemma~\ref{lem:self-bound-greater-1}, we have
    $\eln_{t,t} \geq 1$, which. similar to the discussion above, implies that
    $\log_2(1+\lceil \eln_{t,t} \rceil) \leq \log_2 \eln_{t,t} + \log_2 3$. Putting all these bounds together, we realize that the
    total number of bits used to encode $\Gn_{t,t}$ is at most
    \begin{equation*}
      \log_2 \eln_{t,t} + O ( n / \log n). 
    \end{equation*}
    We conclude that the total number of bits we use to encode partition graphs
    is at most
    \begin{equation}
      \label{eq:encode-partition-graphs-bit-count}
      \sum_{(t,\pp{t}) \in \mEn_{\leq}} \log_2 \eln_{t,\pp{t}} + O ( n / \log n).
    \end{equation}
  \end{itemize}

  Putting the bounds
  \eqref{eq:encode-Vns-bit-count},
  \eqref{eq:star-edges-bits-count},
  \eqref{eq:encode-vertex-types-bit-count},
  and \eqref{eq:encode-partition-graphs-bit-count}
  together, and multiplying by $\log 2$ to convert
  bits to nats, we have
  \begin{align*}
    \nat(\fn_{h, \delta}(\Gn)) &\leq \log \binom{n}{|\Vns|} + |\edgemark|^2 \times |\Vns| \log 2 + 2 \mn_{\star} \log 2 + \mn_{\star} \log n \\
    &\qquad +\log \left( \frac{n!}{\prod_{(\theta, D) \in \mA_\delta} |\{ v \in [n]: (\thetan_v, \Dn(v)) = (\theta, D)\}|!}  \right) + \sum_{(t,\pp{t}) \in \mEn_{\leq}} \log \eln_{t,\pp{t}} +  o(n),
  \end{align*}
which completes the proof.
\end{proof}

\editfinish


\editstart

\section{Proof of Lemmas in Section~\ref{sec:optimality-proof}}
\label{sec:opt-lemmas}

In this section, we give the proof of Lemmas~\ref{lem:opt-bb13-Vns} through
\ref{lem:sum-Snc-mn-mn-star} in
Section~\ref{sec:optimality-proof}. We have repeated  the lemma statements for
convenience.

\begin{lem*}[Lemma~\ref{lem:sum-Snc-mn-mn-star} in Section~\ref{sec:optimality-proof}]
  For $h \geq 1$ and $\delta \geq 1$, we have
  \begin{equation*}
    \frac{1}{2} \sum_{c \in \mCdeltah} \Sn_c = \mn - \mn_\star.
  \end{equation*}
\end{lem*}

\begin{proof}[Proof of Lemma~\ref{lem:sum-Snc-mn-mn-star}]
  We have
  \begin{align*}
    \frac{1}{2} \sum_{c \in \mCdeltah} \Sn_c &= \frac{1}{2} \sum_{c \in \mCdeltah} \sum_{v=1}^n \Dn_c(v) \\
                                             &= \frac{1}{2} \sum_{v=1}^n \sum_{c \in \mCdeltah} \Dn_c(v) \\
                                             &= \frac{1}{2} \sum_{v=1}^n \sum_{c \in \mCdeltah} |\{w \sim_{\Gn} v: \type{v}{w} = c \}| \\
                                             &= \frac{1}{2} \sum_{v=1}^n |\{ w \sim_{\Gn} v: \type{v}{w} \in \mCdeltah\}| \\
                                             &= \frac{1}{2} \sum_{v=1}^n \left( \deg_{\Gn}(v) - |\{w \sim_{\Gn} v: \type{v}{w} \notin \mCdeltah\}| \right) \\
                                             &= \frac{1}{2} \sum_{v=1}^n \left( \deg_{\Gn}(v)  - |\{w \sim_{\Gn} v: \type{v}{w} = (\xi_{\Gn}(w,v), \xi_{\Gn}(v,w))\}| \right) \\
    &= \mn - \mn_\star. \qedhere
  \end{align*}
\end{proof}

\begin{lem*}[Lemma~\ref{lem:opt-bb13-Vns} in Section~\ref{sec:optimality-proof}]
  With the assumptions in the first part of Theorem~\ref{thm:optimality-complexity-main},
  and with $h \geq 1$ and $\delta \geq 1$ fixed, we have 
  \begin{equation*}
  \lim_{n \rightarrow \infty} \frac{|\Vns|}{n} 
  = \eta_1(\mu; h , \delta),
\end{equation*}
and
\begin{equation*}
  \limsup_{n \rightarrow \infty} \frac{1}{n} \log \binom{n}{|\Vns|} \leq \eta_1(\mu;h, \delta)(1 - \log \eta_1(\mu; h, \delta)).
\end{equation*}
Recall that $\eta_1(\mu; h, \delta) = \prwrt{\mu}{\deg_\star^{(h, \delta)}([T,o]) > 0}$.
\end{lem*}

\begin{proof}[Proof of lemma~\ref{lem:opt-bb13-Vns}]
Define the function $F: \mGb_*^h \rightarrow \{0,1\}$ such that
\begin{equation*}
  F([G,o]) =
  \begin{cases}
    1 & \exists v \sim_G o: \UC_o(G)[v,o]_{h-1} \notin \mFdeltah \text{ or } \UC_o(G)[o,v]_{h-1} \notin \mFdeltah \\
    & \qquad \qquad \qquad \text{ or } \deg_T(o) > \delta \text{ or } \deg_T(v) > \delta \\
0 & \text{otherwise}
  \end{cases}
\end{equation*}
Recalling the definition of $\Vns$ and using~\eqref{eq:type-v-w-equivalent}, 
we have 
\begin{equation*}
  \frac{|\Vns|}{n} = \frac{1}{n} \sum_{v=1}^n F([\Gn,v]).
\end{equation*}
But $F$ is bounded and continuous and $U(\Gn) \Rightarrow \mu$, hence
\begin{equation*}
  \lim_{n\rightarrow \infty} \frac{1}{n} \sum_{v=1}^n F([\Gn,v]_h) = \evwrt{\mu}{F([T,o])}.
\end{equation*}
Using the fact that $\mu$ is supported on $\mTb_*$, and for a rooted tree $[T,o]
\in \mTb_*$, we have $[\UC_o(T),o] = [T,o]$, we have
\begin{align*}
  \evwrt{\mu}{F([T,o])} &= \prwrt{\mu}{\exists v \sim_T o: T[o,v]_{h-1} \notin \mFdeltah \text{ or } T[v,o]_{h-1} \notin \mFdeltah \text { or } \deg_T(o) > \delta \text{ or } \deg_T(v) > \delta} \\
  &= \prwrt{\mu}{\deg_\star^{(h, \delta)}([T,o]) > 0} = \eta_1(\mu; h, \delta),
\end{align*}
where the second line uses the definition of $\deg_\star^{(h, \delta)}([T,o])$
from~\eqref{eq:deg-star-deg}.
Putting the above together, we
conclude that 
\begin{equation*}
  \lim_{n \rightarrow \infty} \frac{|\Vns|}{n} 
  = \eta_1(\mu; h , \delta).
\end{equation*}
Using this together with the inequality $\log \binom{r}{s} \leq s \log
\frac{re}{s}$, we get 
\begin{equation*}
  \limsup_{n \rightarrow \infty} \frac{1}{n} \log \binom{n}{|\Vns|} \leq \lim_{n \rightarrow \infty} -\frac{|\Vns|}{n} \log \frac{|\Vns|}{ne} = \eta_1(\mu;h, \delta)(1 - \log \eta_1(\mu; h, \delta)).
\end{equation*}
This completes the proof.
\end{proof}

\begin{lem*}[Lemma~\ref{lem:opt-cc14-general-bounds-for-Dn} in Section~\ref{sec:optimality-proof}]
  With the assumptions in the first part of Theorem~\ref{thm:optimality-complexity-main},
  and with $h \geq 1$ and $\delta \geq 1$ fixed,  for $c \in \mCdeltah =
  \mFdeltah \times \mFdeltah$, we have
  \begin{equation*}
  \lim_{n \rightarrow \infty} \frac{1}{n} \sum_{v=1}^n \Dn_c(v) = \evwrt{\mu}{D_c([T,o])},
\end{equation*}
and
\begin{equation*}
  \lim_{n \rightarrow \infty} \frac{1}{n} \sum_{v=1}^n \log \Dn_c(v)! = \evwrt{\mu}{\log D_c([T,o])!}.
\end{equation*}
\end{lem*}

\begin{proof}[Proof of Lemma~\ref{lem:opt-cc14-general-bounds-for-Dn}]
  Recalling~\eqref{eq:thetan-Dn--UC-Gn}, 
  for $c \in \mCdeltah$,
  we have
\begin{equation*}
  \Dn_c(v) = D_c([\UC_v(\Gn), v]).
\end{equation*}
{\color{peditcolor}
Note that the mapping $[G,o] \mapsto [\UC_o(G), o]$ is continuous. Indeed, for
$[G,o], [G',o'] \in \mGb_*$ and $h \in \nats$, if
$[G,o]_h = [G',o']_h$ then $[\UC_o(G), o]_h = [\UC_{o'}(G'), o']_h$.}
This together
with~\eqref{eq:D-T-o-delta} implies that the mapping $[G,o] \mapsto
D_c([\UC_o(G), o])$ is bounded and continuous. Thereby, since $U(\Gn)
\Rightarrow \mu$, we have
\begin{equation*}
  \lim_{n \rightarrow \infty} \frac{1}{n} \sum_{v=1}^n \Dn_c(v) = \evwrt{\mu}{D_c([\UC_o(T),o])} = \evwrt{\mu}{D_c([T,o])},
\end{equation*}
where the last equality uses the facts that $\mu$ is supported on rooted trees,
and for a rooted tree $[T,o] \in \mTb_*$, we have $[\UC_o(T),o] = [T,o]$. A
similar argument implies that
\begin{equation*}
  \lim_{n \rightarrow \infty} \frac{1}{n} \sum_{v=1}^n \log \Dn_c(v)! = \evwrt{\mu}{\log D_c([T,o])!}. \qedhere
\end{equation*}
\end{proof}


\begin{lem*}[Lemma~\ref{lem:opt-dd15-mnstar-bound} in Section~\ref{sec:optimality-proof}]
  With the assumptions in the first part of Theorem~\ref{thm:optimality-complexity-main},
and  with $h \geq 1$ and $\delta \geq 1$ fixed, we have
  \begin{equation*}
    \lim_{n \rightarrow \infty} \frac{\mn_\star}{n} = \eta_2(\mu; h, \delta).
  \end{equation*}
  Recall that $\eta_2(\mu; h, \delta) =  \frac{1}{2} \evwrt{\mu}{\deg_\star^{(h, \delta)}([T,o])} $.
\end{lem*}

\begin{proof}[Proof of Lemma~\ref{lem:opt-dd15-mnstar-bound}]
From Lemma~\ref{lem:sum-Snc-mn-mn-star}, we may write
\begin{equation*}
  \mn_\star = \mn - \frac{1}{2} \sum_{c \in \mCdeltah} \Sn_c.
\end{equation*}
This together with~\eqref{eq:lim-sum-Dn-c-E-D-To} in Lemma~\ref{lem:opt-cc14-general-bounds-for-Dn},
and the assumption that $\mn / n \rightarrow \deg(\mu) / 2$, implies that 
  \begin{align*}
    \lim_{n \rightarrow \infty} \frac{\mn_\star}{n} &= \frac{\deg(\mu)}{2} - \frac{1}{2}  \sum_{c \in \mCdeltah} \evwrt{\mu}{D_c([T,o])}\\
    &= \frac{1}{2} \evwrt{\mu}{\deg_T(o) - \sum_{c \in \mCdeltah} D_c([T,o])} \\
                                                    &= \frac{1}{2} \mathbb{E}_{\mu}\Big [\Big|\Big\{v \sim_T o : T[o,v]_{h-1} \notin \mFdeltah \text{ or } T[v,o]_{h-1} \notin \mFdeltah \\
    &\qquad \qquad \text{ or } \deg_T(o) > \delta \text{ or } \deg_T(v) > \delta \Big \}\Big| \Big ]\\
    &= \frac{1}{2} \evwrt{\mu}{\deg_\star^{(h, \delta)}([T,o])} = \eta_2(\mu; h, \delta).
  \end{align*}
This concludes the proof.
\end{proof}


\begin{lem*}[Lemma~\ref{lem:theta-D-fraction-limit} in Section~\ref{sec:optimality-proof}]
  With the assumptions in the first part of Theorem~\ref{thm:optimality-complexity-main},
  and with $h \geq 1$ and $\delta \geq 1$ fixed, for $(\theta, D) \in \mA_\delta =
  \{(\theta([T,o]), D([T,o])): [T,o] \in \mTb_*\}$, we have 
  \begin{equation*}
    \lim_{n \rightarrow \infty} \frac{1}{n} \sum_{v=1}^n \one{\thetan_v = \theta, \Dn(v) = D} = \prwrt{\mu}{\theta([T,o]) = \theta, D([T,o]) = D}.
  \end{equation*}
\end{lem*}

\begin{proof}[Proof of Lemma~\ref{lem:theta-D-fraction-limit}]
We may write
\begin{align*}
  \lim_{n \rightarrow \infty} \frac{1}{n} \sum_{v=1}^n \one{\thetan_v = \theta, \Dn(v) = D} &\stackrel{(a)}{=} 
  \lim_{n \rightarrow \infty} 
  \color{black}
  \frac{1}{n} \sum_{v=1}^n \one{\theta([\UC_v(\Gn), v]) = \theta, D([\UC_v(\Gn), v]) = D} \\
                                                                                            &\stackrel{(b)}{=} \prwrt{\mu}{\theta([\UC_o(T),o]) = \theta, D([\UC_o(T),o]) = D} \\
                                                                                            &\stackrel{(c)}{=} \prwrt{\mu}{\theta([T,o]) = \theta, D([T,o]) = D} \\
\end{align*}
where $(a)$ uses~\eqref{eq:thetan-Dn--UC-Gn}, $(b)$ uses $U(\Gn) \Rightarrow
\mu$,
and in $(c)$, we have used the fact that $\mu$ is supported on $\mTb_*$, and for
$[T,o] \in \mTb_*$, we have $[\UC_o(T), o] = [T,o]$. This completes the proof.
\end{proof}

\begin{lem*}[Lemma~\ref{lem:opt-ee16-general-bound-for-ln-ttp} in Section~\ref{sec:optimality-proof}]
  With the assumptions in the first part of Theorem~\ref{thm:optimality-complexity-main},
  and with $h \geq 1$ and $\delta \geq 1$ fixed, we have 
  \begin{equation*}
  \sum_{(t,\pp{t}) \in \mEn_{\leq}} \log \eln_{t,\pp{t}} \leq \frac{1}{2} \sum_{c \in \mCdeltah} \left( \Sn_c \log n + \Sn_c \log \frac{\Sn_c}{n} - \Sn_c - 2 \sum_{v=1}^n \log \Dn_c(v)! \right) + O( \log n).
\end{equation*}
\end{lem*}

\begin{proof}[Proof of Lemma~\ref{lem:opt-ee16-general-bound-for-ln-ttp}]
We define $\mCdeltah_{\neq} := \{(t,\pp{t}) \in \mCdeltah: t \neq \pp{t}\}$, and
$\mCdeltah_= := \{(t,\pp{t}) \in \mCdeltah: t = \pp{t}\}$.
Observe that for $c \in \mCdeltah \setminus \mEn$, we have $\Dn_c(v) = 0$ for
all $v \in [n]$, and  $\Sn_c = 0$.
Therefore, we may write
\begin{equation}
  \label{eq:opt-sum-log-el-sum-log-Snc}
  \sum_{(t,\pp{t}) \in \mEn_{\leq}} \log \eln_{t,\pp{t}} = \frac{1}{2} \sum_{c \in \mCdeltah_{\neq}} \log \frac{\Sn_c!}{\prod_{v=1}^n \Dn_c(v)! \prod_{v=1}^n \Dn_{\bar{c}}(v)!} + \sum_{v \in \mCdeltah_=} \log \frac{(\Sn_c-1)!!}{\prod_{v=1}^n \Dn_c!}.
\end{equation}
Note that the factor $1/2$ in front of the first summation appears since
the pair $(t,\pp{t}) \in \mEn_{\leq}$ with $t \neq \pp{t}$ appears both as
$(t,\pp{t})$ and $(\pp{t}, t)$ in the summation over $\mCdeltah_{\neq}$.
 Using Stirling's
approximation, for $c \in \mCdeltah_{\neq}$ such that $\Sn_c > 0$, we have 
\begin{equation*}
\begin{aligned}
  \log \frac{\Sn_c!}{\prod_{v=1}^n \Dn_c(v)! \prod_{v=1}^n \Dn_{\bar{c}}(v)!} &\leq \Sn_c \log \Sn_c - \Sn_c +1 + \frac{1}{2} \log \Sn_c\\
                                                                              &\qquad - \sum_{v=1}^n \log \Dn_c(v)! - \sum_{v=1}^n \log \Dn_{\bar{c}}(v)!\\
                                                                               & = \Sn_c \log n + \Sn_c \log \frac{\Sn_c}{n} - \Sn_c \\
          &\qquad  - \sum_{v=1}^n \log \Dn_c(v)! - \sum_{v=1}^n \log \Dn_{\bar{c}}(v)! + O(\log n),
\end{aligned}
\end{equation*}
where in the second step, we have used the fact that
from~\eqref{eq:Dn-bounded-delta}, we have $\log \Sn_c = \log (\sum_{v=1}^n
\Dn_c(v)) \leq \log (n \delta) = O(\log n)$.
Note that if $\Sn_c = 0$, the above bound automatically holds. Hence, for all $c
\in \mCdeltah_{\neq}$, we have
\begin{equation}
  \label{eq:opt-Sn-c-<-bound}
  \begin{aligned}
  \log \frac{\Sn_c!}{\prod_{v=1}^n \Dn_c(v)! \prod_{v=1}^n \Dn_{\bar{c}}(v)!}  &\leq \Sn_c \log n + \Sn_c \log \frac{\Sn_c}{n} - \Sn_c \\
  &\qquad  - \sum_{v=1}^n \log \Dn_c(v)! - \sum_{v=1}^n \log \Dn_{\bar{c}}(v)! + O(\log n),
  \end{aligned}
\end{equation}
Likewise, for $c \in \mCdeltah_=$, we have
\begin{equation}
  \label{eq:opt-Sn-c-=-bound}
\begin{aligned}
  \log \frac{(\Sn_c-1)!!}{\prod_{v=1}^n \Dn_c(v)!} &\leq \frac{\Sn_c}{2} \log \Sn_c - \frac{\Sn_c}{2}  + O(1) - \sum_{v=1}^n \log \Dn_c(v)! \\
                                                   &= \frac{1}{2} \Sn_c \log n + \frac{1}{2} \Sn_c \log \frac{\Sn_c}{n} - \frac{1}{2} \Sn_c - \sum_{v=1}^n \log \Dn_c(v)! + O(1).
\end{aligned}
\end{equation}
In the above bounds, we interpret $0 \log 0$ as $0$.
Putting~\eqref{eq:opt-Sn-c-<-bound} and~\eqref{eq:opt-Sn-c-=-bound} back in
\eqref{eq:opt-sum-log-el-sum-log-Snc},
we have
\begin{equation*}
  \sum_{(t,\pp{t}) \in \mEn_{\leq}} \log \eln_{t,\pp{t}} \leq \frac{1}{2} \sum_{c \in \mCdeltah} \left( \Sn_c \log n + \Sn_c \log \frac{\Sn_c}{n} - \Sn_c - 2 \sum_{v=1}^n \log \Dn_c(v)! \right) + O( \log n).
\end{equation*}
This concludes the proof.
\end{proof}

\editfinish


\newcommand{\etalchar}[1]{$^{#1}$}

\end{document}